\documentclass[12pt,twoside,a4paper,normalheadings,headsepline]{scrbook}

\usepackage{feynmp,emp}
\usepackage{axocolor}
\usepackage{graphicx}
\usepackage{amsmath,amssymb}
\usepackage{pennames}
\usepackage{bibspacing}
\usepackage{tdr_master}
\usepackage{tdrdefinitions}

\newcommand{\href}[2]{\hspace*{-3mm}{#2}}

\begin{document}

\setcounter{part}{3}
\setcounter{chapter}{0}
\setcounter{secnumdepth}{3}
\setcounter{tocdepth}{2}

%\setcounter{page}{1}
%\part{PHYSICS AT A {${\rm e}^+{\rm e}^-$} LINEAR COLLIDER}
%\titlehead{\LARGE\begin{center}TESLA Technical Design Report\end{center}}
%\title{Part \thepart \\ Physics at an e$^+$ e$^-$ Linear Collider}
%\date{March 2001\\
%\begin{center}
%Embargo: March 23, 2001, 12:00
%\end{center}}
%\publishers{Editors: R.-D.Heuer, D.Miller, F.Richard, P.M.Zerwas}
%\maketitle
\thispagestyle{empty}
\pagestyle{empty}
\begin{picture}(150,180)\unitlength 1mm

{\sfb
\put(0,40){\Huge TESLA}
\put(0,15){\begin{minipage}{14cm}
       \LARGE\bf
        \begin{flushleft}{\sfb\LARGE
        The Superconducting
        Electron Positron \\Linear Collider
        with an Integrated\\
        X-Ray Laser Laboratory\\
        }\end{flushleft}\end{minipage}}
\put(0,-10){\Huge Technical Design Report}
\put(5,-40){\LARGE Part III Physics at an e$^+$e$^-$ Linear Collider}
\put(-5,-150){{\sfb\bf\large DESY-2001-011, ECFA-2001-209}}
\put(-5,-155){{\sfb\bf\large TESLA-2001-23, TESLA-FEL-2001-05}}

\put(131,-150){\sfb \large March}
\put(135,-155){\sfb \large 2001}}
\end{picture}

\newpage

\thispagestyle{empty}

\begin{picture}(150,180)\unitlength 1mm
  \put(-10,-65){\includegraphics[height=2.5cm]{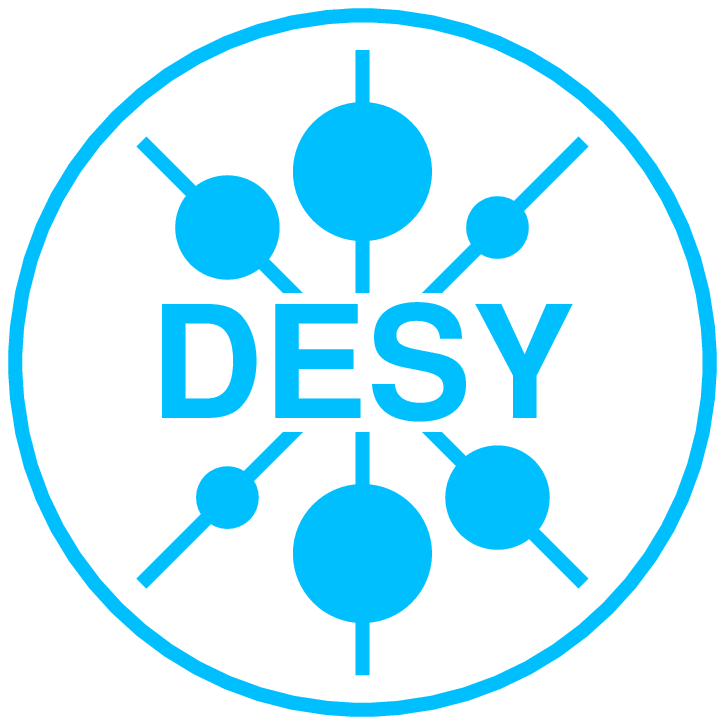}}
  \put(-10,-120){\begin{minipage}{10cm}{
   {\bf\noindent Publisher:} \\
   \noindent DESY\\
   \noindent Deutsches Elektronen-Synchroton\\
   \noindent Notkestra{\ss}e 85, D-22607 Hamburg\\
   \noindent Germany\\
   \noindent \url{http://www.desy.de}\\
   \noindent E-mail: desyinfo@desy.de\\

   \noindent Member of the Hermann von Helmholtz Association\\
   \noindent of National Research Centers (HGF)\\

   \noindent Reproduction including extracts is permitted \\
   \noindent subject to crediting the source.\\

%   {\noindent \bf Printing}
%   \noindent Dierk Heigener Druckerzeugnisse GmBH,\\
%   \noindent Hamburg, Germany\\

   {\bf \noindent Copy deadline:} March 2001 \\
   ISBN 3-935702-00-0\\
   ISSN 0418-9833
}
\end{minipage}}

\end{picture}

%\clearpage

%\begin{center}
%\begin{picture}(150,180)\unitlength 1mm
%  \put(-50,-100){\includegraphics[height=10cm]{wiik.ps}}
%  \put(-20,-140){Dedicated to the memory of Bj{\o}rn H. Wiik (1937-1999)}
%\end{picture}
%\end{center}

\cleardoublepage

\pagestyle{empty}
\begin{picture}(150,180)\unitlength 1mm

{\sfb
\put(0,40){\Huge TESLA}
\put(0,15){\begin{minipage}{14cm}
       \LARGE\bf
        \begin{flushleft}{\sfb\LARGE
        The Superconducting
        Electron Positron \\Linear Collider
        with an Integrated\\
        X-Ray Laser Laboratory\\
        }\end{flushleft}\end{minipage}}
\put(0,-10){\Huge Technical Design Report}

\put(-5,-150){{\sfb\bf\large DESY-2001-011, ECFA-2001-209}}
\put(-5,-155){{\sfb\bf\large TESLA-2001-23, TESLA-FEL-2001-05}}

\put(131,-150){\sfb \large March}
\put(135,-155){\sfb \large 2001}}
\end{picture}

\newpage
\thispagestyle{empty}
{\large
\begin{picture}(150,180)\unitlength 1mm
\put(-5,20)   {{\bf\Large PART I: Executive Summary }}
      \put(25,12){Editors: \begin{minipage}[t]{10cm}{F.Richard, J.R.Schneider, D.Trines, A.Wagner}\end{minipage}}
\put(-5,-10) {{\bf \Large PART II: The Accelerator}}
      \put(25,-18){Editors: \begin{minipage}[t]{9cm}{R.Brinkmann,
      K.Fl\"ottmann, J.Rossbach,\hfil P.Schm\"user, N.Walker, H.Weise}\end{minipage}}
\put(-5,-40) {{\bf \Large PART III: Physics at an e$^+$e$^-$ Linear Collider}}
      \put(25,-48){Editors: \begin{minipage}[t]{10cm}{R.D.Heuer,
      D.Miller, F.Richard, P.Zerwas}\end{minipage} }
\put(-5,-70) {{\bf \Large PART IV: A Detector for TESLA}}
      \put(25,-78){Editors: \begin{minipage}[t]{10cm}{T.Behnke, S.Bertolucci,
      R.D.Heuer, R.Settles }\end{minipage}}
\put(-5,-100) {{\bf \Large PART V: The X-Ray Free Electron Laser Laboratory}}
      \put(25,-108){Editors:
      \begin{minipage}[t]{10cm}{G.Materlik, T.Tschentscher}\end{minipage}}
\put(-5,-130) {{\bf \Large PART VI: Appendices}}
      \put(25,-138){Editors: \begin{minipage}[t]{10cm}{
        R.Klanner\\
        Chapter 1: V.Telnov\\
        Chapter 2: U.Katz, M.Klein, A.Levy\\
        Chapter 3: R.Kaiser, W.D.Nowak\\
        Chapter 4: E.DeSanctis, J.-M.Laget, K.Rith
}\end{minipage}}
\end{picture}
}
\clearpage

\begin{picture}(150,180)\unitlength 1mm

  \put(10,50){\sfb \LARGE Part III: Physics at an e$^+$e$^-$ Linear Collider }
  \put(10,42){\sfb \large Editors:}
  \put(10,34){\sfb \large R.-D.Heuer, D.Miller,}
  \put(10,26){\sfb \large F.Richard, P.Zerwas}

\end{picture}

\clearpage

\pagestyle{headings}
\pagenumbering{roman}
\addsec*{Authors}
\renewcommand{\baselinestretch}{0.8}
%
%AUTHORLIST
%

\noindent \begin{flushleft} J.A.~Aguilar--Saavedra$^{39}$$^{ }$,
J.~Alcaraz$^{59}$$^{ }$,
A.~Ali$^{26}$$^{ }$,
S.~Ambrosanio$^{19}$$^{ }$,
A.~Andreazza$^{68}$$^{ }$,
J.~Andruszkow$^{48}$$^{ }$,
B.~Badelek$^{111,113}$$^{ }$,
A.~Ballestrero$^{110}$$^{ }$,
T.~Barklow$^{102}$$^{ }$,
A.~Bartl$^{114}$$^{ }$
% ...

%
%INSTITUTELIST
%
\newdimen\authorskip
\setlength{\authorskip}{-1mm}
\bigskip 
\vspace*{-5mm} 
\addsec*{Convenors}
\noindent   A.~Bartl, 
            M.~Battaglia,
            W.~Bernreuther,
            G.~Blair,
            A.~Brandenburg,
            P.~Burrows,
            K.~Desch, 
            A.~Djouadi,
            W.~de~Boer,
            A.~De~Roeck, 
            G.~Gounaris, 
            E.~Gross,
            C.A.~Heusch, 
            S.~Jadach,
            F.~Jegerlehner,
            S.~Katsanevas,
            M.~Kr\"{a}mer,
            B.A.~Kniehl,
            J.~K\"uhn,
            W.~Majerotto,
            M.~Martinez, 
            H.-U. Martyn, 
            R.~Miquel, 
            K.~M\"onig,
            T.~Ohl, 
            M.~Pohl, 
            R.~R\"{u}ckl, 
            M.~Spira, 
            V.~Telnov,
            G.~Wilson

\bigskip 
\end{flushleft} 
%\vspace*{-5mm} 
{\addtolength{\baselineskip}{-2mm} 
{\footnotesize \noindent
 \vspace*{\authorskip}$\mbox{\,} $ $ ^{1}$ \begin{minipage}[t]{16cm} RWTH Aachen, Germany \end{minipage}\ \\ 
 \vspace*{\authorskip}$\mbox{\,} $ $ ^{2}$ \begin{minipage}[t]{16cm} Laboratoire d'Annecy--le--Vieux de Physique des Particules, France \end{minipage}\ \\ 
 \vspace*{\authorskip}$\mbox{\,} $ $ ^{3}$ \begin{minipage}[t]{16cm} Universiteit Antwerpen, The Netherlands \end{minipage}\ \\ 
 \vspace*{\authorskip}$\mbox{\,} $ $ ^{4}$ \begin{minipage}[t]{16cm} ANL, Argonne, IL, USA \end{minipage}\ \\ 
 \vspace*{\authorskip}$\mbox{\,} $ $ ^{5}$ \begin{minipage}[t]{16cm} Demokritos National Centre for Scientific Research, Athens, Greece \end{minipage}\ \\ 
 \vspace*{\authorskip}$\mbox{\,} $ $ ^{6}$ \begin{minipage}[t]{16cm} Universit\'e Blaise Pascal, Aubi\`ere, France \end{minipage}\ \\ 
 \vspace*{\authorskip}$\mbox{\,} $ $ ^{7}$ \begin{minipage}[t]{16cm} Indian Institute of Science, Bangalore, India \end{minipage}\ \\ 
 \vspace*{\authorskip}$\mbox{\,} $ $ ^{8}$ \begin{minipage}[t]{16cm} Universitat Autonoma de Barcelona, Spain \end{minipage}\ \\ 
 \vspace*{\authorskip}$\mbox{\,} $ $ ^{9}$ \begin{minipage}[t]{16cm} Universitat de Barcelona, Spain \end{minipage}\ \\ 
$ ^{10}$ \begin{minipage}[t]{16cm} Universit\"at Basel, Switzerland \end{minipage}\ \\ 
$ ^{11}$ \begin{minipage}[t]{16cm} University of Bergen, Norway \end{minipage}\ \\ 
$ ^{12}$ \begin{minipage}[t]{16cm} Universit\"{a}t Bern, Switzerland \end{minipage}\ \\ 
$ ^{13}$ \begin{minipage}[t]{16cm} Bialystok University Bialystok, Poland \end{minipage}\ \\ 
$ ^{14}$ \begin{minipage}[t]{16cm} BNL, Upton, NY,  USA \end{minipage}\ \\ 
$ ^{15}$ \begin{minipage}[t]{16cm} INFN, Sezione di Bologna, Italy \end{minipage}\ \\ 
$ ^{16}$ \begin{minipage}[t]{16cm} Universit\`a degli Studi di Bologna, Italy \end{minipage}\ \\ 
$ ^{17}$ \begin{minipage}[t]{16cm} Universit\"at Bonn, Germany \end{minipage}\ \\ 
$ ^{18}$ \begin{minipage}[t]{16cm} BINP, Novosibirsk, Russia \end{minipage}\ \\ 
$ ^{19}$ \begin{minipage}[t]{16cm} CERN, Gen\`eve, Switzerland \end{minipage}\ \\ 
$ ^{20}$ \begin{minipage}[t]{16cm} The Institute od Mathematical Sciences, CIT Campus, Chennai, India \end{minipage}\ \\ 
$ ^{21}$ \begin{minipage}[t]{16cm} Chonbuk National University, Chonju, Korea \end{minipage}\ \\ 
$ ^{22}$ \begin{minipage}[t]{16cm} University of Cincinnati, OH, USA \end{minipage}\ \\ 
$ ^{23}$ \begin{minipage}[t]{16cm} University of Colorado, Boulder, CO, USA \end{minipage}\ \\ 
$ ^{24}$ \begin{minipage}[t]{16cm} Technische Universit\"at Darmstadt, Germany \end{minipage}\ \\ 
$ ^{25}$ \begin{minipage}[t]{16cm} University of California, Davis, CA, USA \end{minipage}\ \\ 
$ ^{26}$ \begin{minipage}[t]{16cm} DESY, Hamburg and Zeuthen, Germany \end{minipage}\ \\ 
$ ^{27}$ \begin{minipage}[t]{16cm} JINR, Dubna, Russia \end{minipage}\ \\ 
$ ^{28}$ \begin{minipage}[t]{16cm} University of Durham, UK \end{minipage}\ \\ 
$ ^{29}$ \begin{minipage}[t]{16cm} University of Edinburgh, UK \end{minipage}\ \\ 
$ ^{30}$ \begin{minipage}[t]{16cm} Friedrich--Alexander--Universit\"at Erlangen--N\"urnberg, Germany \end{minipage}\ \\ 
$ ^{31}$ \begin{minipage}[t]{16cm} FNAL, Batavia, IL, USA \end{minipage}\ \\ 
$ ^{32}$ \begin{minipage}[t]{16cm} INFN, Sezione di Ferrara, Italy \end{minipage}\ \\ 
$ ^{33}$ \begin{minipage}[t]{16cm} Universit\`a degli Studi di Ferrara, Italy \end{minipage}\ \\ 
$ ^{34}$ \begin{minipage}[t]{16cm} Universit\`a di Firenze, Italy \end{minipage}\ \\ 
$ ^{35}$ \begin{minipage}[t]{16cm} INFN Laboratori Nazionali di Frascati, Italy \end{minipage}\ \\ 
$ ^{36}$ \begin{minipage}[t]{16cm} Albert--Ludwigs--Universit\"at Freiburg, Germany \end{minipage}\ \\ 
$ ^{37}$ \begin{minipage}[t]{16cm} Universit\'e de Gen\`eve, Switzerland \end{minipage}\ \\ 
$ ^{38}$ \begin{minipage}[t]{16cm} Gomel Technical University, Belarus \end{minipage}\ \\ 
$ ^{39}$ \begin{minipage}[t]{16cm} Universidad de Granada, Spain \end{minipage}\ \\ 
$ ^{40}$ \begin{minipage}[t]{16cm} Universit\"at Hamburg, Germany \end{minipage}\ \\ 
$ ^{41}$ \begin{minipage}[t]{16cm} Universit\"{a}t Heidelberg, Germany  \end{minipage}\ \\ 
$ ^{42}$ \begin{minipage}[t]{16cm} Helsinki Institute of Physics, Finland \end{minipage}\ \\ 
$ ^{43}$ \begin{minipage}[t]{16cm} University of Helsinki, Finland \end{minipage}\ \\ 
$ ^{44}$ \begin{minipage}[t]{16cm} Hyogo University, Japan \end{minipage}\ \\ 
$ ^{45}$ \begin{minipage}[t]{16cm} Indiana University, Bloomington, USA \end{minipage}\ \\ 
$ ^{46}$ \begin{minipage}[t]{16cm} Universit\"at Karlsruhe, Germany \end{minipage}\ \\ 
$ ^{47}$ \begin{minipage}[t]{16cm} KEK, Tsukuba, Japan \end{minipage}\ \\ 
$ ^{48}$ \begin{minipage}[t]{16cm} INP, Krak\'ow, Poland \end{minipage}\ \\ 
$ ^{49}$ \begin{minipage}[t]{16cm} Jagellonian University, Krak\'ow, Poland \end{minipage}\ \\ 
$ ^{50}$ \begin{minipage}[t]{16cm} Universidad Nacional de La Plata, Argentina \end{minipage}\ \\ 
$ ^{51}$ \begin{minipage}[t]{16cm} LBNL,  University of California, Berkeley, CA, USA \end{minipage}\ \\ 
$ ^{52}$ \begin{minipage}[t]{16cm} INFN, Sezione di Lecce, Italy \end{minipage}\ \\ 
$ ^{53}$ \begin{minipage}[t]{16cm} Universiteit Leiden, The Netherlands \end{minipage}\ \\ 
$ ^{54}$ \begin{minipage}[t]{16cm} Universit\"at Leipzig, Germany \end{minipage}\ \\ 
$ ^{55}$ \begin{minipage}[t]{16cm} University College London, UK \end{minipage}\ \\ 
$ ^{56}$ \begin{minipage}[t]{16cm} Royal Holloway and Bedford New College, University of London, UK \end{minipage}\ \\ 
$ ^{57}$ \begin{minipage}[t]{16cm} University of Lund, Sweden \end{minipage}\ \\ 
$ ^{58}$ \begin{minipage}[t]{16cm} IPN, Lyon, France \end{minipage}\ \\ 
$ ^{59}$ \begin{minipage}[t]{16cm} CIEMAT, Madrid, Spain \end{minipage}\ \\ 
$ ^{60}$ \begin{minipage}[t]{16cm} Universidad Aut\'{o}noma de Madrid, Spain \end{minipage}\ \\ 
$ ^{61}$ \begin{minipage}[t]{16cm} CSIC, IMAFF, Madrid, Spain \end{minipage}\ \\ 
$ ^{62}$ \begin{minipage}[t]{16cm} Johannes--Gutenberg--Universit\"at Mainz, Germany \end{minipage}\ \\ 
$ ^{63}$ \begin{minipage}[t]{16cm} University of Manchester, UK \end{minipage}\ \\ 
$ ^{64}$ \begin{minipage}[t]{16cm} CINVESTAV-IPN, Merida, Mexico \end{minipage}\ \\ 
$ ^{65}$ \begin{minipage}[t]{16cm} Michigan State University, East Lansing, MI, USA \end{minipage}\ \\ 
$ ^{66}$ \begin{minipage}[t]{16cm} Universit\`a degli Studi Milano--Bicocca, Italy \end{minipage}\ \\ 
$ ^{67}$ \begin{minipage}[t]{16cm} INFN, Sezione di Milano, Italy \end{minipage}\ \\ 
$ ^{68}$ \begin{minipage}[t]{16cm} Universit\`a degli Studi di Milano, Italy \end{minipage}\ \\ 
$ ^{69}$ \begin{minipage}[t]{16cm} Universit\'e de Montpellier II, France \end{minipage}\ \\ 
$ ^{70}$ \begin{minipage}[t]{16cm} ITEP, Moscow, Russia  \end{minipage}\ \\ 
$ ^{71}$ \begin{minipage}[t]{16cm} M.V.~Lomonosov Moscow State University, Russia \end{minipage}\ \\ 
$ ^{72}$ \begin{minipage}[t]{16cm} Ludwigs-Maximilians--Universit\"at M\"unchen, Germany \end{minipage}\ \\ 
$ ^{73}$ \begin{minipage}[t]{16cm} Max Planck Institut f\"ur Physik, M\"unchen, Germany \end{minipage}\ \\ 
$ ^{74}$ \begin{minipage}[t]{16cm} Katholieke Universiteit Nijmegen, The Netherlands \end{minipage}\ \\ 
$ ^{75}$ \begin{minipage}[t]{16cm} NIKHEF, Amsterdam, The Netherlands \end{minipage}\ \\ 
$ ^{76}$ \begin{minipage}[t]{16cm} University of Notre Dame, IN, USA \end{minipage}\ \\ 
$ ^{77}$ \begin{minipage}[t]{16cm} Institute of Mathematics SB RAS, Novosibirsk, Russia \end{minipage}\ \\ 
$ ^{78}$ \begin{minipage}[t]{16cm} Ochanomizu University, Tokyo, Japan \end{minipage}\ \\ 
$ ^{79}$ \begin{minipage}[t]{16cm} Universit\'e Paris XI, Orsay, France \end{minipage}\ \\ 
$ ^{80}$ \begin{minipage}[t]{16cm} LAL, Orsay, France \end{minipage}\ \\ 
$ ^{81}$ \begin{minipage}[t]{16cm} Carleton University, Ottawa, Canada \end{minipage}\ \\ 
$ ^{82}$ \begin{minipage}[t]{16cm} Oxford University, UK \end{minipage}\ \\ 
$ ^{83}$ \begin{minipage}[t]{16cm} Ecole Polytechnique, Palaiseau, France \end{minipage}\ \\ 
$ ^{84}$ \begin{minipage}[t]{16cm} Universit\'es Paris VI et VII, France \end{minipage}\ \\ 
$ ^{85}$ \begin{minipage}[t]{16cm} Universit\`a degli Studi di Parma, Italy \end{minipage}\ \\ 
$ ^{86}$ \begin{minipage}[t]{16cm} INFN, Sezione di Pavia, Italy \end{minipage}\ \\ 
$ ^{87}$ \begin{minipage}[t]{16cm} Universit\`a di Pavia, Italy \end{minipage}\ \\ 
$ ^{88}$ \begin{minipage}[t]{16cm} Pennsylvania State University, Mont Alto, PA, USA \end{minipage}\ \\ 
$ ^{89}$ \begin{minipage}[t]{16cm} Pennsylvania State University, University Park, PA, USA \end{minipage}\ \\ 
$ ^{90}$ \begin{minipage}[t]{16cm} PSI, Villigen, Switzerland \end{minipage}\ \\ 
$ ^{91}$ \begin{minipage}[t]{16cm} RAL, Oxon, UK \end{minipage}\ \\ 
$ ^{92}$ \begin{minipage}[t]{16cm} Universit\"{a}t Regensburg, Germany \end{minipage}\ \\ 
$ ^{93}$ \begin{minipage}[t]{16cm} Weizmann Institute of Science, Rehovot, Israel \end{minipage}\ \\ 
$ ^{94}$ \begin{minipage}[t]{16cm} RIKEN-BNL, Upton, NY, USA \end{minipage}\ \\ 
$ ^{95}$ \begin{minipage}[t]{16cm} University of Rochester, NY, USA \end{minipage}\ \\ 
$ ^{96}$ \begin{minipage}[t]{16cm} INFN, Sezione di Roma I, Italy \end{minipage}\ \\ 
$ ^{97}$ \begin{minipage}[t]{16cm} Universit\`a degli Studi di Roma La Sapienza,  Italy \end{minipage}\ \\ 
$ ^{98}$ \begin{minipage}[t]{16cm} DAPNIA--CEA, Saclay, France \end{minipage}\ \\ 
$ ^{99}$ \begin{minipage}[t]{16cm} University of California, Santa Cruz, CA, USA \end{minipage}\ \\ 
$ ^{100}$ \begin{minipage}[t]{16cm} Shoumen University Bishop K.\ Preslavsky, Bulgaria \end{minipage}\ \\ 
$ ^{101}$ \begin{minipage}[t]{16cm} University of Silesia, Katowice, Poland \end{minipage}\ \\ 
$ ^{102}$ \begin{minipage}[t]{16cm} SLAC, Stanford, CA, USA \end{minipage}\ \\ 
$ ^{103}$ \begin{minipage}[t]{16cm} Stanford University, CA, USA \end{minipage}\ \\ 
$ ^{104}$ \begin{minipage}[t]{16cm} IReS, Strasbourg, France \end{minipage}\ \\ 
$ ^{105}$ \begin{minipage}[t]{16cm} Tblisi State University, Georgia \end{minipage}\ \\ 
$ ^{106}$ \begin{minipage}[t]{16cm} Aristotle University of Thessaloniki, Greece \end{minipage}\ \\ 
$ ^{107}$ \begin{minipage}[t]{16cm} Tohoku University, Sendai, Japan \end{minipage}\ \\ 
$ ^{108}$ \begin{minipage}[t]{16cm} INFN, Sezione di Trieste, Italy \end{minipage}\ \\ 
$ ^{109}$ \begin{minipage}[t]{16cm} Universit\`a degli Studi di Trieste, Italy \end{minipage}\ \\ 
$ ^{110}$ \begin{minipage}[t]{16cm} INFN, Sezione di Torino, Italy \end{minipage}\ \\ 
$ ^{111}$ \begin{minipage}[t]{16cm} University of Uppsala, Sweden \end{minipage}\ \\ 
$ ^{112}$ \begin{minipage}[t]{16cm} Universitat de Val\`{e}ncia, Spain \end{minipage}\ \\ 
$ ^{113}$ \begin{minipage}[t]{16cm} Warsaw University, Poland \end{minipage}\ \\ 
$ ^{114}$ \begin{minipage}[t]{16cm} Universit\"at Wien, Austria \end{minipage}\ \\ 
$ ^{115}$ \begin{minipage}[t]{16cm} Universit\"at W\"urzburg, Germany \end{minipage}\ \\ 
$ ^{116}$ \begin{minipage}[t]{16cm} ETH Z\"urich, Switzerland \end{minipage}\ \\ 
} }\bigskip 
%
%ANNOTATIONLIST
%

{\setlength{\baselineskip}{0.2mm}\footnotesize \noindent$ ^{a}$ now at CERN \\ 
 } 

\renewcommand{\baselinestretch}{1.0}
\clearpage
\tableofcontents
\cleardoublepage
\pagenumbering{arabic}
\setcounter{page}{1}

%------------------------------------------------------------------
\chapter{Introduction}
\label{physics_introduction}
%------------------------------------------------------------------

  \section{Particle Physics Today}

The Standard Model of particle physics was built up through decades of 
intensive
dialogue between theory and experiments at both hadron and electron machines.  
It has become increasingly coherent as experimental analyses have established 
the basic physical concepts.
Leptons and quarks were discovered as the fundamental constituents 
of matter. The 
photon, the $W$ and $Z$ bosons, 
and the gluons were identified as the carriers of
the electromagnetic, weak and strong forces. Electromagnetic and weak 
forces have been unified within the electroweak gauge field theory.  
The QCD gauge field theory has been confirmed as the 
theory of strong interactions. 

In the last few years many aspects of the model have been stringently tested, 
some to the per-mille level, 
with $ e^+ e^-$, $ e p$ and $ p \bar{p}$ 
machines making complementary contributions, especially to the 
determination of the electroweak parameters.  
With the $ e^+ e^-$ data from LEP1 and SLC measurements of the 
lineshape and couplings of the $Z$ boson became so precise that the mass of 
the top quark was already tightly constrained by quantum level 
calculations before it was directly measured in $p \bar{p}$ at the 
Tevatron.  Since then LEP2 and the Tevatron  have extended the 
precision measurements
to the properties of the $W$ bosons.  
Combining these results with neutrino scattering data 
and low energy measurements, the experimental analysis is in
excellent concordance with the electroweak part of the Standard Model.  \par

At the same time the predictions of QCD have also been thoroughly tested. 
Notable among the QCD results from LEP1 and SLC were precise 
measurements of the strong coupling $\alpha_s$.  
At HERA the proton structure is being probed to the shortest 
accessible distances.
HERA and the Tevatron have been able to explore a wide range of QCD phenomena 
at small and large
distances involving both the proton and the photon, supplemented by
data on the photon from $\gamma \gamma$ studies at LEP.  \par

Despite these great successes there are many gaps in our understanding.  
The clearest gap of all is the present
lack of any direct evidence for the microscopic
dynamics of electroweak symmetry breaking and the generation of
the masses of gauge bosons and fermions.  
These masses are generated in the Standard Model by the Higgs mechanism.
A fundamental field is introduced, the Higgs boson field,
whose non--zero vacuum expectation value 
breaks the electroweak symmetry spontaneously. 
Interaction with this field generates the $W$ and $Z$ boson masses 
 while leaving the photon massless; the masses of the quarks and leptons
are generated by the same mechanism.
The precision electroweak analysis
favours a Higgs boson mass which is in the region 
of the limit which has been reached in searches at LEP2.  The LEP experiments have 
reported a tantalising hint of a 
Higgs signal at $M_h \simeq 115$\GeV\  but, even if that is a mirage, the 
95\% confidence level limit on the mass is just above 200\GeV.  If the  electroweak 
sector of the Standard Model is an accurate description of Nature then such a light Higgs boson must be accessible
both at the LHC and at TESLA. \par

Many other puzzles remain to be solved.
We have no explanation for the
wide range of masses of the fermions (from $<$\,eV 
for neutrinos to $\simeq 175$\GeV\  for the top quark).  CP violation
is not understood at the level required to account for the excess of matter
over antimatter
in the universe. The grand unification between the two gauge 
theories, QCD and electroweak, is not realised and gravity has not been brought
into any close relationship to the other forces.
Thus, the Standard Model leaves many deep physics questions unanswered.  \par

Some alternative scenarios have been developed for the physics which may 
emerge beyond the Standard Model as energies are
increased, ranging from
supersymmetric theories - well motivated theoretically and incorporating 
a light Higgs boson - to theories in which 
the symmetry breaking is generated by new strong interactions. 
Supersymmetry opens a new particle world characterised in its standard form by energies
of order 100\GeV\  to order 1\TeV. 
On the other hand, new strong interactions, a dynamical alternative
to the fundamental Higgs mechanism for electroweak symmetry breaking, 
give rise to strong
forces between $W$ bosons at high energies. Quite general arguments suggest that such
new phenomena must appear below a scale of $\simeq$ 3\TeV. \par

There are two ways of approaching the new scales.  The LHC tackles 
them head-on by going to the highest available centre of mass energy,  
but this brings experimental complications 
from the composite quark/gluon nature of the colliding protons.  
Events at TESLA will be much more 
cleanly identified and much more precisely measured.  These
advantages, together with the large statistics which come from 
its high luminosity, will 
allow TESLA to carry out a comprehensive and conclusive physics
programme, identifying the physical nature of the new
new final states, 
and reaching up to high effective scales to recognise
new physics scenarios through its quantum level effects.  
%There is a ``no lose theorem'' for TESLA;
For all the wide range of new and complementary
scenarios that have been studied there 
are ways in which
TESLA can detect their effects, directly or indirectly.

\section{The TESLA Physics Programme}

The physics programme for $e^+ e^-$ linear colliders in the TeV range has been 
developed through numerous theoretical analyses, summarised in 
\cite{CDRphysics}, and in a
decade of experimentally based feasibility studies 
(see Refs.~\cite{CDR,DESY123,LCWS}). The essential 
elements are summarised here and a more comprehensive overview
is given in the following chapters.

\subsection{The Higgs mechanism}

LEP and SLC have established a precise picture of the electroweak interactions 
between matter particles and they have confirmed the structure of the forces. But the
third component of the Standard Model, the Higgs mechanism which breaks
the electroweak symmetry and generates the masses of the particles, 
has not so far been firmly established. \par

Should a Higgs boson exist, then TESLA will be able to measure the full set of 
its properties with high precision, establishing that the
Higgs mechanism is responsible for electroweak symmetry breaking and testing the self consistency of the picture.
The initial question is simple; does the observed Higgs boson have the profile 
predicted by the Standard Model: the mass, the lifetime, the production cross 
sections,
the branching ratios to quarks of different flavours, to leptons and to bosons, the Yukawa coupling to the top quark, the self coupling? 
TESLA will achieve a precision of 50\,(70)\MeV\  on the mass of
a 120\,(200)\GeV\  Higgs, and will measure many of the branching ratios to a 
few percent.
The top-Higgs Yukawa coupling will be measured to 5\%.  
The Higgs self-potential can be established from the $Z\!H\!H$ 
final state, where the self-coupling will be measurable to 20\%. \par

If the Higgs boson does have the Standard Model profile, the next stage of the programme will be to refine even further the 
existing precision measurements which constrain the model at the quantum level.
TESLA can measure the mass of the top quark to a precision of about 100\MeV.
Other important constraints come from the mass of the $W$ boson and the size 
of the electroweak mixing angle which can
be measured very precisely with TESLA's GigaZ option at 90 to 200\GeV.  Lack of
concordance between the parameters of the Higgs sector and the 
parameters derived from precision measurements in the electroweak boson sector
could give direct information about physics scenarios beyond the Standard Model. 
The photon collider option will supplement the picture by precise 
measurements of the Higgs coupling to $\gamma \gamma$, 
an important probe of the quantum loops which would be sensitive to new 
particles with masses beyond direct reach. \par

The Higgs mechanism in the Standard Model needs only one Higgs
doublet, but an extended Higgs sector is required by 
many of the theories in which the Standard Model may be embedded.
In supersymmetric theories, for example, at least two Higgs doublets must be
introduced giving rise to five or more physical Higgs particles. Many
experimental aspects can be inferred from the analysis of the light SM
Higgs boson, though the spectrum of heavy Higgs particles requires new
and independent experimental analyses.  
Examples are given of how these Higgs particles 
can be investigated at TESLA, exploiting the whole energy range up to 800\GeV.

\subsection{Supersymmetry}

Supersymmetry is the preferred candidate for extensions beyond the Standard 
Model.
It retains small Higgs masses in the context of large scales in a natural way.
Most importantly, 
it provides an attractive route towards unification of the 
electroweak and strong interactions.  When embedded in a 
grand-unified theory, it makes a very precise 
prediction of the size of the electroweak mixing parameter $\sin^2 \theta_W$
which has been confirmed experimentally at LEP at the per-mille level.  
In supersymmetric theories electroweak symmetry breaking may be 
generated radiatively.
Last but not least, supersymmetry is deeply related to gravity, the fourth of 
the fundamental
forces. 
The density of dark matter needed in astrophysics and cosmology can be 
accomodated well in supersymmetric theories, where the lightest
supersymmetric particles are stable in many scenarios.

Supersymmetric models give an unequivocal prediction that 
the lightest Higgs boson mass should be below 200\GeV, or even
135\GeV\  in 
the minimal model.
Testing the properties of this particle can reveal its origin in a 
supersymmetric world and can shed light on the other heavy particles in the
Higgs spectrum which may lie outside the range covered by TESLA
(and LHC) directly. However,
if the other SUSY Higgs bosons are within TESLA's mass reach then 
in almost every conceivable SUSY scenario TESLA will be able to measure 
and identify them.  \par

If supersymmetry is realised in Nature there are several alternative 
schemes for the breaking of the symmetry, 
many of which could give rise to superpartners of the 
normal particles with a rich spectrum falling within the reach of TESLA. 
The great variety of TESLA's precision measurements 
can be exploited to tie down the parameters of the
supersymmetric theory with an accuracy which goes well beyond the LHC.  
Polarisation of the electron beam is
shown to be particularly important for these analyses, and polarisation of 
the positrons is desirable, both to increase analysis power in particle 
diagnostics and to reduce backgrounds.
Because TESLA can scan its well defined centre of mass energy across the 
thresholds for new particle production 
it will be able to identify the individual objects one by one and to measure 
supersymmetric particle masses to very high precision. 
It could be demonstrated at LHC 
that supersymmetry is present, and part of its spectrum 
could be resolved.
But overlapping final states will complicate LHC's reconstruction of the
whole set of supersymmetric particles.

The highest possible precision is needed so that the supersymmetric parameters 
measured at the TESLA energy scale can be extrapolated to higher 
energy scales where the 
underlying structure of supersymmetry breaking may be explored and the structure  
of the grand unified supersymmetric theory may be revealed.  
This may be the only way to link particle physics with gravity in
controllable experiments - a most important aspect of TESLA's physics 
potential.

\subsection{Alternative new physics}

Numerous alternatives have been developed to the above picture which
incorporates a fundamental Higgs field to generate electroweak symmetry
breaking and which can be extrapolated to high scales near the 
Planck energy. Out of the important families of possibilities, two
different concepts and their consequences for the TESLA experiments have been
analysed at some detail.

Recent work has shown that
the unification of gravity with the other forces may be 
realised at much lower energy scales than thought previously, 
if there are extra space
dimensions which may be curled-up, perhaps even at semi-macroscopic length scales. 
This could generate new effective spin-2 forces and missing energy events 
which TESLA
would be well equipped to observe or, in alternative scenarios, it could give a new spectroscopy 
at a scale which TESLA could probe. 
Thus TESLA can tackle fundamental problems of the structure of 
space and time. \par

The second analysis addresses the problem of dynamical electroweak 
symmetry breaking induced by new strong
interactions. In this no-Higgs scenario
quantum-mechanical unitarity requires the interactions 
between $W$ bosons to become strong at energies close to 1\TeV. The new effects
would be reflected in anomalous values of the couplings between the
electroweak bosons and in the quasi-elastic $WW$ scattering amplitudes,
from which effective scales for the new strong interactions can be extracted.
Precision measurements of $e^+ e^-$ annihilation to $WW$ pairs at 500\GeV and
$WW$ scattering with TESLA's high luminosity at 800\GeV 
are shown to have the sensitivity required to explore
the onset of these strong interactions in a range up to the limit of
$\sim$3\TeV\ 
for resonance formation. 
If the strong vector-vector boson interactions are characterised by a lower scale
of 1 to 2\TeV, there could be a spectacular spectrum of new
composite bosons at LHC. 
TESLA will be able to extend this scale 
further than the LHC can.    \par

\subsection{Challenging the Standard Model}

Although the SM has been strenuously tested in many directions it still has 
important aspects which require
experimental improvement.  A prime target will be
to establish the non-abelian gauge symmetry of the electroweak forces
by studying the $WW$ self-couplings to the sub per-mille level.
This will close the chapter on one
of the most successful ideas in particle physics. \par

Other improvements will come from running the machine in the
GigaZ mode. The size of the electroweak mixing
angle and the mass of the $W$-boson will be measured much more precisely 
than they have been at
LEP/SLC if TESLA can make dedicated runs with high luminosity at low
energies; close to the $Z$ resonance, around 92\GeV, and above
the $W^+W^-$ threshold, 161 to 200\GeV. \par

Moreover, TESLA in the GigaZ mode can supplement the analyses performed 
at beauty factories by studying the CKM matrix elements directly in
$W$ decays and CP violating B meson decays. \par

If symmetries in grand-unified theories are broken down to the symmetry of
the Standard Model in steps, remnants of those higher symmetries 
may manifest themselves in
new types of vector bosons and extended spectra of leptons and quarks
at the \TeV\  scale and below. These scenarios can be probed in high precision
analyses of SM processes at TESLA, taking advantage of its high luminosity and
polarised beams. Limits close to 10\TeV\  for most kinds of $Z'$ bosons from
TESLA, though indirect, go significantly beyond the discovery limits at LHC.
For the heavy $W'$ bosons the photon collider in its $\gamma e^-$ mode
is particularly sensitive. The $e^-e^-$ option is especially suited
to the search for heavy Majorana neutrinos, exchanged as virtual particles
in lepton-number violating processes. \par

The detailed profile of the top quark is another important goal for
TESLA; its mass (measured to about 100\MeV), 
its width, its decay modes, its static electroweak parameters - charges and magnetic and electric dipole moments. It is anticipated that the 
highest possible precision will be required to constrain the 
future theory of flavour physics in which
the top quark, the heaviest Standard Model fermion, 
will surely play a key role. \par

The QCD programme of TESLA will include a range of new measurements and improvements.
Event shape studies will further test the theory by looking at the way the strong 
coupling runs up to the highest TESLA energy. 
The re-analysis of hadronic $Z$ decays in the GigaZ mode will improve the 
measurement of the QCD coupling to the per-mille level.  
A new class of precise QCD measurements will be made  with the top quark,
particularly at the threshold of top-pair production where the excitation
curve demands new theoretical techniques. 
At the photon collider, QCD in $\gamma \gamma$ physics can be studied for the
first time with relatively well determined energies for the incoming particles.  In particular, the growth of the total $\gamma \gamma$ cross section
can be compared with predictions based on $pp$ and $\gamma p$, up to much higher energies than before.  
The photon structure function $F_{2}^{\gamma}$ can be measured in $\gamma e^-$ to much higher $Q^2$ and lower $x_{Bj}$ than at LEP, 
testing one of the few fundamental predictions of QCD.

\section{Technical Requirements}

\begin{figure}[tbh]
  \begin{center}
    \href{pictures/0/crosssection.pdf}{{\includegraphics[height=10cm]{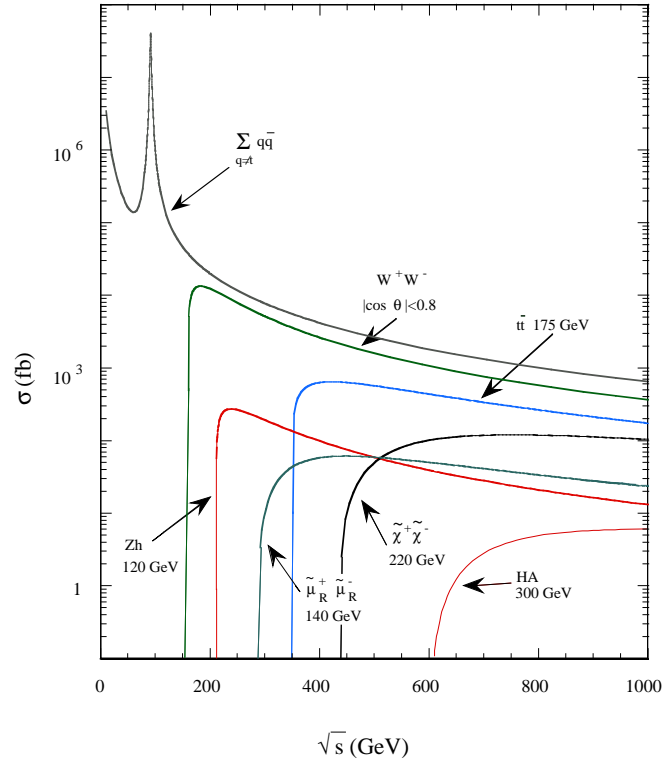}}}
  \end{center}
\vspace*{-5mm}
\caption{Cross secions for some interesting processes at a linear
collider.\label{fig:crossectionplot}}
\vspace*{2mm}
\end{figure}

The physics programme described above demands a large amount of
integrated luminosity for $e^+e^-$ collisions in the energy range 
between 90\GeV\  and $\sim$1\TeV. 
The distribution of luminosity over this energy range will be driven
by the physics scenario realised by Nature but it is obvious that
independent of any scenario a few ab$^{-1}$ will be required.
Most of the interesting cross sections
are of a size 
typical for the electroweak scale (see Fig.~\ref{fig:crossectionplot}), 
for instance $\simeq$ 100\,fb for $Z$ + light Higgs at 500\GeV\  centre
of mass energy 
($\simeq$ 200\,fb at 350\GeV), and event rates in identified channels will
need to be measured to a few percent if the profile is to be established 
unambiguously. 
Important topics which 
motivate running at 800\GeV\  have lower cross sections and require even 
more integrated luminosity, typically $1000\,{\rm fb}^{-1}$ for the 
measurement of the top-Higgs Yukawa coupling 
or to see the effects of new physics in strong $WW$ scattering.  
Supersymmetry, if present, requires the highest possible energy to
reach as many sparticles as possible, and high luminosity to scan
production thresholds in order to measure their masses precisely.
A typical scan requires some 100\,fb$^{-1}$. \par

The absolute luminosity delivered by the machine can be measured to a 
precision of 0.1\% using the high cross section QED process of 
Bhabha scattering in the forward region. This is 
much better than the statistical precision in most 
physics channels, except for the 
GigaZ studies.

The beam-beam interaction at the interaction point will be very intense.
This leads to a focusing of the bunches resulting in  
a luminosity enhancement factor of $\sim$2.
On the other hand beamstrahlung spreads the luminosity spectrum
towards lower centre of mass energies. 
However, about 60\%
of the total luminosity is still produced at energies higher than 99.5\%
of the nominal centre of mass energy. 
For many analyses like threshold scans or high precision measurements in the 
continuum a good knowledge of the luminosity spectrum is required.
This spectrum can be measured from the acolinearity of Bhabha events in the 
forward region. In the same analysis also the beam 
energy spread can be measured.  
The precision with which the beamstrahlung and the beamspread can
be measured is good enough that it will not affect any 
physics analysis. \par

For several measurements, in particular threshold scans, the absolute 
energy of the TESLA beams 
will be determined and monitored with a 
special spectrometer which can give $\Delta E/E \leq 10^{-4}$. \par

SLC demonstrated the power of using polarised electrons
in electroweak studies, 
and the same technologies will be available to TESLA.
Throughout these studies we assume that 80\% electron polarisation can be 
achieved.  In a number of analyses, especially for supersymmetry, 
positron polarisation will also be important.  An outline design exists for 
the production of 45 to 60\% polarised positrons. 
The expected precision for the measurement of the polarisation
is 0.5\%, sufficient for most analyses. For high
precision analyses like $\sin^2 \theta_W$ at the GigaZ
positron polarisation is essential.

The range of physics to be done at TESLA can be significantly extended 
by operating the machine either as an $e^- e^-$ collider, or 
with one or both of the $e^-$ beams converted to
real high energy photons by Compton back-scattering of laser light from 
the incoming $e^-$ bunches. 
The $\gamma \gamma$ and $e^- \gamma$ modes need a non-zero beam-crossing 
angle, which should be foreseen in the
layout of the intersection region for a second collision point. \par

Many of the feasibility studies presented here have been carried out either 
with full simulation of the TESLA detector 
or with a fast simulation, tuned by comparison with the full 
simulation.  The physics processes have been simulated with the 
full suite of available Monte Carlo generators, some of which now include 
beam polarisation.
The experimental precision which TESLA can achieve must be matched
by the theoretical calculations. A continued programme of studies is
needed to improve precision on higher order corrections and to
understand the indirect contributions from new physics.

\section{Conclusions}
This volume describes the most likely
physics scenarios to be explored
at TESLA and describes a detector 
optimised to carry out that programme.  It justifies an immediate commitment 
to the construction
of the collider in its $e^+ e^-$ mode, going up to 500\GeV\  in the 
centre of mass
initially,  with a detector that can be designed and built 
using existing technologies assisted by some well defined R\&D. 

Increasing the centre of mass energy to 800\GeV\  (or higher, if the technology 
will allow) brings important 
physics benefits and should be regarded as an essential continuation of 
the programme.  
The detector can cope easily with this increase.  \par

To carry out the 
programme the collider must achieve high luminosity 
and the electron beam must be polarised.  Polarisation of the positron
beam will also be very useful.  \par

When TESLA has completed its programme of precision measurements at 
high energies up to 800\GeV, matching
improvements will be demanded on
some of the electroweak parameters measured at LEP and SLC.
The TESLA design should make provision for the possibility of high 
luminosity running at these low energies (90 to 200\GeV, the GigaZ 
option). \par

The other options for colliding beams at TESLA ($e^- e^-$,  $\gamma \gamma$ 
or  $\gamma e^-$),
add important extra components to the physics programme.  
Making two polarised electron beams is not difficult.  
The ``photon collider'' is more of
a challenge, but space should be 
left in the TESLA layout for a second interaction region with non-zero 
beam crossing angle where a second detector could be added, 
either to allow for 
$\gamma \gamma$ and $\gamma e^-$ or to give a second facility for 
$e^+ e^-$ physics. 

\bigskip
%The physics programme of TESLA is complementary to that of the
%LHC in many aspects.   \par
%

The present status of the Standard Model could not have been achieved 
without inputs from both hadron and electron accelerators and 
colliders.  This should continue into the era of TESLA and the LHC; 
the physics programme of TESLA is complementary to that of the 
LHC, 
they both have complementary strengths and both are needed.  
TESLA, with its high luminosity over the whole range of energies from 
90\GeV\  to $\sim$1\TeV, will make precise measurements
of the important quantities, masses, couplings, branching ratios, 
which will be needed to reveal the origin of electroweak symmetry
breaking and to understand the new physics, whatever it will be.
There is no scenario in which no new signals would be observed. \par

In the most likely scenarios 
with a light Higgs boson the linear collider's unique ability 
to perform a comprehensive set of
clean precision measurements will allow TESLA to establish the theory
unequivocally.
In the alternative scenario where the electroweak bosons interact
strongly at high energies, TESLA will map out the threshold region
of these new interactions.
In supersymmetric theories the great experimental potential of the
machine will allow us to perform extrapolations to scales near the 
fundamental Planck scale where particle physics and gravity 
are linked -- a unique opportunity to explore the physics area where all four 
fundamental forces of Nature will unify.

\begin{flushleft}

\end{flushleft}

%  \bibliography{physics_intro}
  \cleardoublepage
%------------------------------------------------------------------
\chapter{Higgs Physics}
\label{physics_higgs}
\input higgsdef

%The Higgs mechanism is one of the building blocks of the present picture of
%electroweak interactions. 
The fundamental particles: leptons, quarks and heavy
gauge bosons, acquire mass through their interaction with a scalar field of
non-zero field strength in its ground state~\cite{Higgs,HHG}. To accommodate
the well--established electromagnetic and weak phenomena, the Higgs mechanism
requires the existence of at least one weak isodoublet scalar field.  After
absorbing three Goldstone modes to build up the longitudinal polarisation
states of the $W^\pm/Z$ bosons, one degree of freedom is left over,
corresponding to a real scalar particle. The discovery of this Higgs boson and
the verification of its characteristic properties is crucial for the
establishment of the theory of
the electroweak interactions, not only in the canonical formulation, the 
Standard Model~(SM)~\cite{SM}, but also in supersymmetric extensions of the 
SM~\cite{SUSY,MSSM}. 

If a Higgs particle exists in Nature, the accurate study of
its production and decay properties in order to establish experimentally
the Higgs mechanism
as the mechanism of electroweak symmetry breaking can be performed in the 
clean environment of $e^+ e^-$ linear colliders~\cite{DESY-Reports}. The study
of the profile of the Higgs particles will therefore represent a central theme 
of the TESLA physics programme.  

In Sections~\ref{sec:2.1.1} and~\ref{sec:2.1.2} we review the main 
scenarios considered in this study and their implications for the Higgs sector 
in terms of the experimental Higgs signatures. 
These scenarios are the Standard Model (SM), its minimal supersymmetric 
extension (MSSM) and more general supersymmetric extensions. 
The expected accuracies for the determination of the 
Higgs boson production and decay properties are then presented 
in Section~\ref{sec:2.2} for the SM Higgs boson, in Section~\ref{sec:2.3} for
supersymmetric Higgs bosons and in Section~\ref{sec:2.4} in extended 
models together with a discussion of their 
implications for the Higgs boson profile and its nature.
Finally the complementarity of the TESLA potential to that of the LHC is 
discussed in Section~\ref{sec:2.5}.

%Results from analyses of simulated data are given for total statistics of
%1~ab$^{-1}$ to be collected at $\sqrt{s}$ = 350, 500 and 800\GeV. Results
%on the light Higgs boson refer to an integrated luminosity of 500\,fb$^{-1}$ 
%at $\sqrt{s}$ = 350\GeV or 500\GeV. 

\section{Higgs Boson Phenomenology \label{sec:2.1}}

\subsection{The Standard Model \label{sec:2.1.1}}

In the SM the Higgs sector consists of one doublet of complex 
scalar fields. Their self--interaction leads to a non-zero field strength 
$v=(\sqrt{2} G_F)^{-1/2} \approx 246$\GeV\  of the ground state, inducing the
breaking of the electroweak ${\rm SU(2)_L\times U(1)_Y}$ symmetry down
to the electromagnetic ${\rm U(1)_{EM}}$ symmetry.
Among the four initial degrees of
freedom, three will be absorbed in the $W^\pm$ and $Z$ boson states and the
remaining one corresponds to the physical  $H^0$ particle~\cite{Higgs}. In
addition, the scalar doublet couples to fermions through Yukawa interactions
which, after electroweak symmetry breaking, are responsible for the fermion
masses.
The couplings of the Higgs boson to fermions and gauge bosons are then
proportional to the masses $m_f$ and $M_V$ of these particles and completely 
determined by known SM parameters:
\begin{equation} 
g_{ffH}= m_f/v \ \ , \  \  g_{VVH}=  2 M_V^2/ v.   \label{Hcoup}
\end{equation}

\subsubsection{Constraints on the Higgs boson mass}
 
The only unknown parameter in the SM Higgs sector is the Higgs boson mass, 
$M_H$. Its value is a free parameter of the theory.
However, there are several theoretical and experimental indications 
that the Higgs boson of the SM should be light. In fact, this conclusion
holds quite generally.
\begin{figure}[ht!]
\begin{picture}(150,20)\unitlength 1mm
  \put(65,-5){a)}
  \put(142,-5){b)}
\end{picture}
\vspace{-1.5cm}
\begin{center}
\begin{tabular}{c c}
\href{pictures/1/fig2101a.pdf}{{\epsfig{file=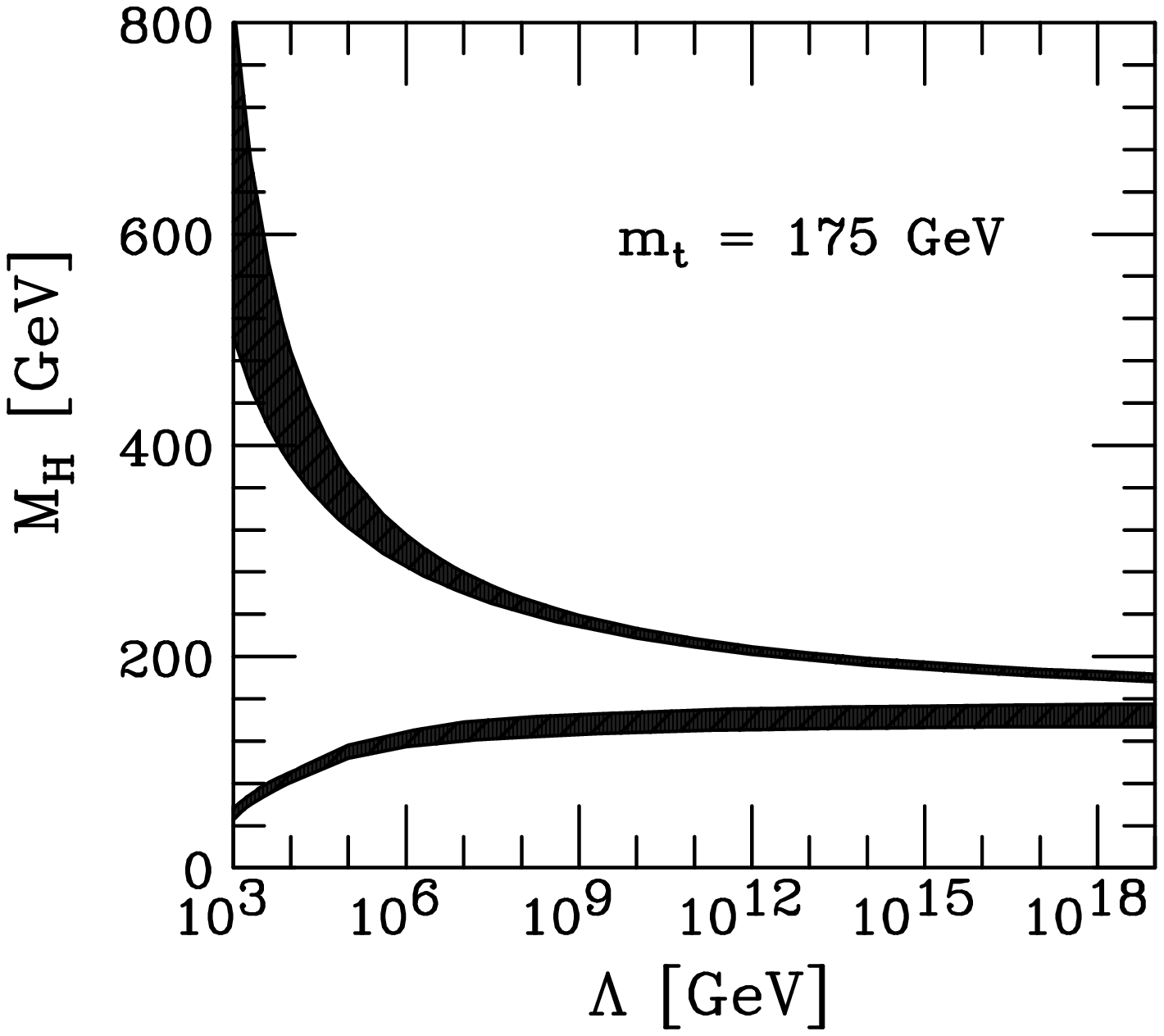,width=0.48\linewidth}}} &
\href{pictures/1/fig2101b.pdf}{{\epsfig{file=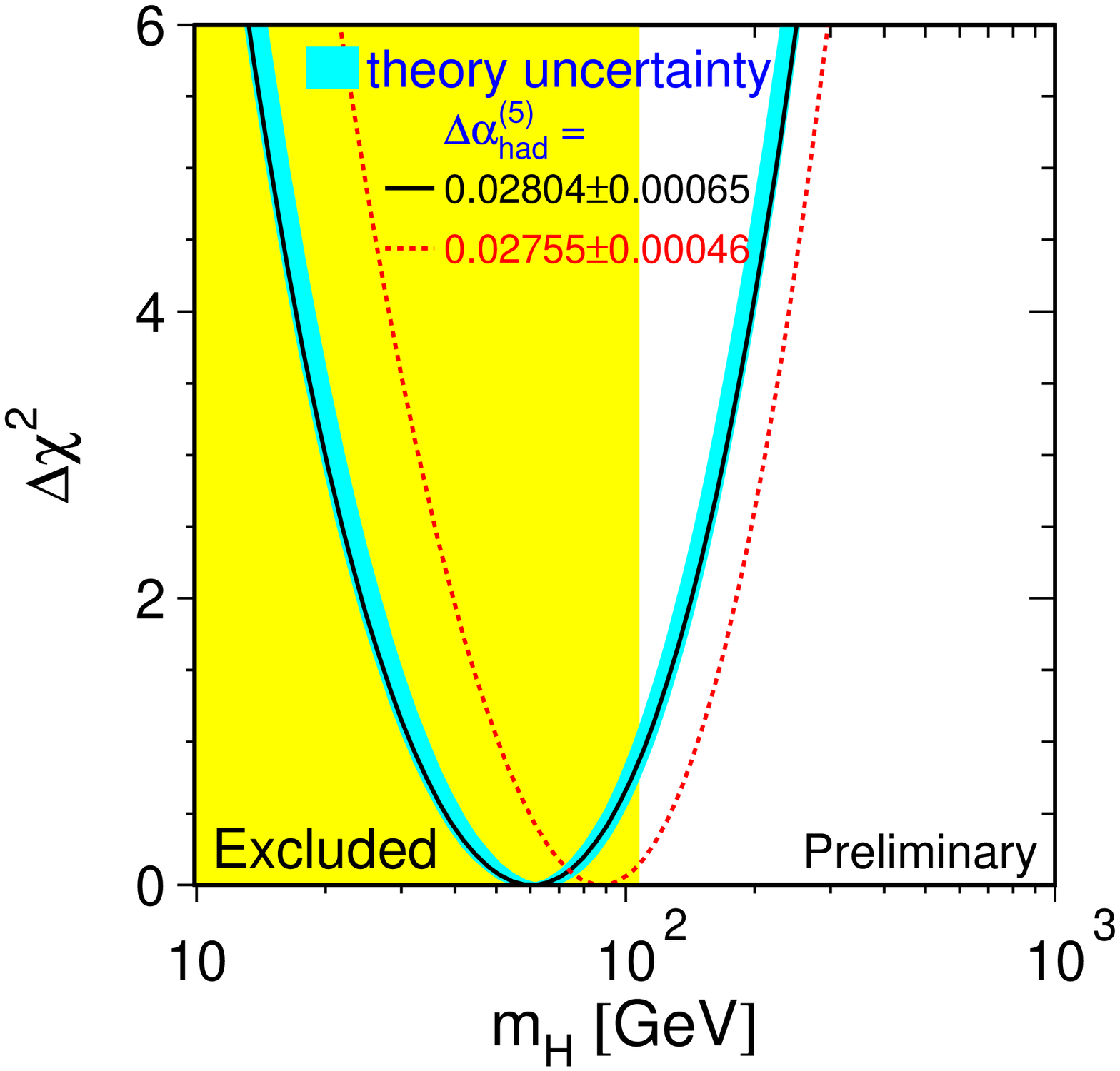,width=0.48\linewidth}}} \\
\end{tabular}
\vspace{-0.8cm}
\caption{ \label{fig:mhew} a): The triviality and vacuum stability bounds on 
the SM Higgs boson mass shown $M_H$ 
by the upper and lower curves as a function of 
the scale of new physics $\Lambda$ (from~\cite{Riesselmann}).
b): The $\Delta\chi^2$ of the electroweak fit to the LEP, SLD and 
Tevatron data as a function of $M_H$ 
(from~\cite{lepewwg}).}
\end{center}

\end{figure} 

For large values of the Higgs boson mass, $M_H \sim {\cal O}(1\; {\rm TeV})$,
the electroweak gauge bosons would have to interact strongly to 
insure unitarity in 
their scattering processes and perturbation theory would not be valid anymore.
Imposing the 
unitarity requirement in the elastic scattering of longitudinal $W$ bosons at 
high--energies, for instance, leads to the bound $M_H \lsim 870$\GeV\  at the 
tree level~\cite{strong}.
% Stringent constraints on $M_H$ can be derived from the scale $\Lambda$ up to
% which the SM is assumed to be valid before the gauge and 
% Higgs particles become
% strongly interacting and new physics phenomena may emerge~\cite{strong}. 

The
strength of the Higgs self-interaction is determined by the Higgs boson mass 
itself at the scale $v$ which characterises the spontaneous breaking of the
electroweak gauge symmetry. As the energy scale is increased, the quartic
self-coupling of the Higgs field increases logarithmically, similarly to the
electromagnetic coupling in QED.  If the Higgs boson mass is small, the energy
cut-off $\Lambda$, at which the coupling diverges, is 
large; conversely, if the Higgs boson mass is large, this $\Lambda$ becomes 
small. The upper band in Fig.~\ref{fig:mhew}\,a) 
shows the upper limit on the
Higgs boson mass as a function of $\Lambda$~\cite{triviality}.  
It has been shown in lattice analyses, which account properly for the onset of
the strong interactions in the Higgs sector, that this condition leads to an
estimate of about 700\GeV\  for the upper limit on $M_H$~\cite{lattice}.
 However, if the Higgs mass is less
than 180 to 200\GeV, the SM can be extended up to the grand unification scale,
$\Lambda_{\rm GUT} \sim 10^{16}$\GeV, or the Planck scale, 
$\sim 10^{19}$\GeV,
while all particles remain weakly
interacting [an hypothesis which plays a key role in explaining the 
experimental value of the mixing parameter $\sin^2 \theta_W$]. 
% However, if the SM has to remain valid up to the grand unification scale 
% $\Lambda_{\rm GUT} \sim 10^{16}$\GeV, the upper limit on $M_H$ becomes
% $\simeq~180$\GeV. This hypothesis, that the particle interactions remain weak 
% up to the GUT scale, is independently supported by the analysis of the value 
% of the electroweak mixing parameter $\sin^2 \theta_W$.

Lower bounds on $M_H$ can be derived from the requirement of vacuum stability.
Indeed, since the coupling of the Higgs boson to the heavy top quark
is fairly large,
corrections to the Higgs potential due to top quark loops can drive the scalar
self--coupling to negative values, leading to an unstable electroweak vacuum.
These loop contributions can only be balanced if $M_H$ is sufficiently 
large~\cite{stability}. Based on the triviality and the vacuum stability arguments,
the SM Higgs boson mass is expected in the window $130 \lsim M_H \lsim 180$\GeV~\cite{Riesselmann} for a top mass value of about 175\GeV, if the SM is extended
to the GUT scale (see Fig.~\ref{fig:mhew}\,a).
% A lower bound on $M_H$ can be obtained by the requirement of vacuum 
% stability.  Negative loop corrections to the Higgs potential due to 
% heavy top quarks can only be balanced if the Higgs boson mass is sufficiently 
% large. Based on these arguments, the SM Higgs boson mass is expected in the 
% window $130 \lsim M_H \lsim 180$\GeV~\cite{Riesselmann} for a top quark mass of 
% $m_t = 175$\GeV.  

The SM Higgs contribution to the electroweak observables, mainly through 
corrections
of the $W^\pm$ and $Z$ propagators, 
provides further information on its mass.
While these corrections only vary logarithmically, $\propto\log{(M_H/M_W)}$,
the accuracy of the electroweak data obtained at LEP, SLC and the Tevatron 
provides sensitivity to $M_H$. The most recent analysis~\cite{lepewwg} yields 
$M_H = 60^{+52}_{-29}$\GeV, corresponding to a 95\% CL upper limit of 
162\GeV.
This result depends on the running of the fine-structure constant $\alpha$. 
Recent improved measurements of $\alpha$ in the 
region between 2 and 5\GeV~\cite{bes} 
which are compatible with QCD--based calculations~\cite{hoecker}
yield $M_H = 88^{+60}_{-37}$\GeV\  corresponding to an upper limit of
206\GeV\ (see Fig.~\ref{fig:mhew}\,b). 
%% Zerwas special
Even using more conservative estimates on the theoretical 
errors~\cite{h_okun}, the upper limit on the Higgs boson mass is well within 
the reach of a 500\GeV\  linear collider.

Since this result is extracted in the framework of the SM, it can be
considered as an effective low-energy approximation to a more fundamental 
underlying theory. It is interesting to verify how this 
constraint on $M_H$ may be modified by the effect of new physics beyond the SM.
This new physics can be parameterised generically, by extending the SM 
Lagrangian with effective operators of mass dimension five and higher,
weighted by inverse powers of a cut-off scale $\Lambda$, representing the 
scale of new physics. In this approach, the SM result corresponds to 
$\Lambda = \infty$. By imposing the necessary symmetry 
properties on these operators and by fixing their dimensionless coefficients to
be $\pm1$, compatibility with the electroweak precision data can be preserved
only with $M_H \lsim$ 400\GeV, if the operators are 
not restricted to an unplausibly small set~\cite{barstr}.
Though slightly above the SM limit, the data nevertheless require a light 
Higgs boson even in quite general extended scenarios.
% An alternative analysis, introducing an {\it ad hoc}
% cancellation of the Higgs mass contribution to the $T$ observable, sets a
% comparable upper limit $M_H <$ 500\GeV~\cite{chihol}.

Direct searches for the Higgs boson at LEP yield a lower bound of 
$M_H \geq 113.5$\GeV\  at the 95\% confidence level~\cite{h_LEPHiggs}.
The LEP collaborations have recently reported a $2.9 \sigma$ excess of 
events beyond the expected SM background in the combination of
their Higgs boson searches~\cite{h_LEPHiggs}. 
This excess is consistent with the production of a
SM--like Higgs boson with a mass $M_H=115 ^{+1.3}_{-0.9}$\GeV.

In summary, the properties of the SM Higgs sector and the experimental data
from precision electroweak tests favour a light
Higgs boson, as the manifestation of symmetry breaking 
and mass generation within the Higgs mechanism.\footnote{For
comments on no--Higgs scenarios and their theoretically very complex 
realisations see Section~\ref{sec:phys-ewsb-strong} on strong WW interactions.}

\subsubsection{Higgs boson production processes}

\begin{figure}[h!t!]
\begin{center}
\href{pictures/1/fig2102.pdf}{{\epsfig{file=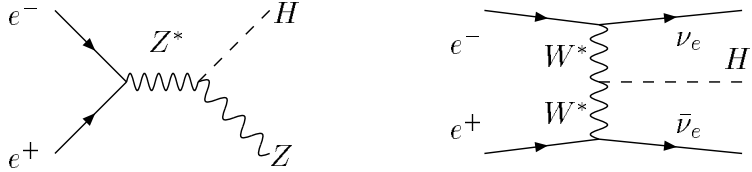,width=0.7\linewidth,clip}}}
\end{center}
\caption{\label{diagH}
Main production mechanisms of the SM Higgs boson at $e^+e^-$ 
colliders.}
\end{figure}

The main production mechanism of this SM Higgs boson 
in $e^+e^-$ collisions at TESLA
are the Higgs-strahlung process~\cite{strahlung}, $e^+e^- \to ZH^0$, 
and the $WW$ fusion process~\cite{fusion}, $e^+ e^- \to W^* W^*\to \bar{\nu}_e 
\nu _e H$; Fig.~\ref{diagH}.
The cross section for the Higgs-strahlung process scales as $1/s$ and 
dominates at low energies:
\begin{equation}
\label{eq:hstr}
\sigma(e^+e^- \to ZH)= \frac{g_{ZZH}^2}{4\pi} \, \frac{G_F (v_e^2+ a_e^2)}{96 
\sqrt{2} s} \, \beta_{HZ} \, \frac{\beta_{HZ}^2 + 12 M_Z^2 /s}
{(1-M_Z^2 /s)^2} ,
\end{equation}
where $\beta_{ij}^2=\left[ 1-(M_i+M_j)^2/s\right ]
\left[1-(M_i-M_j)^2/s\right]$, $v_e=-1+4\sin^2\theta_W$ and $a_e=-1$.
The cross--section for the $WW$ fusion process~\cite{fusion}, 
$e^+e^- \to \nu_e \bar{\nu_e} H^0$, rises $\propto \log(s/M_H^2)$ 
and dominates at high energies:
\begin{equation}
\sigma(e^+e^- \to \bar{\nu}_e \nu _e H) \to 
\frac{g_{WWH}^2}{4\pi} \, \frac{G_F^2 } {8\pi^2} \left[
\left( 1+\frac{M_H^2}{s} \right) \log \frac{s}{M_H^2} -2
\left( 1-\frac{M_H^2}{s} \right)
\right] .
\end{equation}

\vspace*{-5mm}\noindent The $ZZ$ fusion mechanism, $e^+e^- \rightarrow Z^*Z^* e^+ e^- \to e^+e^-H$,
also contributes to Higgs production, with a cross section  suppressed by an 
order of magnitude compared to that for $WW$ fusion, due to the ratio of the CC
to NC couplings, $16 \cos^4 \theta_W \sim 9.5$. In contrast to Higgs-strahlung
and $WW$ fusion, this process is also possible in $e^-e^-$ collisions with 
approximately the same total cross section as in $e^+e^-$ collisions.

\begin{figure}[ht!]
\begin{center}
\href{pictures/1/fig2103.pdf}{{\epsfig{file=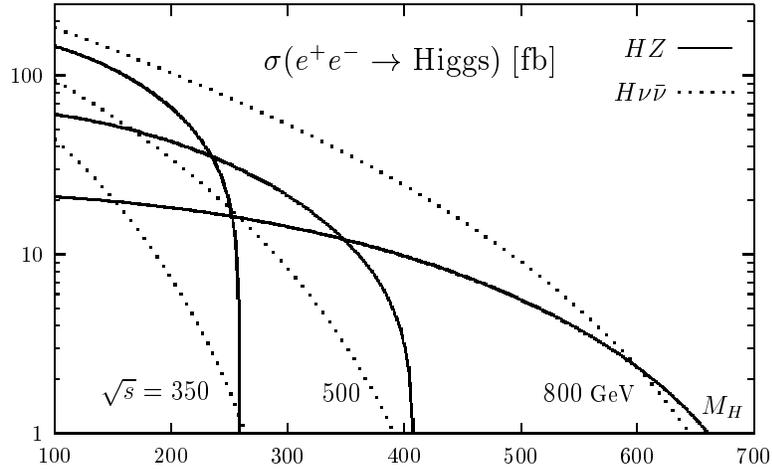,width=0.65\linewidth}}}
\caption{\label{fig:xsec}
The Higgs-strahlung and WW fusion production cross--sections 
vs.~$M_H$ for $\sqrt{s}$ = 350\GeV, 500\GeV\  and 800\GeV.}
\end{center}

\end{figure} 

The cross--sections for the Higgs-strahlung and the $WW$ fusion processes are 
shown in Fig.~\ref{fig:xsec} for three values of $\sqrt{s}$.
\begin{figure}[ht!]
\begin{center}
\href{pictures/1/fig2104.pdf}{{\epsfig{file=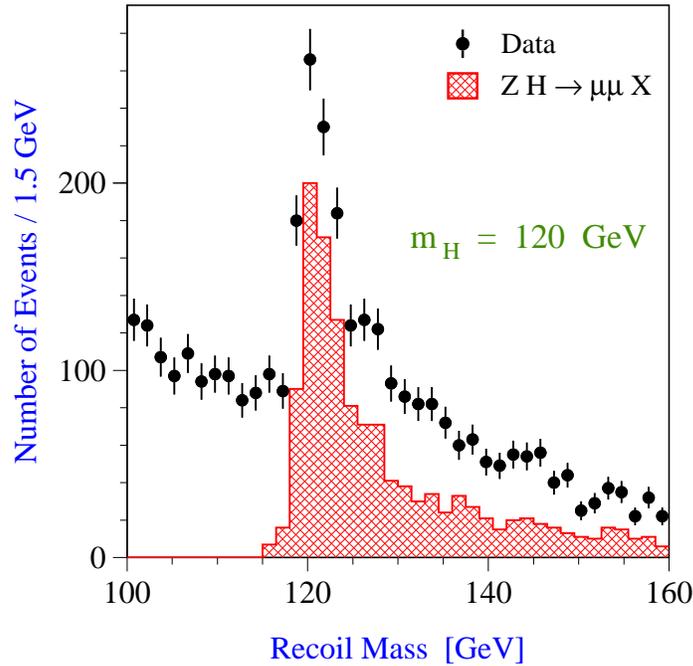,width=0.6\linewidth}}}
\end{center}
\caption{\label{fig:recm}
The $\mu^+\mu^-$ recoil mass distribution in the process
$e^+ e^- \rightarrow H^0 Z \rightarrow X \mu^+\mu^-$ for 
$M_H$ = 120\GeV and 500\,fb$^{-1}$ at $\sqrt{s}$ = 350\GeV. The dots with
error bars are Monte Carlo simulation of Higgs signal and background. 
The shaded histogram represents the signal only.}
\end{figure}
At $\sqrt{s} = 350$\GeV, a sample of $\sim$ 80.000 Higgs bosons
is produced, predominantly through Higgs-strahlung,
for $M_H = 120$\GeV\  with an integrated luminosity 
of 500\,fb$^{-1}$, corresponding to one to two years of 
running. The Higgs-strahlung process, $e^+e^-\to Z H^0$, 
with $Z \rightarrow \ell^+ \ell^-$, offers a very distinctive
signature~(see Fig.~\ref{fig:recm}) ensuring the observation of the SM Higgs 
boson up to the production kinematical limit independently of its decay 
(see Table~\ref{tab:discovery}).
At $\sqrt{s} = 500$\GeV, the Higgs-strahlung and the 
$WW$ fusion processes have approximately the same cross--sections, 
${\cal O}$(50 fb) for 100\GeV\  $\lsim M_H \lsim$ 200\GeV.  
\begin{table}[ht!]
\begin{center}
\begin{tabular}{|c|c|c|c|}
\hline
$M_H$ (GeV) & $\sqrt{s}$ = 350\GeV & 500\GeV & 800\GeV \\
\hline
\hline
120     & 4670 & 2020 & ~740 \\
140     & 4120 & 1910 & ~707 \\
160     & 3560 & 1780 & ~685 \\
180     & 2960 & 1650 & ~667 \\
200     & 2320 & 1500 & ~645 \\
250     & ~230 & 1110 & ~575 \\ \hline
Max $M_H$ (GeV)  & 258 & 407 & 639 \\
\hline
\end{tabular}
\caption{\label{tab:discovery}
 Expected number of signal events for 500\,fb$^{-1}$ for the 
Higgs-strahlung channel with di-lepton final states $e^+e^- \rightarrow
Z H^0 \rightarrow \ell^+ \ell^- X$, ($\ell = e,~\mu$) at different $\sqrt{s}$
values and maximum value of $M_H$ yielding more than 50 signal events in this
final state.}
\end{center}
\end{table}

\vspace*{-2mm} \noindent
At a $\gamma \gamma$ collider, Higgs bosons can be produced 
in the resonant 
s--channel process $\gamma \gamma \to H$ which proceeds predominantly
through a loop of $W$ bosons and top quarks~\cite{gamma}. 
This process provides the
unique opportunity to measure precisely the di--photon partial width 
$\Gamma_{\gamma\gamma}$ of the
Higgs boson which represents one of the most important measurements to be
carried out at a $\gamma \gamma$ collider. Deviations of 
$\Gamma_{\gamma\gamma}$ from its predicted SM value are a probe of any 
new charged heavy particle exchanged in the loop such as charged Higgs
bosons and supersymmetric particles even if they are too heavy to be
directly observed at TESLA or the LHC.
The large backgrounds from
the continuum process $\gamma\gamma\to q\bar{q}, q=(c,b)$ are theoretically
and experimentally under control~\cite{melles2, hgamma}.

\subsubsection{Higgs boson decays}

In the SM, the Higgs boson branching ratios are completely 
determined~\cite{hdecay}, once the Higgs boson mass is fixed. For values of the 
Higgs boson mass in the range $M_Z \leq M_H \lsim 140$\GeV, 
the Higgs boson dominantly decays
to fermion pairs, in particular $b\bar{b}$ final states since the Higgs 
fermion couplings are proportional to the fermion masses. The partial width 
for a decay of the SM Higgs boson into a fermion pair is given by:
\begin{equation} 
\Gamma (H^0 \to f\bar{f} ) = \frac{g^2_{ffH} (M_H^2)}{4\pi} \, \frac{N_C}{2} 
\,  M_H \left( 1 - \frac{4m_f^2}{M_H^2} \right)^{\frac{3}{2}} ,
\end{equation} 
with $N_C = 1(3)$ for leptons (quarks).
For $M_H \lsim 140\GeV$,
the decays $H^0 \to \tau ^+ \tau ^-, c\bar{c}$ and $gg$ remain significantly 
suppressed compared to $b \bar b$ but they are important to test the relative 
Higgs couplings to up-type and down-type fermions and the scaling of these 
couplings with the fermion masses. The precise value of the running quark mass
at the Higgs boson scale $m_q (M_H)$ represents a significant source of 
uncertainty 
in the calculation of the rates for these decays. QCD corrections to 
the hadronic decays, being quite substantial, introduce an additional 
uncertainty. At present, the $c$-quark mass and the $\alpha_s$ uncertainties 
limit the accuracy for rate predictions for the $c \bar c$ and $g g$ channels 
to about $\pm 14\%$ and $\pm 7\%$ respectively. Improvements on $m_b$ and 
$m_b - m_c$, possibly by a factor $\simeq 2$, can be envisaged after the study
of the data on $B$~decays from the $B$~factories and the {\sc LHC}. On the 
contrary, the $b \bar b$ and $\tau^+ \tau^-$ predictions can be obtained with 
accuracies comparable to, or better than, the experimental uncertainties 
discussed later in this chapter.          

Above the $ZZ$ threshold and except in a mass range above the $t\bar{t}$ 
threshold, the Higgs boson decays almost exclusively into 
the $WW$ or $ZZ$ channels, with widths
\begin{equation} 
\Gamma (H^0 \to VV) = \frac{g^2_{VVH}}{4\pi} \, \frac{3 \delta_V}{8 M_H}  \, 
\left(1 - \frac{M_H^2}{3M_V^2} + \frac{M_H^4}{12M_V^4} \right)
\left( 1 -\frac{4M_V^2}{M_H^2} \right)^{\frac{1}{2}} 
, \;  \delta_{W/Z}=2/1 .
\end{equation}

\begin{figure}[ht!]
\begin{picture}(150,20)\unitlength 1mm
  \put(65,-45){a)}
  \put(131,-45){b)}
\end{picture}
\vspace{-1.5cm}
\begin{center}
\href{pictures/1/fig2105.pdf}{{\epsfig{file=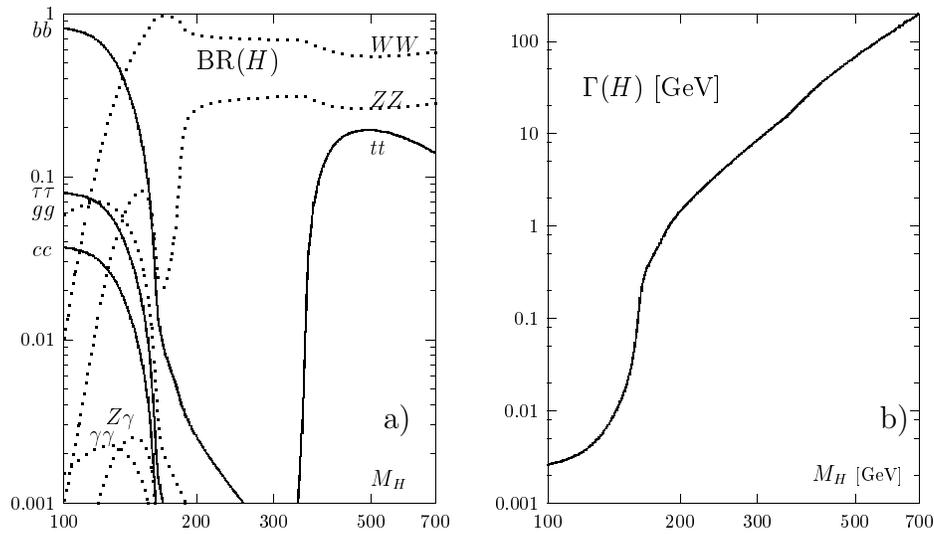,width=0.8\linewidth}}}
\end{center}
\caption{ \label{fig:hbrth}
The branching ratios (a) and the total decay width (b) 
of the SM Higgs boson as a function of its mass.}
\end{figure}

Decays into $WW^*$ pairs, with one of the two gauge bosons being 
virtual, become comparable to the $b\bar{b}$ mode at $M_H \simeq 140$\GeV. 
The Higgs boson branching ratios 
are shown in Fig.~\ref{fig:hbrth}\,a) 
as a function of $M_H$. QCD corrections to the
hadronic decays have been taken into account as well as the virtuality of the 
gauge bosons, and of the top quarks. 
The top quark and $W$ boson mediated loop decays into
$\gamma \gamma$ and $Z \gamma$ final states have 
small branching ratios, reaching a
maximum of $\sim 2.5 \times 10^{-3}$ at 125 and 145\GeV, respectively.
However, they lead to clear signals and are 
interesting because they are sensitive to new heavy particles. 

By adding up all possible decay channels, we obtain the total Higgs 
boson decay width, as shown in Fig.~\ref{fig:hbrth}\,b) for $m_t= $~175\GeV.  
Up to masses of 140\GeV, the Higgs particle is very narrow, 
$\Gamma (H) \le $~10\,MeV.  After opening the mixed real/virtual gauge 
boson channels, the state becomes rapidly wider, reaching $\sim
$~1\GeV\  
at the $ZZ$ threshold.

\subsection{Supersymmetric Extension of the Standard Model
\label{sec:2.1.2}}

Several extensions of the SM introduce additional Higgs doublets and
singlets. In the 
simplest of such extensions the Higgs sector consists of two doublets 
generating five physical Higgs states: $h^0$, $H^0$, $A^0$ and $H^{\pm}$.
The $h^0$ and $H^0$ states are ${\cal CP}$ even 
and the $A^0$ is ${\cal CP}$ odd. Besides the masses,
two mixing angles define the properties of the Higgs bosons and their 
interactions with gauge bosons and fermions, namely the ratio of the 
vacuum expectation values $v_2/v_1 = \tan \beta$ and a mixing angle $\alpha$
in the neutral ${\cal CP}$-even sector. 
These models are generally referred to as
2HDM and they respect the SM phenomenology at low energy. In particular, the 
absence of flavour changing neutral currents is guaranteed by either 
generating the mass of both up- and down-like quarks through the same 
doublet (Model~I) or by coupling the up-like quarks to the first doublet and
the down-like quarks to the second doublet (Model~II). 
Two Higgs field doublets naturally arise in the context of the minimal 
supersymmetric extension of the SM (MSSM).

One of the prime arguments for introducing Supersymmetry~\cite{SUSY0, SUSY} 
is the solution of the hierarchy problem. By assigning fermions and 
bosons to common
multiplets, quadratically divergent radiative corrections to the Higgs boson
mass can be cancelled in a natural way~\cite{HHG,MSSM} by adding up bosonic and
opposite--sign fermionic loops. As a result of the bosonic--fermionic
supersymmetry, Higgs bosons can be retained as elementary spin--zero particles
with masses close to the scale of the electroweak symmetry breaking even in the
context of very high Grand Unification scales. These supersymmetric
theories are strongly supported by the highly successful prediction of the
electroweak mixing angle: $\sin^2{\theta_W^{\mathrm{SUSY}}} = 
0.2335 \pm 0.0017$, $\sin^2{\theta_W^{\mathrm{exp}}} = 
0.2310 \pm 0.0002$. 
% \footnote{The additional degrees
% of freedom of this theory will also account for the small discrepancy between
% the predicted value of the electroweak mixing angle, 
% $\sin^2 \theta_W \sim 0.2$
% in the SM, and the experimentally measured value, $\sin^2\theta_W \simeq
% 0.23$.}. 
In addition, the breaking of the
electroweak symmetry may be generated in supersymmetric models
in a natural way via radiative corrections associated with
the heavy top quark. The MSSM serves as a useful guideline into 
this area, since only a few phenomena are specific to this
model and many of the characteristic patterns are realized also in 
more general extensions.  

%%%%%%%%%%%%%%%%%%%%%%%%%%%%%%%%%%%%%%%%%%%%%%%%%%%%%%%%%%%%%%%%%%%%%%%%%%
\subsubsection{The Higgs spectrum in the MSSM}

In the MSSM, two doublets of Higgs fields are needed to break the electroweak
symmetry, leading to a Higgs spectrum consisting of five particles~\cite{GH}:
two ${\cal CP}$--even bosons $h^0$ and $H^0$, 
a ${\cal CP}$--odd boson $A^0$ and two
charged particles $H^\pm$. 
% Besides the four masses, two mixing angles define
% the properties of the scalar particles and their 
% interactions with gauge bosons
% and fermions: the ratio of the two vacuum expectation values $\tb = v_2/v_1$
% and a mixing angle $\alpha$ in the neutral ${\cal CP}$--even sector. 
Supersymmetry leads to several relations among these parameters and, in fact,
only two of them are independent at the tree level.  
These relations impose
a strong hierarchical structure on the mass spectrum $ [ M_h<M_Z , M_A < M_H$
and $M_W <M_{H^\pm}]$ some of which are, however, broken by radiative 
corrections.  

\begin{figure}[ht!]
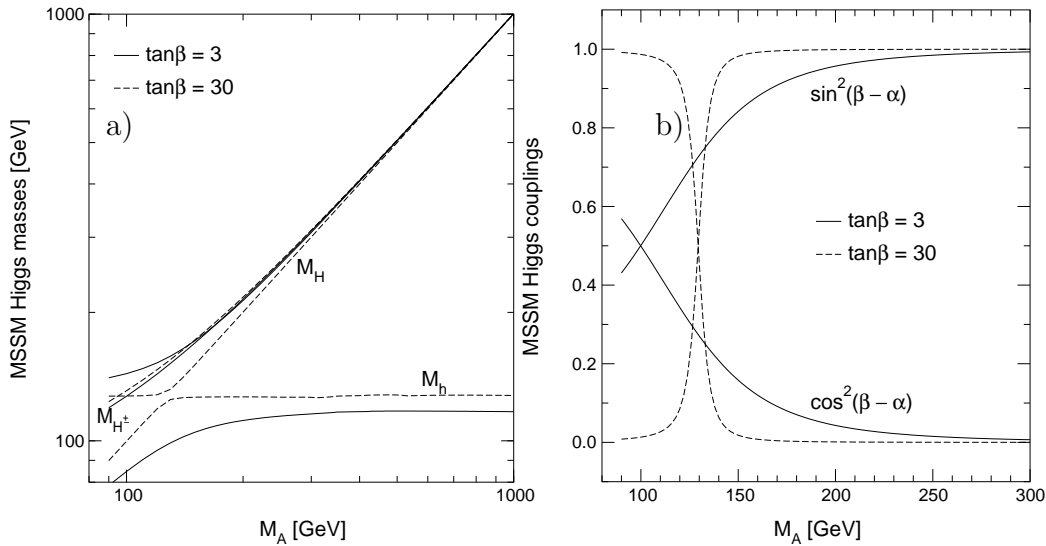

\begin{picture}(150,20)\unitlength 1mm
  \put(20,-6){a)}
  \put(93,-6){b)}
\end{picture}
\vspace{-1.5cm}
\begin{center}
\begin{tabular}{c c}
\href{pictures/1/fig2106a.pdf}{{\epsfig{file=fig2106a.eps,width=0.45\linewidth}}}
\href{pictures/1/fig2106b.pdf}{{\epsfig{file=fig2106b.eps,width=0.45\linewidth}}} \\
\end{tabular}
\caption{ \label{fig:mass}
The masses of the Higgs bosons in the MSSM (a) and their squared 
couplings to the gauge bosons (b) for two representative values of 
$\tan \beta$ = 3 and 30~\cite{FeynHiggs}.}
\end{center}
\vspace*{-1cm}
\end{figure}

The leading part of these radiative corrections~\cite{radcor1,radcor2,mhiggsletter} to the
Higgs boson masses and couplings grows as the fourth power of the top quark mass
% [defined in the ${\rm \overline{MS}}$ scheme to account for QCD corrections]
and logarithmically with the SUSY scale or common squark 
mass $M_S$~\cite{radcor1}; mixing in the stop sector~$\tilde{A}_t$ has also
to be taken into account.  The radiative corrections push the maximum value of 
the lightest $h$ boson mass upwards by several ten\GeV~\cite{radcor2,mhiggsletter}; a recent
analysis, including the dominant two--loop contributions gives an upper bound
$M_{h} \lsim 135$\GeV~\cite{FeynHiggs}; c.f.  Fig.~\ref{fig:mass}\,a)
where the MSSM
Higgs masses are shown for $M_S=1$\,TeV and $\tilde{A}_t=\sqrt{6} M_S$ 
% [the so--called maximal mixing scenario].  
This upper bound is obtained for large
values of $M_A \sim 1$\,TeV and $\tb \sim m_t/m_b \sim 30$ and crucially
depends on the value of the top quark mass. The precise determination of
$M_t$ possible at TESLA is instrumental for precision physics in the MSSM
Higgs sector.

The couplings of the MSSM Higgs bosons to fermions and gauge bosons depend
strongly on the angles $\alpha$ and $\beta$. The pseudo-scalar and charged 
Higgs boson couplings to down (up) type fermions are (inversely) proportional 
to $\tb$; the pseudo-scalar $A^0$ has no tree level couplings to two gauge 
bosons. 
For the ${\cal CP}$--even Higgs bosons, 
the couplings to down (up) type fermions 
are enhanced (suppressed) compared to the SM Higgs couplings [for values $\tb>
1$]; the couplings to gauge bosons are suppressed by $\sin/\cos(\beta-\alpha)$ 
factors (see Tab.~\ref{tab:coupl2hdm} and Fig.~\ref{fig:mass}\,b)).  
\begin{table}[ht!]
\begin{center}
\renewcommand{\arraystretch}{1.2}
\begin{tabular}{|c|c|c|c|c|} \hline
$\ \ \ \Phi \ \ \ $ &$ g_{\Phi \bar{u}u} $      & $ g_{\Phi \bar{d} d} $ &
$g_{ \Phi VV} $ \\ \hline\hline 
$h^0$  & \ $\; \cos\alpha/\sin\beta \rightarrow 1  \; $ \ & \ $ \; -\sin\alpha/
\cos\beta \rightarrow 1 \; $ \ & \ $ \; \sin(\beta-\alpha) \rightarrow 1 \; 
$ \ \\
$H^0$  & \   $\; \sin\alpha/\sin\beta \rightarrow 1/\tb \; $ \ & \ $ \; 
\cos\alpha/ \cos\beta \rightarrow \tb \; $ \ & \ $ \; \cos(\beta-\alpha) 
\rightarrow 0 \; $ 
\ \\ $A^0$  & \ $\; 1/ \tb \; $\ & \ $ \; \tb \; $ \   & \ $ \; 0 \; $ \ 
\\   \hline
\end{tabular}
\end{center}
\vspace*{-.2cm}
\caption{\label{tab:coupl2hdm}
MSSM neutral Higgs boson couplings to fermions and gauge bosons 
normalized to the SM Higgs couplings, and their limit for $M_A \gg M_Z$ 
[decoupling regime].}
\vspace*{-.2cm}
\end{table}

If $M_h$ is very close to its upper limit for a given value of $\tb$, the 
couplings of the $h$ boson to fermions and gauge bosons are SM like, while the 
couplings of the heavy $H$ boson become similar to that of the pseudoscalar 
$A^0$ boson; Tab.~\ref{tab:coupl2hdm}. 
This decoupling limit~\cite{Decoupling} is 
realized when $M_A \gg M_Z$
% , but is reached in practice if the pseudoscalar 
% mass $M_A$ exceeds 300\GeV for small $\tb$,  or exceeds the maximum allowed 
% $M_h$ value for large $\tb$. 
and in this regime, the $A^0,H^0$ 
and $H^\pm$ bosons are almost degenerate in mass.

%%%%%%%%%%%%%%%%%%%%%%%%%%%%%%%%%%%%%%%%%%%%%%%%%%%%%%%%%%%%%%%%%%%%%%%%%%
\subsubsection{MSSM Higgs production}

\begin{figure}[ht!]
% \vspace*{-0.8cm}
\begin{center}
\hspace*{5mm}
\href{pictures/1/fig2107.pdf}{{\epsfig{file=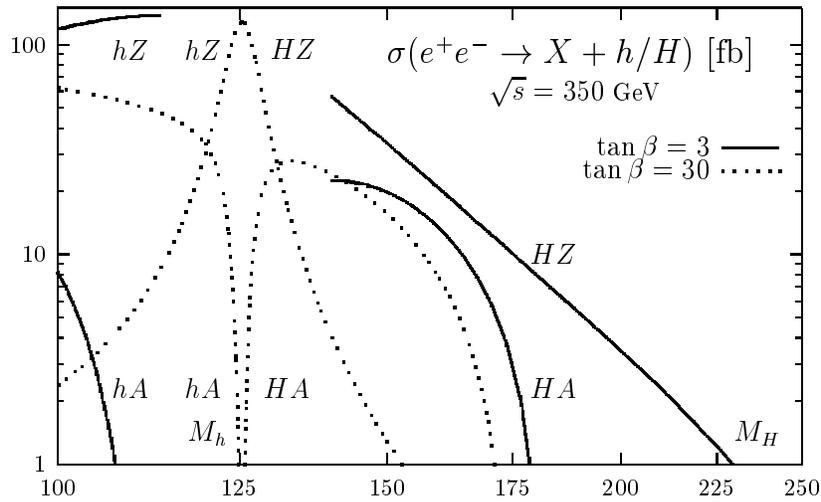,width=0.7\linewidth}}}
\end{center}
% \vspace{-13.8cm}
\caption[]{Production cross sections of the MSSM neutral Higgs bosons at 
$\sqrt{s}=350$\GeV\  in the Higgs-strahlung and pair production processes; 
$\tan \beta=3$ and 30.  \protect\label{xs-350} }
% \vspace*{-5mm}
\end{figure}

\begin{figure}[ht!]
% \vspace*{-1.2cm}
\begin{center}
\hspace*{5mm}
\href{pictures/1/fig2108.pdf}{{\epsfig{file=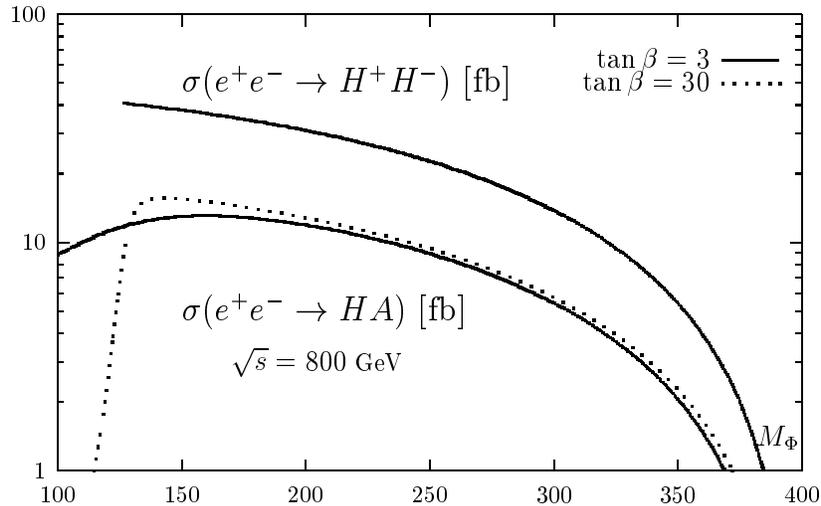,width=0.7\linewidth}}}
\end{center}
% \vspace{-13.8cm}
\caption[]{Production cross sections for the associated $H^0A^0$ and the 
$H^+H^-$ production mechanisms at $\sqrt{s}=800$\GeV\  as functions of the $A$ and
$H^\pm$ masses, respectively, for $\tan \beta=3$ and 30.  
\protect\label{fig:2108} }
% \vspace*{-5mm}
\end{figure}

In addition to the Higgs-strahlung and $WW$ fusion production processes
for the ${\cal CP}$--even Higgs particles $h^0$ and $H^0$, 
$e^+e^- \to Z+h^0/H^0$ and
$e^+e^- \to \nu_e \bar{\nu}_e +h^0/H^0$, the associated pair production process,
$e^+ e^- \to A^0+h^0/H^0$, also takes place 
in the MSSM or in two--Higgs doublet
extensions of the SM. The pseudoscalar $A^0$ cannot be produced in the
Higgs-strahlung and fusion processes to leading order. The cross sections for 
the Higgs-strahlung and pair production processes can be expressed as~\cite{MSSMprod}
\begin{eqnarray}
\sigma(e^+ e^-  \rightarrow Z + h^0/H^0) & =& \sin^2/\cos^2(\beta-\alpha) \ 
\sigma_{\rm SM} \nonumber \\
\sigma(e^+ e^-  \rightarrow A^0 + h^0/H^0) & =& \cos^2/\sin^2(\beta-\alpha) \
\bar{\lambda} \  \sigma_{\rm SM}
\end{eqnarray}
where $\sigma_{\rm SM}$ is the SM cross section for Higgs-strahlung and the
coefficient $\bar{\lambda}$, given by 
$\bar{\lambda} = \beta^3_{Aj} / [ \beta_{Zj} ( 12 M_Z^2 + \beta^2_{Zj})]$
($\beta_{ij}$ is defined below eq.~\ref{eq:hstr} and $j = h$~or $H$, 
respectively)
accounts for the suppression of the ${\cal P}$--wave $A^0h^0/A^0H^0$ 
cross sections near threshold. Representative examples of the 
cross sections in these
channels are shown as a function of the Higgs masses in Fig.\ref{xs-350} at a
c.m. energy $\sqrt{s}=350$\GeV\  for $\tb= 3$ and 30.  The cross sections for the
Higgs-strahlung and for the pair production, likewise the cross sections for
the production of the light and the heavy neutral Higgs bosons 
$h^0$ and $H^0$, are
mutually complementary to each other, coming either with coefficients
$\sin^2(\beta-\alpha)$ or $\cos^2(\beta-\alpha)$.  As a result, since
$\sigma_{\rm SM}$ is large, at least the lightest ${\cal CP}$--even Higgs boson
must be detected.
For large $M_A$ values, the main production mechanism 
for the heavy neutral Higgs bosons is the associated $H^0A^0$ process when 
kinematically allowed; the cross section is shown for a c.m. energy 
$\sqrt{s}=800$\GeV\  in Fig.~\ref{fig:2108}. 


Charged Higgs bosons, if lighter than the top quark, can be produced in top
decays, $t  \rightarrow b + H^+$, with a branching ratio varying between $2\%$
and $20\%$ in the kinematically allowed region. Charged Higgs particles can 
also be directly pair produced in $e^+ e^-$ collisions, $e^+ e^- \to H^+H^-$, 
with a  cross section which depends mainly on the $H^\pm$ mass~\cite{MSSMprod}.
It is of ${\cal 
O}$(50 fb) for small masses at $\sqrt{s} = 800$\GeV, but it drops very quickly 
due to the ${\cal P}$--wave suppression $\sim \beta^3$ near the threshold 
( see Fig.~\ref{fig:2108}). For $M_{H^{\pm}} = 375$\GeV, the cross section falls to a 
level of $\sim 1\,$\,fb, which for an integrated luminosity of $500\,{\rm 
fb}^{-1}$ corresponds to $\sim 500$ events.  

The MSSM Higgs bosons can also be produced in $\gamma \gamma$ collisions, 
$\gamma \gamma \to H^+ H^-$ and $\gamma \gamma \to h^0,H^0, A^0$, 
with favourable 
cross sections~\cite{MSSMgamma}. For the neutral $H^0$ and $A^0$ 
bosons, this mode 
is interesting since one can probe higher masses than at the $e^+ e^-$ collider,
$M_{H,A} \sim 400$\GeV\  for a 500\GeV\  initial c.m. $e^+ e^-$ energy. 
Furthermore, an energy scan could resolve the small $A^0$ and $H^0$ 
mass difference near the decoupling limit.  
%%%%%%%%%%%%%%%%%%%%%%%%%%%%%%%%%%%%%%%%%%%%%%%%%%%%%%%%%%%%%%%%%%%%%%%%%%%%%%%%%
\subsubsection{MSSM Higgs decays}

\begin{figure}[ht!]
\begin{center}
% \vspace*{-3cm}
\hspace*{-1cm}
\href{pictures/1/fig2109.pdf}{{\epsfig{file=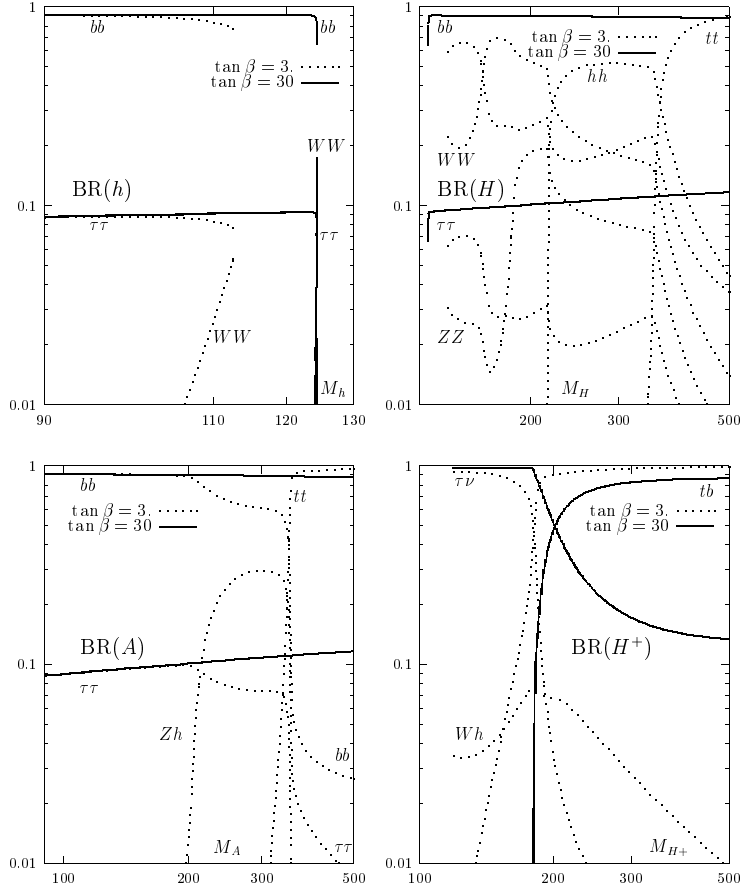,width=0.65\linewidth}}} 
\end{center}
% \vspace*{-4.5cm}
\caption{\label{fig:2107}
MSSM Higgs decay branching ratios as a function of the Higgs masses}
\vspace*{3mm}
\end{figure}

The decay pattern of the Higgs bosons in the MSSM~\cite{MSSMdecays} is more
complicated than in the SM and depends strongly on the value of $\tb$
( see Fig.~\ref{fig:2107}).  

The lightest neutral $h^0$ boson will decay  mainly into fermion pairs since its
mass is smaller than $\sim$~130\GeV.  This is, in general, also the dominant
decay mode of the pseudo-scalar boson $A^0$.  
For values of $\tb$ much larger than
unity, the main decay modes of the three neutral Higgs bosons are decays into
$b \bar{b}$ and $\tau^+ \tau^-$ pairs; the branching ratios being of order $
\sim 90\%$ and $10\%$, respectively. For large masses, the top decay channels
$H^0, A^0 \rightarrow t\bar{t}$ open up, yet for large $\tb$ this mode remains
suppressed.  If the masses are high enough, the heavy $H^0$ boson can decay into
gauge bosons or light $h^0$ boson pairs and the pseudo-scalar $A^0$ 
particle into $h^0Z$ 
final states; these decays are strongly suppressed for $\tb \gsim 5$. 
The charged Higgs particles decay into fermions pairs: mainly $t\bar{b}$ and
$\tau \nu_{\tau}$ final states for $H^\pm$ masses, respectively, above and below
the $tb$ threshold.  If allowed kinematically, the $H^\pm$ bosons decay also
into $h^0W^\pm$ final states. 
Adding up the various decay modes, the Higgs bosons widths
remain narrow, being of order 10\GeV\  even for large masses.
However, the total width of the $h^0$ boson may become much larger than
that of the SM $H^0$ boson for large $\tan{\beta}$ values.

Other possible decay channels for the MSSM bosons, in particular the heavy $H^0,
A^0$ and $H^\pm$ states, are decays into supersymmetric particles~\cite{SUSYdecays}. In addition to light sfermions, decays into charginos and
neutralinos could eventually be important if not dominant.  Decays of the
lightest $h^0$ boson into the lightest neutralinos (LSP) or sneutrinos can be
also important, exceeding 50\% in some parts of the SUSY parameter space, in
particular in scenarios where the gaugino and sfermion masses are not unified at
the GUT scale~\cite{LSPdecays}.  These decays strongly affect experimental
search techniques. In particular, invisible neutral Higgs decays could
jeopardise the search for these states at hadron colliders where these 
modes are very difficult to detect.  

\subsubsection{Non--minimal SUSY extensions} 
A straightforward extension of the MSSM is the addition of an iso--singlet 
scalar field $N$~\cite{NMSSM1,NMSSM2}. This next--to-minimal extension of the 
SM or (M+1)SSM has been advocated to solve the so--called $\mu$ problem, i.e. to
explain why the Higgs--higgsino mass parameter $\mu$ is of ${\cal O}(M_W)$. 
The Higgs spectrum of the (M+1)SSM includes
in addition one extra scalar and pseudo-scalar Higgs particles.  The
neutral Higgs particles are in general mixtures of the iso--doublets, which
couple to $W, Z$ bosons and fermions, and the iso--singlet, decoupled from the
non--Higgs sector. Since the two trilinear couplings involved in the potential,
$H_1 H_2 N$ and $N^3$, increase with energy, upper bounds on the mass of the
lightest neutral Higgs boson $h_1$ can be derived, in analogy to the SM, from
the assumption that the theory be valid up to the GUT scale: $M_{h_1} \lsim 150
$\GeV~\cite{NMSSM2}.  If $h_1$ is (nearly) pure iso--scalar and decouples, its
role is taken by the next Higgs particle with a large isodoublet component,
implying the validity of the mass bound again.  

The couplings of the ${\cal CP}$--even neutral Higgs boson $h_i$ to the $Z$
boson, $g_{ZZh_i}$, are defined relative to the usual SM coupling.  If  $h_1$
is primarily isosinglet, the coupling $g_{ZZh_1}$ is small and the particle
cannot be produced by Higgs-strahlung.  However, in this case $h_2$ is
generally light and couples with sufficient strength to the $Z$ boson; if not,
$h_3$ plays this role. Thus, despite the additional interactions, the
distinct pattern of the minimal extension remains valid also in this SUSY
scenario~\cite{NMSSM3}. 

In more general SUSY scenarios, one can add an arbitrary number of Higgs doublet
and/or singlet fields without being in conflict with high precision data. The
Higgs spectrum becomes then much more complicated than in the MSSM, and much
less constrained. However, the triviality argument always imposes a bound on
the mass of the lightest Higgs boson of the theory as in the case of the 
(M+1)SSM. In the most general SUSY model, with arbitrary
matter content and gauge coupling unification near the GUT scale, an absolute
upper limit on the mass of the lightest Higgs boson, $M_h \lsim 200$\GeV, has
been recently derived~\cite{NMSSM4}. 

Even if the Higgs sector is extremely complicated, there is always a light
Higgs boson which has sizeable couplings to the $Z$ boson. This Higgs particle
can thus be produced in the Higgs-strahlung process, 
$e^+ e^-  \rightarrow Z+$``$h^0$",
and using the missing mass technique this ``$h^0$" particle can be detected
independently of its decay modes [which might be rather different from those of
the SM Higgs boson]. Recently a powerful ``no lose theorem" has been 
derived~\cite{NMSSM5}:  a Higgs boson in SUSY theories can always be detected 
at a 500\GeV\  $e^+ e^-$ collider with a luminosity of $\int {\cal L} \sim 500$\,fb$^{-1}$
in the Higgs-strahlung process, regardless of its 
decays and of the complexity of the Higgs sector of the theory.  

To summarise: Experiments at  $e^+ e^-$ colliders are in a no--lose 
situation~\cite{NMSSM3,NMSSM5} for detecting the Higgs particles in general SUSY theories
for energies $\sqrt{s} \sim 500 $\GeV, if integrated luminosities 
$\int {\cal L} \sim {\cal O}(100$\,fb$^{-1})$ are available. 

\section{Study of the Higgs Boson Profile \label{sec:2.2}}

\subsection{Mass measurement \label{sec:2.2.1}}

Since the SM Higgs boson mass $M_H$ is a fundamental parameter of the theory,
the measurement is a very important task. Once $M_H$
is fixed, the profile of the Higgs particle is uniquely determined in the SM. 
In theories with extra Higgs doublets, the measurement of the masses of
the physical boson states is crucial to predict their production and 
decay properties as a function of the remaining model parameters and thus
perform a stringent test of the theory.

At the linear collider, $M_H$ can be measured best by exploiting the 
kinematical characteristics of the Higgs-strahlung production process
$e^+e^- \rightarrow Z^* \rightarrow H^0 Z$, where the $Z$ boson can be 
reconstructed in both its hadronic and leptonic decay modes~\cite{hmass1}.

For the case of SM-like couplings, a neutral Higgs boson with mass 
$M_H \le$ 130\GeV\  decays predominantly to $b \bar b$. Thus, $H^0Z$
production gives four jet $b \bar b q \bar q$ and two jet plus two lepton 
$b \bar b \ell^+ \ell^-$ final states. 

\begin{figure}[ht!]
\begin{center}
\begin{tabular}{c c}
\href{pictures/1/fig2201a.pdf}{{\epsfig{file=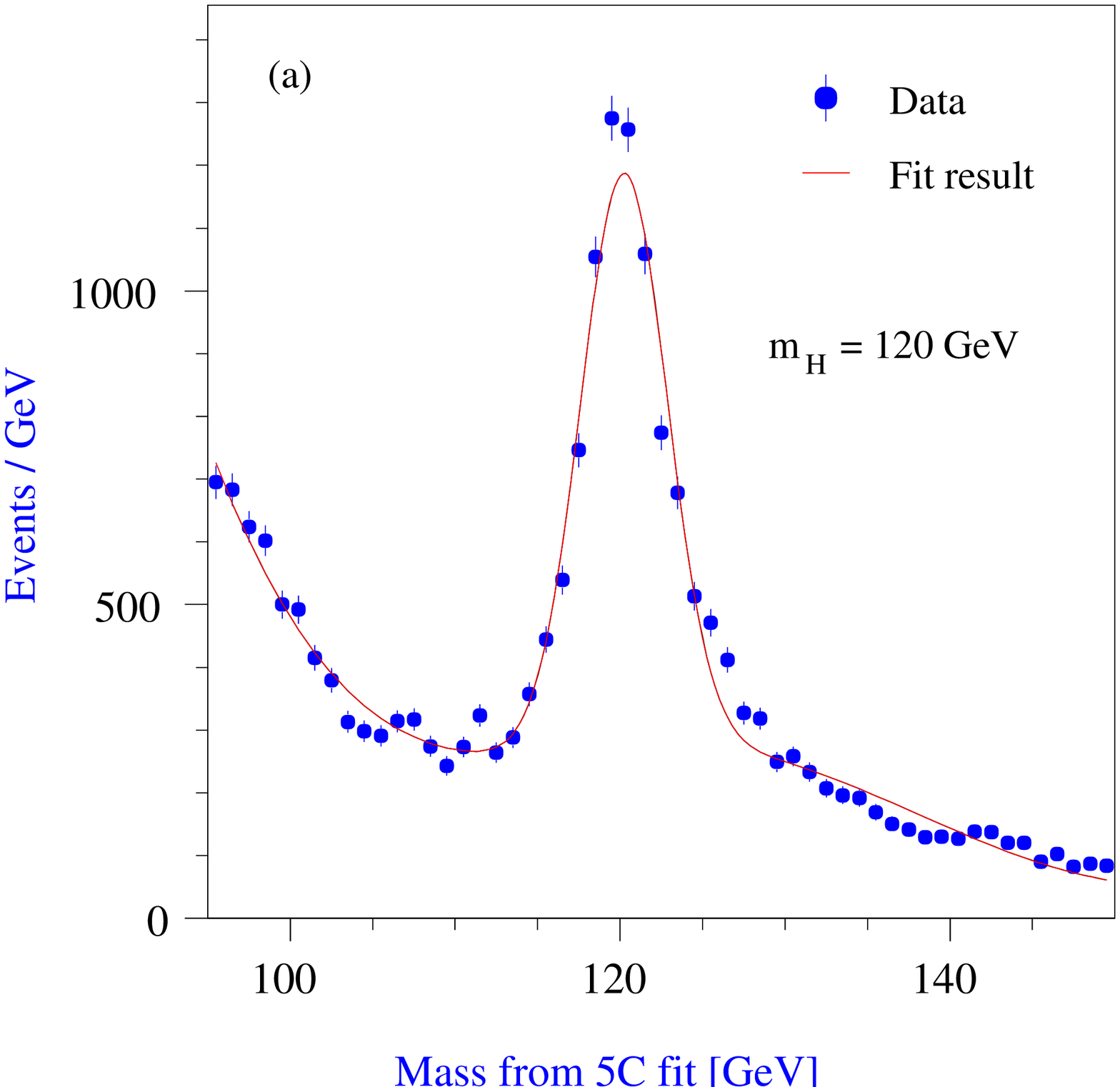,width=0.45\linewidth}}} &
\href{pictures/1/fig2201b.pdf}{{\epsfig{file=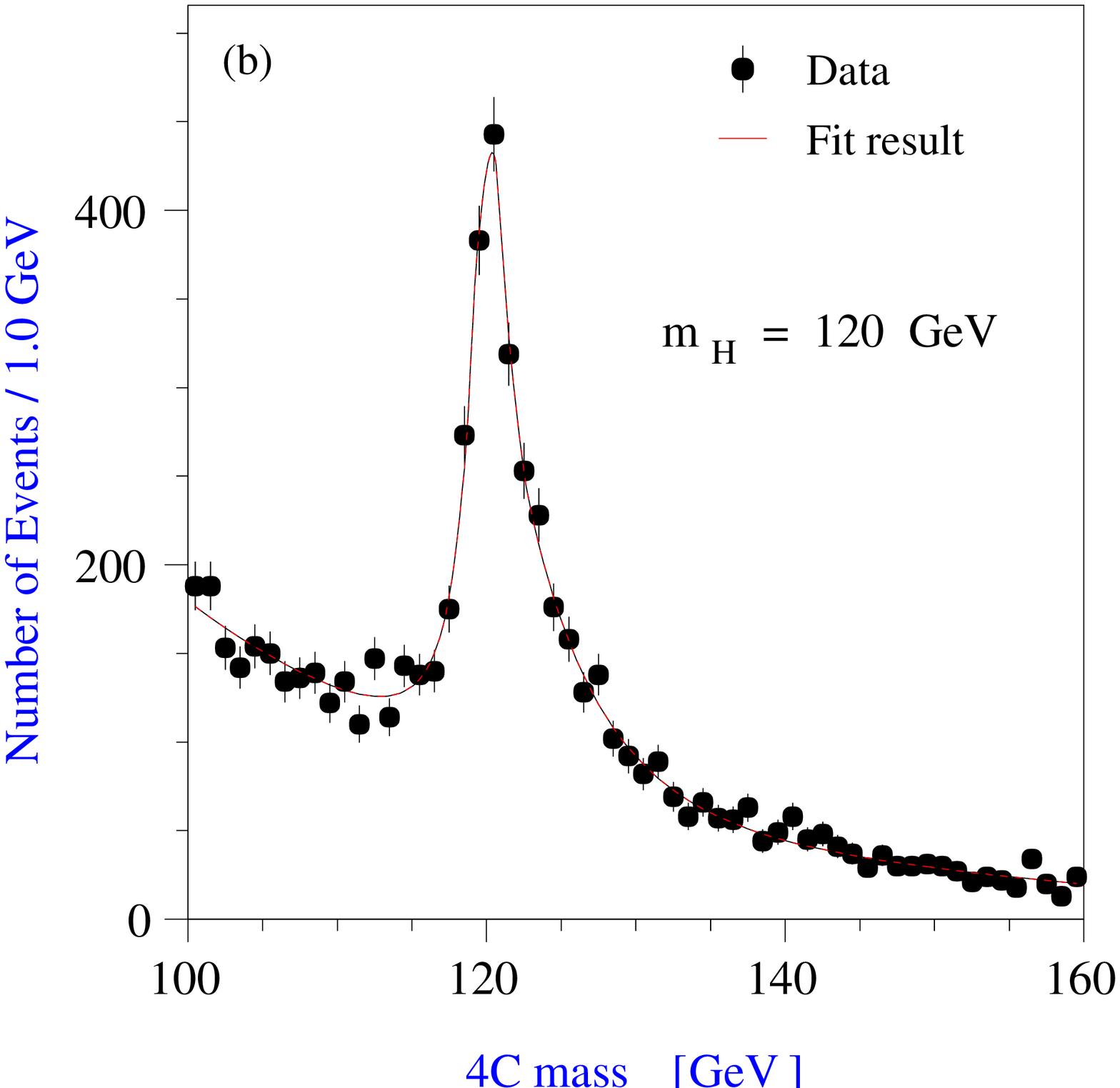,width=0.45\linewidth}}} \\
\href{pictures/1/fig2201c.pdf}{{\epsfig{file=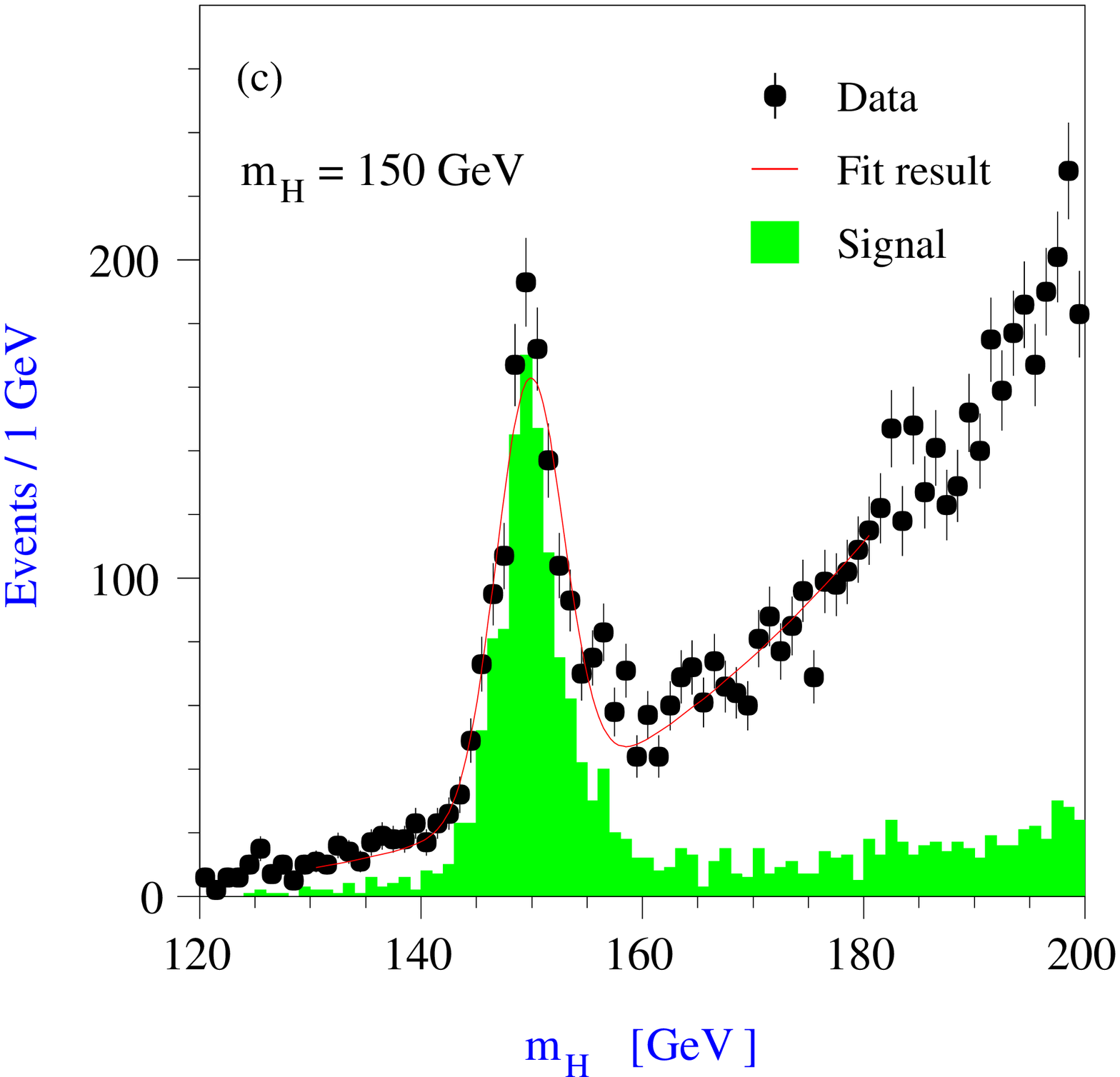,width=0.45\linewidth}}} &
\href{pictures/1/fig2201d.pdf}{{\epsfig{file=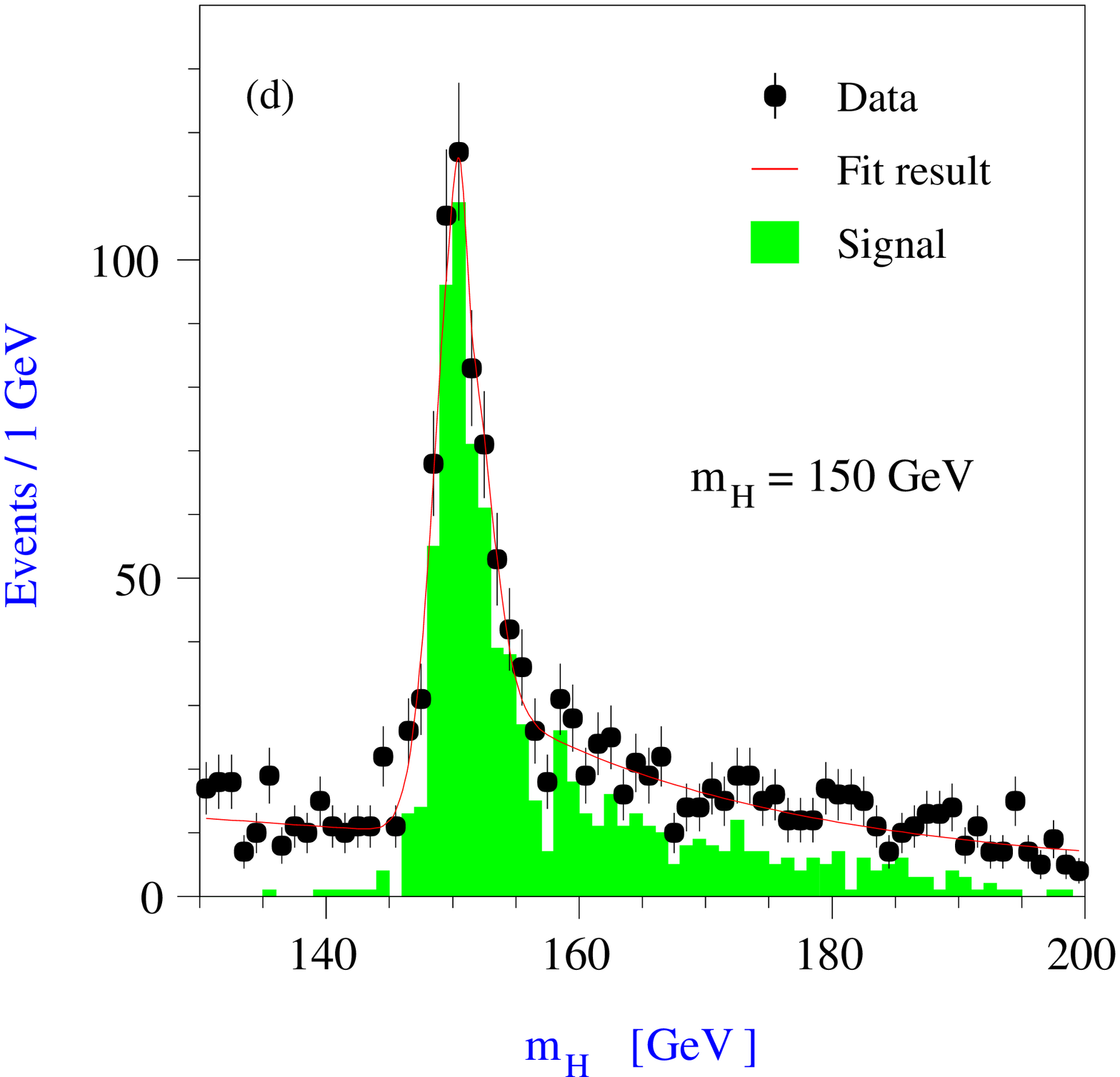,width=0.45\linewidth}}} \\
\end{tabular}
\caption[]{\label{fig:hmass}  The Higgs boson mass peak reconstructed in 
different channels with constrained fits for two values of $M_H$.
(a): $H^0Z \rightarrow b \bar b q \bar q$ at $M_H$ = 
120\GeV;
(b): $H^0Z \rightarrow q \bar q \ell^+ \ell^-$ at $M_H$ =
120\GeV;
(c): $H^0Z \rightarrow W^+W^- q \bar q$ at $M_H$ = 
150\GeV;
(d): $H^0Z \rightarrow W^+W^- \ell^+ \ell^-$ at $M_H$ = 150\GeV. 
The figures are for an integrated luminosity of 500\,fb$^{-1}$ at
$\sqrt{s} = $~350\GeV.}
\end{center}
\end{figure}

In the four--jet channel, the Higgs 
boson is reconstructed through its decay to $b \bar b$ with the $Z$ boson
decaying into a $q \bar q$ pair. 
The Higgs boson mass determination relies on a kinematical 5-C fit 
imposing energy and momentum conservation and 
requiring the mass of the jet pair 
closest to the $Z$ mass to correspond to $M_Z$.
This procedure gives a mass resolution of approximately 2\GeV\  
for individual events.
A fit to the resulting mass distribution, shown in
Fig.~\ref{fig:hmass}\,a), 
gives an expected accuracy of 50\,MeV~\cite{hmass2} for $M_H = 120
$\GeV\ 
and an integrated luminosity of 500\,fb$^{-1}$ at $\sqrt{s} = 350 $\GeV.

The leptonic $Z$ decays $Z \rightarrow e^+e^-$ and $Z \rightarrow 
\mu^+ \mu^-$ offer a clean signature in the detector, and the lepton momenta 
can be measured with high accuracy in the large tracking volume of the TESLA 
detector. In the case of $Z \to e^+e^-$ backgrounds are larger than in
the $Z\to \mu^+\mu^-$ channel due to large cross section for 
Bhabha scattering. Bhabha events with double ISR can be efficiently 
suppressed using a likelihood technique~\cite{hmass3}.
In order to further improve the resolution of the recoil mass, a 
vertex constraint is applied in reconstructing the lepton trajectories. 
Signal selection efficiencies in excess of 50\% are achieved for both the 
electron and the muon channels, with a recoil mass resolution of 
1.5\GeV\  for single events. 
The recoil mass spectrum is fitted with the Higgs boson mass, the mass 
resolution and the signal fraction as free parameters. The shape of the 
signal is parametrised using a high statistics simulated $H^0Z$ sample 
including initial state radiation and beamstrahlung effects while the 
background shape is fitted by an exponential. The shape of the luminosity 
spectrum can be directly measured, with high accuracy, using Bhabha events. 
The estimated precision on $M_H$ is 110\,MeV for a luminosity of 
500\,fb$^{-1}$ at $\sqrt{s}$ = 350\GeV, without any requirement on the nature of
the Higgs boson decays. By requiring the Higgs boson to decay hadronically and 
imposing a 4-C fit, the precision can be improved to 70\,MeV~\cite{hmass2}
(see Fig.~\ref{fig:hmass}\,b)).

As $M_H$ increases above 130\GeV, the $W W^*$ channel 
becomes more important and eventually dominates for masses from
150\GeV\  
up to the $ZZ$ threshold. In this region, the Higgs boson decay can be fully 
reconstructed by selecting hadronic $W$ decays leading to 
six jet (Fig.~\ref{fig:hmass}\,c)) and four jet plus two leptons 
(Fig.~\ref{fig:hmass}\,d)) final states~\cite{hmass2}. 

The recoil mass technique, insensitive to the actual Higgs boson decay 
channel, is also exploited and provides a comparable mass determination 
accuracy, the smaller statistics being compensated by the better mass 
resolution.

Table~\ref{tab:mass} summarises the expected accuracies on the Higgs boson 
mass determination. 
If the Higgs boson decays predominantly into invisible final states,
as predicted by some models mentioned earlier, but its total width remains
close to that predicted by the SM, the recoil mass technique is still 
applicable and determines the achievable accuracy on the mass determination.
\begin{table}[ht!]
\begin{center}
\begin{tabular}{|c|c|c|}
\hline
$M_H$  & Channel & $\delta M_H$ \\
(GeV) &  & (MeV) \\ \hline \hline
 120              & $\ell \ell q q$            & ~$\pm$70 \\
 120              & $qqbb$                     & ~$\pm$50 \\ \hline
 120              & Combined                   & ~$\pm$40 \\ \hline \hline
 150              & $\ell \ell$ Recoil         & ~$\pm$90 \\
 150              & $qqWW$                     & $\pm$130 \\ \hline
 150              & Combined                   & ~$\pm$70 \\ \hline \hline
 180              & $\ell \ell$ Recoil         & $\pm$100 \\
 180              & $qqWW$                     & $\pm$150 \\ \hline
 180              & Combined                   & ~$\pm$80 \\ \hline
\end{tabular}
\end{center}
\caption{ \label{tab:mass}
Summary of Higgs boson mass determination accuracies for 
500\,fb$^{-1}$ at $\sqrt{s} = $ 350\GeV.}
\end{table}

\subsection{Couplings to massive gauge bosons \label{sec:2.2.2}}

The couplings of the Higgs boson to massive gauge bosons is probed best in the
measurement of the production cross--section for Higgs-strahlung 
($e^+e^- \rightarrow Z^* \rightarrow H^0 Z$ probing $g_{HZZ}$) 
and $WW$ fusion ($e^+e^- \rightarrow H^0 \nu_e \bar \nu_e$ probing $g_{HWW}$).
%with cross--section $\sigma_{HZ} \propto 1/s$, $WW$ fusion  with cross--section
%$\sigma_{H^0 \nu \bar \nu} \propto \log(s/M^2_H)$ and  $ZZ$ fusion 
%($e^+e^- \rightarrow Z^*Z^* e^+ e^- \rightarrow H^0 e^+ e^-$) with 
%cross--section $\sigma_{Hee}$ suppressed by an order of magnitude compared to
%that for $WW$ fusion. These production cross--sections provide information on 
%the Higgs couplings $g_{HZZ}$ and $g_{HWW}$. 
The measurement of these cross--sections is also needed to extract the 
Higgs boson branching ratios from the observed decay rates and 
provide a determination of the Higgs boson total width when matched with the 
$H^0 \rightarrow WW^*$ branching ratio as discussed later. 

The cross--section for the Higgs-strahlung process can be measured by analysing 
the mass spectrum of the system recoiling against the $Z$ boson as already 
discussed in Section~\ref{sec:2.2.1}. This method provides a cross--section 
determination independent of the Higgs boson decay modes. From the 
number of signal events fitted to the di-lepton recoil mass spectrum, the 
Higgs-strahlung cross--section is obtained with a statistical accuracy of 
$\pm$ 2.8\%, combining the $e^+e^-$ and $\mu^+\mu^-$ channels. The systematics
are estimated to be $\pm$ 2.5\%,  mostly due to the uncertainties on the 
selection efficiencies and on the luminosity spectrum~\cite{hmass1}.
The results are summarised in Table~\ref{table:hxsec}.

\begin{table*}[ht!]
\begin{center}
\begin{tabular}{|c|c|c|}
\hline
$M_H$ & Fit $\sigma_{H^0 Z \rightarrow H^0 \ell^+ \ell^-}$ & 
$\delta\sigma/\sigma$ 
(stat) \\
(GeV) & (fb) & \\ \hline \hline
120 & 5.30$\pm$0.13(stat)$\pm$0.12(syst) & $\pm$0.025 \\
140 & 4.39$\pm$ 0.12(stat)$\pm$0.10(syst) & $\pm$0.027 \\
160 & 3.60$\pm$ 0.11(stat)$\pm$0.08(syst) & $\pm$0.030 \\ \hline
\end{tabular}
%\vspace{0.1cm}
\caption{ \label{table:hxsec}
The fitted Higgs-strahlung cross--sections for different 
values of $M_H$ with
500\,fb$^{-1}$ at $\sqrt{s}$ = 350\GeV. The first error is statistical and 
the second due to systematics. The third column gives the relative statistical
accuracy.} 
\end{center}
\vspace{-0.2cm}
\end{table*}

\begin{figure}[ht!]
\begin{center}
\begin{tabular}{cc}
\href{pictures/1/fig2202a.pdf}{{\epsfig{file=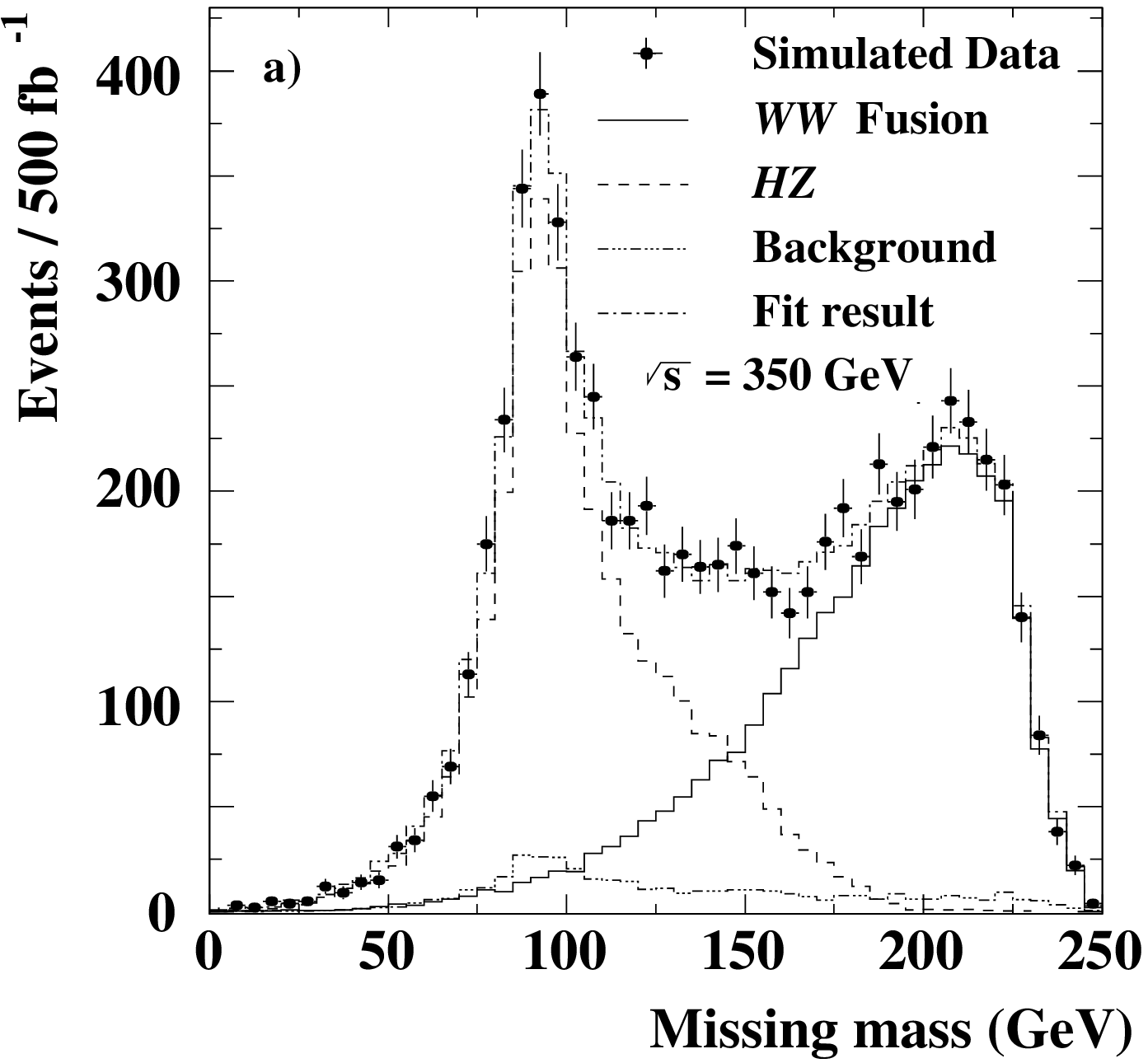,height=0.51\linewidth,width=0.48\linewidth}}} &
\href{pictures/1/fig2202b.pdf}{{\epsfig{file=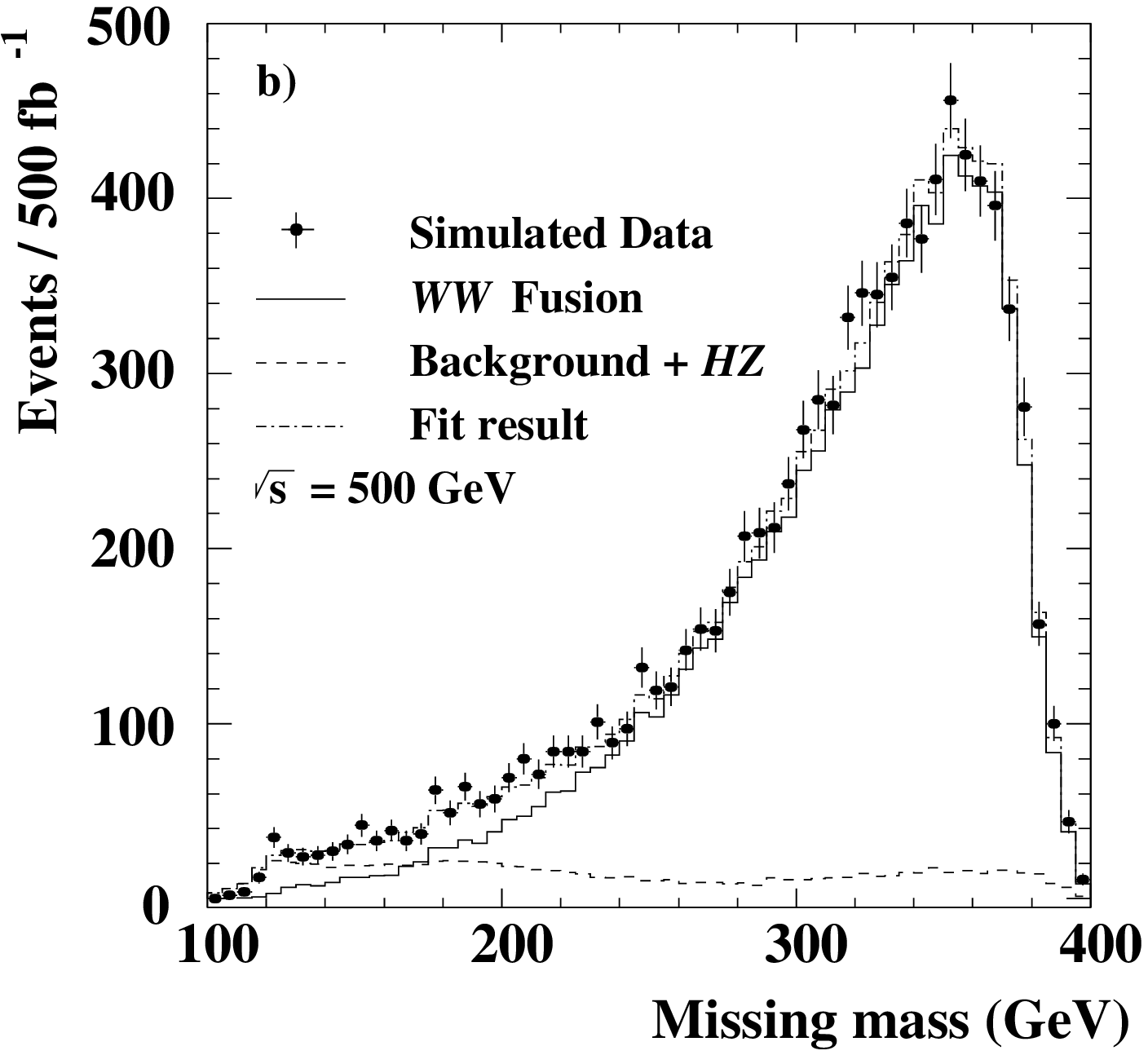,height=0.51\linewidth,width=0.48\linewidth}}} \\
\end{tabular}
\caption{ \label{fig:wwfusion}
Simulation of the missing mass distribution in
$b\bar{b}\nu\bar{\nu}$ events for 500~fb$^{-1}$ at $\sqrt{s} = 350 $\GeV\; (a)
and 500\GeV\; (b). The contributions from $WW$~fusion, Higgs-strahlung
and background can be disentangled using a fit to the shape of their
distributions. }
\end{center}
\end{figure}

The cross--section for $WW$~fusion can be determined in the 
$b\bar{b}\nu\bar{\nu}$ final state, where these events can be well separated 
from the corresponding Higgs-strahlung final state, 
$H^0Z \to b \bar b \nu \bar \nu$, and the background processes 
by exploiting their 
different spectra for the $\nu\bar{\nu}$ invariant mass 
(Fig.~\ref{fig:wwfusion}). There could be serious contamination of $H^0 \nu 
\bar \nu$ events from overlapping $\gamma \gamma \rightarrow 
{\mathrm{hadrons}}$ events, but the good spatial resolution of the vertex
detector will make it possible to resolve the longitudinal displacement of the
two separate event vertices, within the TESLA bunch length~\cite{gghad}. 
The precision to which the cross--section for $WW$~fusion can be measured with 
500 $\mathrm{fb}^{-1}$ at $\sqrt{s} = 500$ \GeV\  is given in 
Table~\ref{tab:sigsummary}~\cite{xs-hww}. Further, by properly choosing 
the beam polarisation configurations, the relative contribution of
Higgs-strahlung and $WW$ fusion can be varied and systematics arising from the 
contributions to the fitted spectrum from the two processes and their effect 
can be kept smaller than the statistical accuracy~\cite{polarisation}.

An accurate determination of the branching ratio for the decay 
$H^0/h^0 \rightarrow W W^*$ can be obtained in the Higgs-strahlung process by
analysing semi-leptonic~\cite{borisov} and fully hadronic~\cite{hjs} $W$ 
decays. The large $W^+W^-$ and
$t \bar t$ backgrounds can be significantly reduced by imposing the 
compatibility of the two hadronic jets with the $Z$ mass and that of their 
recoil system with the Higgs boson mass. Further background suppression is 
ensured by an anti-$b$ tag requirement that rejects the remaining $ZZ$ 
and $t \bar t$ events. The residual $WW^*$ background with one off-shell $W$ 
can be further suppressed if the electron beam has right-handed 
polarisation.
\begin{table}[ht!]
\begin{center}
\begin{tabular}{|c|c|c|c|}
\hline
Channel & $M_H$ = 120\GeV & 140\GeV & 160\GeV \\
\hline \hline
$\sigma (e^+e^- \rightarrow H^0Z)$       & $\pm$ 0.025 & $\pm$ 0.027 &
$\pm$ 0.030 \\        
$\sigma (e^+e^- \rightarrow WW \rightarrow H^0 \nu \bar \nu)$ & $\pm$ 0.028 &
$\pm$ 0.037 &  $\pm$ 0.130 \\ \hline
$H^0  \rightarrow W W^*$        & $\pm$ 0.051 & $\pm$ 0.025 & 
$\pm$ 0.021 \\
$H^0 \rightarrow Z Z^*$      & \mbox{ }    &    & $\pm$ 0.169\\
\hline
\end{tabular}
\caption{ Relative accuracy in the determination of the SM Higgs boson 
production cross--sections and decay rates into gauge bosons for 500\,fb$^{-1}$
at $\sqrt{s} = $~350\GeV\  and 500\GeV.
\label{tab:sigsummary}}
\end{center}
\end{table}

\subsection{Coupling to photons \label{sec:2.2.3}}

The Higgs effective coupling to photons is mediated by loops. These are
dominated, in the SM, by the contributions from the $W$ boson and the
top quark but are also 
sensitive to any charged particles coupling directly to the Higgs 
particle and to the photon. 

\begin{figure}[ht!]
\begin{picture}(150,20)\unitlength 1mm
  \put(20,-6){a)}
  \put(98,-6){b)}
\end{picture}
\vspace{-1.5cm}
\begin{center}
\begin{tabular}{c c}
\href{pictures/1/fig2203a.pdf}{{\epsfig{file=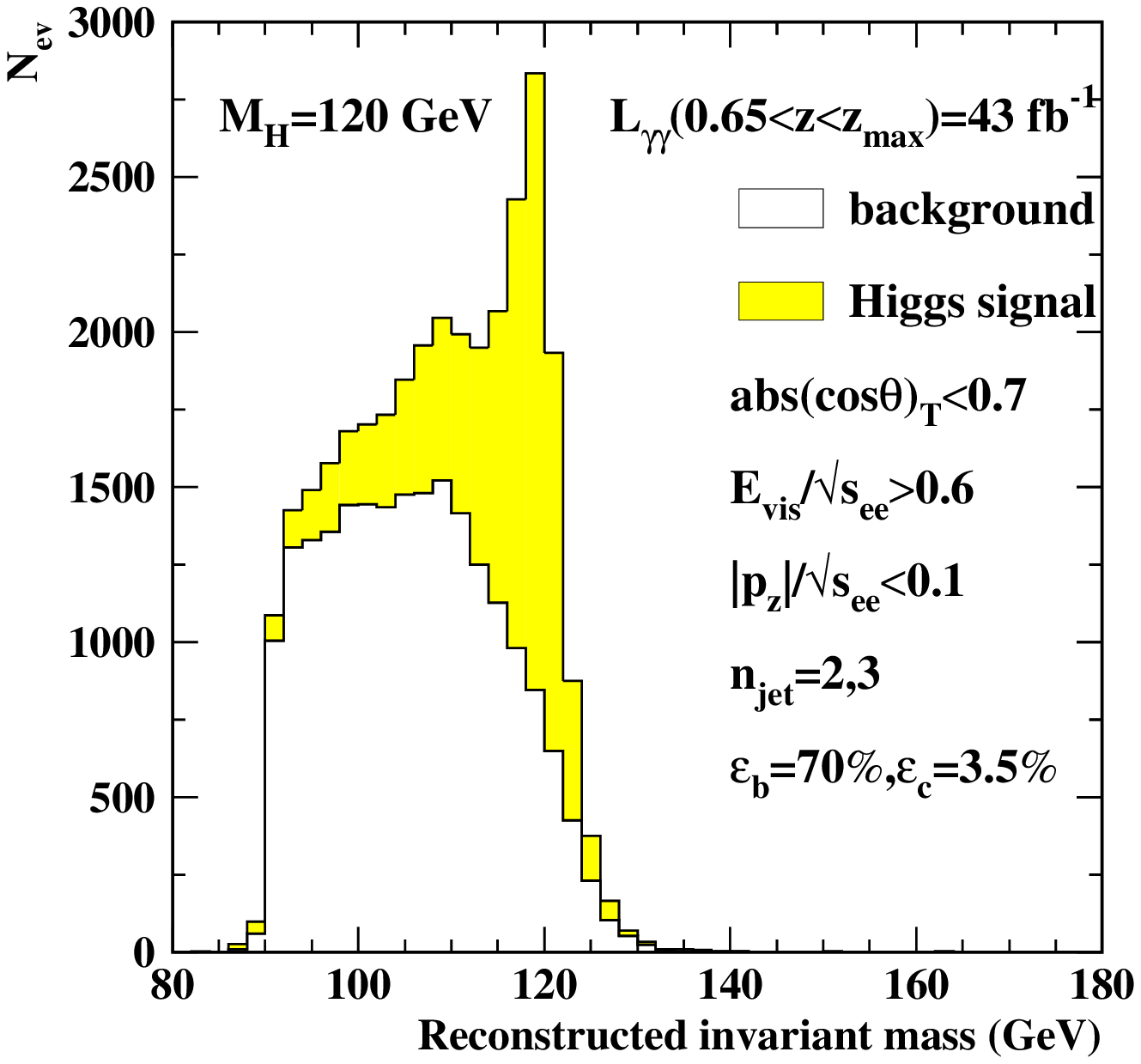,height=0.45\linewidth,width=0.45\linewidth,clip}}} &
\href{pictures/1/fig2203b.pdf}{{\epsfig{file=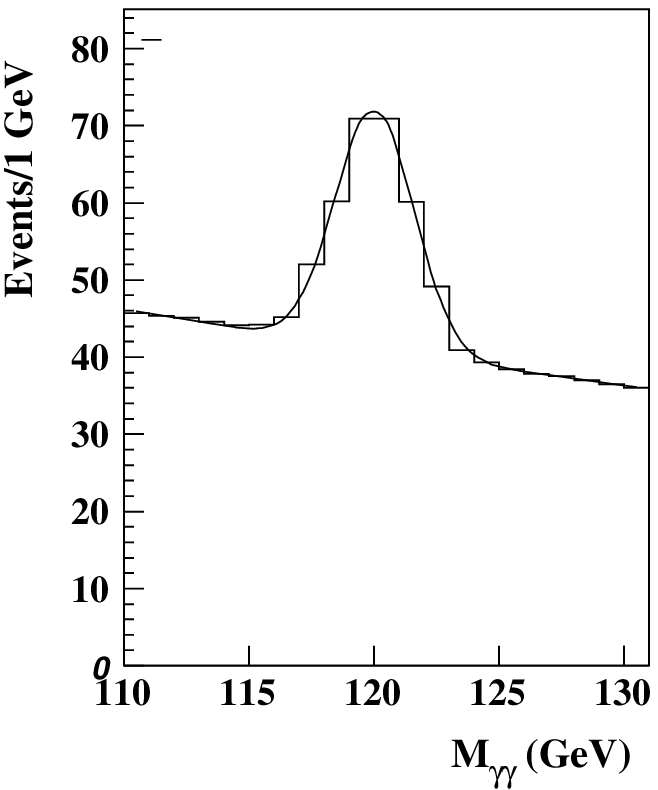,height=0.45\linewidth,width=0.45\linewidth,clip,bb=0
0 187 230}}} \\
\end{tabular}
\caption{\label{fig:hgamma}
(a): The Higgs signal reconstructed at the $\gamma \gamma$ 
collider for $M_H$ = 120\GeV\  with 43\,fb$^{-1}$ $\gamma\gamma$ 
luminosity in the hard part of the spectrum. 
(b): The signal for $e^+e^- \rightarrow \nu\bar{\nu}H^0 
\rightarrow \nu \bar{\nu} \gamma \gamma$ for 
$M_H$ = 120\GeV\  at $\sqrt{s}$ = 500\GeV\  and an integrated 
luminosity of 1000\,fb$^{-1}$.} 
\end{center}
\end{figure}

At the $\gamma \gamma$ collider, the process $\gamma \gamma \to H$ has a very 
substantial cross--section. The observation of
the Higgs signal through its subsequent decay $H^0 \to b \bar b$ requires an 
effective suppression of the large non--resonant 
$\gamma \gamma \to c \bar c$ and $\gamma \gamma \to b \bar b$ backgrounds.
Profiting from the effective $b$/$c$ jet flavour discrimination of the TESLA 
detector, it is possible to extract the Higgs signal with good background 
rejection (see Fig.~\ref{fig:hgamma}\,a)). Assuming $M_H$ = 120\GeV\  and an
integrated $\gamma\gamma$ luminosity of 43\,fb$^{-1}$ in the hard part
of the spectrum,
an accuracy of about 2\% on
$\sigma(\gamma \gamma \to H)$ can be achieved~\cite{hgamma, melles}
(see Part VI, Chapter 1.).

The Higgs coupling to photons is also accessible through the $H^0 \to \gamma 
\gamma$ decay. The measurement of its branching ratio
together with the production cross--section at the TESLA 
$\gamma \gamma$~collider is important 
for the extraction of the Higgs boson width.
The branching ratio analysis is performed 
in both the $e^+e^-\to\gamma\gamma\nu\bar{\nu}$ 
and the $e^+e^-\to\gamma\gamma$ +~jets final states, 
corresponding to the sum of
the $WW$ fusion, $ZH^0\to \nu \bar{\nu} H^0$, and $ZH^0\to q\bar{q}H^0$,
respectively~\cite{reid1}. The most important background in both channels 
comes from the double-bremsstrahlung $Z \gamma\gamma$ process. This background
and the smallness of the $H^0 \to \gamma \gamma$ partial decay width make the
analysis a considerable experimental challenge. However the signal can be 
discriminated from this irreducible background, since the photons in the signal
have a spectrum peaked at high energy and rather isotropic production contrary
to the background process which has photons produced at large polar angles and
with lower energies. Efficiency values in the range  50\% to 65\% are obtained
for the $\nu \bar \nu \gamma \gamma$ and $q \bar q \gamma \gamma$ final 
states. Combining both channels, the relative accuracy for the measurement
of ${\mathrm{BR}}(H^0 \to \gamma \gamma)$ for $M_H = 120 $\GeV\  is 26\% (23\%), 
for an integrated luminosity of 500\,fb$^{-1}$ at 
$\sqrt{s}$ = 350\GeV\  (500\GeV). For 1000\,fb$^{-1}$, an accuracy of
18\% (16\%) can be reached (see Fig.~\ref{fig:hgamma}\,b)).

\subsection{The Higgs boson total decay width \label{sec:2.2.4}}

The SM Higgs boson total width, $\Gamma_H$, is extremely small 
for light mass values and 
increases rapidly once the $WW^*$ and $ZZ^*$ decay channels become accessible, 
reaching a value of 1\GeV\  at the $ZZ$ threshold. Therefore, for 
$M_H \gsim $~200\GeV\ 
the total decay width becomes directly accessible from the reconstruction of 
the Higgs boson line-shape. However at the linear collider and for lower
masses, it can be obtained semi--directly, in a nearly model--independent way,
from the combination of the 
measurements of a Higgs coupling constant with the corresponding 
branching ratio.

Absolute measurements of coupling constants can be obtained
(i) for $g_{\mathrm{HZZ}}$ through the Higgs-strahlung cross--section, for 
$g_{\mathrm{HWW}}$ through (ii) the $WW$~fusion cross--section or, more
model-dependently, (iii) by using the symmetry
 $g^2_{\mathrm{HWW}}/g^2_{\mathrm{HZZ}} 
= \cos^2{\theta_W}$ and, in the  $\gamma \gamma$ collider 
option, for 
$g_{\mathrm{H}\gamma\gamma}^{\mathrm{effective}}$ through (iv) the 
cross--section for $\gamma\gamma\rightarrow H^0$.

For a mass below 160\GeV, the best method is to use the $WW$ fusion 
process. Combined with the measurement of the branching ratio for 
$\mathrm{H} \rightarrow \mathrm{WW}^*$ (see section \ref{sec:2.2.2}) an 
accuracy ranging from 4\% to 13\% can be obtained for $\Gamma_H$,
as shown in Table~\ref{tab:width}.

\begin{table}[ht!]
\begin{center}
\begin{tabular}{|c|c|c|c|c|}
\hline
$\Gamma_{H \rightarrow X}$ & BR($H \rightarrow X$) & $M_H$ = 120\GeV & 
140\GeV & 160\GeV \\
\hline \hline
 $WW = WW \nu \nu$ & $H^0 \rightarrow WW$ & $\pm$0.061  & $\pm$0.045 
& $\pm$0.134 \\
 $WW = HZ$       & $H^0 \rightarrow WW$ & $\pm$0.056  & $\pm$0.037 
& $\pm$0.036 \\
 $\gamma \gamma \rightarrow H^0$ & $H^0 \rightarrow \gamma \gamma$ & $\pm$0.23 
& -  & - \\
\hline
\end{tabular}
\end{center}
\caption{ \label{tab:width} 
Relative accuracy on the determination of the total Higgs boson 
decay width $\Gamma_H$ for 500\,fb$^{-1}$ using the three methods described 
in the text.}
\end{table}

An alternative method is to exploit the effective 
$\mathrm{H}\gamma\gamma$ coupling through the measurement of the
cross--section for $\gamma\gamma\rightarrow\mathrm{H}\rightarrow
\mathrm{b} \bar{\mathrm{b}}$ using the $\gamma \gamma$ collider option.
This cross--section and hence the partial width $\Gamma_{\gamma\gamma}$ can be 
obtained to 2\% accuracy for $m_\mathrm{H} \lsim 140\GeV$\  and to better 
than 10\% for $m_\mathrm{H} \lsim 160\GeV$. The derivation of 
the total width however needs the measurement of the branching ratio 
$\mathrm{H}\rightarrow\gamma\gamma$ as input. As it was shown in 
Sec.~\ref{sec:2.2.3}, this can only be achieved to 23\% precision
for 500\,fb$^{-1}$ and thus 
dominates the uncertainty on the total width reconstructed from the 
H$\gamma\gamma$ coupling.
% (unless more integrated luminosity becomes available).

\subsection{Couplings to fermions \label{sec:2.2.5}}

The accurate determination of the Higgs couplings to fermions is important
as a proof of the Higgs mechanism and to establish the nature of the Higgs 
boson. The Higgs-fermion couplings being proportional to the fermion masses, 
the SM Higgs boson branching ratio into fermions are 
fully determined once the Higgs boson and the fermion masses are fixed.

Deviations of these branching ratios from those predicted for the SM 
Higgs boson can be the signature of the lightest 
supersymmetric $h^0$ boson. Higgs boson 
decays to $gg$, like those to $\gamma \gamma$, proceed through loops,
dominated in this case by the top contribution. The measurements of these 
decays are sensitive to the top Yukawa coupling in the SM and the existence
of new heavy particles contributing to the loops.

The accuracy on the Higgs boson branching ratio measurements at the 
linear collider has been the subject of several studies~\cite{hildreth}. 
With the high resolution Vertex Tracker, the more advanced jet flavour tagging 
techniques, the experience gained at LEP and SLC
(see Part IV, Chapter~9), 
and the large statistics available at the 
TESLA collider, these studies move into the domain of precision 
measurements.

In the hadronic Higgs boson decay channels at TESLA, the fractions of 
$b \bar b$, $c \bar c$ and $g g$ final states are extracted by a binned 
maximum likelihood fit to the jet flavour tagging probabilities for the Higgs 
boson decay candidates~\cite{mba}. The background is estimated over a wide
interval around the Higgs boson mass peak and subtracted. It is also possible 
to study the flavour composition of this background directly in the real data 
by using the side-bands of the Higgs boson mass peak. 
The jet flavour tagging response can be checked by using low energy runs at
the $Z$ as well as $ZZ$ events at full energy, thus reducing
systematic uncertainties from the simulation.

For the case of $H^0/h^0 \rightarrow \tau^+ \tau^-$, a global $\tau \tau$ 
likelihood is defined by using the response of discriminant variables
such as charged multiplicity, jet invariant mass and track impact 
parameter significance. 
\begin{figure}[ht!]
\begin{center}
\href{pictures/1/fig2204.pdf}{{\epsfig{file=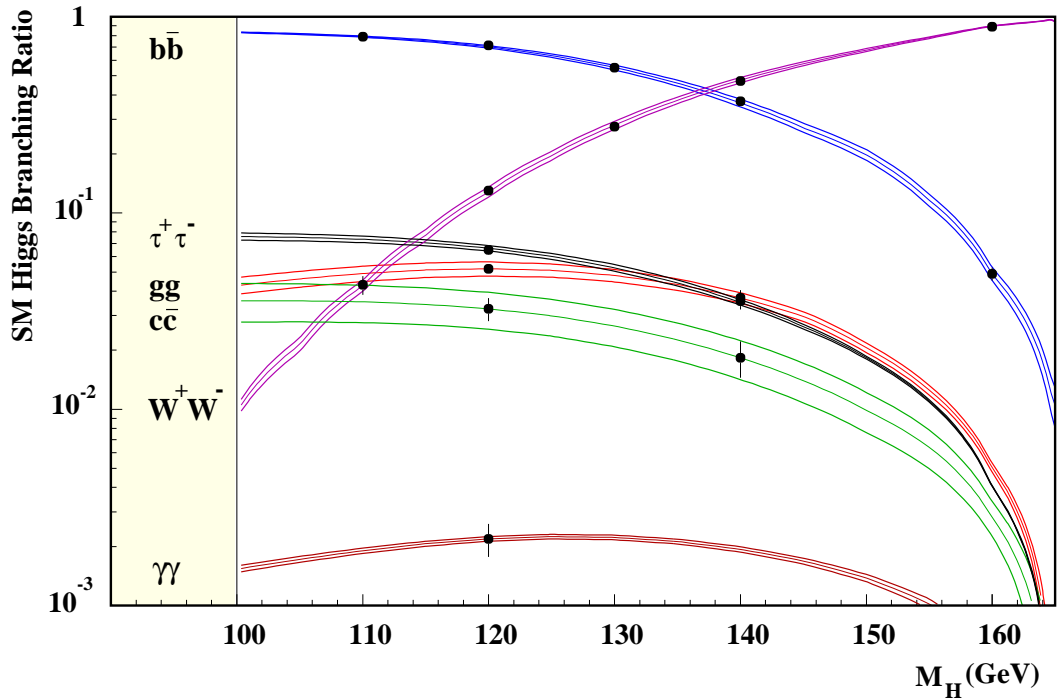,width=0.9\linewidth}}}
\caption{ The predicted SM Higgs boson branching ratios. 
Points with error bars show the expected experimental accuracy, while the
lines show the estimated uncertainties on the SM predictions.\label{fig:hbr}
} 
\end{center}
\end{figure}             
These measurements are sensitive to the product $\sigma_{H^0Z, H^0
 \nu \bar \nu}
\times {\mathrm{BR}} (H^0 \rightarrow f \bar f)$. Using the results discussed
above for the production cross--sections 
$\sigma_{H^0Z, H^0 \nu \bar \nu}$, the 
branching ratios can be determined to the accuracies summarised in 
Table~\ref{table:brsummary} and shown in Fig.~\ref{fig:hbr}~\cite{mba}.

\begin{table}[ht!]
\begin{center}
%\vspace{0.4cm}
\begin{tabular}{|c|c|c|c|}
\hline
Channel & $M_H$ = 120\GeV & $M_H$ = 140\GeV &
$M_H$ = 160\GeV \\
\hline \hline
$H^0 / h^0 \rightarrow b \bar b$        & $\pm$ 0.024 & $\pm$ 0.026 & 
$\pm$ 0.065\\
$H^0 / h^0 \rightarrow c \bar c$        & $\pm$ 0.083 & $\pm$ 0.190 &   \\
$H^0 / h^0 \rightarrow g g$             & $\pm$ 0.055 & $\pm$ 0.140 &   \\
$H^0 / h^0 \rightarrow \tau^+ \tau^-$   & $\pm$ 0.050 & $\pm$ 0.080 &   \\
\hline
\end{tabular}
\end{center}
\caption{ Relative accuracy in the determination of Higgs 
boson branching ratios for 500\,fb$^{-1}$ at $\sqrt{s}$ = 350\GeV.
\label{table:brsummary}}
\end{table}

\subsection{Higgs top Yukawa coupling \label{sec:2.2.6}}   

The Higgs Yukawa coupling to the top quark is the largest coupling in 
the SM ($g_{ttH}^2 \simeq 0.5$ to be compared with $g_{bbH}^2 \simeq 
4 \times 10^{-4}$). If $M_H < 2 m_{t}$ this coupling is directly accessible 
in the process $e^+e^- \rightarrow t \bar t H$~\cite{ttH}. This process, with 
a cross--section of the order of 0.5\,fb for $M_H \sim 120$\GeV\  at 
$\sqrt{s}$ = 500\GeV\  and 2.5\,fb at $\sqrt{s}$ = 800\GeV, including
QCD corrections~\cite{ttHth}, leads to a distinctive signature consisting 
of two $W$ bosons and four $b$-quark jets.
\begin{figure}[ht!]
\begin{center}
\href{pictures/1/fig2205.pdf}{{\epsfig{file=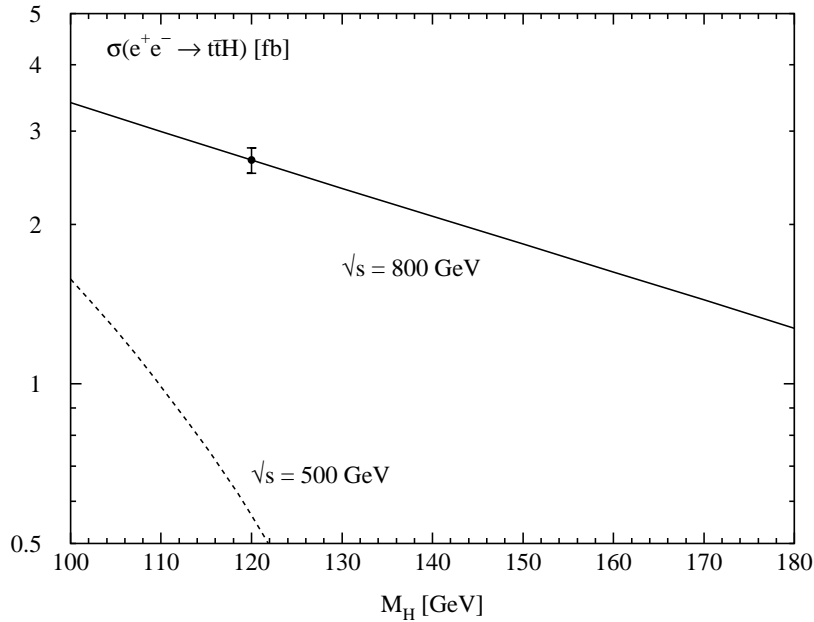,width=0.7\linewidth}}}
\caption{ \label{fig:tth}
The $t \bar t H^0$ cross--section, computed at next-to-leading 
order, as a function of  $M_H$ for $\sqrt{s}$ = 500\GeV\  and 
800\GeV\ with 
the expected experimental accuracy for $M_H$ = 120\GeV\  shown by the dot with
error bar for an integrated luminosity of 1000\,fb$^{-1}$. }
\end{center}
\end{figure} 

% The QCD corrections to this process have been calculated recently
% up to next-to-leading order~\cite{ttHth}
% and found to be large and positive at $\sqrt{s} \sim 500$\GeV because of 
% threshold effects, while they become small and negative 
% at $\sqrt{s} \sim 1$~TeV. The expected sensitivity reaches a plateau for
% $\sqrt{s} \geq 700$\GeV. % (see Figure~\ref{fig:tth}). 
% \begin{table}[htpb]
% \begin{center}
% \begin{tabular}{|l|c|c|c|c|c|}
% \hline
%  & $\epsilon\:(\%)$ & S/B & $\delta_{\rm {stat.}}$ (\%) & 
% $\delta_{\rm{syst.}}$ (\%) & $\delta_{\rm{tot.}}$  (\%)    \\
% \hline \hline
% Semileptonic & & & & & \\ \hline
%   {\rm After Preselection } & $54.1$ & $0.03$ & $11.5$ & $63.1$ & $64.1$ \\
%  {\rm Fit to NNSel (50 bins)} & $54.1$ & $0.03$ & $4.4$ & $9.1$ & $10.1$ \\
%  {\rm NNSel$>$0.9 (optimal cut)} & $27.1$ & $0.51$ & $5.1$ & $3.8$ & $6.3$ \\ 
%  \hline \hline 
% Hadronic & & & & & \\ \hline
%  {\rm After Preselection} & $77.1$ & $0.03$ & $9.8$ & $83.5$ & $83.5$ \\
%  {\rm Fit to NNSel (50 bins)} & $77.1$ & $0.03$ & $4.2$ & $13.7$ & $14.3$ \\
%  {\rm NNSel$>$0.95 (optimal cut)} & $8.5$ & $0.90$ & $7.3$ & $3.0$ & $7.9$ \\ 
% \hline
% \end{tabular}
% \end{center}
% caption{\label{sl_errors}\protect
% Statistical and systematic uncertainties in the Higgs top Yukawa coupling
% for $M_H$ = 120\GeV and an integrated luminosity of 1000\,fb$^{-1}$ at 
% $\sqrt{s}$ = 800\GeV.}
% \label{tab:tthres}
% \end{table}   

The experimental accuracy on the determination of
the top Yukawa coupling has been studied for $\sqrt{s}$ = 800\GeV\  and 
$L = 1000$\,fb$^{-1}$ in both the semileptonic and fully hadronic 
channels~\cite{juste}.
The main sources of efficiency loss are from failures of the jet-clustering 
and of the $b$-tagging due to hard gluon radiation and to large multiplicities.
The analysis uses a set of highly efficient pre-selection criteria and
a Neural Network trained to separate the signal from the remaining backgrounds.
Because of the large backgrounds, it is crucial that they are well 
modelled both in normalisation and event shapes. A conservative estimate
of 5\% uncertainty in the overall background normalisation has been
used in the evaluation of systematic uncertainties.
% which determine how 
% they are treated by the Neural Network.
% Table~\ref{tab:tthres} summarises the results of the 
% analysis assuming a 5\% systematic uncertainty in the knowledge of 
% the background cross--section after the event selection. 
%Figure~\ref{fig:nnout_sel} illustrates the importance of an accurate 
%knowledge of the signal-like background.
For an integrated luminosity of 1000\,fb$^{-1}$ the statistical uncertainty in 
the Higgs top Yukawa coupling after combining the semileptonic and the 
hadronic channels is $\pm$ 4.2\%~(stat). This results in an uncertainty
of 5.5\%~(stat.+syst.)~\cite{juste} (see Fig.~\ref{fig:tth}). 

% Anomalous top Yukawa couplings may be extracted from the angular
% distribution of $e^+e^-\to t\bar t\ H$ with the help of the
% optimal-observable method~\cite{GGH} discussed in the next section.

If $M_H > 2 m_t$, the Higgs top Yukawa couplings can be measured from the 
$H^0 \rightarrow t \bar t$ branching ratio, similarly to those of the other
fermions discussed in the previous section. A study has been performed for
the $WW$~fusion process $e^+e^- \rightarrow \nu_e \bar \nu_e H^0 
\rightarrow \nu_e \bar \nu_e t \bar t$ for 350\GeV\  $< M_H <$
500\GeV\ 
at $\sqrt{s}$ = 800\GeV~\cite{ttheavy}. 
The $e^+e^- \rightarrow t \bar t$ and the 
$e^+e^- \rightarrow e^+e^- t \bar t$ backgrounds are reduced by the event 
selection based on the characteristic event signature with six jets, two of 
them from a $b$ quark, on the missing energy and the mass.
Since the S/B ratio is expected to be large,
the uncertainty on the top Yukawa coupling is dominated by the statistics and 
corresponds to 5\% (12\%) for $M_H$ = 400 (500) \GeV\  for an integrated
luminosity of 1000\,fb$^{-1}$~\cite{ttheavy}.

\subsection{Extraction of Higgs couplings \label{sec:2.2.7}}

The Higgs boson production and decay rates discussed above, can be used to 
measure the Higgs couplings to gauge bosons and fermions. 
After the Higgs boson is discovered, this is the first crucial step in
establishing experimentally the Higgs mechanism for mass generation.
Since some of the couplings
of interest can be determined independently by different observables while 
other determinations are partially correlated, it is interesting to perform
a global fit to the measurable observables and to extract the Higgs couplings
in a model--independent way.
This method optimises the available information and can take properly
into account the experimental correlation between different measurements.

\begin{figure}[ht!]
\begin{center}
\begin{tabular}{c c}
\href{pictures/1/fig2206a.pdf}{{\epsfig{file=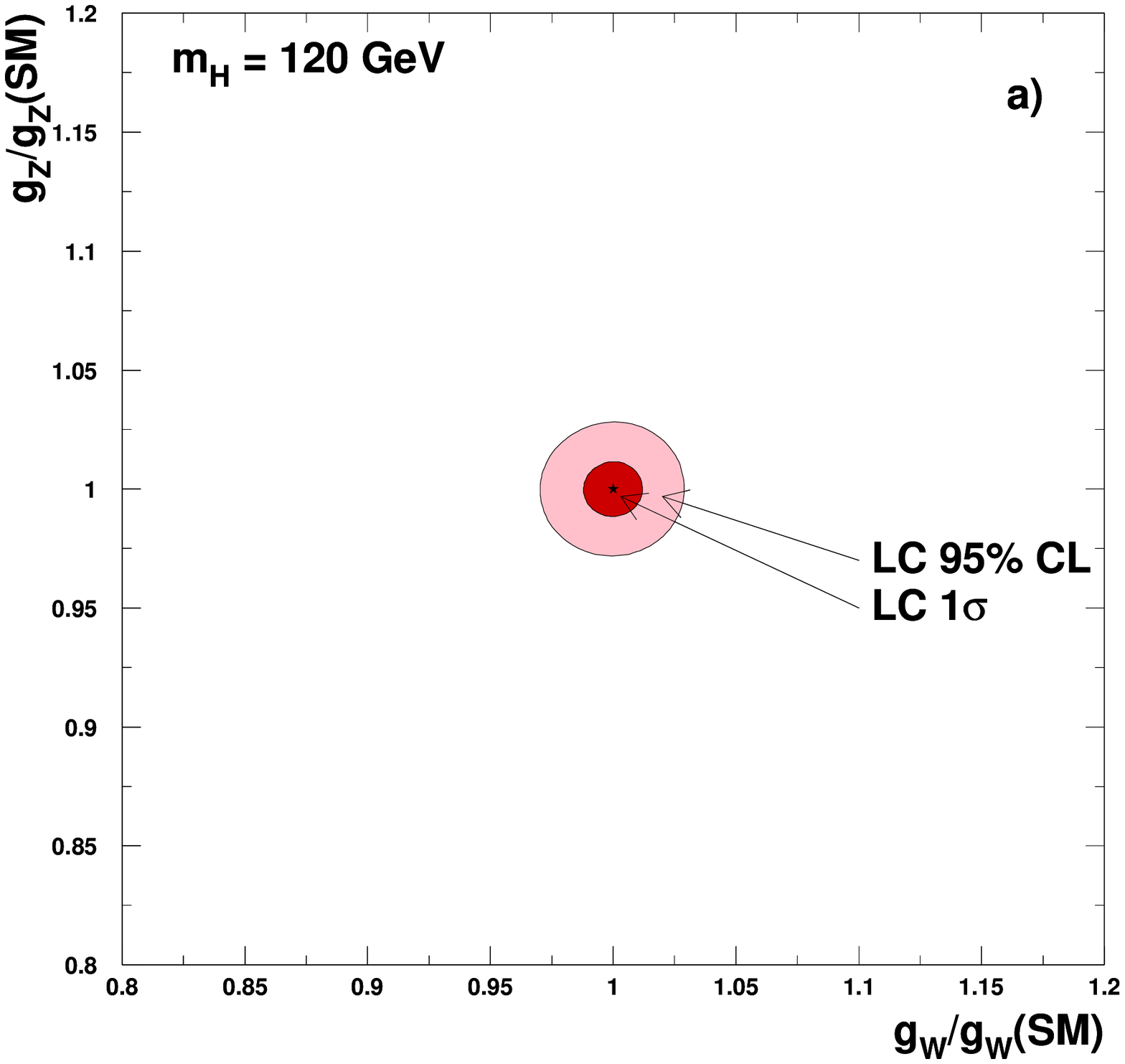,width=0.45\linewidth}}} &
\href{pictures/1/fig2206b.pdf}{{\epsfig{file=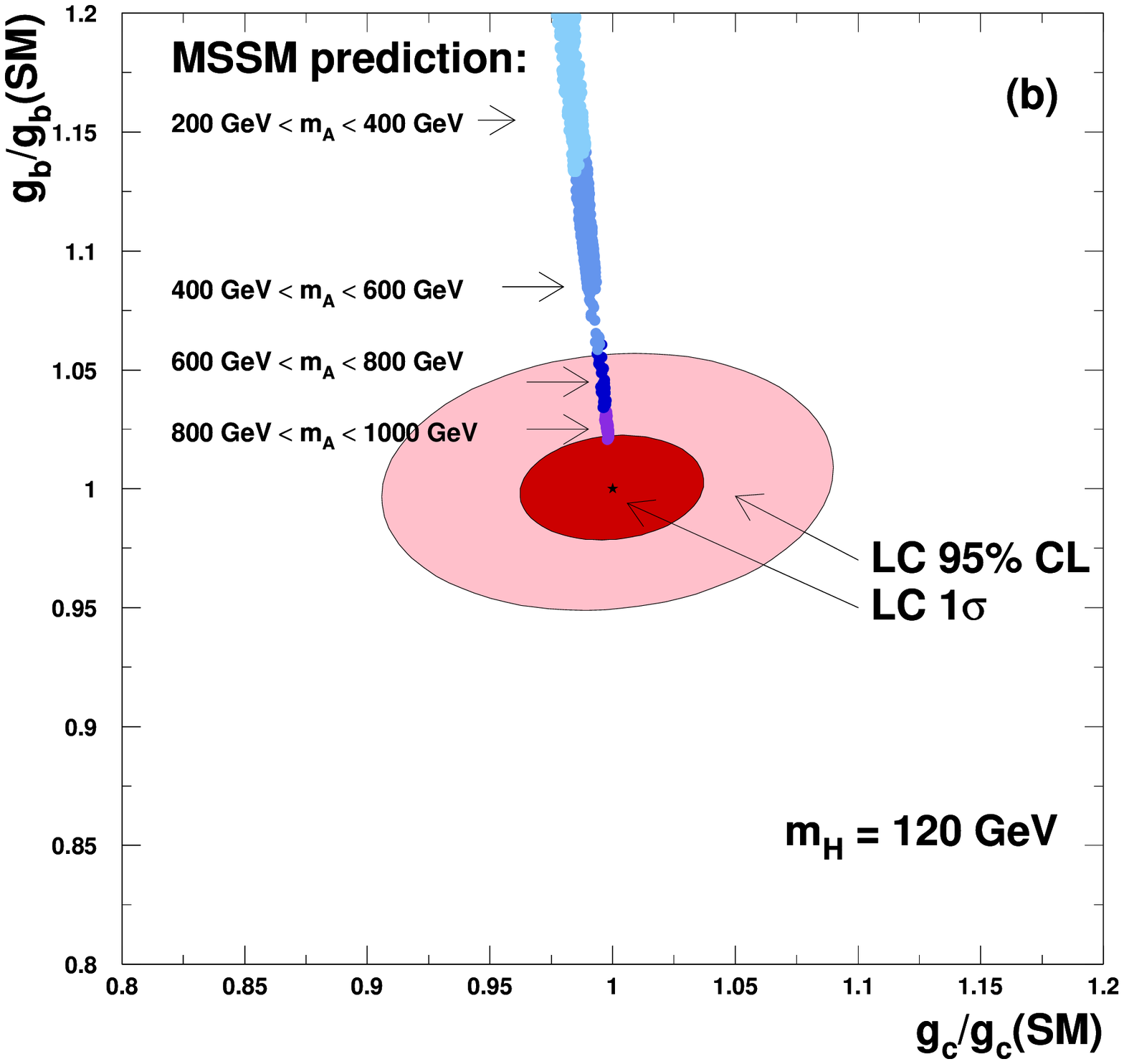,width=0.45\linewidth}}} \\
\href{pictures/1/fig2206c.pdf}{{\epsfig{file=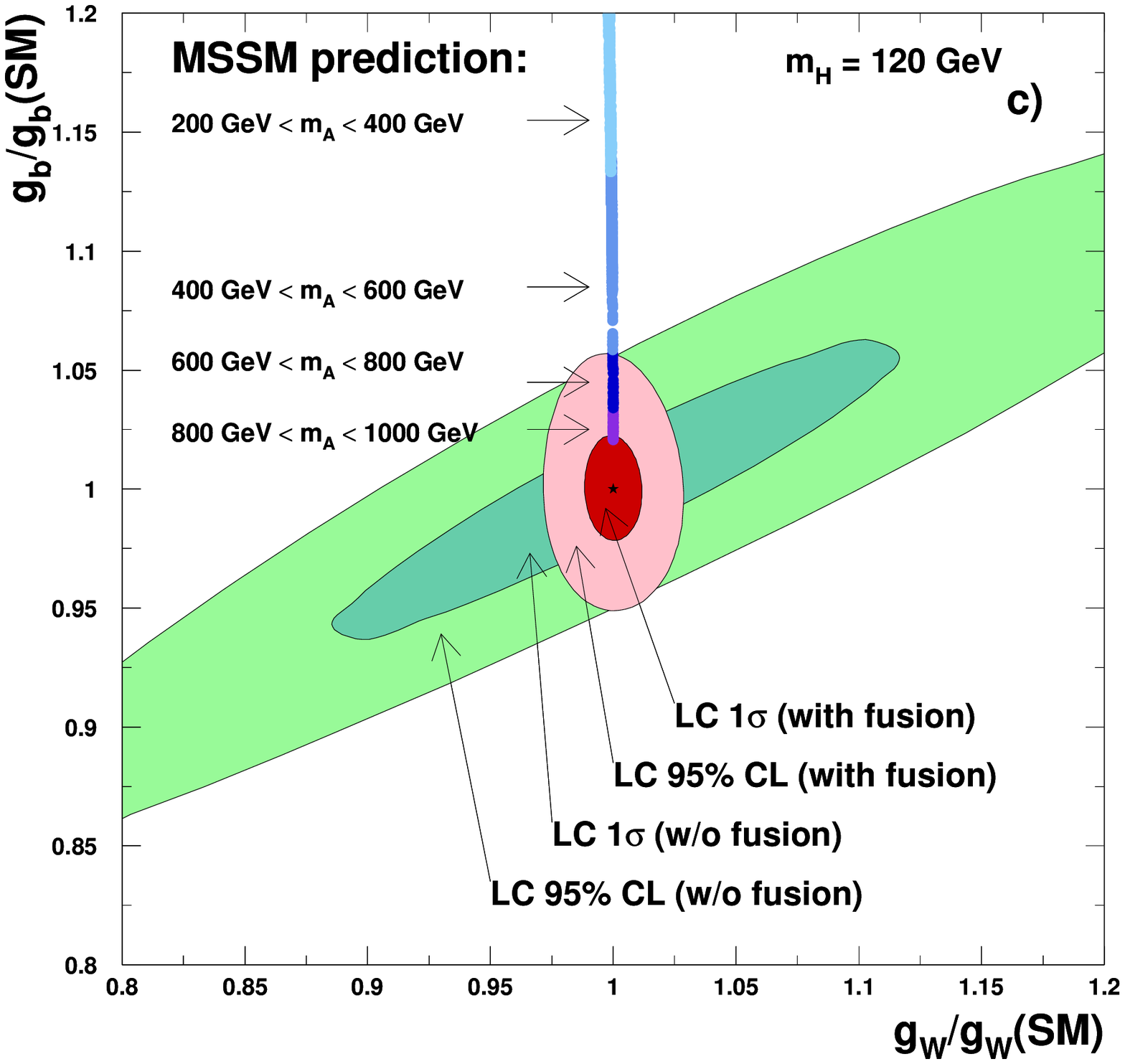,width=0.45\linewidth}}} &
\href{pictures/1/fig2206d.pdf}{{\epsfig{file=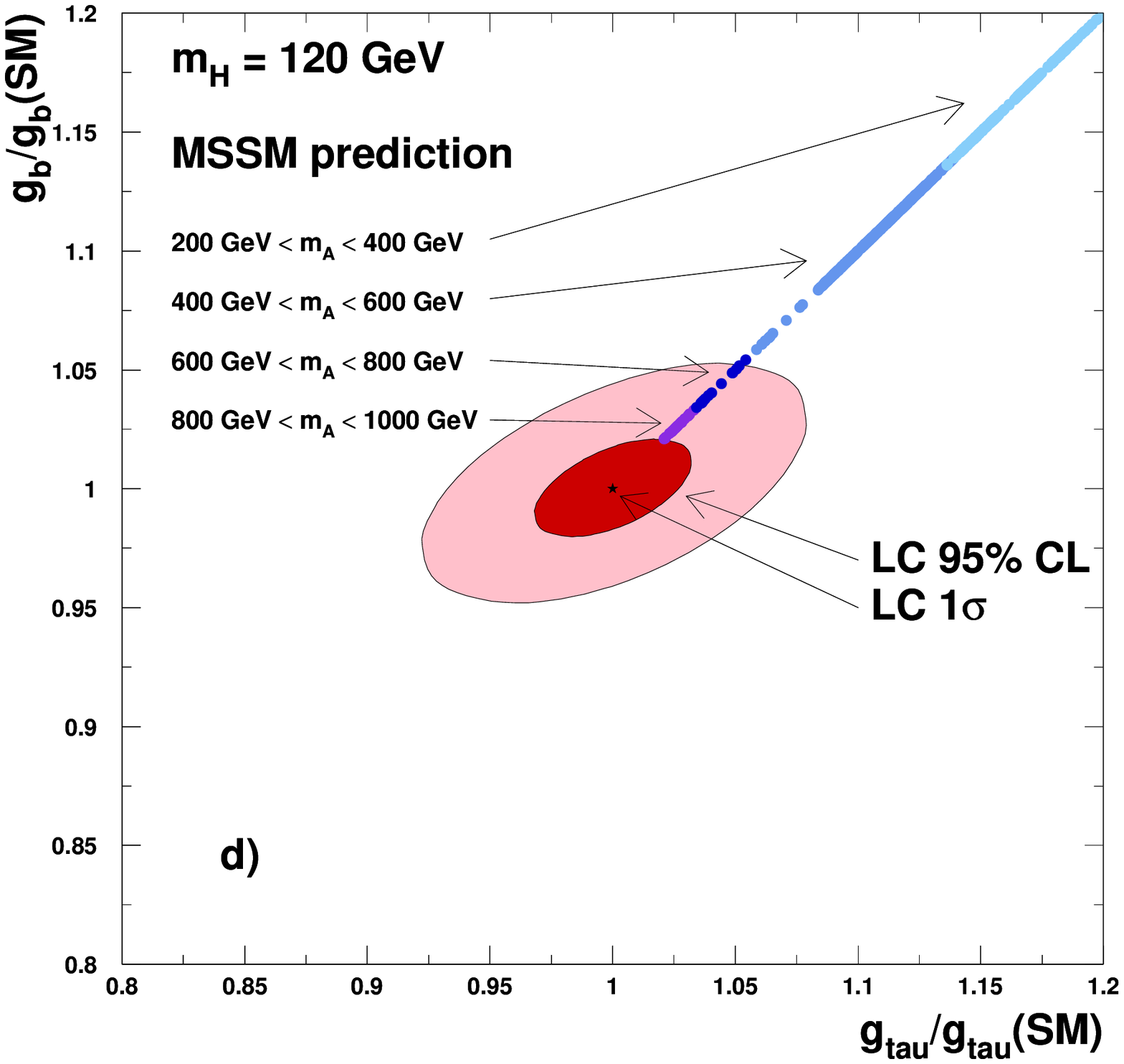,width=0.45\linewidth}}} \\
\end{tabular}
\caption{\label{fig:hfitter}
Higgs coupling determinations at TESLA. The contours for the 
$g_{HZZ}$ vs. $g_{HWW}$ (a), $g_{Hbb}$ vs. $g_{Hcc}$ (b), 
$g_{Hbb}$ vs. $g_{HWW}$ (c) and $g_{Hbb}$ vs. $g_{H\tau\tau}$ (d) 
couplings for a 120\GeV\  Higgs boson as measured with 500\,fb$^{-1}$ of data.}
\end{center}
\end{figure}

\begin{table}[h!]
\begin{center}
\begin{tabular}{|l|c|c|}
\hline
Coupling & $M_H$ = 120\GeV & 140\GeV \\
\hline \hline
$g_{HWW}$ & $\pm$ 0.012       &  $\pm$ 0.020           \\
$g_{HZZ}$ & $\pm$ 0.012       & $\pm$ 0.013            \\ \hline
$g_{Htt}$ & $\pm$ 0.030       & $\pm$ 0.061            \\
$g_{Hbb}$ & $\pm$ 0.022       & $\pm$ 0.022             \\
$g_{Hcc}$ & $\pm$ 0.037       & $\pm$ 0.102            \\ \hline
$g_{H\tau\tau}$ & $\pm$ 0.033       & $\pm$ 0.048            \\ \hline\hline
$g_{HWW} / g_{HZZ}$ & $\pm$ 0.017        & $\pm$ 0.024           \\\hline
$g_{Htt} / g_{HWW}$ & $\pm$ 0.029        & $\pm$ 0.052           \\
$g_{Hbb} / g_{HWW}$ & $\pm$ 0.012        & $\pm$ 0.022           \\
$g_{H\tau\tau} / g_{HWW}$ & $\pm$ 0.033  & $\pm$ 0.041       \\ \hline
$g_{Htt} / g_{Hbb}$ & $\pm$ 0.026        & $\pm$ 0.057            \\
$g_{Hcc} / g_{Hbb}$ & $\pm$ 0.041        & $\pm$ 0.100            \\
$g_{H\tau\tau} / g_{Hbb}$ & $\pm$ 0.027  & $\pm$ 0.042       \\ \hline
\end{tabular}
\end{center}
\caption{\label{tab:hfitter}
Relative accuracy on Higgs couplings and their ratios obtained from a
global fit (see text).
An integrated luminosity of 500\,fb$^{-1}$ at $\sqrt{s} = 500 $\GeV\  is assumed
except for the measurement of $g_{Htt}$, which assumes 1000\,fb$^{-1}$ at
$\sqrt{s} = $~800\GeV\  in addition.}
\end{table}

A dedicated program, {\sc HFitter}~\cite{hfitter} has been developed based on 
the {\sc Hdecay}~\cite{hdecay} program for the calculation of the Higgs boson
branching ratios. The following inputs have been used: $\sigma_{HZ}$, 
$\sigma_{H\nu\bar\nu}$, BR($H^0 \rightarrow WW$), BR($H^0 \rightarrow \gamma 
\gamma$), BR($H^0 \rightarrow b \bar b$), BR($H^0 \rightarrow \tau^+ \tau^-$),
BR($H^0 \rightarrow c \bar c$), BR($H^0 \rightarrow g g$), $\sigma_{t \bar t H}$.
For correlated measurements the full covariance matrix has been used.
The results are given for $M_H$ = 120\GeV\  and 140\GeV\  and 
500\,fb$^{-1}$. Table~\ref{tab:hfitter} shows the accuracy which can be achieved
in determining the couplings and their relevant ratios.
Fig.~\ref{fig:hfitter} shows 1$\sigma$ and 95\% 
confidence level contours for the fitted values of various pairs of ratios of
couplings, with comparisons to the sizes of changes expected from the MSSM. 

\subsection{Quantum numbers of the Higgs boson\label{sec:2.2.8}}

The spin, parity, and charge-conjugation quantum numbers $J^{PC}$ of the Higgs
bosons can be determined at TESLA in a model-independent way~\cite{BCDKZ}.
The observation of Higgs boson production at the $\gamma\gamma$ collider 
or of the $H^0 \rightarrow \gamma\gamma$ decay would rule out $J=1$ and require
$C$ to be positive. 
\begin{figure}[ht!]
\begin{center}
\href{pictures/1/fig2207.pdf}{{\epsfig{file=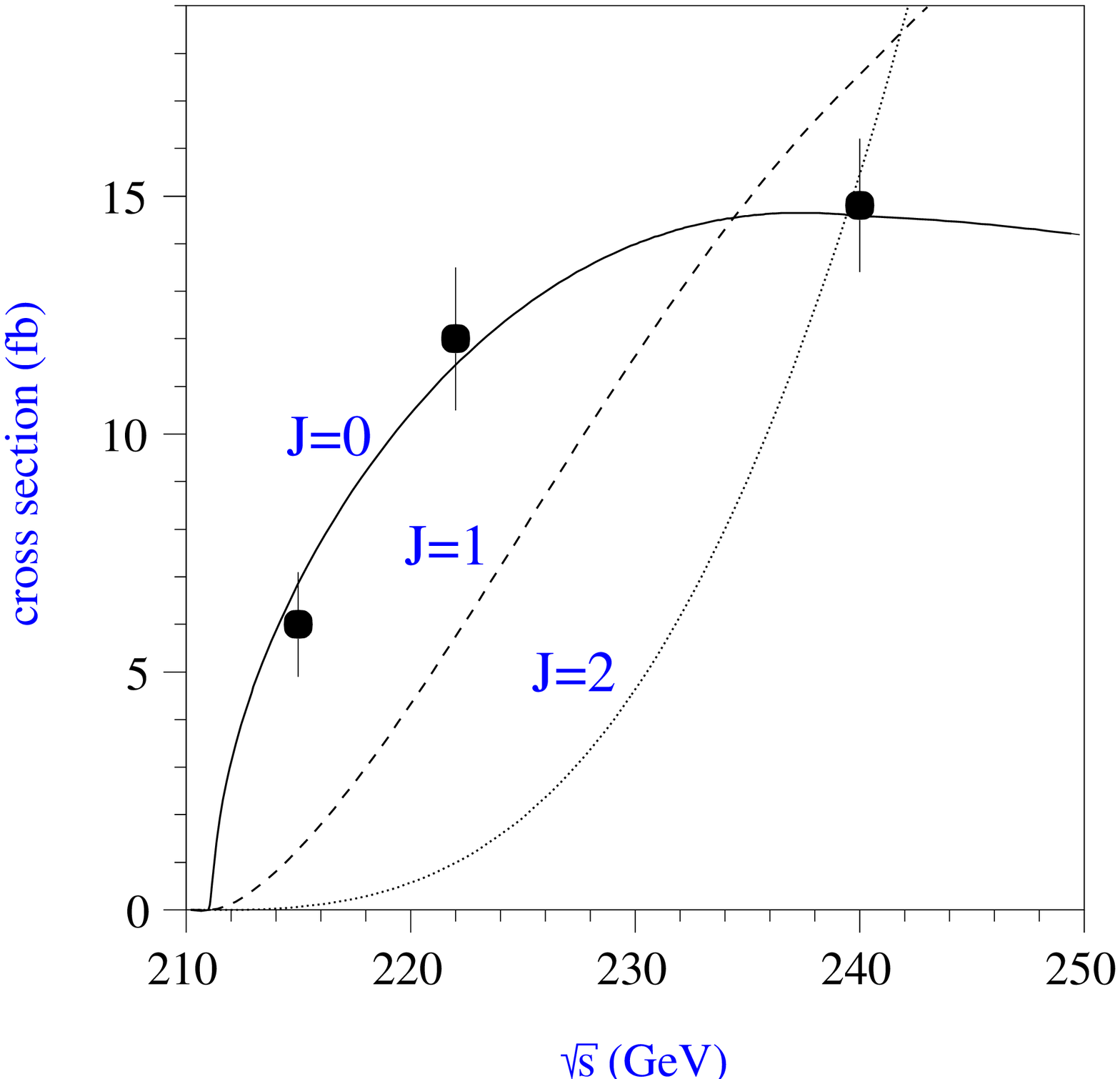,width=0.53\linewidth,clip}}}
\end{center}
\caption{\label{fig:htresh}
Simulated measurement 
of the $e^+e^- \rightarrow H^0Z$ cross section for 
$M_H$ = 120\GeV\  with 20\,fb$^{-1}$/point at three centre-of-mass energies 
compared to the predictions for a spin-0 (full line) and 
examples of spin-1 (dashed line) and spin-2 (dotted line) particles.}
\end{figure}
The measurement of the rise of the total Higgs-strahlung cross section 
at threshold and the angular dependence of the cross--section in the continuum 
allow $J$ and $P$ to be uniquely determined.

The threshold rise of the process $e^+e^-\rightarrow ZX$ for a boson $X$
of arbitrary spin $J$ and normality $n = (-1)^J P$ has been 
studied in~\cite{dmiller}. While for $J=0$ the cross section at threshold
rises $\propto \beta_{ZX}$ (see eq.~\ref{eq:hstr}), for higher spins
the cross section rises generally with higher powers of $\beta_{ZX}$ except
for some scenarios with which can be distinguished through the angular
dependence in the continuum. A threshold scan with a luminosity of
20\,fb$^{-1}$ at three centre--of--mass energies is sufficient to distinguish
the different behaviours (see Fig.~\ref{fig:htresh})~\cite{lohmannspin}.

\begin{figure}[ht!]
\begin{picture}(150,10)\unitlength 1mm
  \put(12,2){a)}
  \put(87,2){b)}
\end{picture}
\vspace*{-10mm}
\begin{center}
\begin{tabular} {c c}
{\href{pictures/1/fig2208a.pdf}{{\epsfig{figure=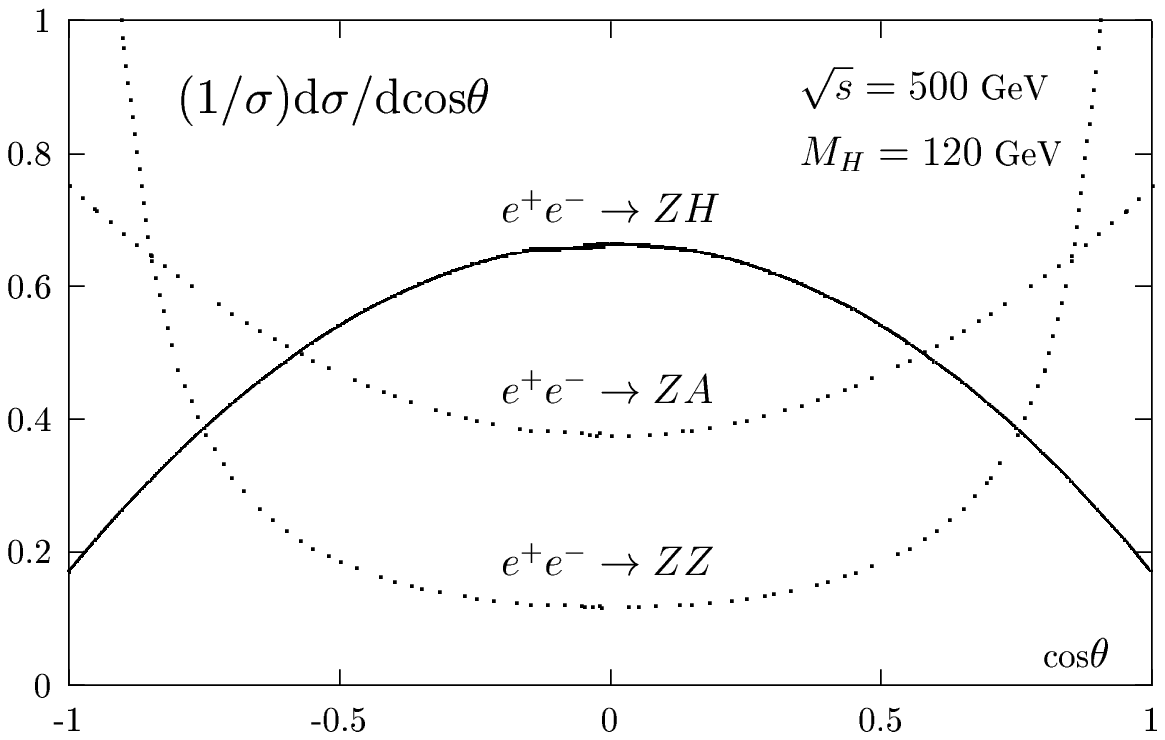,height=0.43\linewidth,width=0.42\linewidth}}}} &
{\href{pictures/1/fig2208b.pdf}{{\epsfig{figure=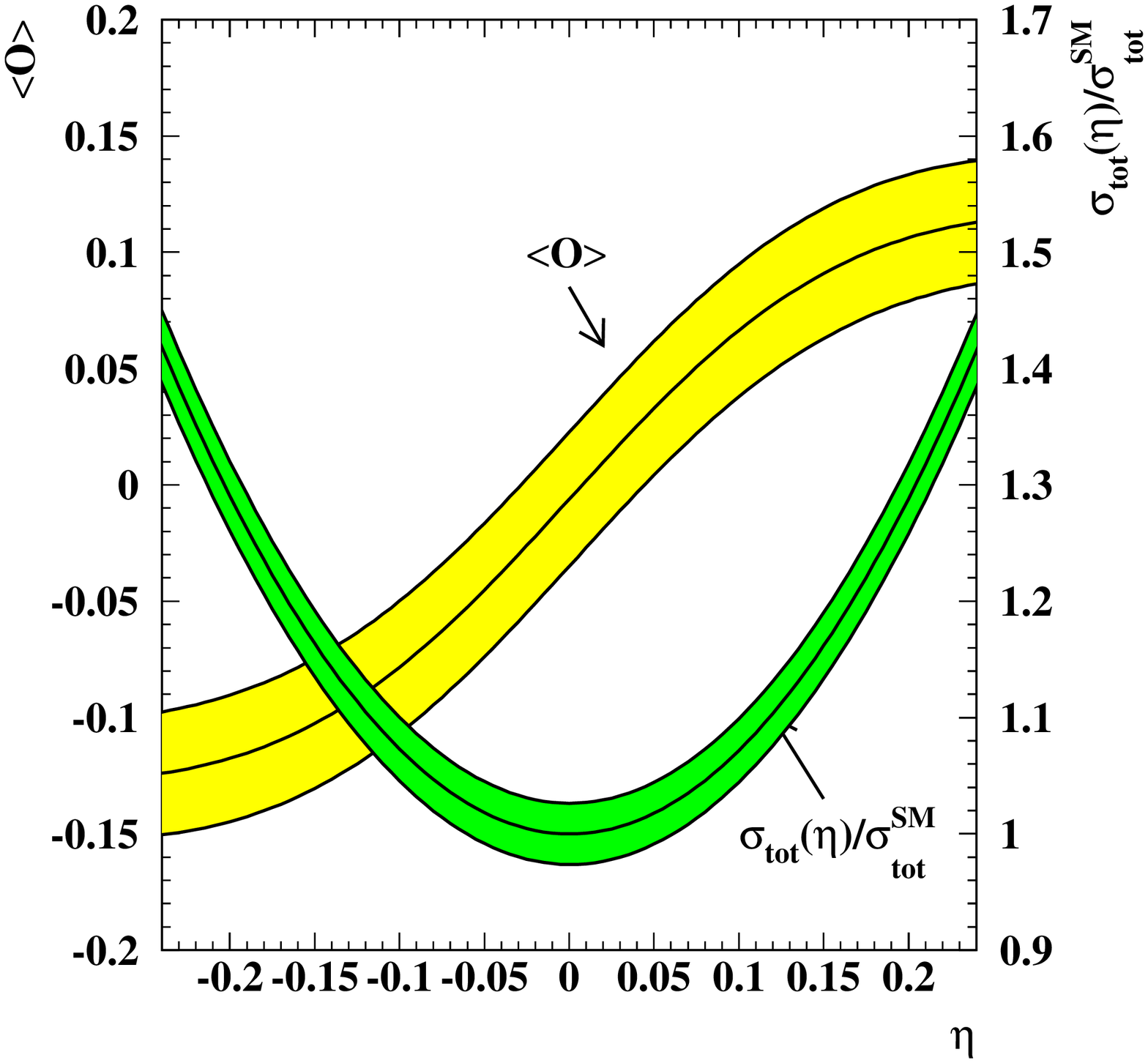,width=0.465\linewidth}}}} \\
\end{tabular}
\caption{\label{fig:bak1}
a): The $\cos\theta$ dependence of $e^+e^-\to ZH$, $e^+e^-\to ZA$,
$e^+e^-\to ZZ$ for $\protect\sqrt s=500$\GeV, assuming 
$M_H = M_A = $~120\GeV~\protect\cite{BCDKZ} and b): the dependence of the 
expectation value of the optimal observable and the total cross--section
on $\eta$ for $M_H = 120$\GeV,$\sqrt{s}$ = 350\GeV\ and $\cal{L}$ = 
500\,fb$^{-1}$ after applying the selection cuts. The shaded bands show the 
$1\sigma$ uncertainty in the determination of $\langle{\cal{O}}\rangle$ 
and the total cross section.}
\end{center}
\end{figure}

In the continuum, one can distinguish the SM Higgs $0^{++}$ boson from 
a ${\cal CP}$-odd $0^{-+}$ state $A^0$, or a ${\cal CP}$--violating mixture 
of the two (generically denoted by $\Phi$ in the following). 
The $J=0$ nature of the Higgs bosons can be established by comparing the cross
section angular dependence with that of the $e^+e^-\rightarrow ZZ$ process, 
which exhibits a distinctly different angular 
momentum structure (see Fig.~\ref{fig:bak1}\,a)) due to the t-channel
electron exchange. However, in a general 2HDM model the three neutral 
Higgs bosons correspond to arbitrary mixtures of ${\cal CP}$ 
eigenstates, and their 
production and decay exhibit ${\cal CP}$ violation. In this case, the 
amplitude for the Higgs-strahlung process can be described by adding a 
ZZA coupling with strength $\eta$ to the SM matrix element 
${\cal{M}} = {\cal{M}}_{ZH} + i \eta {\cal{M}}_{ZA}$. In general the parameter
$\eta$ can be complex, we assume it to be real in the following.
If $\eta=0$, we recover the coupling of SM Higgs boson $H$.
However, in a more general scenario, $\eta$ need not be loop suppressed as in 
the MSSM, and it is useful to allow for $\eta$ to be arbitrary in the 
experimental data  analysis. 
%The squared amplitude for the Higgs-strahlung 
%process $Z\rightarrow Z \phi$ is then given by~\cite{DK}:
%\begin{equation}
%|{\cal{M}}|^2 = |{\cal{M}}_{ZH}^{SM}|^2 + \eta |{\cal{M}}_{\cal{CP}}|^2 + 
%\eta^2 |{\cal{M}}_{ZA}|^2 \label{matel} 
%\end{equation}
%The first term in $|{\cal{M}}|^2$ corresponds to the SM cross--section, the 
%interference term $|{\cal{M}}_{\cal{CP}}|^2 = 2 Re({\cal{M}}_{ZH}^*
%{\cal{M}}_{ZA})$, linear in $\eta$, generates a forward-backward asymmetry, 
%which would represent a distinctive signal of $CP$ violation, while the 
%CP-even contribution  $|{\cal{M}}_{ZA}|^2$ increases the total $e^+e^- 
%\rightarrow Z\phi$ cross--section. 
The most sensitive single kinematic variable
to distinguish these different contributions to Higgs boson production is the
production angle $\theta_Z$ of the $Z$ boson w.r.t.\ to the beam axis, in the 
laboratory frame. The differential cross--section for the process
$e^+e^-\to Z\Phi$ is given by:
\begin{equation}\nonumber
\label{eq:cos}
\frac{d\sigma}{d\cos\theta_Z} \propto \beta_{\Phi Z}
\left[1+\frac{s\beta_{\Phi Z}^2}{8M_Z^2}\sin^2\theta_Z
% +\eta\frac{2s\beta_{\Phi Z}}{M_Z^2}\frac{v_ea_e}{v_e^2+a_e^2}\cos\theta_Z
+\eta\frac{2s\beta_{\Phi Z}}{M_Z^2}\kappa\cos\theta_Z
+\eta^2\frac{s^2\beta_{\Phi Z}^2}{8M_Z^4}(1+\cos^2\theta_Z)\right],
\end{equation}
where $\kappa = v_e a_e / (v_e^2+a_e^2)$ and $v_e$, $a_e$ and 
$\beta_{\Phi Z}$ are defined below equation~\ref{eq:hstr}.
The angular distribution of $e^+e^-\to ZA$, $\propto (1+\cos^2\theta_Z)$,
corresponding to transversely polarised $Z$ bosons, is 
therefore very distinct from that of $ZH$ in the SM, $\propto \sin^2\theta_Z$,
for longitudinally polarised $Z$ bosons 
in the limit~$\sqrt s\gg M_Z$~\cite{BCDKZ}. 
In the above equation, the interference
term, linear in $\eta$, generates a forward-backward 
asymmetry, which would represent a distinctive signal of ${\cal CP}$ 
violation, while 
the term proportional to $\eta^2$ increases the total 
$e^+e^- \rightarrow Z\phi$ cross--section. 

The angular distributions of the accompanying $Z \rightarrow f \bar f$ decay 
products are also sensitive to the Higgs boson ${\cal CP}$ parity 
and spin as well as 
to anomalous couplings~\cite{DK}. 
In fact, at high energies, the $Z$ bosons from $e^+e^-\to ZH$ are dominantly
longitudinally polarised, while those from $e^+e^-\to ZA$
($e^+e^-\to ZZ$) are fully (dominantly) transversely polarised~\cite{BCDKZ}.
These distributions can be described in terms of the angles $\theta^*$ and 
$\phi^*$, where $\theta^*$ is the  polar angle between the flight direction of
the decay fermion $f$ in the $Z$-boson rest frame and that of the 
$Z$-boson in
the laboratory frame and $\phi^*$ is the corresponding azimuthal angle w.r.t.\ 
the plane defined by the beam axis and the $Z$-boson flight direction.

The information carried by these three angular distributions can be 
analysed using the optimal observable formalism~\cite{oom}, in terms of a 
single variable ${\cal{O}}$ defined as the ratio of the ${\cal CP}$-violating 
contribution to the SM cross--section, 
${\cal{O}} = 2{\mathrm{Re}}(\cal{M}_{\it{ZA}}^* \cal{M}_{\it{ZH}}) / 
|\cal{M}_{\it{ZH}}|^{\it{2}}$.
If the Higgs boson production respects 
${\cal CP}$ symmetry, 
the expectation value of this ${\cal CP}$--odd observable must vanish, 
i.e. $\langle{\cal{O}}\rangle = 0$. Any significant deviation of 
$\langle{\cal{O}}\rangle$ from 0 implies the existence of ${\cal CP}$ violation,
independent of the specific model.

This analysis has been performed for 
$M_H$ = 120\GeV\ assuming an integrated luminosity of 500\,fb$^{-1}$ at 
$\sqrt{s}$ = 350\GeV, following the criteria of the $H^0Z$ reconstruction 
discussed above. However, in order not to bias the analysis towards specific 
Higgs boson decay modes, only cuts on $Z$ decay products are applied. 
The resulting sensitivity is shown in Fig.~\ref{fig:bak1}\,b) for the
case of $Z\to\mu^+\mu^-$.
The accuracy in the determination of $\eta$, obtained using 
the expectation value of the optimal observable is 0.038, and it
improves to 0.032 when the total cross--section dependence is exploited
in addition~\cite{markus}.

\begin{table}[ht!]
\begin{center}
\smallskip
{\renewcommand{\arraystretch}{1.1}
\begin{tabular}{|c|c|c|c|}
\hline
$\epsilon_\tau$ & --- & 0.5 & 0.5 \\
$\epsilon_b$    & --- & 0.6 & 0.6 \\
$|P_{e^-}|$     & --- & --- & 0.8 \\
$|P_{e^+}|$     & --- & --- & 0.45 \\
\hline \hline
Re$\,(b_Z)$ & $\pm$0.00055 & $\pm$0.00029 & $\pm$0.00023 \\
Re$\,(c_Z)$ & $\pm$0.00065 & $\pm$0.00017 & $\pm$0.00011 \\
Re$\,(b_\gamma)$ & $\pm$0.01232 & $\pm$0.00199 & $\pm$0.00036 \\
Re$\,(c_\gamma)$ & $\pm$0.00542 & $\pm$0.00087 & $\pm$0.00008 \\
Re$\,(\tilde b_Z)$ & $\pm$0.00104 & $\pm$0.00097 & $\pm$0.00055 \\
Re$\,(\tilde b_\gamma)$ & $\pm$0.00618 & $\pm$0.00101 & $\pm$0.00067 \\
%Im$\,(b_Z-c_Z)$ & 0.01055 & 0.00049 & 0.00046 \\
%Im$\,(b_\gamma-c_\gamma)$ & 0.00206 & 0.00057 & 0.00054 \\
%Im$\,\left(\tilde b_Z\right)$ & 0.00521 & 0.00022 & 0.00022 \\
%Im$\,\left(\tilde b_\gamma\right)$ & 0.00101 & 0.00026 & 0.00026 \\
\hline
\end{tabular}
}
\end{center}
\caption{ \label{tab:bak1}
Accuracy on general $ZZ\Phi$ and $Z\gamma\Phi$ couplings for
various values for the $\tau$ helicity reconstruction and $b$ charge 
identification efficiencies ($\epsilon_\tau$ and $\epsilon_b$) and
beam polarisations ($|P_{e^-}|$ and $|P_{e^+}|$). The numbers correspond
to 300\,fb$^{-1}$ of data at $\sqrt{s}$ = 500\GeV. Detector resolution effects
are not simulated.}
\end{table}
\begin{figure}[ht!]
\begin{picture}(150,10)\unitlength 1mm
  \put(20,-5){a)}
  \put(135,-5){b)}
\end{picture}
\vspace*{-8mm}
\begin{center}
\begin{tabular}{cc}
\href{pictures/1/fig2209a.pdf}{{\epsfig{file=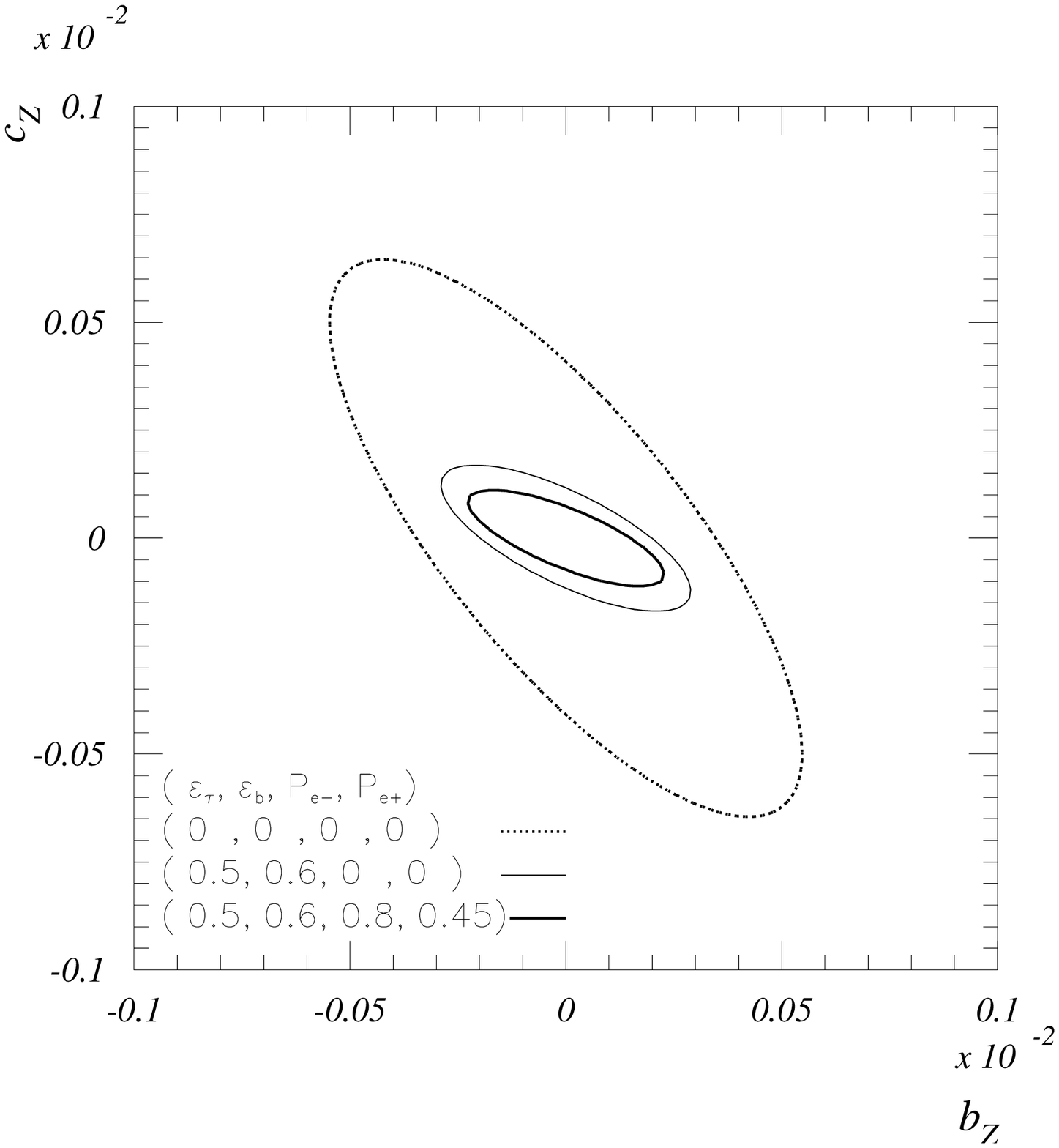,width=0.45\linewidth}}} &
\href{pictures/1/fig2209b.pdf}{{\epsfig{file=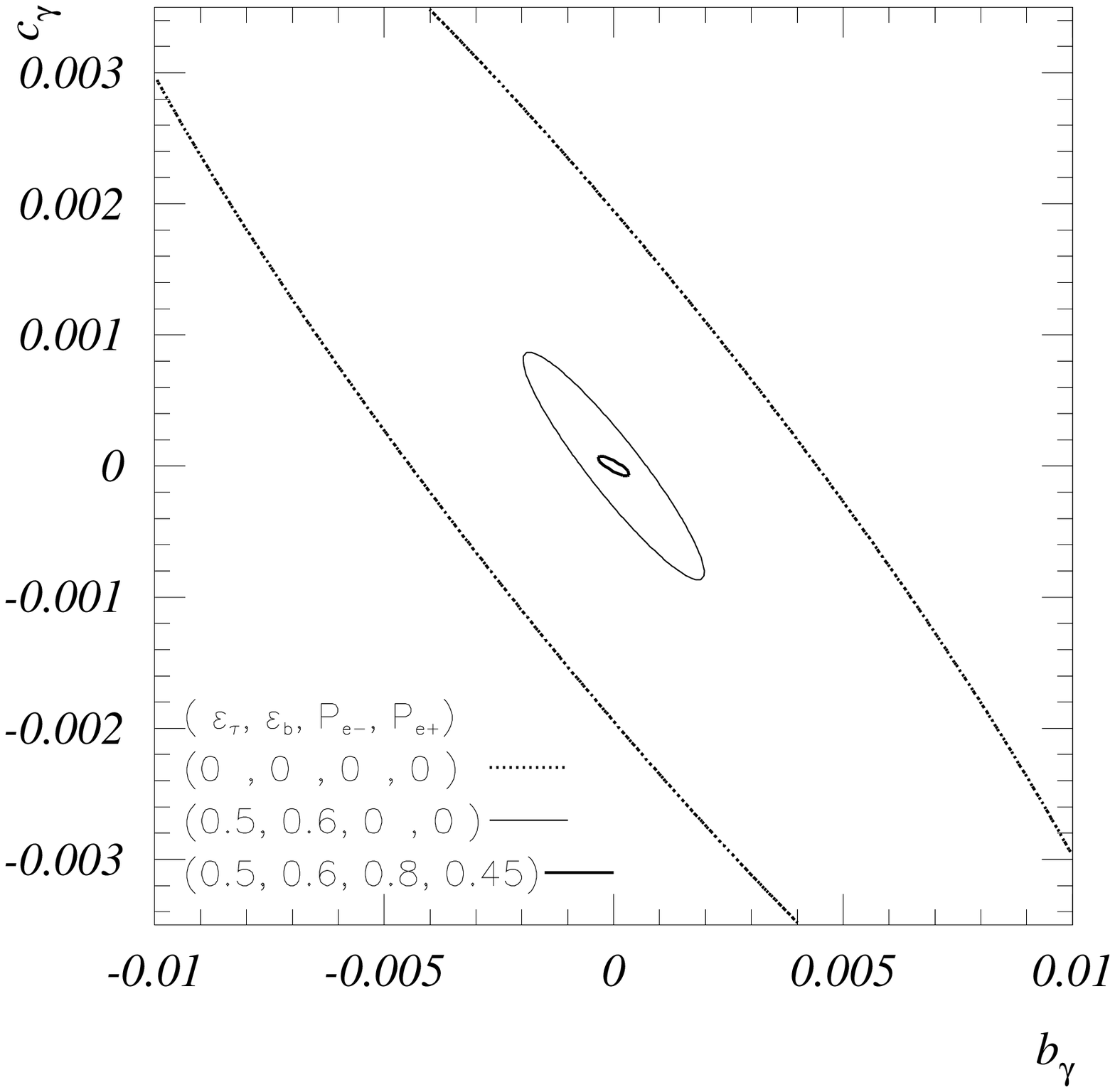,width=0.45\linewidth}}} \\
\end{tabular}
\caption{ \label{fig:bak2}
a): The 68\% C.L. contours in the $(b_Z,c_Z)$  and b):
$(b_\gamma,c_\gamma)$ (right) planes. In each case, the 
other degrees of freedom have been integrated out.
The contours correspond
to 300\,fb$^{-1}$ of data at $\sqrt{s}$ = 500\GeV. Detector resolution effects
are not simulated.}
\end{center}
\end{figure}
In the effective-Lagrangian approach, the most general $ZZ\Phi$ coupling 
can have two more independent ${\cal CP}$--even terms~\cite{HIKK}.
Similarly, there may also be an effective $Z\gamma\Phi$ coupling, generated by
two ${\cal CP}$--even and one ${\cal CP}$--odd terms~\cite{HIKK} making a total of seven 
complex couplings, $a_Z$, $b_Z$, $c_Z$, $\tilde b_Z$, $b_\gamma$, $c_\gamma$, 
and $\tilde b_\gamma$, where the ${\cal CP}$--odd 
couplings are indicated by a tilde. 
With sufficiently high luminosity, accurate $\tau$ helicity and good $b$ 
charge identification and electron and positron beam polarisation it will 
be possible to determine these couplings from the angular distributions of
$e^+e^-\to Z\Phi\to\left(f\bar f\right)\Phi$~\cite{polarisation}
A global analysis of these angular distributions, based on the
optimal observable method~\cite{GGH,oom} and assuming $\sqrt s=500$\GeV, 
${\cal L} = 300$\,fb$^{-1}$, $\tau$ helicity and $b$ charge identification
efficiencies $\epsilon_\tau=50\%$ and $\epsilon_b=60\%$, and beam polarisations
$P_{e^-}=\pm80\%$, $P_{e^+}=\mp45\%$ gives the results summarised in
Table~\ref{tab:bak1} and in Fig.~\ref{fig:bak2} for fixed $a_Z$.
The coupling $a_Z$ can be determined by repeating
the analysis at two different values of $\sqrt s$, such as 350\GeV\ and 500\GeV.
%We observe that, for $\epsilon_\tau=\epsilon_b=P_{e^-}=P_{e^+}=0$, 
% the $ZZ\Phi$ couplings are generally well constrained, even
%for $\epsilon_\tau=\epsilon_b=P_{e^-}=P_{e^+}=0$, while the $Z\gamma\Phi$
%couplings are not.
We observe that the $ZZ\Phi$ couplings are generally well constrained, even
for $\epsilon_\tau=\epsilon_b=P_{e^-}=P_{e^+}=0$.
The constraints on the $Z\gamma\Phi$ couplings may be improved by
approximately a factor of 6 through $\tau$ and $b$ tagging and by another
factor of 1.5 to 10 through beam polarisation.

\subsection{Higgs potential \label{sec:2.2.9}}

To establish the Higgs mechanism experimentally in an unambiguous way, the
self potential of the Higgs field:
\begin{equation} 
V = \lambda \left(|\varphi|^2 -\textstyle{\frac{1}{2}} v^2 \right)^2, 
\end{equation} 
with a minimum at $\langle \varphi \rangle_0 = v/\sqrt{2}$, must be
reconstructed. This can be accomplished by measuring the
self-couplings of the physical Higgs boson $H$~\cite{triple, triple2} as
predicted by the potential:
\begin{equation}
V = \lambda v^2 H^2 + \lambda v H^3 + \textstyle{\frac{1}{4}} \lambda H^4.
\end{equation}
The coefficient of the bilinear term in the Higgs field defines the
mass $M_H=\sqrt{2\lambda} v$ so that the trilinear and quadrilinear
couplings can be predicted unambiguously in the SM.
\vspace{0.2cm}
\begin{figure}[ht!]
\begin{center}
\href{pictures/1/fig2210.pdf}{{\epsfig{file=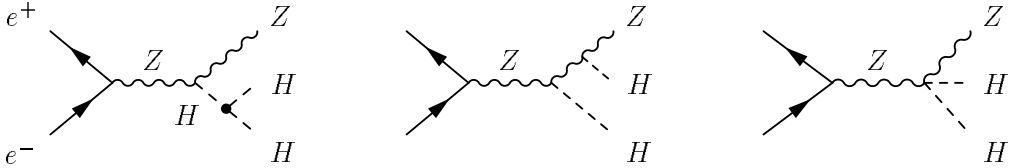,width=0.9\linewidth}}}
\end{center}
\caption{\label{fig:diag}
Double Higgs boson associated production with a $Z$~boson.
}
\end{figure}

\begin{figure}
\begin{center}
{\epsfig{figure=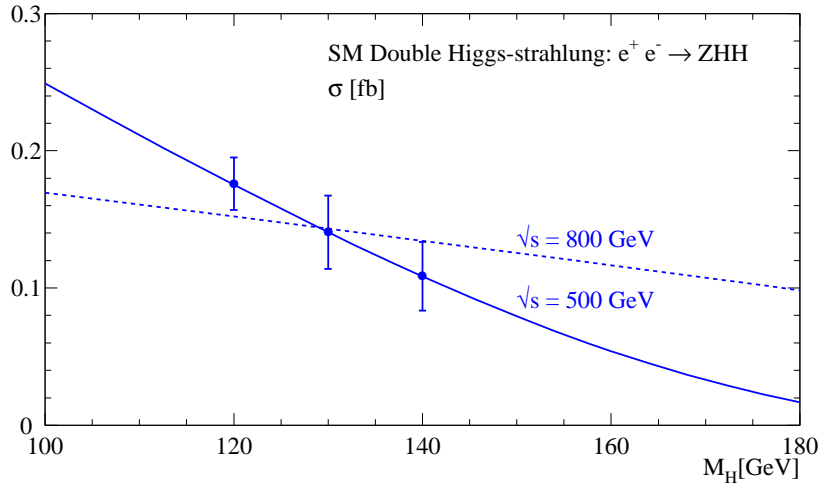,width=0.7\linewidth}}
\vspace*{-0.25cm}
\end{center}
\caption{ \label{fig:hhz}
The cross--section for double Higgs-strahlung
  $e^+e^-\to ZHH$ in the Standard Model at two collider energies:
  $\sqrt{s}=500$\GeV\ and 800\GeV. The dots with error bars show
the achievable experimental accuracies for 1000\,fb$^{-1}$ (see text).}
\end{figure}

The trilinear Higgs coupling  $\lambda_{H\!H\!H} = 6\sqrt{2}\lambda$, 
in units of $v/\sqrt{2}$, can be measured directly in pair-production of Higgs
particles at high-energy $e^+ e^-$ colliders~\cite{triple, triple2, triple3}.
The most interesting process at TESLA centre--of--mass energies
is the associated production of two Higgs bosons 
with a $Z$~boson, $e^+e^- \to H^0H^0Z$. 
As evident from Fig.~\ref{fig:diag}, this process is built up by the
amplitude involving the trilinear Higgs coupling superimposed on the
two other mechanisms which lead to the same final state but do not
involve $\lambda_{H\!H\!H}$.
The cross--section for double Higgs production, which
is therefore a binomial in the coupling $\lambda_{H\!H\!H}$,
is of the order of 0.20\,fb for $M_H$ = 120\GeV\ at $\sqrt{s}$ = 500\GeV\ 
and 0.15\,fb at $\sqrt{s}$ = 800\GeV\ (see Fig.~\ref{fig:hhz}). 
The quadrilinear Higgs coupling can in principle be measured 
in triple Higgs boson production, but the cross--section is 
suppressed by an additional electroweak factor, and is therefore 
too small to be observable at TESLA energies~\cite{triple2}.  

A detailed analysis of
the reconstruction of double Higgs-strahlung events has been 
performed~\cite{lutz}. The large four and six fermion background and the tiny
signal cross--section make  this analysis a genuine experimental challenge. 
However, by profiting from the characteristic signature with four $b$ jets and 
a $Z$~boson, reconstructed either in its leptonic or hadronic decay modes, 
and from the excellent tagging and energy flow reconstruction capabilities 
of the TESLA detector (see Part IV, Chapter~9), 
this process can be isolated from backgrounds.

\begin{table}[ht!]
\begin{center}
\begin{tabular}{|l|c|c|c|}
\hline
$M_H$ (GeV)              & 120    & 130 & 140 \\
\hline \hline
$N_{HHZ}$                & 80     & 64      & 44  \\ 
Efficiency               & 0.43   & 0.43    & 0.39 \\ \hline
$\delta \sigma / \sigma$ & $\pm$0.17   & $\pm$0.19    & $\pm$0.23 \\ \hline 
\end{tabular}
\end{center}
\caption{\label{tab:hhz}
Number of selected signal $H^0H^0Z$ events, 
selection efficiency and relative uncertainty on
the double Higgs-strahlung cross--section for 1000\,fb$^{-1}$ of TESLA data 
at $\sqrt{s}$ = 500\GeV\ for a cut--based selection (see text).}
\end{table}

In the hadronic channel, after kinematical cuts,
the events are forced into six jets and the jet
pair most consistent with the $Z$ hypothesis is identified. In the 
leptonic channel two identified leptons consistent with a $Z$ boson are
required instead. Then the
jets recoiling against the reconstructed $Z$ boson are required to
contain identified b--quarks. With this selection, accuracies of
approximately 20\% on the $H^0H^0Z$ cross--section can be obtained
for $M_H$ between 120 and 140\GeV\ and 1000\,fb$^{-1}$ 
(see Table~\ref{tab:hhz} and Fig.~\ref{fig:hhz}).
The sensitivity can be further improved when a multi-variable selection based 
on a neural network is applied, reducing the uncertainty from 17\% to 13\% 
for $M_H = 120$\GeV\ and yielding a signal significance $S/\sqrt{B} \sim 6$.

The sensitivity to $\lambda_{H\!H\!H}$ is diluted 
due to the additional diagrams shown in Fig.~\ref{fig:diag}.
Taking this into account, the trilinear Higgs coupling $\lambda_{H\!H\!H}$ 
can be obtained at TESLA with a statistical accuracy 
of 22\% for $M_H=120$\GeV\ with an integrated luminosity of 1000\,fb$^{-1}$,
using the neural network selection~\cite{lutz}.
This measurement crucially depends on the high luminosity anticipated 
for the TESLA operation and the accurate
decay reconstruction provided by the optimised detector.
It represents an essential element for the reconstruction of the 
characteristic Higgs potential which leads to the
non-zero value of the Higgs field in the vacuum, the physical basis of the
Higgs mechanism for breaking the electroweak symmetry and 
generating the masses of the fundamental particles.

% In the SM extensions with extra Higgs doublets, additional trilinear Higgs 
% couplings are also present such as $\lambda_{hhH}$, $\lambda_{hhA}$, 
% $\lambda_{hhh}$ and $\lambda_{HAA}$. 
% While these depend also on the $\tan \beta$ and $M_A$ 
% parameters, the topologies analysed for the case of the SM also apply to
% that of the $\lambda_{hhh}$ except for the limited region of parameters
% where the $h \rightarrow b \bar b$ decay is suppressed. 
% The corresponding analysis can be repeated for trilinear
% Higgs couplings in the supersymmetric extension of the Standard Model.
% Between the five physical Higgs states of the MSSM a large ensemble of
% couplings can be realized. Their interplay renders Higgs boson 
% pair-production quite complex~\cite{triple}. The trilinear coupling of the 
% light ${\cal CP}$--even neutral Higgs boson $h$ can be expressed as:
% \begin{equation}
% \lambda_{hhh} = 3 \cos2\alpha \sin (\beta+\alpha) 
% + {\cal O}(G_F M_t^4/M_Z^2)\, ,
% \end{equation}
% in terms of the mixing angles $\alpha$ and $\beta$
% Fig.~2b. The figure presents the $2\sigma$ sensitivity area in the
% $[M_A,\tan\beta]$ plane at an integrated luminosity of 2~ab$^{-1}$. 
% The sensitive area in the $[M_A,\tan\beta]$ plane depends on the states that
% can be analysed as described in detail in~\cite{triple}.
% the heavy MSSM Higgs bosons $H,A$ in the final state are analysed, the
% sensitivity areas are reduced significantly as described in detail in
% Ref.~\cite{triple}.

\section{Study of SUSY Higgs Bosons \label{sec:2.3}}

If supersymmetry exists in Nature, a major goal of TESLA will be the 
measurement of its parameters. 
In this way, the underlying SUSY-breaking mechanism 
could be determined and thorough consistency checks of the model itself could 
be performed. The TESLA potential in the investigations of the supersymmetric
particle partners is described in detail in Section~\ref{physics_SUSY}.
Here the perspectives of the 
study of the extended Higgs sector as predicted in supersymmetry is discussed.

The study of the lightest neutral MSSM Higgs boson $h^0$ follows closely that
of the SM-like $H^0$ discussed above, and similar results, in terms of the 
achievable experimental accuracies, are valid. This light Higgs boson, $h^0$,
can be found at $e^+e^-$ colliders easily.
The ability of TESLA to 
distinguish the SM/MSSM nature of a neutral Higgs boson is discussed below.

In SUSY models, additional decay channels may open for the Higgs bosons if 
supersymmetric particles exist with light enough masses. The most interesting
scenario is that in which the lightest Higgs boson decays in particles escaping 
detection giving a sizeable $H^0 \rightarrow {\mathrm{invisible}}$ decay width. 
While the Higgs boson observability in the di-lepton recoil mass in the 
associated $H^0Z$ 
production channel is virtually unaffected by this scenario, such an invisible
decay width can be measured by comparing the number of 
$e^+e^- \rightarrow ZH^0 \rightarrow \ell^+\ell^- {\mathrm{anything}}$ events 
with the sum over the visible decay modes corrected by the $Z \rightarrow
\ell^+ \ell^-$ branching ratio: BR($Z \rightarrow \ell^+\ell^-$) $\times
(\sum_{i=b,c,\tau,...} N_{ZH \rightarrow f_i \bar f_i} + 
\sum_{j=W,Z,\gamma} N_{ZH \rightarrow B_j \bar B_j}$). Using the accuracies
on the determination of the individual branching ratios discussed above,
the rate for the $H^0 \rightarrow {\mathrm{invisible}}$ decay can be determined to
better than 20\% for BR($H^0 \rightarrow {\mathrm{invisible}}$) $>$ 0.05.

\subsection[Study of the $H^0$, $A^0$ and $H^{\pm}$ bosons]{Study of the $\bH^0$, $\bA^0$ and $\bH^{\pm}$ bosons \label{sec:2.3.1}}

A most distinctive feature of extended models such as supersymmetry, or general
2HDM extensions of the SM, is the existence of additional Higgs bosons.
Their mass and coupling patterns vary with the model parameters. However in
the decoupling limit, the $H^{\pm}$, $H^0$ and $A^0$ bosons are expected to be
heavy and to decay predominantly into quarks of the third generation.
Establishing their existence and the determination of their masses and of their
main decay modes will represent an important part of the TESLA physics
programme at centre-of-mass energies exceeding 500\GeV.

For a charged Higgs boson mass $M_{H^\pm}$ larger than 
$M_t$\footnote{The case $M_{H^\pm} < M_t$ with the decay $t\to H^\pm b$
is discussed in Section~\ref{physics_top_profile}}, the dominant
production mode is pair production, $e^+e^-\to H^+ H^-$ with the dominant 
decay modes being $H^+ \to t \bar b$ with contributions 
from $H^+ \to \tau^+ \bar{\nu_\tau}$ and $H^+ \to W^+ h^0$
(see Fig.~\ref{fig:2107}).
The cross--section depends mainly on the charged Higgs boson
mass $M_{H^{\pm}}$ and is of the order of 15\,fb for $M_{H^{\pm}}$ = 
300\GeV\ at $\sqrt{s}$ = 800\GeV\ (see Fig.~\ref{fig:2108}).
The radiative corrections do significantly 
change these results~\cite{hpmguasch}.

As an example of the performance of TESLA, a study has been made of the
$e^+e^- \rightarrow H^+H^- \rightarrow t \bar b \bar t b$ and the 
$e^+e^- \rightarrow H^+H^- \rightarrow W^+ h^0 W^- h^0$, 
$h \rightarrow b \bar b$ processes with $M_{H^{\pm}}$ = 300\GeV,
$M_h$ = 120\GeV\ and $\sqrt{s}$ = 800\GeV~\cite{hpm300}.
\begin{figure}[ht!]
\begin{center}
\begin{tabular}{c c}
\href{pictures/1/fig2301a.pdf}{{\epsfig{file=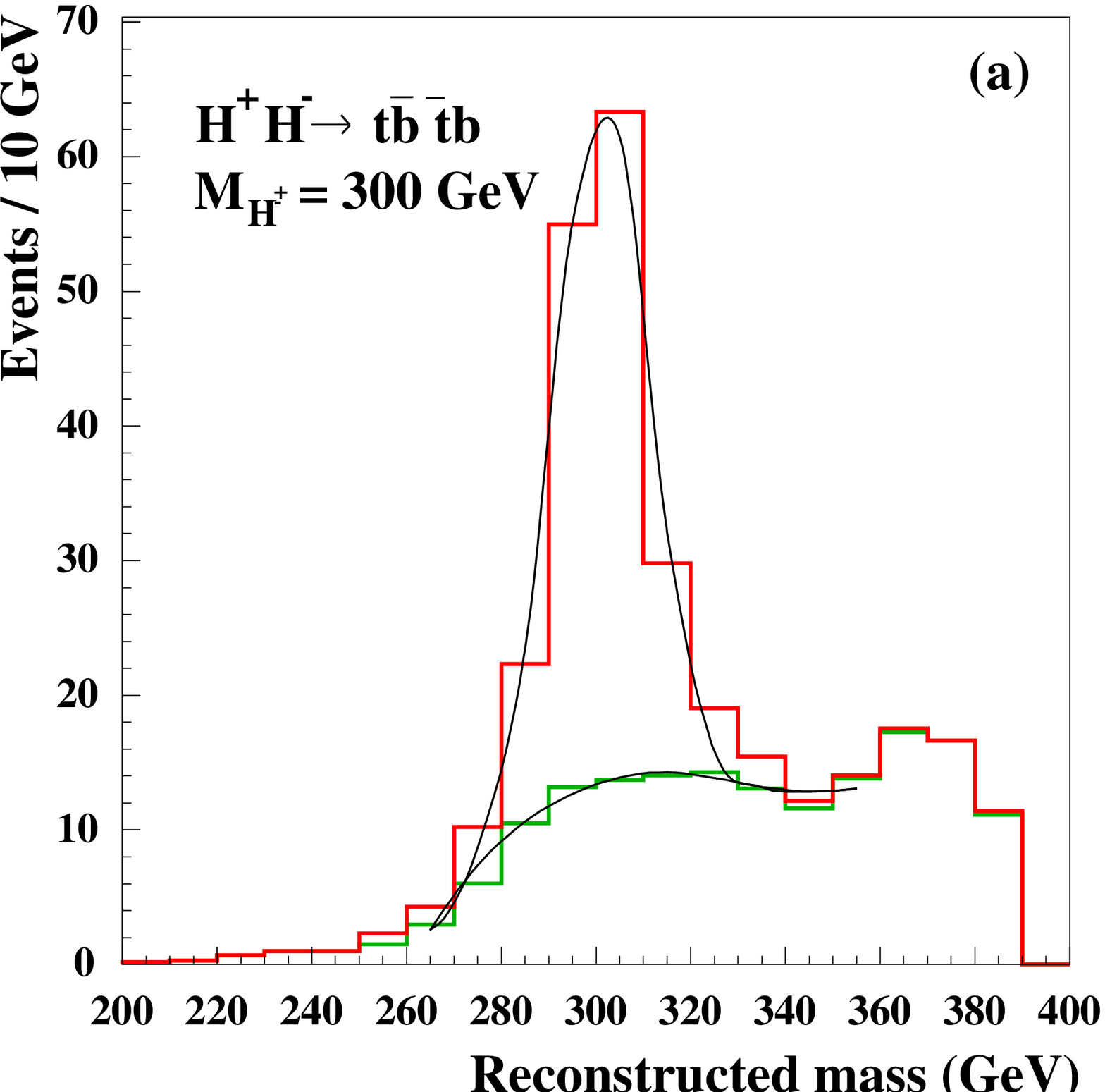,height=0.45\linewidth,width=0.45\linewidth,clip}}} &
\href{pictures/1/fig2301b.pdf}{{\epsfig{file=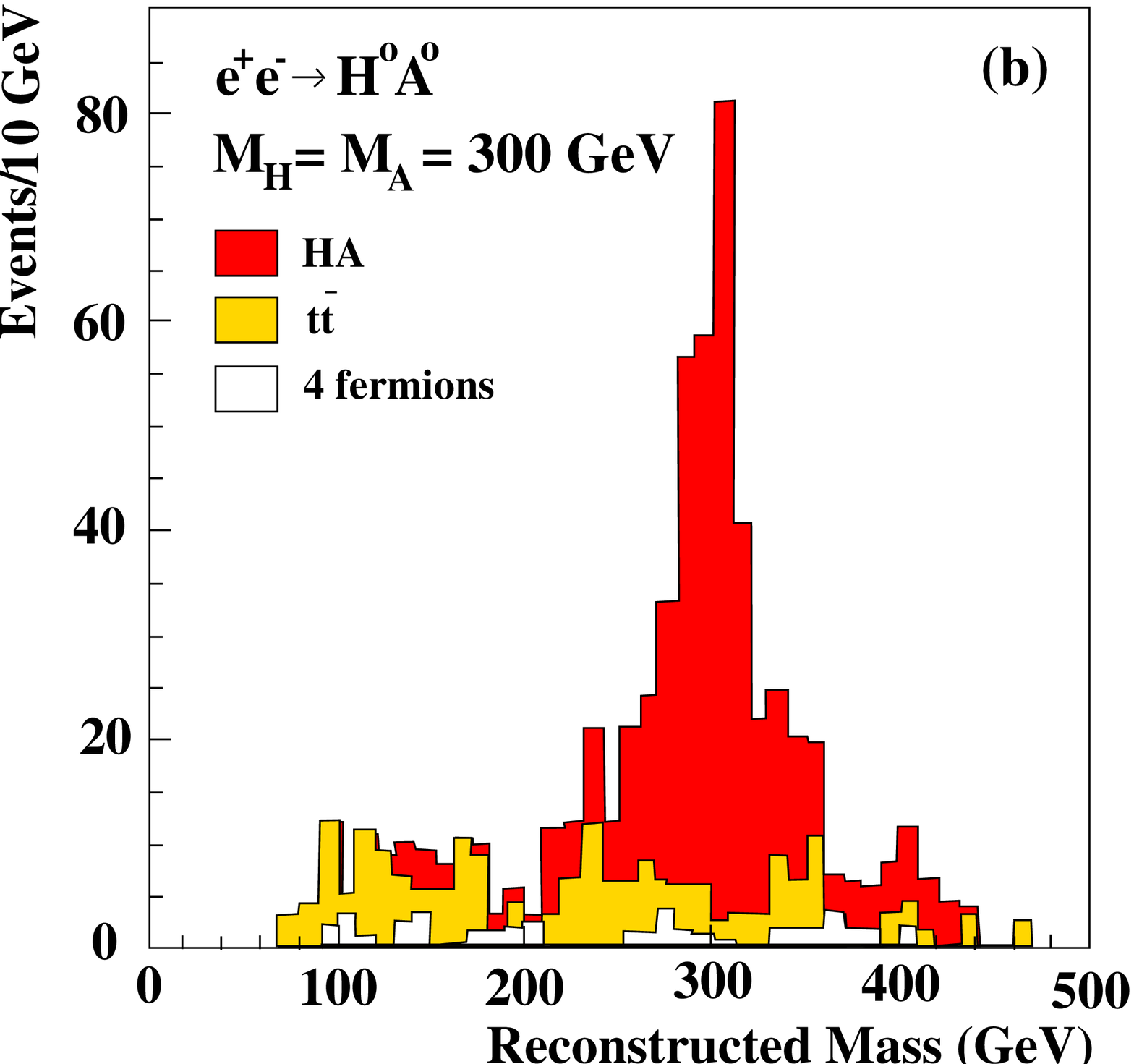,height=0.49\linewidth,width=0.49\linewidth,clip}}}\\
\end{tabular}
\caption{ \label{fig:hpm}
a): the di--jet invariant mass distribution for (left) $e^+ e^- 
\rightarrow H^+H^- \rightarrow t \bar b \bar t b$ candidates after 
applying the intermediate $W$ and $t$ mass and the equal mass final state 
constraints for 500\,fb$^{-1}$ at $\sqrt{s} = $ 800\GeV. b):
Mass peak for $e^+ e^- \rightarrow H^0A^0 \rightarrow b \bar b b \bar b$
for 50\,fb$^{-1}$ at $\sqrt{s} = $ 800\GeV.}
\end{center}
\end{figure} 
In the resulting 8~jet final state with 4~$b$-quark jets it is possible to 
beat down the backgrounds to a low level by using $b$~tagging and mass
constraints on the intermediate $t$, or $h^0$, and $W$. The combinatorial 
background due to jet-jet pairing ambiguities in signal events can be 
resolved, since $b$-tagged jets can not come from the $W$ decays. Using the 
$t$ and $W$ mass constraints, the estimated resolution on the charged Higgs 
boson mass is 10\GeV. 
Assuming an integrated luminosity of 500\,fb$^{-1}$, the analysis gives
120~signal events on an estimated background of 50 
misreconstructed events (see Fig.~\ref{fig:hpm}\,a)). 
The product $\sigma(e^+e^- \rightarrow H^+H^-) \times 
{\mathrm{BR}}(H^+H^- \rightarrow t \bar b \bar t b)$ or ($W^+ h^0 W^- h^0$) 
and the charged Higgs boson mass $M_{H^{\pm}}$ are obtained from a likelihood 
fit to the reconstructed mass distribution with the number of signal events,
the mass $M_{H^{\pm}}$ and the mass resolution as free parameters. 
The resulting
statistical uncertainty on the mass is $\pm$1\GeV, and that on
the product of the production cross--section with the branching ratio 
$\sigma(e^+e^- \rightarrow H^+H^-) \times {\mathrm{BR}}(H^+H^- \rightarrow t 
\bar b \bar t b)$ or ($W^+ h^0 W^- h^0$) is smaller than 15\%.  

The two neutral heavy Higgs bosons in SM extensions with an additional
doublet can be produced in the pair production process $e^+e^- \rightarrow 
H^0 A^0 \rightarrow b \bar b b \bar b$. This has been studied for $\sqrt{s}$ =
800\GeV\ in the decoupling limit where their masses become almost 
degenerate~\cite{ha340}.
The tagging of the characteristic four $b$-jet final state reduces the large 
$e^+ e^- q \bar q g g$ and $t \bar t$ backgrounds significantly. The $HA$ 
production is already observable for masses up to 340\GeV\ with only 
50\,fb$^{-1}$ (see Fig.~\ref{fig:hpm}). 
A determination of their mass with a relative accuracy of 
$\delta M_A / M_A$ = 0.2--0.4\% and of the product 
$\sigma(e^+e^- \rightarrow H^0 A^0) \times \mathrm{BR}(A^0 \rightarrow b \bar b) \times
\mathrm{BR}(H^0 \rightarrow b \bar b)$ = 5\% - 11\% can be obtained with 200\,fb$^{-1}$ 
for $260\GeV < M_A < 340\GeV$.

The heavy Higgs bosons $A^0, H^0, H^\pm$ are produced at $e^+e^-$ colliders
primarily in pairs and they can be discovered for masses close to the beam
energy. The range for $A^0, H^0$ can be extended into regions not accessible
at the LHC at the $\gamma\gamma$ collider where they are formed as 
single resonances.

\subsection[Indirect determination of the SM/MSSM nature of a light Higgs]{Indirect determination of the SM/MSSM nature of a light Higgs 
boson \label{sec:2.3.2}}

The discovery of a neutral Higgs boson, with mass in the range 114\GeV\ $<
M_H \lsim $ 140\GeV, will raise the question of whether the observed particle
is the SM Higgs or the lightest boson from the Higgs sector of a SM extension.
It has been shown that, for a large fraction of the $\tan \beta - M_A$ 
parameter plane in the MSSM, this neutral boson is the only Higgs state
observable at the LHC. 
Supersymmetric particles will most probably be observed at both the LHC
and TESLA. However, it is difficult to shade light on 
the structure of the supersymmetric Higgs sector with only one Higgs
boson observed. In this situation the precision measurements of the 
Higgs boson couplings are powerful to obatin information about additional
Higgs doublets, their structure and even the masses of the heavier Higgs
boson states. This will be exemplified in the context of the MSSM in the
following.

If the $H^0ZZ$ coupling, measured by the 
Higgs-strahlung production cross--section 
independently from the Higgs boson decay mode, turns out to be significantly 
smaller than the SM expectation, this will signal the existence of extra Higgs
doublets or other new physics.

The determination of the Higgs boson branching ratios with the accuracy
anticipated by these studies can be employed to identify the SM or MSSM 
nature of a light neutral Higgs boson. 
The Higgs boson decay widths $\Gamma^{MSSM}$ to a specific final state
are modified as follows with respect to the SM $\Gamma^{SM}$:
$\Gamma^{MSSM}_{b \bar b} \propto \Gamma^{SM}_{b \bar b} 
(\sin^2 \alpha/\cos^2 \beta)$ and
$\Gamma^{MSSM}_{c \bar c} \propto \Gamma^{SM}_{c \bar c} 
(\cos^2 \alpha/\sin^2 \beta)$.
Therefore, deviations in the ratios of branching ratios such as
{$\mathrm{BR}(h \rightarrow W W^*)/\mathrm{BR}(h \rightarrow b \bar b)$~\cite{borisov},
$\mathrm{BR}(h \rightarrow c \bar c)/\mathrm{BR}(h \rightarrow b \bar b)$ and
$\mathrm{BR}(h \rightarrow g g)/\mathrm{BR}(h \rightarrow b \bar b)$~\cite{kamoshita}
from their SM 
expectations can reveal the MSSM nature of the Higgs boson and also provide 
indirect information on the mass of the $\cp$-odd $A^0$ Higgs boson, even when
it is so heavy that it can not be directly observed at $\sqrt{s}$ = 500\GeV. 

In particular, it has been shown that the accuracy obtained at TESLA for 
$\mathrm{BR}(h \rightarrow W W^*)/\mathrm{BR}(h \rightarrow b \bar b)$, 
implies a 
statistical sensitivity to the MSSM up to $M_A \simeq$ 1\,TeV~\cite{borisov}. 
This may also be complemented by the high precision electroweak data from the 
GigaZ operation (see Section~\ref{sec:gaugebosons}).

To fully account for the sensitivity provided by different accessible 
branching ratios as well as the theoretical uncertainties on the SM branching 
ratio predictions, a complete scan of MSSM parameter phase space has been 
performed~\cite{mba}.

For each set of parameters, the $h^0$ mass has been computed using the
diagrammatic two-loop result~\cite{mhiggsletter}. Solutions corresponding to 
$M_{h^0} = (120 \pm 2)$\GeV\ have been selected and used to compute the 
$h^0$ branching ratios taking into account the dominant loop corrections
(including those arising from supersymmetric particles)~\cite{hdecay}.
The deviations from the SM predictions for
BR($h \rightarrow b \bar b$)/BR($h \rightarrow$ hadrons),
BR($h \rightarrow c \bar c$)/BR($h \rightarrow$ hadrons),
BR($h \rightarrow g g$)/BR($h \rightarrow$ hadrons) and
BR($h \rightarrow b \bar b$)/BR($h \rightarrow W W^*$) have been used to 
investigate the SM/MSSM discrimination. For $M_A \lsim 750\GeV$, 68\%
of all MSSM solutions can be distiguished from the SM and for 
$M_A \lsim 600\GeV$, 95\% of all MSSM solutions can be distiguished
from SM at the 95\% confidence level. This confidence level is derived
from a $\chi^2$ test which compares the deviation of the above-mentioned 
ratios in the MSSM from their SM values and accounts for their
uncertainties.

If a significant deviation from the SM has been observed, 
% which implies that $M_A$ is within the limit of Figure~\ref{fig:matgb}, 
it is possible to go further and use the accurate measurements of 
the Higgs boson decays estimate $M_{A^0}$ in the framework of the MSSM. 
By varying the $A^0$ mass together with the other MSSM parameters within the 
range compatible with experimental and theoretical uncertainty on the 
branching ratios. The range of values of $M_A$ for the accepted MSSM solutions 
corresponds to an accuracy of 70\GeV\ to 100\GeV\ for the indirect
determination of $M_A$ in the mass range 300\GeV\ $ < M_A <$ 600\GeV~\cite{mba}.

The SUSY contributions considered above enter via the dependence on 
$\tan\beta$ and the mixing angle $\alpha$ and affect the Higgs couplings to 
all up-type fermions and to all down-type fermions in a universal way. 
However, for large values of $\tan\beta$ and/or of the Higgs mixing parameter 
$\mu$, gluino and higgsino loop corrections can also induce important SUSY 
effects. They affect the tree-level relations between 
the fermion masses and the Yukawa couplings~\cite{deltab}, thus inducing 
further deviations in particular for the ratio of the $g_{h b \bar b}$ to 
the $g_{h \tau^+ \tau^-}$ couplings. 
A determination of BR($h \to \tau^+ \tau^-$) to the accuracy 
anticipated at TESLA can probe these effects and will thus enhance the 
sensitivity to differences between the SM and the MSSM in this region of the 
SUSY parameter space. 

If a light Higgs boson is observed and found to correspond to the decay 
properties expected for the lightest neutral Higgs boson in MSSM, and if a 
light $A^0$ boson exists, the associated production with 
$e^+e^- \to b\bar{b} A^0$ could be observed with a significant cross--section 
in the MSSM at large values of $\tan \beta$. In such a case, this process 
allows for a direct determination of the important $\tan \beta$ parameter. 
An experimental study has been
performed for 500\,fb$^{-1}$ at $\sqrt{s}$ = 500\GeV\ using an iterative 
discriminant analysis technique.
% This analysis is quite delicate
% because it relies on a small signal to be separated from the $b \bar b b 
% \bar b$ background and the interference effect. 
The resulting uncertainty on 
$\tan \beta$ for $\tan \beta$ = 50 has been estimated to be 7\% 
for $M_A =$~100\GeV~\cite{bbh}. 

\section{Non SUSY Extension of the SM \label{sec:2.4}}

\subsection{Higgs detection in 2HDM \label{sec:2.4.1}}

\label{2hdm}
Models abound in which the 2HDM extension of the SM (without
supersymmetry) is the effective theory, 
correct up to some new physics scale, $\Lambda$~\cite{okada}.
We focus on the ${\cal CP}$--conserving 2HDM of type II, as defined earlier, 
with eigenstates $\hl$, $\hh$, $\ha$ and $\hpm$.
The phenomenology of the neutral Higgs bosons in the 2HDM is essentially 
determined by the parameters $\tan{\beta}$ and $\alpha$ and the Higgs
boson masses. 
In constrast to the supersymmetric models, these parameters are not 
correlated, so that the no loose theorem does not apply.
If $h^0$ production is kinematically accessible through
the $e^+e^-\to h^0Z$ process, it can be observed at TESLA unless the 
$ZZh^0$ coupling is heavily supressed, i.e.~$\sin(\beta-\alpha)$
is very small (${\cal O}(10^{-3})$). 
Precision measurements of the $h^0$ properties, in particular the 
$e^+e^-\to h^0Z$ and $\gamma\gamma\to h^0$ production cross--sections, can
reveal the nature of the model~\cite{mariatwogam}.
In this situation, should either $A^0$ or $H^0$ be
light enough, Higgs bosons will be observed in the 
$e^+e^-\to h^0A^0$ and $e^+e^-\to H^0Z$ processes, since 
the relevant couplings are proportional
to $\cos{(\beta-\alpha)}$ and thus large.

The most difficult situations to probe at TESLA are:
(a) that of small $\sin(\beta-\alpha)$ and the lightest Higgs boson 
is $h^0$ while  
$A^0$ and $H^0$ too heavy to be produced via the above mentioned processes, 
(b) the lightest Higgs boson is  $A^0$ while $h^0$ and $H^0$ too heavy. 
The case (b) is discussed in the following as an example, while the case (a) 
is discussed in~\cite{gckz}.
In this situation alternative production processes have 
to be considered since the loop--induced $ZA^0A^0$ coupling will also be too 
small~\cite{ghsmall}. The most improtant process is the
Yukawa processes $e^+e^-\to f\bar{f} A^0$. 
The Yukawa couplings are proportional 
to $\tan{\beta}$ for down-type quarks and charged leptons and to $\cot{\beta}$
for up-type quarks in the $\sin(\beta-\alpha)\to 0$ limit, hence they can not 
both be supressed simultaneously~\cite{twolight}. Furthermore, the 
$e^+e^-\to A^0A^0Z$ and $\gamma\gamma\to A^0$ channels can contribute.

\begin{figure}[ht!]
%\begin{picture}(150,10)\unitlength 1mm
%  \put(68,-6){a)}
%  \put(125,-7){b)}
%\end{picture}
%\vspace*{-8mm}
\begin{center}
\href{pictures/1/fig2401.pdf}{{\epsfig{file=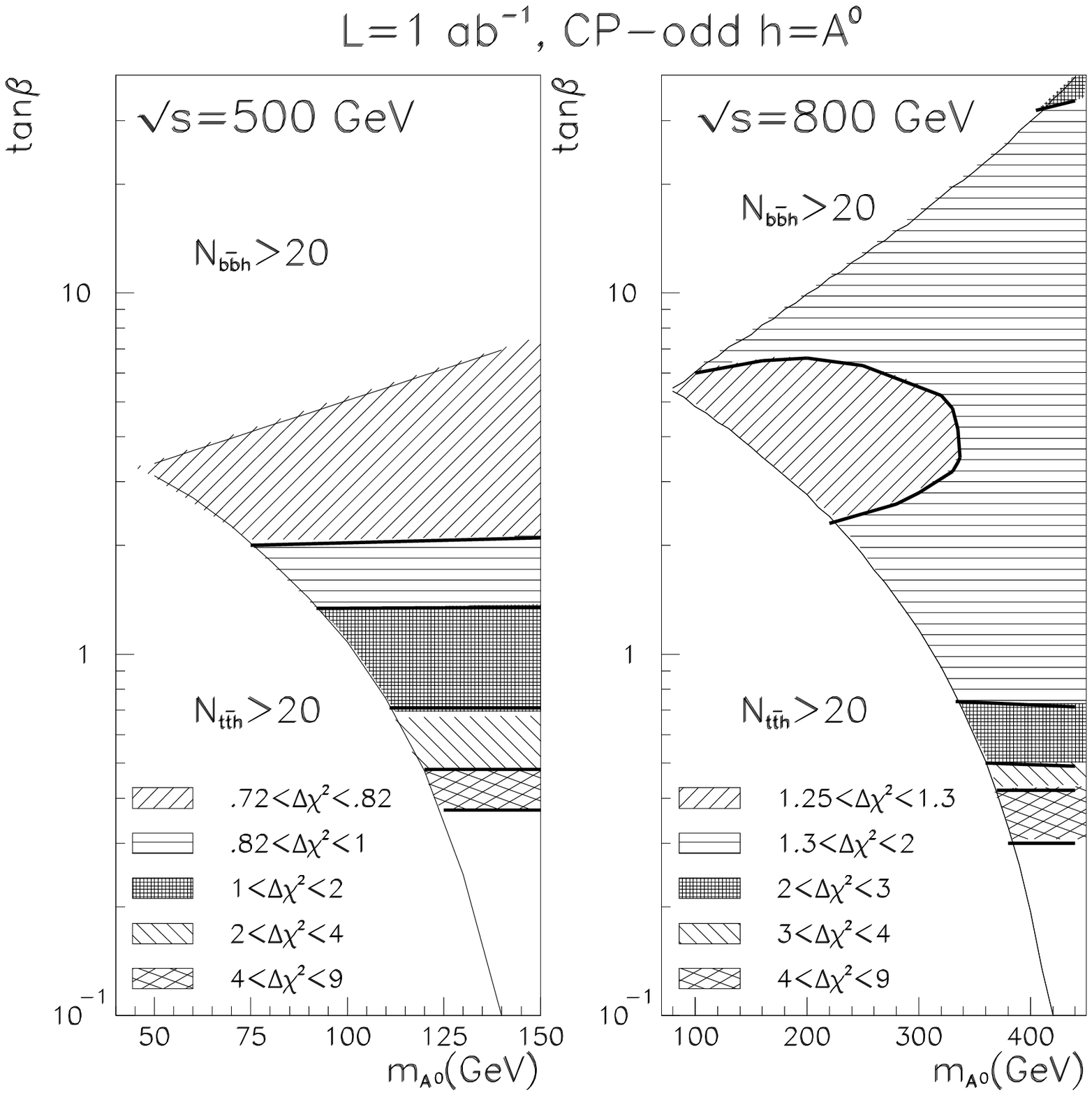,height=0.47\linewidth,width=0.75\linewidth}}}
\caption{\label{lctdr_regions}
For $\protect\rts=500\gev$  and $\protect\rts=800\gev$,
the solid lines show
as a function of $M_A$ the maximum and minimum $\tanb$
values between which $t\anti t \ha$, $b\anti b \ha$ 
final states will both have fewer than 20 events assuming
$L=1\abi$. The different regions indicate
the best $\dchisq$ values (relative to the best SM $\chi^2$) 
obtained for fits to the present precision electroweak
data after scanning: a) over the masses of the remaining Higgs bosons 
subject to the constraint they are 
too heavy to be directly produced; and b) over the mixing angle
in the ${\cal CP}$--even sector. Results are shown only for
$M_A<\protect\rts-2\mt$, but extrapolate to higher $M_A$ in obvious fashion.}
\end{center}
\end{figure}

Studies have been performed to investigate to what extent this particular 
scenario could be observed or excluded with TESLA running in the 
GigaZ mode~\cite{kzm} and at high energy~\cite{gckz}. While significantly 
larger regions in the $m_A$--$\tan{\beta}$--plane are accessible than at LEP, 
there are regions remaining
for which a luminosity of 1000\,fb$^{-1}$ is not sufficient 
to guarantee discovery.

In Fig.~\ref{lctdr_regions} the regions for $A^0$ which cannot be accessed at 
$\sqrt{s} = 500$\GeV\ and 800\GeV\ with 1000\,fb$^{-1}$, respectively, are shown. 
A minimum production of 20 events in either the $b\bar{b}A^0$ 
or the $t\bar{t}A^0$ 
process is assumed as an optimistic observability criterion~\cite{twolight}.
In the problematic regions $\gam\gam\to A^0$ production is also
unlikely to produce a detectable signal for the expected luminosities.

The $A^0A^0 Z$ and $\wp\wm\to A^0A^0$ processes~\cite{gang}
are sensitive up to $M_A<155\gev$ ($<250\gev$)
at $\rts=500\gev$ ($800\gev$) for  $1<\tanb<50$ with $L=1\abi$, 
assuming that 20 events will be adequate for observation.
The other Higgs boson masses are assumed to be larger than $\sqrt{s}/2$ such 
that they can not be pair produced.

%An observation in the remaining parameter region would only be possible at 
%either much higher luminosity or at a multi-TeV collider, where the heavy 
%Higgs states become kinematically accessible.

Surprisingly, in these scenarios, 
the parameters for the other (heavy) Higgs bosons can be chosen so
that the fit to the present precision electroweak observables is nearly as good
as that obtained with a light SM Higgs boson, despite the
fact that the ${\cal CP}$--even Higgs boson with substantial
$WW,ZZ$ couplings is heavier than $\rts$~\cite{gckz}.
This is illustrated
in Fig.~\ref{lctdr_regions} for the case of $A^0$ being the lightest
Higgs boson. The $\Delta\chi^2$ values between the best 2HDM and SM
precision electroweak fits are seen to obey 
$\Delta\chi^2< 2$  in the $\rts=500\gev$ and
$\rts=800\gev$ $L=1\abi$  `no-discovery' wedges 
when $\tanb>0.7$. With increased precision of the electroweak data
from a GigaZ run, the sensitivity to this scenario increases significantly
(see Section~\ref{sec:gaugebosons}).

A third generation linear collider with sufficient centre--of--mass energy could
then completely reveal the Higgs states by observing not only $ZH^0$
and/or $W^+W^-\to H^0$ production 
but also $\hl\ha$ production (regardless of which 
is light) and possibly $\hp\hm$ production.  
        
\subsection{Higgs boson detection in the Stealth Model \label{sec:2.4.2}}

A possible extension of the SM consists of the introduction of singlet Higgs 
particles. Since these particles do not couple directly to ordinary matter,
their existence is unconstrained by the precision electroweak data.
Owing to these characteristics, such particles are a suitable candidate for
dark matter and may also play an important role in the phenomenology of 
technicolor models or theories with higher dimensions. 
\begin{figure}[ht!]
\begin{picture}(150,10)\unitlength 1mm
  \put(29,5){a)}
  \put(87,5){b)}
\end{picture}
\vspace*{-8mm}
\begin{center}
\href{pictures/1/fig2402.pdf}{{\epsfig{file=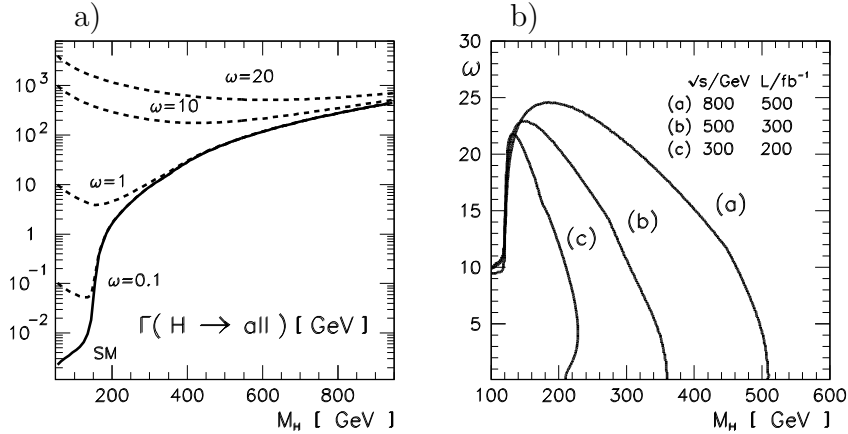,width=0.72\linewidth}}}
\caption{ \label{fig:stealth}
The Higgs boson width for several values of the coupling strength 
$\omega$ compared to (a) the SM width  and  (b) the exclusion limits achievable 
at TESLA in the $M_H - \omega$ plane.}
\end{center}
\end{figure}
Since the SM Higgs boson has direct interactions of strength $\omega$ 
with these singlet Higgs 
particles, it can decay into a pairs of Higgs singlets generating experimentally
invisible decay modes. Further, the invisible decay width of the Higgs boson
may be 
sizeable, generating a wide Higgs state that would not be detected as a narrow 
peak in the recoil mass spectrum discussed earlier and would also escape 
detection at the LHC. 
Still the signal is observable at TESLA as an excess of events over the 
precisely known SM backgrounds. Such a scenario is known as the stealth Higgs 
model~\cite{hstealth}. Fig.~\ref{fig:stealth} shows the range of Higgs boson
mass values detectable at TESLA for different values of the Higgs coupling 
$\omega$ to light invisible matter particles.

\section{The Complementarity with the LHC \label{sec:2.5}}

In proton--proton collisions at $\sqrt{s}=14$\,TeV at the LHC, Higgs bosons
are mainly produced through the loop induced gluon-gluon fusion mechanism;
the contributions from the associated $WH$, $ZH$, $t\bar{t} H$
and $WW$/$ZZ$ fusion production processes are also relevant. 
The ATLAS~\cite{h_atlas} and
CMS~\cite{h_cms} experiments have shown, that they are sensitive to
the SM Higgs boson over the whole mass range of 100 -- 1000\GeV.
In the  range 100\GeV\ $ < M_H < $ 130\GeV\ Higgs bosons will be searched 
for in the $H^0\to b\bar{b}$ and $H^0\to\gamma\gamma$ decay modes, 
while for larger masses the $H^0\to ZZ^{0(*)}$ and 
and $H^0\to W W^{(*)}$ will take over.
After combining different channels and results 
from two experiments in the whole mass range a 5$\sigma$ significance 
can be reached already with an integrated luminosity of 10\,fb$^{-1}$ 
per experiment. 
The expected number of Higgs bosons varies from 50 events for 
$t\bar{t}H^0$, $H^0\to b\bar{b}$ and 30\,fb$^{-1}$ 
to about 1000 events for $H^0\to\gamma\gamma$ and 100\,fb$^{-1}$, 
expected in each experiment. 

% In proton--proton collisions at $\sqrt{\mathrm{s}} = 14 $ TeV at the LHC,
% light Higgs bosons are mainly produced through the loop induced gluon--gluon 
% fusion mechanism with additional important 
% contributions from associated production mechanisms
%   $WH$, $ZH$ and $t\bar{t}H$ and from $WW$ fusion. 
% The ATLAS and CMS experiments have shown that the 
% SM Higgs boson can be discovered over the 
% whole theoretically allowed mass range
% 110\GeV/$ < M_H <$ 700\GeV
% with convincing significance~\cite{caner} with an integrated luminosity of 
% 30\,fb$^{-1}$. 

There is a variety of channels in which the MSSM Higgs boson can be
discovered. The lightest Higgs boson can be discovered in the same 
decay modes as the SM Higgs boson of the same mass. It 
might be also observed in the cascade decays of SUSY particles, namely 
$\chi^0_2\to h^0 (\to b\bar b) \chi^0_1$.
%  As
% the heavy Higgs H and A couplings to ZZ/WW bosons get suppressed,
% observability of the new decay modes became possible eg. H/A-->tautau,
% mumu for large tanb. The charged Higgs boson will be observable in the
% H-->tau nu and H-->tb decay modes for low and/or tanb beta. For low tanb,
% almost already excluded by LEP, also H-->hh and A-->Zh modes would be
% accessible.  
At least one Higgs
boson can be discovered for the whole parameter range. In a fraction
of the parameter space, more than one Higgs boson is accessible. However,
there is a region (see Fig.~\ref{fig:2501}),
in which the
extended nature of the supersymmetric Higgs sector might not be observable, 
unless cascade decays of supersymmetric particles into the Higgs bosons
are accessible, since only the lightest Higgs boson can be seen 
in SM--like production processes.

Beyond its discovery, a limited number of measurements 
of Higgs boson properties can be carried out at the LHC.
Combining results from both experiments with
an integrated luminosity of 100\,fb$^{-1}$, the Higgs boson mass can be
measured with an accuracy of few permil over the whole mass range
and the total decay
width with an accuracy of about 10\% only for large masses, $M_H > 300 $\GeV.
% Precision better than 1\% can be reached for the measurements 
% of the Higgs masses in the MSSM sector and below 10\% 
% for the determination of tanb. 
Further perspectives for an indirect measurement of the total Higgs 
boson width at lower $M_H$ have been recently 
proposed~\cite{zeppenfeld, sopczak}, and their experimental
feasibility is presently under investigation by both LHC collaborations. 
Beyond that, the ratio of couplings $g_{HWW}/g_{HZZ}$ can be measured
for $M_H \gsim 160 $\GeV\ and $g_{Htt}/g_{HWW}$ for $M_H \lsim 120 $\GeV
(see Table~\ref{tab:lhc1} and Fig.~\ref{fig:lhclcgh}).

% However beyond its discovery, measurements of the Higgs boson properties at 
% the LHC are difficult due to limited signal statistics and/or 
% large backgrounds and systematic uncertainties on parton densities.
% While the LHC may provide some ratios of branching ratios
% and couplings, as listed in Table~\ref{tab:lhc1} for $M_H = 120$\GeV,
% precision measurements of the absolute branching ratios and couplings can only
% be addressed at a high luminosity linear collider, such as TESLA as
% exemplified in Figure~\ref{fig:lhclcgh}.

% Further perspectives for measurements of the total Higgs boson width and of 
% the $g_{HWW}$ coupling have been recently 
% proposed for the LHC~\cite{zeppenfeld} but are still waiting to be confirmed 
% by both experiments, with full account taken of the backgrounds and detector 
% response.

% A summary of MSSM Higgs boson measurements which can be carried out at LHC
% is summarised in the ATLAS TDR~\cite{atlastdr}. Over a large range of the 
% parameter space, only the lightest ${\cal CP}$--even Higgs boson $h^0$ is
% observable. In this region, the precise measurement of the Higgs boson 
% branching ratios at TESLA is essential to determine the parameters of the 
% Higgs sector and reveal its supersymmetric nature (see Fig.~\ref{fig:2501}).
\begin{figure}[htbp]
\begin{center}
\href{pictures/1/fig2501.pdf}{{\epsfig{file=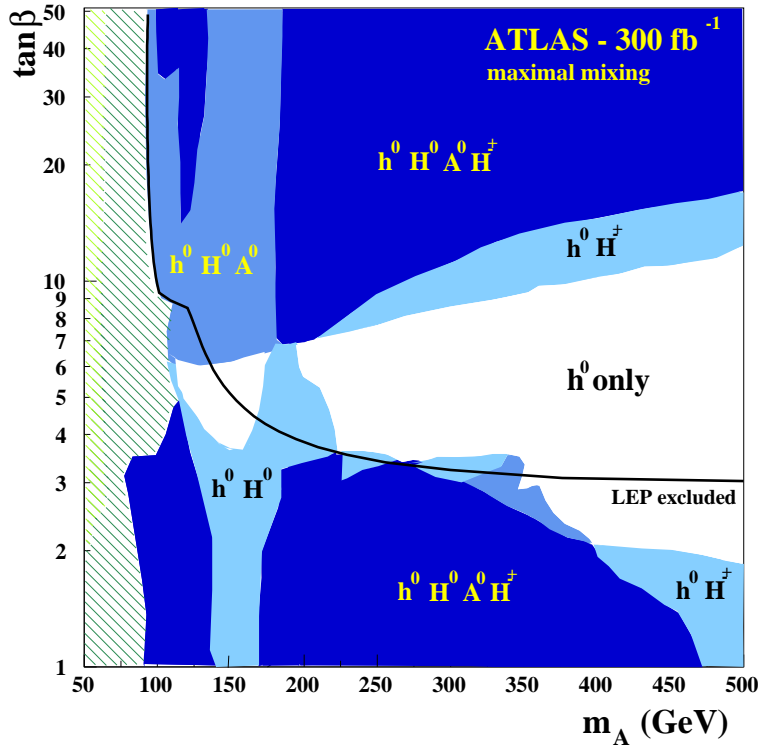,width=0.65\linewidth,clip}}}
\caption{\label{fig:2501}
Higgs bosons which are observable in the ATLAS experiment with 300\,fb$^{-1}$
in the maximal mixing scenario of the MSSM in the plane of
$\tan\beta$ vs. $M_A$. In the white region only the lightest $h^0$ boson 
is observable at the LHC if only SM--like decays are accessible.
With TESLA, the $h^0$ boson can be distinguished from
the SM Higgs boson through the accurate determination of its couplings
and thus reveal its supersymmetric nature.}
\end{center}
\end{figure}
\begin{figure}[hbtp]                           
\begin{center}
\href{pictures/1/fig2502.pdf}{{\epsfig{file=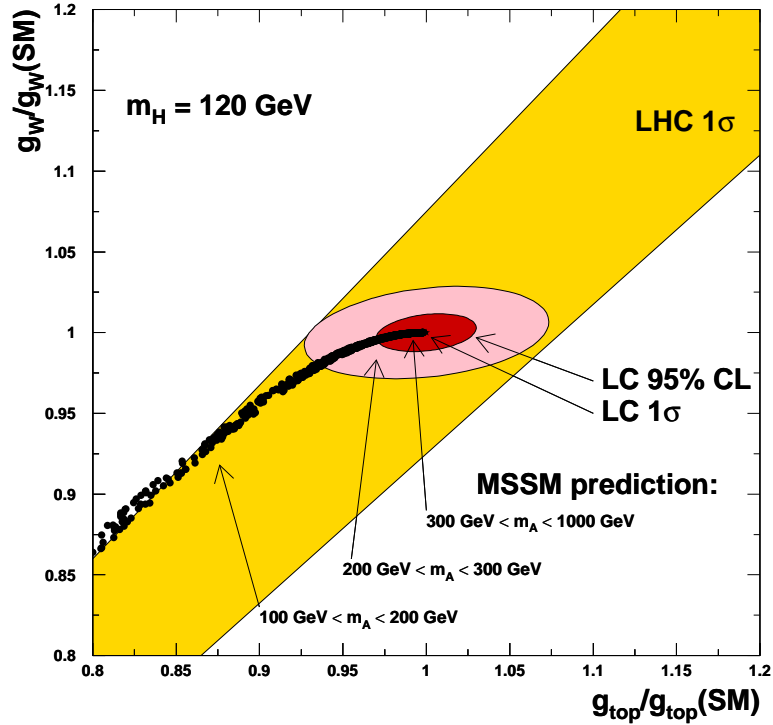,width=0.65\linewidth}}}
\caption{ \label{fig:lhclcgh}
A comparison of the accuracy in the determination of the 
$g_{ttH}$ and
$g_{WWH}$ Higgs couplings at the LHC and at TESLA compared to the predictions
from MSSM for different values of the $M_A$ mass.}
\end{center}
\end{figure}

It is very clear that
the precise and absolute measurement of all relevant Higgs boson
couplings can only be performed at TESLA.
Furthermore, the unambiguous determination of the quantum numbers of
the Higgs boson and the high sensitivity to ${\cal CP}$--violation
represent a crucial test. The measurement of the Higgs self coupling
gives access to the shape of the Higgs potential.
These measurements together will allow to establish the Higgs meachnism as the
mechanism of electroweak symmetry breaking.

At TESLA, extended Higgs sectors as present in supersymmetric 
thoeries can be distinguished from the SM Higgs sector with 
little assumptions about
their precise structure. As an example, in the MSSM the $h^0$ boson can be 
distingished from SM Higgs boson over the whole parameter region shown
in Fig.~\ref{fig:2501}. Heavy Higgs bosons can studied if kinematically
accessible with high precision.

Almost any conceivable extended Higgs boson scenario can be seen at
TESLA. 
In particular any Higgs boson, which couples to the $Z$ boson can be
observed in $ZH^0$ production through the recoil mass method, 
independent of its decay. Therefore,
TESLA is able to close possible loopholes, if they exist in the LHC
discovery potential (e.g. the accessibility of invisibly decaying Higgs
bosons was not confirmed so far by the LHC collaborations).

In summary, TESLA has the unique
opportunity to study Higgs bosons with high precision in all essential
aspects. These measurements will provide the information necessary to
reveal the mechanism of electroweak symmetry breaking and mass generation.

\clearpage

\begin{table}[htb!]
\begin{center}
{\renewcommand{\arraystretch}{1.2}
\begin{tabular}{|l|c|c|c|}
\hline 
         & $M_H$            & $\delta(X)/X$ & $\delta(X)/X$ \\
         & (GeV)            & LHC & LC \\
         &                  & 2 $\times$ 300\,fb$^{-1}$ & 500\,fb$^{-1}$ \\ 
 \hline \hline
  $M_H$      & 120           & ~9 $\times 10^{-4}$ & 3 $\times 10^{-4}$ \\
  $M_H$      & 160           & 10 $\times 10^{-4}$ & 4 $\times 10^{-4}$ \\
  $\Gamma_{tot}$ & 120-140 &   -     & 0.04 - 0.06 \\ \hline
  $g_{Hu\bar u}$ & 120-140 & - & 0.02 - 0.04 \\
  $g_{Hd \bar d}$ & 120-140 & - & 0.01 - 0.02 \\
  $g_{HWW}$ & 120-140 & - & 0.01 - 0.03 \\ \hline
  $\frac{g_{Hu \bar u}}{g_{H d \bar d}}$ & 120-140 & - & 0.023-0.052  \\
  $\frac{g_{Hb \bar b}}{g_{HWW}}$ & 120-140 & - & 0.012-0.022  \\
  $\frac{g_{Ht \bar t}}{g_{HWW}}$ & 120 & 0.070 & 0.023  \\
\rule[-3mm]{0mm}{5.5mm}  $\frac{g_{HZZ}}{g_{HWW}}$ & 160 & 0.050 & 0.022  \\ \hline
  ${\cal CP}$ test & 120   & - &  0.03  \\
  $\lambda_{H\!H\!H}$ & 120 & - & 0.22 \\ \hline 
 \end{tabular}
}
\caption[]{ \label{tab:lhc1}
Comparison of the expected accuracy in the determination of the 
SM-like Higgs profile at the  LHC and at TESLA. The mass, width, couplings to
up-type and down-type quarks and to gauge bosons, several of the ratios of 
couplings, the triple Higgs coupling and the sensitivity to a ${\cal CP}$-odd 
component are considered.}
\end{center}
\end{table}

% In summary, the complementarity of TESLA to the LHC in Higgs physics is
% threefold:
% \begin{enumerate}
% 
% \item{The absolute measurements of all relevant Higgs boson couplings 
%  (including the Higgs self coupling) at the percent level can only be 
% performed at TESLA.} 
% \item{The accuracy of the measurements which are possible at the LHC can be
% significantly increased.} 
% \item{Almost all conceivable extended Higgs boson scenarios can be observed 
% at TESLA. In particular any Higgs boson which couples to the $Z$ boson can
% be observed in $ZH^0$ production through the recoil mass method. 
% In case LHC should not observe any Higgs boson, TESLA will be able to 
% close the loopholes such as invisible Higgs decays.}
% \end{enumerate}

{\raggedright}

  \cleardoublepage
%------------------------------------------------------------------
\chapter{Supersymmetry}}
\label{physics_SUSY}
%------------------------------------------------------------------
  
  %
% TESLA TDR - SUSY chapter
% latex definitions
% last update 25 Jan 2001
%

\def\noi {\noindent}

\def\a               {\alpha}
\def\b               {\beta}
\def\g               {\gamma}
\def\d               {\delta}
\def\l               {\lambda}
\def\s               {\sigma}
\def\t               {\theta}
\def\x               {\chi}
\def\G               {\Gamma}
\def\D               {\Delta}

\def\ti    {\tilde}
\def\sf    {{\ti f}}
\def\sq    {{\ti q}}
\def\st    {{\ti t}}
\def\sb    {{\ti b}}
\def\stau  {{\ti\tau}}
\def\snu   {{\ti\nu}}
\def\sl    {{\ti \ell}}
\def\ch    {\ti \x}
\def\nt    {\ti \x^0}
\def\sg    {\ti g}

\def\cth   {\cos\theta}
\def\sth   {\sin\theta}
\def\tsf   {\theta_{\ti f}}
\def\tst   {\theta_{\ti t}}
\def\tsb   {\theta_{\ti b}}
\def\tstau {\theta_{\ti\tau}}
\def\csf   {\cos\theta_{\ti f}}
\def\cst   {\cos\theta_{\ti t}}
\def\csb   {\cos\theta_{\ti b}}
\def\cstau {\cos\theta_{\ti\tau}}

\providecommand{\msf}[1]   {m_{\ti{f_{#1}} }}
\providecommand{\mst}[1]   {m_{\ti{t_{#1}} }}
\providecommand{\msb}[1]   {m_{\ti{b_{#1}} }}
\newcommand{\mstau}[1] {m_{\ti {\tau_{#1} }}}
\newcommand{\mnt}[1]   {m_{\ti {\chi^0_{#1}} }}
\newcommand{\mch}[1]   {m_{\ti {\chi^+_{#1}} }}
\newcommand{\msnu}     {m_{\ti \nu}}
\newcommand{\msg}      {m_{\ti g}}

\def\sz  {\sqrt{2}}
\def\Pm  {{\cal P}_-^{}}
\def\Pp  {{\cal P}_+^{}}
\def\fbi {{\rm fb}^{-1}}
\providecommand{\cL}{{\cal L}}
\providecommand{\cP}{{\cal P}}

\newcommand{\ra}{\rightarrow}
\newcommand{\Ra}{\Rightarrow}

% Def. for figure references
\newcommand{\fig}[1]   {Fig.\,\ref{fig:#1}}
\newcommand{\figs}[1]  {Figs.\,\ref{fig:#1}}
           
% --- UM ---

\def\smu   {{\ti\mu}}
\def\smul  {{\ti\mu}_L}
\def\smur  {{\ti\mu}_R}
\def\se    {{\ti e}}
\def\sel   {{\ti e}_L}
\def\ser   {{\ti e}_R}

\newcommand{\rpv}{\slash\hspace{-2.5mm}{R}_{p}}
\renewcommand{\hdick}{\noalign{\hrule height1.4pt}}

% --- JFG ---
\newcommand{\dmchi}{\Delta m_{\tilde\chi_1}}
\newcommand{\mslash}{/ \hskip-8pt M}

% --- moreau ---
\def \Eslash {E \kern-.5em\slash }

%
% TESLA TDR - SUSY chapter
% Introduction to SUSY  
% last update 24 Jan 2001
%

\noindent Despite the enormous success of the Standard Model (SM) this cannot be the
ultimate wisdom to understand nature for many reasons. 
The introduction of Supersymmetry (SUSY) is considered the most attractive 
extension of the Standard Model~\cite{nilles}. 
Firstly, there are important theoretical motivations. 
It is the only non--trivial extension of the Poincar\'e group in 
quantum field theory. SUSY as a {\it local} symmetry becomes a 
supergravity (SUGRA) theory, incorporating gravity. SUSY appears in 
superstring theories, which may lead to the final theory of all fundamental 
interactions. From a phenomenological point of view, the most 
important feature of SUSY is that it can explain the hierarchy 
between the electroweak scale of $\sim$~100\,GeV, responsible for the 
$W$ and $Z$ masses, and the unification scale $M_{\rm GUT}$ $\simeq$ 
$10^{16}$\,GeV or the Planck scale $M_{\rm Pl}$ $\simeq$ 
$10^{19}$\,GeV. It also stabilises the Higgs mass with respect to 
radiative corrections, if $m_{\rm SUSY} \leq {\cal O}(1)$\,TeV. 
Moreover, the Minimal Supersymmetric Standard Model (MSSM) allows the 
unification of the gauge couplings of electroweak and strong 
interactions and yields precisely the measured value of $\sin^2\theta_W$. 
Furthermore, in the MSSM electroweak symmetry breaking is a natural result 
of renormalisation  group evolution. 
Another attractive feature of SUSY is that the lightest supersymmetric 
particle is a good cold dark matter candidate. 
Furthermore, a supersymmetric theory naturally contains extra sources 
of CP violation to ensure baryogenesis assuming an initially 
matter--antimatter symmetric universe. 

\begin{figure}[t] \centering
  \href{pictures/2/spectrum.pdf}{\epsfig{file=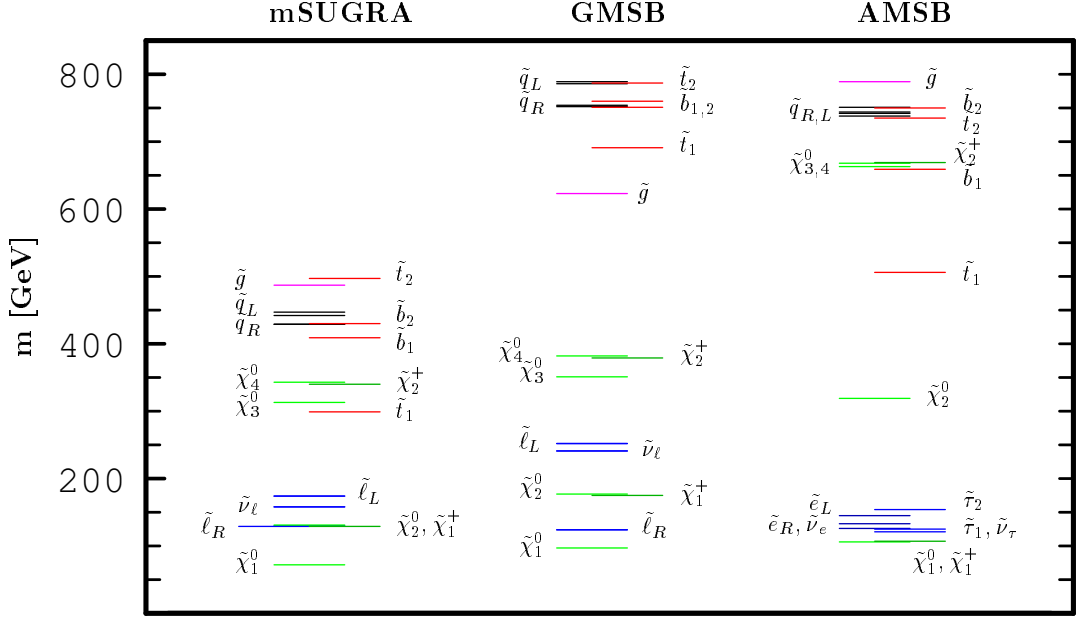,angle=0,width=.9\textwidth}}
  \caption{Examples of mass spectra in mSUGRA, GMSB and AMSB models for 
    $\tan\b=3$, sign$\,\mu > 0$. The other parameters are  
    $m_0=100$~\,eV, $m_{1/2}=200$\,GeV for mSUGRA; 
    $M_{\rm mess}=100$\,TeV, $N_{\rm mess}=1$, $\Lambda=70$\,TeV for GMSB; and 
    $m_0=200$\,GeV, $m_{3/2}=35$\,TeV for AMSB.}
  \label{fig:susy_mass-spectra}
\end{figure}

Most motivations for supersymmetry lead us to expect that SUSY particles will 
be found at the next increase in energy, at {\sc Tevatron} and/or LHC. If 
gluinos and squarks have masses below $2.5$\,TeV, they will be seen at 
LHC. In most scenarios some SUSY particles, especially the 
partners of $W$ and $Z$, the charginos and neutralinos, are expected 
to be lighter and should lie in the energy region of {\sc Tesla}.
Examples of mass spectra for three SUSY breaking mechanisms
are shown in \fig{susy_mass-spectra}.

The discovery potential of {\sc Tesla} for SUSY particles has been extensively
studied in the literature and in previous workshops~\cite{intro1}. 
Two important new issues have been addressed at the 2nd Joint ECFA/DESY 
Study~\cite{intro2}: The availability of high luminosity, 
${\cal L} \simeq 500\,\fbi$ per year, 
and of polarised electron {\it and} positron beams. 
The high luminosity makes precision experiments possible. 
We will therefore discuss in detail accurate measurements of the masses of
SUSY particles and the determination of the couplings and mixing properties of
sleptons, charginos, neutralinos and scalar top quarks.
A precise knowledge of the sparticle spectrum and the SUSY parameters is 
necessary to reveal the underlying supersymmetric theory. 

If kinematically accessible the complete sparticle spectrum can be studied in 
detail with the high luminosity available at {\sc Tesla}.
It is vital to have highly polarised electrons %($\cP_{-} \gsim 80\%$)
and it is very desirable to have polarised positrons %($\cP_{+} \gsim 60\%$) 
as well.
It is assumed that polarisations of $\cP_{-} = 80\%$ for electrons and 
$\cP_{+} = 60\%$ for positrons are achievable.
A proper choice of polarisations and center of mass energy helps disentangle
the various production channels and suppress background reactions. 
Electron polarisation is essential to determine the weak quantum numbers, 
couplings and mixings. Positron polarisation provides additional important 
information~\cite{moortgat}: 
(i) an improved precision on parameter measurements by exploiting all 
combinations of polarisation;
(ii) an increased event rate (factor 1.5 or more) resulting in a higher 
sensitivity to rare decays and subtle effects;
and (iii) discovery of new physics, {\em e.g.} spin 0 sparticle exchange.
In general the expected background is dominated by decays of other 
supersymmetric particles, while the Standard Model processes like
$W^+ W^-$ production can be kept under control at reasonably low level.

The most fundamental open question in SUSY is how supersymmetry is broken and 
in which way this breaking is communicated to the particles. 
Here three different schemes are considered:
the minimal supergravity (mSUGRA) model, gauge mediated (GMSB) and anomaly 
mediated (AMSB) supersymmetry breaking models.
The phenomenological implications are worked out in detail.
The measurements of the sparticle properties, like masses, 
mixings, couplings, spin-parity and other quantum numbers, do not 
depend on the model chosen.

In a kind of `bottom--up' approach a study demonstrates how the SUSY 
parameters, determined at the electroweak scale with certain errors, can be 
extrapolated to higher energies. In this way model assumptions made at higher
energies, for example at the GUT scale, can be tested. 

$R$--parity conservation is an additional assumption in most 
SUSY models. However, there is no fundamental reason for this, and a section 
is devoted to the phenomenology of $R$--parity violation. 

Finally, a comparison is made between {\sc Tesla} and LHC concerning the 
determination of the SUSY particle spectrum and the SUSY parameters.

%
% TESLA TDR - SUSY chapter
% section on MSSM   
% last update 29 Jan 2001
%

\section{The Minimal Supersymmetric Standard Model \label{SUSY_mssm}}
    
    The Minimal Supersymmetric Standard Model (MSSM) is the minimal extension 
    of the Standard Model (SM) to incorporate supersymmetry~\cite{nilles}. 
    In addition to the particles of the 
    SM, the MSSM contains their supersymmetric partners: 
    sleptons $\tilde{\ell}^{\pm}$, $\tilde{\nu}_{\ell}$ ($\ell = e,\,
    \mu ,\, \tau$), squarks $\tilde{q}$ and gauginos 
    $\tilde{g},\, \tilde{W}^{\pm},\, \tilde{Z}, \,\tilde{\gamma}$. Two Higgs 
    doublets are necessary: $H_{1} = (H^{0}_{1}, H^{-}_{1}$) and 
    $H_{2} = (H_{2}^{+}, H^{0}_{2})$, together 
    with their superpartners, the higgsinos ($\tilde{H}^{0}_{1,2}, 
    \tilde{H}^{\pm}$). The two doublets lead to five physical 
    Higgs bosons $h^{0},\, A^{0}$ ($CP = -1)$, $H^{0},\, H^{\pm}$. 
    
    The non--strongly interacting gauginos mix with the higgsinos to 
    form corresponding mass eigenstates: two pairs of charginos
    $\tilde{\chi}^{\pm}_{i}\ (i = 1,2)$ and four neutralinos 
    $\tilde{\chi}^{0}_{i}\,(i = 1, \ldots , 4)$. The masses and couplings 
    of the charginos and neutralinos are determined by the 
    corresponding mass matrices, which depend on the parameters $M_{1}, \,
    M_{2},\, \mu$ and $\tan\beta = v_{2}/v_{1}$. Usually, the GUT 
    relation $M_{1} / M_{2} = (5/3)\, \tan^{2} \theta_{W}$ is taken.
    
    Corresponding to the two chirality states of the %charged
    leptons and quarks one has the left and right scalar partners 
    $\tilde{\ell}_{R}, \,\tilde{\ell}_{L}$, and $\tilde{q}_{R}, \,
    \tilde{q}_{L}$. In the third generation one expects
    mixing between the $R$ and $L$ states with mass eigenstates 
    called $\tilde{t}_{1},\, \tilde{t}_{2},\, \tilde{b}_{1},\, 
    \tilde{b}_{2}$ and $\tilde{\tau}_{1},\, \tilde{\tau}_{2}$.
    
    In the MSSM the multiplicative quantum number $R$--parity is 
    conserved, $R_p = +1$ for SM particles and $R_p = -1$ for the 
    supersymmetric partners. This implies that there is a lightest 
    supersymmetric particle (LSP), which is stable and into which all 
    SUSY particles eventually decay. Usually, the 
    neutralino $\tilde{\chi}^{0}_{1}$ is assumed to be the LSP.
    
    In the most general case, the MSSM contains 105 parameters in 
    addition to the SM parameters.
    This number can be considerably reduced by invoking specific models, 
    which allow a systematic study of the whole parameter space. 
    In the so--called {\em minimal supergravity} (mSUGRA) model,
    due to universal unification conditions at 
    $M_{\rm{GUT}} \simeq 10^{16}$\,GeV,
    one has only five parameters $m_{0}$, $m_{1/2}$, $A_{0}$, $\tan\beta$ 
    and ${\rm{sign}}\,\mu$, 
    where $m_{0}$ and $m_{1/2}$ are the common scalar mass and gaugino
    mass at $M_{\rm{GUT}}$ and $A_{0}$ is the universal trilinear coupling
    parameter.
    
    In the following studies two mSUGRA scenarios, labeled 
    RR1 and RR2, are used to calculate masses, cross sections, 
    branching ratios and other physical quantities~\cite{intro2}. 
    The model parameters for RR1 are: 
    $m_{0} = 100$\,GeV, $m_{1/2} = 200$\,GeV, $A_{0} = 0$\,GeV, 
    $\tan\beta = 3$,  ${\rm sign}\,\mu  > 0$; 
    the corresponding mass spectrum is shown in \fig{susy_mass-spectra}.
    The parameters for RR2 are: $m_{0} = 160$\,GeV, 
    $m_{1/2} = 200$\,GeV, $A_{0} = 600$\,GeV, $\tan\beta = 30$, 
    ${\rm sign}\,\mu > 0$. 
    The low $\tan\beta$ scenario gives a Higgs mass of $ m_h=98$\,GeV,
    which is ruled out by LEP searches. 
    The $h^{0}$ mass does not affect the studies discussed here, 
    it only enters in cascade decays of the higher $\tilde{\chi}$ states.
    Moreover, a Higgs mass of, for instance, 115\,GeV can be achieved
    within mSUGRA by shifting the trilinear coupling 
    $A_0 \rightarrow -600$\,GeV and $\tan\beta \rightarrow 4.5$, which has 
    only little influence on the slepton, chargino and neutralino masses 
    and properties.

%
% TESLA TDR - SUSY chapter
% section on sleptons
% last update 29 Jan 2001
%

\section{Sleptons \label{SUSY_sleptons}}

Scalar leptons are the superpartners of the right-handed and left-handed 
leptons. They are produced in pairs 
\begin{eqnarray}
   e^+e^- & \to &  \ser  \ser, \ \sel \sel, \ \ser \sel,\  \snu_e\bar{\snu}_e 
   \nonumber  \\
   e^+e^- & \to &  \smur  \smur, \ \smul \smul, \ \snu_\mu\bar{\snu}_\mu \\
   e^+e^- & \to &  \stau_1  \stau_1, \ \stau_2 \stau_2, \ \stau_1 \stau_2, \
                   \snu_\tau\bar{\snu}_\tau
   \nonumber
\end{eqnarray}
via $s$-channel $\gamma / Z$ exchange. 
In addition the $t$-channel contributes in selectron production via neutralinos
and in electron-sneutrino production via charginos.
The two-body kinematics of the decays
$\sl^- \to \ell^- \ti\chi^0_i $ and $\snu_\ell \to \ell^- \ti\chi^+_i$
allows a clean identification and accurate measurements of 
the sparticle masses involved and other slepton properties like spin,  
branching ratios, couplings and mixing parameters.  
Polarisation is indispensible to determine the weak quantum numbers $R,\, L$  
of the sleptons. 
Detailed simulations of slepton production based on the {\sc Tesla} detector  
design are reported by~\cite{martyn,martyn2}, where 
it is assumed that beam polarisations of $\cP_{-}=0.8$ and $\cP_{+}=0.6$  
are available.

\subsection{Mass determinations} 

\begin{figure}  {\setlength{\unitlength}{1mm}
\begin{picture}(155,52)
%\put(0,0){\framebox(155,52)}
\put(-6,0){\mbox{\href{pictures/2/sl_mur132_emu.pdf}{{\epsfig{file=sl_mur132.emu.eps,angle=90,width=.55\textwidth}}}}}
\put(81,1){\mbox{\href{pictures/2/sl_mur132_min.pdf}{{\epsfig{file=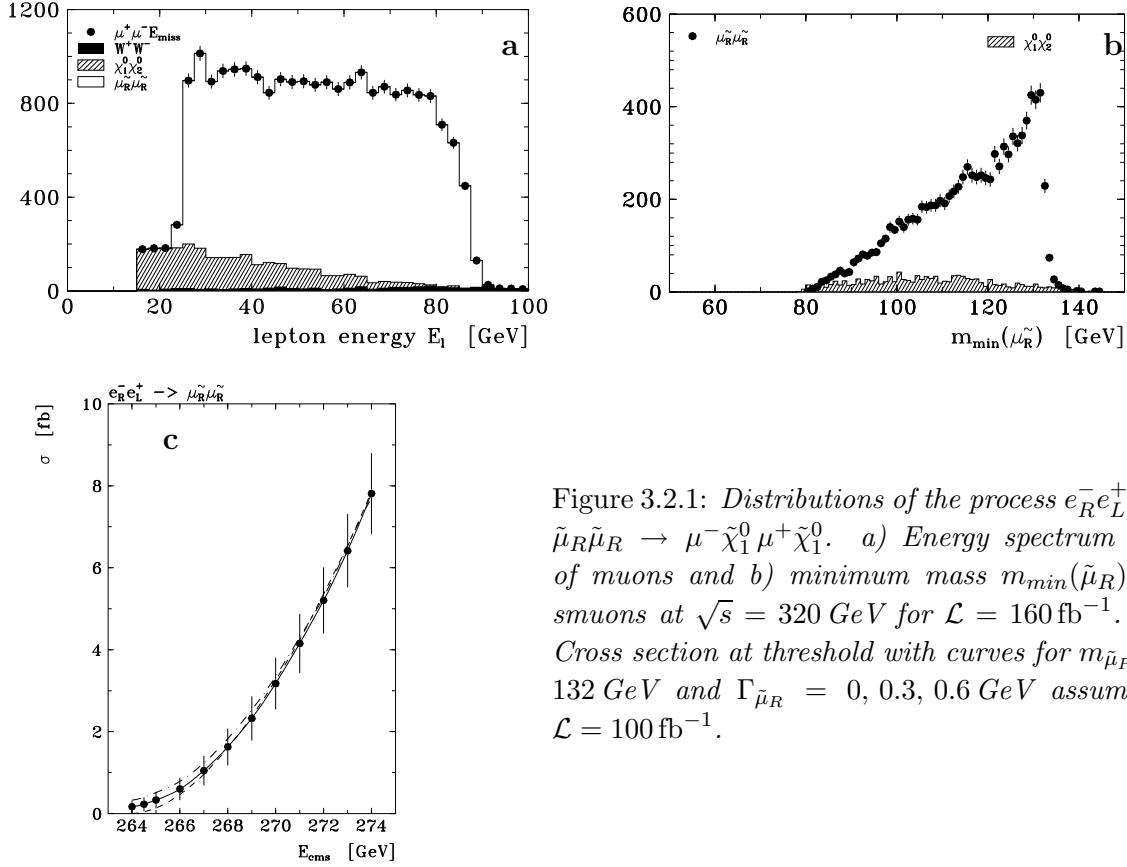,angle=0,width=.4\textwidth}}}}}
\put(65,42){\mbox{\bf a}}
\put(145,42){\mbox{\bf b}}
\end{picture}\\
\begin{minipage}[t]{55mm} 
\begin{picture}(55,66)
%\put(0,0){\framebox(55,66)}
\put(5,0){\mbox{\href{pictures/2/sl_mur132_scan.pdf}{{\epsfig{file=sl_mur132.scan.eps,%
    bbllx=0pt,bblly=0pt,bburx=350pt,bbury=450pt,height=6.8cm}}} }}
\put(20,56){\mbox{\bf c}}
\end{picture}
\end{minipage} \hfill
\begin{minipage}[t]{85mm}  \vspace{-55mm}
\caption{ Distributions of the process
  $e^-_R e^+_L \to \smur\smur \to \mu^-\ti\chi^0_1\, \mu^+ \ti\chi^0_1$.
  a) Energy spectrum $E_\mu$ of muons and 
  b) minimum mass $m_{min}(\smur)$ of smuons at $\sqrt{s} = 320$\,GeV 
     for $\cL = 160\,\fbi$.
  c) Cross section at threshold with curves for $m_{\tilde{\mu}_R}=132$\,GeV 
  and $\Gamma_{\tilde{\mu}_R} = 0,\,0.3,\,0.6$\,GeV assuming $\cL=100\,\fbi$.}
\label{fig:sl_smuonR}
\end{minipage}\\
}\end{figure}

The potential of an $e^+e^-$ collider will be illustrated for the second 
generation of sleptons $\smu$ and $\snu_\mu$. %~\cite{martyn,martyn2}.
The simplest case is the production and decay of right smuons
$e^-_R e^+_L \to \tilde{\mu}_R \tilde{\mu}_R
             \to \mu^- \ti\chi^0_1 \, \mu^+ \ti\chi^0_1$.
The results of a simulation are shown in~\fig{sl_smuonR}\,a.
The dominant background from $\ti\chi^0_2 \ti\chi^0_1$ production can be kept 
small.
The energy spectrum of the decay muons is flat apart from beamstrahlung,
initial state radiation and resolution effects at the high edge.
The end points can be related to the masses $m_{\tilde{\mu}_R}$ and 
$m_{\ti\chi^0_1}$ of the primary and secondary particles with an accuracy of 
about 3 per mil.
If the neutralino mass can be fixed by other measurements one can exploit the 
momentum correlations of the two observed muons~\cite{feng}
and construct the minimum kinematically allowed smuon mass $m_{min}(\smur)$.
From the `end point' or maximum of this distribution, 
shown in \fig{sl_smuonR}\,b, the accuracy on $m_{\smur}$ can be improved
by a factor of two.

Higher accuracy can be achieved by measuring the pair production cross 
section around threshold, which rises as $\sigma_{\smu\smu}\sim\beta^3$,
%$\sigma(\smu\smu)\sim\beta^3$, 
see~\fig{sl_smuonR}c.
For an integrated luminosity of $\cL = 100\,\fbi$, 
to be collected within a few month at {\sc Tesla},
a mass resolution $\delta m_{\smur} < 0.1$\,GeV can be reached.
With such a sensitivity finite width effects cannot be 
neglected~\cite{martyn2}. From a fit to the excitation curve one expects to
measure the width, here $\Gamma_{\smur} = 0.3$\,GeV, with an error of the
same order.
It is also important to include in the threshold cross section calculations
sub-dominant diagrams which lead to the same final state, because they may 
mimic a mass shift comparable to the resolution~\cite{freitas}.

\begin{figure}[ht]  {\setlength{\unitlength}{1mm}
\begin{picture}(155,53)
%\put(0,0){\framebox(155,53)}
\put(-1,0){\mbox{ \hspace{-25pt} 
  \href{pictures/2/sl_mul176_emu.pdf}{{\epsfig{file=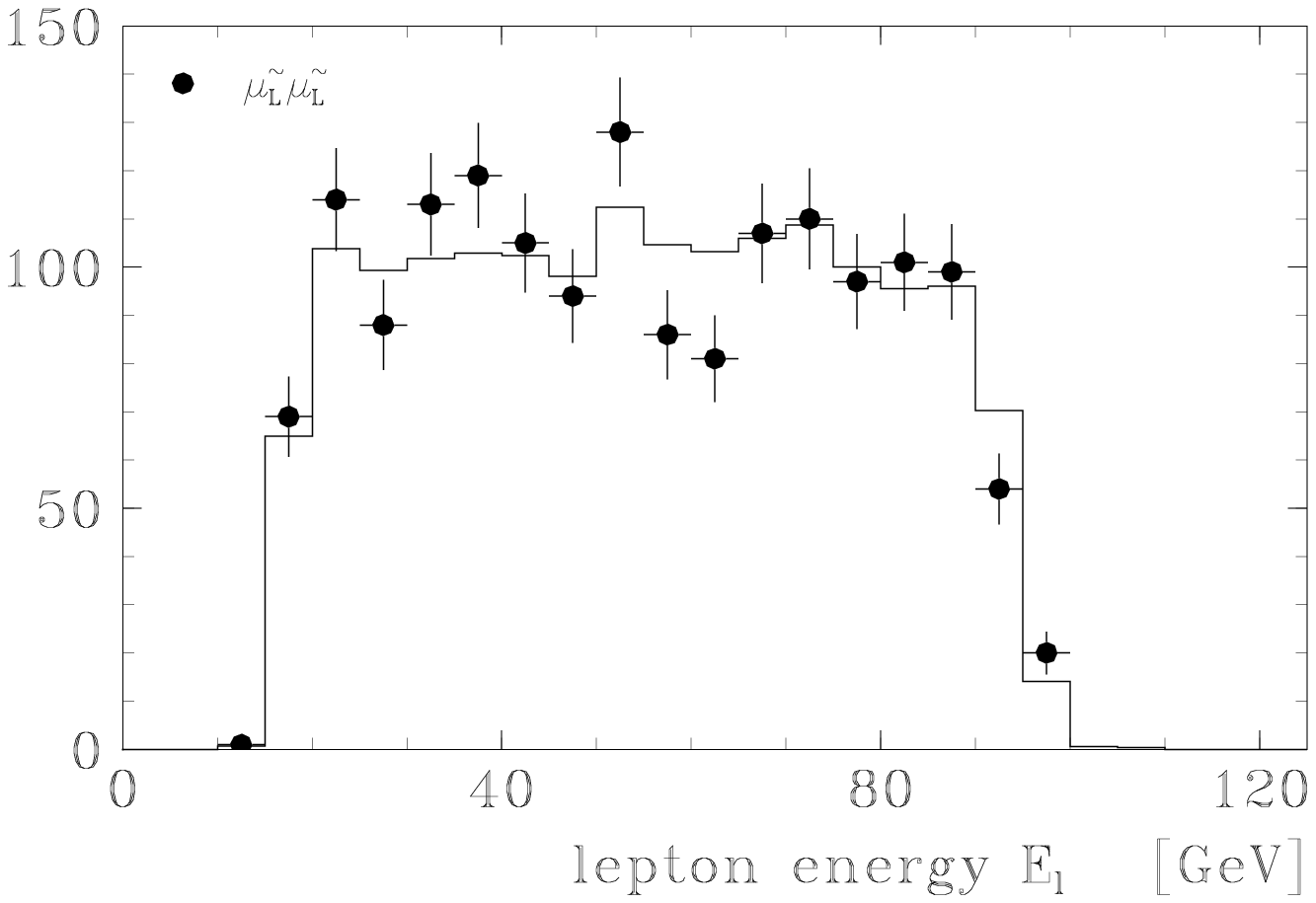,angle=90,width=.55\textwidth}}} \hspace{-25pt}
  \href{pictures/2/sl_nul161_emu.pdf}{{\epsfig{file=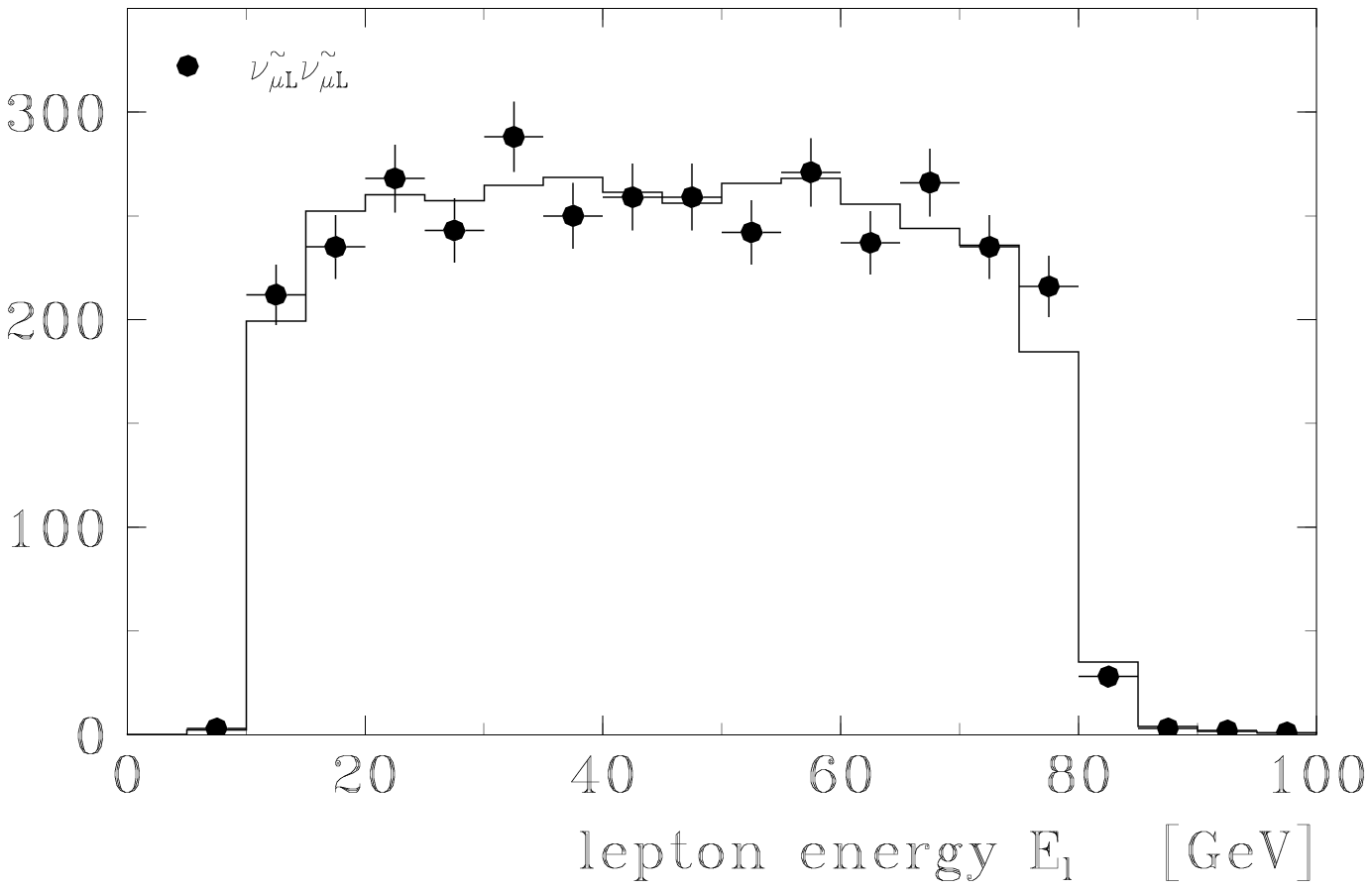,angle=90,width=.55\textwidth}}} }}
\put(64,41){\mbox{\bf a}}
\put(143,41){\mbox{\bf b}}
\end{picture}
\caption{Energy spectra $E_\mu$ of muons from the reactions
  a) $e^-_L e^+_R \to \smul\smul\to\mu^-\ti\chi^0_2 \,\mu^+\ti\chi^0_2$ and 
  b) $e^-_L e^+_R \to \snu_{\mu}\bar{\snu}_{\mu} 
                   \to \mu^-\ti\chi^+_1 \, \mu^+\ti\chi^-_1$ 
  at $\sqrt{s} = 500$\,GeV for $\cL = 250\,\fbi$.}
\label{fig:sl_smuonL}
}\end{figure}

The left partner $\tilde{\mu}_L$ is more difficult to detect due to background
from $W^+W^-$ pairs and SUSY cascade decays.
It can be identified via a unique $\mu^+\mu^-\,4\ell^\pm\, \Eslash$ signature: 
$e^-_L e^+_R \to \smul \smul \to \mu^- \ti\chi^0_2 \, \mu^+ \ti\chi^0_2$
followed by $ \ti\chi^0_2 \to \ell^+ \ell^- \ti\chi^0_1$.
Despite the low cross section, $\sigma\,{\cal B} \simeq 4$\,fb in scenario RR1,
such a measurement is feasible at {\sc Tesla}, see~\fig{sl_smuonL}\,a,
providing the  masses  $m_{\tilde{\mu}_L}$ and $m_{\ti\chi^0_2}$ 
with a precision of 2 per mil.
Another example is sneutrino production, where the flavour is tagged via
its charged decay
$e^-_L e^+_R \to \snu_\mu \bar{\snu}_\mu \to \mu^- \ch^+_1 \mu^+ \ch^-_1$.
The subsequent decays
$\ti\chi^\pm_1 \to  \ell^\pm \nu_l  \ti\chi^0_1$ and $ q \bar{q}' \ti\chi^0_1$
lead to a clean $\mu^+ \mu^- \, \ell^\pm \, 2 jet \, \Eslash$ topology.
The spectrum of the primary muons in~\fig{sl_smuonL} can be used to 
determine $m_{\snu_{\mu}}$ and $m_{\ti\chi^\pm_1}$ to better than 2 per mil.

Even more accurate mass measurements can be done for the first
generation of sleptons $\se$ and $\snu_e$,
due to much larger cross sections from additional $t$-channel contributions.
Of particular interest is associated selectron production 
$e^-e^+ \to \ser \sel$ via $\nt$ exchange.% in the $t$-channel.
The cross section rises as $\sigma_{\ser \sel} \sim \beta$, contrary to other 
slepton pairs, which is an advantage for mass determination
via threshold scans.
In case of polarised beams the charge of the observed lepton can be directly 
related to the $L,\, R$ quantum number of the produced selectron, 
$e^-_{L,\,R} \to \se^-_{L,\,R}$ and $e^+_{L,\,R} \to \se^+_{L,\,R}$.
This elegant separation of the selectron decay spectra can be considerably
improved if not only the $e^-$ beam but also the $e^+$ beam is polarised.

Assuming that the incoming electrons and positrons have the same helicity 
only the $t$--channel production $e^-_L e^+_L \to \se^-_L \se^+_R$ and 
$e^-_R e^+_R \to \se^-_R \se^+_L$ is possible. 
This allows one to easily identify $\se_L^{}$ and $\se_R^{}$ separately.

\begin{table}[htb]  
    \begin{center}  \mbox{
      \begin{tabular}{l c c c l}
        \hdick \\[-1.5ex]
        $ \sl, \snu $ & $m\, [\mathrm{GeV}]$ & $\delta m_{c}\, [\mathrm{GeV}]$ 
                                    & $\delta m_{s}\, [\mathrm{GeV}]$ 
        \\[1ex] \hdick  
\addtop $\tilde{\mu}_R$   & 132.0 & 0.3 & 0.09 \\
        $\tilde{\mu}_L$   & 176.0 & 0.3 & 0.4 \\
        $\tilde{\nu}_\mu$ & 160.6 & 0.2 & 0.8 \\  %\hline
        $\tilde{e}_R$     & 132.0 & 0.2 & 0.05 \\
        $\tilde{e}_L$     & 176.0 & 0.2 & 0.18 \\
        $\tilde{\nu}_e$   & 160.6 & 0.1 & 0.07 \\ %\hline
        $\tilde{\tau}_1$  & 131.0 &     & 0.6 \\
        $\tilde{\tau}_2$  & 177.0 &     & 0.6 \\
        $\tilde{\nu}_\tau$& 160.6 &     & 0.6 %\\ \hline
      \end{tabular}  \hspace{2mm}
      \begin{tabular}{l c c c l}
        \hdick \\[-1.5ex]
        $\ti\chi$ & $m\, [\mathrm{GeV}]$ & $\delta m_{c}\, [\mathrm{GeV}]$ 
                                 & $\delta m_{s}\, [\mathrm{GeV}]$ 
        \\[1ex] \hdick  
\addtop $\ti\chi^\pm_1$      & 127.7 & 0.2  & 0.04\\
        $\ti\chi^\pm_2$      & 345.8 &      & 0.25 \\ %\hline
        $\ti\chi^0_1$        &  71.9 & 0.1 & 0.05 \\
        $\ti\chi^0_2$        & 130.3 & 0.3  & 0.07 \\
        $\ti\chi^0_3$        & 319.8 &      & 0.30 \\
        $\ti\chi^0_4$        & 348.2 &      & 0.52 
        \\ \\ \\ \\
      \end{tabular}               }
    \end{center}  \vspace{-5mm}
\caption{Expected precision on masses, scenario RR1, 
    using polarised $e^\pm$ beams ($\cP_{-}=0.8,\,\cP_{+}=0.6$).
    $\delta m_{c}$ from decay kinematics measured in the continuum
    (${\cal L} = 160\,(250)\,{\rm fb}^{-1}$ at $\sqrt{s} = 320\,(500)$\,GeV)
    and $\delta m_{s}$ from threshold scans (${\cal L} = 100\,{\rm fb}^{-1}$).}
\label{tab:masses}
\end{table}

Measurements of $\stau$ and $\snu_\tau$ of the third slepton generation
are less favourable.
While identification via decays $\tau\ch$ will be easy and efficient,
the background is large ($W^+W^-$ production)
and a mass determination via energy spectra is much less accurate,  
of the order of a few per cent~\cite{nojiri}.
But from cross section measurements at threshold one may obtain mass 
resolutions around half a per cent.
The expected accuracies on slepton masses for mSUGRA model RR1
are given in Table~\ref{tab:masses}.

\subsection{Slepton properties}

\begin{figure}  
  \vspace*{3mm}
  \begin{minipage}[t]{70mm} 
    \href{pictures/2/sl_mur132_cost.pdf}{{\epsfig{file=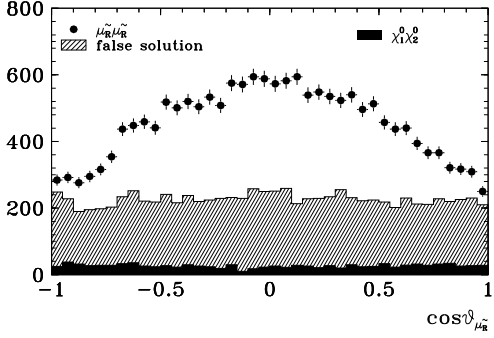,width=.85\textwidth}}}
  \end{minipage} \hfill
  \begin{minipage}[t]{70mm}  \vspace{-41mm}
    \caption{Angular distribution of smuons 
      (two entries per event) in the reaction
      $e^-_R e^+_L \to \smur\smur \to \mu^-\ti\chi^0_1 \, \mu^+\ti\chi^0_1$.
      The hatched histogram represents the false solution.}
    \label{fig:sl_cosmuonR}
  \end{minipage}\\
  %\vspace{-5mm}
\end{figure}

A very important topic is the determination of the quantum numbers.
Sleptons carry spin~0, but otherwise the SM quantum numbers of leptons.
The differential cross section for $s$-channel exchange is proportional
to $\beta^3\,\sin^2\vartheta$. % with $\beta = \sqrt{1 - 4\,m_{sl}/s}$.
A consistency check, although not unique, can be obtained from the $\beta$
dependence of the cross section scan at threshold. 
A more direct method is to measure the angular distribution of the sleptons.
Using the masses of the particles involved the event kinematics allows the 
slepton directions to be reconstructed up to a twofold ambiguity. 
The wrong solution turns out to be flat in $\cos\vartheta$ and can be 
subtracted.
The smuon angular distribution for $e^-_R e^+_L \to \smur\smur$ production is 
displayed in~\fig{sl_cosmuonR} and clearly exhibits the expected behaviour
of a scalar spin~0 particle.
The association of $\sl_{R}$ and  $\sl_{L}$
to their right-handed and left-handed    
SM partners can be unambiguously done by studying the dependence of 
the production cross section on the electron and/or positron beam polarisation.

Precise mass measurements allow the flavour dependence of the underlying 
supersymmetry model to be checked at the level of one per mil for the first 
two slepton generations and to a few per mil for the stau family.
An important application is to test general SUSY mass relations.
The tree level prediction 
\begin{eqnarray}
  m^2_{\sl_L} - m^2_{\snu_\ell} & = & - M^2_W\,\cos  2\,\beta
\end{eqnarray}
offers a model independent determination of $\tan\beta$
%in the low and intermediate range 
from the slepton sector.
Using typical measurements as given in Table~\ref{tab:masses} one finds
$\tan\beta = 3.0 \pm 0.1$. The sensitivity degrades at larger $\tan\beta$
values to $\tan\beta \simeq 10 \pm 5$.

In the case of large $\tan\b \sim 30$ the slepton analyses of the first
and second generation remain essentially unaffected. 
Major differences occur in the stau sector where a large mass 
splitting between $\stau_R$ and $\stau_L$ is expected. 
The physical eigenstates are mixed,
$\stau_1^{} = \stau_L^{} \cth_{\stau} + \stau_R^{}\sin\theta_{\stau}$ and
$\stau_2^{} = \stau_R^{} \cth_{\stau} - \stau_L^{}\sin\theta_{\stau}$,
and are no longer degenerate with the selectron and smuon masses.
These properties allow $\tan\b$ to be accessed via the relation 
\begin{eqnarray}
  \mu\,\tan\beta & = & 
  A_\tau-\frac{(\mstau{1}^2-\mstau{2}^2)\,\sin 2\,\tstau}{2\,m_\tau} \ , 
  \label{sl_eq3}
\end{eqnarray}
which follows from the diagonalisation of the $\stau$ mass matrix. 
If the directly measurable quantities $m_{\stau_1},\,m_{\stau_2}$ and
$\tstau$ can be determined to \mbox{$\sim\,1\%$} 
and $\mu$ to \mbox{$\sim\,1\%$} (from the chargino sector),
one can extract $\tan\b$ with an accuracy of ${\cal O}(10\%)$,
dominated by large uncertainties on the value of $A_\tau$.

It has been noted that the polarisation $P_\tau$ of tau's in the 
decay $\stau_1 \to \tau\nt_1$ is very sensitive to $\tan\beta$ if it is large  
or if $\nt_1$ has a large higgsino component~\cite{nojiri}.
%The reason is that the chirality flipping higgsino interaction becomes 
%comparable to the chirality conserving gaugino interaction.
The $P_\tau$ measurement is based on the characteristic energy distributions 
of the decay products of the polarised $\tau$. % and can be done to $5- 10\%$.
In a combined analysis of $\stau_1\stau_1$ and $\ser\ser$ pair production for 
${\cal L}=100\,\fbi$, one obtains an accuracy of $\tan\beta \simeq 15 \pm 2$, 
depending slightly on the gaugino parameter $M_1$.

%
% TESLA TDR - SUSY chapter
% section on charginos/neutralinos
% last update 29 Jan 2001
%

\section{Charginos and Neutralinos \label{SUSY_gauginos}}

Charginos and neutralinos are produced in pairs 
\begin{eqnarray}   
  e^+e^- & \to & \ti\chi^+_i \ti\chi^-_j \qquad\qquad [i,j = 1,2] 
  \label{eq:cijproduction} \\
  e^+e^- & \to & \ti\chi^0_i \ti\chi^0_j \qquad\qquad \ \, [i,j = 1, \ldots,4] 
  \label{eq:nijproduction} 
\end{eqnarray}
via $s$-channel $\gamma / Z$ exchange and $t$-channel selectron or sneutrino
exchange. 
They are easy to detect via their decays into lighter charginos/neutralinos
and gauge or Higgs bosons or into sfermion-fermion pairs.
If these two-body decays are kinematically not possible, 
typically for the lighter chargino and neutralino,
they decay via virtual gauge bosons and sfermions,
{\em e.g.}  $\ti\chi^+_1 \to f \bar{f}' \ti\chi^0_1$ or 
$\ti\chi^0_2 \to f \bar{f} \ti\chi^0_1$.
In $R$-parity conserving MSSM scenarios the lightest neutralino $\ti\chi^0_1$ 
is stable.% (LSP).
The experimental signatures are multi-lepton and/or multi-jet events with large
missing energy. Detailed {\sc Tesla} detector simulations of chargino and 
neutralino production assuming beam polarisations of $\cP_{-}=0.8$ and 
$\cP_{+}=0.6$ are performed in~\cite{martyn}.

\subsection{Mass determinations}

\begin{figure}[ht]  {\setlength{\unitlength}{1mm}
\begin{picture}(155,65)
%\put(0,0){\framebox(155,65)}
\put(-4,0){\mbox{  
  \href{pictures/2/chi_n22spectra.pdf}{{\epsfig{file=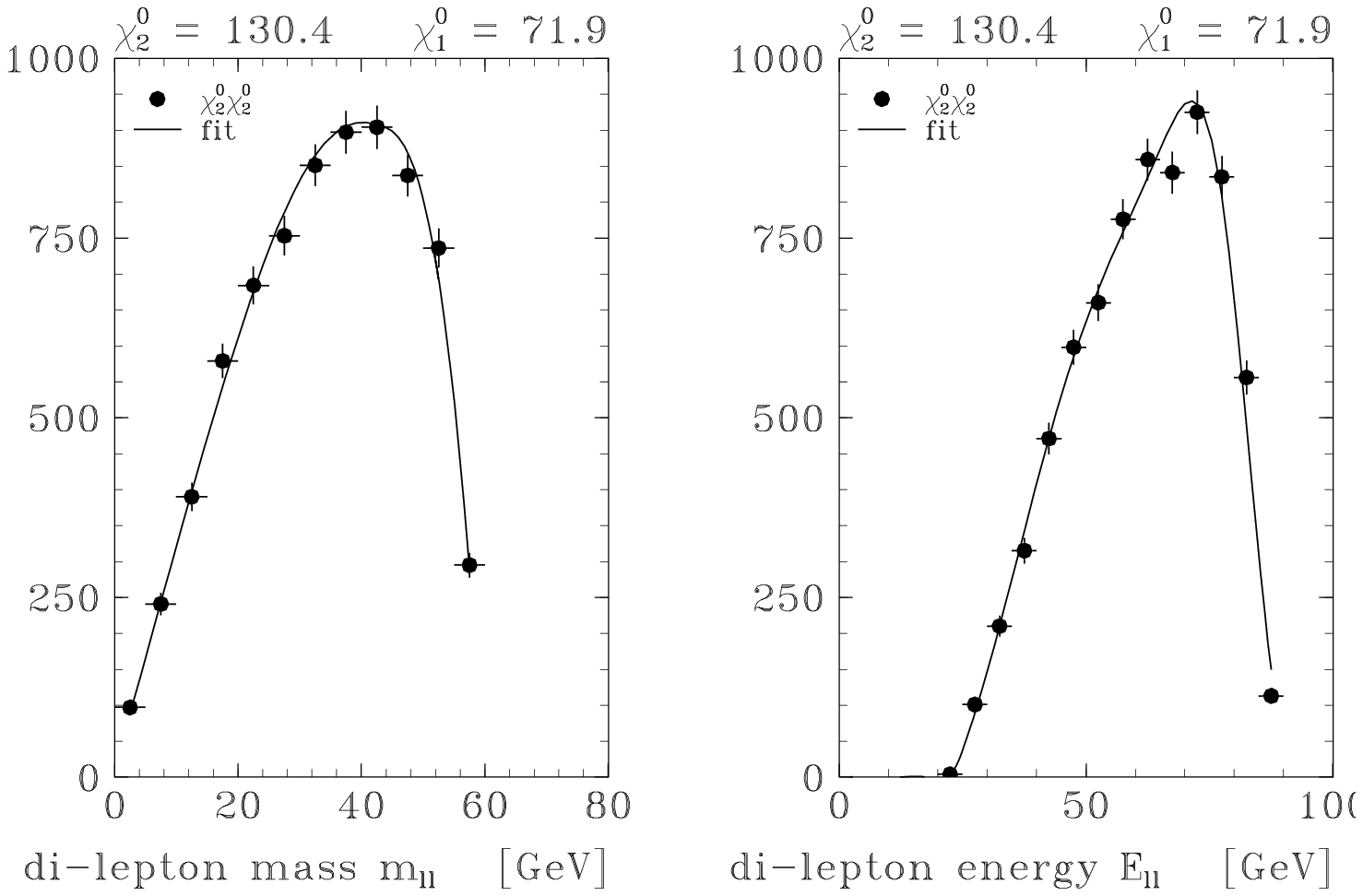,angle=0,width=.66\textwidth}}}\hspace{1mm}
  \href{pictures/2/chi_n22scan.pdf}{{\epsfig{file=chi_n22scan.eps,%
    bbllx=0pt,bblly=400pt,bburx=500pt,bbury=800pt,clip=,%
    angle=90,height=6.4cm}}} }}  
\put(41,53){\mbox{\bf a}}
\put(92,53){\mbox{\bf b}}
\put(142,55.5){\mbox{\bf c}}
\end{picture}
\caption{Distributions of the reaction
    $e^-_L e^+_R \to \ti\chi^0_2 \ti\chi^0_2 \to 
    l^+ l^-\ti\chi^0_1\, l^+ l^-\,\ti\chi^0_1$.
    a) Di-lepton mass and b) di-lepton energy spectra 
    at $\sqrt{s} = 320$\,GeV for $\cL = 160\,\fbi$.
    c) Cross section near threshold assuming 
    ${\cal L} = 10\,\fbi$ per point.}
\label{fig:chi_neutralino22}
}\end{figure}

The lightest observable neutralino can be detected via its 3-body decay
$\ti\chi^0_2 \rightarrow l^+ l^-\,\ti\chi^0_1$. 
In direct production
%$e^-_L e^+_R \to \ti\chi^0_2 \ti\chi^0_2 \to 4 l^\pm + \Eslash$
$e^-_L e^+_R \to \ti\chi^0_2 \ti\chi^0_2 \to 2(l^+l^-)\, \Eslash$
the energy spectra of the di-lepton systems,~\fig{chi_neutralino22},
can be used to determine the masses of the primary and secondary neutralinos
with typical uncertainties of 2~per mil. %  $\sim 2$~per mil.
From the di-lepton mass spectrum one gets additional information on
the mass difference $\Delta m (\ti\chi^0_2 - \ti\chi^0_1)$.
Moreover, $\ti\chi^0_2$'s are abundantly produced in decay chains of other SUSY
particles. By exploiting all di-lepton modes it will be possible to measure the
mass difference with a precision of better than $50$\,MeV, limited only by 
the resolution of the detector.

\begin{figure}[ht]  {\setlength{\unitlength}{1mm}
\begin{picture}(155,65)
%\put(0,0){\framebox(155,65)}
\put(-4,0){ \mbox{\href{pictures/2/chi_c11spectra.pdf}{{\epsfig{file=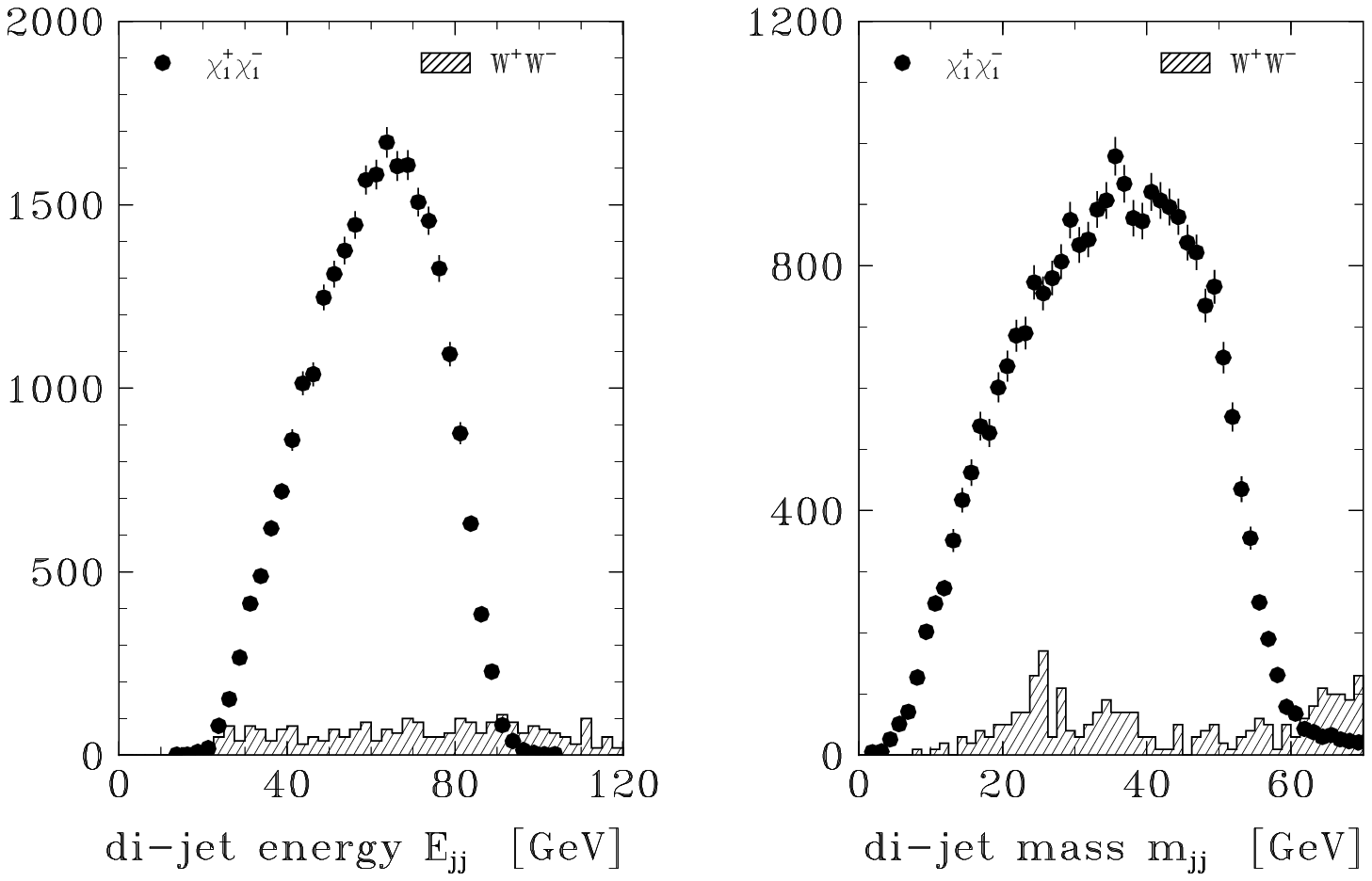,width=.66\textwidth}}} }} 
\put(103,1){\mbox{\href{pictures/2/chi_c11scan.pdf}{{\epsfig{file=chi_c11scan.eps,%
    bbllx=0pt,bblly=400pt,bburx=500pt,bbury=800pt,clip=,%
    angle=90,height=6.4cm}}} }}  
\put(41,52){\mbox{\bf a}}
\put(91,52){\mbox{\bf b}}
\put(146,52){\mbox{\bf c}}
\end{picture}
\caption{Distributions of the reaction
    $e^-_L e^+_R \to \ti\chi^-_1 \ti\chi^+_1
    \to l^\pm \nu\,\ti\chi^0_1\, q \bar{q}'\ti\chi^0_1$.
    a) Di-jet energy and b) di-jet mass spectra 
    at $\sqrt{s} = 320$\,GeV for $\cL = 160\,\fbi$.
    c) Cross section near threshold assuming 
    ${\cal L} = 10\,\fbi$ per point.}
\label{fig:chi_chargino11}
}\end{figure}

Charginos will be copiously produced, for example
$e^-_L e^+_R \to \ti\chi^-_1 \ti\chi^+_1 
             \to l^\pm \nu\,\ti\chi^0_1 \ q \bar{q}'\ti\chi^0_1$,
see \fig{chi_chargino11}. Using the same techniques as for neutralinos,
the di-jet energy distribution gives an accuracy of
$\delta m_{\ti\chi^\pm_1} = 0.2$\,GeV.
Similarly, the di-jet mass spectrum allows to get the
chargino-neutralino mass difference $\Delta m (\ti\chi^\pm_1 - \ti\chi^0_1)$ 
to better than $50$\,MeV, when using all possible cascade decays.

The cross sections for neutralino and chargino 
pair production rise as $\sigma_{\chi\chi} \propto \beta$.
This leads to steep excitation curves around threshold, 
see~\figs{chi_neutralino22}\,c and~\ref{fig:chi_chargino11}\,c, 
from which excellent mass resolutions of ${\cal O}(50\,\mathrm{MeV})$ 
with an integrated luminosity of $\cL = 100\,\fbi$
can be obtained for the light chargino/neutralinos, degrading for the heavier 
$\ch$ states to the per mil level.
At the same time, the shape of the cross section at threshold provides a 
consistency check of a spin $1/2$ assignment to neutralinos and charginos.
The expected accuracies of various mass determinations are summarised in
Table~\ref{tab:masses} for mSUGRA scenario RR1.

For large $\tan\beta$ the chargino and neutralino decays may be very 
different. Depending on the SUSY parameters the mass splitting of the $\stau$ 
sector, which rises with $\tan\beta$, see eq.~(\ref{sl_eq3}), may result in a 
situation where $m_{\stau_1} < m_{\ch^\pm_1},\, m_{\nt_2}$.
As a consequence 
the chargino decay $\ch^+_1\to \stau^+_1\nu \to \tau^+\nu\nt_1$ 
and the neutralino decay $\nt_2 \to \stau^+_1 \tau^- \to \tau^+ \tau^-\nt_1$ 
dominate over all other decay modes via lepton or quark pairs. 
Although $\tau$'s are easy to detect, their energy cannot be reconstructed 
(missing neutrinos) and their decay products provide much less information on 
masses and mass differences of the $\chi$ states.
A simulation of 
$e^+e^-_L\to \ch^+_1\ch^-_1 \to \stau_1^+\nu\ \stau_1^-\nu 
         \to \tau^+\nu\nt_1 \ \tau^-\nu\nt_1$ 
at $\sqrt{s}=400$\,GeV 
with $m_{\ch^\pm_1}=172.5$\,GeV, $m_{\stau_1}=152.7$\,GeV,
$m_{\nt_1}=86.8$\,GeV and $\tan\beta=50$ is reported by~\cite{kamon}.
Fitting the energy distribution of hadronic $\tau$ decays 
results in resolutions of about $4\%$ for the $\ch^\pm_1$ and $\stau_1$ masses.
Note that chargino and neutralino cross section measurements, in particular 
around threshold, are much less affected by $\tau$ topologies 
and become more important for precise mass determinations in large 
$\tan\beta$ scenarios.

\subsection{Chargino properties}

Charginos are composed of Winos and Higgsinos.
%$\ti\chi^\pm_i = \alpha\tilde{W}^\pm + \beta\tilde{H}^\pm$.
An easy way to access the Wino component is via $t$-channel $\snu_e$ exchange,
which couples only to left-handed electrons.
Thus the mixing parameters of the chargino system 
can be determined by varying the beam polarisation.
Such studies have been presented in detail by~\cite{cdgksz,mbfm1}. 
The chargino mass matrix in the $(\tilde{W}^-,\tilde{H}^-)$ basis 
\begin{eqnarray}
  {\cal M}_C & = &
  \left(\begin{array}{cc}
            M_2                & \sqrt{2}\,m_W\cos\beta  \\
         \sqrt{2}\,m_W\sin\beta  &      \mu   
                  \end{array}\right)\
  \label{eq:massmatrix}
\end{eqnarray}
depends on the parameters $M_2$, $\mu$ and $\tan\beta$.
Two mixing angles $\phi_L$ and $\phi_R$ are needed to diagonalise the 
mass matrix.
If the collider energy is sufficient to produce all chargino states of
reaction~(\ref{eq:cijproduction}) the underlying SUSY parameters 
can be extracted in a model independent way from the masses
and production cross sections~\cite{cdgksz}. 

\begin{figure}[htb] \centering
  \href{pictures/2/chi_cijcontour.pdf}{{\epsfig{file=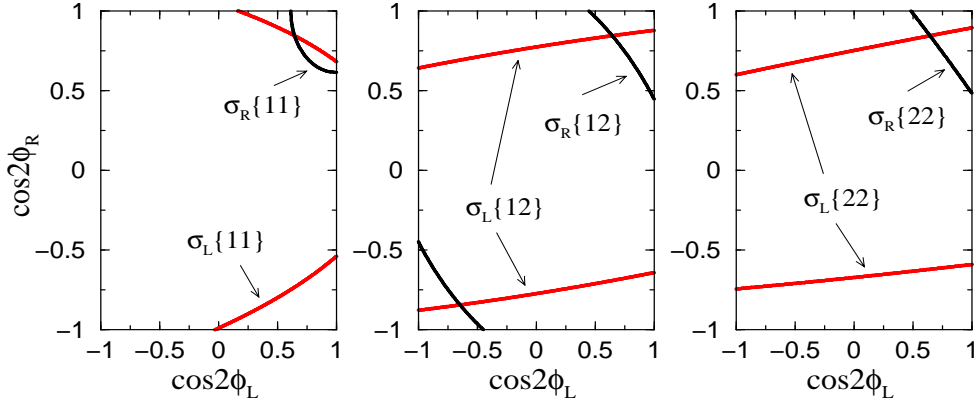,%
    bbllx=30pt,bblly=110pt,bburx=590pt,bbury=360pt,clip=,%
    width=13.2cm,height=5.5cm}}} \vspace{-5mm}
\caption {Contours of $\sigma_L\{ij\}, \ \sigma_R\{ij\}$
  for $\ti\chi^+_i \ti\chi^-_j$  production
  with completely polarised $e^-_{L, R}$ 
  in the $\cos2\phi_L - \cos2\phi_R$ plane,
  scenario RR1 at $\sqrt{s} = 800$\,GeV.}
\label{fig:chi_charginoij}
\end{figure}

Figure~\ref{fig:chi_charginoij} shows lines of constant cross sections for 
$\ti\chi^-_i\ti\chi^+_j$ pair production with completely polarised electrons  
in the $\cos2\phi_L - \cos2\phi_R$ plane.
Ambiguities can be resolved and the expected accuracy on the mixing angles is
$\cos2\phi_L=0.645 \pm 0.02$ and  $\cos2\phi_R=0.844 \pm 0.005$
for a total luminosity ${\cal L} = 2\times 500\,\fbi$.
Given precise measurements of $m_{\ti\chi^\pm_{1,2}}$ 
and the exchanged sneutrino mass $m_{\ti\nu_e}$,
the fundamental supersymmetry parameters 
$M_2$, $\mu$, and (low to medium) $\tan\beta$ can be 
accurately determined independently of the structure of the neutralino sector,
as illustrated in Table~\ref{tab:chi_cijrr12}.
If $\tan\beta$ is large, this parameter is difficult to extract,
only a significant lower bound can be derived. 
In this case the $\stau$ sector provides a higher sensitivity,
see section~\ref{SUSY_sleptons}.

\begin{table}[hbt] 
\begin{center}
  \begin{tabular}{l c c c c}
  \hdick \\[-1.5ex]
  & input RR1 & fit value & input RR2 & fit value  
  \\[1ex] \hdick  \\[-1.5ex]
  $M_2$    & 152\, GeV & $152\pm 1.8$\, GeV & 150\, GeV & $150\pm 1.2$\, GeV \\
  $\mu$    & 316\, GeV & $316\pm 0.9$\, GeV & 263\, GeV & $263\pm 0.7$\, GeV \\
  $\tan\beta$& 3       & $3\pm 0.7$ \ \ \   & 30        &  $> 20$          \\
  $M_1$  & 78.7\, GeV & $78.7\pm 0.7$\, GeV & 78.0\, GeV & $78.0\pm 0.4$\, GeV\\
  \end{tabular} \vspace*{-5mm}
\end{center}
\caption{ Estimated accuracy for the parameters $M_2$, $\mu$ and
    $\tan\beta$ from chargino masses and %production cross sections 
    and $M_1$ from neutralino production
    for mSUGRA scenarios RR1 and RR2
    (statistical errors based on ${\cal L}=500\,\fbi$ 
    per $e^-$ polarisation).} 
\label{tab:chi_cijrr12}
\end{table}

\begin{figure}[ht]  {\setlength{\unitlength}{1mm}
\begin{picture}(155,52)
\put(7,-3){\mbox{  
  \href{pictures/2/chi_c11sigma.pdf}{{\epsfig{file=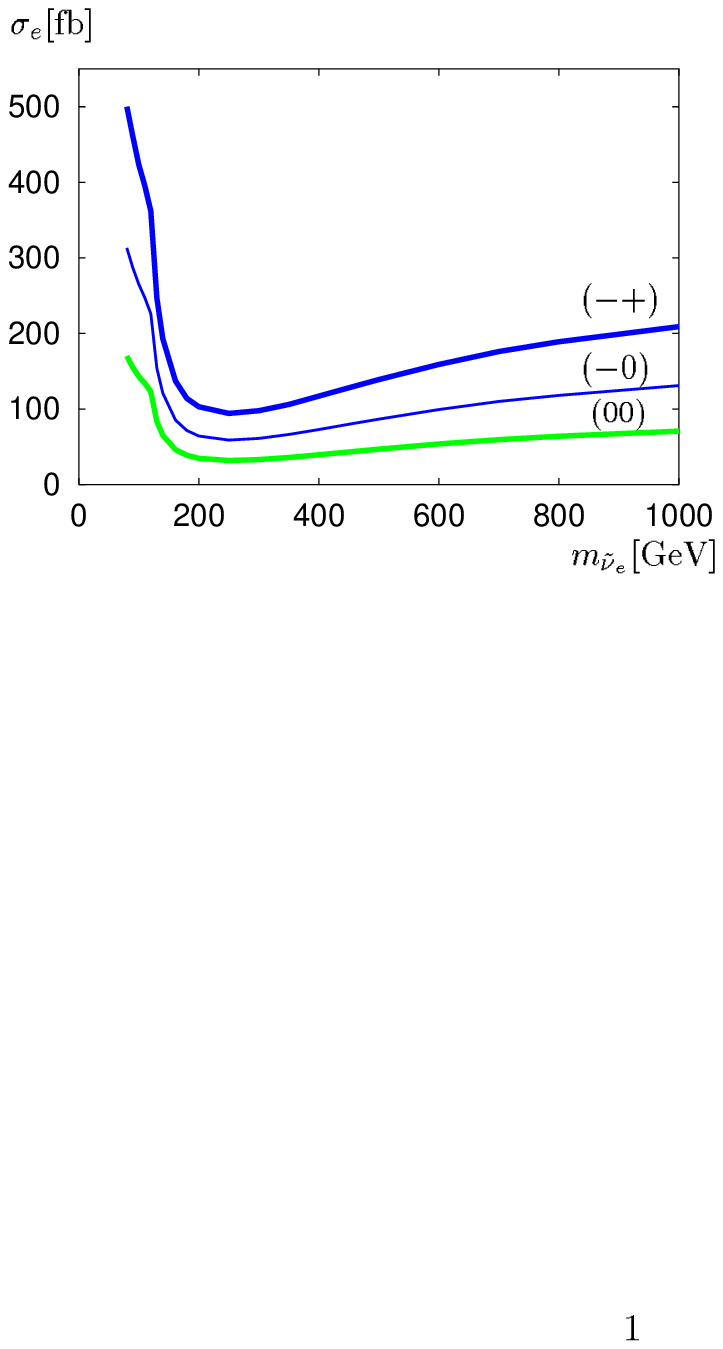,%
    bbllx=120pt,bblly=356pt,bburx=332pt,bbury=536pt,clip=,%
    width=64mm}}} \hspace{11mm}
  \href{pictures/2/chi_c11asym.pdf}{{\epsfig{file=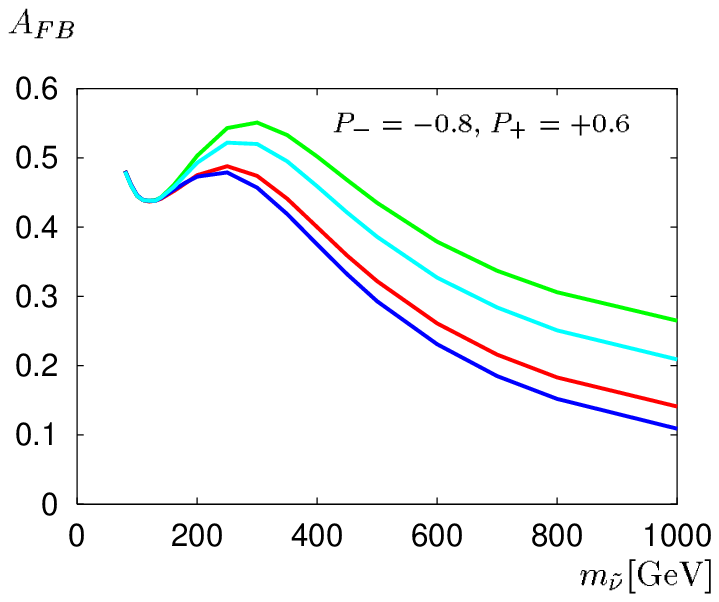,%
    bbllx=200pt,bblly=356pt,bburx=412pt,bbury=536pt,clip=,%
    width=64mm}}} }}  
\put(2,44){\mbox{\bf a)}}
\put(79,44){\mbox{\bf b)}}
\end{picture}
\caption{Dependence of $e^+e^-\to\ti\chi^+_1\ti\chi^-_1$ 
  with $\ti\chi^-_1\to e^-\bar{\nu_e}\ti\chi^0_1$ 
  at $\sqrt{s}=500$\,GeV on the $\snu_e$ mass, scenario RR1.
  a) Cross sections for various beam polarisations ($\cP_-, \cP_+$) and   
  b) forward--backward asymmetry $A_{FB}$ of the decay electron  
  for $m_{\tilde{e}_L}=130$\,GeV (top curve), $150$\,GeV, $200$\,GeV 
  and using $m^2_{\tilde{e}_L} = m^2_{\tilde{\nu}_e} - m^2_W\cos2\beta$
  (bottom curve) assuming polarisations $\cP_-=-0.8$ and $\cP_+=+0.6$.}
\label{fig:chi_c11asym}
}\end{figure}

The analysis of the chargino system depends via the cross sections $\sigma_L$
on the mass of the exchanged sneutrino which may not be directly accessible, 
{\em e.g.} if $m_{\snu_e} > \sqrt{s}/2$.
The sensitivity to the sneutrino mass
can be considerably enhanced by a proper choice of polarisations and
by making use of spin correlations between production and decay %~\cite{mbfm1} 
in the reaction $e^+ e^-\to\ti\chi^+_1\ti\chi^-_1$ and 
$\ti\chi^-_1\to e^-\bar{\nu_e}\ti\chi^0_1$~\cite{mbfm1}. 
From the cross section $\sigma\cdot {\cal B}_e$
and the forward-backward asymmetry $A_{FB}$ of the decay electrons,
shown in \fig{chi_c11asym},
one can determine sneutrino masses up to $1$\,TeV 
with a precision of $\sim 10$\,GeV.

For final precision measurements the inclusion of electroweak radiative 
corrections will be important, as they have been calculated for example 
in~\cite{blank}.

\subsection{Neutralino properties}

\begin{figure}[ht]  {\setlength{\unitlength}{1mm}
\begin{picture}(155,52)
\put(7,-2){\mbox{  
  \href{pictures/2/chi_n12sigma.pdf}{{\epsfig{file=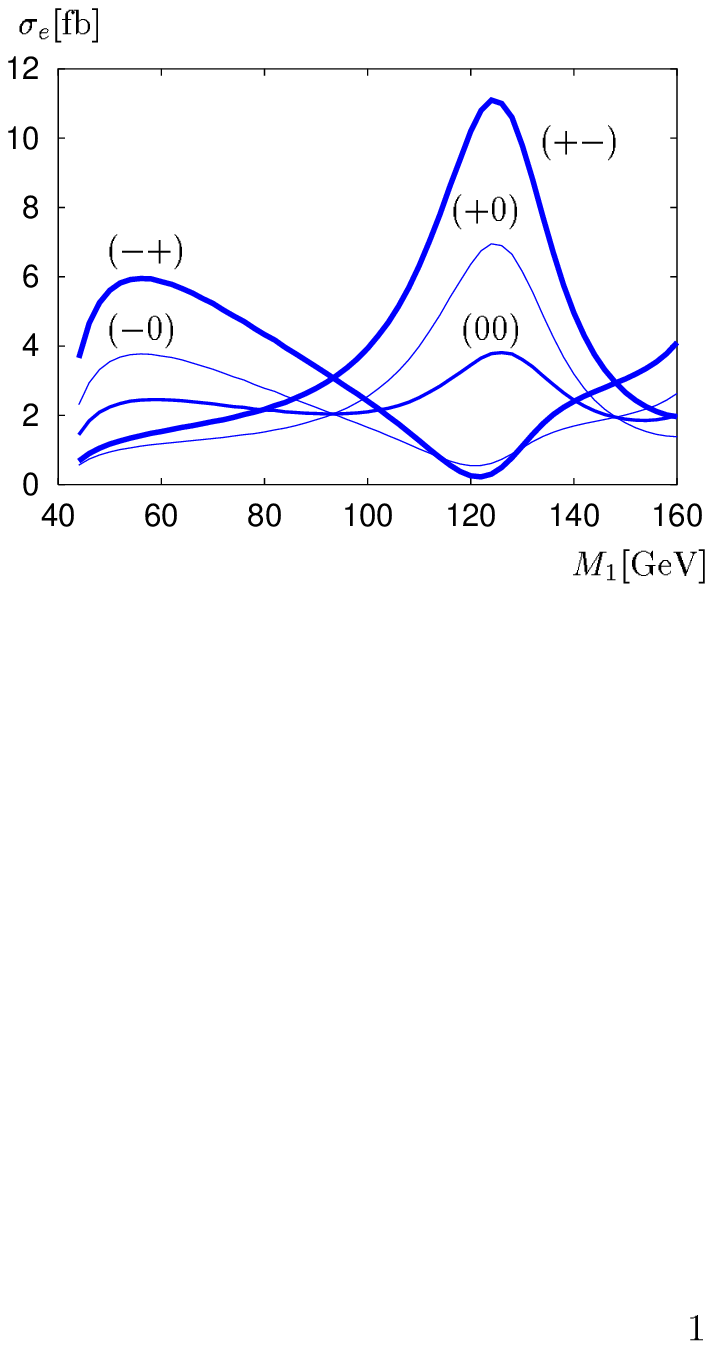,%
    bbllx=100pt,bblly=357pt,bburx=310pt,bbury=530pt,clip=,%
    width=64mm}}} \hspace{9mm}
  \href{pictures/2/chi_n12asym.pdf}{{\epsfig{file=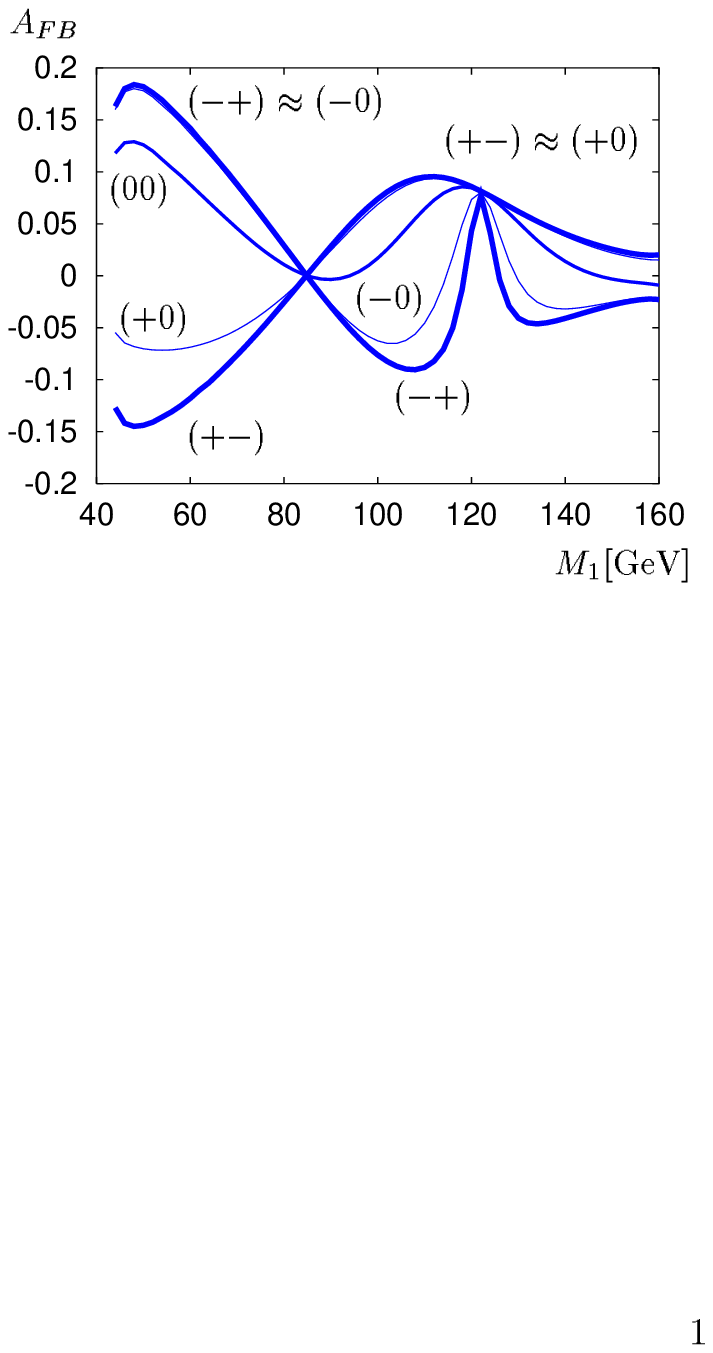,%
    bbllx=100pt,bblly=357pt,bburx=310pt,bbury=530pt,clip=,%
    width=64mm}}} }}  
\put(2,44){\mbox{\bf a)}}
\put(79,44){\mbox{\bf b)}}
\end{picture}
\caption{Dependence of $e^+ e^-\to\ti\chi^0_2\ti\chi^0_1$ 
  with $\ti\chi^0_2\to e^+e^-\ti\chi^0_1$
  at $\sqrt{s}= m_{\ti\chi^0_1}+m_{\ti\chi^0_2}+30$\,GeV  
  on the gaugino parameter $M_1$, scenario RR1.
  a) Cross sections and b) forward--backward asymmetry $A_{FB}$ 
  of the decay electron  
  for various polarisations ($|\cP_-|=0.8, |\cP_+|=0.6$).}
\label{fig:chi_n12asym}
}\end{figure}

In a similar way the properties of the neutralino system,
which is a mixture of Bino, Wino and two Higgsino fields,
have been investigated. 
In a general MSSM model the neutralino sector depends  
in addition to $M_2$, $\mu$ and $\tan\beta$ on the gaugino parameter $M_1$.
Very useful analysis tools are angular distributions of leptons in the 
reaction 
$e^+e^- \to \ti\chi^0_2 \ti\chi^0_1 \to \ell^+\ell^- \ti\chi^0_1 \ti\chi^0_1$
exploiting spin correlations~\cite{mbfm2}.
Figure~\ref{fig:chi_n12asym} shows the sensitivity of the production 
cross section and the forward-backward asymmetry of the decay electron 
to the parameter $M_1$.
Again the importance of $e^\pm$ beam polarisations to
determine the neutralino mixing parameters is clearly borne out.

A richer neutralino spectrum with quite different properties is expected if
supersymmetry is extended by an additional Higgs superfield, 
like in the NMSSM or $E_6$ inspired models. 
With the high luminosity available at {\sc Tesla} neutralinos with a dominant 
singlino component can be easily detected and studied 
over large regions in the parameter space~\cite{hesselbach}.
A characteristic feature of these scenarios is that in certain parameter 
regions the second lightest supersymmetric particle may have a long life time
leading to displaced vertices. Polarisation of both beams is important to 
enhance the production cross sections and to determine the underlying SUSY 
model~\cite{moortgat}.

Quite generally, the parameters $M_1$, $M_2$ and $\mu$ can be {\em complex},
which also leads to $CP$ violation.
It is, however, possible to take $M_2$ real, so that only two phases remain,
$\mu = |\mu|\, e^{i\, \phi_\mu}$ and $M_1 = |M_1|\, e^{i\, \phi_{M_1}}$. 
A method to extract $\cos\phi_\mu$ from chargino production is described 
in~\cite{cdgksz} giving $\Delta\cos\phi_\mu = \pm 0.1$. 
A rather simple algebraic algorithm has been proposed 
by~\cite{kneur} to determine $\mu$, $M_1$, $M_2$, $\phi_\mu$, 
$\phi_{M_1}$ for given $\tan\beta$ in terms of the masses of both 
charginos and two neutralinos and one of the chargino mixing angles 
as physical input. The remaining twofold ambiguity
in $|M_1|$ and $\phi_{M_1}$ can be resolved by a measurement of the
$e^+e^- \to \ti\chi^0_1 \ti\chi^0_2$ production cross section.

%
% TESLA TDR - SUSY chapter
% section on stops
% last update 26 Jan 2001
%

\section{Stop Particles \label{SUSY_stops}}

Supersymmetry requires the existence of scalar partners
$\ti f_L$ and $\ti f_R$ to each fermion $f$. In case of the scalar 
partners of the top quark one expects a large mixing between 
$\ti t_L$ and $\ti t_R$ due to the large top quark mass thus making 
the lighter mass eigenstate $\ti t_1$ presumably lighter than the 
squark states of the first two generations. In $e^+e^-$
collisions stops can be pair produced by  $\g/Z$ exchange
\begin{equation} \label{stop_eq1}
  e^+e^-\to\st_i \st_j \qquad\qquad [i,j = 1,2] \ .
\end{equation}
The cross sections
have a very characteristic dependence on the stop mixing angle 
$\theta_{\st}$, where
$\ti t_1^{} = \ti t_L^{} \cth_{\ti t} + \ti t_R^{}\sin\theta_{\ti t}$ and
$\ti t_2^{} = \ti t_R^{} \cth_{\ti t} - \ti t_L^{}\sin\theta_{\ti t}$.

\begin{figure}[htb]
  \begin{minipage}[t]{65mm} %\centering 
    \mbox{\href{pictures/2/stop_fig1.pdf}{{\epsfig{file=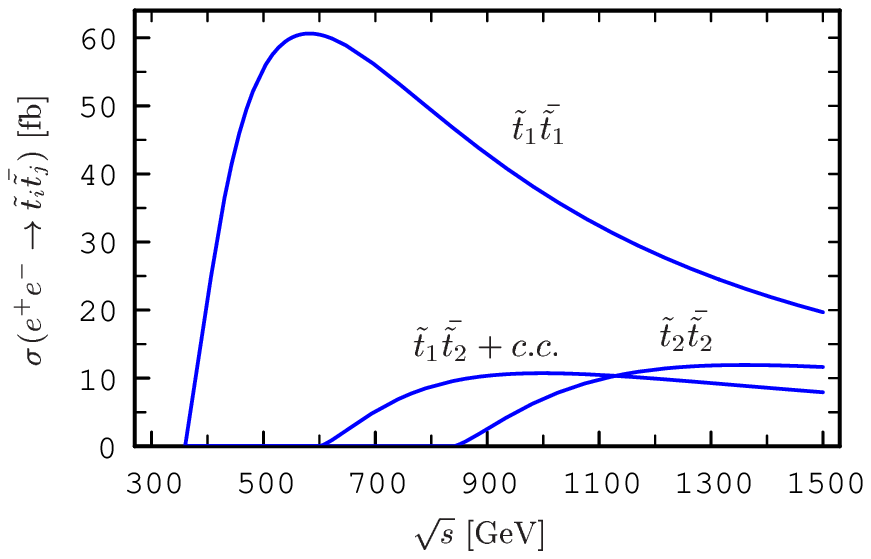,%
        bbllx=165pt,bblly=599pt,bburx=420pt,bbury=765pt,clip=,%
        height=5.5cm }}}}
  \end{minipage} \hfill
  \begin{minipage}[t]{65mm}  \vspace{-45mm}
    \caption{
      Energy depen\-dence of $\st_i \st_j$ production cross sections 
      with unpolarised beams
      for $m_{{\ti t}_1}=180$\,GeV, $m_{{\ti t}_2}=420$\,GeV, $\cst=0.66$.}
    \label{fig:stop_fig1}
  \end{minipage}
  %\vspace{-8mm}
\end{figure}

The phenomenology of stop production and decay at a linear collider
has been discussed in detail in~\cite{bartl1}.  
Figure~\ref{fig:stop_fig1} shows the energy dependence of the $\st_i \st_j$ 
pair production cross sections.
Initial state radiation, supersymmetric QCD~\cite{glucorr,helmut} 
and Yukawa coupling corrections~\cite{yuk} are included.

\begin{figure}[ht]  {\setlength{\unitlength}{1mm}
\begin{picture}(155,60)
%\put(0,0){\framebox(155,60)}
\put(2,0){\mbox{  
  \href{pictures/2/stop_fig2.pdf}{{\epsfig{file=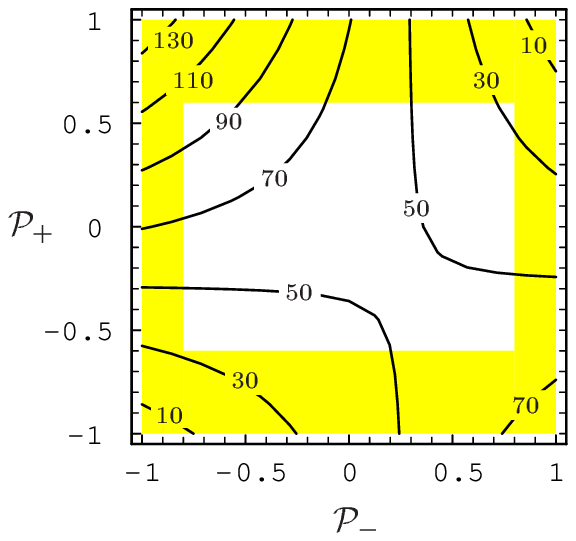,
      bbllx=208pt,bblly=599pt,bburx=380pt,bbury=762pt,height=6cm}}} 
  \hspace{14mm}
  \href{pictures/2/stop_fig3.pdf}{{\epsfig{file=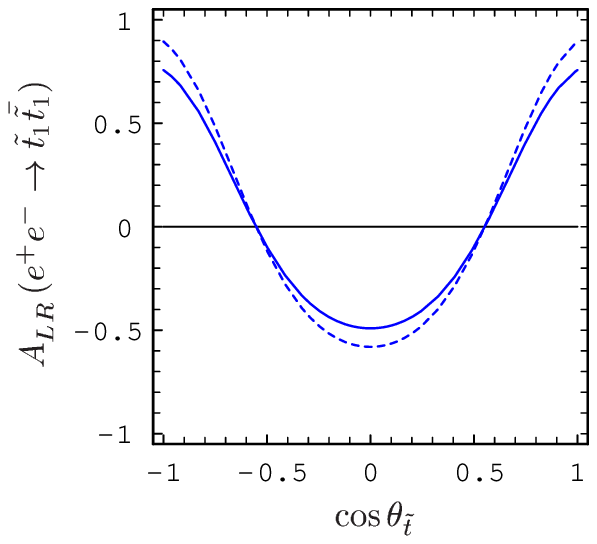,
      bbllx=208pt,bblly=599pt,bburx=380pt,bbury=762pt,height=6cm}}} }}  
\put(5,54){\mbox{\bf a)}}
\put(82,54){\mbox{\bf b)}}
\end{picture}
\caption{a) Contours of $\sigma(e^+e^-\to\tilde t_1\bar{\tilde t_1})$ 
  in [f\/b] as function of $e^\mp$ polarisations
  ${\cal P}_-$ and ${\cal P}_+$ for $\sqrt{s}=500$\,GeV,
  $m_{\tilde t_1}=180$\,GeV, $\cos\theta_{\tilde t}=0.66$. 
  The white area shows the accessible range at {\sc Tesla}. 
  b) Left--right asymmetry $A_{LR}(e^+e^-\to\tilde t_1\bar{\tilde t_1})$
  as function of $\cos\theta_{\tilde t}$ 
  with $|\cP_-| = 0.8$ and $\cP_+ = 0$ (solid curve) 
  or $|\cP_+| = 0.6$ (dashed curve).}
\label{fig:stop_fig2}
}\end{figure}

Figure~\ref{fig:stop_fig2}\,a shows the contour lines of the cross section
$\sigma (e^+e^- \ra \ti t_1 \bar{\ti t_1})$ as a function of the $e^-$
and $e^+$ beam polarisation.
The cross section can be significantly increased by 
choosing the maximally possible $e^-$ {\it and} $e^+$ polarisation.
Using polarised beams one can also measure the left--right polarisation
asymmetry  
\begin{equation}\label{stop_eq2}
  A_{LR} \equiv \frac{\s_L - \s_R}{\s_L+\s_R}
\end{equation}
where $\s_L^{} = \s\,(-\cP_-,\,\cP_+)$ and $\s_R^{} = \s\,(\cP_-,\,-\cP_+)$. 
This observable is very sensitive to the stop mixing $\cst$ , 
as shown in \fig{stop_fig2}\,b.

\subsection{Parameter determination}

Owing to the large luminosity and the availability of polarised 
beams, it is possible to determine the mass and the mixing angle of 
the stop very precisely. One method consists 
of measuring production cross sections $\sigma_R$ and $\sigma_L$ of 
different polarisations.

\begin{figure}[ht] {\setlength{\unitlength}{1mm}
\begin{picture}(155,70)
%\put(0,0){\framebox(155,70)}
%%%\put(0,6){\mbox{ \href{pictures/2/stop_fig4.pdf}{{\epsfig{file=stop_fig4.eps,height=7.1cm}}} }}  
\put(-2,0){\mbox{ \href{pictures/2/stop_fig4N.pdf}{{\epsfig{file=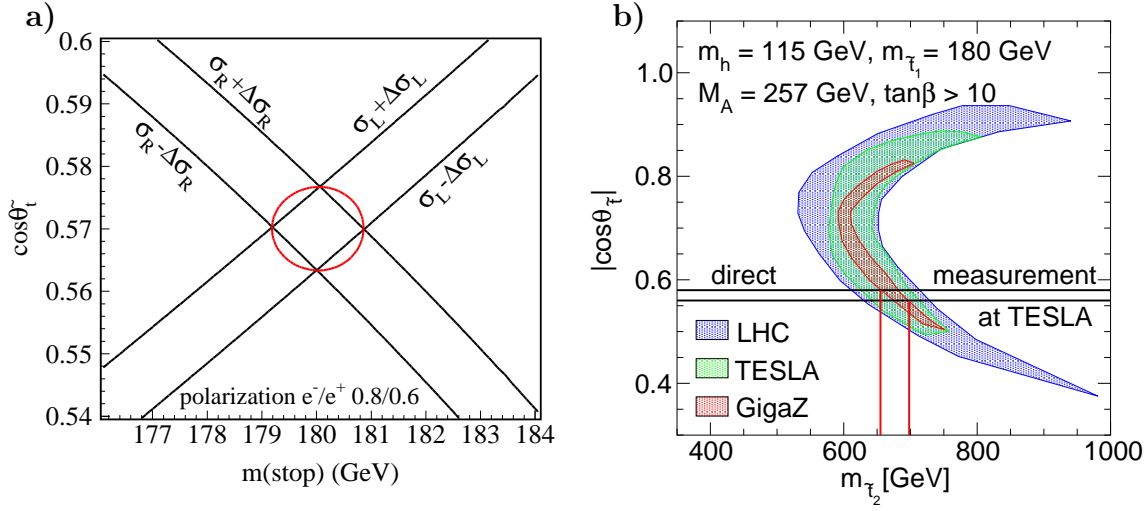,height=7.8cm}}} }}  
\put(80,0){\mbox{\href{pictures/2/stop_fig5.pdf}{{\epsfig{file=stop_fig5.eps,height=6.5cm}}} }}  
\put(2,64){\mbox{\bf a)}}
\put(80,64){\mbox{\bf b)}}
\end{picture}
\caption{ a) Contours of 
    $\sigma_{R}(\tilde{t}_1 \overline{\tilde{t}}_1)$ and
    $\sigma_{L}(\tilde{t}_1 \overline{\tilde{t}}_1)$, %  with
    $\tilde{t}_1\to c\,\ch^0_1$ as a function of
    $m_{\tilde{t}_{1}}$ and $\cos\theta_{\tilde{t}}$
    for $\sqrt{s} = 500$\,GeV, $\cL=2 \times  500\,\fbi$,
    $|\cP_-|= 0.8$ and $|\cP_+|=0.6$.
    b) Indirect constraints on $m_{{\ti t}_2}$ and $\cos\theta_{\tilde{t}}$
    expected from future high precision measurements at LHC and {\sc Tesla}.}
\label{fig:stop_fig4}
}\end{figure}

A simulation of $e^{+}e^{-} \to \tilde{t}_{1}\overline{\tilde{t}}_{1}$ 
with decay modes $\tilde{t}_{1} \to \ch^{0}_{1} c$ and 
$\tilde{t}_{1} \to \ch^{+}_{1} b$ is performed in~\cite{keranen} 
including full SM background.
The decay $\tilde{t}_{1} \rightarrow \ch^{0}_{1} c$ results in a
signature of two jets and large missing energy, while
$\tilde{t}_{1} \rightarrow \ch^{+}_{1} b$ leads to two $b$ jets, 
further jets from the $\ch^{+}_{1}$ decay and missing energy.
The study, based on a fast simulation of the {\sc Tesla} detector, is done for 
$m_{\tilde{t}_{1}} = 180$\,GeV and $\cos\theta_{\tilde{t}} = 0.57$ with
$m_{\ch^{0}_{1}} = 100$\,GeV  for the neutralino decay channel and
$m_{\tilde{\chi}^{+}_{1}} = 150$\,GeV, $m_{\tilde{\chi}^{0}_{1}} = 60$\,GeV 
for the chargino channel.
%%%%%%%%%
%A measurement of the stop production cross section with a statistical error 
%of 2\% for $\tilde{t}_{1} \rightarrow \ch^{0}_{1} c$ and about 1\%
%for $\tilde{t}_{1} \rightarrow \ch^{+}_{1} b$ appears feasible.
%%%%%%%%%
For 80\% $e^-$ and 60\% $e^+$ polarization and 
assuming $\cL=2 \times  500\,\fbi$, 
a measurement of the stop production cross section with a statistical 
error of 1.5\% for $\tilde{t}_{1} \rightarrow \ch^{0}_{1} c$ 
and about 0.75\% for $\tilde{t}_{1} \rightarrow \ch^{+}_{1} b$ 
appears feasible. 
The systematical error is of the order of 1\%. 
Figure~\ref{fig:stop_fig4}\,a shows the corresponding $\sigma_R$, $\sigma_L$
measurements in the $m_{\tilde{t}_{1}} - \cos\theta_{\tilde{t}}$ plane.
%%%%%%%%%
%For the channel $\tilde{t}_{1} \rightarrow \ch^{0}_{1} c$
%one obtains accuracies of $\delta m_{\tilde{t}_{1}} = 1.25$~GeV and 
%$\delta \cos\theta_{\tilde{t}} = 0.01$. The errors for the decay 
%$\tilde{t}_{1} \rightarrow \chi^{+}_{1} b$ are about half the size.
%%%%%%%%%
For the channel $\tilde{t}_{1} \rightarrow \ch^{0}_{1} c$
one obtains accuracies of $\delta m_{\tilde{t}_{1}} = 0.8$\,GeV and 
$\delta \cos\theta_{\tilde{t}} = 0.008$. The errors for the decay 
$\tilde{t}_{1} \rightarrow \chi^{+}_{1} b$ are about half the size.

A more precise determination of the stop mass can be achieved by a 
method proposed in~\cite{feng}. 
The mass is obtained as `end-point' of the minimum kinematically allowed 
squark mass distribution ${\rm d}\s/{\rm d}m_{min}(\st)$
(in analogy to the smuon case, \fig{sl_smuonR}b).
Taking into account initial state radiation of photons and gluon radiation, 
a mass of $m_{\st}=300$\,GeV can be determined with a statistical error of 
$\sim 0.3\%$ with an integrated luminosity of $500\,\fbi$~\cite{degk}.

A very good way to measure the stop mixing angle is to use 
the left--right asymmetry $A_{LR}$, eq.~(\ref{stop_eq2}), where
kinematic effects and uncertainties should largely cancel. For 
the parameter point above one gets an error on $\cst$ at the per mil level
with an integrated luminosity of $250\,\fbi$ for each beam 
polarisation~\cite{bartl1}, cf.~\fig{stop_fig2}\,b.

If the energy is high enough the heavier stop $\st_2$ can also be 
produced via $e^+e^-\to\st_1\bar{\st_2}$, and its mass $m_{{\ti t}_2}$ can 
be determined. The stop masses $m_{{\ti t}_1}$, $m_{{\ti t}_2}$
%$\mst{1}$, $\mst{2}$ 
and the mixing $\cst$ are 
related to the basic  soft SUSY breaking parameters $M_{\ti Q},M_{\ti U}$ 
and $A_t$, which can be calculated if $\tan \beta$ and $\mu$ are known. 
With $m_{{\ti t}_2}$ measured to an error of $\sim 2\%$ and if 
$\tan\b$ and $\mu$ are known from other measurements to $\lsim 10\%$, then
$M_{\ti Q}, \ M_{\ti U}$ and $A_t$ can be determined with an accuracy 
of few percent~\cite{bartl1}.

Direct information on the stop parameters can be combined with indirect 
information by requiring consistency of the MSSM with precise measurements 
of the Higgs-boson mass $m_{h^0}$, and the electroweak observables $M_W$ 
and $\sin^2\theta_{\mathrm{eff}}$~\cite{heineweig}.
This is shown in \fig{stop_fig4}\,b, where the allowed parameter space 
expected from future measurements at LHC and {\sc Tesla}
is displayed in the $m_{\tilde t_2} - |\cos\theta_{\tilde t}|$ plane,  
for the stop parameters of \fig{stop_fig4}\,a and the other MSSM parameters 
chosen according to the RR2 scenario.
The allowed region %in the $m_{\tilde t_2} - |\cos\theta_{\tilde t}|$ plane 
is significantly reduced by data from {\sc Tesla}, in particular in the GigaZ 
scenario. 
Using $\cos\theta_{\tilde t}$ from polarised cross section 
measurements one gets $m_{\tilde t_2}$ with a precision of $\sim 5\%$. 
A comparison with direct mass measurements would test
the MSSM at its quantum level in a sensitive and highly 
non-trivial way.

In general, the stop can decay in a variety of ways 
depending on its mass and  those of the other SUSY particles~\cite{bartl1}. 
There are additional two--body decays: 
$\st_i\to \nt_k t$, $\ch^+_j b$, $\sg t$, $\sb_j W^+ (H^+)$ and 
$\st_2\to \st_1 Z^0 (h^0,H^0,A^0)$. 
If these decays are kinematically not possible, 
the loop--decays $\st_1\to\nt_{1,2}\,c$~\cite{hikasa} as well as
three--~\cite{werner} and four--~\cite{djouadi} particle decays can be 
important.

One should point that a light stop of
$m_{{\ti t}_1} \lsim 250$\,GeV may escape detection at the hadron colliders 
{\sc Tevatron} and LHC and may only be discovered at {\sc Tesla}.
 
%
% TESLA TDR - SUSY chapter
% section on mSUGRA   
% last update 29 Jan 2001
%

\section{The Minimal Supergravity (mSUGRA) Model \label{SUSY_sugra}}

In supergravity supersymmetry is broken in a `hidden' 
sector and the breaking is transmitted to the `visible' sector 
by gravitational interactions. %~\cite{nilles}.  
In the more specific minimal supergravity (mSUGRA) model 
all scalar particles (sfermions and Higgs bosons) have a 
common mass $m_{0}$ at the unification point $M_{\rm{GUT}} 
\approx 10^{16}$\,GeV. The gaugino masses $M_{1},\, M_{2},\, M_{3}$ 
(corresponding to $U(1),\, SU(2)$ and $SU(3)$, respectively) unify to a 
common gaugino mass $m_{1/2}$ and all trilinear coupling 
paramters $A_{ijk}$ have the same value $A_{0}$ at $M_{\rm{GUT}}$. 
One also has unification of the electroweak and strong coupling 
parameters $\alpha_{i}\, (i = 1,2,3)$. A further reduction of the 
parameters is given by invoking `radiative electroweak symmetry 
breaking'. As a consequence, one has only the following input 
parameters: $m_{0},\, m_{1/2},\, A_{0},\, \tan\beta,\, {\rm{sign}}\,\mu$. 
The whole SUSY particle spectrum can then be calculated by making use 
of renormalization group equations.

In mSUGRA it turns out quite generally that $|\mu|>M_2$, so that 
$\mnt{2}\simeq\mch{1}\simeq 2\mnt{1}\sim M_2$. 
Both $\nt_1$ and $\nt_2$ are gaugino--like, $\nt_1$ is almost 
a pure $B$--ino and  $\nt_2$ is almost a pure $W^3$--ino. 
The slepton masses of the first and second generation are given by: 
$m_{\ti\ell_R}^2 = m_0^2 + 0.15\,m_{1/2}^2 - \sin^2\t_W M_Z^2\cos2\b$, 
$m_{\ti\ell_L}^2 = m_0^2 + 0.52\,m_{1/2}^2 
                   - (\frac{1}{2}-\sin^2\t_W) M_Z^2\cos2\b$, and 
$m_{\ti\nu_\ell}^2 = m_0^2 + 0.52\,m_{1/2}^2 + \frac{1}{2}M_Z^2\cos2\b$. 
Analogous equations hold for squarks. 
It is also noteworthy that in mSUGRA the lightest neutralino $\nt_1$ 
is most naturally a good dark matter candidate if 
$\mnt{1},\,m_{\ti\ell_R}\leq 200$\,GeV~\cite{drees_cdm}. 

The precise mass measurements of sleptons, neutralinos and charginos
described in sections \ref{SUSY_sleptons} and \ref{SUSY_gauginos} 
(see Table~\ref{tab:masses})
constitute an over--constrained set of observables which allow 
to determine the structure and parameters of the underlying SUSY 
theory~\cite{martyn}.
A widely employed strategy, for example at the LHC,
is to assume a SUSY breaking scenario and then fit to the corresponding
low--energy particle spectrum including experimental uncertainties.
Applying such a model dependent top--down approach to scenario RR1, 
one expects accuracies on the mSUGRA parameters as given 
in Table~\ref{tab:msugraerr2}.

\begin{table}[htb]
\begin{center}
\begin{tabular}{l c c}
  parameter &  \ input RR1 \ & error\\
  \hdick
\addtop  $m_0$             &\            100\,GeV \ & \ 0.09\,GeV\\
  $m_{1/2}$         &\            200\,GeV \ & \ 0.10\,GeV\\
  $A_0$             &\ \hphantom{00}0\,GeV \ & \ 6.3\hphantom{0}\,GeV\\
  $\tan\beta$       &\ 3                  \ & \ {0.02}\\
  ${\rm sign}(\mu)$ &\ +                  \ & \ no fit
\end{tabular}
\end{center}
\caption{Estimated accuracy on the mSUGRA parameters.}
\label{tab:msugraerr2}
\end{table}

The common scalar and gaugino masses $m_0$ and $m_{1/2}$ can be determined
to better than one per mil, $\tan\beta$ to better than a percent,  
and there is even some sensitivity to the trilinear coupling $A_0$
(coming from the higher mass sparticles).
The magnitude of $\mu$ is obtained implicitly by the requirement
of electroweak symmetry breaking.

While this method is a useful illustration of the SUSY measurement
potential, the scenario assumptions are effectively constraints in the fit. 
This is particularly dangerous for models
with pseudo-fixed point structures, where the low energy predictions will
be quite similar for a large range of fundamental parameters.
Also, new intermediate scales below the GUT scale will not be immediately
apparent in a top--down approach.
The advantages of {\sc Tesla} to perform a model independent analysis of SUSY
parameters will be discussed in section~\ref{SUSY_hiE}.

%
% TESLA TDR - SUSY chapter
% section on GMSB
% last update 24 Jan 2001
%

\section{Gauge--Mediated SUSY Breaking (GMSB) \label{SUSY_gmsb}}

In supergravity models the typical fundamental scale of SUSY breaking 
is ${\cal O}(10^{11}\,\mathrm{GeV})$. 
%Such an approach is also suggested by string theory.
An alternative possibility is that supersymmetry breaking occurs at lower 
energies with gauge interactions serving as the messengers, referred to as 
`gauge mediated supersymmetry breaking' (GMSB)~\cite{dine}. 
It avoids some potential problems of SUGRA, {\em e.g.}
flavour changing neutral currents and CP violation. 
GMSB models are also very predictive as 
%they involve just a few parameters:
the MSSM spectrum depends on just a few parameters:
\begin{equation}\label{gmsb_eq3}
  M_{\rm mess},\; N_{\rm mess},\; \Lambda,\; \tan\beta,\; {\rm sign}\,\mu,
\end{equation}
where $M_{\rm mess}$ is the messenger scale and $N_{\rm mess}$ is the 
messenger index parameterising the structure of the messenger sector. 
$\Lambda$ is the universal soft SUSY breaking scale felt by the low energy 
sector.

The MSSM parameters and the sparticle spectrum (at the weak scale) 
are determined from renormalisation group equation evolution starting from 
boundary conditions at the messenger scale $M_{\rm mess}$. 
The gaugino masses at $M_{\rm mess}$ are given by 
$M_{i} = N_{\rm mess}\Lambda\,g(\Lambda/M_{\rm mess})\,(\alpha_i/4\pi),$
\ and\,  the\, squark/slepton masses by \hfill
${m_{\tilde f }^2 = 2 N_{\rm mess}}$$\Lambda^2 $ $f(\Lambda/M_{\rm mess})$
$\sum_i\,(\alpha_i/4\pi)^2\,C_i$, where $g$ and $f$ are one-- and two--loop 
functions and $C_i$ are known constants. 
If $M_2 = 100-300$\,GeV at the electroweak scale, then 
${10\,\mathrm{TeV} \lsim \Lambda \lsim}$ $120$\,TeV. 
As an illustration a GMSB sparticle mass spectrum is shown in 
\fig{susy_mass-spectra}. The charginos, neutralinos and sleptons are much
lighter than the gluino and squarks. 

\begin{figure}[htb] {\setlength{\unitlength}{1mm}
  \begin{minipage}[t]{70mm} 
    \begin{picture}(70,72)
      \put(-1,0){\mbox{\href{pictures/2/gmsb_fig15.pdf}{{\epsfig{file=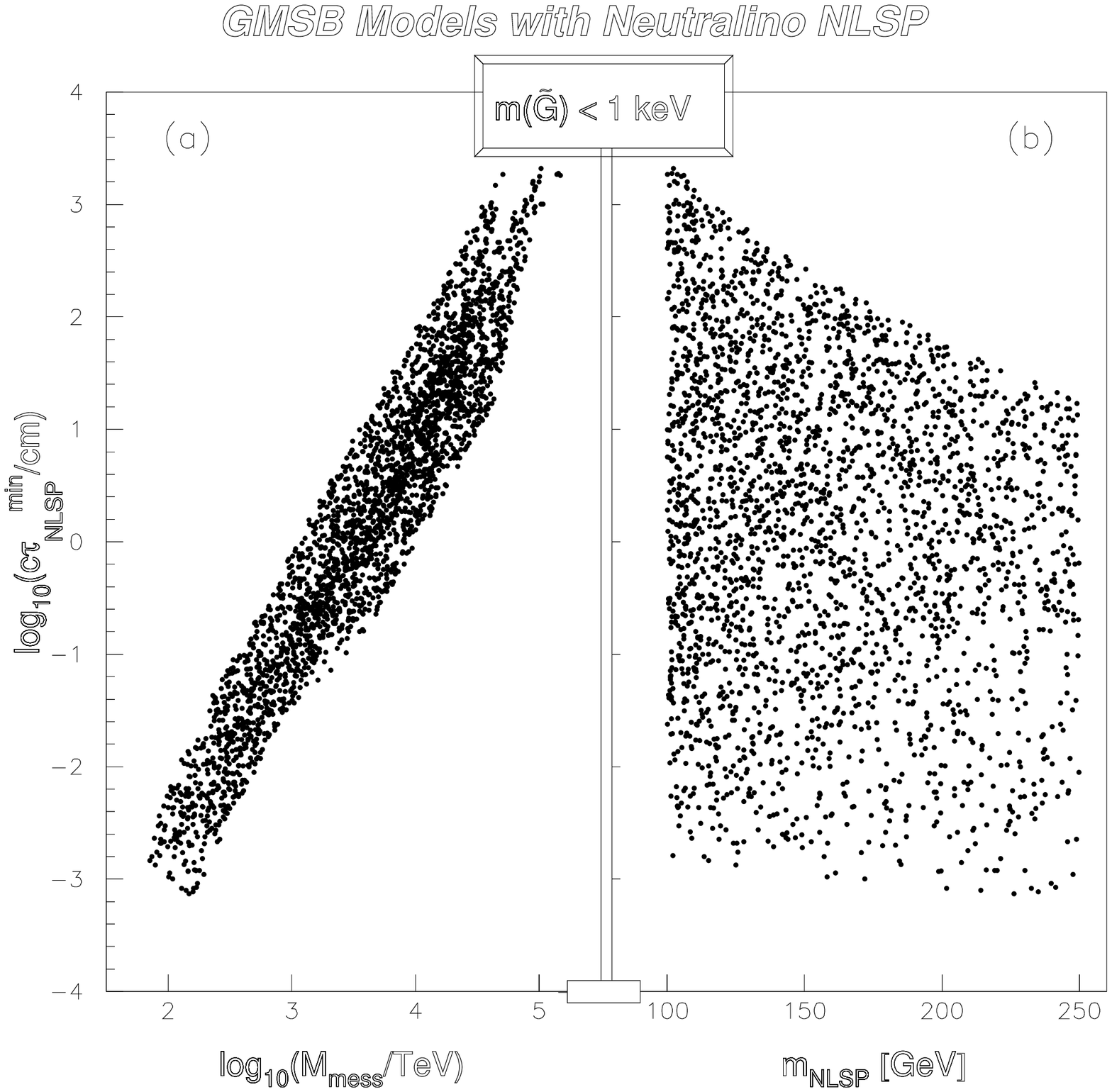,height=7cm}}}}}
    \end{picture} 
    \refstepcounter{figure}
    {\footnotesize Figure~\thefigure: 
      {\it Neutralino NLSP lifetime %(lower limit) 
        as a function 
        of a) the messenger scale $M_{\rm{mess}}$ and 
        b) the NLSP mass $m_{\ch^{0}_{1}}$.
        Each dot represents a different choice
        of GMSB model parameters. }
      \label{fig:gmsb_fig4}}
  \end{minipage} \hspace{5mm}
  \begin{minipage}[t]{70mm} 
    \begin{picture}(70,72)
      \put(3,0){\mbox{\href{pictures/2/gmsb_fig9a.pdf}{{\epsfig{file=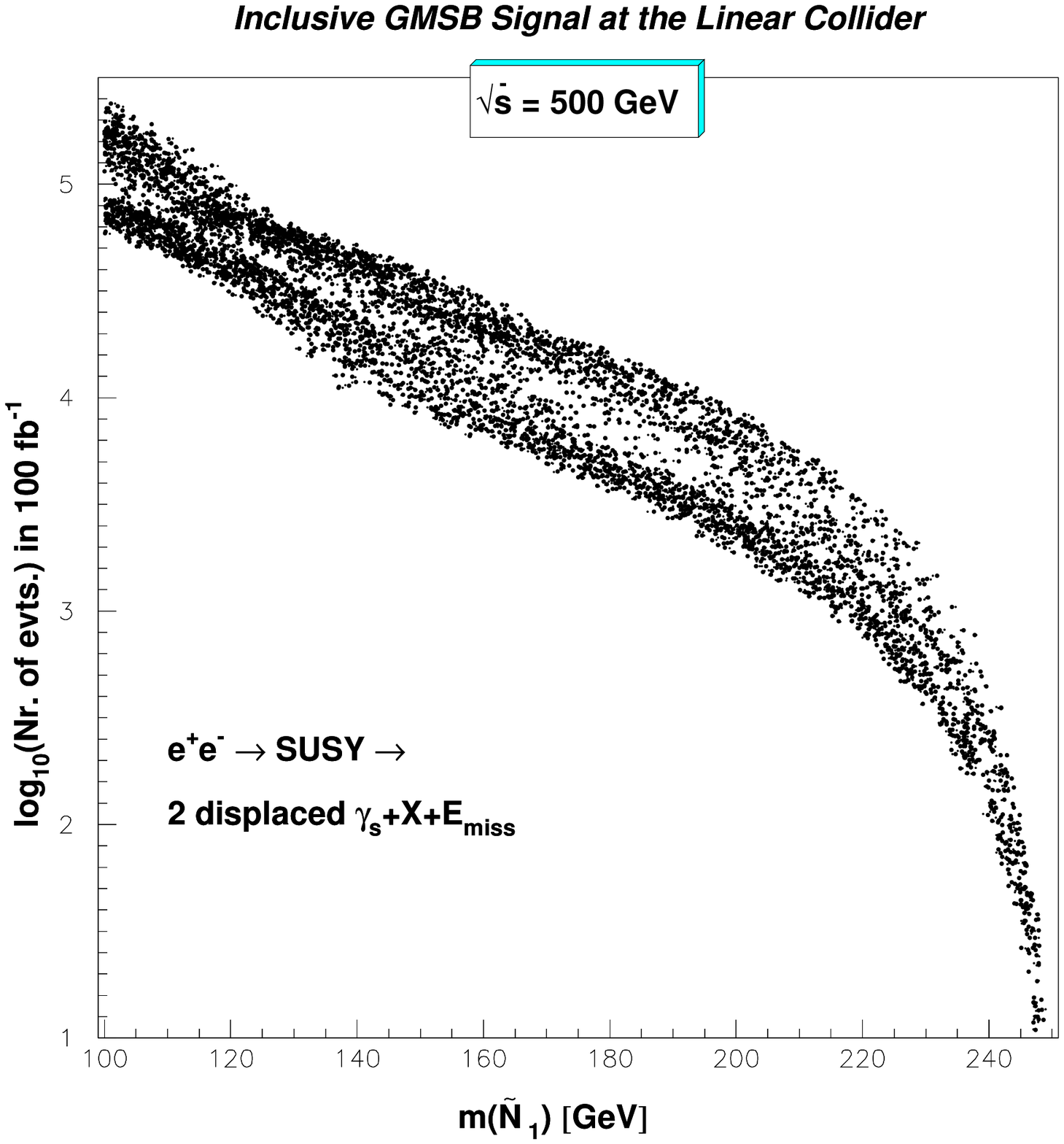,height=7cm}}}}}    
    \end{picture}    
    \refstepcounter{figure}
    {\footnotesize Figure~\thefigure: 
      {\it Event rate for displaced photon signatures from
        $e^{+}e^{-} \to \nt_1\, \nt_1\,X$, $\nt_1 \to \gamma \tilde G$
        as a function of the NLSP mass
        at $\sqrt{s} =  500$\,GeV, ${\cal L} = 100\,\fbi$.}
      \label{fig:gmsb_fig5}}
  \end{minipage}
}\end{figure}

A very interesting feature of GMSB is that the lightest 
supersymmetric particle is the gravitino
\begin{equation}\label{gmsb_eq4}
  m_{3/2} \equiv m_{\tilde{G}} = \frac{F}{\sqrt{3}\,M_{P}^{'}}\simeq
  {\left(\frac{\sqrt{F}}{100\,{\rm TeV}} \right)}^{2} 2.37\,{\rm eV} \, ,
\end{equation}
where $M_{P}^{'} = 2.4 \cdot 10^{18}$\,GeV is the reduced Planck mass
and $\sqrt{F}$ is the fundamental scale of SUSY breaking with a typical
value of $100$\,TeV. Therefore, the phenomenology is strongly 
determined by the next-to-lightest SUSY particle (NLSP), which decays 
into the gravitino $\tilde G$.
The NLSP can be the neutralino, which decays dominantly via 
$\tilde{\chi}^{0}_{1} \to \gamma\, \tilde G, \ f\, \bar f\, \tilde G$. 
The lifetime is given by
\begin{equation}\label{gmsb_eq5}
    c\,\tau_{NLSP} \simeq \frac{1}{100\,\cal{B}} 
    \left(\frac{\sqrt{F}}{100\,{\rm TeV}}\right)^{4} \left( 
    \frac{m_{NLSP}}{100\,{\rm GeV}}\right)^{-5}\,{\rm cm} \, ,
\end{equation}
where $\cal{B}$ is of order unity depending on the nature of the NLSP. 
Assuming $m_{\ti G}<1$\,keV as favoured by cosmology, 
typical decay lengths range from micro-meters to tens of meters. 
Figure~\ref{fig:gmsb_fig4} shows the neutralino NLSP lifetime as a function of the 
messenger scale and $m_{\tilde{\chi}_{i}^{0}}$ for various sets of GMSB 
parameters~\cite{ambrosanio}.

A detailed simulation of inclusive $\nt_1$ production and decays
$\nt_1 \to \gamma\,\ti G, \ f \bar f \,\ti G$ is presented 
in~\cite{ambrosanio}.
The proposed {\sc Tesla} detector is capable of identifying neutralino decays 
and measuring its mass to within a few per mil from the endpoints of the 
$E_\gamma$ spectrum. 
The event rate for displaced photons, not pointing to the interaction vertex, 
can be large even for NLSP masses close to the production limit, 
see~\fig{gmsb_fig5}.
Various techniques, such as tracking, pointing calorimetry and statistical 
photon counting methods, provide accurate measurements of the decay length 
$c\tau$ over a large range of $30\,\mu{\rm m} - 40\,{\rm m}$ to better than 10\%.
Such data would allow one to extract the scale $\sqrt{F}$ with an accuracy 
of $\sim5\%$. 
Together with a knowledge of the SUSY particle spectrum, a determination of 
the other fundamental GMSB parameters is feasible with high precision: at the 
level of per mil for $\Lambda$ and $N_{\rm mess}$ and per cent for $\tan\beta$
and $M_{\rm mess}$.

Other scenarios with a slepton as NLSP have also been 
studied~\cite{ambrosanio}, {\em e.g.}
$\stau_1$  decaying to $\stau_1 \to \tau \ti G $, producing
long-lived, heavy particles or $\tau$ pairs, possibly coming from secondary 
decay vertices.

\clearpage

%
% TESLA TDR - SUSY chapter
% section on AMSB
% last update 24 Jan 2001
%

\section{Anomaly--Mediated SUSY Breaking (AMSB) \label{SUSY_amsb}}

SUSY breaking may not be directly communicated from the hidden
to the visible sector. This is the case in the
so-called anomaly mediated SUSY breaking models (AMSB), where gauginos masses
are generated at one loop and scalar masses at two loops as a consequence
of the 'super--Weyl (superconformal) anomaly'~\cite{randall,giudice}.
The gaugino masses are no more universal, but are given by
\begin{equation}
  M_{i} = \frac{\beta_{i}}{g_i}\, m_{3/2} \ ,
\end{equation}
where $\beta_i$ are the one--loop beta functions.
In the simplest form, however, the squared masses of the sleptons turn out to
be negative (tachyonic).  To avoid this, it suffices phenomenologically
to introduce a universal scalar mass $m_0^2$ at the GUT scale.  
The parameters of the model are then
$m_0$, $m_{3/2}$, $\tan\beta$ and $\rm{sign}\,\mu$.
An example of an AMSB mass spectrum is shown in~\fig{susy_mass-spectra}.

The most characteristic feature is the relation $M_1 \sim 3\, M_2$
in contrast to SUGRA scenarios, where $M_1 \simeq 0.5\, M_2$. 
Therefore, in the AMSB framework the wino is the lightest supersymmetric
particle.  Furthermore, one has near degeneracy of the lighter chargino
$\ch^\pm_1$ and the wino--like neutralino $\ch^0_1$ masses, 
which has important phenomenological implications.
Another property of the mass spectrum is the near degeneracy of sleptons 
$\sl_R$ and $\sl_L$, 
which can be tested at {\sc Tesla} very precisely.  

\begin{figure}[htb] 
  \begin{minipage}[b]{75mm} %\vspace{-5mm}  
    \href{pictures/2/amsb_deltam.pdf}{\epsfig{file=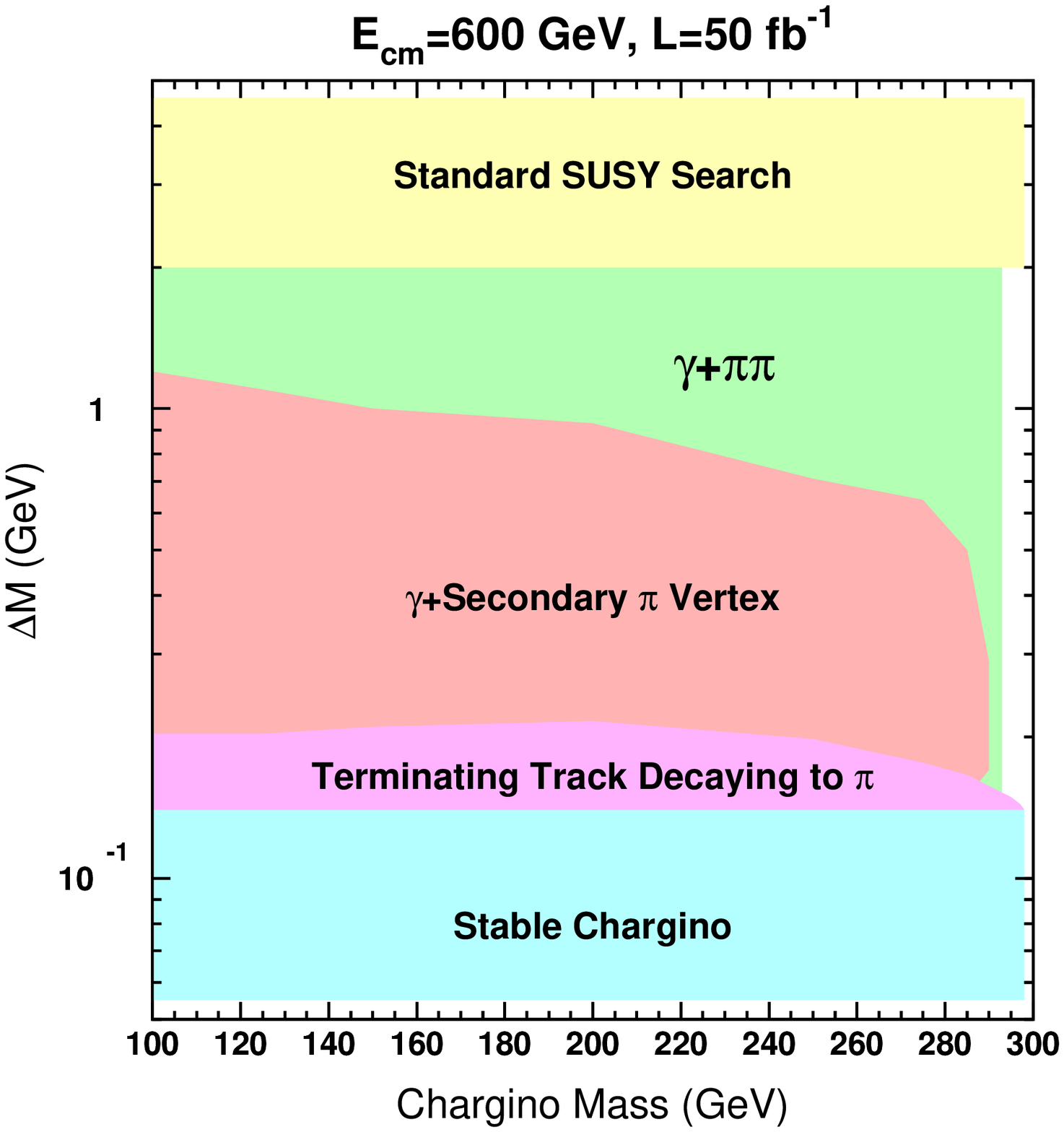,height=9cm}}
  \end{minipage} \hfill
  \begin{minipage}[b]{55mm}  %\vspace{30mm}
  \caption{Discovery reach of AMSB scenarios as a function of 
    $\dmchi \equiv m_{\tilde\chi^\pm_1} - m_{\tilde\chi^0_1}$ and 
    $\mch{1}$ at $\sqrt{s}=600$\,GeV for $\cL=50\,\fbi$.
    The signature $\gamma + \mslash$ extends up to chargino masses of 
    $\sim200$\,GeV.\vspace*{16mm}}
  \label{fig:amsb_deltam}
  \end{minipage}
\end{figure}

Search strategies for $e^+e^- \to \ch^+_1 \ch^-_1(\gamma)$ production with
almost degenerate chargino and neutralino masses are discussed 
in~\cite{gunion}.
The most critical ingredients are the lifetime and the decay modes of 
$\ch^\pm_1$, which depend almost entirely on  the small mass difference
$\dmchi \equiv m_{\ch^{\pm}_{1}} - m_{\ch^{0}_{1}}$. 
(i) 
For $\dmchi< m_\pi$ the chargino may exit the detector as heavily ionising 
stable particle, or decay to a soft, but visible $e$ or $\mu$ yielding a 
secondary vertex.
(ii)
If $m_\pi < \dmchi < 0.2$\,GeV the $\tilde\chi^{\pm}_1$ may decay inside 
the tracking system to a soft $\pi^\pm$, which need not be visible.
The signature is a terminating track.  
(iii) 
For $ 0.2\,\mathrm{GeV} < \dmchi \lsim 2-3$\,GeV the decay pion(s) will be detected, 
possibly associated to a secondary vertex.
The large background from $\gamma\gamma \to \pi\pi$ may be suppressed
by requiring an additional tagged photon.
If the pions have too low an energy to be detected, then one relies on a single
photon $\gamma + \mslash$ signature from $e^+e^-\to\gamma\ch^+_1 \ch^-_1$ 
production, which, however, has a large $\gamma\nu\bar\nu$ background.
(iv) 
Once $\dmchi\gsim 2-3$\,GeV, the $\tilde\chi^{\pm}_1$ decay products have 
sufficient energy to be detected and resemble the usual MSSM topologies.

The AMSB discovery potential is shown in~\fig{amsb_deltam} as a function of 
$\dmchi$ and $\mch{1}$ for $\sqrt{s} = 600$\,GeV.
With $\cL=50\,\fbi$ a large $\dmchi$ region can be covered almost to the 
kinematic limit.
The discovery regions increase only slightly with higher luminosity, except
for the $\gamma + \mslash$ channel, which would be extended beyond 
$\sim 200$\,GeV accesible with low luminosity.
Since $\dmchi = 0.2 - 2$\,GeV is typical of models with 
loop-dominated gaugino masses, the tagged $\gamma$ signals are very important.

%
% TESLA TDR - SUSY chapter
% section on R-parity violation
% last update 29 Jan 2001
%

\section[Supersymmetry with { $R$}--Parity Violation]{Supersymmetry with { $\bR$}--Parity Violation \label{SUSY_rpv}}

So far it has been assumed that the multiplicative quantum number 
$R$--parity is conserved. Under this symmetry all standard model 
particles have $R_p = +1$ and their superpartners $R_p = -1$. As a 
consequence, SUSY particles are only produced in pairs with the 
lightest of them (LSP) being stable, giving rise to missing energy in 
an experiment. In the MSSM, this is the neutralino $\tilde{\chi}^{0}_{1}$. 

$R$--parity conservation has, however, no strong theoretical justification. 
The superpotential admits explicit $R$--parity violating ($\rpv$) 
terms such as~\cite{hall}
\begin{equation}\label{rpv_eq1}
  W_{\rpv} = 
    \sum_{i,j,k} \left ( \frac{1}{2} \lambda_{ijk} L_{i} L_{j} 
    \bar{E}_{k} + \lambda^{'}_{ijk} L_{i} Q_{j} \bar{D}_{k} + 
    \frac{1}{2} \lambda^{''}_{ijk} \bar{U}_{i} \bar{D}_{j} \bar{D}_{k} 
    \right) \ ,
\end{equation}
where $L,Q$ are the left--handed lepton and squark 
superfield and $\bar{E}, \bar{D}, \bar{U}$ are the corresponding 
right--handed fields. If both lepton--number violating 
($\lambda_{ijk}$ and $\lambda^{'}_{ijk}$) and baryon--number 
violating ($\lambda^{''}_{ijk}$) couplings were present, they would 
give rise to fast proton decay. 
This is avoided by assuming at most one coupling to be finite.

$R$--parity violation changes the SUSY phenomenology drastically. 
The lightest supersymmetric particle decays, so the typical missing energy 
signature in the $R_p$ conserving MSSM is replaced by  multi-lepton and/or 
multi-jet final states.

\subsection{Single SUSY particle production}

If $R$--parity is violated, then single SUSY particle production is possible,
for instance 
$e^{+}e^{-}  \to \tilde{\nu} \to \ell \bar\ell, \
 \nu\tilde{\chi}^{0}, \ \ell^\pm \tilde{\chi}^{\mp}$,
which extends the accessible mass reach considerably.
For sneutrino masses $m_{\snu} < \sqrt{s}$ one expects spectacular 
resonances~\cite{lola,heyssler}. 
Since the exchanged sneutrino carries spin 0, the $\rpv$ signal can be further 
enhanced by polarising both the incoming electron and positron beams with the 
same helicities and thereby reducing any background mediated through 
$\gamma / Z$ exchange substantially~\cite{moortgat}.

\begin{figure}[ht] \vspace*{-8mm}
  \begin{minipage}[t]{70mm} \centering
    \mbox{{\href{pictures/2/rpv_ee.pdf}{{\epsfig{figure=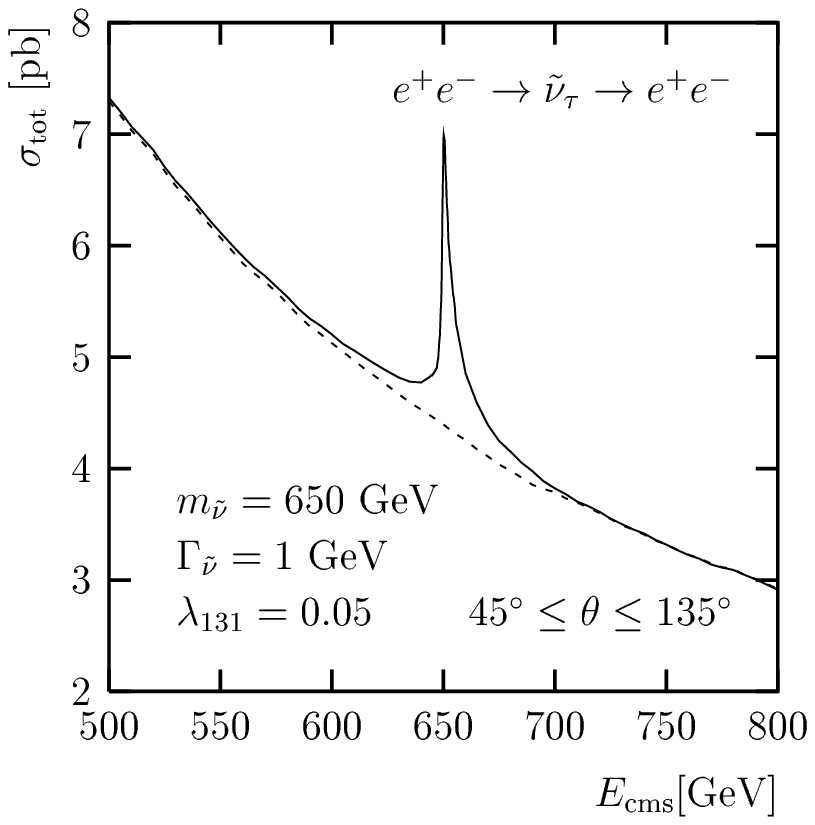,width=7.cm}}}}}
    \caption  {\footnotesize 
      $R_p$ violation in
      resonance production of $e^+e^- \to \snu_\tau \to e^+e^-$ interfering
      with Bhabha scattering.}
    \label{fig:rpv_fig1}
  \end{minipage} \hfill %\hspace{5mm}
  \begin{minipage}[t]{70mm} \centering
    \mbox{{\href{pictures/2/rpv_muchi.pdf}{{\epsfig{figure=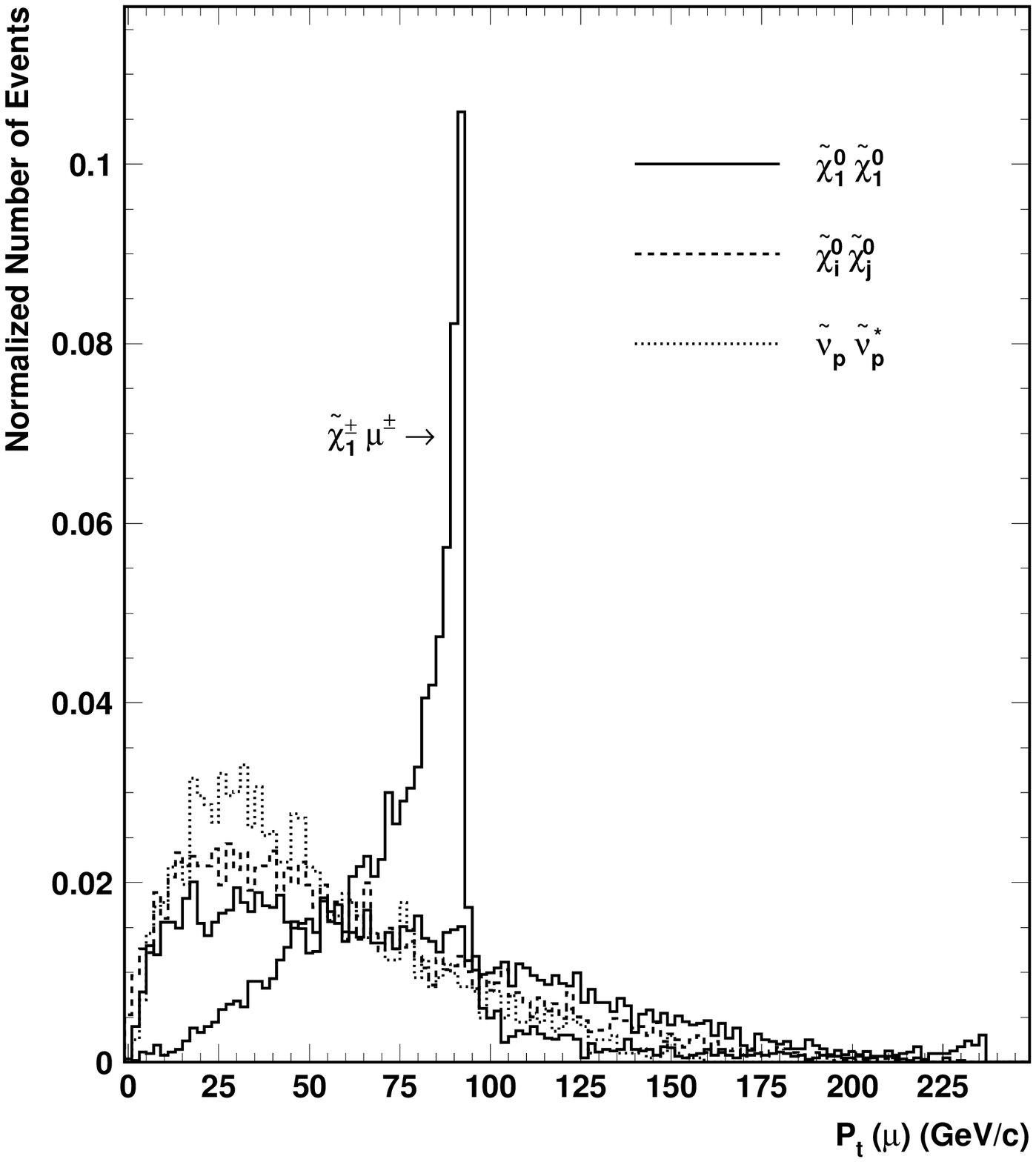,width=7.cm}}}}}
    \caption {\footnotesize 
      Highest $\mu$
      transverse momentum spectrum $P_t(\mu)$ in the $\rpv$ process
      $e^+e^-\to\tilde \chi^\pm_1 \mu^\mp\to4 \ell + \Eslash $ 
      at $\sqrt{s}= 500$\,GeV for
      $\l_{121}=0.05$, $m_{\tilde \nu}=240$\,GeV,
      $m_{\chi^+_1} = 115.7$\,GeV.}
    \label{fig:rpv_fig2}
  \end{minipage} 
\end{figure}

If both production and decay occur via $\lambda_{1j1}$ couplings, Bhabha 
scattering $e^+e^- \to e^+e^-$ is particular sensitive to the interference
with heavy sneutrino exchange diagrams~\cite{heyssler}. 
This is illustrated in~\fig{rpv_fig1} in case of $s$--channel resonance 
production. Masses beyond the center of mass energy are accessible via contact
interactions.
The effects scale as $(\lambda/m_{\snu})^2$ and one is sensitive to masses of 
$m_{\snu}\simeq 1.8$\,TeV for a coupling of $\lambda=0.1$
at the highest {\sc Tesla} energy.

A detailed study of single chargino production 
$e^{+}e^{-} \rightarrow \tilde{\chi}^\pm \mu^\mp$ 
with subsequent decays $\ch^+_1 \rightarrow \ell^+\nu \ch^0_1$ 
and $\ch^0_1 \rightarrow e e \nu_\mu$, $ \mu e \nu_e$
leading to a $4\ell + \Eslash $ final state 
is presented in~\cite{moreau}.
The production proceeds via $\tilde{\nu}_{\mu}$ exchange in the 
$s$--channel and $\tilde{\nu}_{e}$ exchange in the $t$--channel. 
Due to the $\lambda_{121}$ coupling which flips helicity, 
$\tilde{\chi}^{-}_{1}$ production occurs through 
$e^-_Le^+_L \to \tilde{\chi}^-_1 \mu^+$ and 
$\tilde\chi^+_1$ production through $e^-_R e^+_R \to \tilde\chi^+_1 \mu^-$. 
Figure~\ref{fig:rpv_fig2} shows the distribution of the highest 
muon transverse momentum.
Single chargino production can be easily separated from the background and
the pronounced peak in $P_{t} (\mu)$ can be used to reconstruct the
$\ch^\pm_1$ mass.
With ${\cal{L}} = 500\,\fbi$ at $\sqrt{s} = 500$\,GeV,
values of the $\rpv$ coupling $\lambda_{121}$ much smaller than 
present low--energy bounds can be probed for 
$150\,\mathrm{GeV} \leq m_{\snu} \leq 600\,\mathrm{GeV}$. 
The sensitivity strongly increases when approaching a resonance, 
{\em e.g.} $m_{\tilde{\nu}} = \sqrt{s'}$ via ISR radiation, 
where $\lambda_{121} \sim 10^{-4}$ can be reached.

Other $\rpv$ couplings are accessible via the LSP decays. 
A simulation of $e^+e^- \to \nt_1\nt_1$ production with hadronic 
$\nt_1$ decays via $\lambda^{''}_{233}$ coupling, %see eq.~(\ref{rpv_eq1}), 
which lead to 6 jets including 2 $b$--quark jets, is presented 
in~\cite{besancon}.
Exploiting the overconstrained kinematics of the final state, the SM
and SUSY background can be efficiently reduced and the %lightest 
neutralino 
$\nt_1$ can be reconstructed with a mass resolution of $\sim15\%$ 
for $m_{\nt_1} = 90 - 140$\,GeV.

A classification of $R_p$ violating signals in 
$e^+e^- \to \ch^+_1 \ch^-_1, \ \ch^0_i \ch^0_j$ % $(i,j = 1,2)$  
production,
where the LSP decays via $\lambda_{ijk}$ or $\lambda_{ijk}^{'}$ couplings,
is performed in~\cite{godbole}.
The $\rpv$ signature is an excess of events with at least three
leptons plus missing energy or jets, which should be easily recognisable
over the $R_p$ conserving MSSM and SM expectation.

\subsection[Bilinear violation of {$R$}--parity]{Bilinear violation of { $\bR$}--parity}

A particularly simple form of $R$--parity breaking is realised by additional 
bilinear couplings in the superpotential~\cite{epsrad}
\begin{equation}
    W'_{\rpv} = 
    \epsilon_{i}\, {L}_{i} {H}_{2} \ ,
\end{equation}
where ${L}_{i}$ and ${H}_{2}$ are the lepton and Higgs superfields.
The electroweak symmetry is
broken when the two Higgs doublets $H_{1}$ and $H_{2}$ and the neutral
components of the slepton doublets $L_{i}$ acquire vacuum expectation
values.  The model breaks lepton number and generates non--zero
Majorana neutrino masses, % \cite{Schechter80}, 
thus providing an elegant mechanism for the origin of neutrino masses.  
At tree--level only one of the neutrinos gets a mass
by mixing with neutralinos, leaving the other two neutrinos massless. 
While this can explain the atmospheric neutrino problem, to reconcile
it with the solar neutrino data requires going beyond the tree--level
approximation.  A full one--loop calculation of the neutralino--neutrino
mass matrix consistent with solar and atmospheric neutrino data
was performed in~\cite{Hirsch00}.

\begin{figure}[htb]
  \begin{minipage}[t]{70mm} \centering
    \mbox{{\href{pictures/2/rpv_bilinear.pdf}{{\epsfig{figure=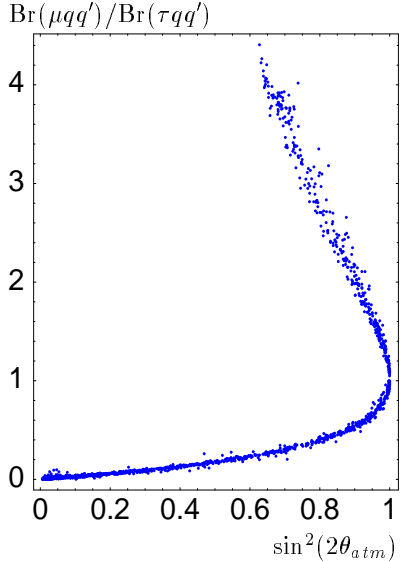,height=7.5cm}}}}}
  \end{minipage} \hspace{10mm}
  \begin{minipage}[t]{70mm} \vspace{-50mm}
    \caption{
        The branching ratios 
        Br($\nt_1\to\mu q q'$)/Br($\nt_1\to\tau q q'$) as a function of 
        the atmospheric neutrino mixing $\sin^2(2\,\theta_{atm})$  
        for various bilinear $\rpv$ models.}
    \label{fig:rpv_BiLinear}
  \end{minipage} 
\end{figure}

An interesting feature of this model is, that the semileptonic
branching ratios of the neutralino decays
$\tilde{\chi}^{0}_{1} \rightarrow \mu qq'$ and 
$\tilde{\chi}^{0}_{1} \rightarrow \tau qq'$ can be related to
the atmospheric neutrino mixing $\sin^2(2\,\theta_{\rm{atm}})$,
shown in \fig{rpv_BiLinear} for a variety of  $\rpv$ model parameters.
Note that in this class of theories neutrino
mixing angles can be probed at accelerators. 

Another property of this model is that the light stop decays 
may indicate $R$--parity violation~\cite{restrepo}. 
The $\rpv$ decay $\tilde{t}_{1} \to b\tau$
can be as important as the $R_p$ conserving three-body 
decays $\tilde{t}_{1} \to b W \tilde{\chi}^{0}_{1},\; 
be^{+} \tilde{\nu}_{e},\; b\tilde{e}^{+} \nu_e$ and the loop-decay 
$\tilde{t}_{1} \rightarrow c \tilde{\chi}^{0}_{1,2} $.
The main reason is that for $\epsilon_{3} \neq 0$ the chargino mixes with the 
$\tau$ lepton. The corresponding mass region is 
$m_{\tilde{t}_{1}} \leq 250$\,GeV, where it might be difficult to detect the 
stop at the LHC.

%
% TESLA TDR - SUSY chapter
% section on e-e-, e-g, gg options
% last update 9 Jan 2001
%

\section[ $e^-e^-$, $e^-\gamma$ and $\gamma\gamma$ Options ]{
 $\bel^-\bel^-$, $\bel^-\bgamma$ and $\bgamma\bgamma$ Options 
  \label{SUSY_egamgam}}

Additional information on the supersymmetry particle spectrum may
be obtained when operating {\sc Tesla} in the $e^-e^-$, $e^-\gamma$ and 
$\gamma\gamma$ modes, each with highly polarised beams.

Supersymmetry in $e^-e^-$ collisions is limited to selectron pair production
$e^-e^- \to \se^-\se^-$ via neutralino exchange. 
The main interest lies in mass determinations through threshold scans.
Selectrons associated to the same helicity, $e^- e^- \to \se^-_R\se^-_R$ 
and $e^- e^- \to \se^-_L\se^-_L$, are produced with angular momentum $J=0$
leading to a $\beta$ dependence of the cross section~\cite{feng2},
in contrast to the less steep $\beta^3$ behaviour in $e^+e^-$ collisions. 
This apparent advantage, however, is depleted by initial state radiation and
beamstrahlung effects, which severely degrade the shape 
(flattening of the steep rise)
and magnitude of the excitation curve at threshold~\cite{heusch}.
Given the considerably lower luminosity,
it is questionable whether on gets competitive or even more precise mass 
measurements in comparable running times.
An interesting possibility is to search for mixing in the slepton sector
(analogous to neutrino mixing) via lepton number violating decays
$\se \to \mu \,\nt$, where electron collisions provide a very clean 
environment.

Higher selectron masses beyond the $e^+ e^-$ kinematic limit can be probed 
by associated production of $e^-\gamma \to \se^- \nt_1 \to e^- \nt_1 \nt_1$.
Further, this process offers an interesting possibility to access 
the gaugino mass parameter $M_1$~\cite{bloechinger}.
Using highly polarised electron and photon beams the cross sections
are large and any ambiguities are easily resolved by measuring
polarisation asymmetries.
With moderate luminosities, the parameter $M_1$ can be determined within a
per cent or better depending on the MSSM scenario.
Such measurements combined with those from the chargino sector,
see Table~\ref{tab:chi_cijrr12},
allow a stringent test of the GUT relation
$M_1 / M_2 = \frac{5}{3}\, \tan^2\theta_W$.

In photon collisions the production of charged sfermions, sleptons or squarks,
%$\gamma\gamma \to \ti f \bar{\ti f}$~\cite{klasen}
$\gamma\gamma \to \sl^+ \sl^-, \ \ti q \bar{\ti q}$~\cite{klasen}
and charginos $\gamma\gamma \to \ti\chi^+\ti\chi^-$~\cite{mayer} 
are pure QED processes and depend essentially on the sparticle masses
and charges (interesting for squarks).  
Therefore, in contrast to $e^+e^-$ annihilation, the decay mechanisms can be
separated from the production, which simplifies an analysis of the relevant 
SUSY parameters.
In general the polarised cross sections are larger than in $e^+e^-$ 
annihilation up to the kinematic limit, thus allowing to study more subtle
effects.
Another interesting possibility is resonant stoponium 
production~\cite{gorbunov}. 
The cross section for $\gamma\gamma\to S$ collisions with total helicity 0
may be quite large and the dominant decay modes to gluons or Higgs pairs
easily detectable. 
Such a resonance would be observable in $e^+e^-$ annihilation at an
appreciably lower rate only if the decay to Higgs bosons is dominant.

%
% TESLA TDR - SUSY chapter
% section on high E extrapolation
% last update 26 Jan 2001
%

\section{Extrapolation of SUSY Parameters to High Energy Scales 
  \label{SUSY_hiE}}

In most studies of SUSY models assumptions are made at a high energy 
scale. In the minimal supergravity model with the input 
parameters $m_{0}$, $m_{1/2}$, $\tan\beta$, $A_{0}$, ${\rm sign}\,\mu$ at the 
GUT scale $M_{U} \simeq 2\cdot 10^{16}$\,GeV, all gauge couplings 
$\alpha_{1,2,3}$, all gaugino masses $M_{1,2,3}$ and all scalar 
masses unify at $M_{U}$. In the  GMSB model one starts from boundary 
conditions at the messenger scale $M_{mess}$ for the gaugino and 
scalar masses. The evolution of the parameters down to the electroweak scale 
is described by the renormalisation group equations. % (RGE).

In order to test these assumptions and models one can also start from 
the particle spectrum measured at lower energies and extrapolate the 
corresponding SUSY parameters by RGE to higher energies. Such a `bottom--up' 
approach is presented in~\cite{blair}.  They analyse in detail the 
mSUGRA point $m_{0} = 200$\,GeV, $m_{1/2} = 190$\,GeV $A_{0} = 550$\,GeV, 
$\tan\beta = 30$ and ${\rm sign}\,\mu < 0$, which determines the particle 
spectrum at low energy. 
From fits to the mass spectrum and cross sections,
as given by simulations of {\sc Tesla}~\cite{martyn} 
and LHC~\cite{hinchliffe,cms} experiments,
one extracts the SUSY parameters including their correlated errors.
Typical mass errors are given in Table~\ref{tab:hiE_masserrors}.

\begin{table}[htb]
  \begin{center}
    \begin{tabular}{l c c c }
      \hdick \\[-1.5ex]
       particle          & \ m [GeV]\ & \multicolumn{2}{c}{$\delta$m [GeV]}\\
               &        & \ LHC \    & LHC+LC  \\[1ex] \hdick  
\addtop  $h^0$             & 109    & 0.2 & 0.05  \\
      $A^0$             & 259    &  3  & 1.5   \\ \hline
      $\chi^+_1$        & 133    &  3  & 0.11  \\
      $\chi^0_1$        & 72.6   &  3  & 0.15  \\ \hline
      $\tilde{\nu_e}$   & 233    &  3  & 0.1   \\
      $\tilde{e_1}$     & 217    &  3  & 0.15  \\
      $\tilde{\nu_\tau}$& 214    &  3  & 0.8   \\
      $\tilde{\tau_1}$  & 154    &  3  & 0.7   \\ \hline
      $\tilde{u_1}$     & 466    &  10 & 3     \\
      $\tilde{t_1}$     & 377    &  10 & 3     \\ \hline
      $\tilde{g}$       & 470    &  10 & 10    \\ \hline
    \end{tabular}
  \end{center}
  \caption{Representative masses and experimental errors used in 
           mSUGRA fits to the mass spectra.} 
  \label{tab:hiE_masserrors}
\end{table}

\begin{figure} {\setlength{\unitlength}{1mm}
\begin{picture}(155,70)
%\put(0,0){\framebox(155,70)}
\put(0,-1){\mbox{\href{pictures/2/hiE_sugra1.pdf}{{\epsfig{file=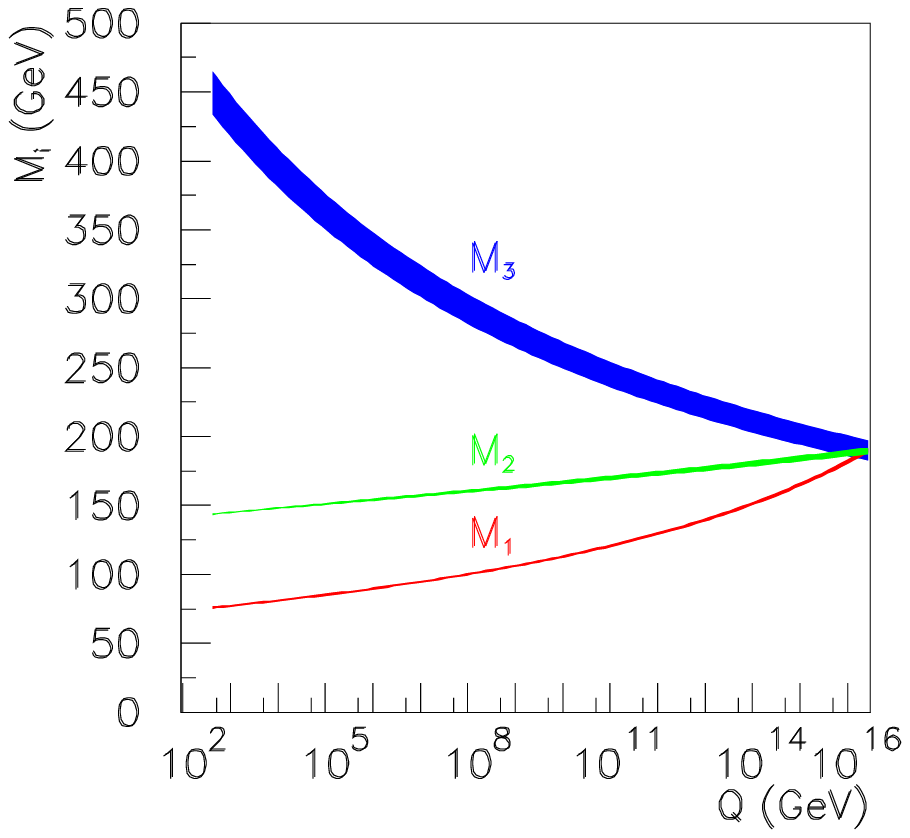,
   bbllx=0pt,bblly=270pt,bburx=270pt,bbury=520pt,height=7cm}}}}}
\put(76,-1){\mbox{\href{pictures/2/hiE_sugra2.pdf}{{\epsfig{file=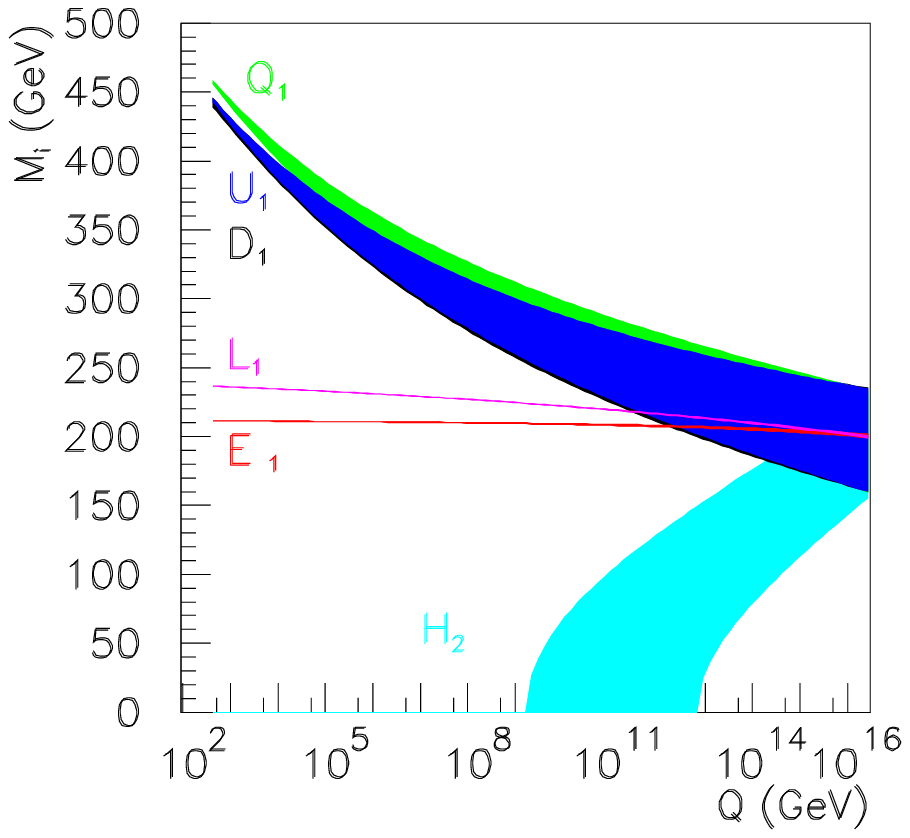,
   bbllx=0pt,bblly=270pt,bburx=270pt,bbury=520pt,height=7cm}}}}}
\end{picture}
%  \centering  \vspace{-10mm}
%%%  \mbox{\href{pictures/2/hiE_sugra.pdf}{{\epsfig{file=hiE_sugra.eps,height=8cm }}}} \vspace{-10mm}
%  \mbox{\href{pictures/2/hiE_sugra1.pdf}{{\epsfig{file=hiE_sugra1.eps,height=8cm }}} \hspace{4mm}
%        \href{pictures/2/hiE_sugra2.pdf}{{\epsfig{file=hiE_sugra2.eps,height=8cm }}}} 
%  \vspace{-10mm}
  \caption{
    Evolution of gaugino and sfermion mass parameters in mSUGRA for
    $m_{0} = 200$\,GeV, $m_{1/2} = 190$\,GeV, $A_{0} = 500$\,GeV, 
    $\tan\beta = 30$ and ${\rm sign}\,\mu <0 $. 
    The bands indicate 95\% CL contours.}
  \label{fig:hiE_fig1}
}\end{figure}

The extrapolation of the corresponding SUSY parameters from the weak 
scale to the GUT scale within the mSUGRA scenario are shown 
in~\fig{hiE_fig1}. It can be seen that the gaugino mass parameters $M_{1,2,3}$ 
and the slepton mass parameters $M_{L_{1}}, M_{E_{1}}$ for the first 
and second generation are in excellent agreement with unification,
due to the precise measurements in the slepton and chargino/neutralino sectors.
Using only LHC information would give uncertainties on the unification scale 
worse by more than an order of magnitude.
The squark parameters $M_{Q_{1}}, M_{U_{1}}, M_{D_{1}}$ and the Higgs 
parameter $M_{H_{2}}$, being less well known, still allow to test unification.

\begin{figure}[htb]
  \begin{minipage}[t]{70mm} \centering \vspace{-10mm}
    \mbox{\href{pictures/2/hiE_gmsb.pdf}{{\epsfig{file=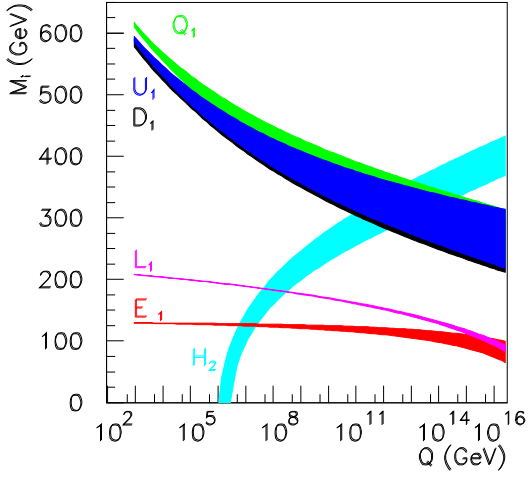,height=8cm}}}}   
  \end{minipage} \hspace{10mm}
  \begin{minipage}[t]{70mm}  \vspace{20mm}
    \caption{
        Evolution of sfermion mass parameters in a GMSB model for 
        $M_{\rm mess} = 2\cdot 10^{5}\,\mathrm{TeV}$, $\Lambda = 28\,\mathrm{TeV}$, 
        $N_{\rm mess} = 3$, 
        $\tan\beta = 30$ and ${\rm sign}\,\mu < 0$. 
        The bands indicate 95\% CL contours.}
      \label{fig:hiE_fig2}
  \end{minipage}
\end{figure}

To confront the mSUGRA scenario with an alternative one, the analysis 
was also done for the GMSB model with the parameters $M_{mess} = 
2\cdot10^{5}$\,TeV, $\Lambda = 28$\,TeV, $N_{5} = 3$, $\tan\beta = 30$ 
and ${\rm sign}\,\mu < 0$. The results are shown in~\fig{hiE_fig2}. 
Note that $M_{H_2}$ approaches the parameter for $M_{L_1}$ 
at the GMSB scale around $10^8$\,GeV 
as both belong to weak isodoublet fields which do not have strong interaction.
As one can see one gets a very different picture at high energy scales 
compared to the mSUGRA model, and obviously both scenarios cannot be confused. 
Moreover, from both~\fig{hiE_fig1} and~\fig{hiE_fig2} 
one can see that precision data are essential for stable 
extrapolations to high energy scales.

%
% TESLA TDR - SUSY chapter
% section on LHC comparison
% last update 29 Jan 2001
%

\section{Comparison of TESLA with LHC \label{SUSY_compLHC}}

If supersymmetry is realized at low energies, it will be discovered at 
the LHC. In particular, squarks and gluinos -- if they exist --
will be produced abundantly because of their strong interaction. 
It will be possible to 
discover gluinos and squarks up to a mass of 2.5\,TeV by a variety of 
distinctive signatures (multiple jets, multi-leptons, etc. + missing 
energy)~\cite{hinchliffe,cms}. However, at LHC all kinematically accessible 
SUSY particles (charginos, neutralinos, sleptons) are produced 
simultaneously either directly or in cascade decays of gluinos 
and squarks. 
It is extremely difficult and often impossible to separate the many SUSY 
processes which can occur. 
A model independent experimental analysis, aiming at a measurement of the 
masses and other properties of the particles, is essentially precluded.
In most cases, one has to invoke model assumptions and to compare the 
predictions with the experimental distributions. 
By selecting decay chains, {\em e.g.} $\tilde{g} \rightarrow 
q\tilde{q} \rightarrow qq\tilde{\chi}^{0}_{2} \rightarrow qq 
\ell^{+} \ell^{-} \chi^{0}_{1}$, it is possible to construct enough 
kinematic constraints to determine the masses of the primary and 
daughter particles. Such studies have been performed in a variety of 
SUSY scenarios~\cite{hinchliffe,cms},
covering a large range of model parameters.

Concerning the mass reach, LHC has of course a larger discovery potential 
for almost all  SUSY particles than {\sc Tesla} due to the high centre of mass 
energy and will be able to determine particle masses in given scenarios
with an accuracy of a few percent. 
Other sparticle properties, however,
remain almost inaccessible.
 
{\sc Tesla}, on the other hand, offers far superior measurements of the SUSY
particle spectrum wihin its energy range, specifically:
\begin{itemize}
    \item a precise determination of particle masses:\\[1mm]
      \begin{tabular}{lcllcl}
         $\Delta m_{\tilde{\chi}^{\pm, 0}}$     & = & 0.1 -- 1\,GeV,    & 
         $\Delta m_{\tilde{\ell}, \tilde{\nu}}$ & = & 0.05 -- 0.3\,GeV, \\
         $\Delta m_{\tilde{\tau}, \tilde{\nu}_{\tau}}$ & = & 0.6\,GeV,  & 
         $\Delta m_{\tilde{t},\tilde{b}}$              & = & 1\,GeV.
      \end{tabular}\\[1mm]    
      The high accuracy of masses allows the extrapolation to very high 
      energies (GUT scale) revealing the origin of SUSY breaking.    
    \item precise measurement of the widths and branching ratios      
    \item precise determination of the couplings 
    \item determination of the mixing parameters in the 
          chargino/neutralino sectors
    \item measurement of the mixing angles in the $\tilde{t}$ 
          and $\tilde{\tau}$ sectors 
    \item determination of $\tan\beta$ in the $\tilde{\tau}$ sector 
          if $\tan\beta > 10$   
    \item determination of the spin and the quantum numbers
    \item model independent determination of SUSY parameters
    \item measurement of CP violating phases 
\end{itemize}

It should be emphasised that for all these precision measurements the 
use of {\em polarised beams} is important. Polarised $e^+$ in addition to 
polarised $e^-$ are especially useful for the separation of $\se_L^{}$ 
and $\se_R^{}$ in $\se\se$ production, and in $R$--parity violating analyses.

Only via the precision measurements which are possible at {\sc Tesla} can 
the underlying SUSY model be revealed and its parameters determined. 
The input of findings from the LHC will of course be valuable for 
experimentation at {\sc Tesla}.
It is worth pointing out that precise measurements from {\sc Tesla},
particularly of the masses of the LSP and sleptons, would greatly improve
the quality of the information which LHC can derive from multiple
decay chains.

To summarise, from these comparisons it is obvious that the LHC and 
Linear Collider programmes complement each other. 
The LHC may discover supersymmetry and constrain its gross features. 
However, only high precision measurements at the {\sc Tesla} Linear Collider 
will be able to pin down the detailed structure of the underlying 
super\-symmetry theory.

\clearpage

  \clearpage
  {\raggedright{
}}
  \cleardoublepage
%------------------------------------------------------------------
\chapter{Alternative Theories}
\label{physics_alternatives}
%------------------------------------------------------------------

  \section{Introduction}
%\section{Alternative Theories: Introduction}

%HEP MAP:  standard standard physics      standard non-standard physics
%          non-standard standard physics  non-standard non-standard physics

Microscopic physics is characterized in the standard 
formulation by two scales, the electroweak scale of order
$10^2$ GeV at which the Standard Model is defined, and the 
Planck scale of order $10^{19}$ GeV where particle physics
and gravity are linked. The large gap between the two scales 
can be stabilized by supersymmetry. This picture of Nature 
is strongly supported by the successful prediction of the 
electroweak mixing angle; however, alternative scenarios 
are not ruled out.

{\bf 1.)} Extending the Minkowski--Einstein Universe by extra 
space dimensions not near the Planck scale but at semi-macroscopic
scales may change the picture
dramatically [Antoniadis; Arkani-Hamed, Dimopoulos, Dvali].
Gravity may become strong in extended space already at the
TeV scale and the hierarchy problem, present in the standard picture,
is non-existent. Towers of Kaluza--Klein states are realized
on the compactified extra dimensions which affect high-energy
processes, giving rise to missing-energy signals
and new contact interactions, or novel resonances with masses 
in the TeV range. 

The large gap between the scales may also be generated
by localizing gravity on a wall different from the wall of the 
Standard Model in higher space-time dimensions [Randall-Sundrum].
The projection of gravity down to the SM wall is weak,
the large Planckian energy scale being reduced exponentially. 
Kaluza--Klein graviton resonances should be observed 
at the TeV scale in such a scenario. 
 
Even though all these ideas are highly hypothetical, they open new 
vistas, in particular on the unsolved theoretical problems in 
gravity. Observation of effects in high-energy experiments as
described above, would revolutionize the basic space-time picture 
of the world.

{\bf 2.)} So long as the Higgs mechanism is not established firmly,
rival theories for generating the masses of the fundamental
particles in the Standard Model must be considered seriously.
In the standard Standard Model the masses are generated by 
interactions of the particles with the fundamental Higgs field,
being of non-zero field strength in the ground state as a result 
of spontaneous symmetry breaking. Alternatively, the symmetry 
breaking could be of dynamical origin as realized in theories 
of new strong interactions at the TeV scale [Susskind, Weinberg].
The masses of the gauge fields are generated by absorption 
of Goldstone bosons associated with the breaking of global symmetries. 

In such scenarios, the $W$ bosons become strongly interacting particles 
at high energies. This will affect the production of $WW$ pairs in
$e^+e^-$ annihilation, and the amplitudes for $WW$ scattering in the 
threshold region of the strong interactions can be predicted. 
From both effects the scale of the new strong interactions can be 
determined at a sub-TeV  collider. New $WW$ resonances will be observed 
in the (multi-)TeV mass range. Extending these ideas to fermions 
generates quite a number of serious difficulties, inflicted by the 
necessary coexistence of disgruent large scales. They require rather
complex theoretical constructs in attempts to solve these problems.

{\bf 3.)} Strong interactions between particles have signalled quite often 
in the past, hidden composite structures. Solving the problem
of mass by new strong interactions naturally raises the question
of non-pointlike structures of electroweak gauge bosons, leptons
and quarks. Analyzing contact interactions in high-energy 
$e^+e^-$ scattering experiments will probe or set bounds on the radii 
of these particles. The same compositeness picture suggests 
leptoquarks as novel bound states.

Experimentation at TESLA may thus open vistas to new physics areas, 
``unexpected'' in the standard form of non-standard physics.   

\section{Extra Dimensions}

A novel approach which exploits the geometry of extra spatial
dimensions has recently been proposed \cite{add,ant,rs,seiwit} as a means
of addressing the gauge hierarchy (for a different approach to the link
between electroweak symmetry breaking and gravity see Ref.~\cite{bij}).
These models 
make use of the fact that gravity has yet to be probed at energy
scales much above $10^{-3}$ eV in laboratory experiments. In the
scenario of Arkani--Hamed, Dimopoulos and Dvali \cite{add}, 
the apparent hierarchy is generated by a large
volume for the extra dimensions, while in the Randall--Sundrum model
\cite{rs}, the observed hierarchy is created
by an exponential warp factor which arises from the localization of
gravity in a 5-dimensional non-factorizable geometry.
Moreover, recent theoretical results have demonstrated that
non-commutative field theories naturally appear within the context 
of string/M-theory \cite{seiwit}. An exciting feature of these three
classes of theories is that they afford concrete and distinctive
phenomenological, as well as astro-physical, tests.
Furthermore, if they truly describe the source of the observed
hierarchy, then their signatures should appear in experiments at
the TeV scale. We now review these models and
discuss their signatures at the TESLA collider.

\subsection{Gravity at large dimensions}
In the scenario of Ref.~\cite{add}, gravitational interactions become strong
near the weak scale and take place mainly in $\delta$ new large spatial 
dimensions, known as the bulk.  Since it is known experimentally
that the Standard Model fields do not feel the effects of 
extra dimensions with a compactification scale of
less than a few TeV, they are constrained to lie on a $3+1$-dimensional 
brane, or wall, in the higher dimensional space.  Gravity thus 
appears weak  in ordinary 4-dimensional space-time as we only observe 
its projection onto the wall.  The relation between the scales
where gravity becomes strong in the $4+\delta$ and 4-dimensional theories
can be derived from Gauss's Law and is given by
$M^2_{Pl}=V_\delta M_D^{2+\delta}$,
where $M_D$ denotes the fundamental Planck scale in the higher
dimensional space, and $V_\delta$ is the volume of the compactified
dimensions.  Setting $M_D\sim 1$ TeV thus determines the size of the
extra dimensions for a given value of $\delta$.  The case of $\delta=1$ 
is ruled out by astronomical data. Cavendish-type experiments
have excluded departures from the gravitational inverse square law 
for length scales exceeding 190 $\mu$m \cite{adel}.
For $\delta=2$ this rules out $M_D < 1.6$ TeV using the mass-scale
convention of \cite{grw}.  
In addition, 
astro-physical and cosmological considerations \cite{astro}, such 
as the rate of supernova cooling and the $\gamma$-ray flux spectrum, 
disfavor a value of $M_D$ near the TeV scale for $\delta=2$.

The Feynman rules for this scenario \cite{grw,hlz} are obtained by 
considering a linearized theory of gravity in the bulk.
Upon compactification, the bulk gravitational field expands into 
Kaluza-Klein (KK) towers of gravitons, which are equally spaced and 
have masses of $n/R$ where $n$ labels the KK excitation level and $R$
denotes the radius of the compactified extra dimensions.  
Taking $M_D=1$ TeV, we see that the KK state mass splittings are
equal to $5\times 10^{-4}$ eV, 20 keV and 7 MeV for $\delta=2,4$
and 6, respectively.  Note that due to the form of the action
the spin-1 KK states do not interact with the wall fields,
and that the scalar states are
phenomenologically irrelevant for most processes at the TESLA collider.
Each state in the spin-2 KK tower, $G_n$, couples identically 
to the Standard Model wall fields 
via the stress energy tensor
and the strength of the couplings
is given by the inverse 4-d Planck scale, $M_{Pl}^{-1}$.

\subsubsection{Graviton emission}
There are two classes of collider signatures for gravity at large
dimensions, with the 
first we discuss being that of graviton KK tower emission in scattering
processes \cite{grw,emit}.  The signal process at the TESLA collider 
is $e^+e^-\to\gamma/Z+G_n$, where the graviton appears as missing
energy in the detector as it behaves as if it were a massive, 
non-interacting, stable particle.  The cross section is computed for
the production of a single massive graviton excitation, and then summed
over the full tower of KK states.  Since the mass splittings of the
KK excitations are quite small compared to the collider center of mass 
energy, this sum can be replaced by an integral weighted by the density 
of KK states and which is cut off by the specific process kinematics.
This has the effect of removing the 4-d Planck scale suppression; the
$M^{-2}_{Pl}$ factor which appears from the graviton couplings is
exactly cancelled by the $M^2_{Pl}$ dependence of the phase space
integration.  The process now scales as simple powers of $\sqrt s/M_D$.
It is important to note that due to the effective density of states,
the emitted graviton appears to have a continuous mass distribution;
this corresponds to the probability of emitting gravitons with
different momenta in the extra dimensions.  
The differential cross-section of $e^+e^-\to \gamma G$ is given~\cite{grw} by
\begin{equation}
  \frac{d^2\sigma}{dx_{\gamma}\,d\cos\theta} = 
  \frac{\alpha S_{\delta-1}} {64 {M_D}^2 }
  \left(\frac{\sqrt{s}}{M_D} \right)^{\delta} 
   f_{\gamma G}(x_{\gamma},\cos\theta)
  \label{eq:Gg1}
\end{equation}
where $x_{\gamma}=E_{\gamma} / \Ebeam$, $\delta$ is the number of 
extra dimensions and
$S_{\delta-1}$ is the surface area of a $\delta$-dimensional
sphere of unit radius, 
with
\begin{equation}
   f_{\gamma G}(x,\cos\theta) = 
        \frac{2 (1-x)^{\frac{\delta}{2} - 1}}{x(1-\cos^2\theta)}
        \left[(2-x)^2 (1-x+x^2) - 3x^2 (1-x) \cos^2\theta - 
                x^4 \cos^4\theta \right]
  \label{eq:ggamma}
\end{equation}

   The discovery reach of the TESLA collider for direct graviton 
production in $e^+e^- \rightarrow \gamma \mathrm{G} $ 
is estimated for 
an integrated luminosity of 1 ab$^{-1}$ at $\sqrt{s}=800$~GeV.
Details of these studies are given in \cite{LCNOTE_Wilson_XD}
and \cite{LCNOTE_Vest_XD}.
The signature is a relatively soft photon and missing energy.
The major background is
$e^+e^- \rightarrow \nng$ and it is largely irreducible.
   The following kinematic acceptance cuts are imposed on the photon:
\begin{itemize}
\item{Within the acceptance of the electromagnetic calorimeter,
$\sin{\theta_{\gamma}} > 0.1$. }
\item{$p_T > 0.06 \Ebeam$ in order to reject 
events with no genuine missing $p_T$ such as 
$e^+e^- \rightarrow e^+e^-\gamma$ where electrons at polar angles below 
the mask calorimeter
acceptance of 27.5~mrad mimic missing $p_T$.
}
\item{$x_{\gamma} < 0.625$ in order to reject the energetic photons
from  $e^+e^- \rightarrow \nng$ which arise
from  $e^+e^- \rightarrow \mathrm{Z} \gamma$.
}
\end{itemize}
   With these cuts, the accepted cross-sections including ISR 
and beamstrahlung 
from $e^+e^- \rightarrow \nng$ for 100\% electron polarisation 
are
$\sigma_{{e_L^-} e^+} = 1.90$~pb 
and 
$\sigma_{{e_R^-} e^+} = 23$~fb,
evaluated using
NUNUGPV \cite{NUNUGPV}.
The cross section for left-handed electrons is much enhanced due to the
dominance of $W$ exchange contributions in this kinematic region.
Other backgrounds have so far been neglected; they will be small but 
should not be ignored, e.g.~$e^+e^- \rightarrow \nnnng$. 
For the signal, as an example for $M_D=5$~TeV and $\delta=2$, the 
unpolarised
accepted Born cross-section without beamstrahlung is 12~fb. 
 
\begin{figure}[hbt]
\vspace*{-0.5cm}
\begin{center}
  \includegraphics[height=10cm]{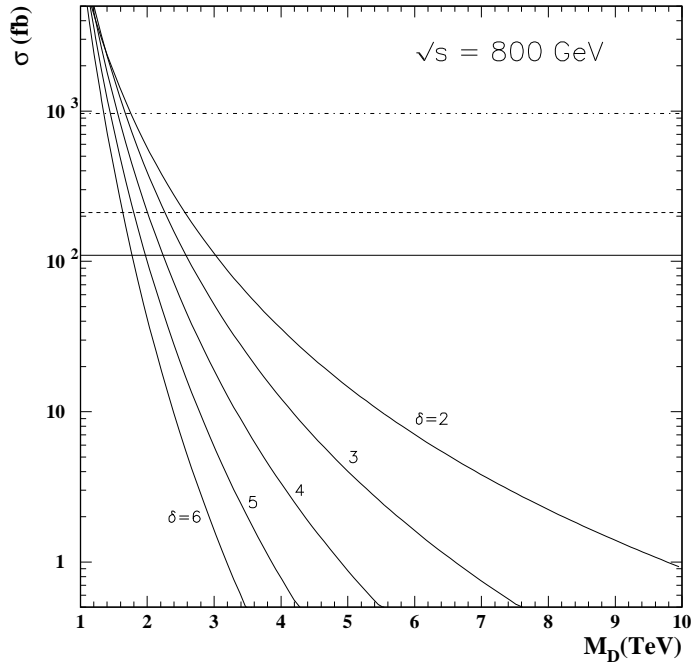}
\vspace*{-0.5cm}
  \caption{\it \label{fg:xdim} Total cross sections for $e^+e^-\to \gamma
G$ at $\sqrt{s}=800$ GeV
as a function of the scale $M_D$ for different numbers $\delta$ of extra
dimensions. These signal cross-sections take into account 80\% electron 
and 60\% positron polarisation \cite{pospol}.
% (enhancement
%factor of 1.48).
The three horizontal lines indicate the background cross-sections from
$e^+ e^- \rightarrow \nng$ 
for both beams polarised (solid), only electron beam polarisation (dashed) 
and no polarisation (dot-dashed). Signal cross-sections
are reduced by a factor of 1.48 for the latter two scenarios.
}
\end{center}
\vspace*{-0.8cm}
\end{figure}

Given the near maximal polarisation asymmetry of 
the background, polarised 
beams of appropriate helicity are extremely effective in 
suppressing the background and therefore extending the 
reach of the TESLA collider in
the quest for evidence of extra dimensions.
Fig.~\ref{fg:xdim} compares the signal cross-sections with the background
cross-sections for several polarisation assumptions.
Numerical sensitivity estimates shown here are based on a 
normalisation uncertainty of 0.3\%.
%The signal and background include an estimated 
%inefficiency of 10\% from additional cuts required 
%against multi-particle final states.
%The signal cross-sections have been
%reduced by 15\% corresponding to the estimated
%cross-section reduction from ISR and beamstrahlung. 
For completeness, the studies in \cite{LCNOTE_Wilson_XD,LCNOTE_Vest_XD} 
have also considered normalisation uncertainties varying from
0.1\% (optimistic) to 1.0\% (conservative). 
Many sources of
systematic error will have to be controlled at quite challenging levels
of precision: theoretical error on background cross-section, absolute
luminosity, selection efficiency, energy scale and polarisation. 
However the availability of large control data-sets such 
as 
$\mathrm{Z} \rightarrow e^+e^-$ and $e^+e^- \rightarrow \gamma \gamma$ 
should allow detector related systematics to be kept 
under sufficient control. 

The inclusive cross-section 
measurement is used to estimate the sensitivity.
A modest improvement is expected
if information on the energy and polar angle 
distributions is also included.
The sensitivity estimates are shown 
in Table~\ref{tab:xdim} for numbers of extra dimensions
ranging from 2 to 6. 
By polarising both beams to a high degree, the TESLA collider potential for
exploring this physics is maximised.

%\begin{table}[hbt]
%\centering
% \renewcommand{\arraystretch}{1.5}
%\begin{tabular}{|l|c|c|c|c|c|}
%\hline
%$\delta$                       &   2  &  3  &  4  &  5  &  6  \\ \hline
%$M_D (P_{-,+}=0)$              &  4.5 & 3.5 & 2.9 & 2.5 & 2.2 \\ 
%$M_D (P_{-}=0.8)$              &  6.3 & 4.6 & 3.6 & 3.0 & 2.6 \\
%$M_D (P_{-}=0.8, P_{+}=0.6)$   &  7.9 & 5.6 & 4.2 & 3.4 & 2.9 \\ \hline
%\end{tabular}
%\caption[]{\it Discovery ($5\sigma$) reach in mass scale $M_D$ in TeV 
%for direct graviton production in $e^+e^- \rightarrow \gamma \mathrm{G} $ for 
%various values of $\delta$ taking a
%0.3\% normalisation error.
%The change in the discovery reach for 
%alternative assumptions 
%on the normalisation error of 0.1\% (optimistic) and 1.0\% (conservative)
%is indicated by the upper and lower value. 
%}
%\label{tab:xdim}
%\end{table}

\begin{table}[hbt]
\centering
 \renewcommand{\arraystretch}{1.5}
\begin{tabular}{|l|c|c|c|c|c|}
\hline
$\delta$                       &   2  &  3  &  4  &  5  &  6  \\ \hline
$M_D (P_{-,+}=0)$              &  5.9 & 4.4 & 3.5 & 2.9 & 2.5 \\ 
$M_D (P_{-}=0.8)$              &  8.3 & 5.8 & 4.4 & 3.5 & 2.9 \\
$M_D (P_{-}=0.8, P_{+}=0.6)$   & 10.4 & 6.9 & 5.1 & 4.0 & 3.3 \\ \hline
\end{tabular}
\caption[]{\it Sensitivity (95\% CL) in mass scale $M_D$ in TeV 
for direct graviton production in $e^+e^- \rightarrow \gamma \mathrm{G} $ for 
various values of $\delta$ taking a
0.3\% normalisation error.
%The change in the discovery reach for 
%alternative assumptions 
%on the normalisation error of 0.1\% (optimistic) and 1.0\% (conservative)
%is indicated by the upper and lower value. 
}
\label{tab:xdim}
\end{table}

   At the LHC, direct graviton production can be explored using 
the signature of $pp \rightarrow jet \; G$.
However for certain values of ($M_D,\delta$), the partonic centre-of-mass
energy can exceed $M_D$ and the effective low energy theory approach 
breaks down at LHC.  A recent study \cite{Vacavant} of the LHC potential with 
the ATLAS experiment shows that in 100 fb$^{-1}$, 
direct graviton production can be discovered at at least $5\sigma$
for $M_D$ in the ranges shown in Table~\ref{tab:xdim_comp} for 
$\delta=2,3$ and 4.
% 4--7.5 TeV ($\delta=2$), 4.5--5.9 TeV ($\delta=3$)
%and 5.0--5.3~TeV ($\delta=4$).
However the effective theory 
approach breaks down at LHC for $\delta \ge 5$
and 
for $M_D$ values below the given ranges 
when $\delta=$2,3 and 4.
At TESLA, if $\sqrt{s}_{e^+e^-} \ll M_D$ the effective theory approach 
should
be valid and the measured single photon cross-section can be
used to constrain ($M_D,\delta$). 
TESLA offers a more
model-independent test of this theory while the LHC may be
in the string theory regime whose phenomenology is 
perhaps rich but presently unknown.
For regions 
which can be compared, as shown in Table~\ref{tab:xdim_comp},
the 5-$\sigma$ discovery reach in $M_D$ for 
TESLA and the LHC is similar.

\begin{table}[hbt]
\centering
 \renewcommand{\arraystretch}{1.5}
\begin{tabular}{|l|c|c|c|c|c|}
\hline
$\delta$ & 2  &  3  &  4  &  5  &  6  \\ \hline
LHC      & 4.0---7.5 & 4.5---5.9 & 5.0---5.3 & none & none \\
TESLA    & 0.5---7.9 & 0.5---5.6 & 0.5---4.2 & 0.5---3.4 & 0.5---2.9 \\ \hline 
\end{tabular}
\caption[]{\it The range of $M_D$ values in TeV
which can lead to a
discovery at at least $5\sigma$
for direct graviton production at LHC (ATLAS study) 
and TESLA with both beams polarised.
}
\label{tab:xdim_comp}
\end{table}

    Anomalous single photon signatures at TESLA and monojet 
signatures at LHC can both arise from many types of new physics
other than extra dimensions. Therefore 
measurement of processes sensitive to 
direct graviton production with complementary initial and final states 
would help to confirm whether the correct diagnosis had been made.
At TESLA, the process $e^+e^- \rightarrow Z G$ 
can be explored in $e^+e^-$ collisions; however the
sensitivity relative to $e^+e^- \rightarrow \gamma G$ is 
rather limited \cite{CK}.
A more promising channel is 
$e^- \gamma \rightarrow e^- G$ \cite{Iowa}.

\begin{figure}[hbt]
\vspace*{-0.5cm}
\begin{center}
  \includegraphics[height=10cm]{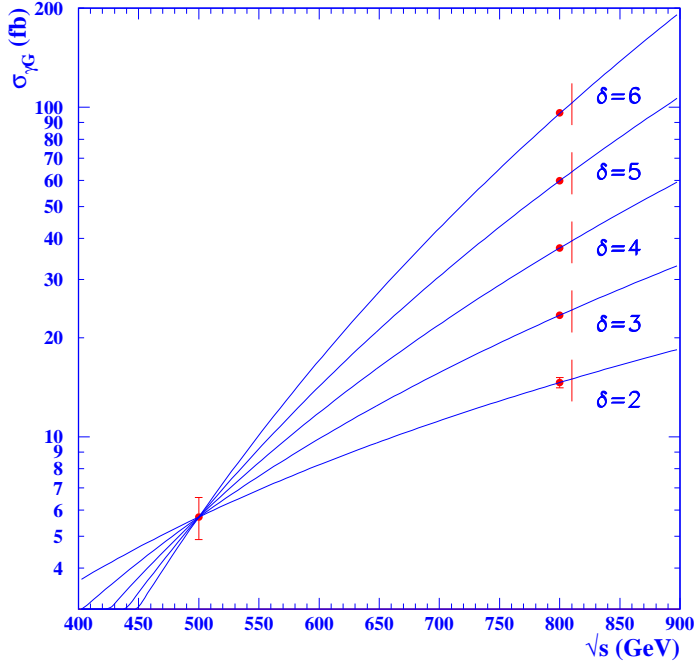}
\vspace*{-0.5cm}
  \caption{\it \label{fg:xdim_del} 
   Determining $\delta$ from
anomalous single photon cross-section measurements at $\sqrt{s}=500$~GeV
and 800~GeV.
The sensitivity shown corresponds to integrated luminosities 
of 500$fb^{-1}$ at $\sqrt{s}=500$~GeV and 1 $ab^{-1}$ at
$\sqrt{s}=800$~GeV with 80\% electron and 60\% positron polarisation
with a cross-section at 500 GeV equivalent to $M_D=5$~TeV if $\delta=2$.
The points with error bars show the measurements one could expect.
The smooth curves show the cross-section dependence on $\sqrt{s}$ for 
the central value  
of the 500 GeV cross-section measurement under the
hypotheses of $\delta=$~2,3,4,5 and 6. The vertical lines adjacent to 
the 800 GeV measurements indicate the range that would be 
consistent within $\pm 1\sigma$ with the 500 GeV measurement.
}
\end{center}
\vspace*{-0.8cm}
\end{figure}

    If extra dimensions are the cause
of the anomalous single photon rate, the $\sqrt{s}$ dependence of
the cross-section should follow 
$\sigma \propto \left({\sqrt{s}} \right)^{\delta}$.
Fig.~\ref{fg:xdim_del} illustrates how a measurement of an
excess of single photon events at $\sqrt{s}=500$~GeV
together with a measurement at $\sqrt{s}=800$~GeV
can be used to determine the number of extra dimensions.
%The measured excess cross-section with polarised beams at 
%$\sqrt{s}=500$~GeV is taken to be $5.7 \pm 0.8$~fb. 
%For $\delta=2$ this would be consistent 
%with $M_D=5.0\pm 0.2$~TeV.
%The uncertainty corresponds to 500 fb$^{-1}$ and a normalisation 
%systematic uncertainty of 0.3\%.
%Cross-section measurements at 800~GeV with 1 $ab^{-1}$
%are illustrated for values of $M_D$ and $\delta$
%consistent with the 500 GeV cross-section measurement.
%The full lines and dashed lines 
%illustrate the $\sqrt{s}$ dependence for the 
%central value and $\pm 1 \sigma$ 
%bands of the 500 GeV cross-section measurement under the
%hypotheses of $\delta=2,3,4,5,6$.
%As an example, if $\delta$ were 2, the $\delta=3$ hypothesis
%could be excluded on average at 2.8 standard deviations.
%With $\delta$ determined, $M_D$ can be inferred unambiguously: in 
%particular for the above example with $\delta=2$, one would measure
%$M_D=5.0\pm 0.04$~TeV where the sensitivity is dominated by the
%800 GeV data.
   Determination of the number of extra dimensions
is possible with this data-taking scenario for 
excess cross-sections at 500 GeV down to $5.3\pm0.8$~fb.
This cross-section is equivalent 
to $M_D=5.1$~TeV for $\delta=2$ and for these values, one would
exclude $\delta=3$ on average at 99\% CL.
Inconsistency with the expected $\sqrt{s}$
dependence, i.~e.~excluding integer values of $\delta$,
would exclude the extra dimensions interpretation.

    The LHC and TESLA therefore have valuable complementary roles
to play in experimentally testing theories with extra space dimensions.

\subsubsection{Virtual effects}
\begin{figure}[hbt]
\vspace*{-0.0cm}
\begin{minipage}[t]{7.4cm} {
\begin{center}
  \includegraphics[height=7cm]{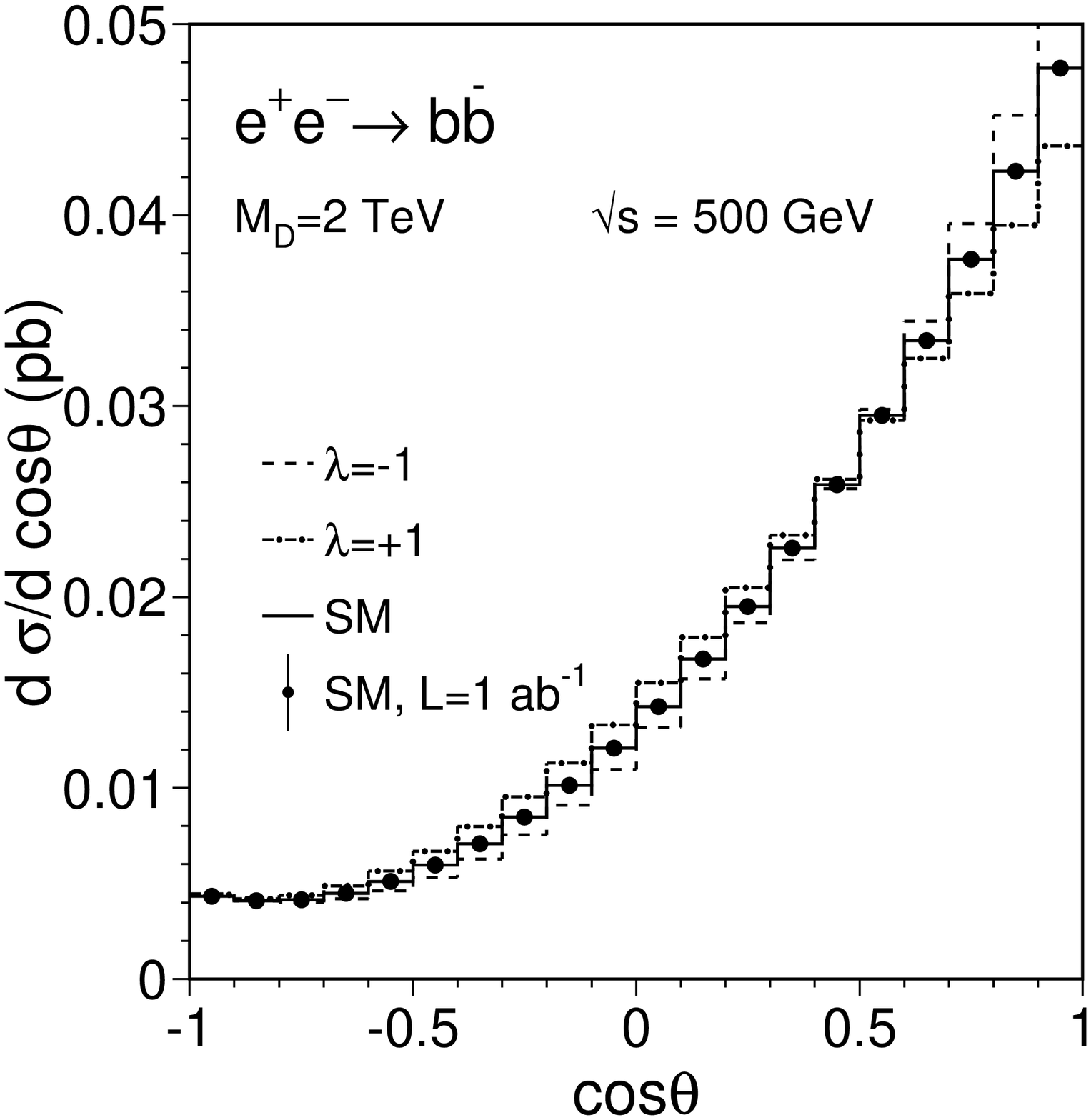}
\vspace*{-0.5cm}
\caption{\it \label{exch1} Angular distribution for $e^+e^-\to b\bar b$.
The solid histogram represents the Standard Model, while the dashed ones
are for $M_D=2$ TeV with $\lambda=\pm 1$ \cite{riemanns}.}
\end{center} }
\end{minipage}
\hspace*{0.5cm}
\begin{minipage}[t]{7.4cm} {
\begin{center}
\includegraphics[height=7cm]{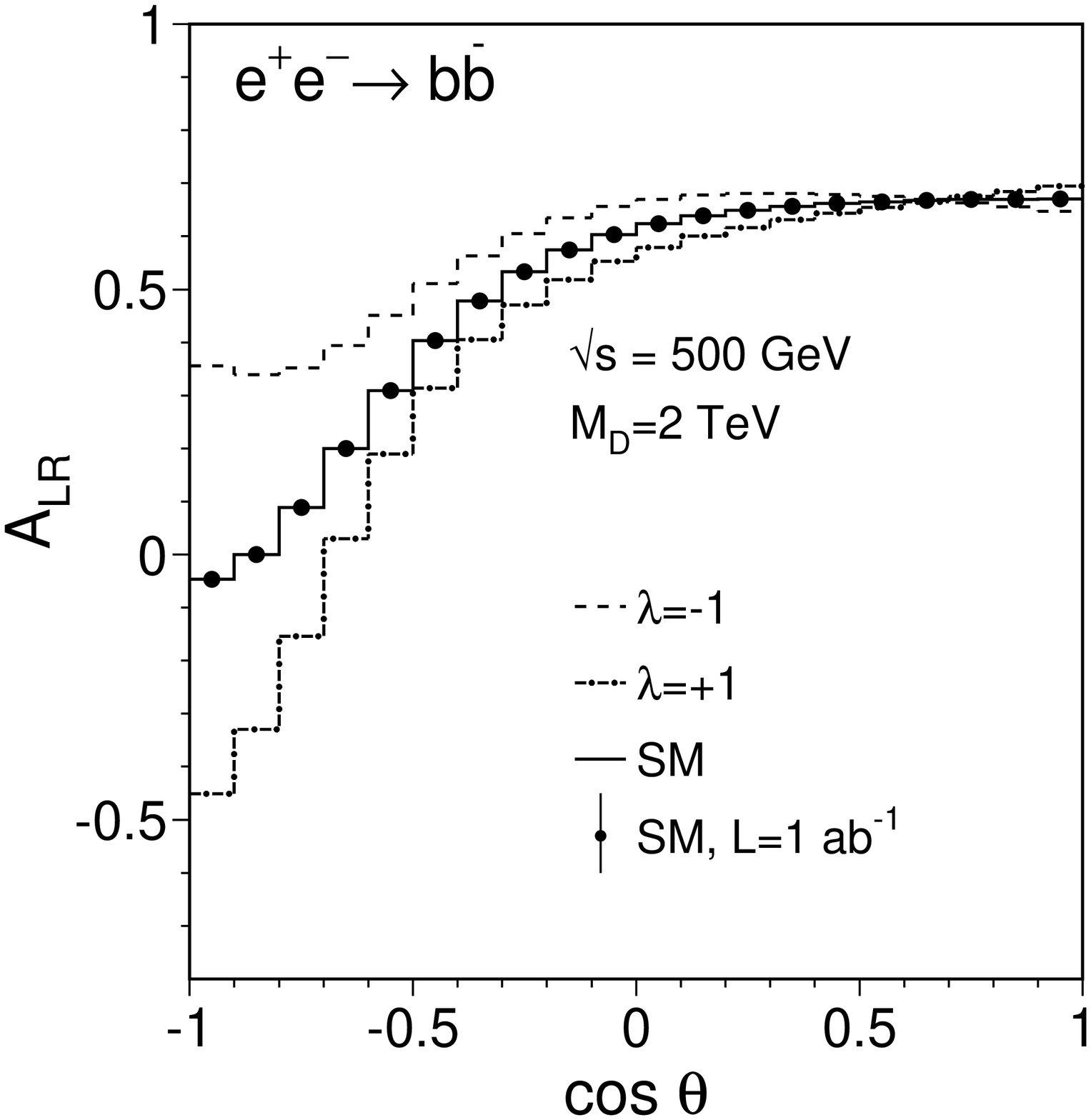}
\vspace*{-0.5cm}
\caption{\it \label{exch2} Left-Right asymmetry for $e^+e^-\to b\bar b$.
The solid histogram represents the Standard Model, while the dashed ones
are for $M_D=2$ TeV with $\lambda=\pm 1$ \cite{riemanns}.}
\end{center} }
\end{minipage}
\vspace*{-0.5cm}
\end{figure}

The second class of processes considered here is that of graviton 
exchange \cite{grw,hlz,exch} in $2\to 2$ scattering.  This virtual 
exchange mechanism leads to deviations in cross sections and asymmetries
in Standard Model processes, such as $e^+e^-\to f\bar f$, and can 
also mediate new processes which are not present at tree-level in
the Standard Model, such as $e^+e^-\to hh,$ or $\tilde g\tilde g$.
The exchange amplitude is proportional to the sum over the
propagators for the entire graviton KK tower and, again, can be 
converted to an integral over the density of states.  However, in this 
case the integral is divergent for $\delta>1$ and thus introduces a 
sensitivity to the unknown ultraviolet physics.  Several approaches have
been proposed to regulate this integral.
%been proposed to regulate this integral: {\it (i)} a naive 
%cut-off scheme \cite{grw,hlz,exch}, {\it (ii)} the tension of the 
%brane \cite{bando},
%{\it (iii)} the finite extent of the fermion wave-function in the 
%bulk \cite{martin}, or {\it (iv)} the inclusion of full weakly coupled
%TeV-scale string theory in the scattering process \cite{dudas}.
Here, we adopt the most model independent approach, that of a naive
cut-off \cite{grw,hlz,exch}, and set the cut-off equal to $M_D$.
Assuming that the integral is dominated by
the lowest dimensional local operator, which is dimension-8, this
results in a contact-type interaction limit for graviton exchange.
This is described in the matrix element for $s$-channel $2\to 2$ 
scattering by the replacement
\begin{equation}
\frac{1}{8\overline{M}_{Pl}^2}\, \sum^\infty_{n=1} 
\frac{1}{s-m_n^2}\to \frac{\lambda}{M_D^4}
\end{equation}
with corresponding substitutions for $t$- and $u$-channel scattering.
Here $\overline{M}_{Pl}$ represents the reduced Planck scale
$\overline{M}_{Pl}=M_{Pl}/\sqrt{8\pi}$, $m_n$ 
is the mass of the n$^{th}$ graviton KK excitation, and $\lambda$ 
is a model dependent factor, which we take to be of order unity and
of either sign.  This substitution is universal for all final states.  
The resulting angular distributions for fermion pair production
are quartic in $\cos\theta$ and thus provide a unique signal for
spin-2 exchange. We present an example of this in Figs.~\ref{exch1} and
\ref{exch2} which display the angular distribution and Left-Right asymmetry
in $b\bar b$ production for $M_D=2$ TeV and $\sqrt{s}=500$ GeV. The two
dashed histograms correspond to the two choices of sign for $\lambda$.
Table~\ref{tb:edlim} presents the sensitivities in $M_D$ in
$\mu^+\mu^-, b\bar b, c\bar c$ final states. Combining all final states
TESLA will be sensitive up to $M_D=8$ TeV at $\sqrt{s}=800$ GeV
\cite{riemanns}.
\begin{table}[hbt]
 \renewcommand{\arraystretch}{1.2}
  \begin{center}
    \begin{tabular}{|l|cc|}
\hline
   &\multicolumn{2}{|c|}{$M_D$ [TeV]}               \\
   & $\sqrt{s}=0.5$ TeV &$\sqrt{s}=0.8$ TeV \\
  \hline 
          $e^+e^-\rightarrow \mu^+ \mu^-$ &  4.1   & 5.8  \\
          $e^+e^-\rightarrow b\bar b$     &  5.0   & 7.1  \\
          $e^+e^-\rightarrow c\bar c$     &  5.1   & 7.1  \\ \hline
          combined                        &  5.6   & 8.0  \\ \hline
    \end{tabular}
  \end{center}
\vspace*{-0.6cm}
\caption{\label{tb:edlim} \it Sensitivity (95\% CL) in $M_D$ for indirect
effects in $\mu^+\mu^-,b\bar b$ and $c\bar c$ production \cite{riemanns}.}
\end{table}

It should be noted that the best limits, as phenomenologically calculated,
on $M_D$ in the model of Ref.~\cite{add} may arise from virtual graviton
exchange in $\gamma \gamma \to W^+W^-$. The reach in $M_D$ at a
$\gamma\gamma$ collider can be
estimated to be about $11\sqrt{s_{e^+e^-}}$ \cite{gagaww}.

In a certain class of models with extra dimensions compactified at the
TeV scale the SM fields can also explore the extra dimensions \cite{ant,ant2}.
For $\delta=6$ extra dimensions this translates into a fundamental quantum
gravity scale $M_D$ of about 8000 TeV. This scenario can lead to superparticle
mass of the order of the electroweak scale. Since the scale $M_D$ is
large, direct quantum effects of gravity will be inaccessible at TESLA.
However, these models exhibit spin 1 excitation states of the SM gauge
bosons $\gamma,Z,W,g$, called ``Kaluza--Klein recurrences''. By means of
the process $e^+e^-\to \mu^+\mu^-$ TESLA can reach a sensitivity in the
compactification scale $1/R$ beyond 10 TeV at high energy and high
luminosity \cite{ant2} [corresponding to scales $M_D\gsim 45000$ TeV for
$\delta=6$]. This turns out to be significantly larger than the
sensitivity $1/R\sim 6$ TeV, which can be reached at the LHC by means of
dilepton production $pp\to \ell^+\ell^- X$ \cite{ant2}.

\subsection{Randall--Sundrum model}
In the Randall--Sundrum model the
hierarchy is generated by an exponential function of the 
compactification radius.  In its simplest form, this model consists 
of a 5-dimensional non-factorizable geometry based on a slice of
Anti-de-Sitter (AdS$_5$) space of length $\pi r_c$, where $r_c$ denotes the
compactification radius.  Two 3-branes, with equal and opposite
tensions, reside rigidly at $S_1/Z_2$ orbifold fixed points at the
boundaries of the AdS$_5$ slice.  The 5-dimensional Einstein's 
equations permit a solution which preserves 4-d Poincar\' e
invariance with the metric
\begin{equation}
ds^2=e^{-2kr_c|\phi|}\eta_{\mu\nu}dx^\mu dx^\nu-r_c^2d\phi^2\,,
\end{equation}
where $0\leq |\phi|\leq\pi$.  Here, $k$ is the AdS$_5$ curvature scale
which is of order the Planck scale and is determined by the bulk
cosmological constant $\Lambda_5=-24M_5^3k^2$, where $M_5$ is the 5-d
Planck scale.  Examination of the 4-d effective action yields
$\overline{M}_{Pl}^2=(1-e^{-2kr_c\pi})M_5^3/k$.  The scale of
physical phenomena as realized by the 4-d flat metric transverse to
the 5$^{th}$ dimension is specified by the exponential
warp factor.  TeV scales can naturally be attained on the 3-brane
at $\phi=\pi$ if gravity is localized on the Planck brane at
$\phi=0$ and $kr_c\simeq 11-12$.  The scale of physical processes
on this TeV-brane is then $\Lambda_\pi\equiv\overline{M}_{Pl}
e^{-kr_c\pi}\sim 1$ TeV.  It has been demonstrated \cite{gw} that this 
value of $kr_c$ can be stabilized  without the fine tuning of 
parameters.

\begin{figure}[hbt]
\vspace*{0.0cm}
\begin{center}
  \includegraphics[height=9cm,angle=90]{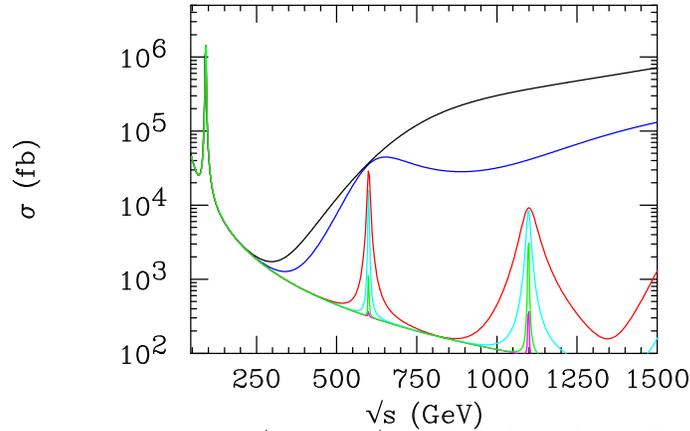}
\vspace*{-0.5cm}
  \caption{\it \label{fg:kkspect} 
The cross section for $e^+e^-\to\mu^+\mu^-$ including the
exchange of a KK tower of gravitons with $m_1=600$ GeV.  From top
to bottom the curves correspond to $k/\overline{M}_{Pl}=1.0\,, 0.7\,,
0.5\,, 0.3\,, 0.2\,$ and $0.1$.
}
\end{center}
\vspace*{-0.8cm}
\end{figure}

The 4-d phenomenology of this model is governed by 2 parameters,
$\Lambda_\pi$ and the ratio
$k/\overline M_{Pl}$, where constraints on the 5-d curvature
$|R_5|=20k^2<M_5^2$ suggest that $k/\overline M_{Pl}<0.1$.  The
Feynman rules are also obtained by a linear expansion of the flat
metric, which in this case includes the warp factor.
On the TeV-brane, the
resulting KK tower of gravitons now have masses given by $m_n=x_nk
e^{-kr_c\pi}=x_n\Lambda_\pi k/\overline M_{Pl}$ with the $x_n$ being
the roots of the first-order Bessel function, {\it i.e.} $J_1(x_n)
=0$.  Note that the first excitation is naturally of order a few
hundred GeV and that the KK states are not evenly spaced.
Due to the explicit form of the interactions of the KK tower with the
Standard Model fields on the TeV-brane the zero-mode decouples and the
excitation state couplings are now an inverse TeV.  This results in a
strikingly different phenomenology than in the above case of large 
extra dimensions, as now the graviton KK tower states can undergo
single, direct, resonance production.  
To exhibit how the tower of 
graviton excitations may appear at the TESLA
collider, Fig. \ref{fg:kkspect} displays the cross section for $e^+e^-\to
\mu^+\mu^-$ as a function of $\sqrt s$, assuming $m_1=600$ GeV and taking
various values of $k/\overline M_{Pl}$ for purposes of demonstration.
%The tower of 
%graviton excitations may appear as resonances in the process $e^+e^-\to
%\mu^+\mu^-$ at the TESLA collider.
Searches for the first KK resonance in Drell-Yan and di-jet data
at the Tevatron already place non-trivial constraints \cite{dhr1}
on the parameter space of this model.  If the KK gravitons are too 
massive to be produced directly, their contributions to fermion
pair production may still be felt via virtual exchange.  Since in this
case there
is only one additional dimension, the uncertainties associated with the
introduction of a cut-off do not appear, as opposed to the model of
Ref.~\cite{add} as discussed above. As shown in Ref. \cite{dhr1},
scales of order $\Lambda_\pi=1-10$ TeV may be excluded for
$k/\overline M_{Pl}=0.01-1.0$ at a 500 GeV TESLA collider with 500 
fb$^{-1}$ of integrated luminosity.
Lastly, we note that if the Standard Model fields are also allowed to
propagate in the bulk \cite{bulkph}, the phenomenology can 
be markedly different, and is highly dependent 
upon the value of the 5-dimensional fermion mass, which is of order of
the Planck mass and thus different from the effective fermion mass in 4
dimensions.

\subsection{Non-commutative quantum field theory (NCQFT)}
In models of NCQFT the usual 
$\delta$-dimensional space associated with commuting space-time 
coordinates is generalized to one which is non-commuting.  In such
a space the conventional coordinates are represented by operators
which no longer commute,
\begin{equation}
[\hat X_\mu, \hat X_\nu]=i\theta_{\mu\nu}\equiv {i\over\Lambda^2_{NC}}
c_{\mu\nu} \,.
\end{equation}
Here, we have parameterized the effect in terms of an overall scale
$\Lambda_{NC}$, which characterizes the threshold where non-commutative
(NC) effects become important, and a real antisymmetric matrix 
$c_{\mu\nu}$, whose dimensionless elements are presumably of order
unity. The most likely value of $\Lambda_{NC}$ is probably near the
string or Planck scale, however, given the possibility of the onset of
stringy effects at the TeV scale, and that the fundamental Planck
scale may be lower due to the existence of large extra dimensions
as discussed above, it is feasible that NC effects could also set
in at a TeV.  There is a clear relation between the matrix $c_{\mu\nu}$
and the Maxwell field strength tensor $F_{\mu\nu}$ as NCQFT arises
in string theory through the quantization of strings as described by
the low energy excitations of D-branes in the presence of background
electromagnetic fields\footnote{NCQFT has been formulated so far only
for QED. Extensions to SU(N) theories including quarks are studied
presently.}$^,$\footnote{Astrophysical bounds have not been derived so
far in a consistent way.}. $c_{\mu\nu}$ is identical in all reference
frames, defining a preferred NC direction in space, and hence Lorentz
invariance is violated at energies of order $\Lambda_{NC}$.  However,
due to the rotation of the Earth and its revolution about the Sun, the
violation of Lorentz invariance is only observable in processes which 
are quadratic (or higher order even powers) in $\theta_{\mu\nu}$.

\begin{figure}[hbt]
\vspace*{-0.2cm}
\centerline{
\psfig{figure=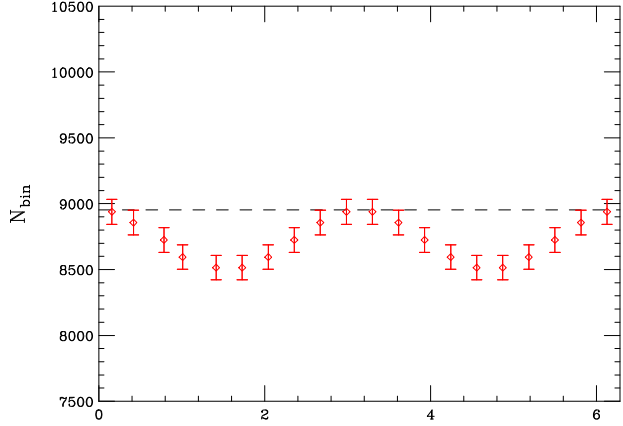,height=6cm,angle=0}}
\vspace*{-0.5cm}
\caption{\it $\phi$ dependence of the 
$e^+ e^- \rightarrow \gamma \gamma$ cross section, taking 
$\Lambda_{NC} = {\sqrt s}= 500$ GeV and a luminosity 
of $500 \, {\rm fb}^{-1}$.  A cut of 
$|\cos\theta| < 0.5$ has been employed.  
The dashed line corresponds to the SM 
expectations and the `data' points represent the NCQED results.}
\label{azim}
\vspace*{-0.5cm}
\end{figure}

A striking consequence of NCQFT is that the
NC version of QED takes on a non-abelian nature in that both 3-point
and 4-point photon couplings are generated.  In addition, all QED
vertices pick up additional phase factors which are dependent upon
the momenta flowing through the vertex.
NCQED thus has striking
effects in QED processes at the TESLA collider.  The modifications
to pair annihilation, Bhabha and M{\o}ller scattering, as well as
$\gamma\gamma\to\gamma\gamma$ have been studied in Ref. \cite{jpr}.
Pair annihilation and $\gamma\gamma$ scattering both receive new
diagrammatic contributions due to the non-abelian couplings, and
all four processes acquire a phase dependence due to the relative
interference of the vertex kinematic phases.  The lowest order
correction to the Standard Model to these processes occurs at
dimension 8.  The most interesting result is that a $\phi$
angular dependence is induced in $2\to 2$ scattering processes
due to the existence of the NC preferred direction in space-time.
This azimuthal dependence in pair annihilation is illustrated in 
Fig. \ref{azim} for the case where the NC direction is perpendicular
to the beam axis.  The results of Ref. \cite{jpr} are summarized in
Table \ref{nonctab} which displays the $95\%$ CL search reach for
the NC scale in these four reactions.  We see that these processes
are complementary in their ability to probe different structures
of non-commuting space-time.  These results indicate that NCQED
can be probed to scales of order a TeV, which is where one would
expect NCQFT to become relevant if stringy effects of the
fundamental Planck scale are also at a TeV.
\begin{table}[hbt]
\centering
\begin{tabular}{|c|c|c|} \hline
Process & Structure Probed  & Bound on $\Lambda_{NC}$   \\ \hline
$e^+e^-\to\gamma\gamma$   & Space-Time & $740-840$ GeV \\
M{\o}ller Scattering & Space-Space &  1700 GeV\\ 
Bhabha Scattering & Space-Time & 1050 GeV \\
$\gamma\gamma\to\gamma\gamma$ & Space-Time & $700-800$ GeV \\ 
   & Space-Space & 500 GeV \\ \hline
\end{tabular}
\caption{\it Summary of the $95\%$ CL search limits on the NC scale
$\Lambda_{NC}$ from the various processes considered above at
a 500 GeV $e^+e^-$ TESLA collider with an integrated luminosity
of 500 fb$^{-1}$.}
\label{nonctab}
\end{table}

  %%%%%%%%%%%%%%%%%%%%%%%%%%%%%%%%%%%%%%%%%%%%%%%%%%%%%%%%%%%%%%%%%%%%%%%%
\section{Strong Electroweak Symmetry Breaking}
\label{sec:phys-ewsb-strong}
\label{physics_alternatices_WW}

In the absence of a light Higgs particle or other low lying
resonances, unitarity requires that the interaction among gauge bosons
becomes strong at high energies. In this case, the physics of
electroweak symmetry
breaking~(EWSB) below the symmetry breaking
scale~$\Lambda_{\text{EWSB}}=4\pi v\approx3\,\mathrm{TeV}$ is
described by the most general effective Lagrangian for the Goldstone
bosons required by the spontaneous
$\mathrm{SU}(2)_L\otimes\mathrm{U}(1)_Y\to\mathrm{U}(1)_Q$ breaking.
This Lagrangian describes the physics of longitudinal gauge bosons and
its parameters can be probed in their interactions.  Effective field
theory allows to explore the multidimensional parameter space
systematically, where the course of this exploration is laid out using
power counting, dimensional analysis and symmetry.

All deviations of the $\rho$-parameter from the minimal Standard Model
tree level value of unity that have been observed by the LEP1/SLC
experiments can naturally be accounted for by loop corrections in the
minimal Standard Model itself.  Therefore, the EWSB sector appears to
have an approximate global $\mathrm{SU}(2)_c$ custodial symmetry,
protecting the $\rho$-parameter from renormalisation, that
is only broken by the hypercharge couplings of the Goldstone bosons.
Hence it is reasonable to look for possible
$\mathrm{SU}(2)_c$-conserving deviations from the Standard Model
predictions first.

The luminosity and centre of mass energy of TESLA will allow
experiments to probe the interactions of weak gauge bosons with
unprecedented precision, so that the nature of EWSB can be determined even in a scenario without light
resonances.  For this purpose, the couplings of three gauge bosons
$\mathrm{WWZ}$ and $\mathrm{WW}\gamma$ can be studied comprehensively
in the production of four-fermion final states, using methods
established at LEP2.  The interactions of four gauge bosons can be
analysed by disentangling the final states
$\nu\bar\nu\mathrm{W}^+\mathrm{W}^-$,
$\bar\nu\mathrm{e}^-\mathrm{W}^+\mathrm{Z}$,
$\mathrm{e}^+\nu\mathrm{Z}\mathrm{W}^-$, and
$\nu\bar\nu\mathrm{Z}\mathrm{Z}$ in six fermion
production $\mathrm{e}^+\mathrm{e}^-\to6\mathrm{f}$.  A convenient
basis for the analysis of weak gauge boson scattering is provided by
the weak isospin channels.  This analysis can be completed
by measuring the purely $I=2$ channel
$\mathrm{e}^-\mathrm{e}^-\to\nu\nu\mathrm{W}^-\mathrm{W}^-$, which is
accessible in the $\mathrm{e}^-\mathrm{e}^-$ mode of TESLA.  Finally,
GigaZ will contribute substantially improved limits on LEP1/SLC
oblique corrections, which will provide the best constraints in one
direction of parameter space.

In addition to model independent analyses of effective Lagrangians for
the EWSB sector, specific models for the EWSB can be studied.  These
models typically include vector and scalar resonances, whose
properties can be determined at TESLA.

%%%%%%%%%%%%%%%%%%%%%%%%%%%%%%%%%%%%%%%%%%%%%%%%%%%%%%%%%%%%%%%%%%%%%%%%
%%% \subsection{Strong $\mathrm{WW}$ Interactions}
\subsection{Strong WW Interactions}

Assuming only spontaneous
$\mathrm{SU}(2)_L\otimes\mathrm{U}(1)_Y\to\mathrm{U}(1)_Q$ symmetry
breaking, the most general ($C$ and $P$ conserving) effective
Lagrangian contains 10 dimension-four
interactions~\cite{Appelquist/Longhitano/Wu}.  As mentioned above,
$\mathrm{SU}(2)_c$ appears 
to be conserved in EWSB and a natural strategy for multiparameter fits
is to start the systematic exploration with the $\mathrm{SU}(2)_c$
invariant\footnote{The operators $L_{4,5}$ are explicitely
$\mathrm{SU}(2)_c$ invariant, but they do contribute to the
renormalisation of the $\mathrm{SU}(2)_c$ violating dimension-two
operator~\cite{Appelquist/Longhitano/Wu,Dawson/Brunstein/vanderBij}.
This induces a deviation of the $\rho$-parameter from unity at loop
level, but the resulting limits on~$\alpha_{4,5}$ from LEP1/SLC are
not competitive with the direct measurements discussed below.}  and
linearly breaking operators. In the notation
of~\cite{Appelquist/Longhitano/Wu}, these are
\begin{subequations}
\label{eq:strong-ww}
\begin{align}
  L_{1} & = \frac{\alpha_{1}}{16\pi^2} \frac{gg'}{2} B_{\mu\nu}
              \mathop{\mathrm{tr}} \Bigl( \sigma_3 W^{\mu\nu} \Bigr) \\
  L_{2} & = \frac{\alpha_{2}}{16\pi^2} \mathrm{i}g' B_{\mu\nu}
              \mathop{\mathrm{tr}} \Bigl( \sigma_3 V^\mu V^\nu \Bigr) \\
  L_{3} & = \frac{\alpha_{3}}{16\pi^2} 2\mathrm{i}g
              \mathop{\mathrm{tr}} \Bigl( W_{\mu\nu} V^\mu V^\nu \Bigr) \\
  L_{4} & = \frac{\alpha_{4}}{16\pi^2}
              \mathop{\mathrm{tr}} \Bigl( V_\mu V_\nu \Bigr)
              \mathop{\mathrm{tr}} \Bigl( V^\mu V^\nu \Bigr) \\
  L_{5} & = \frac{\alpha_{5}}{16\pi^2}
              \mathop{\mathrm{tr}} \Bigl( V_\mu V^\mu \Bigr)
              \mathop{\mathrm{tr}} \Bigl( V_\nu V^\nu \Bigr)\,,
\end{align}
\end{subequations}
where $W_{\mu\nu}=(\partial^{\vphantom{i}}_{\mu}W^i_\nu
 - \partial^{\vphantom{i}}_{\nu}W^i_\mu
 + g \epsilon^{ijk} W^j_\mu W^k_\nu ) \frac{\sigma^i}{2}$
and $V_\mu$ simplifies to $-\mathrm{i}g\frac{\sigma^i}{2}W^i_\mu
+\mathrm{i}g'\frac{\sigma^3}{2}B_\mu$ in unitarity gauge.  The
remaining five operators consist of four that break $\mathrm{SU}(2)_c$
quadratically with two explicit $\sigma_3$s and one that breaks it
quartically with four explicit $\sigma_3$s.  Since
$\mathrm{SU}(2)_c$ appears to be approximately conserved in EWSB, its
breaking must be governed by a higher
scale~$\Lambda_{\text{F}}>\Lambda_{\text{EWSB}}$, probably related to
flavour physics.  Then each explicit $\sigma_3$ is naturally suppressed
by one power of a small ratio
$\Lambda_{\text{EWSB}}/\Lambda_{\text{F}}$.  This observation
justifies to start the exploration with the
operators~(\ref{eq:strong-ww}).
The coefficients $\alpha_i$ in~(\ref{eq:strong-ww}) are related to
scales of new physics~$\Lambda^*_i$ by naive dimensional analysis
(NDA)~\cite{Weinberg_Georgi_NDA}
\begin{equation}
  \frac{\alpha_i}{16\pi^2} = \left(\frac{v}{\Lambda^*_i}\right)^2\,,
\end{equation}
where the Fermi scale $v = 246\,\mathrm{GeV}$ is fixed by low energy
weak interactions.  In the absence of resonances that are lighter than
$4\pi v$, one expects from NDA in a strongly interacting
symmetry breaking sector
\begin{equation}
\label{eq:ewsb-benchmark}
  \Lambda^*_i \approx \Lambda_{\text{EWSB}}
     = 4\pi v \approx 3\,\mathrm{TeV}\;,\;\;\;\text{i.e.}\;\;\;
  \alpha_i \approx \mathcal{O}(1)\,,
\end{equation}
unless some couplings are naturally suppressed by symmetries
(e.\,g.~$\mathrm{SU}(2)_c$).  Therefore, the crucial benchmark for a
linear collider from strong EWSB physics is given by this natural size
of the couplings, in order to be able to probe the EWSB sector in any
realistic scenario.

%%%%%%%%%%%%%%%%%%%%%%%%%%%%%%%%%%%%%%%%%%%%%%%%%%%%%%%%%%%%%%%%%%%%%%%%
\subsubsection{Final states with four fermions}
Three of the operators in~(\ref{eq:strong-ww}) contribute to triple
gauge couplings~(TGCs) at tree level
\begin{equation}
  L_{\mathrm{TGC}} = L_{1} + L_{2} + L_{3}\,.
\end{equation}
Of these, $L_{3}$ conserves the approximate $\mathrm{SU}(2)_c$, while
$L_{1,2}$ break it linearly.  The customary
anomalous TGCs~\cite{Hagiwara:1987vm} are related to
the coefficients of the effective Lagrangian via
\begin{subequations}
\label{eq:alpha2tgc}
\begin{align}
  g_1^{\mathrm{Z}} &= 1 + \frac{e^2}{\cos^2\theta_w(\cos^2\theta_w-\sin^2\theta_w)}
               \frac{\alpha_{1}}{16\pi^2}
             + \frac{e^2}{ \sin^2\theta_w\cos^2\theta_w}
               \frac{\alpha_{3}}{16\pi^2} \\
  \kappa_{\mathrm{Z}} &= 1 + \frac{2e^2}{\cos^2\theta_w-\sin^2\theta_w}
                  \frac{\alpha_{1}}{16\pi^2}
                - \frac{e^2}{\cos^2\theta_w}\frac{\alpha_{2}}{16\pi^2}
                + \frac{e^2}{\sin^2\theta_w}\frac{\alpha_{3}}{16\pi^2}\\
  \kappa_\gamma &= 1
                   - \frac{e^2}{\sin^2\theta_w}\frac{\alpha_{1}}{16\pi^2}
                   + \frac{e^2}{\sin^2\theta_w}\frac{\alpha_{2}}{16\pi^2}
                   + \frac{e^2}{\sin^2\theta_w}\frac{\alpha_{3}}{16\pi^2}\,.
\end{align}
\end{subequations}
The transformation~(\ref{eq:alpha2tgc}) is singular and the resulting
anomalous couplings satisfy
\begin{equation}
\label{eq:tgc-constraint}
  (\Delta g^1_{\mathrm{Z}} - \Delta\kappa_{\mathrm{Z}}) \cdot \cos^2\theta
      = \Delta\kappa_\gamma \cdot \sin^2\theta.
\end{equation}
Only two dimensions of the $\alpha_{1,2,3}$ parameter space
can be determined directly in processes with TGCs, such
as four-fermion production. The blind direction
\begin{equation}
\label{eq:alpha123-blind}
  (\alpha_{1},\; \alpha_{2},\; \alpha_{3})_{\text{blind}}
    \propto
  (\cos^2\theta_w-\sin^2\theta_w,\; \cos^2\theta_w,\; -\sin^2\theta_w)
\end{equation}
in the parameter space can not be constrained from TGCs alone.

\begin{table}
  \begin{center}
    \renewcommand{\arraystretch}{1.2}
    \begin{tabular}{c|cc|cc}
      $\alpha_1=0$     & \multicolumn{2}{|c}{$P_{\mathrm{e}^-}=80\%, P_{\mathrm{e}^+}=0\%$}
                       & \multicolumn{2}{|c}{$P_{\mathrm{e}^-}=80\%, P_{\mathrm{e}^+}=60\%$}
         \\\hline
      $\sqrt{s}$       & 500\,GeV & 800\,GeV & 500\,GeV & 800\,GeV \\
      $\int\!\mathcal{L}\mathrm{d}t$
                       & 500\,fb${}^{-1}$ & 1000\,fb${}^{-1}$
                       & 500\,fb${}^{-1}$ & 1000\,fb${}^{-1}$ \\\hline
      $\Delta\alpha_2$ & 0.329    & 0.127     & 0.123     & 0.090 \\
      $\Delta\alpha_3$ & 0.143    & 0.071     & 0.083     & 0.048 \\\hline
      $\Lambda^*_2$    & 5.4\,TeV &  8.7\,TeV &  8.8\,TeV & 10.3\,TeV \\
      $\Lambda^*_3$    & 8.2\,TeV & 11.6\,TeV & 10.7\,TeV & 14.1\,TeV
    \end{tabular}
  \end{center}
  \caption{\label{tab:alpha23}%
    68\%~C.L. sensitivities for the strong EWSB
    parameters~$(\alpha_2,\alpha_3)$, assuming $\alpha_1=0$, in a
    study of TGCs at a TESLA experiment, with and without positron
    polarisation~\cite{Menges:2000:TDR-backup}.  Results without the
    constraint $\alpha_1=0$ are presented in
    Fig.~\ref{fig:alpha_123} and in the text on
    page~\pageref{text:alpha1-nonzero}.}
\end{table}

\begin{figure}
  \begin{center}
    %%% Please don't change the aspect ratio and try to avoid
    %%% rescaling as well (the bitmap fonts would look ugly).
    \href{pictures/4/sewsb_fig1.pdf}{{\includegraphics[width=10cm]{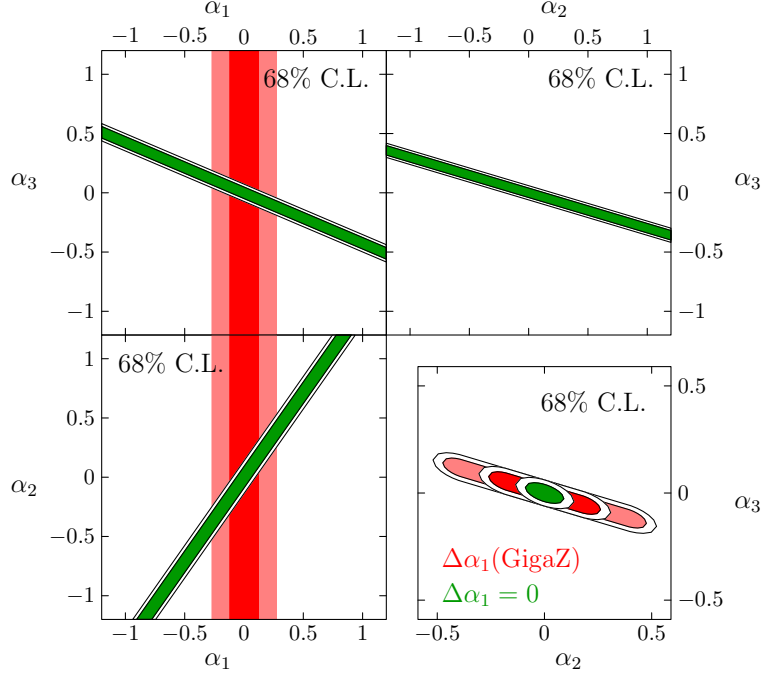}}}
  \end{center}
  \caption{\label{fig:alpha_123}%
    Sensitivity for the strong EWSB parameters~$\alpha_{1,2,3}$ in a
    study of TGCs at a TESLA experiment (800\,GeV,
    $1000\,{fb}^{-1}, P_{\mathrm{e}^-}=80\%,
    P_{\mathrm{e}^+}=60\%$)~\cite{Menges:2000:TDR-backup}.
    The~$\alpha_{1,2,3}$ are normalised such that their natural size
    from dimensional analysis is~$\mathcal{O}(1)$.  The inner shaded
    diagonals correspond to~$\Delta\chi^2=1$, while the outer diagonals
    correspond to 68\%~C.L.{} in two dimensions,
    i.e.~$\Delta\chi^2=2.3$.  The dark and light vertical bands
    correspond to 68\%~C.L.{} limits on $\alpha_1$ from fitting
    $\epsilon_3$ at GigaZ with and without the constraint
    $\epsilon_2=\epsilon_2(\mathrm{SM})$ (cf.~Fig.~\ref{fig:eps_lc}).
    The blowup in the lower right corner shows the allowed region in
    $(\alpha_{2},\alpha_{3})$ for $\alpha_1$ fixed and constrained at
    GigaZ, respectively.}
\end{figure}

The simulation of TGC measurements at TESLA summarised in
Fig.~\ref{fig:fiveparfit}
and~\cite{Menges:2000:TDR-backup} can be used to perform fits of pairs
of parameters with the third parameter fixed.  The results
for~$(\alpha_{2},\alpha_{3})$ with $\alpha_1=0$ are collected in
Table~\ref{tab:alpha23}, showing that the
benchmarks~(\ref{eq:ewsb-benchmark}) are reached easily.  The results
of Table~\ref{tab:alpha23} can be combined with the parameters of the
blind direction~(\ref{eq:alpha123-blind}) to obtain the allowed volume
in three dimensional parameter space. Fig.~\ref{fig:alpha_123} shows
the projections for 1000\,fb${}^{-1}$ polarised scattering at
800\,GeV.  The limits scale with integrated luminosity as
$\Delta\alpha_i \propto (\int\mathcal{L})^{-1/2}$ since the
measurement is dominated by statistics~\cite{Menges:2000:TDR-backup}.

%%% The most conservative estimate of the experimental sensitivity
%%% includes correlations of the coupling parameters that violate the
%%% constraint~(\ref{eq:tgc-constraint}).  This corresponds to
%%% \emph{projecting} the allowed region on the
%%% hyperplane~(\ref{eq:tgc-constraint}), while the two dimensional fit
%%% corresponds to taking a \emph{slice} along~(\ref{eq:tgc-constraint}).  The
%%% blowup in the lower right corner of Fig.~\ref{fig:alpha_123} shows
%%% the result $\Delta\alpha_2=0.281$ and~$\Delta\alpha_3=0.357$,
%%% i.\,e.~$\Lambda^*_2=5.8\,\text{TeV}$
%%% and~$\Lambda^*_3=5.2\,\text{TeV}$.  While deviations
%%% from~(\ref{eq:tgc-constraint}) could only be caused by
%%% unexpected new physics beyond EWSB, this result demonstrates that
%%% TESLA retains its sensitivity for the EWSB sector even in this case.

Additional independent measurements constrain the parameters
further.  $L_1$ contributes to LEP1/SLC oblique electroweak
corrections at tree level as $\Delta S=-\alpha_1/(2\pi)$ resulting
in~$\Delta\alpha_1=0.69$ at 68\,\%\,C.L.{}
(cf.~Fig.~\ref{fig:eps_lc}). The constraint on $S$ (or $\epsilon_3$,
respectively) can be improved by more than a factor of two at GigaZ
(cf.~Fig.~\ref{fig:eps_lc}). This observation motivates the blowup
of the $(\alpha_{2},\alpha_{3})$-plane in the lower right corner of
Fig.~\ref{fig:alpha_123}, where the blind direction is removed by
the expected limits on~$\alpha_1$ from the measurement of $\epsilon_3$
at GigaZ (the dark and light areas correspond to fits with and
without the constraint~$\epsilon_2=\epsilon_2(\mathrm{SM})$,
respectively). The resulting
conservative limits $\Delta\alpha_2=0.5$ and~$\Delta\alpha_3=0.2$,
\label{text:alpha1-nonzero}
i.\,e.~$\Lambda^*_2=4.4\,\text{TeV}$ and~$\Lambda^*_3=6.9\,\text{TeV}$,
still probe the EWSB parameters at their natural
size~(\ref{eq:ewsb-benchmark}).
The size of these constraints is of the order of electroweak radiative
corrections and further theoretical studies of the systematics of non
leading corrections will be useful.

In addition, $L_3$ contributes to quartic gauge boson interactions.
However, limits on~$\alpha_{3}$ derived from measurements of quartic
couplings can not be expected to improve the limits from
the other two measurements~(cf.~Table~\ref{tab:alpha45-LHC}).
Instead, they will provide important consistency checks.

In summary, the limits on the TGCs translate to a physics reach of
\begin{equation}
  \Lambda^*_i \approx 5\,\mathrm{TeV}
     > \Lambda_{\text{EWSB}} \approx 3\,\mathrm{TeV}
\end{equation}
for the EWSB sector in $\mathrm{W}$ pair
production at TESLA.  However, this naive translation should only be
understood as confirmation that \emph{any reasonable scenario} for the
symmetry breaking sector can be probed in detail, since new physics is
to be expected in the symmetry breaking sector \emph{below} these scales
at~$\Lambda_{\text{EWSB}}$.

It is worth pointing out that the measurements at the linear collider
probe the coefficients of the effective Lagrangian directly and do
\emph{not} depend on \emph{ad-hoc} unitarisation prescriptions.  In
particular, all momenta remain in the region where the momentum
expansion in $\sqrt{s}/\Lambda_i<1$ converges.
Fig.~\ref{fig:tgccomp} shows that a TESLA experiment
has an advantage over LHC for~$\kappa_{\gamma,\mathrm{Z}}$, while being
competitive for~$\lambda_{\gamma,\mathrm{Z}}$.  Therefore TESLA is
particularly powerful for constraining the strong EWSB
parameters~$\alpha_{1,2,3}$.

%%%%%%%%%%%%%%%%%%%%%%%%%%%%%%%%%%%%%%%%%%%%%%%%%%%%%%%%%%%%%%%%%%%%%%%%
\subsubsection{Final states with six fermions}
Two of the $\mathrm{SU}(2)_c$ conserving operators
in~(\ref{eq:strong-ww}) contribute solely to quartic gauge
couplings~(QGCs)
\begin{equation}
  L_{\text{QGC}} = L_{4} + L_{5}\,,
\end{equation}
while $L_{\text{TGC},\text{QGV}} = L_{3}$
contributes to both TGCs and QGCs.

\begin{figure}
  \begin{center}
    %%% The following is output of FeynMP that has been made selfcontained.
    \setlength{\unitlength}{1mm}
    \def\fmfL(#1,#2,#3)#4{\put(#1,#2){\makebox(0,0)[#3]{#4}}} 
    \begin{picture}(40,30)
      \put(0,0){\href{pictures/4/sewsb_fig2a.pdf}{{\includegraphics{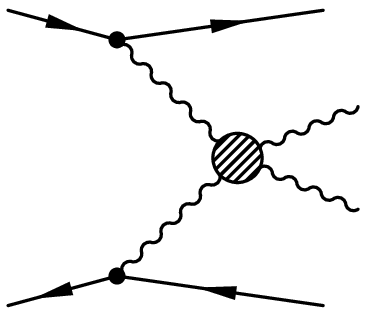}}}}
      % tesla-tdr-ewgb-pics.t1 -- generated from tesla-tdr-ewgb-pics.mp
      \fmfL(2.4618,-1.44228,rt){$\mathrm {e}^+$}%
      \fmfL(2.4618,31.44228,rb){$\mathrm {e}^-$}%
      \fmfL(37.5382,-1.44228,lt){$\mathrm {e}^+/\bar \nu $}%
      \fmfL(41.58289,9.24266,l){$\mathrm {Z}/\mathrm {W}^+$}%
      \fmfL(41.58286,20.7576,l){$\mathrm {Z}/\mathrm {W}^-$}%
      \fmfL(37.5382,31.44228,lb){$\mathrm {e}^-/\nu $}%
    \end{picture}
    \qquad\qquad
    \begin{picture}(40,30)
      \put(0,0){\href{pictures/4/sewsb_fig2b.pdf}{{\includegraphics{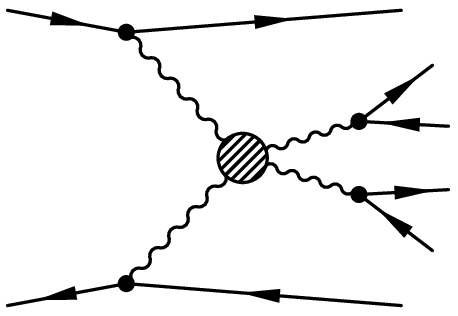}}}}
      % tesla-tdr-ewgb-pics.t2 -- generated from tesla-tdr-ewgb-pics.mp
      \fmfL(3.31328,-1.26527,rt){$\mathrm {e}^+$}%
      \fmfL(3.31328,31.26527,rb){$\mathrm {e}^-$}%
      \fmfL(46.68672,-1.26527,lt){$\mathrm {e}^+/\bar \nu $}%
      \fmfL(50.11566,4.79782,l){$\bar {\mathrm {f}}_1$}%
      \fmfL(51.8827,11.50629,l){$\mathrm {f}_2$}%
      \fmfL(51.8828,18.4932,l){$\bar {\mathrm {f}}_3$}%
      \fmfL(50.11584,25.2017,l){$\mathrm {f}_4$}%
      \fmfL(46.68701,31.26483,lb){$\mathrm {e}^-/\nu $}%
    \end{picture}
  \end{center}
  \caption{\label{fig:ee->ffVV}%
    Gauge boson scattering with on-shell gauge boson final states,
    studied comprehensively with irreducible backgrounds
    in~\cite{Boos_etal:1998:WWWW} (left).  Gauge boson scattering
    subprocess in six-fermion production (right).}
\end{figure}

In~\cite{Boos_etal:1998:WWWW}, unpolarised on-shell vector boson
production processes
$\mathrm{e}^+\mathrm{e}^-\to\nu\bar\nu\mathrm{W}^+\mathrm{W}^-$,
$\mathrm{e}^+\mathrm{e}^-\to\nu\bar\nu\mathrm{ZZ}$, and
$\mathrm{e}^-\mathrm{e}^-\to\nu\nu\mathrm{W}^-\mathrm{W}^-$ (see the
left hand side of Fig.~\ref{fig:ee->ffVV}) have been studied
comprehensively.  It has been demonstrated that the natural
size~(\ref{eq:ewsb-benchmark}) of all quartic couplings
can be probed at a high luminosity 800\,GeV TESLA.

Realistic studies including detector simulation must handle the decays
of the gauge bosons into the fermions that are observed (see the right
hand side of Fig.~\ref{fig:ee->ffVV}), including all irreducible and
reducible background.  Detailed simulations of six fermion production
have been performed~\cite{Chierici_Kobel_Rosati:2000:TDR-backup}, using an
unweighted event generator for the six particle phase
space~\cite{Kilian:2000:TDR-backup-WHIZARD,Ohl:1999:VAMP} and complete
tree level amplitudes for signal and irreducible
backgrounds~\cite{Moretti_Ohl_etal:2000:TDR-backup-O'Mega}.

\begin{figure}
  \begin{center}
    \href{pictures/4/alpha45-separate.pdf}{{\includegraphics[width=0.45\textwidth,height=0.45\textwidth]{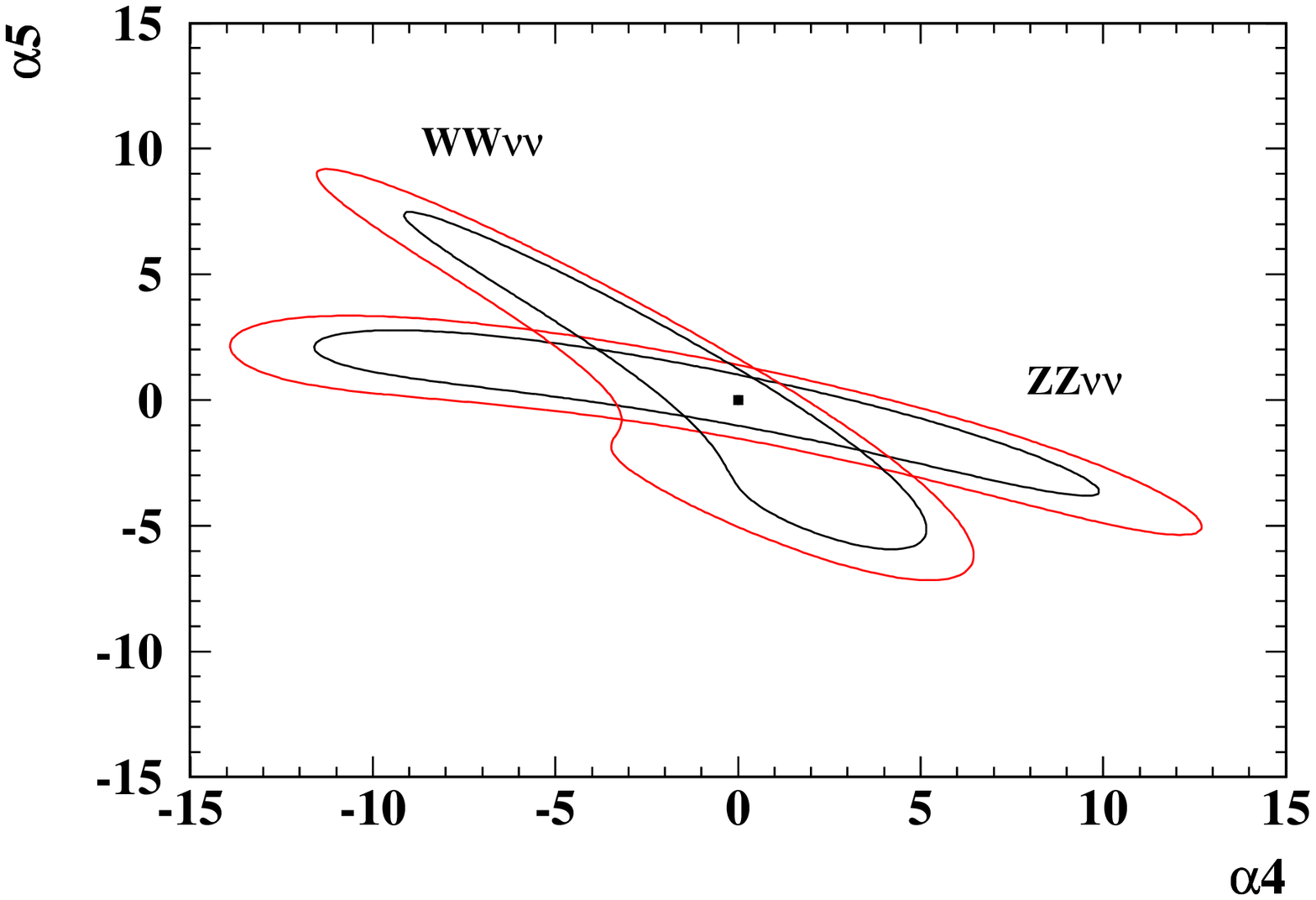}}}\qquad
    \href{pictures/4/alpha45-combined.pdf}{{\includegraphics[width=0.45\textwidth,height=0.45\textwidth]{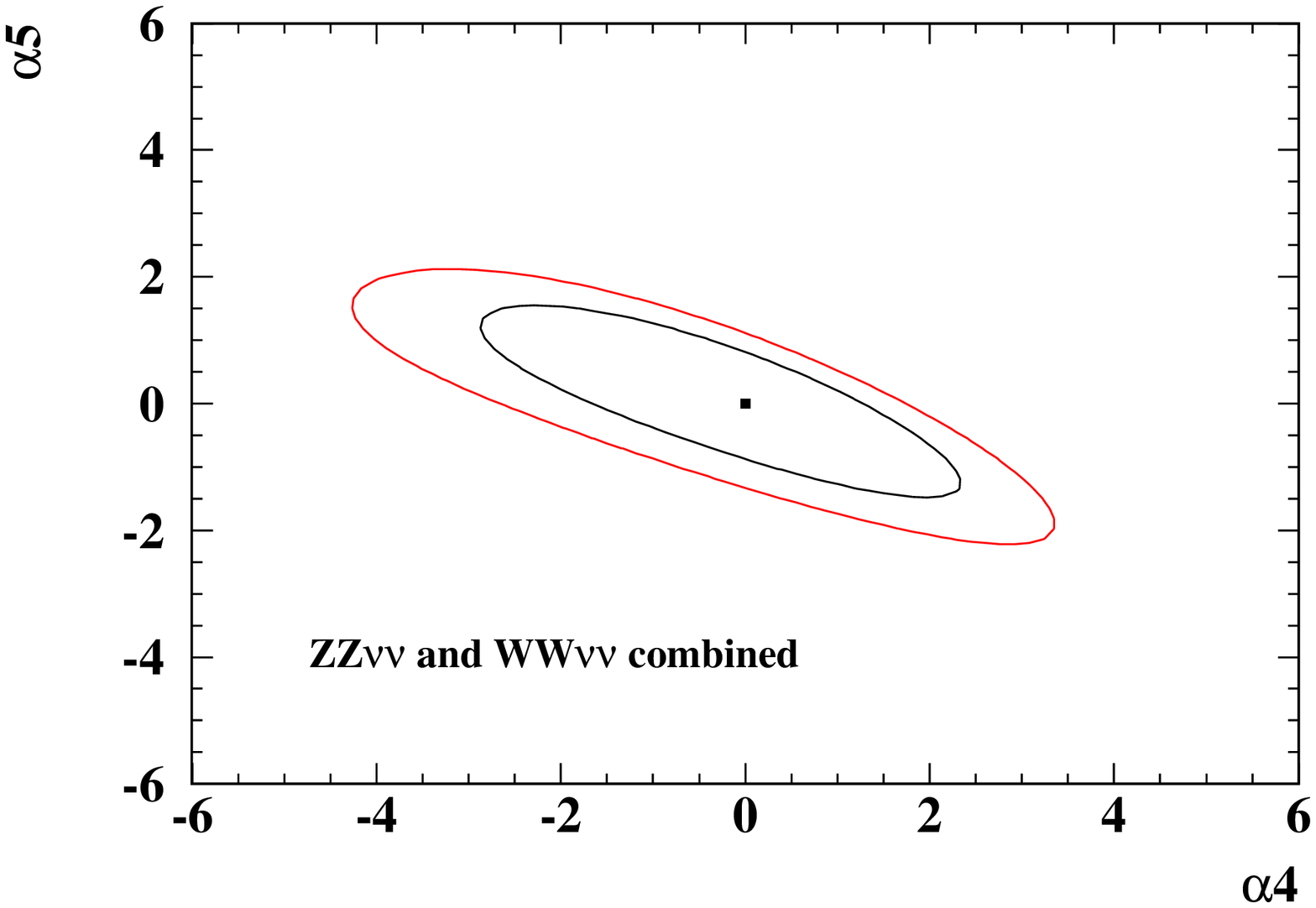}}}
  \end{center}
  \caption{\label{fig:alpha4/5}%
    Sensitivity for the strong EWSB parameters~$\alpha_4$ and~$\alpha_5$
    in the processes $\mathrm{e}^+\mathrm{e}^-\to
    \nu\bar{\nu}\mathrm{W}^+\mathrm{W}^-\to\nu\bar{\nu}jjjj$
    and~$\mathrm{e}^+\mathrm{e}^-\to\nu\bar{\nu}\mathrm{ZZ}\to\nu\bar{\nu}jjjj$
    in a TESLA experiment with $1000\,\textrm{fb}^{-1}$
    at 800\,GeV and~$P_{\mathrm{e}^-}=80\%$,
    $P_{\mathrm{e}^+}=40\%$~\cite{Chierici_Kobel_Rosati:2000:TDR-backup}.
    The inner and outer contours represent 68\,\%\,C.L.{}
    (i.\,e.~$\Delta\chi^2=2.3$) and 90\,\%\,C.L.{} limits, respectively.}
\end{figure}

\begin{table}
  \begin{center}
    \renewcommand{\arraystretch}{1.2}
    \begin{tabular}{c|c|c}
      $\sqrt{s}$       & \multicolumn{2}{c}%
                         {$800\,\textrm{GeV}$,
                          $P_{\mathrm{e}^-}=80\%$,
                          $P_{\mathrm{e}^+}=40\%$}  \\
      $\int\!\mathcal{L}\mathrm{d}t$
                       & $1000\,\textrm{fb}^{-1}$ & $2000\,\textrm{fb}^{-1}$ \\ \hline
      $\alpha_4$       & $-1.8$\;\ldots\;$+1.5$ & $-1.3$\;\ldots\;$+1.1$ \\
      $\alpha_5$       & $-0.9$\;\ldots\;$+1.0$ & $-0.6$\;\ldots\;$+0.7$ \\\hline
      $\Lambda^*_4$    & 2.3\,TeV               & 2.7\,TeV \\
      $\Lambda^*_5$    & 3.1\,TeV               & 3.7\,TeV
    \end{tabular}
  \end{center}
  \caption{\label{tab:alpha45}%
    68\%~C.L.{} sensitivities for
    the strong EWSB parameters~$(\alpha_4,\alpha_5)$ in a realistic
    study of QGCs at a TESLA
    experiment~\cite{Chierici_Kobel_Rosati:2000:TDR-backup}.}
\end{table}

The experimental sensitivity of an experiment at TESLA to the
$\mathrm{SU}(2)_c$~conserving parameters~$(\alpha_{4},\alpha_{5})$ has
been estimated from a likelihood fit to the angular distributions of
the gauge boson production angles and the angles of the decay fermions
in the $\nu\bar{\nu}jjjj$
channel, after detector simulation (using SIMDET~V1.3~\cite{SIMDET})
and event selection~\cite{Chierici_Kobel_Rosati:2000:TDR-backup}.  The
results are summarised in Fig.~\ref{fig:alpha4/5} and
Table~\ref{tab:alpha45}.  The $\mathrm{WW}/\mathrm{ZZ}$ separation and
background suppression in hadronic final states use cuts on transverse
momenta and recoil masses, utilising the excellent energy resolution
and hermiticity, including the forward region, of the proposed TESLA
detector. Three to four years of operation will cover $\Lambda^*_{4,5}$
up to the EWSB scale
$\Lambda^*_{4,5}\lesssim\Lambda_{\text{EWSB}}\approx 3\,\textrm{TeV}$ 

The exotic $I=2$~channel
$\mathrm{W}^-\mathrm{W}^-\to\mathrm{W}^-\mathrm{W}^-$ can be accessed
by operating TESLA in the $\mathrm{e}^-\mathrm{e}^-$~mode.  The
different angular distributions in this channel allow a further
reduction of correlations among $\mathrm{SU}(2)_c$~conserving
interactions.  In particular, there can be no contaminations from a
$\mathrm{SU}(2)_c$~violating sector in this
channel~\cite{Boos_etal:1998:WWWW}.

In the Standard Model with a very heavy Higgs, two loop diagrams
create a sizeable contribution to the parameter~$\alpha_5$, while the
contributions to the other parameters remain
small~\cite{vanderBij:1985fi}.  Using $\alpha_5\approx-g^2/(16\pi^2)
\cdot m_{\mathrm{H}}^2/m_{\mathrm{W}}^2$~\cite{vanderBij:1985fi}, the
limits on~$\alpha_5$ from the two-dimensional fit in
Table~\ref{tab:alpha45} translate to a Higgs mass of
$m_{\mathrm{H}}\approx 1.8\,\mathrm{TeV}$, demonstrating that virtual
effects of an extremely heavy Higgs can be observed in vector boson
scattering at a TESLA experiment.

\begin{table}
  \begin{center}
    \renewcommand{\arraystretch}{1.2}
    \begin{tabular}{c|c|c}
      $\sqrt{s}$       & LHC
                       & TESLA 800\,GeV \\
      $\int\!\mathcal{L}\mathrm{d}t$
                       & $100\,\textrm{fb}^{-1}$
                       & $1000\,\textrm{fb}^{-1},
                            P_{\mathrm{e}^-}=80\%, P_{\mathrm{e}^+}=40\%$ \\ \hline
      $\alpha_4$       & $-0.17$\;\ldots\;$+1.7$ & $-1.1$\;\ldots\;$+0.8$ \\
      $\alpha_5$       & $-0.35$\;\ldots\;$+1.2$ & $-0.4$\;\ldots\;$+0.3$ \\\hline
      $\Lambda^*_4$    & 2.3\,TeV                & 2.9\,TeV \\
      $\Lambda^*_5$    & 2.8\,TeV                & 4.9\,TeV
    \end{tabular}
  \end{center}
  \caption{\label{tab:alpha45-LHC}%
    Comparison of 68\%~C.L.{}
    sensitivities from one dimensional fits of
    the strong EWSB parameters~$(\alpha_4,\alpha_5)$
    at LHC~\cite{Belyaev_etal:LHC-Limits} and at a TESLA
    experiment~\cite{Chierici_Kobel_Rosati:2000:TDR-backup}.}
\end{table}

The one dimensional 68\%~C.L.{} limits from $100\,\mathrm{fb}^{-1}$ at
LHC~\cite{Belyaev_etal:LHC-Limits} are compared with the
analogous prediction for a TESLA experiment in
Table~\ref{tab:alpha45-LHC}.  Even though all backgrounds are included in
the simulation of a TESLA experiment, TESLA exceeds the physics reach of
LHC for QGCs and reaches the strong EWSB
benchmark~(\ref{eq:ewsb-benchmark}).  As anticipated above, the limits for
the QGCs are sligthly worse than the limits for the TGCs.  As in the
case of TGCs, no unitarisation prescriptions are required.

In summary, it has been demonstrated with realistic simulations that a
TESLA experiment can probe the $\mathrm{SU}(2)_c$ invariant and
linearly breaking parameters of a strong EWSB sector exhaustively in
the threshold region of strong $\mathrm{WW}$~interactions
up to the EWSB scale $\Lambda^*_i \approx 3\,\mathrm{TeV}$.

%%%%%%%%%%%%%%%%%%%%%%%%%%%%%%%%%%%%%%%%%%%%%%%%%%%%%%%%%%%%%%%%%%%%%%%%
\subsection{Vector resonances and Pseudo-Goldstone bosons}

\begin{figure}
  \begin{center}
    \begin{minipage}{0.5\textwidth}
      \href{pictures/4/barklow-let.pdf}{{\includegraphics[height=\textwidth,angle=-90]{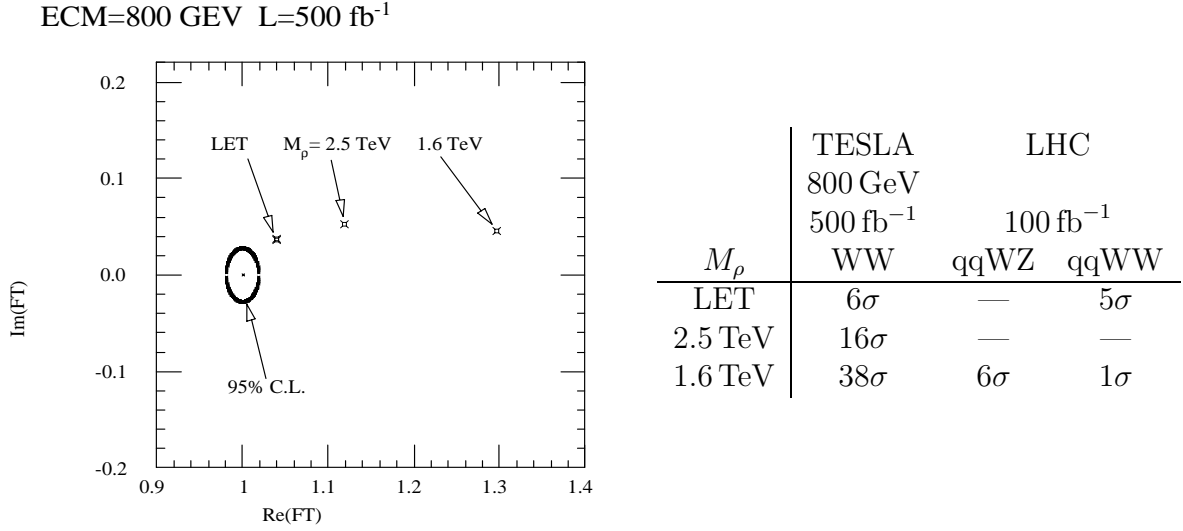}}}
    \end{minipage}
    \quad
    \begin{tabular}{c|ccc}
                 & TESLA                 & \multicolumn{2}{c}{LHC} \\
                 & 800\,GeV              &                         \\
                 & $500\,\text{fb}^{-1}$ & \multicolumn{2}{c}{$100\,\mathrm{fb}^{-1}$} \\
      $M_{\rho}$ & $\mathrm{WW}$         & $\mathrm{qqWZ}$ & $\mathrm{qqWW}$ \\ \hline
      LET        & $6\sigma$             & ---             & $5\sigma$  \\
      2.5\,TeV   & $16\sigma$            & ---             & ---  \\
      1.6\,TeV   & $38\sigma$            & $6\sigma$       & $1\sigma$
    \end{tabular}
  \end{center}
  \caption{\label{fig:barlow-let}%
    Sensitivity for a resonance form
    factor~(\ref{eq:barklow-form-factor}) at TESLA (assuming perfect
    charm tagging) and the LHC~\cite{Barklow:2000:LCWS}.}
\end{figure}

If the EWSB sector includes a resonance below the EWSB scale, the
vector boson pair production amplitude is unitarised by a Omn\`es
rescattering factor 
\begin{equation}
\label{eq:barklow-form-factor}
  F_{\mathrm{TC}}(s) = \exp \left(\frac{s}{\pi}
      \int_0^\infty\!\frac{\mathrm{d}s'}{s'}\,
      \frac{\delta_{\mathrm{LET}}(s')+\delta_\rho(s')}{s'-s-\mathrm{i}\epsilon}\right)\,,
\end{equation}
with one contribution reproducing the low energy theorem
$\delta_{\mathrm{LET}}(s)=s/(8\Lambda_{\mathrm{EWSB}}^2)$
for Goldstone boson scattering at threshold far below any resonance
and a second contribution from a resonance
$\delta_\rho(s)=3\pi/8\cdot(\tanh(s-M_\rho^2)/(M_\rho\Gamma_\rho)+1)$.
A phenomenological study~\cite{Barklow:2000:LCWS} shows that
$500\,\mathrm{fb}^{-1}$ of $\mathrm{W}^+\mathrm{W}^-$ production at a
800\,GeV TESLA is competitive with $100\,\mathrm{fb}^{-1}$ at LHC, as
shown in Fig.~\ref{fig:barlow-let}.  The~$6\sigma$ exclusion limit
for LET (also excluding any $I=1$ resonance) at TESLA assumes perfect
charm tagging, which is a realistic
approximation for the proposed TESLA detector.  Without any charm
tagging, the significance is reduced to~$4.6\sigma$.

An example of a concrete model for the EWSB sector without a Higgs
particle is the BESS model~\cite{Casalbuoni:1987vq}, which includes most
technicolour models.  The model assumes a triplet of new vector
resonances~$V^{\pm,0}$, similar to the $\rho$ or techni-$\rho$.  These
vector bosons mix with the electroweak gauge bosons and the mixing angle
is proportional to the ratio $g/g''$, where $g''$ is the self-coupling
of the~$V^{\pm,0}$.  The coupling of the~$V^{\pm,0}$ to fermions is
determined by a second parameter $b$.  The so called degenerate BESS
model~\cite{Casalbuoni_etal:Degenerate-BESS} is a special case, in which axial and vector
resonances are almost degenerate in mass.  Models for dynamical EWSB
typically predict many pseudo Nambu-Goldstone bosons~(PNGBs) from the
breaking of a large global symmetry group $G$ and the lightest neutral
PNGB $P^0$ calls for special attention.

\begin{figure}
  \begin{center}
    \href{pictures/4/bess-limits.pdf}{{\includegraphics[width=0.5\textwidth]{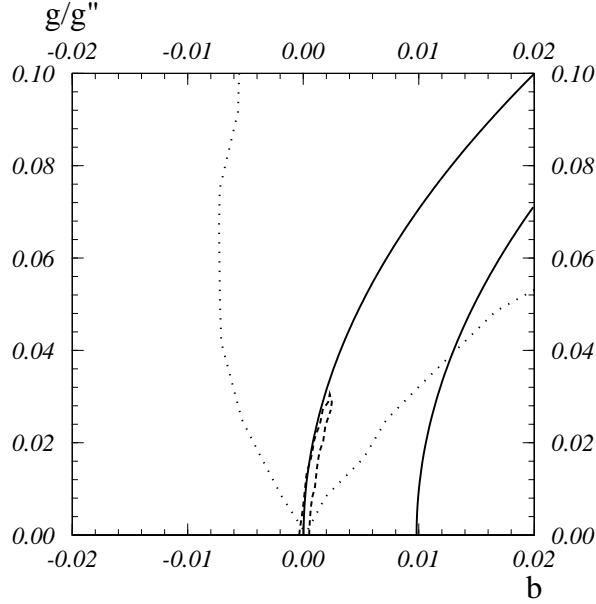}}}
    \caption{\label{fig:bess-limits}%
       The 95\,\%\,C.L. bounds for the BESS model parameters from
       a TESLA experiment with $\int\mathcal{L}=1000~\mathrm{fb}^{-1}$
       at~$\sqrt{s}=800\,\mathrm{GeV}$, assuming $M_V=2\,\mathrm{TeV}$
       (interior of the dashed boundary).  This is compared with
       present electroweak data (interior of the solid lines) and
       with the expected limits from LHC for $M_V=2\,\mathrm{TeV}$
       (inside of the dotted wedge).}
  \end{center}
\end{figure}

The $\mathrm{W}^+\mathrm{W}^-$-channel is the preferred channel for
the discovery of the vector resonances of the general BESS model,
while the $\bar{\mathrm{f}}\mathrm{f}$-channel is preferred for the
degenerate BESS model~\cite{Casalbuoni_etal:Degenerate-BESS}.  The analysis in the fermionic
channels is based on the observables $\sigma^{\mu}$,
$\sigma^{\mathrm{hadr.}}$, $A_{\mathrm{FB}}^{\mathrm{e}^+\mathrm{e}^-
\to \mu^+\mu^-,\bar{\mathrm{b}}\mathrm{b}}$,
$A_{\mathrm{LR}}^{\mathrm{e}^+\mathrm{e}^- \to
\mu^+\mu^-,\bar{\mathrm{b}}\mathrm{b}}$ and
$A_{\mathrm{LR}}^{\mathrm{e}^+\mathrm{e}^- \to \mathrm{hadr.}}$ (all
with $P_{\mathrm{e}^-}=0.8$).  In the $\mathrm{W}^+\mathrm{W}^-$
channel, the observables
$\mathrm{d}\sigma^{\mathrm{e}^+\mathrm{e}^-\to
\mathrm{W}^+\mathrm{W}^-}/\mathrm{d}\cos\theta$,
$A_{\mathrm{LR}}^{\mathrm{e}^+\mathrm{e}^- \to
\mathrm{W}^+\mathrm{W}^-}$ and the longitudinally and transversely
polarised differential $\mathrm{W}$ cross sections and asymmetries
have been used.  The expected bounds from a TESLA experiment for the
BESS model, obtained by combining all the observables, are shown in
Fig.~\ref{fig:bess-limits}.  In particular at large mixing, the
sensitivity of a TESLA experiment is much higher than the combination
of current electroweak data and expected LHC results.

In the case of the degenerate BESS, the LHC has the better discovery
potential.  However, if a neutral resonance with a mass below 1\,TeV
were discovered at the LHC, a TESLA experiment could study it in
detail and attempt to split the two nearly degenerate resonances and
measure their widths~\cite{Casalbuoni:1999mm}.

% For example, considering a spread in the c.m.~energy given by
% $\sigma_E\sim 0.007 E$, one can split the two
% resonances for $g/g''\geq 0.16$~\cite{Casalbuoni:1999mm}.

% At the LHC, the $\mathrm{gg}\to P^0\to \gamma\gamma$ channel provides
% a robust detection mode for the lightest neutral PNGB~$P^0$. In a
% $\mathrm{SU}(N_{\mathrm{TC}})$ technicolor theory with
% $N_{\mathrm{TC}}=4$, the~$P^0$ can be discovered in the mass range
% $30\ldots200\,\mathrm{GeV}$ (the upper mass reach probably being
% considerably higher) at high luminosities
% $\int\mathcal{L}=150\,\mathrm{fb}^{-1}$.  Even for the unrealistically
% low value of $N_{\mathrm{TC}}=1$, discovery at the LHC is possible for
% $m_{P^0}=70\ldots150\,\mathrm{GeV}$.  At the Tevatron, the same mode
% allows, for $N_{\mathrm{TC}}=4$, discovery at luminosities
% $\int\mathcal{L} \gtrsim 6\,\mathrm{fb}^{-1}$ in the mass range
% $60\ldots200\,\mathrm{GeV}$. Both at the LHC and Tevatron, $P^0$
% detection in the dominant $\mathrm{b}\bar{\mathrm{b}}$ decay mode is
% hindered by large backgrounds.

\begin{figure}
  \begin{center}
    \href{pictures/4/bess-pngb-ee.pdf}{{\includegraphics[width=0.5\textwidth]{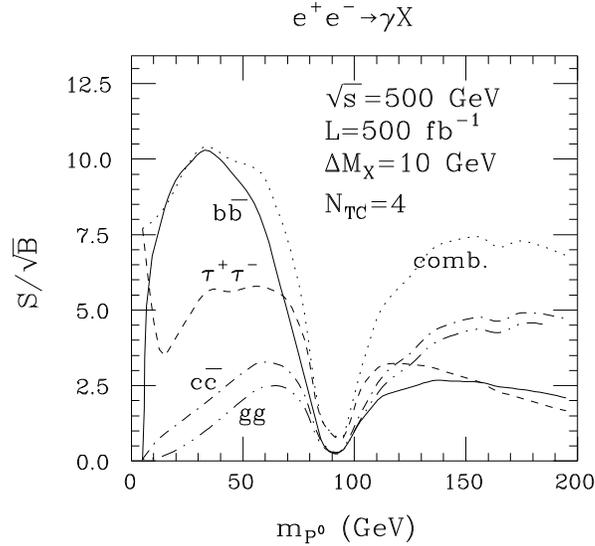}}}
    \caption{\label{fig:bess-pngb-ee}%
      The statistical significances $S/\sqrt B$ for a $P^0$ signal
      in various tagged channels as a function of $m_{P^0}$ at a
      $500\,\mathrm{GeV}$ collider for  an integrated luminosity of
      $500\,\mathrm{fb}^{-1}$.  The significance from the optimal
      combination of the channels is shown as a dotted line.}
  \end{center}
\end{figure}

The best mode for $P^0$ production at TESLA is
$\mathrm{e}^+\mathrm{e}^-\to\gamma P^0$. Results for the significance
$S/\sqrt B$, in the various tagged channels, for a
$\mathrm{SU}(N_{\mathrm{TC}})$ technicolour model with
$N_{\mathrm{TC}}=4$ and integrated luminosity $\int {\cal
L}=500\,\mathrm{fb}^{-1}$ at $\sqrt{s}=500\,\mathrm{GeV}$, are plotted
in Fig.~\ref{fig:bess-pngb-ee}~\cite{Casalbuoni:1999fs}.  Also shown
is the significance that can be achieved with the optimal combination
of the $\mathrm{gg}$, $\mathrm{c}\bar{\mathrm{c}}$,
$\mathrm{b}\bar{\mathrm{b}}$ and $\tau^+ \tau^-$ channels.  A strong
signal is visible for all but $m_{P^0}=80\ldots110\,\mathrm{GeV}$.
From the scaling law $S/\sqrt B\propto N_{\mathrm{TC}}^2$, one sees that
a $P^0$ discovery would be possible for $N_{\mathrm{TC}}\gtrsim 2.5$
for $m_{P^0}$ near 35\,GeV.  Unlike the LHC, for high enough
luminosities, a TESLA experiment can probe quite low values of
$m_{P^0}$ and could measure ratios of a number of interesting $P^0$
decay modes.  In the scenario of Fig.~\ref{fig:bess-pngb-ee} and in
the case of $m_{P^0}\approx 35\,\mathrm{GeV}$, an accuracy of $\approx
11\,\%$ could be achieved for the product
$\Gamma(P^0\to\gamma\gamma)B(P^0\to \mathrm{b}\bar{\mathrm{b}})$.

\begin{figure}
  \begin{center}
    \href{pictures/4/bess-pngb-gg.pdf}{\includegraphics[width=0.7\textwidth]{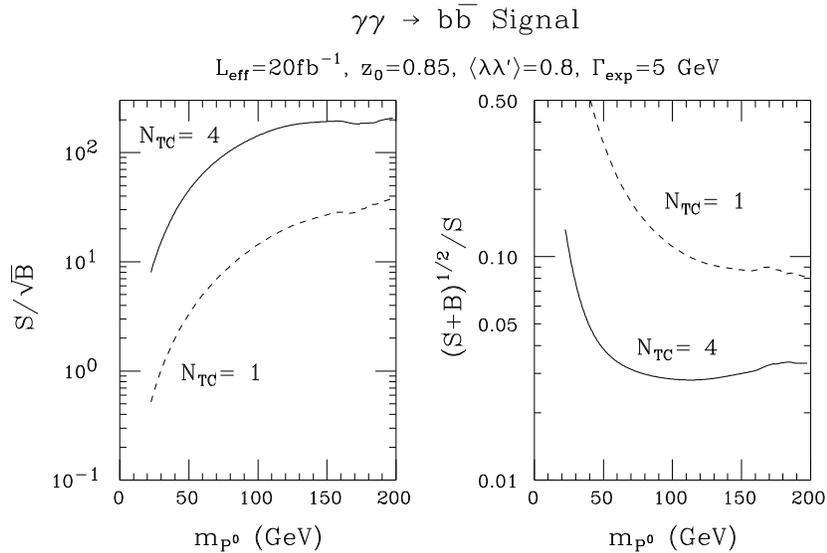}}
    \caption{\label{fig:bess-pngb-gg}%
      The statistical significance $S/\sqrt B$ and error $(S+B)^{1/2}/S$
      for the $\gamma\gamma\to P^0\to \mathrm{b}\bar{\mathrm{b}}$ signal as a function
      of $m_{P^0}$ at a $500\,\mathrm{GeV}$ collider for minimal
      effective luminosity of $20\,\mathrm{fb}^{-1}$.}
  \end{center}
\end{figure}

The $\gamma\gamma$ option at an $\mathrm{e}^+\mathrm{e}^-$ collider is
actually a more robust tool for discovering the $P^0$ than the
$\mathrm{e}^+\mathrm{e}^-$ collision mode.  For $N_{\mathrm{TC}}=4$
and using a setup with a broad $E_{\gamma\gamma}$ spectrum, a $P^0$
signal should be easily detectable in $\gamma\gamma\to P^0\to
\mathrm{b}\bar{\mathrm{b}}$ for masses up to 70\% of the
$\mathrm{e}^+\mathrm{e}^-$ CMS energy with minimal luminosity ($\int
\mathcal{L}_{\text{eff.}}\approx20\,\mathrm{fb}^{-1}$) and a
reasonably accurate measurement of $\Gamma(P^0\to \gamma\gamma)B
(P^0\to \mathrm{b}\bar{\mathrm{b}})$ would be obtained, as illustrated
in Fig.~\ref{fig:bess-pngb-gg}.

Note that the $\gamma\gamma$ discovery mode is strongest at larger
$m_{P^0}$, i.\,e.~where $\mathrm{e}^+\mathrm{e}^-\to \gamma P^0$
becomes less robust.  Once $m_{P^0}$ is known, the $\gamma\gamma$
collision setup can be configured to yield a luminosity distribution
that is strongly peaked at $E_{\gamma\gamma}\approx m_{P^0}$ and, for
much of the mass range of $m_{P^0}\leq 200$~GeV, a measurement of
$\Gamma(P^0\to \gamma\gamma)B (P^0\to \mathrm{b}\bar{\mathrm{b}})$ can
be made with statistical accuracy in the $\mathcal{O}(1\%)$
range~\cite{Casalbuoni:1999fs}, which is competitive with the LHC
accuracy for measuring $\Gamma(P^0\to \mathrm{gg}) B(P^0\to
\gamma\gamma)$.

%%%%%%%%%%%%%%%%%%%%%%%%%%%%%%%%%%%%%%%%%%%%%%%%%%%%%%%%%%%%%%%%%%%%%%%%

  \section{Compositeness}

As one among other physical scenarios, strongly interacting electroweak bosons
at energies of order 1\TeV\  could be interpreted as a signal of composite
substructures of these particles at a scale of $10^{-17} $~cm.  Moreover,
the proliferation of quarks and leptons could be taken as evidence for
possible substructures of the matter particles \cite{701}.  In this
picture, masses and mixing angles are a consequence of the
interactions between a small number of elementary constituents -- in
perfect analogy to the quark/gluon picture of hadrons.  No theoretical
formalism has been set up so far which would reconcile, in a
satisfactory manner, the small masses in the Standard Model with the tiny radii
of these particles which imply very large kinetic energies of the
constituents.  However, the lack of theoretical formalism does not
invalidate the physical picture or its motivation.

\subsection{Contact interactions}

In this agnostic approach, stringent bounds have been derived from
high energy scattering experiments on possible non-zero radii of
leptons, quarks and gauge bosons from $Z$ decay data \cite{702} and
Bhabha scattering \cite{703} in $e^+e^-$ collisions, as well as from
electron-quark and quark-quark scattering at HERA \cite{147A} and
the Tevatron \cite{704}, respectively.  From these analyses the
compositeness scale has been bounded to less than $10^{-17}$\,cm.

\begin{figure}[hbt]
\vspace*{0cm}
\begin{minipage}[t]{7.4cm} {
\begin{center}
\vspace*{-10.05cm}
  \href{pictures/5/ciqq.pdf}{{\includegraphics[height=10cm]{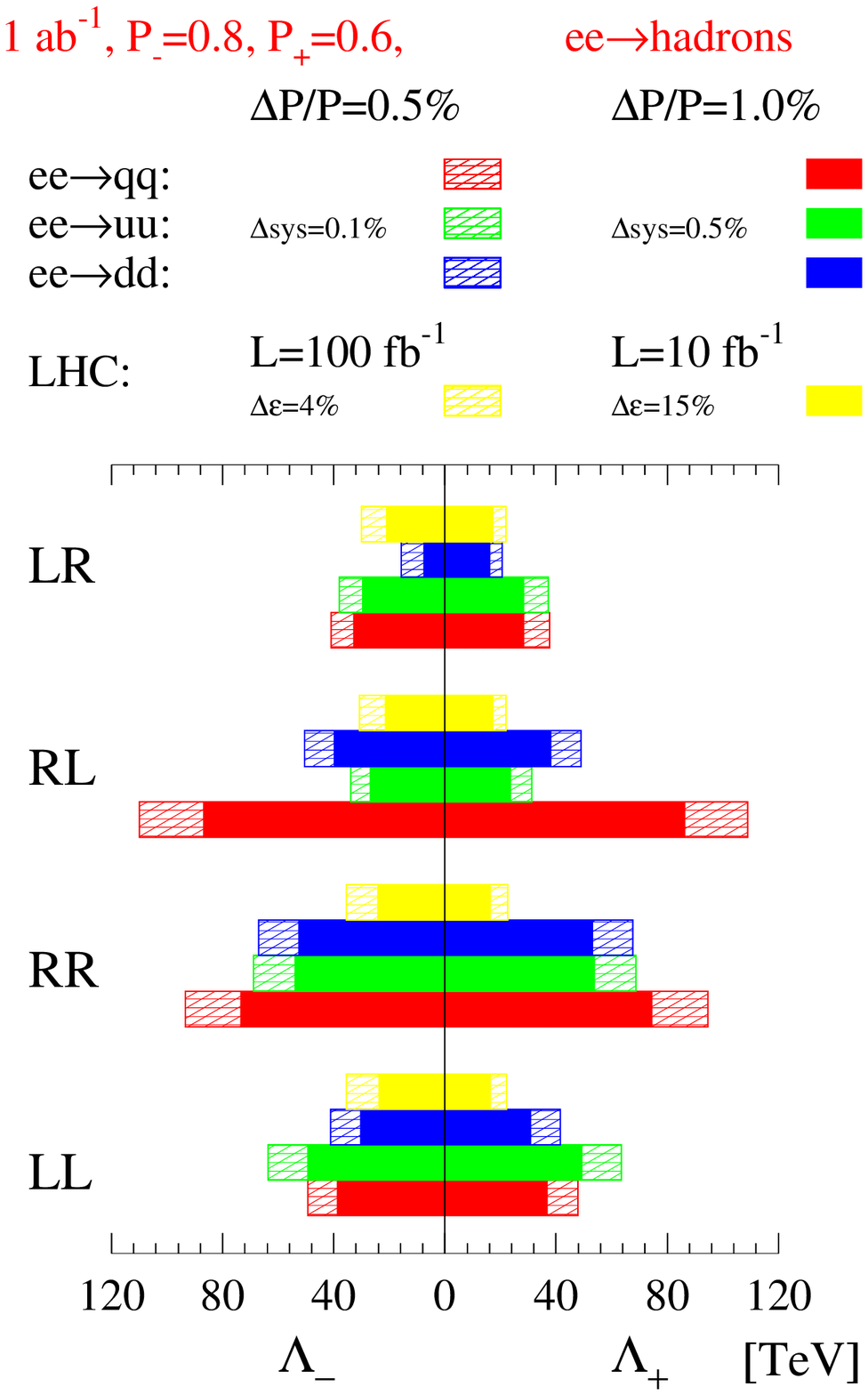}}}
\vspace*{-1.2cm}
\end{center} }
\end{minipage}
\hspace*{0.5cm}
\begin{minipage}[t]{7.4cm} {
\begin{center}
  \href{pictures/5/cimm.pdf}{{\includegraphics[height=10cm]{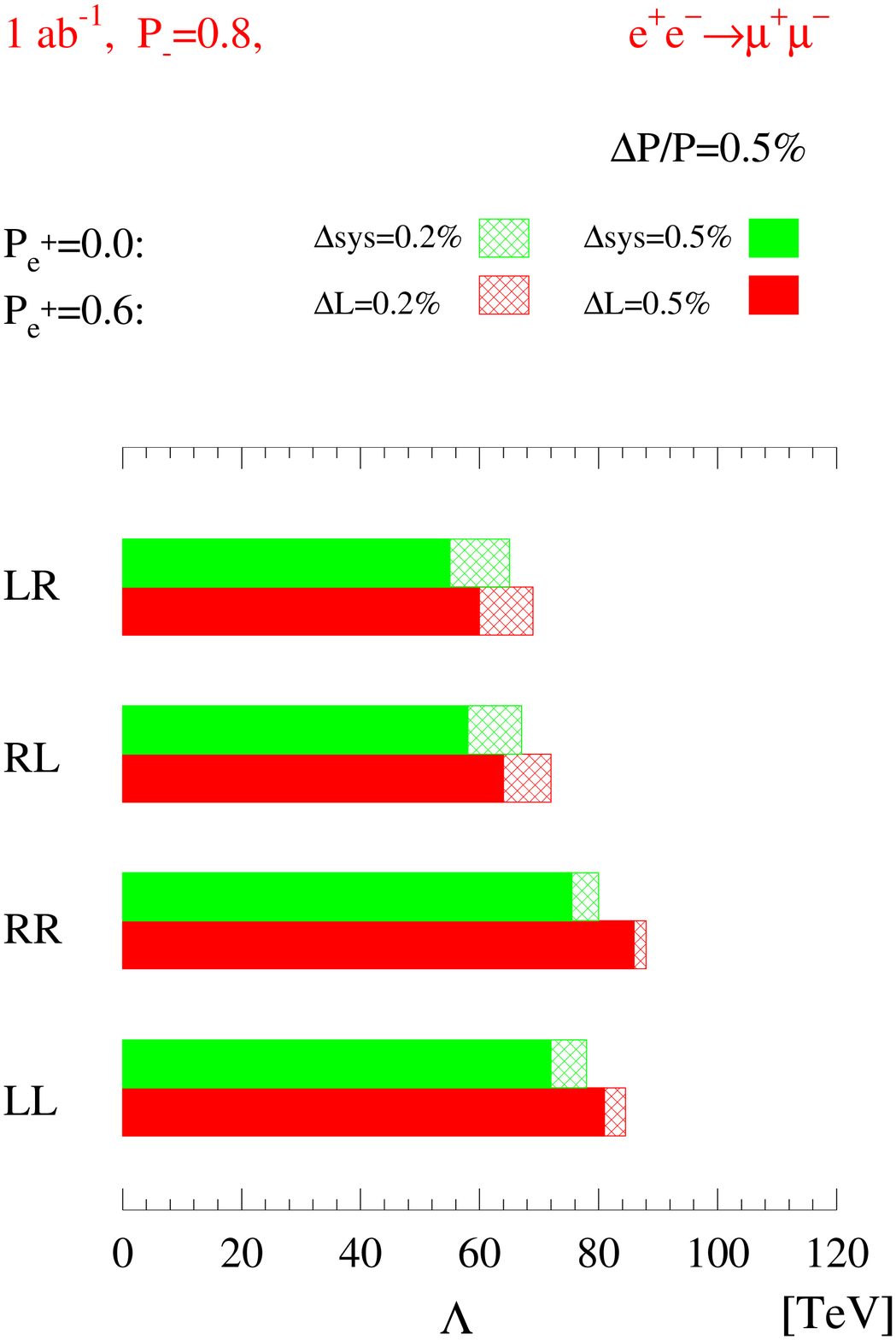}}}
\vspace*{-1.2cm}
\end{center} }
\end{minipage}
\caption{\it \label{fg:ci} Sensitivities (95\% CL) of TESLA to contact
interaction scales $\Lambda$ for different helicities in
$e^+e^-\to$ hadrons (left) and
$e^+e^-\to \mu^+\mu^-$ (right) including polarization of both beams
at $\sqrt{s} = 500$\GeV~\cite{riemanns} [At $\sqrt{s} = 800$\GeV the
limits will be about 30\% larger.].
For hadronic final states the corresponding results of the LHC are shown,
while the LHC cannot probe $e^+e^-\mu^+\mu^-$ couplings.}
\end{figure}

Fermion pair production $e^+ e^- \to f \bar f$ at high energies
provides a very powerful instrument to set limits on fermion
compositeness.  This problem has been studied, based
on four-fermion contact interactions which can be generated by the
exchange of subconstituents \cite{706}:
\begin{equation}
{\cal L}_{eff} = \sum_{i,j=L,R} \eta_{ij} \frac{4 \pi}{\Lambda_{ij}^2}
\bar{e}_i \gamma^{\mu} e_i \cdot \bar{f}_j \gamma_{\mu} f_j
\end{equation}
The strength of the interaction has been set to $g_*^2 / 4\pi = 1$.
The (inverse) contact scales $\Lambda_{ij}$ can be identified, within an
uncertainty of a factor of order 3, with the radius of the fermions.
Detailed experimental simulations have shown that fermion pair
production at TESLA provides a larger sensitivity to compositeness
scales than the LHC, which reaches sensitivities up to about 20--35\TeV~
\cite{lhcci}.
The high polarization that can be achieved for $e^+$ and $e^-$ beams,
gives the TESLA collider another advantage. At c.m.~energies of 500\GeV, the
bounds on fermion compositeness are presented in Fig.~\ref{fg:ci} for
the production of hadrons and muon pairs
for an integrated luminosity of $\int {\cal L} \sim$~$1$\,ab$^{-1}$
\cite{riemanns} (see also \cite{riemanns2,paver}).
For muon pair production the significant effect of positron polarization
\cite{pospol} is also shown. The dependence on the sign of the interference term
between composite and SM contributions is negligible in muon pair
production, and only the average of $\Lambda_+$ and $\Lambda_-$ is
presented in Fig.~\ref{fg:ci}. Increasing the c.m.~energy to 800\GeV
results in an increase of the sensitivities to $\Lambda_+$ and
$\Lambda_-$ by about 30\% \cite{riemanns}.

\subsection{Leptoquarks}

A very exciting prediction of fermion compositeness is the existence
of leptoquarks \cite{708}.  They are novel bound states of
subconstituents which build up leptons and quarks in this scenario.
While the size of the couplings to $\gamma$ and $Z$ bosons follows from the
electroweak symmetries, the Yukawa couplings to leptons and quarks are
bound by experiment \cite{yukawa}. In the interesting mass range, these Yukawa
couplings are expected to be weak. Currently leptoquark masses below about
250, 200 and 100\GeV\  are excluded for 1st, 2nd and 3rd generation
leptoquarks \cite{lqbounds}.

These particles can also occur in grand unified theories.  Moreover, in
supersymmetric theories in which the $R$ parity is broken, 
squarks may be coupled to quarks and leptons, giving rise
to production mechanisms and decay signatures analogous to leptoquarks.
However, whereas leptoquarks {\it sui generis}
disintegrate solely to leptons and quarks, a wide variety of decay
modes is in general expected for squarks, including the
large ensemble of standard supersymmetric decay channels, see e.g. \cite{f163A}.
Since leptoquark bound states in the compositeness picture build up a
tower of states with non-zero spins, the phenomenology of the two
scenarios is clearly distinct.

Leptoquarks can exist in a large variety of states carrying
$[l_iq_j]$ or $[l_i\overline{q}_j]$ quantum numbers $(i, j = L, R)$
and being scalar or vectorial in the simplest representations
\cite{709}.
They can be produced in $e^+e^-$ collisions pairwise, $e^+e^- \to LQ +
\overline{LQ}$,
through $s$--channel $\gamma, Z$ exchange and partly through
$t$--channel $q$ exchange \cite{710,711}.
The particles decay to a charged lepton, or a neutrino, and a jet,
giving rise to visible {\it (a)} $l^+l^-jj$, {\it (b)} $l^{\pm}jj$, and
{\it (c)} $jj$
final states.  Since leptoquarks generate a peak in the invariant
($lj$) mass, they are easy to detect in the cases {\it (a)} and {\it
(b)} up to mass values close to the kinematical limit \cite{712}.

Since leptoquarks carry color, they are produced copiously
\cite{713} in hadron collisions through the subprocesses $gg,
q\bar{q}, qq \to LQ + LQ'$ and $gq \to LQ + l$.  Leptoquarks can
therefore be generated at the LHC with masses up to about
1.5\TeV~\cite{lhclq}. Experiments at $e^+e^-$ colliders are nevertheless
important to identify the electroweak properties of these novel states.
Taking into account statistical errors for leptoquark pair production,
their electroweak couplings to $\gamma$ and $Z$ can be measured at the
level of ${\cal O}(1-10\%)$ \cite{bluemlein}. Combining the processes
$e^+e^-\to LQ~\overline{LQ} \to e^+e^-jj, e\nu jj$ and $\gamma e\to LQ +
X \to ej + X$ [with brems- and beamstrahlung photons], the Yukawa
couplings $\lambda_{L,R}$ can be determined with an accuracy of ${\cal
O} (5\%)$, as can be inferred from Table~\ref{tb:zarnecki}
\cite{zarnecki}.
\begin{table}[hbt]
\renewcommand{\arraystretch}{1.5}
\begin{center}
\begin{tabular}{|l||l|l|l|l|l|l|}
\hline
& \multicolumn{3}{c|}{$\lambda_L/e$} & \multicolumn{3}{c|}{$\lambda_R/e$}
\\ \hline
$M_{S_{1/2}}$ & TESLA & LHC & current & TESLA & LHC & current \\ \hline \hline
330\GeV & $0.150\pm 0.005$ & $<0.18$ & $<0.31$ &
          $0.150\pm 0.005$ & $<0.17$ & $<0.53$ \\
350\GeV & $0.150\pm 0.006$ & $<0.19$ & $<0.33$ &
          $0.150\pm 0.005$ & $<0.18$ & $<0.56$ \\
370\GeV & $0.150\pm 0.006$ & $<0.20$ & $<0.35$ &
          $0.150\pm 0.006$ & $<0.20$ & $<0.59$ \\
390\GeV & $0.150\pm 0.007$ & $<0.21$ & $<0.37$ &
          $0.150\pm 0.006$ & $<0.21$ & $<0.63$ \\
\hline
\end{tabular}
\end{center}
\renewcommand{\arraystretch}{1}
\caption{\label{tb:zarnecki} \it Expected
results of the log-likelihood fit to the $S_{1/2}$ leptoquark angular
distributions for $\sqrt{s} = 800$\GeV~\cite{zarnecki}. $1\sigma$
uncertainties, resulting from the simultaneous fit to all
considered distributions are compared for different leptoquark masses
accessible at TESLA. Also presented are current $95\%$ CL
exclusion limits and limits expected from the Drell--Yan
$e^+e^-$ pair production at the LHC. Leptoquark production
events were generated assuming $\lambda_L=0.15~e, \lambda_R=0~e$, and
$\lambda_L=0~e, \lambda_R=0.15~e$. Luminosity uncertainty is $1\%$.}
\end{table}

For leptoquark masses beyond the kinematical limit TESLA can study
virtual leptoquark effects in fermion pair production processes
$e^+e^-\to f\bar f$. Assuming Yukawa couplings of electromagnetic
strength, TESLA will be sensitive to scalar leptoquark masses of
$m_{LQ}\lsim 3.5$\TeV\  for $\sqrt{s}=800$\GeV\  at high luminosities
including electron (positron) polarization of 80\% (60\%)
\cite{riemanns}, thus extending the reach of the LHC significantly.

\section{Conclusions}

TESLA provides a rich environment for precision tests of theories beyond
the SM. In general the reach in the masses of new particles that can be
found at LHC will be comparable or larger than at TESLA. However, TESLA
can measure the couplings and properties of the novel particles with
high accuracy contrary to the LHC, thus supporting the complementarity of
both colliders.

A major class of extensions beyond the SM consists of adding extra
dimensions to the conventional Minkowski space, which may turn out to be
large with string scales in the\TeV\  range. The LHC will be
sensitive to Kaluza--Klein-graviton emission up to scales $M_D\sim 7.5$
\TeV\ , which denotes the fundamental Planck scale in the higher
dimensional space. While the reach of TESLA is comparable to the LHC,
TESLA will be able to disentangle the scale $M_D$ and the number $\delta$
of extra dimensions from the energy dependence of the graviton emission
cross section unambiguously in contrast to the LHC. Moreover, TESLA can
probe scales up to $M_D\sim 8$\TeV\  via virtual effects due to KK
exchange, leading effectively to fermionic contact interactions. The
angular distributions of fermion pair production processes can be used to
show that the virtual KK states carry spin 2.

If Higgs bosons do not exist, the onset of strong $WW$ interactions will
become visible at high energies. TESLA will probe the threshold region
for strong $WW$ interactions
up to the cut-off scale $\Lambda_*=4\pi v\sim 3$\TeV\ , which defines the
scale of strong electroweak symmetry breaking in this scenario. The
TESLA sensitivity exceeds the LHC sensitivity. Moreover,
the couplings of novel strong vector resonances and
pseudo Nambu-Goldstone bosons can be determined much more accurately than
at the LHC.

In the framework of conventional compositeness theories TESLA will
exceed the sensitivity to the compositeness scales significantly
compared to the LHC. In addition TESLA will allow an accurate
measurement of leptoquark electroweak and Yukawa couplings, when
leptoquarks will have been discovered at the LHC in the accessible
mass range. However, if the leptoquark masses turn out to be too large
for direct production, TESLA can extend the mass reach indirectly
by means of virtual leptoquark exchange in fermion pair production.

The comparison between TESLA and the LHC is summarized in
Table~\ref{tb:eelhcalt}, which clearly confirms the complementarity of
these two colliders.
\begin{table}[hbt]
\renewcommand{\arraystretch}{1.5}
\begin{center}
\begin{tabular}{|l|l|l|}
\hline
Alternative & TESLA & LHC \\ \hline \hline
$KK$ graviton radiation & $M_D\lsim 8$\TeV & $M_D\lsim 7.5$\TeV \\
$KK$ graviton exchange  & $M_D\lsim 8$\TeV & ? \\ \hline
strong $WW$ interactions & $\Lambda_*\gsim \Lambda_{\rm EWSB}$ (3\TeV) &
$\Lambda_*\lsim \Lambda_{\rm EWSB}$ \\
vector resonance couplings & ${\cal O}(0.1-1\%)$ & ${\cal O}(1-10\%)$ \\
Goldstone couplings & ${\cal O}$ (1\%) & ${\cal O}$ (10\%) \\ \hline
leptoquark Yukawa couplings & ${\cal O}(5\%)$ & upper bounds ${\cal
O}(0.2e)$ \\
compositeness scale & $\Lambda\lsim 110$\TeV & $\Lambda\lsim 35$\TeV \\ \hline
\end{tabular}
\end{center}
\renewcommand{\arraystretch}{1}
\caption{\label{tb:eelhcalt} \it Comparison of TESLA and LHC for several
aspects of alternative scenarios beyond the SM.}
\end{table}
  
  \clearpage
  {\renewcommand {\baselinestretch}{1.2}
  {\raggedright{\def\IJP    #1 #2 #3 {{\em Int. J. Phys.} #1:#2, #3}             % Corrected
\def\PLB    #1 #2 #3 {{\em Phys.~Lett.}~B#1:#2, #3}              % Corrected
\def\PA     #1 #2 #3 {{\em Physica}~A#1:#2, #3}                  % Corrected
\def\PRD    #1 #2 #3 {{\em Phys.~Rev.}~D#1:#2, #3}               % Corrected
\def\PRL    #1 #2 #3 {{\em Phys.~Rev.~Lett.} #1:#2, #3}          % Corrected
\def\ZPC    #1 #2 #3 {{\em Z.~Phys.}~C#1:#2, #3}                 % Corrected
\def\APAS   #1 #2 #3 {{\em Acta~Phys.~Austr.~Suppl.} #1:#2, #3}  % Corrected
\def\EPJC   #1 #2 #3 {{\em Eur.~Phys.~J.} C#1:#2, #3}            % Corrected
\def\NPB    #1 #2 #3 {{\em Nucl.~Phys.} B#1:#2, #3}              % Corrected
\def\CPC    #1 #2 #3 {{\em Comput.~Phys.~Commun.} #1:#2, #3}     % Corrected
\def\NPBPS  #1 #2 #3 {{\em Nucl.~Phys.~Proc.~Suppl.} B#1:#2, #3} % Corrected
\def\JHEP   #1 #2 #3 {{\em JHEP} #1:#2, #3}                      % Corrected
\def\PRP    #1 #2 #3 {{\em Phys.~Rept.} #1:#2, #3}

%% bibliography for alternatives section

}}
   \renewcommand {\baselinestretch}{1.0}}
  \cleardoublepage
%------------------------------------------------------------------
\chapter{Precision Measurements}
\label{physics_precision}
%------------------------------------------------------------------

  %
\providecommand{\swsqeffl}    {\sin^2\!\theta_{\rm{eff}}^\ell}
\providecommand{\ALR}{\mbox{$A_{\rm {LR}}$}}
\providecommand{\cAe} {\mbox{$\cal A_{\rm e}$}}
\providecommand{\cAt} {\mbox{$\cal A_{\tau}$}}
\providecommand{\cAf} {\mbox{$\cal A_{\rm f}$}}
\providecommand{\cAq} {\mbox{$\cal A_{\rm q}$}}
\providecommand{\cAl} {\mbox{$\cal A_{\ell}$}}
\providecommand{\cAb} {\mbox{$\cal A_{\rm b}$}}
\providecommand{\cAc} {\mbox{$\cal A_{\rm c}$}}
\providecommand{\cAtau} {\mbox{$\cal A_{\tau}$}}
\providecommand{\MW}      {m_{\mathrm{W}}}
\providecommand{\MZ}      {m_{\mathrm{Z}}}
\providecommand{\MH}      {m_{\mathrm{H}}}
\providecommand{\Mh}      {m_{\mathrm{h}}}
\providecommand{\MT}      {m_{\mathrm{t}}}
\providecommand{\GZ}      {\Gamma_{\mathrm{Z}}}
\providecommand{\GF}         {G_{\mathrm{F}}}
\providecommand{\ppl}  {{\cal P}_{\rm{e}^+}}
\providecommand{\pmi}  {{\cal P}_{\rm{e}^-}}
\providecommand{\ppm}  {{\cal P}_{\rm{e}^\pm}}
\providecommand{\peff}  {{\cal P}_{\rm{eff}}}
\providecommand{\pol}  {{\cal P}}
\providecommand{\Rb}   {R_{\rm{b}}}
\providecommand{\ee}   {\rm{e^+e^-}}
\providecommand{\bb}   {\rm{b\overline{b}}}
\providecommand{\thw}   {\theta_{W}}
\let\gev=\GeV
\let\mev=\MeV
\providecommand{\nb}{\,\mathrm{nb}}
\providecommand{\pb}{\,\mathrm{pb}}
\providecommand{\fb}{\,\mathrm{fb}}
\providecommand{\pbi}{\,\mathrm{pb}^{-1}}
\providecommand{\fbi}{\,\mathrm{fb}^{-1}}
\providecommand{\Cdgz}{\ensuremath{\Delta g^\mathrm{Z}_1}}
\providecommand{\Cdgg}{\ensuremath{\Delta g^\mathrm{\gamma}_1}}
\providecommand{\Cdkz}{\ensuremath{\Delta \kappa_\mathrm{Z}}}
\providecommand{\Cdkg}{\ensuremath{\Delta \kappa_{\gamma}}}
\providecommand{\Ckg}{\ensuremath{\kappa_{\gamma}}}
\providecommand{\Ckz}{\ensuremath{\kappa_{\mathrm{Z}}}}
\providecommand{\Clg}{\ensuremath{\lambda_{\gamma}}}
\providecommand{\Clz}{\ensuremath{\lambda_{\mathrm{Z}}}}
\providecommand{\Cgv}[1]{\ensuremath{g^V_{#1}}}
\providecommand{\Cgz}[1]{\ensuremath{g^Z_{#1}}}
\providecommand{\Cgg}[1]{\ensuremath{g^{\gamma}_{#1}}}
\providecommand{\Ckzt}{\ensuremath{\tilde{\kappa}_\mathrm{Z}}}
\providecommand{\Clzt}{\ensuremath{\tilde{\lambda}_\mathrm{Z}}}
\providecommand{\Ckgt}{\ensuremath{\tilde{\kappa}_{\gamma}}}
\providecommand{\Clgt}{\ensuremath{\tilde{\lambda}_{\gamma}}}
\providecommand{\phmi}{\phantom{-}}
\section{Electroweak Gauge Bosons}
\label{sec:gaugebosons}
The measurement of gauge boson properties has in the past strongly
influenced our knowledge of electroweak interactions. 
The primary goal is to establish the non-Abelian nature of electroweak 
interactions.
With very precise measurements one can constrain new physics at scales
above the direct reach of the machine through loop effects.
Alternatively, small effects
from operators in an effective Lagrangian, that
are suppressed by $(s/\Lambda)^n$, can be measured, 
where $\Lambda$ is the scale where new physics sets in. 
Also for the extrapolation of couplings to high scales, to test
theories of grand unification, very high precision is needed.
At TESLA there are mainly two ways to study properties of W- and Z-bosons:
\begin{itemize}
\item One can study the couplings amongst gauge bosons.
  These couplings are especially sensitive to models of strong
  electroweak symmetry breaking  and are most precisely measured at the
  highest possible energies.
\item The masses and couplings of the W and Z, especially the
  effective weak mixing angle in Z decays, $\swsqeffl$, can be measured,
  similar to LEP and SLC, however with much higher luminosity and
  polarised beams.
%\item An improved measurement of the couplings of the
%  electroweak gauge bosons to quarks will provide
%  additional insight into the flavour physics of the
%  CKM-matrix.   The experimental methods are complementary
%  to the b-factories and hadron colliders and can provide
%  independent consistency checks.
\end{itemize}
In addition, an improved measurement of the couplings of the
electroweak gauge bosons to quarks will provide
further insight into the flavour physics of the
CKM-matrix. The experimental methods are complementary
to the b-factories and hadron colliders and can provide
independent consistency checks.
\subsection{W-production at high energies}
At high energies W bosons are produced either in pairs,
$\ee \rightarrow {\rm W}^+{\rm W}^-$, or singly via 
$\ee \rightarrow \rm{We}\nu$. W-pair production falls, far above
threshold, like $1/s$ while single W-production rises
logarithmically with the energy. 
At TESLA-energies both cross sections are of about the same size.

\begin{figure}[htb]
\begin{center}
\href{pictures/6/feyn_ww.pdf}{{\includegraphics[height=3cm]{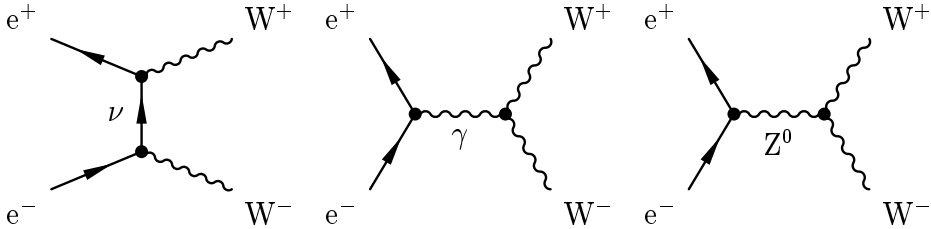}}}
\end{center}
\caption{Feynman graphs for the production of W-pairs in $\ee$-annihilation.}
\label{fig:wwfeyn}
\end{figure}

The Feynman diagrams for on-shell W-pair
production are shown in Fig.~\ref{fig:wwfeyn}. Due
to the $(V\!-\!A)$ nature of the charged current couplings,
the contribution of the $t$-channel $\nu$-exchange
diagram vanishes for right-handed electrons or
left-handed positrons.  Therefore it can be switched off
completely by polarising one of the beams appropriately.
Its contribution can also be enhanced by a
factor two or four by polarising one or both beams in the
opposite way.  
For energies that are much higher
than the weak boson-masses, the combined Z
and~$\gamma$ exchange can be replaced by the neutral
member of the W weak isospin triplet,
because the orthogonal combination corresponding to the
weak hypercharge boson does not couple to the
$\mathrm{W}^\pm$.  Therefore the coupling to the
electrons and positrons is also purely~$(V\!-\!A)$ at high
energies.  Already at TESLA energies, the cross section
for right-handed electrons is suppressed by at least a
factor of ten relative to left-handed electrons for all
polar angles.

\begin{figure}[htb]
\begin{center}
\href{pictures/6/feyn_sw.pdf}{{\includegraphics[height=4cm]{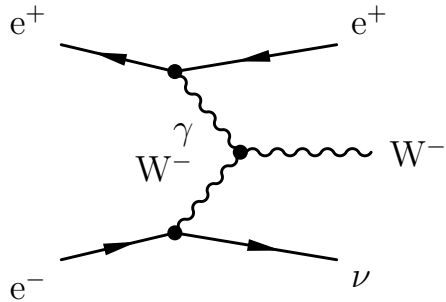}}}
\end{center}
\caption{Dominating Feynman graph for single W production in 
  $\ee$-annihilation.}
\label{fig:swfeyn}
\end{figure}

Single W production is dominated by photon-W fusion (see
Fig. \ref{fig:swfeyn}). Since the helicity only matters for the beam
that radiates the W, varying the electron polarisation can switch
off or double single $\rm{W}^-$ production while varying the positron
polarisation affects single $\rm{W}^+$ production in the same way.

\begin{figure}[htb]
\begin{center}
\href{pictures/6/sigw_yfs.pdf}{{\includegraphics[height=8.cm,bb=22 10 468 517]{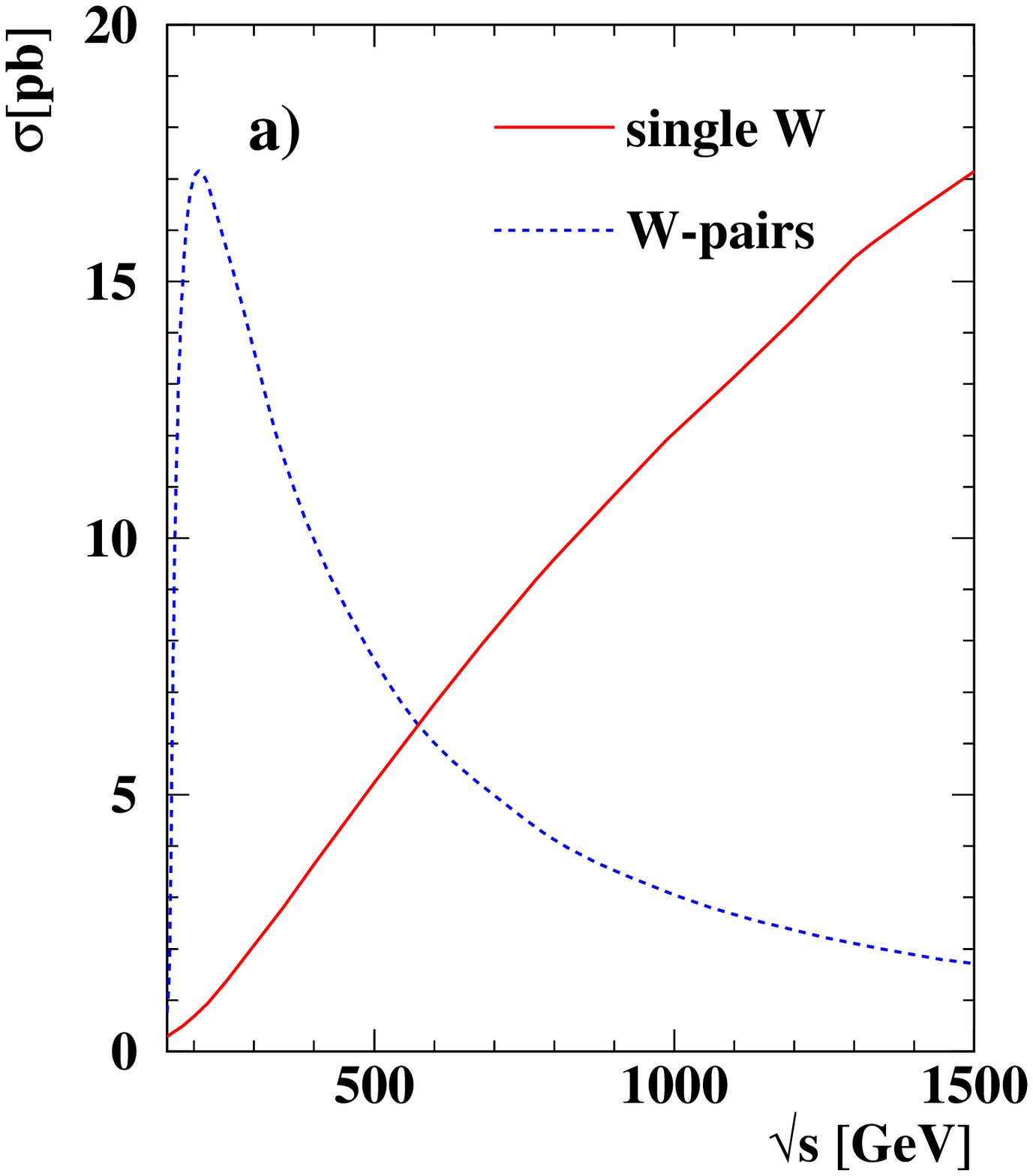}}}
\href{pictures/6/sigw_pol.pdf}{{\includegraphics[height=8.cm,bb=31 10 445 517]{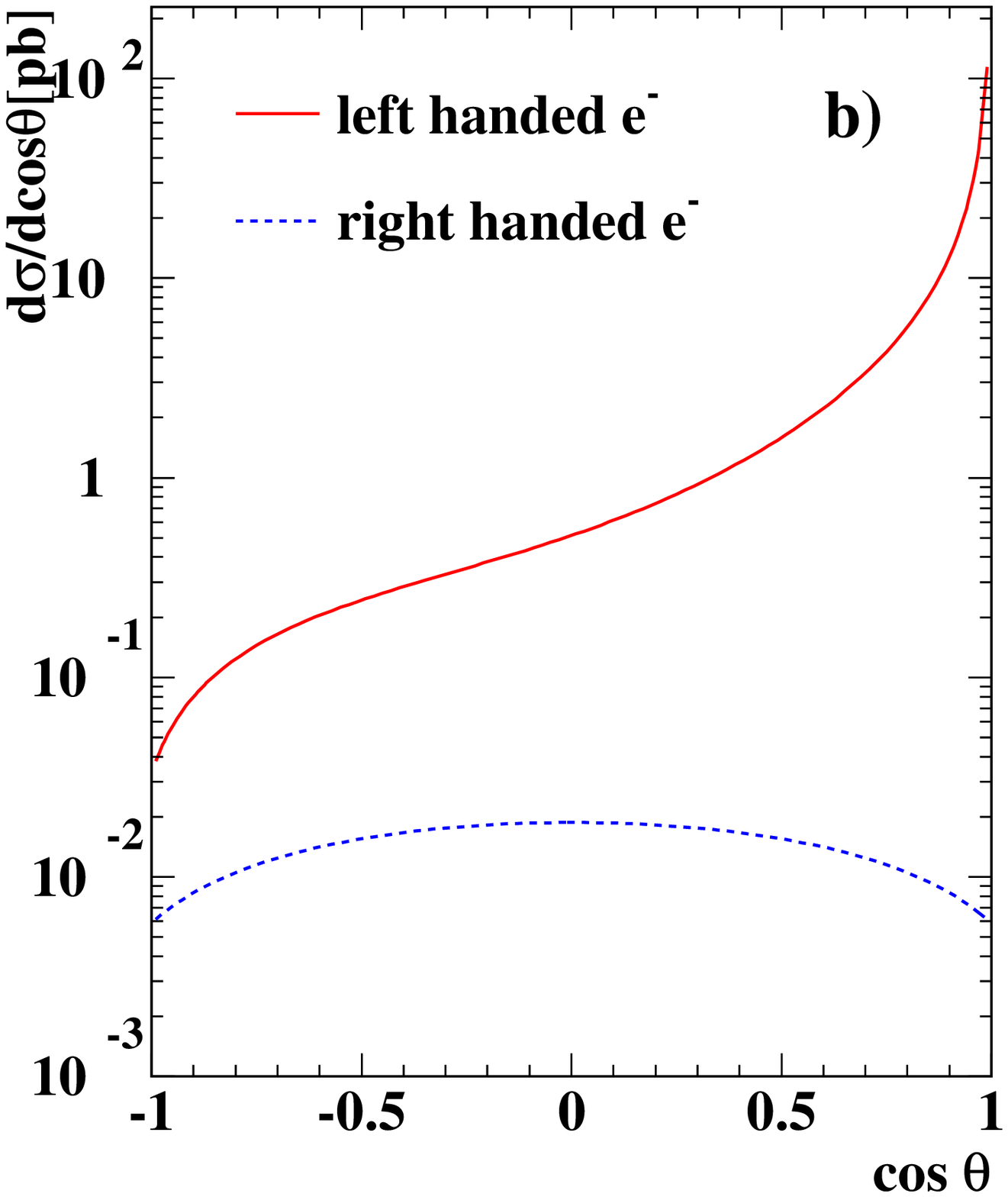}}}
\end{center}
\caption{a): Total cross section for single W \cite{ref:comphep}
  and W pair production \cite{YFSWW3} as
  a function of the centre of mass energy.
  b): Differential cross
  section for W-pair production for different beam polarisation.
  }
\label{fig:wsigma}
\end{figure}

Figure \ref{fig:wsigma} a) shows the total cross section as a function
of the centre of mass energy for both processes and Fig.
\ref{fig:wsigma} b) the differential cross section for W-pair
production for the two electron helicities at $\sqrt{s} = 500\GeV$. 

In addition, at a $\gamma \gamma$- and e$\gamma$-collider the processes 
$\gamma \gamma \rightarrow {\rm W}^+{\rm W}^-$ 
and ${\rm{e}}^- \gamma \rightarrow {\rm W}^- \nu$ are 
accessible. The first process proceeds via W-exchange in the $t$-channel
while the second one involves the vertex shown in
Fig. \ref{fig:swfeyn}. The cross sections for these two processes are large
($\sim 80\,\pb$ for $\gamma\gamma$ and $\sim 30 \, \pb$ for e$\gamma$
at $500\GeV$), however they occur predominantly at a lower scale.

% All processes are sensitive to the triple gauge couplings WWV,
% V=Z,$\gamma$. 
% If operators up to dimension six are considered, the effective
% Lagrangian for the WWV couplings can be written as \cite{ref:wwlag}:
% \begin{eqnarray*}
% i {\cal L}_{\rm{eff}}^{WWV} & = & g_{WWV} \cdot [ \\
%  & & g_1^V V^\mu \left( W^-_{\mu\nu}W^{+\nu} - W^+_{\mu\nu}W^{-\nu} 
% \right) + 
% \kappa_v W^+_{\mu}W^-_{\nu}V^{\mu\nu} + 
% \frac{\lambda_V}{m_W^2} V^{\mu\nu} W_\nu^{+\rho} W^-_{\rho\mu} + \\
%  & & i g_5^V \epsilon_{\mu\nu\rho\sigma} \left(
% \left(\partial^\rho W^{-\mu}\right) W^{+\nu} - %\right.\\
% % & & \ \ \ \ \ \ \ \ \ \ \ \ \ \  
% %\left.
%  W^{-\mu} \left(\partial^\rho W^{+\nu}\right) \right)V^\sigma + \\
%  & & i g_4^V W^-_\mu W^+_\nu 
% \left( \partial^\mu V^\nu + \partial^\nu V^\mu \right) -   
% \frac{\tilde{\kappa}_V}{2} W^-_\mu W^+_\nu 
% \epsilon^{\mu\nu\rho\sigma} V_{\rho\sigma} -   \\
%  & & \frac{\tilde{\lambda}_V}{2m_W^2} W^-_{\rho\mu} W^{+\mu}_{\ \ \ \ \nu}
% \epsilon^{\nu\rho\alpha\beta} V_{\alpha\beta} ]
% \end{eqnarray*}
% with $\ g_{WW\gamma} = e, \ g_{WWZ} = e \cot \thw$, $\thw$ being the weak 
% mixing angle,
% and $V_{\mu\nu} = \partial_\mu V_\nu - \partial_\nu V_\mu$.
% For zero momentum transfer, gauge invariance requires 
% $g_1^\gamma (q^2=0) = 1, \, g_5^\gamma (q^2=0) = 0$.
%
All processes are sensitive to the triple gauge
couplings~$\mathrm{WWV}$, $\mathrm{V}=\mathrm{Z},\gamma$, which are
conventionally parameterised as~\cite{ref:wwlag} 
% \begin{multline}
%   \label{eq:hagiwara}
%   L_{\mathrm{WWV}} / g_{\mathrm{WWV}} = \\
%   \mathrm{i} g_1^{\mathrm{V}}  V^\mu \left( W^{-,\nu} W^{+}_{\mu\nu}
%     - W_{\mu\nu}^- W^{+,\nu} \right)
%   + \mathrm{i} \kappa_{\mathrm{V}} W_\mu^- W_\nu^+ V^{\mu \nu }
%   + \mathrm{i} \frac{\lambda_{\mathrm{V}}}{m_{\mathrm{W}}^2}
%   W_{\lambda\mu}^- W^{+,\mu}_{\hphantom{+,\mu}\nu} V^{\nu\lambda} \\
%   + g_4^{\mathrm{V}} W_\mu^- W_\nu^+
%   \left( \partial^\mu V^\nu + \partial^\nu V^\mu \right)
%   + g_5^{\mathrm{V}} \epsilon^{\mu\nu\lambda\rho}
%   \left( W_\mu^- \partial_\lambda W_\nu^+ -
%     \partial_\lambda W_\mu^- W_\nu^+ \right) V_\rho \\
%   + \mathrm{i} \tilde \kappa_{\mathrm{V}} W_\mu^- W_\nu^+ \tilde V^{\mu\nu}
%   + \mathrm{i} \frac{\tilde\lambda_{\mathrm{V}}}{m_{\mathrm{W}}^2} 
%   W_{\lambda\mu}^-
%   W^{+,\mu}_{\hphantom{+,\mu}\nu} \tilde V^{\nu\lambda}\,,
% \end{multline}
\begin{eqnarray}
  \nonumber
  L_{\mathrm{WWV}} & = & g_{\mathrm{WWV}} [\\
  \nonumber
  & & \mathrm{i} g_1^{\mathrm{V}}  V_\mu \left( W^{-}_{\nu} W^{+}_{\mu\nu}
    - W_{\mu\nu}^- W^{+}_{\nu} \right)
  + \mathrm{i} \kappa_{\mathrm{V}} W_\mu^- W_\nu^+ V_{\mu \nu }
  + \mathrm{i} \frac{\lambda_{\mathrm{V}}}{m_{\mathrm{W}}^2}
  W_{\lambda\mu}^- W^{+}_{\mu\nu} V_{\nu\lambda} \\
  \nonumber
  & & + \, g_4^{\mathrm{V}} W_\mu^- W_\nu^+
  \left( \partial_\mu V_\nu + \partial_\nu V_\mu \right)
  + g_5^{\mathrm{V}} \epsilon_{\mu\nu\lambda\rho}
  \left( W_\mu^- \partial_\lambda W_\nu^+ -
    \partial_\lambda W_\mu^- W_\nu^+ \right) V_\rho \\
  \label{eq:hagiwara}
  & & + \,  \mathrm{i} \tilde \kappa_{\mathrm{V}} W_\mu^- W_\nu^+ 
       \tilde V_{\mu\nu}
  + \mathrm{i} \frac{\tilde\lambda_{\mathrm{V}}}{m_{\mathrm{W}}^2} 
  W_{\lambda\mu}^-
  W^{+}_{\mu\nu} \tilde V_{\nu\lambda}]\,,
\end{eqnarray}
using the antisymmetric combinations $V_{\mu\nu} = \partial_\mu V_\nu -
\partial_\nu V_\mu$ and their duals ${\tilde
  V}_{\mu\nu}=\epsilon_{\mu\nu\rho\sigma}V_{\rho\sigma}/2$.  
The overall coefficients are $g_{\mathrm{WW}\gamma}=e$ and
$g_{\mathrm{WWZ}}=e\cot\theta_W$ with $\theta_W$ being the weak mixing
angle.  
%
% With the couplings as momentum dependent form factors,
% (\ref{eq:hagiwara}) parameterises the most general vertex, that couples
% three vector bosons.  In a systematic analysis
% (\ref{eq:hagiwara}) should be derived from an effective Lagrangian.
% In the lowest orders (dimension four and six) of a systematic
% expansion in the energy, constant values for the coupling arise.
% If the terms in effective Lagrangian are properly organised
% according to their gauge transformation properties, triple
% couplings will in general be related to quartic couplings.
%
With the couplings as momentum dependent form factors,
eq. (\ref{eq:hagiwara}) parameterises the most general vertex, that couples
three vector bosons.  
%In a systematic analysis
%the triple gauge coupling part of the effective Lagrangian
%has to be arranged in the form of eq. (\ref{eq:hagiwara}).
In a systematic analysis the coefficients of the triple gauge
couplings in eq.~(\ref{eq:hagiwara}) are related to the coefficients of an
effective Lagrangian and the latter can be inferred from measurements
of the former.
Keeping only the lowest orders (dimension four and six) of a systematic
expansion in the energy, constant values for the coupling arise.
If the terms in the effective Lagrangian are properly organised
according to their gauge transformation properties, triple
couplings will in general be related to quartic couplings.

Electromagnetic gauge invariance requires that
$g_1^{\gamma}(q^2=0)=1$ and $g_5^{\gamma}(q^2=0)=0$ at zero momentum transfer.
In the Standard Model one has {$g_1^V = \kappa_V = 1$},
all other couplings are equal to zero. 

Amongst the different couplings $g_1,\,\kappa$ and $\lambda$ are C-
and P-conserving, $g_5$ is C and P-violating, but CP-conserving while 
$g_4,\,\tilde{\kappa},\,\tilde{\lambda}$ violate CP.

\begin{figure}[htb]
\begin{center}
\href{pictures/6/tgc_sig.pdf}{{\includegraphics[width=7.cm,bb=18 10 449 515]{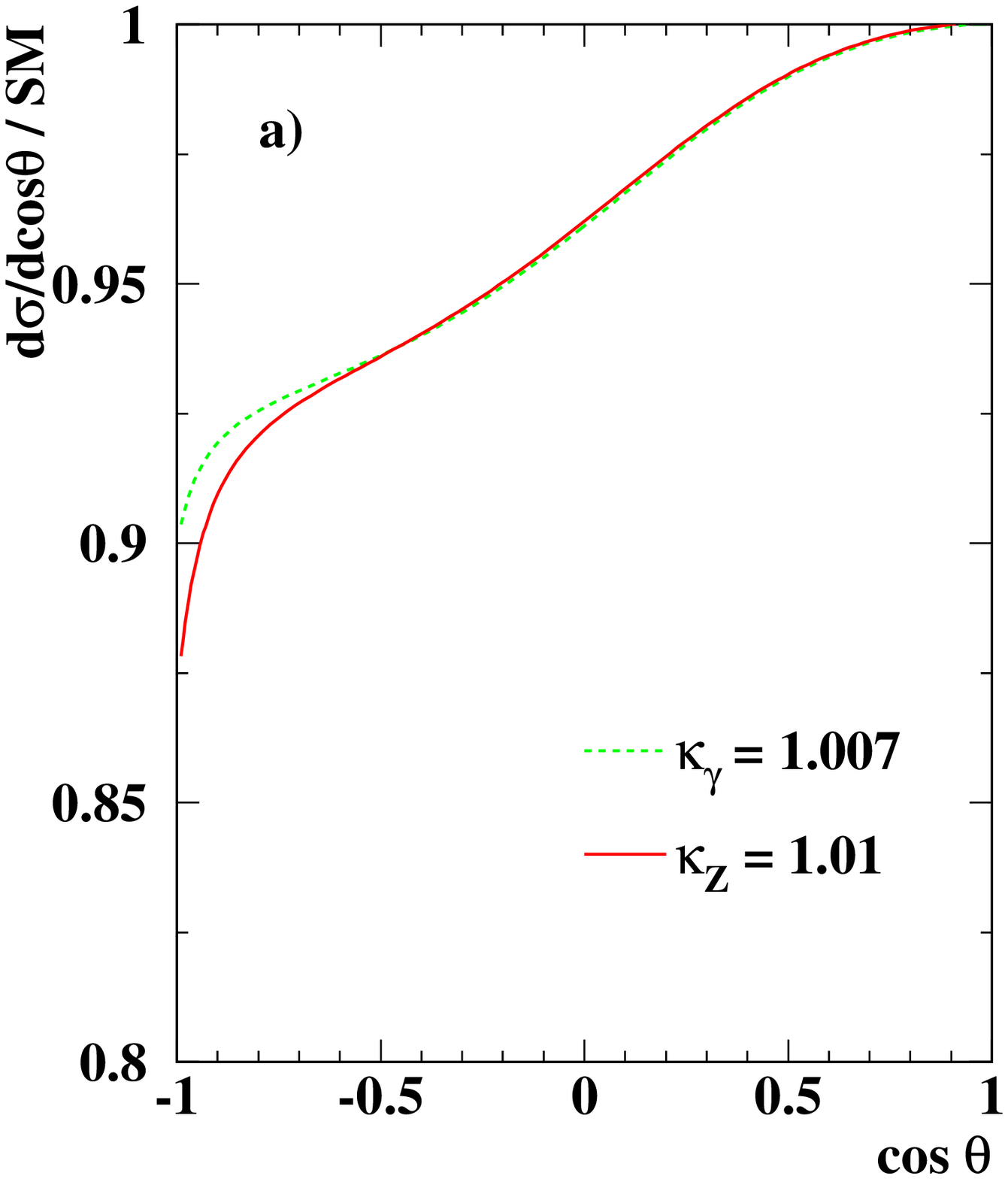}}}
\href{pictures/6/tgc_alr.pdf}{{\includegraphics[width=7.cm,bb=18 10 449 515]{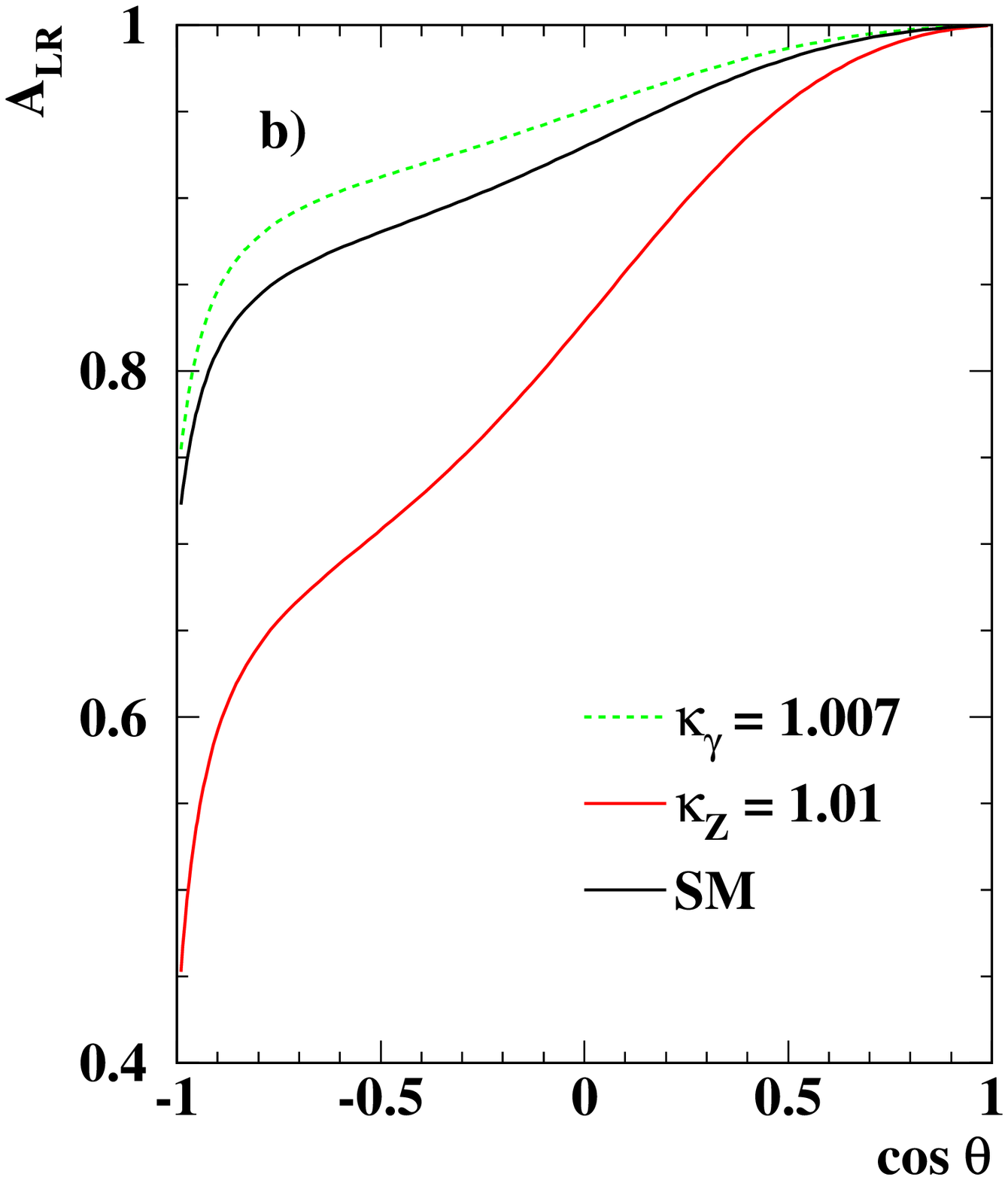}}}
\end{center}
\caption{Ratio of the differential cross section for W-pair production
  to the Standard Model prediction (a) and left-right asymmetry for
  this process
  (b) as a function of the W-production angle for anomalous 
  $\Ckg$ or $\Ckz$.
  }
\label{fig:tgcpol_pr}
\end{figure}

While single W production is basically sensitive to WW$\gamma$
couplings only, W pair production always involves a mixture of
WW$\gamma$ and WWZ couplings. However, as it is demonstrated in Fig.
\ref{fig:tgcpol_pr}, the two types of couplings can be disentangled
with the help of beam polarisation.

%Since the gauge cancellations that are required to maintain unitarity
%are destroyed by any anomalous coupling, these couplings have to
%vanish at high energies. This is often parameterised by a form factor
%of the type $x' = \frac{x} {\left(1 + q^2/\Lambda^2 \right)^n}$ with $n>0.5$ 
%for $\Delta \kappa$ $n>1$ for $\lambda$. For W-pair production the
%scale is fixed to $q^2=s$ so that one does not need to care about the
%form factor in the analysis. For single W production the effective
%scale of the WW$\gamma$-vertex varies but is generally quite low.

For the analysis of triple gauge couplings in W-pair production in principle
five different observables are available:
\begin{itemize}
\item the polar angle of the outgoing $\rm{W}^-$ with respect to the
  incoming $\rm{e}^-$ direction, $\Theta_{\rm W}$;
\item the polar angle of the fermion with respect to the W flight
  direction in the W rest frame for both W-bosons, $\theta^*$, this variable is
  sensitive to the longitudinal polarisation of the W;
\item the azimuthal angle of the fermion in the W-beam plane for both
  Ws, $\phi^*$, sensitive to the transverse polarisation.
\end{itemize}
Not all of the above variables can be determined unambiguously
in all $\mathrm{W}$ decays.
For about 44\% of the W-pairs one W decays leptonically and the other one into
two jets. In these events the ${\rm W}^-$ polar angle can be reconstructed
from the jet momenta and the lepton charge. The decay angles of the
leptonically decaying W can be reconstructed without 
ambiguity and for the hadronically decaying W,
since the quark and the antiquark cannot be distinguished, with the ambiguity 
$(\cos \theta^*,\phi^*) \leftrightarrow (-\cos \theta^*,\phi^*+\pi)$.
This event sample has the by far highest sensitivity
to gauge boson couplings. 46\% of the W-pairs decay into four jets. 
If the correct jet pairing is found one still has
the sign ambiguity for the decay angles of both W bosons plus, since the
W-charges cannot be determined, the ambiguity
$\pm \cos \Theta_{\rm W}$ for the production angle, so that these events add
only little to the sensitivity.
The remaining 11\% decay fully leptonically. In about half of them one
lepton is a $\tau$, so that because of the additional neutrinos, too
little information is available. For the other half all information
can be calculated with a twofold ambiguity. However the additional
statistics from these events is so small that the analysis of the
mixed decays alone gives a good estimate of the total sensitivity.

Since it is inconvenient to work with five independent
variables, always some variable reduction is used. For the
TESLA studies the spin density matrix has been applied which obtains
close to optimal results \cite{ref:spindens}.

At TESLA mixed decays of W-pairs can be selected with very high
efficiency and low background. The large forward peak,
that is partially lost in the beampipe,
is dominated by $t$-channel neutrino exchange and is thus not
sensitive to anomalous couplings (see Fig. \ref{fig:tgcpol_pr}). 
Due to the large boost, the
W-production angle can be measured with significantly higher accuracy
than at LEP. Also the resolution of the W-decay angles is good enough
that detector effects can be almost neglected.

In the spin density formalism the signals from the C,P,CP-violating
couplings are clearly separated from the C,P-conserving ones. For
example the imaginary parts of the off-diagonal elements of the spin-density
matrix are non-zero
only if CP-violating couplings are present. 
Because of the negligible correlations between the different sets of couplings
the fits can be done separately.
\begin{figure}[th!]
\begin{center}
\begin{tabular}{cc}
\href{pictures/6/tgc_g1_kz.pdf}{{\includegraphics[height=7cm,bb=2 8 542 528]{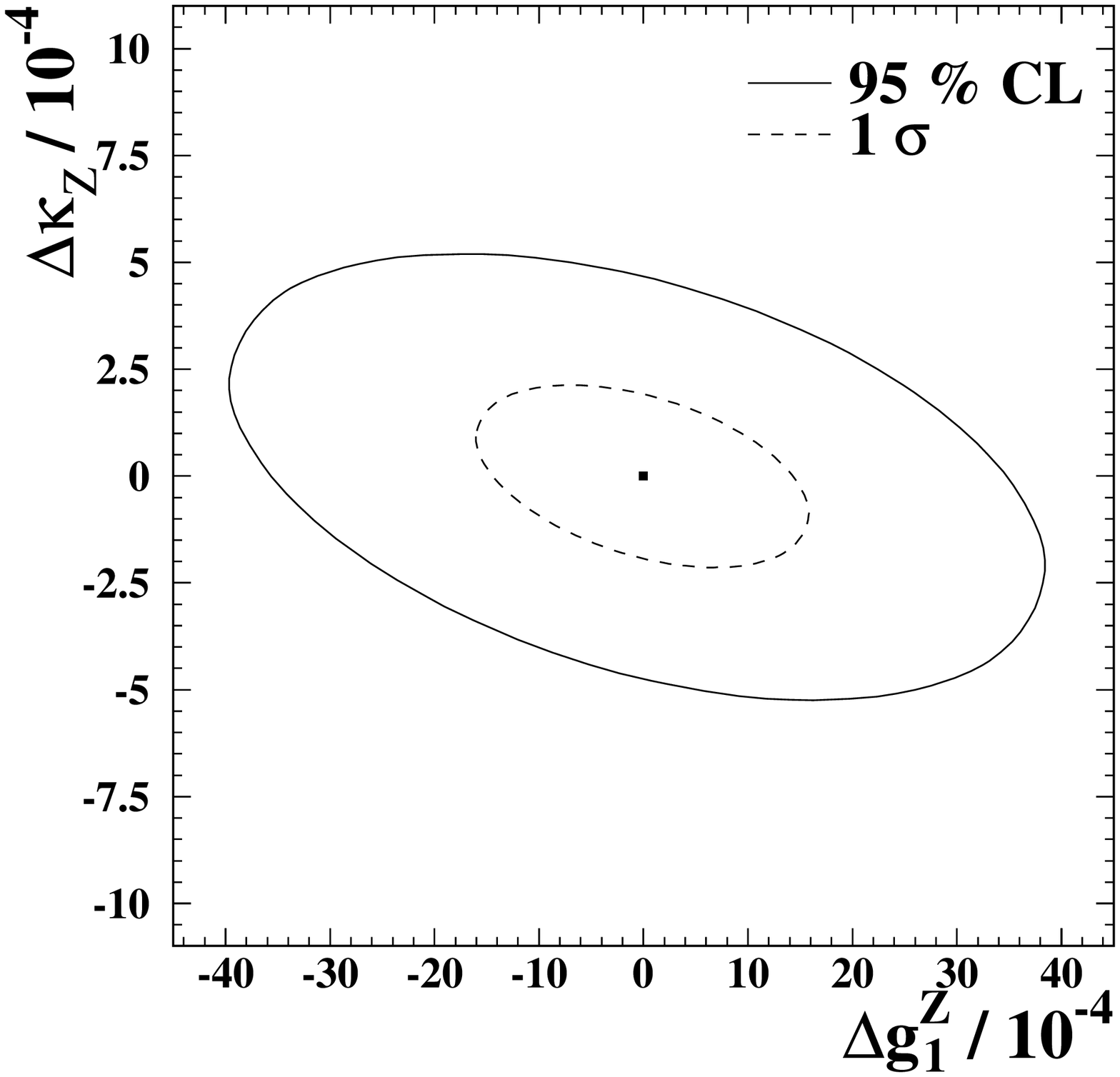}}}&
\href{pictures/6/tgc_g1_lz.pdf}{{\includegraphics[height=7cm,bb=2 8 542 528]{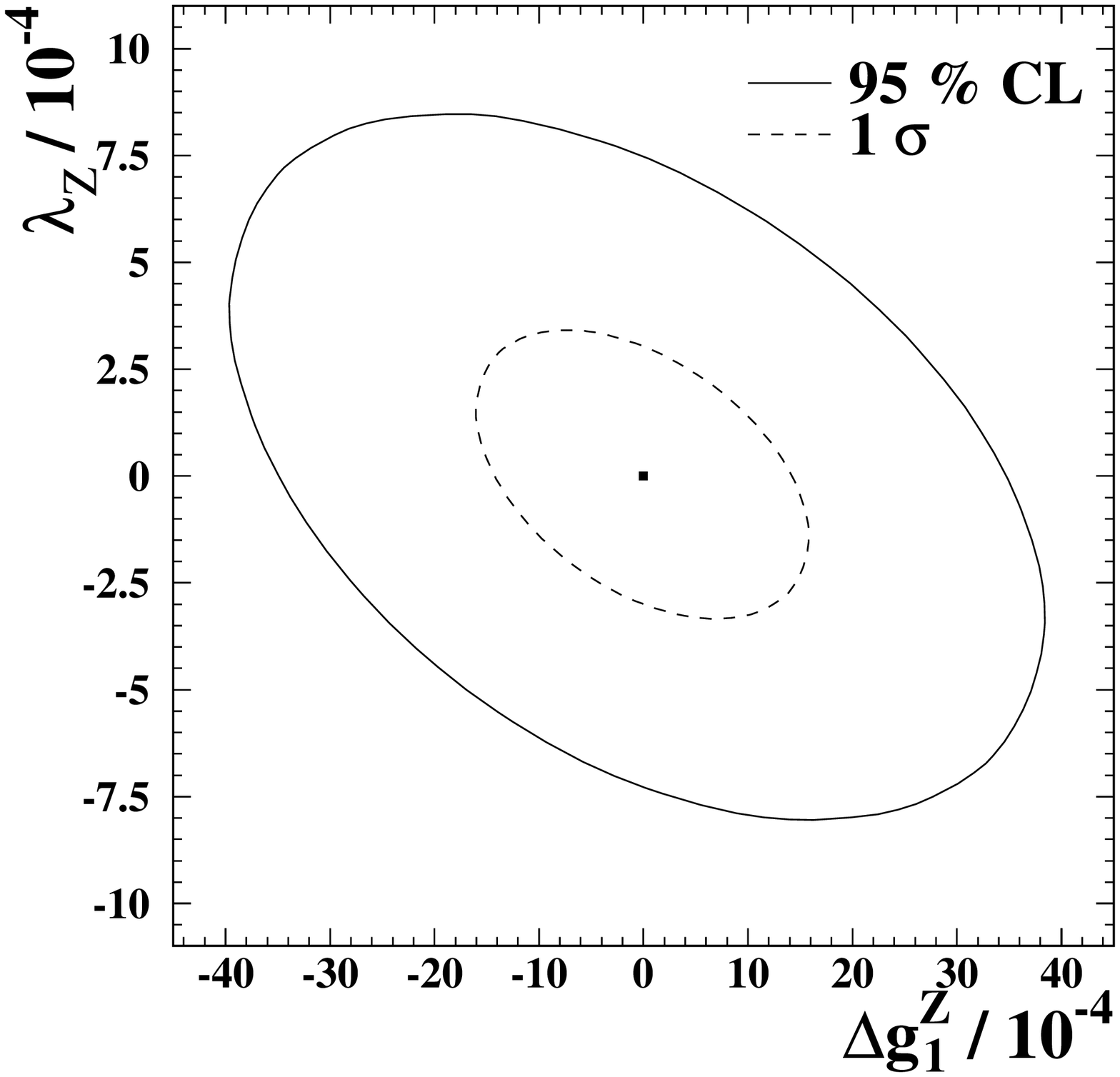}}}\\
\href{pictures/6/tgc_kg_kz.pdf}{{\includegraphics[height=7cm,bb=2 8 542 528]{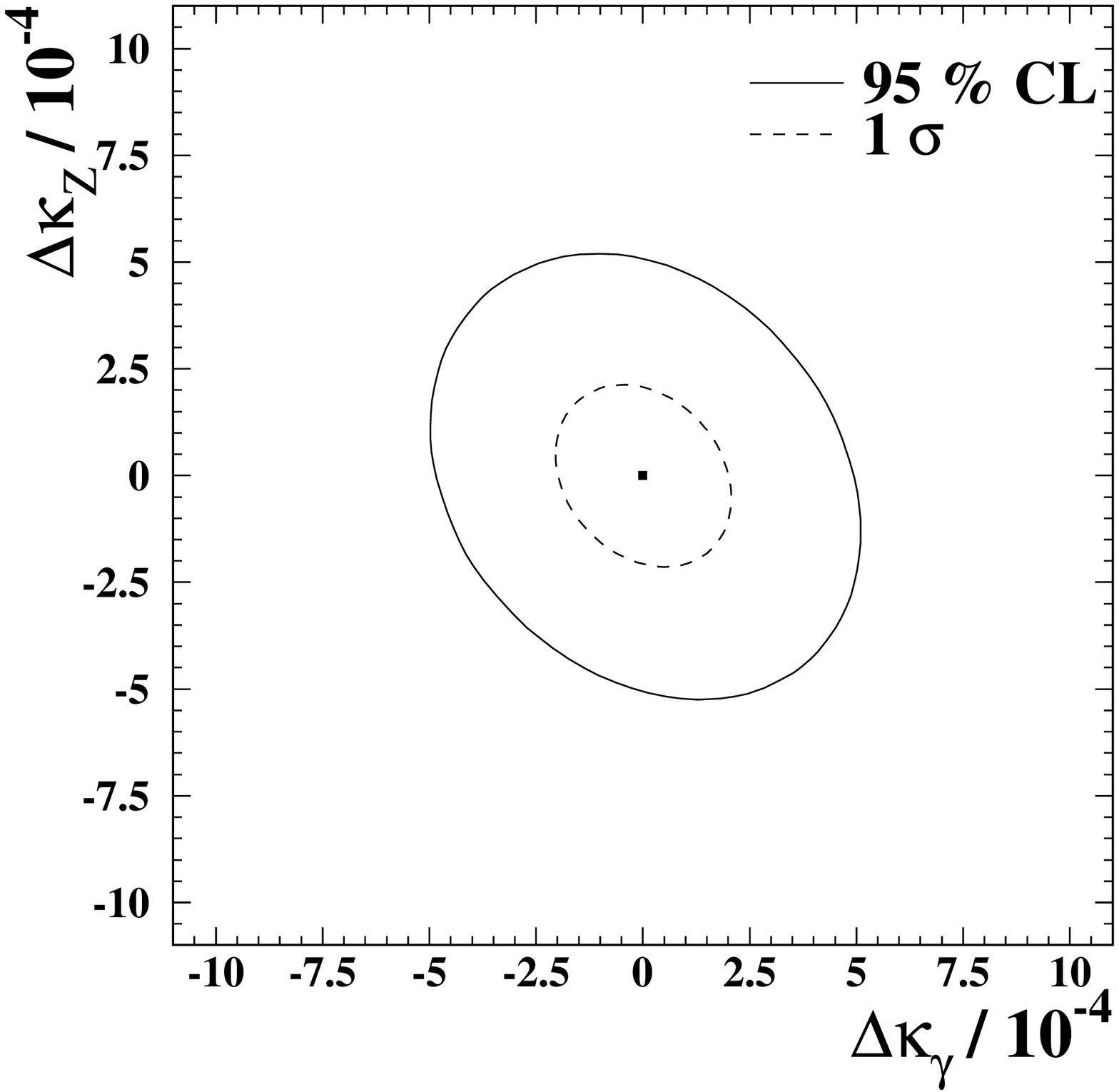}}}&
\href{pictures/6/tgc_lg_lz.pdf}{{\includegraphics[height=7cm,bb=2 8 542 528]{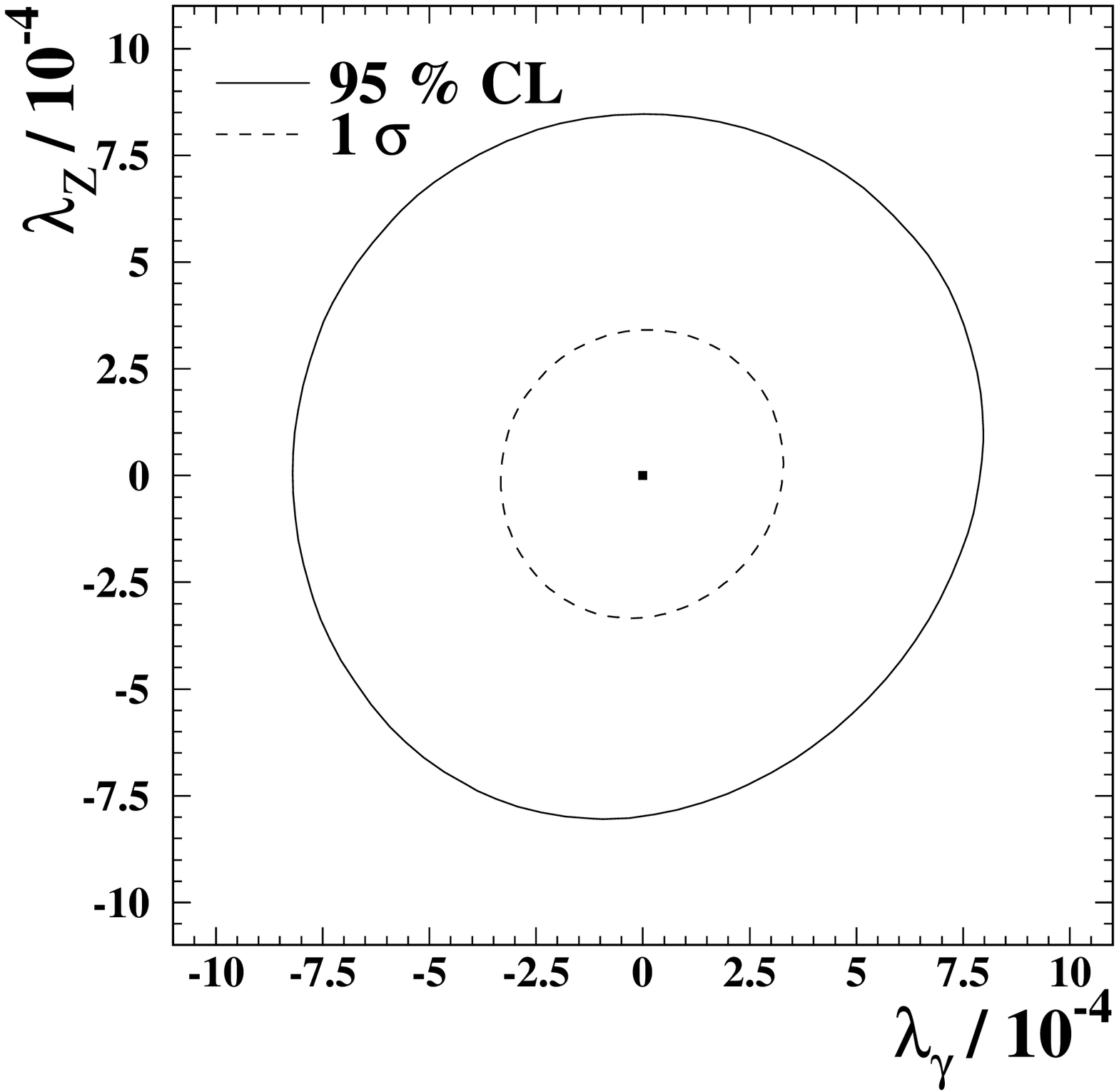}}}
\end{tabular}
\end{center}
\caption{$1 \sigma$ and $95\%$ c.l. (2D) contours for \Cdgz-\Cdkz,
  \Cdgz-\Clz, \Cdkg-\Cdkz and \Clg-\Clz in the 5-parameter fit 
  ($\sqrt{s}=800\GeV, \, {\cal L} = 1000\,\fbi,\, \pmi=0.8,\,\ppl=0.6$).
  For the combinations not shown the correlations are small.
}
\label{fig:fiveparfit} 
\end{figure}

Although with beam polarisation all five C,P-conserving couplings can 
be determined simultaneously, to test certain models it is still useful to 
perform single
parameter fits with all other couplings fixed to the values predicted 
by the Standard Model.
In these fits it is also reasonable to impose the relations amongst the 
parameters suggested by $\rm{SU}(2) \times \rm{U}(1)$ invariance
\cite{ref:tgcsu}:
\begin{eqnarray*}
%\Delta \kappa_\gamma & = & - \frac{\cos^2 \thw}{\sin^2 \thw}
\Delta \kappa_\gamma & = & - \cot^2 \thw
 ( \Delta \kappa_{\rm Z} - g_1^{\rm Z} ) \\
\lambda_\gamma & = & \lambda_{\rm Z}.
\end{eqnarray*}
Table \ref{tab:singtgc} shows the results of the different single parameter
fits including the C or P violating couplings
for $500\,\fbi$ at $\sqrt{s}=500\GeV$ and $1000\,\fbi$ at $\sqrt{s}=800\GeV$.
For both cases an electron polarisation of 80\% and a positron polarisation of
60\% is assumed.
Figure \ref{fig:fiveparfit} shows the results of the five-parameter fit for
$\sqrt{s} = 800\GeV$. Only the combinations with large correlations are shown.

\begin{table}
\begin{center}
\begin{tabular}[c]{|c|c|c|}
\hline
coupling & \multicolumn{2}{|c|}{error $\times 10^{-4}$} \\
\cline{2-3}
         & $\sqrt{s}=500\GeV$ & $\sqrt{s}=800\GeV$ \\
\hline
\multicolumn{3}{|l|}{C,P-conserving, $\rm{SU}(2) \times \rm{U}(1)$ 
relations:}\\
\hline
  \Cdgz  &$  2.8 $&$  1.8 $\\
  \Cdkg  &$  3.1 $&$  1.9 $\\
  \Clg   &$  4.3 $&$  2.6 $\\
\hline
\multicolumn{3}{|l|}{C,P-conserving, no relations:}\\
\hline
  \Cdgz  &$ 15.5 \phantom{0} $&$ 12.6 \phantom{0} $\\
  \Cdkg  &$  3.3 $&$  1.9 $\\
  \Clg   &$  5.9 $&$  3.3 $\\
  \Cdkz  &$  3.2 $&$  1.9 $\\
  \Clz   &$  6.7 $&$  3.0 $\\
\hline
\multicolumn{3}{|l|}{not C or P conserving:}\\
\hline         
  \Cgz{5}&$ 16.5 \phantom{0} $&$ 14.4 \phantom{0} $\\
  \Cgz{4}&$ 45.9 \phantom{0} $&$ 18.3 \phantom{0} $\\
  \Ckzt  &$ 39.0 \phantom{0} $&$ 14.3 \phantom{0} $\\
  \Clzt  &$  7.5 $&$  3.0 $\\
  \hline
\end{tabular}
\end{center}
\caption{Results of the single parameter fits ($1 \sigma$) to the different 
triple gauge couplings. For $\sqrt{s}=500\GeV\ {\cal L}=500\,\fbi$ and for
$\sqrt{s}=800\GeV\ {\cal L}=1000\,\fbi$ has been assumed. For both energies
$\pmi = 80\%$ and $\ppl = 60\%$ has been used.
}
\label{tab:singtgc} 
\end{table}

Systematic uncertainties from detector effects, backgrounds and
beamstrahlung are small. 
The beam polarisation can be determined from a Blondel 
scheme \cite{ref:alain,ref:blondelhe},
so that no additional systematics enter.
If only electron polarisation is available the statistical errors
increase by roughly 50\%. However, since the forward peak in the cross
section is completely dominated by the neutrino $t$-channel exchange,
which is present for left handed electrons only, also in this case the
beam polarisation can be determined from the data alone \cite{ref:blondelhe}.

The radiative corrections need to be
known significantly better than 1\%.
Using the double-pole approximation, the cross section for W-pair production
can currently be predicted to 
better than 0.5\% away from the threshold region up to 500\GeV
with RaccoonWW \cite{ref:racoon} and YFSWW3 
\cite{YFSWW3}. 
%Beyond 500\GeV large logarithmic corrections (electroweak 
%Sudakov logarithms) arise from the virtual exchange of soft 
%and collinear gauge bosons \cite{ref:beenakker}. 
%
Above $500\GeV$ large double logarithmic corrections
(electroweak Sudakov logarithms) arise from the virtual exchange of
soft and collinear gauge bosons.  These corrections are numerically
important, but they have been studied extensively and are
theoretically under control~\cite{Sudakov}.
%\cite{Sudakov_ciafaloni2,Sudakov_fadin,
%Sudakov_melles,Sudakov_beenakker,
%Sudakov_ciafaloni,Sudakov_hori,Sudakov_kuehn}.

In general the total errors on the anomalous couplings are 
${\rm few} \times 10^{-4}$. Loop corrections to the couplings are
expected to be of order $g^2/16 \pi^2$, one order of magnitude larger
than the expected precision. For the case of Supersymmetry it has
been shown that the loop corrections are indeed of that 
size \cite{ref:alam,ref:kneur} and should thus be visible at TESLA.

\begin{figure}[htb]
\begin{center}
\href{pictures/6/tgc_comp_k.pdf}{{\includegraphics[width=7.5cm,bb=33 40 492 470]{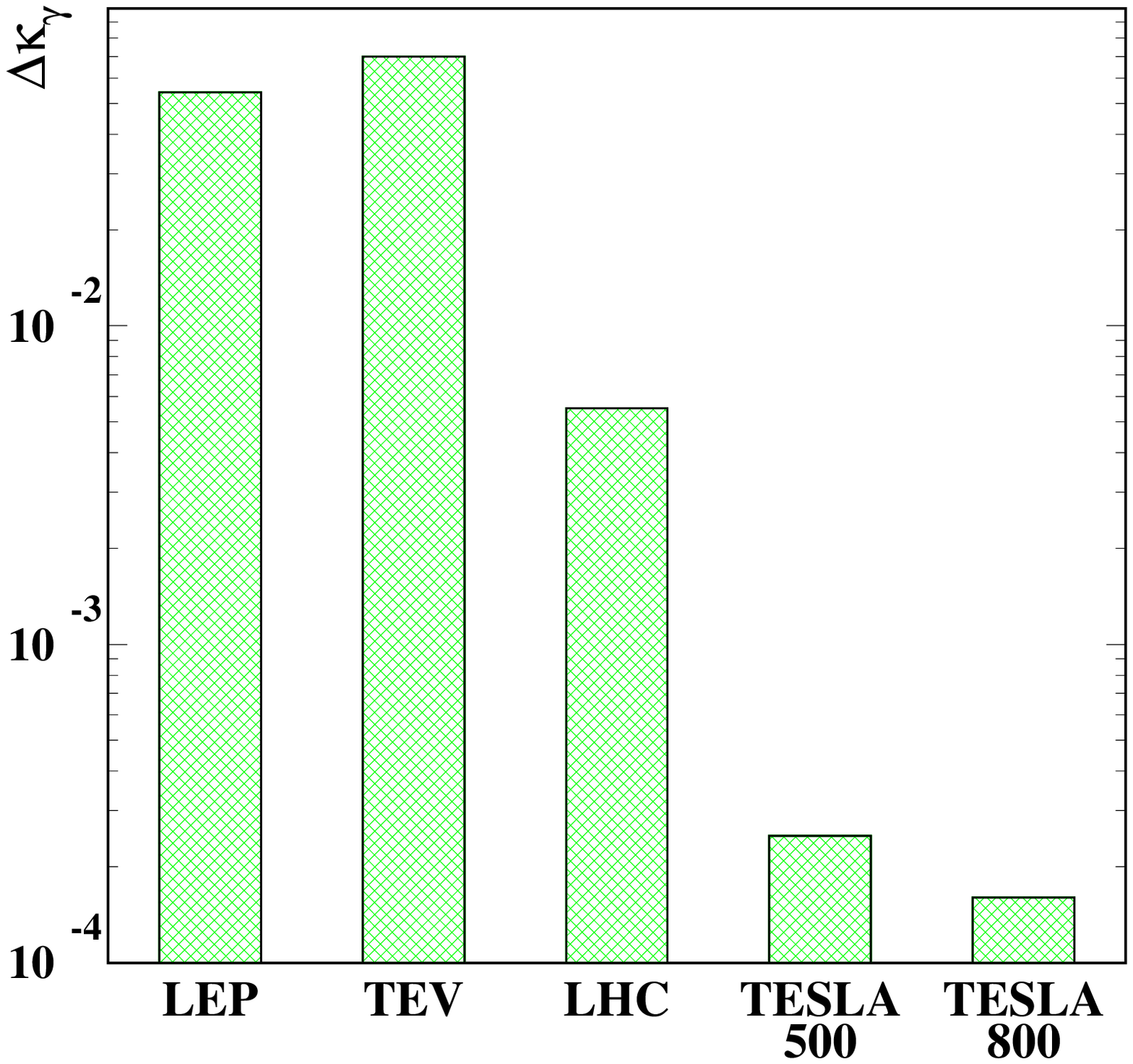}}}
\href{pictures/6/tgc_comp_l.pdf}{{\includegraphics[width=7.5cm,bb=33 40 492 470]{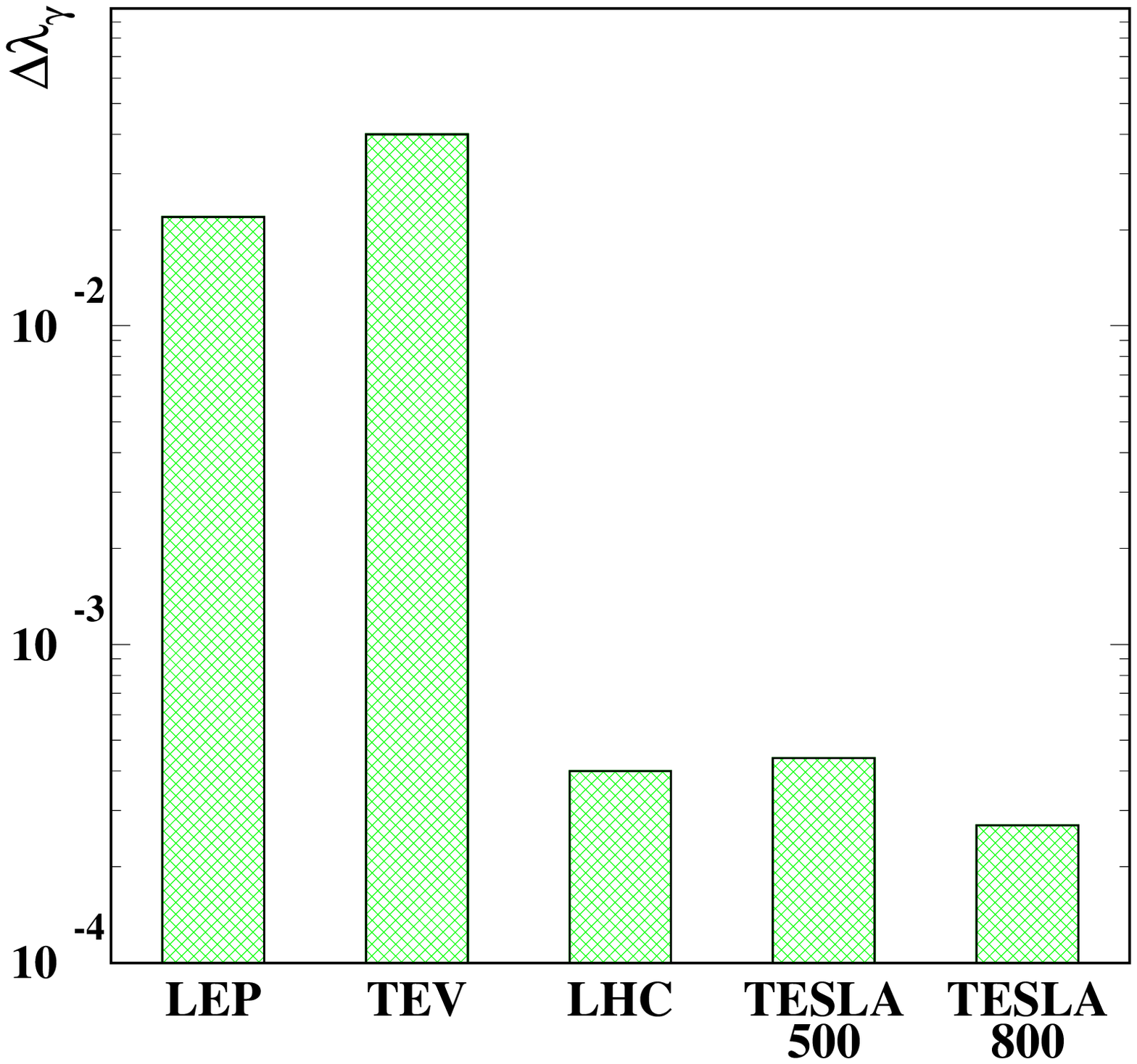}}}
\end{center}
\caption{Comparison of $\Delta \kappa_\gamma$ and $\Delta
  \lambda_\gamma$ at different machines. For LHC and TESLA three years
  of running are assumed (LHC: $300\,\fbi$, 
  TESLA $\sqrt{s}=500\GeV$: $900\,\fbi$, TESLA $\sqrt{s}=800\GeV$: $1500 \fbi$).
  }
\label{fig:tgccomp}
\end{figure}

Figure \ref{fig:tgccomp} compares the obtainable precision of
$\kappa_\gamma$ and $\lambda_\gamma$ at the different machines. 
Especially for $\kappa_\gamma$, where, because of the lower
dimension of the corresponding operator, 
experiments are sensitive at a lower energy to potential new
physics at a high scale,
TESLA has a much higher sensitivity than LHC.

For the additional processes,
$\gamma \gamma \rightarrow {\rm W}^+ {\rm W}^-$ and
${\rm e}^- \gamma \rightarrow \nu {\rm W}^-$,
at the $\gamma \gamma$- and e$\gamma$-collider only theoretical
studies with low luminosity exist \cite{ref:tgcgg}. An extrapolation
to the presently expected luminosity still yields errors that are an
order of magnitude worse than the ones expected from W-pair production
in $\ee$. However, these studies use only the total cross section in
the central region of the detector and additional sensitivity can be
expected from a detailed analysis of the angular dependence and the
W-polarisation. With the same simplifications the expectations for
single W production in $\ee$ collisions are slightly worse than for the 
$\gamma \gamma$- and e$\gamma$-collider.

In addition to single and pair production of gauge bosons also triple
gauge boson production will be visible at TESLA. The cross sections
are, in the heavy Higgs limit, ${\cal O}(50\,\fb)$ for WWZ and 
${\cal O}(1\,\fb)$ for ZZZ \cite{ref:wwzbarger,ref:wwzfawzi}. 
Both processes have their maximum cross section between 500 and 1000\GeV.
Requiring a photon of more than 20\GeV\  energy at a polar angle above
$15^\circ$ the total cross section for ZZ$\gamma$ is of the order $10\,\fb$ 
and for WW$\gamma$ about $100\,\fb$.
Using the latter two channels the anomalous couplings $a_0$ and $a_c$
which modify the $VV\gamma\gamma$-vertex but not the
triple-gauge-couplings can be measured to the 0.2 level with
$\sqrt{s}=500\GeV$, an integrated luminosity of $300\,\fbi$ and
polarised beams, 
corresponding to $\Lambda_{0,c} \approx 1.7\TeV$ in the
operators $1/\Lambda_0^2 \cdot 1/2 \cdot F^{\mu\nu} F_{\mu\nu}
W^i_{\rho} W^{i,\rho}$ and $1/\Lambda_c^2 \cdot 1/2 \cdot F^{\mu\nu}
F_{\nu\rho} W^{i,\rho} W^i_\mu$\cite{ref:anja,ref:wwzfawzi}.

Also these measurements require adequate theoretical
calculations. The present status is summarised in \cite{ref:lep_workshop}.

\subsection{High precision measurements at lower energies}
With a luminosity of ${\cal L}=5 \cdot 10^{33} {\rm cm}^{-2} {\rm s}^{-1}$ 
at energies close to the Z-pole TESLA can produce $10^9$ Z-bosons in about 
50-100 days of running. A similar luminosity is possible close to the W-pair
threshold.
In this scenario, referred to as GigaZ in the following, the
measurements already performed at LEP and SLC can be redone with
increased precision.
\subsubsection{Measurement of the weak mixing angle}
One of the most sensitive quantities to loop corrections from the
Higgs-boson is the effective weak mixing angle in Z-decays $\swsqeffl$.
The most sensitive observable to $\swsqeffl$ is the left-right asymmetry
\[
\ALR = \frac{1}{{\cal P}}\frac{\sigma_L-\sigma_R}{\sigma_L+\sigma_R},
\]
where $\sigma_{L/R}$ is the total cross section for left/right-handed
polarised  electrons and ${\cal P}$ the longitudinal electron polarisation. 
For pure Z-exchange $\swsqeffl$ is then given by 
%$\ALR = \cAe = \frac{2 v_e a_e}{v_e^2 +a_e^2}$,
$\ALR = \cAe = 2 v_{\mathrm{e}} a_{\mathrm{e}} /
(v_{\mathrm{e}}^2 + a_{\mathrm{e}}^2)$,
$v_{\mathrm{e}} (a_{\mathrm{e}})$ being the vector- (axial-vector-) coupling 
of the Z to the electron
and $v_{\mathrm{e}} /a_{\mathrm{e}} = 1 - 4 \swsqeffl$.
$\ALR$ can be measured during GigaZ running from hadronic Z-decays with very 
high efficiency and low background. 
Details on the measurement of $\ALR$ and the other observables can be
found in \cite{ref:zfact}.
The statistical error with $10^9$
events will be of the order $\Delta \ALR = 3 \cdot 10^{-5}$ which has to
be matched by systematics. The polarisation needs to be known to 
$\Delta {\cal P}/{\cal P} < \Delta \ALR/\ALR \sim 2 \cdot 10^{-4}$. 
This is only
possible if polarised electrons and positrons are available, so that 
the polarisation can be measured directly from data using the
Blondel-scheme \cite{ref:alain}. 
The cross section for an electron polarisation $\pmi$ and a positron
polarisation $\ppl$ is given by
\begin{equation}
\sigma \, = \, \sigma_u \left[ 1 - \ppl \pmi + \ALR (\ppl - \pmi) \right],
\end{equation}
where $\sigma_u$ is the cross section for unpolarised beams.

If all four combinations of beam helicities are measured $\ALR$ can be
obtained independently from an external polarisation measurement:
\begin{equation}
  \label{eq:alrblondel}
  \ALR \, = \, \sqrt{\frac{
      ( \sigma_{++}+\sigma_{-+}-\sigma_{+-}-\sigma_{--})
      (-\sigma_{++}+\sigma_{-+}-\sigma_{+-}+\sigma_{--})}{
      ( \sigma_{++}+\sigma_{-+}+\sigma_{+-}+\sigma_{--})
      (-\sigma_{++}+\sigma_{-+}+\sigma_{+-}-\sigma_{--})}}
\end{equation}
where in $\sigma_{ij}$ $i$ denotes the sign of the electron- and $j$ the 
sign of the positron polarisation.

This formula assumes, however, that the absolute polarisation values
of the bunches with opposing helicity states are equal. 
To assure this, or to get the
relevant corrections, polarimeters are still needed. Since only
relative measurements within one beam are needed most systematics cancel, 
so that with
this scheme the polarisation can be measured with the required accuracy.
To obtain optimal statistical precision only one tenth of of the luminosity 
needs to be spent on the small cross sections $(++,--)$. 
For $\ppl > 50 \%$ the statistical error
using the Blondel scheme is only slightly larger than with an external 
polarisation measurement.
For 20\% positron polarisation and $10^9$ Zs the statistical error is
$\Delta \ALR = 8 \cdot 10^{-5}$.

Around the Z peak the change of $\ALR$ with the beam energy is 
${\rm d} \ALR / {\rm d} \sqrt{s} = 2 \cdot 10^{-2}/\GeV$. 
The variation is due to the $\gamma$-Z
interference, so that the difference of $\sqrt{s}$ and $\MZ$
needs to be known. Not to be dominated by the knowledge of the beam
energy one needs a spectrometer with a precision of
1\,MeV that can be calibrated relative to $\MZ$ with a short scan around the 
Z-resonance.
Also because of the energy dependence of $\ALR$, the amount of beamstrahlung
expected for GigaZ running shifts $\ALR$ by 
$\Delta \ALR \, = \, 9 \cdot 10^{-4}$. 
The beamstrahlung thus needs to
be understood to a few percent which seems possible \cite{ref:bs}. 
If the same beamstrahlung as in the $\ALR$ measurement
is also present in the calibration scan the beamstrahl-shift is
absorbed in an apparent shift of the centre of mass energy, so
that in principle no corrections are necessary.
Since all other systematic errors are small, $\Delta \ALR = 10^{-4}$
is a realistic estimate of the final error. 
This corresponds to an error in the weak mixing angle of 
$\Delta \swsqeffl = 0.000013$.

Due to the polarised beams and the excellent b-tagging also the $\cAb$
measurements using the b-quark forward-backward asymmetry can be improved 
by roughly a factor 15 relative to LEP and SLC. GigaZ thus 
can clear up the slight discrepancy between the b-asymmetry at LEP and
SLC and $\ALR$ at SLC \cite{ref:lepew}.

\subsubsection{Measurements of the Z-partial widths}
For the observables sensitive to the partial and total widths of the Z
the situation is less spectacular. The measurement of the
total Z-width will be dominated
by the relative precision of the beam spectrometer. A total precision of 
$\Delta \GZ \approx 1 \MeV$ is thus within reach 
(see Part IV-7.3).%section \ref{detector_mdi_spt}).
%\cite{ref:cmscal}. 
For the selection efficiencies for hadrons, muons and taus a factor
three improvement relative to the best LEP experiment should be
possible \cite{ref:marc}. Also the experimental systematics on the
luminosity might be improved, 
however, this would in addition require an improvement
of the theoretical error, which is 0.05\% at present.

The interesting physics parameters that can be derived from the lineshape
parameters are
\begin{itemize}
\item the mass of the Z $(\MZ)$;
\item the strong coupling constant at the scale of the Z-mass 
  $(\alpha_s(\MZ^2))$;
\item the radiative correction parameter, normalising the strength of
the leptonic Z-couplings to the fermions $(\Delta \rho_\ell)$ \cite{ref:drho};
\item the number of light neutrino species $(N_\nu)$.
\end{itemize}

The possible improvements in these parameters are, together with the
other observables at GigaZ, summarised in Table~\ref{tab:line}. 
The precision on all observables, obtained from the cross section around 
$\MZ$ apart from $\MZ$ itself can be improved by a factor two to three.

\begin{table}
\begin{center}
\begin{tabular}[c]{|c|c|c|}
\hline
 & LEP/SLC/Tev \cite{ref:lepew} & TESLA \\
\hline
$\swsqeffl$ & $0.23146 \pm 0.00017$ & $\pm 0.000013$\\
\hline
\multicolumn{3}{|l|}{lineshape observables:}\\
\hline
$\MZ$ & {$ 91.1875 \pm 0.0021\GeV$} & {$ \pm 0.0021\GeV$} \\
$\alpha_s(\MZ^2)$ & {$ 0.1183 \pm 0.0027 $} & {$ \pm 0.0009$} \\
$\Delta \rho_\ell$ & {$ (0.55 \pm 0.10 ) \cdot 10^{-2}$} 
& {$ \pm 0.05\cdot 10^{-2}$}  \\
$N_\nu$ & {$ 2.984 \pm 0.008 $} & {$ \pm 0.004 $} \\
\hline
\multicolumn{3}{|l|}{heavy flavours:}\\
\hline
$\cAb$ & $0.898 \pm 0.015$   &$\pm 0.001$ \\
$\Rb^0$ &$0.21653 \pm 0.00069$ & $\pm 0.00014$  \\
\hline
$\MW$ & $80.436 \pm 0.036\GeV$ & $\pm 0.006\GeV$\\
\hline
\end{tabular}
\end{center}
\caption{Possible improvement in the electroweak physics quantities 
at TESLA.
For $\alpha_s$ and $\Delta \rho_\ell \ N_\nu=3$ is assumed.}
\label{tab:line}
\end{table}

Due to the extremely good b-tagging capabilities at TESLA, also the
ratio of the Z partial width to $\bb$ to the 
hadronic width, $\Rb$, can be improved by a factor five relative to LEP.
%constraining models of dynamical symmetry breaking \cite{ref:rbtheo}.

\subsubsection{Measurement of the W-mass}
The W-mass can be obtained from a scan around the W-pair production 
threshold \cite{ref:wscan}.
Near threshold the $s$-channel production is suppressed by 
$\beta^3$ while the $t$-channel is
only suppressed by $\beta$, where $\beta$ is the velocity of the W in
units of $c$. Due to the leading, $\beta$-suppressed, contribution, a
scan around the threshold has a high sensitivity to the W-mass. Also
for the $t$-channel only the well known We$\nu$-coupling is involved, so
that the total cross section can be predicted without uncertainties from new 
physics. Any anomalous
triple gauge couplings enter via the $s$-channel and are therefore
suppressed by an additional factor $\beta^2$. It is therefore possible
to measure the W-mass precisely from a scan of the threshold region.

It should however be noted, that the double pole approximation is not
valid in the threshold region.
In order to reach sufficient accuracy in this energy range, 
a full four-fermion 
calculation with radiative corrections is required. The 
necessary improvements should be possible within the coming 
years such that the theoretical accuracy will be no obstacle 
to the precision tests.

\begin{figure}[htb]
\begin{center}
\href{pictures/6/wmassdemonstrate.pdf}{{\includegraphics[height=9cm]{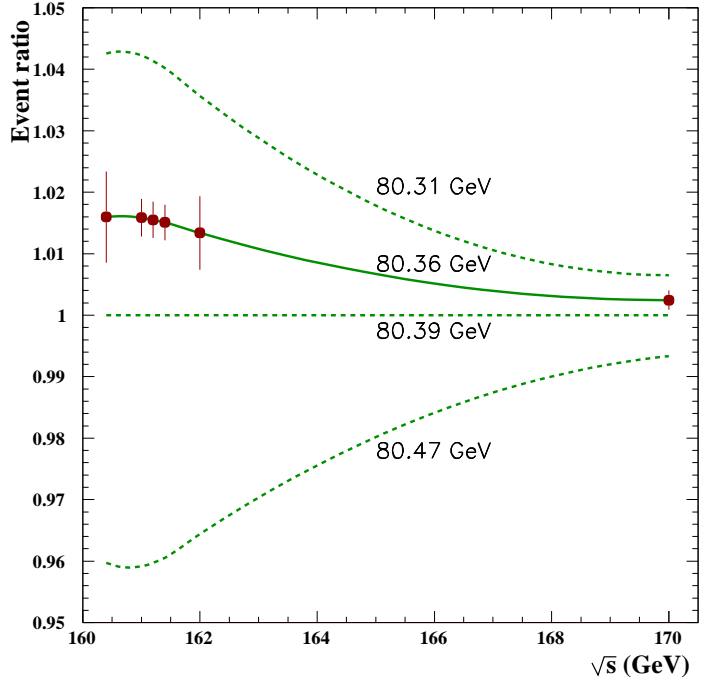}}}
\end{center}
\caption{Sensitivity of the W-pair threshold scan to the W-mass. The
  vertical axis shows the ratio of the cross section to the predicted
  cross section for $\MW=80.39\GeV$. The error bars represent the
  expected errors for the scan described in the text.}
\label{fig:wscan}
\end{figure}

With TESLA one can collect an integrated luminosity of $100\,\fbi$ 
per year at $\sqrt{s} \sim 161\GeV$. The different
polarisation states allow to enhance or suppress the signal helping
to obtain the background directly from the data.
A five point scan with $160.4\GeV \le \sqrt{s} \le 162\GeV$ and an
additional point at $\sqrt{s}=170\GeV$ has been 
simulated \cite{ref:wscan}, assuming the same efficiency and
purity as reached at LEP. With a total error of $0.25\%$ on the luminosity and
on the selection efficiencies $\MW$ can be measured with a total
precision of $6 \MeV$. 
The method is experimentally robust, for example even if the
efficiencies are left free in the fit, the error only increases to $7 \MeV$.
The achievable errors at the scan points are compared with the
sensitivity to the W-mass in Fig. \ref{fig:wscan}.

\subsubsection{Interpretation of the high precision data}
The high precision measurements can be used to test the Standard Model
at the loop level. However one of the most important radiative corrections
is due to the
running of the electromagnetic coupling $\alpha$ from zero momentum
transfer to the Z-scale. 
This running is mainly caused by the contribution of fermion loops.
The lepton loops can be calculated reliably without any
significant uncertainty. However, due to additional QCD corrections the
quark loops are much more uncertain. 
$\alpha(s)$ can be expressed as
$\alpha(s)^{-1} = \left(1-\Delta \alpha_{\rm{lep}}(s) -
\Delta \alpha_{\rm{had}}^{(5)}(s) - \Delta \alpha_{\rm{top}}(s)\right) \cdot
\alpha^{-1}$.
%If the running of~$\alpha$ is calculated in a
If $\Delta \alpha_{\rm{had}}^{(5)}(s)$ is calculated in a
completely model independent fashion from a
convolution of the $\ee \rightarrow \rm{hadrons}$ 
cross section alone using only the optical
%theorem~\cite{ref:aem_data}, one obtains from the low energy data 
%$\Delta \alpha_{\rm{had}}^{(5)}(\MZ^2) = (280.4 \pm 6.4) \cdot 10^{-4}$
%corresponding to an
%uncertainty in the Standard Model prediction
%of~$\swsqeffl$ of~0.00023.  
%Including recent preliminary data from BES \cite{ref:bes} this
%uncertainty can be reduced by a factor two to three 
%\cite{ref:aem_martin,ref:aem_bolek}.
theorem~\cite{ref:aem_data,ref:aem_data_new}, one obtains from the low energy 
data, including the latest BES results \cite{ref:bes},
$\Delta \alpha_{\rm{had}}^{(5)}(\MZ^2) = (279.0 \pm 4.0) \cdot 10^{-4}$
corresponding to an uncertainty in the Standard Model prediction
of~$\swsqeffl$ of~0.00014.  
However,
the sensitivity to the details of the resonance region
can be reduced significantly, if the low energy data
is used to fit the coefficients of a QCD operator
product expansion instead of integrating the total cross
section.
If the hadronic cross section is known to 1\% up to the
$\Upsilon$-resonances the uncertainties are $\Delta \swsqeffl = 0.000017$ 
and $\Delta \MW = 1 \MeV$ \cite{ref:aem_data_new}.

A Z-mass error of 2 MeV from LEP contributes 0.000014 to the
uncertainty of the $\swsqeffl$
prediction, about the same size as the experimental error and the
uncertainty from $\alpha(\MZ^2)$. For $\MW$ the direct uncertainty 
due to $\MZ$ is
2.5 MeV. However, if the beam energy is calibrated relative to the Z-mass, 
so that the relevant observable is
$\MW/\MZ$, the error is smaller by a factor three.

An uncertainty in the top quark-mass of $1\GeV$ results in an uncertainty of
the $\swsqeffl$ prediction of 0.00003 and in the one for $\MW$ of $6 \MeV$. 
For a top-mass error
of $\Delta \MT \approx 100 \MeV$, as it is possible from a top-threshold scan
at TESLA
(see section \ref{physics_top}) this uncertainty is completely 
negligible.

\begin{figure}[htb!]
\begin{center}
  \href{pictures/6/higgs_chi2_lc.pdf}{{\includegraphics[height=6cm,bb=0 0 567 519]{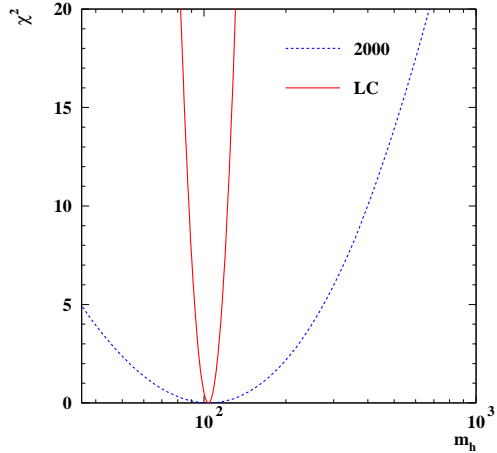}}}
\end{center}
\caption{$\Delta \chi^2$ as a function of 
the Higgs-mass for the electroweak
precision data now and after GigaZ running.
\label{fig:gz_fit} }
\end{figure}

Including the possible improvement on $\alpha(\MZ^2)$ very stringent
tests of the Standard Model are possible. 
Figure \ref{fig:gz_fit}
shows as an example the variation of the fit-$\chi^2$ as a function of
the Higgs-mass for the present data and for TESLA. 
It can be seen that the Higgs-mass can indirectly be constrained at the level 
of 5\% \cite{ref:zfact,ref:gzsven}.

\begin{figure}[htp!]
\vspace{1em}
\begin{center}
\begin{tabular}{cc}
\href{pictures/6/TBMA11b.pdf}{{\includegraphics[height=7cm]{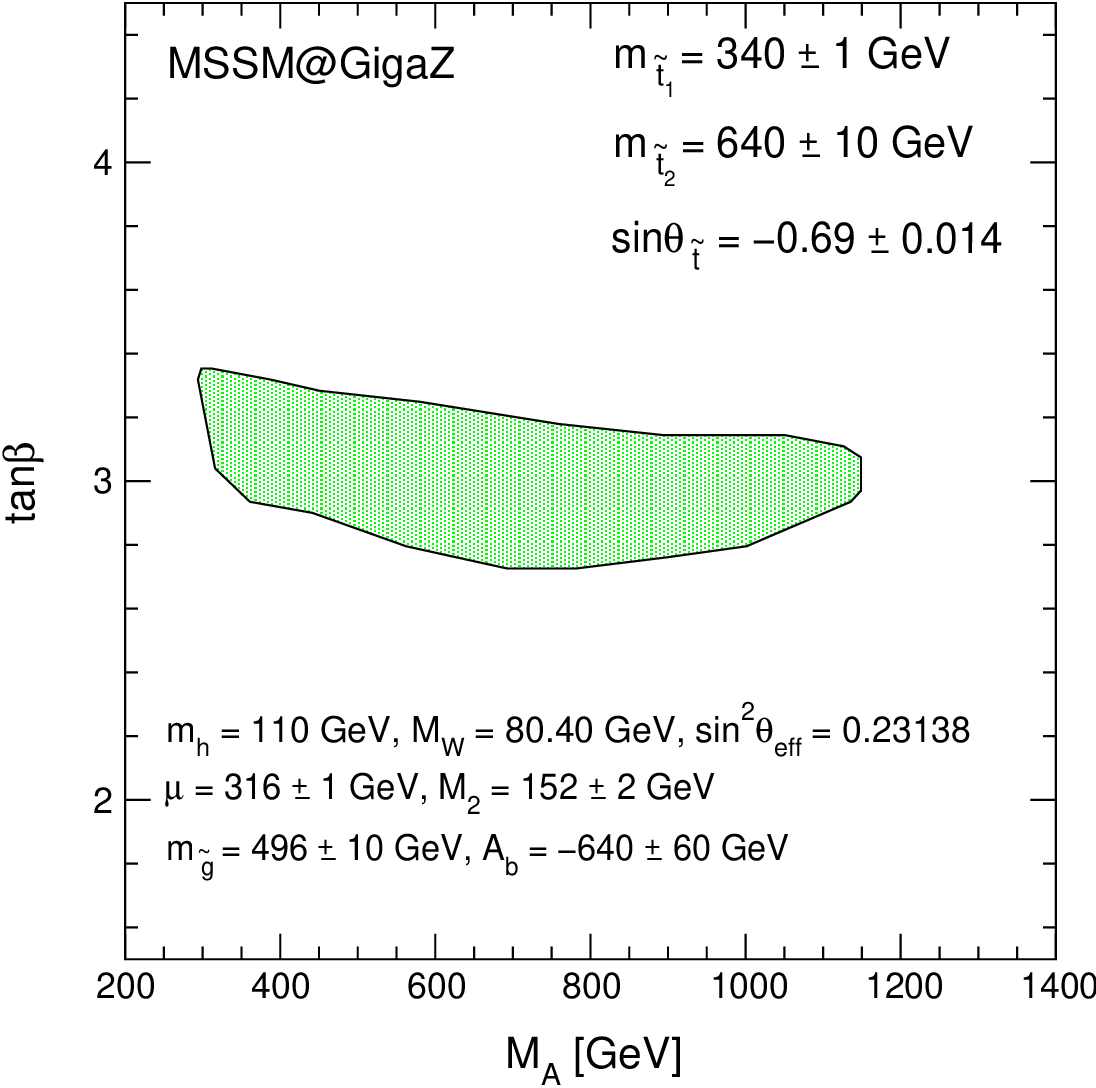}}}&
\href{pictures/6/MSt2MA11c.pdf}{{\includegraphics[height=7cm]{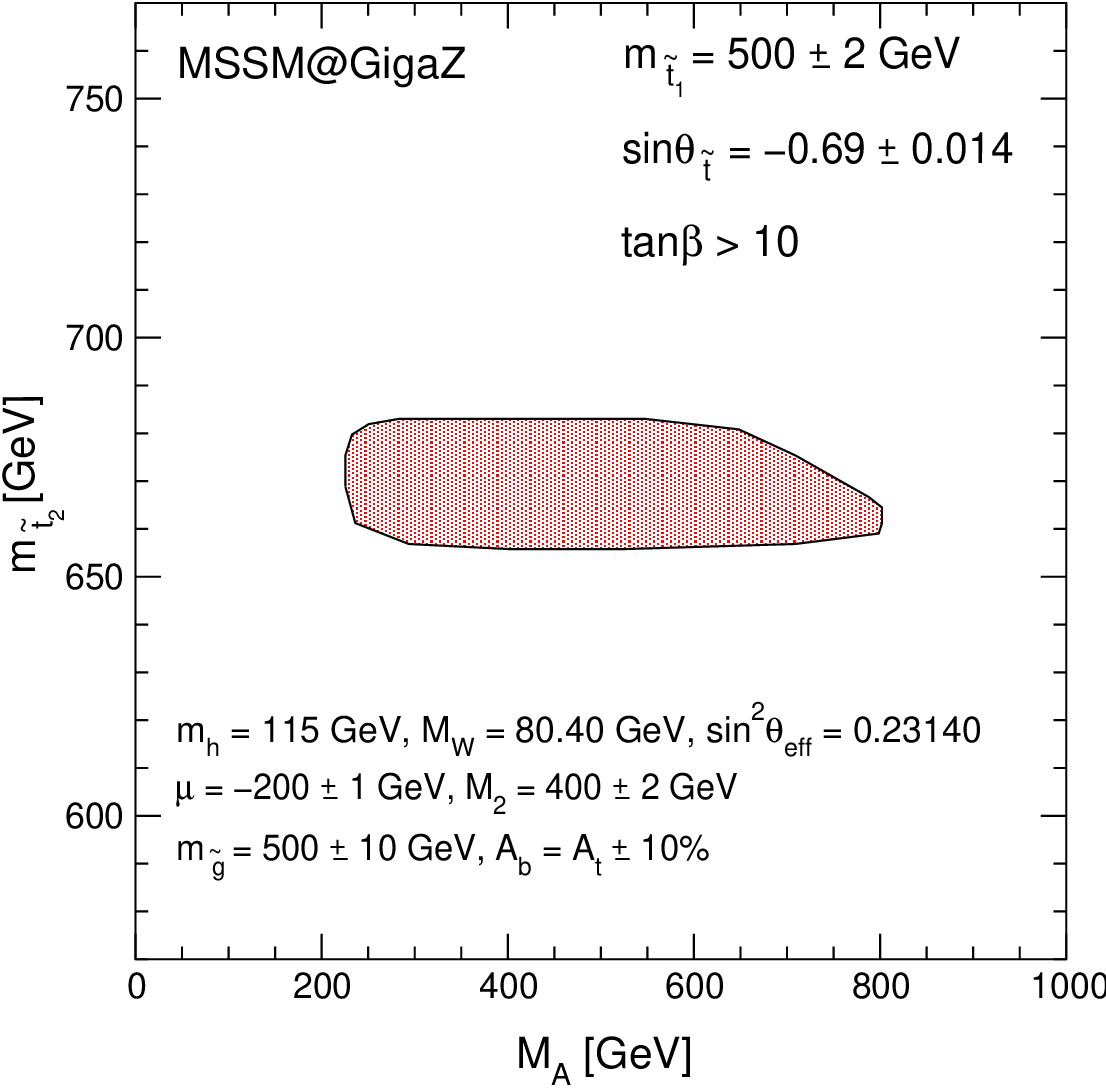}}}
\end{tabular}
\end{center}
\caption[]{
  The regions in the $m_{\rm{A}}- \tan \beta$ and
  $m_{\rm{A}} - m_{\tilde{{\rm t}}_2}$ plane, allowed by
$1\,\sigma$ errors of the measurements of $\Mh$, $\MW$ and $\swsqeffl$.
}
\label{fig:mwsw_susy} 
\end{figure}

If the Higgs-mass is in the range predicted by the current precision
data, the Higgs will have been found at the time of the high precision
electroweak measurements. In this case the data can be used to check
the consistency of the SM or to measure free parameters in by then
established extensions of the model. As an example Fig.
\ref{fig:mwsw_susy} shows the constraints that can be obtained in
$m_{\rm{A}}$ and $\tan \beta$ from the low energy running if other SUSY
parameters, especially the stop sector, are already known or, alternatively,
in $m_{\rm{A}}$ and $m_{\tilde{{\rm t}}_2}$ if $\tan \beta$ and the parameters,
that can be measured from the light stop only, are known
\cite{ref:gzsven}.
Further applications of GigaZ to Supersymmetry are discussed in chapter
\ref{physics_SUSY}.

\begin{figure}[thp!]
\begin{center}
\begin{tabular}{cc}
\href{pictures/6/eps1eps2.pdf}{{\includegraphics[height=7cm,bb=0 0 544 521]{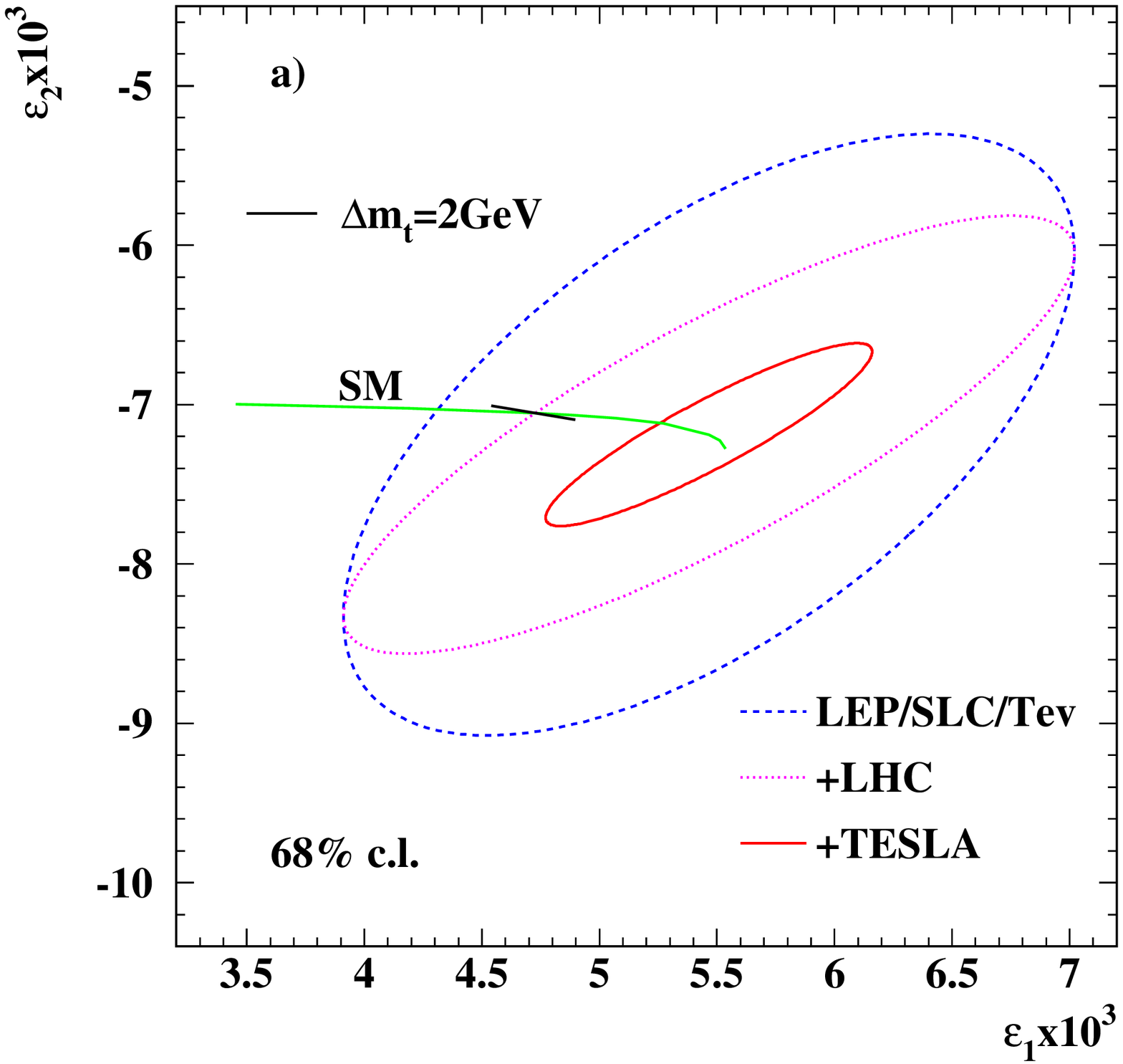}}}&
\href{pictures/6/eps1eps3.pdf}{{\includegraphics[height=7cm,bb=0 0 544 521]{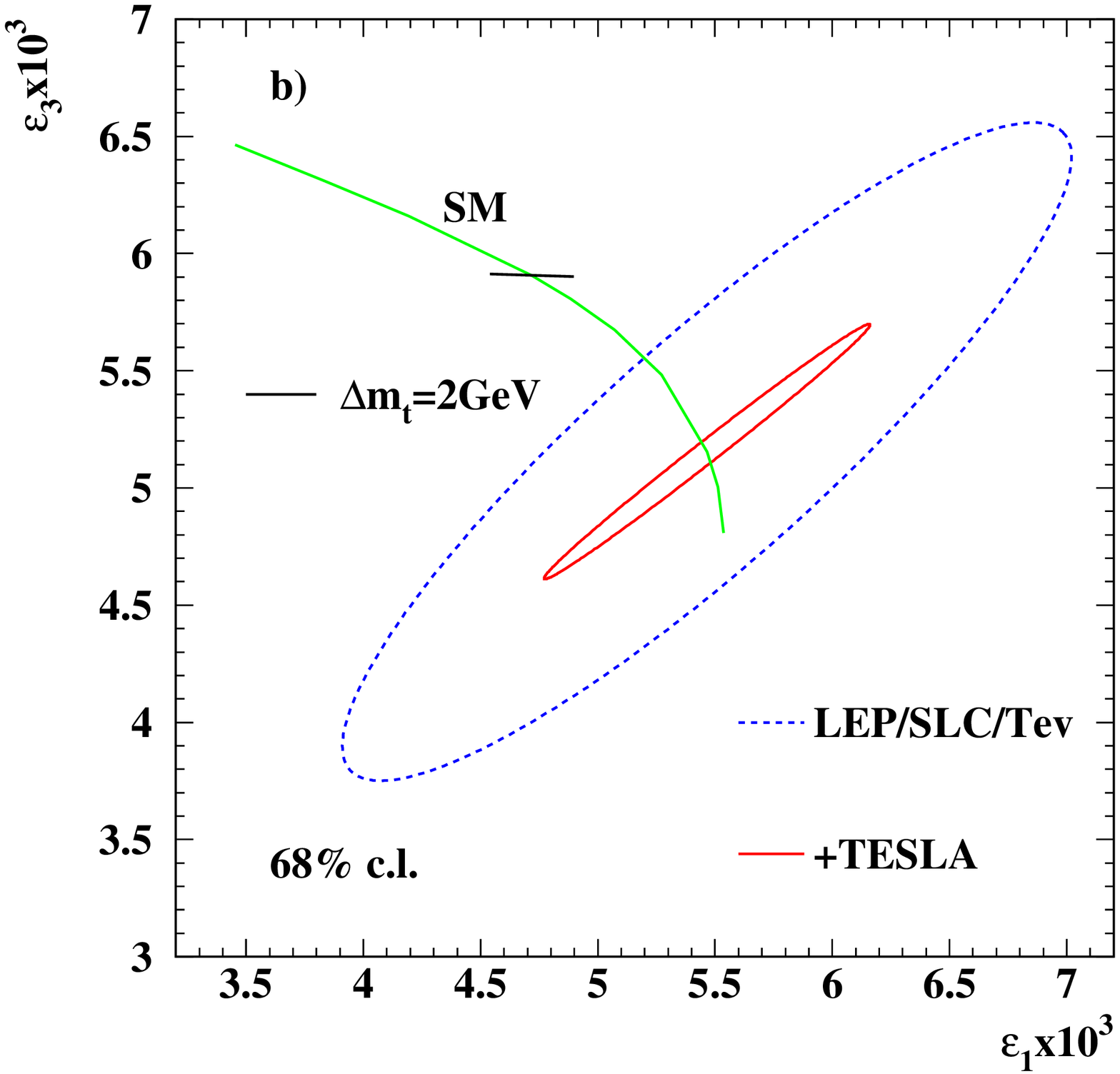}}}\\
\href{pictures/6/eps2eps3.pdf}{{\includegraphics[height=7cm,bb=0 0 544 521]{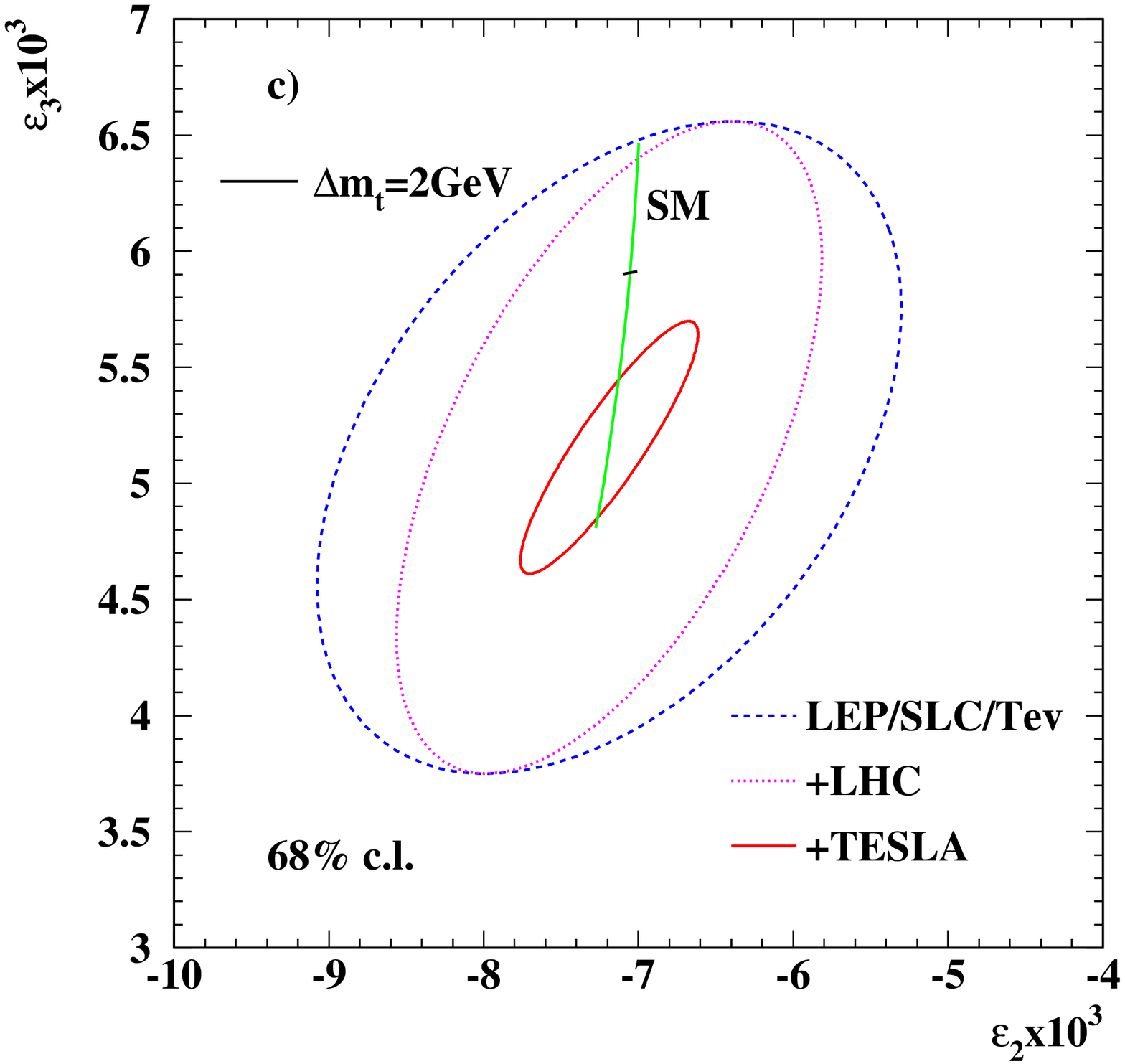}}}&
\href{pictures/6/eps1eps3_e2sm.pdf}{{\includegraphics[height=7cm,bb=0 0 544 521]{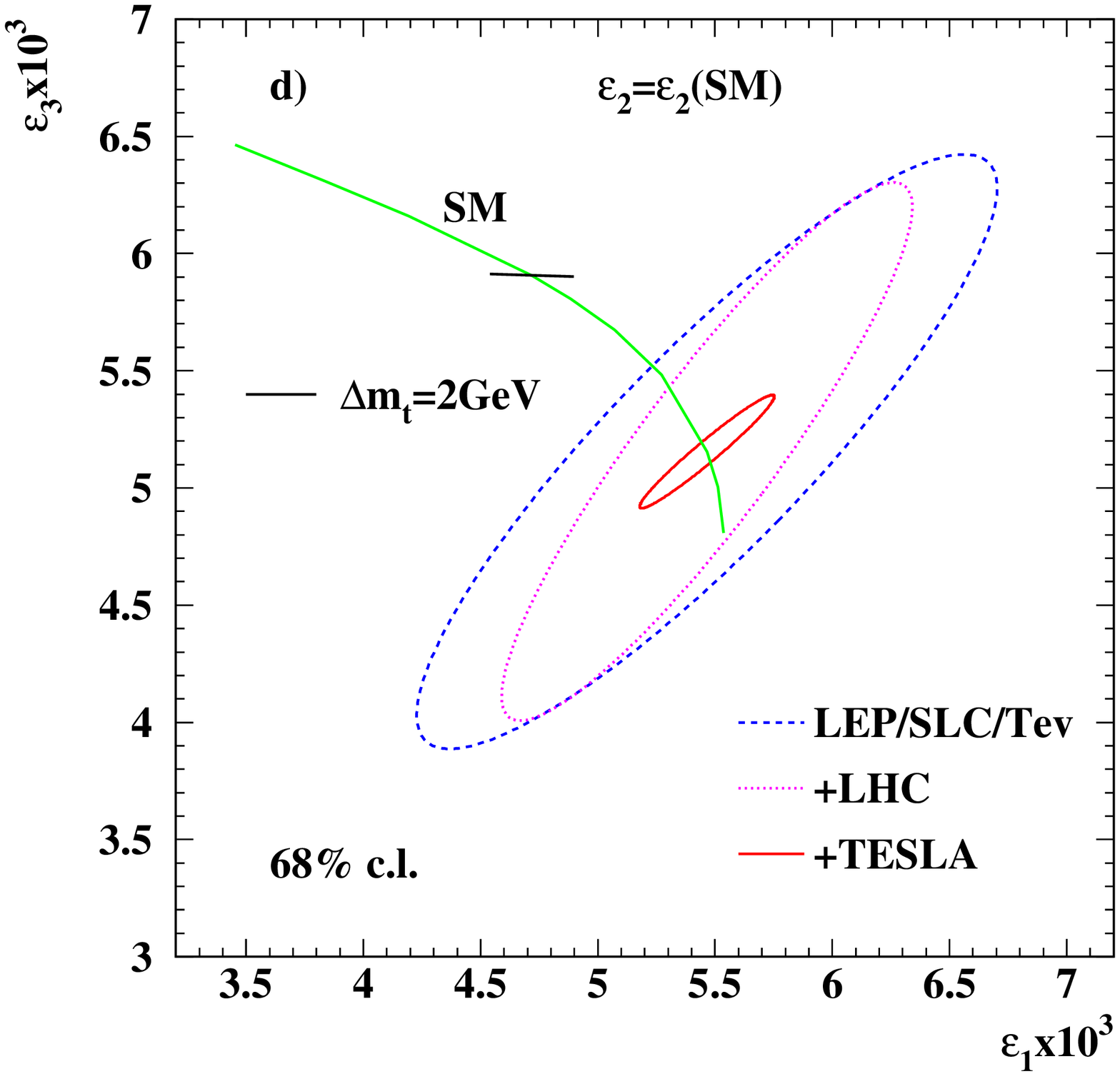}}}
\end{tabular}
\end{center}
\caption{
$\varepsilon_1-\varepsilon_2$ (a),
$\varepsilon_1-\varepsilon_3$ (b), and
$\varepsilon_2-\varepsilon_3$ (c)
for the present data and the expectation at a linear collider. 
The line marked ``SM'' shows
the SM prediction with $\MT=174\GeV$ and varying $\MH$ from 70\GeV\ 
(lower right end) to 1\TeV\  (upper left end).
The effect of an uncertainty in $\MT$ is indicated by the black line.
$\Delta \MT=100\MeV$, expected from TESLA, is inside the SM line width.
In d) $\varepsilon_1-\varepsilon_3$ is shown with $\varepsilon_2$ fixed
to its SM expectation.
}
\label{fig:eps_lc} 
\end{figure}

For more model independent analyses
frequently reparameterisations of the radiative correction parameters
are used where the large isospin-breaking corrections are absorbed
into one parameter, so that the others depend only on the logarithmic 
terms.
One example are the so called $\varepsilon$ parameters \cite{ref:epspar}
\begin{eqnarray*}
\Delta \rho_\ell & = & \varepsilon_1 \\
\swsqeffl & = & \frac{1}{2}\left(1 - \sqrt{1 - 
\frac{4 \pi \alpha(\MZ^2)}{\sqrt{2} \GF \MZ^2}} \right) 
\left( 1 -1.43 \varepsilon_1 + 1.86 \varepsilon_3 \right) \\
\frac{\MW^2}{\MZ^2} & = & \frac{1}{2}\left(1 + \sqrt{1 - 
\frac{4 \pi \alpha(\MZ^2)}{\sqrt{2} \GF \MZ^2}} \right) 
\left( 1 +1.43 \varepsilon_1 -1.00 \varepsilon_2 - 0.86 \varepsilon_3 \right).
%\Delta \rho_{\rm{b}} - \Delta \rho_{\rm{d}} & = & 2 \varepsilon_{\rm{b}}
\end{eqnarray*}
In this parameterisation $\varepsilon_1$ absorbs the large isospin-splitting
corrections,
$\varepsilon_3$ contains only a logarithmic $\MH$ dependence while 
$\varepsilon_2$ is almost constant in the Standard Model and most extensions.
Figure \ref{fig:eps_lc} a)-c) shows the the expectations in the 
$\varepsilon_i-\varepsilon_j$-planes, compared to present data and to the SM
prediction.
Since the prediction for $\varepsilon_2$ is almost constant, in 
Fig. \ref{fig:eps_lc} d) the $\varepsilon_1-\varepsilon_3$-plane is shown, 
if $\varepsilon_2$ is fixed to the predicted value.
In this case the precision along the large 
axis of the ellipse is dominated by the precise measurement of the W-mass.

In many extensions of the Standard Model $\varepsilon_1$ can be
varied freely by adjusting some masses,
so that, 
with the new physics for every Higgs-mass up to a \TeV the prediction
can be brought in agreement with the present data. This is also true to 
some extent with TESLA if the W-mass is not measured precisely.
Only with the very accurate determination of $\MW$, one can 
tightly constrain $\MH$ without the knowledge of $\varepsilon_1$.

\begin{figure}[thp!]
\begin{center}
\href{pictures/6/s-t-2hdm.pdf}{{\includegraphics*[height=13cm,bb=0 0 542 476,clip=]{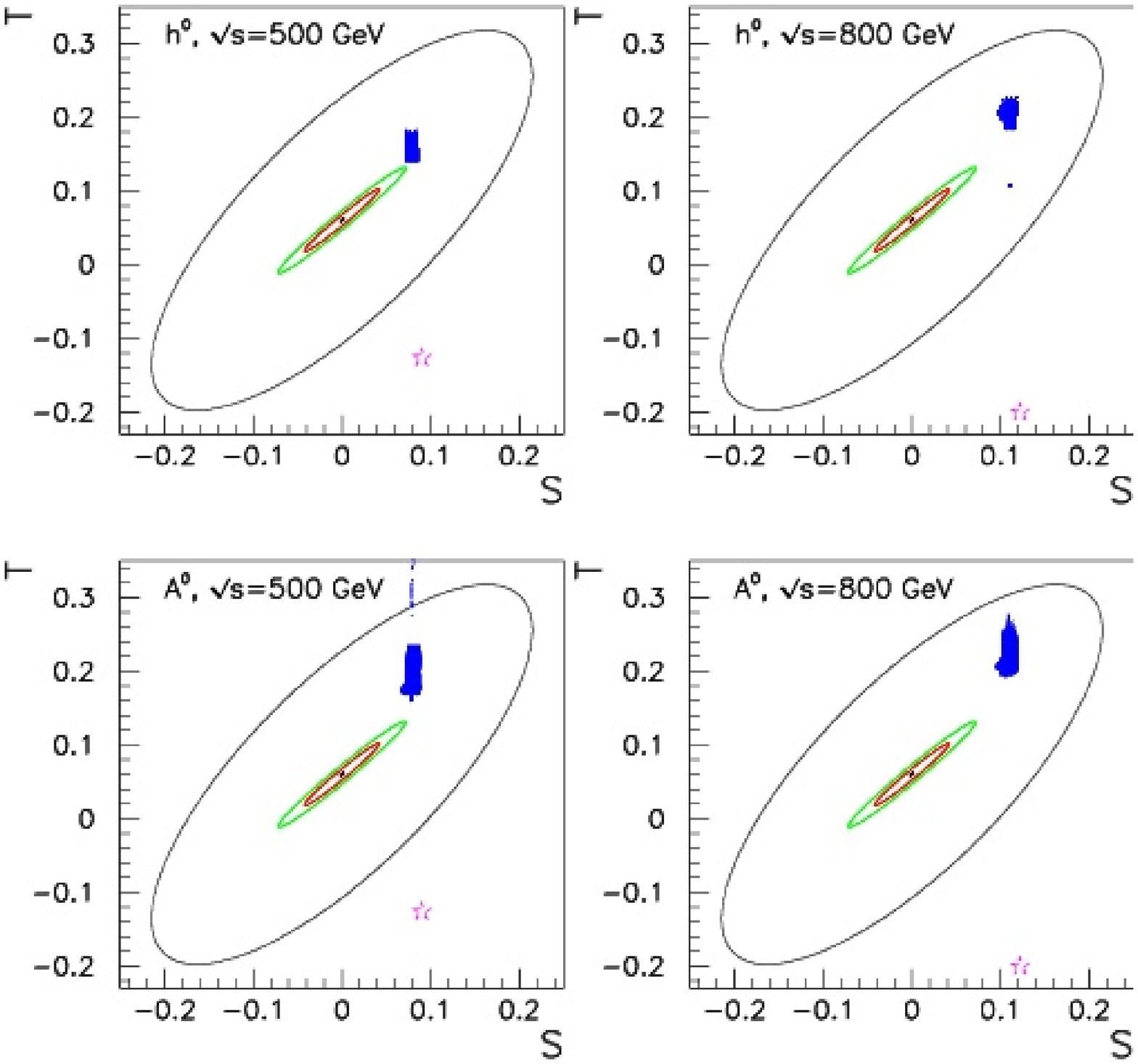}}}
\end{center}
\caption{Prediction for S and T from the 2 Higgs doublet model with a
  light Higgs for the cases where no Higgs is found, compared to the
  current electroweak data and the projection for GigaZ. 
  The outermost (black) ellipse is the 90\% c.l. interval allowed by
  the present data. The green and the red ellipses are the 90\% and
  99\% c.l. expectations for GigaZ. The blue points are the prediction
  of the 2HDM while the open star
  denotes the Standard Model prediction if the Higgs-mass is 
  $\sqrt{s} - 100\GeV$.
  The plots labelled ``$h^0$'' represent the case where the h is
  light, while the ones labelled ``$A^0$'' are for the case where the
  A is light.
}
\label{fig:stthdm} 
\end{figure}

As an example of a possible interpretation of the precision data in
the model independent framework,
Fig. \ref{fig:stthdm} shows the prediction of the 2 Higgs-doublet
model (2HDM) for the ST parameters \cite{ref:stdef}, 
which are basically equivalent
to the $\varepsilon$ parameters, for cases where a light Higgs exists but will
not be seen directly (see section \ref{sec:2.3}) compared
to the present data and the projection of
GigaZ \cite{thdm}. Only with the precision of GigaZ it will be possible to
distinguish between the Standard Model and the 2HDM.

\subsection{Measurements of CKM-matrix elements}

TESLA can also contribute to the measurement of the CKM
matrix in W decays. There are three ways how a linear collider can access these
matrix elements:
\begin{itemize}
\item in principle the absolute values of all elements not involving
  top-quarks can be measured in W-decays;
\item the elements involving b-quarks, including phases, can be
  accessed in B-decays with the GigaZ option;
\item the elements involving top quarks can be obtained from top decays.
%  this is discussed in section \ref{physics_top}.
\end{itemize}

The absolute values of the CKM elements can be obtained from the 
partial width of the W decaying into a specific quark final state.
For this measurement both quarks in a W-decay need to be tagged. In such an
analysis \cite{ref:ckmw} W pairs and single W events can be used. c-
and b-quarks are tagged with the microvertex detector while light
quarks can be separated by identifying the leading charged hadron with
dE/dx in the TPC. 
The tagging efficiencies can be measured free of any 
assumptions on the  hadronisation mechanism during 
Z-running. Based on well know QCD scaling properties
they can be extrapolated from the Z-mass to the W-mass.

If no unitarity of the CKM matrix is imposed
the elements $|V_{\mathrm{u}i}|,\,i=\rm{d,s,b}$ can be determined with
precisions that are comparable to what is currently known \cite{ref:pdg}.
All elements $|V_{\mathrm{c}i}|$ can be determined to a better accuracy 
than present and foreseen measurements. This is in particular
true for $|V_{\rm cs}|$.

Especially $|V_{\rm cb}|$, which is important in the interpretation of 
CP-violation in 
B-decays, is also competitive with the expected precision at the b-factories. 
In any case, a measurement of the CKM matrix elements in
$\mathrm{W}$~decays will be complementary to the measurements
in heavy meson decays and can provide independent
cross-checks.  Furthermore, the errors in $\mathrm{W}$~decays
are dominated by statistical errors and the theoretical
interpretation does not involve the advanced theoretical
machinery required for the reliable evaluation of heavy meson
matrix elements.

In the GigaZ option about $4 \cdot 10^{8}$ b-hadrons are produced. The
statistics is comparable to the $\rm{e^+ e^-}$ b-factories with the
additional advantage that also ${\rm B_s}$-mesons and b-baryons are produced. 
The event sample is much smaller than at the experiments at hadron
machines, BTev and LHCb, but the events are much cleaner and all
b-decays can be triggered. 

The possibilities to measure CP-violation in B-decays have been
studied in \cite{ref:zfact}. 
Due to the high beam polarisation and the large forward-backward
asymmetry for $\rm{Z} \rightarrow \bb$-events the charge of the
produced b-quark can be tagged with high efficiency and purity 
from its polar angle only.
$\sin 2 \beta$ can be measured from the time dependent asymmetry of the decay
${\rm B}^0 \rightarrow {\rm J}/\psi {\rm K}^0_s$ and $\sin 2 \alpha$ from
${\rm B}^0 \rightarrow \pi^+ \pi^-$, where the excellent mass resolution of
the detector largely replaces particle identification.
Table~\ref{tab:cpres} compares the capabilities of TESLA for $10^9$
Z-decays with other machines. TESLA with this statistics will not
provide the best measurement in any channel, but still gives an
interesting cross check.
Furthermore the branching ratios ${\rm B}^0 \rightarrow \pi^0 \pi^0$ and
${\rm B}^+ \rightarrow \pi^+ \pi^0$, which are needed to separate penguin
contributions in the ${\rm B}^0 \rightarrow \pi^+ \pi^-$ analysis can be
measured with similar precision as at BaBar or Belle.

\begin{table}[htbp]
  \begin{center}
    \begin{tabular}[c]{|l|c|c|}
      \hline
      & $\sin 2 \beta$ & ``$\sin 2 \alpha$'' \\
      \hline
      BaBar/Belle \cite{babarbook} & $0.12$ & $0.26$ \\
      CDF \cite{mpaul}   & $0.08$ & $0.10$ \\
      ATLAS \cite{atlasp}  & $0.01$ & $0.09$ \\
      LHCb  \cite{lhcbtp} & $0.01$ & $0.05$ \\
      \hline
      TESLA & $0.04$ & $0.07$ \\
      \hline
    \end{tabular}
    \caption{Accuracy of CP violation measurements in the B system at
      different machines. The error on $\sin 2 \alpha$ is under the
      assumption that no penguin diagrams contribute to the asymmetry.
      }
    \label{tab:cpres}
  \end{center}
\end{table}

In addition to the measurement of CKM phases
the combination of luminosity, polarisation and clean environment
offers some other interesting possibilities in B-physics \cite{GigaZ-B}.

The observation of the rare $\mathrm{b}\to \mathrm{s}\nu\bar\nu$
transitions
requires a clean environment and GigaZ can provide enough luminosity
to make the measurement feasible with $\mathcal{O}(10^3)$ expected
events.  
The transition $\mathrm{b}\to \mathrm{s}\nu\bar\nu$ is of special
interest, since it is
very sensitive to Z-penguins, which receive contributions from new
physics in a wide
class of models like fourth generation, SUSY or models with an
additional ${\rm Z}'$ \cite{ref:buchalla}.
Particularly intriguing would be a deviation from the
Standard Model prediction for
$\mathrm{b} \to \mathrm{s}\nu_\tau\bar\nu_\tau$ as a signature for
anomalous couplings in the third generation.

The Standard Model predicts that $\Gamma(\mathrm{b}_R \to
\mathrm{s}_L\gamma) \gg \Gamma(\mathrm{b}_L\to \mathrm{s}_R\gamma)$,
because the Penguin diagrams for right handed light quarks are
suppressed by $\mathcal{O}(m_{\mathrm{s}}/m_{\mathrm{b}})$.  On the
other hand, contributions from physics beyond the Standard Model can
be comparable for the two decay modes.  
The helicity structure of the underlying quark decay can thus be measured
analysing the decays of polarised $\Lambda_{\rm{b}} \to \Lambda \gamma$. 

%The branching ratio for
%$\mathrm{b} \to \Lambda_{\mathrm{b}}$ of
%about 8\% easily allows the study of rare decays.  Such a
%measurement can only be performed at GigaZ.
%
At the Z-pole b-quarks are polarised with a a polarisation of $-94\%$.
About two third of this gets transferred into the polarisation of 
the $\Lambda_{\rm{b}}$.
At GigaZ about 750 decays $\Lambda_{\rm{b}} \to \Lambda \gamma$ should be seen.
In a detailed analysis it has been shown that with such a sample of fully
reconstructed events the asymmetry of the photon momentum with respect to
the $\Lambda_{\rm{b}}$-spin is sensitive to ratios between left- and
right-handed couplings in the range 0.5 and 1.9 at the
$5 \sigma$ level \cite{ref:gudrun}.

Although theoretically less clean, similar angular asymmetries in rare
hadronic 2-body decays such as  $\Lambda_{\rm{b}} \to \Lambda \Phi$
offer a unique opportunity to probe for new physics contributions to
penguin operators with chiralities opposite to those in the Standard
Model \cite{ref:gudrun}.

Polarised beauty baryons can also be used to measure
novel effects of new physics
through $\mathrm{CP}$-odd correlations in
exclusive as well as inclusive decays.

It should be noted that most results in B-physics, discussed in this
section are statistics limited. If it is found worthwhile it should
thus be possible to decrease the error by a factor of three by collecting
$10^{10}$ Zs, which can be done in a few years of running.

\subsection{Other electroweak tests at GigaZ}
\begin{figure}[htb!]
\begin{center}
\href{pictures/6/zmutau.pdf}{{\includegraphics[height=8cm]{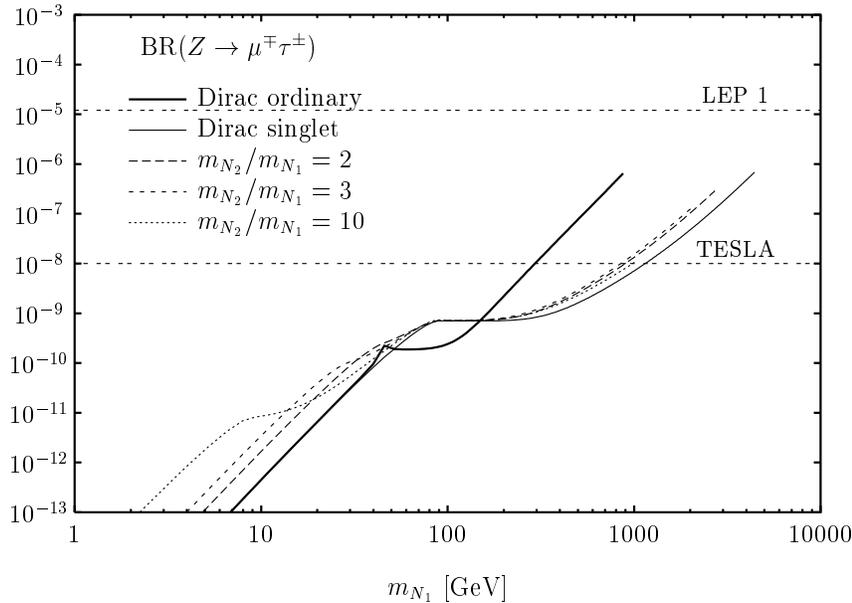}}}
\end{center}
\caption{Upper limit for BR($Z\to\mu^\mp\tau^\pm$) if
the SM is extended with: (i) one heavy ordinary (thick solid) or singlet 
(thin solid) Dirac neutrino of mass $m_{N_1}$; (ii) two heavy right-handed 
singlet Majorana neutrinos (dashed lines) with masses $m_{N_1}$ and $m_{N_2}$.
}
\label{fig:zmutau}
\end{figure}

Taking advantage of the high statistics at GigaZ a couple of other
electroweak tests are possible.
With $10^9$ events rare Z-decays can be tested. Especially for lepton
flavour violating decays
of the type ${\rm Z} \rightarrow \rm{e} \tau$
or ${\rm Z} \rightarrow \mu \tau$ the sensitivity is on the $10^{-8}$ level.
The Standard Model predictions for these decays are completely
negligible \cite{ref:tord}, but, amongst others, models with heavy extra 
neutrinos \cite{ref:tord} or several classes of supersymmetric models
\cite{ref:fcncs1,ref:fcncs2} make predictions that can be tested.
As an example Fig. \ref{fig:zmutau} shows predictions of some models
with extra neutrinos compared to the TESLA sensitivity.
The rise of the Z decay rate is proportional to the fourth power of the
leading neutrino mass scale due to symmetry breaking (neutrinos
of different generations have different masses and mix with each other).
The rise finally gets stopped by unitarity.

%Also in $\tau$-physics some improvements can be expected with $5 \cdot 10^7$
%$\tau$-pairs.
%No quantitative studies exist yet. However, due to the large boost at
%the Z-pole the $\tau$ decay products are easier to identify and due to
%the high beam polarisation also the $\tau$s are highly polarised which
%should help in the determination of the polarisation dependent Michel
%parameters $\xi$ and $\delta \xi$.

\subsection{Conclusions}
Measuring the properties of gauge bosons physics at very high scales
can be tested either through loop corrections or via effective
operators parameterising Born level effects suppressed by large masses.
With TESLA the gauge boson couplings can be measured with good enough
precision that, depending how electroweak symmetry breaking is
realised in nature, either loop effects, for example from
Supersymmetry, can be seen or signals from a strongly interacting
electroweak sector (see section \ref{physics_alternatices_WW}) are visible.

Similarly, the strongly improved precision on the Z-couplings and the
W-mass from high statistics running at lower energies allows stringent
tests of the then-Standard Model. As an example, in Supersymmetry
unmeasured parameters can be predicted in the same way as LEP and SLC
have predicted the mass of the top-quark and later the mass of the Higgs-boson.
%\bibliographystyle{unsrt}
%\bibliography{gauge_bosons_005}

\newcommand{\lrg}{\displaystyle}
\renewcommand{\be}{\begin{equation}}
\renewcommand{\ee}{\end{equation}}
\newcommand{\beq}{\begin{eqnarray}}
\newcommand{\eeq}{\end{eqnarray}}
\newcommand{\nn}{\noindent}
\renewcommand{\non}{\nonumber}
\newcommand{\lra}{\longrightarrow}
\renewcommand{\ra}{\rightarrow}
\newcommand{\tgb}{\tan\beta}
\newcommand{\stw}{\sin^2\vartheta_W}
\renewcommand{\lsim}{\raisebox{-0.13cm}{~\shortstack{$<$ \\[-0.07cm] $\sim$}}~}
\renewcommand{\gsim}{\raisebox{-0.13cm}{~\shortstack{$>$ \\[-0.07cm] $\sim$}}~}
%% Some abbrevs
%\renewcommand{\TeV}{\ensuremath{\textrm{\TeV}}}
%\renewcommand{\GeV}{\ensuremath{\textrm{GeV}}}
\renewcommand{\ab}{\ensuremath{\textrm{ab}}}
\renewcommand{\fb}{\ensuremath{\textrm{fb}}}
\newcommand{\LL}{\mathcal{L}}
\newcommand{\tr}[1]{\textrm{tr}\left[#1\right]}

%%%%%%%%%%%%%%%%%%%%%%%%%%%%%%%%%%%%%%%%%%%%%%%%%%%%%%%%%%%%%%%%%%%%%%%%%%%%%

\section{Extended Gauge Theories}

Despite its tremendous success in describing the experimental data
within the range of energies available today, the Standard Model,
based on the gauge symmetry ${\rm SU}(3) \times$ ${\rm SU}(2) \times
{\rm U}(1)$, cannot be the ultimate theory.  It is expected that in a
more fundamental theory the three forces are described by a single
gauge group at high energy scales.  This grand unified theory would be
based on a gauge group containing ${\rm SU}(3) \times {\rm SU}(2)
\times {\rm U}(1)$ as a subgroup, and it would be reduced to this
symmetry at low energies.

Two predictions of grand unified theories may have interesting
phenomenological consequences in the energy range of a few hundred
GeV:

{$(i)$} The unified symmetry group must be broken at the
unification scale $\Lambda_{\rm GUT} \gsim 10^{16}$ \GeV\  in order to
be compatible with the experimental bounds on the proton lifetime.
However, the breaking to the SM group may occur in several steps and
some subgroups may remain unbroken down to a scale of order 1\TeV.  In
this case the surviving group factors allow for {\it new gauge bosons}
with masses not far above the scale of electroweak symmetry breaking.
Besides ${\rm SU}(5)$, two other unification groups have received much
attention: In ${\rm SO}(10)$ three new gauge bosons $W^\pm_R, Z_R$ may
exist, in E$_6$ a light neutral $Z'$ in the\TeV\ range.

{$(ii)$} The grand unification groups incorporate extended fermion
representations in which a complete generation of SM quarks and
leptons can be naturally embedded.  These re\-pre\-sentations
accommodate a variety of additional {\it new fermions}.  It is
conceivable that the new fermions [if they are protected by
symmetries, for instance] acquire masses not much larger than the
Fermi scale.  This is necessary, if the predicted new gauge bosons are
relatively light.  SO(10) is the simplest group in which the 15 chiral
states of each SM generation of fermions can be embedded into a single
multiplet.  This representation has dimension {\bf 16} and contains a
right-handed neutrino.  The group ${\rm E_6}$ contains $\rm SU(5)$ and $\rm
SO(10)$ as subgroups, and each quark-lepton generation belongs to a
representation of dimension {\bf 27}.  To complete this
representation, twelve new fields are needed in addition to the SM
fermion fields.

\subsection[$Z'$ limits]{$\bZ'$ limits}

\begin{figure}[hbt]
\vspace*{-0.5cm}
\begin{center}
  \href{pictures/7/zpmin.pdf}{{\includegraphics[height=9cm]{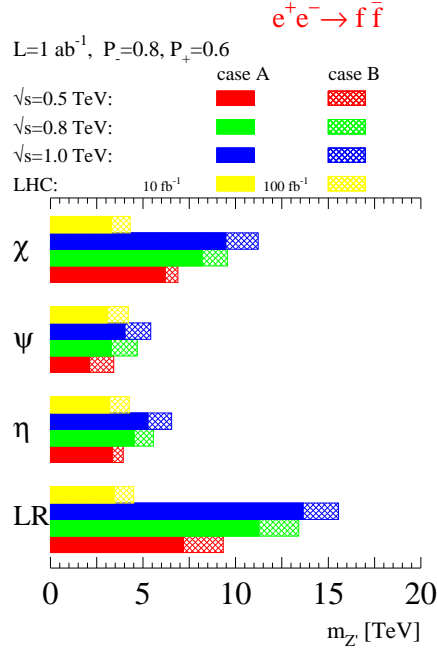}}}
  \vspace*{-0.5cm}
  \caption{\label{fig-zp-mass} \it Sensitivity to lower bounds on the Z$'$,
    Z$_R$ masses (95\% C.L.)
    in E$_6$ ($\chi,~\psi$ and $\eta$ realization) and left--right
    symmetric models \cite{riemannszp}. The integrated
    luminosity is 1000 $fb^{-1}$; the error scenarios A and B are
     described in the text.}
\end{center}
\end{figure}

\begin{figure}[[hbt]
  \vspace*{0.5cm}
\begin{minipage}[t]{7.2cm} {
\begin{center}
%  \href{pictures/7/tdr-lcoupall.pdf}{{\includegraphics[height=8cm]{tdr-lcoupall.eps}}}
  \href{pictures/7/tdr-lcoupcol.pdf}{{\includegraphics[height=8cm]{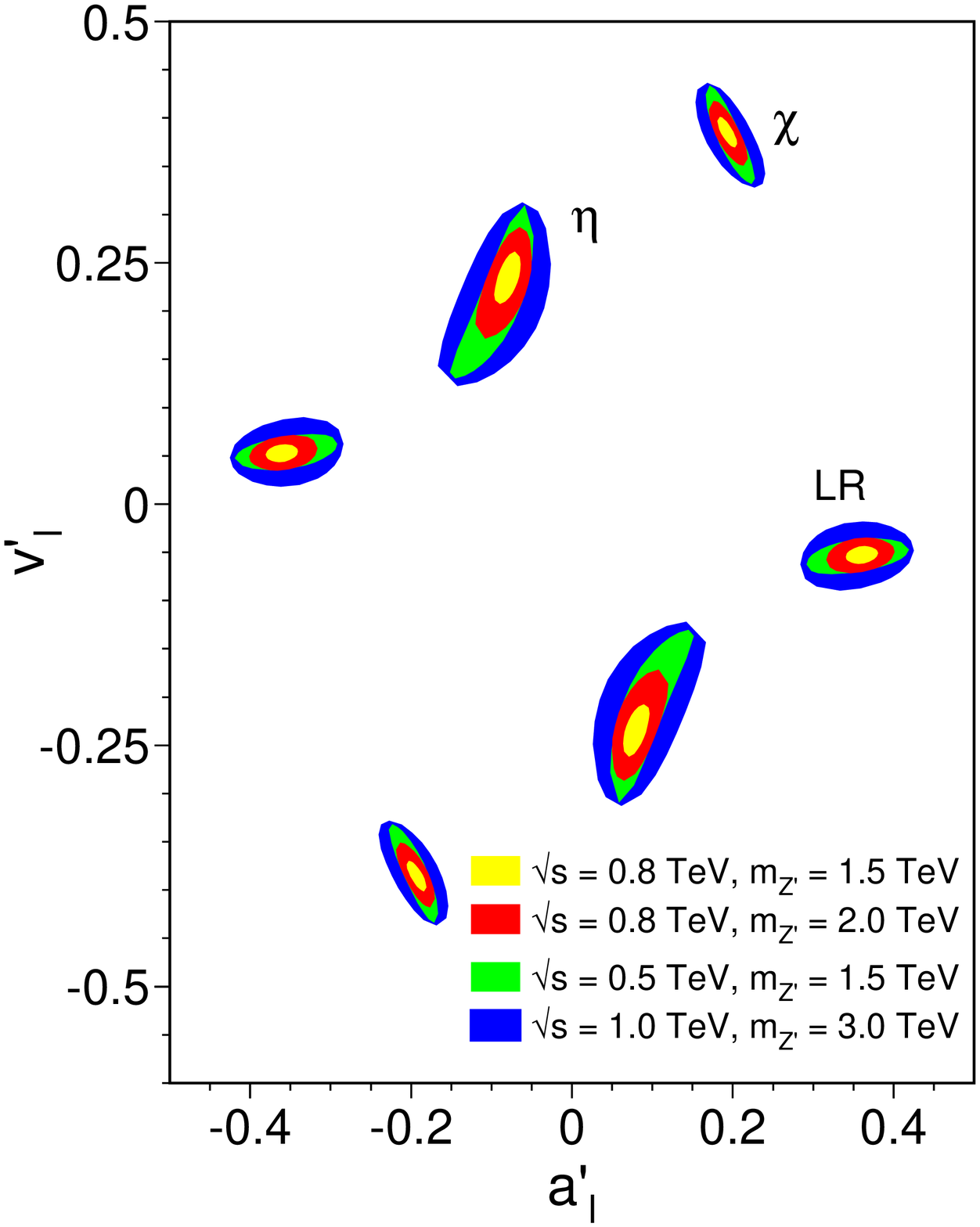}}}
  \vspace*{-0.5cm}
  \caption{\label{fig-zp-coup} \it Resolution power (95\% CL) for different
           $m_{Z'}$ based on measurements of leptonic observables at
           $\sqrt{s}$=500\GeV, 800\GeV, 1\TeV\  with a luminosity
           ${\cal L}_{int}=$1000 fb$^{-1}$ \cite{riemannszp}.
           The leptonic couplings
           of the Z$'$ correspond to the $\chi$, $\eta$ or LR model.}
\end{center} }
\end{minipage}
\hspace*{0.5cm}
\begin{minipage}[t]{7.2cm} {
\begin{center}
  \href{pictures/7/tdr-mav.pdf}{{\includegraphics[height=8cm]{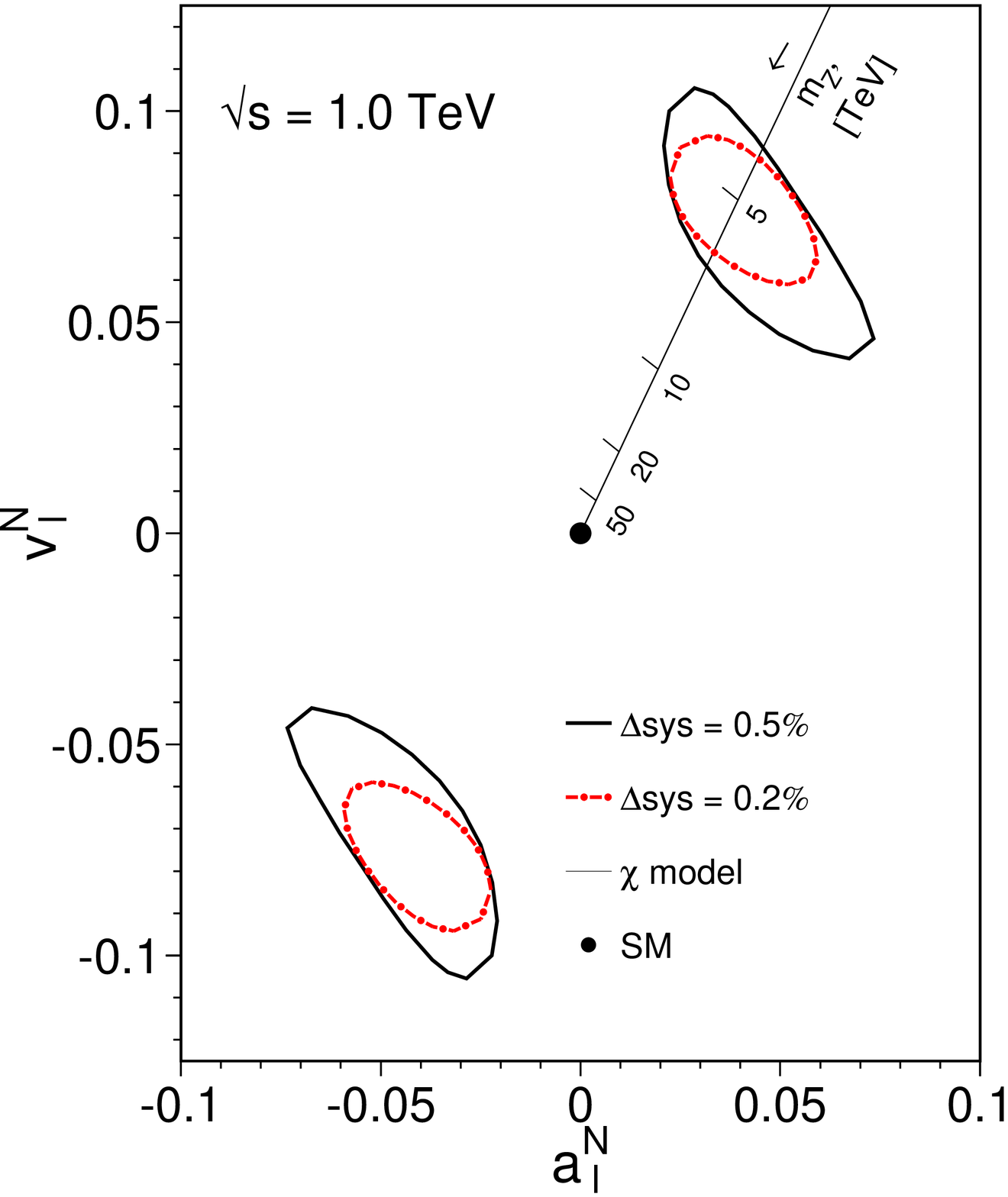}}}
  \vspace*{-0.5cm}
  \caption{\label{fig-zp-mav} \it Resolution power (95\% CL) for a Z$'$
          based on measurements of leptonic observables
          at $\sqrt{s}=1$\TeV and ${\cal L}_{int}=$1000\,fb$^{-1}$
          \cite{riemannszp}.
          The Z$'$ is exemplified in the $\chi$ model
          with $m_{Z'}$=5\TeV; the Z$'$ mass is unknown.}
\end{center} }
\end{minipage}
\end{figure}

The virtual effects of a new $Z'$ or $Z_R$ vector boson
associated with the most general effective theories which arise from
breaking ${\rm E_6} \ra {\rm SU(3) \times SU(2) \times U(1) \times
  U(1)_{Y'}}$ and ${\rm SO(10) \ra}$ ${\rm SU(3) \times SU(2)_L \times
  SU(2)_R \times U(1)}$, have been investigated
Ref.~\cite{extg_riemanns1}.  Assuming that the $Z' (Z_R)$ are heavier than
the available c.m. energy, the propagator effects on various
observables of the process
\[
e^+e^- \stackrel{\gamma, Z, Z'}{\longrightarrow} f\bar{f}
\]
have been analyzed.
Here, the sensitivity reach to detect  Z$'$ bosons is studied for  three
center-of-mass energies ($\sqrt{s}=500\GeV,~800\GeV,~1\TeV$)  
and for different scenarios of accuracy:
\begin{itemize}
\item case A: \\
$\Delta P_{e^{\pm}} =  1.0\%$,
$\Delta \cal L$ = 0.5\%,
$\Delta^{sys} \epsilon_{lepton}= 0.5\%$,
$\Delta^{sys} \epsilon_{hadron}= 0.5\%$;
\item case B: \\
$\Delta P_{e^{\pm}} = 0.5\%$,
$\Delta \cal L$ = 0.2\%,
$\Delta^{sys} \epsilon_{lepton}= 0.1\%$,
$\Delta^{sys} \epsilon_{hadron}= 0.1\%$;
\end{itemize}
An integrated luminosity of 
1000\,fb$^{-1}$ is assumed to be collected at
each centre-of-mass energy.
The polarization of electrons and positrons are 80\% and 60\%,
respectively.
The corresponding lower bounds (95\% CL) on the Z$'$ masses
are given in Figure~\ref{fig-zp-mass} in comparison to the corresponding
numbers at the LHC \cite{atlas}.
Below a Z$'$ resonance measurements of fermion-pair production are
sensitive only to the ratio of Z$'$ couplings and Z$'$ mass.
If a Z$'$ will be detected at the LHC its origin can be found by
determining  the Z$'$ couplings. Figure \ref{fig-zp-coup}
demonstrates the resolution power between Z$'$ models
assuming that the mass of the new boson is
measured at the LHC. Here, leptonic final states are considered and
lepton--universality is assumed.
If the potential Z$'$ is heavier than 4.8\TeV and/or it does not couple to
quarks, there will be no sensitivity to m$_{Z'}$ at the LHC.
Nevertheless, the analysis of fermion--pair  production could detect 
a Z$'$ and resolve the model. Instead of extracting the Z$'$ couplings
to fermions, $v'_f, ~a'_f$,
normalized  Z$'$ couplings, $a^N_f,~v^N_f$, have to be obtained:
\begin{equation}
a^N_f = a'_f \displaystyle{\sqrt{\frac{s}{m_{Z'}^2-s}}};~~~~~~
v^N_f = v'_f \displaystyle{\sqrt{\frac{s}{m_{Z'}^2-s}}}.
\label{equ-normcoup}
\end{equation}
Assuming a Z$'$ with $m_{Z'}$=5\TeV, its detection and identification
will be possible as demonstrated in Figure \ref{fig-zp-mav}.
A Z$'$ is postulated with m$_{Z'}$=5\TeV and with leptonic
couplings as suggested in the $\chi$ model. 
Measurements at $\sqrt{s} $=1\TeV\  with $\cal L$=1000 fb$^{-1}$
and accuracies of $\Delta P_{e^{\pm}}$=0.5\%, $\Delta \cal{L}$=0.5\%
and  $\Delta^{sys} \epsilon_{lepton}$=0.5\% will allow to derive
bounds on $a^N_l$, $v^N_l$ as shown by the solid line.
If  $\Delta^{sys} \epsilon_{lepton}$=0.2\% and $\Delta \cal{L}$=0.2\%
these bounds could be shrinked to the dashed-dotted line.
On the line of the $\chi$ model
the allowed area in the ($a^N_l,v^N_l)-$plane is located around the
point 
$(a^N_{\chi}(m_{Z'}=5\TeV,v^N_{\chi}(m_{Z'}=5\TeV)$.
Fixing the leptonic  Z$'$  couplings
to the $\chi$ model, a mass range of $4.3\TeV \leq m_{Z'}
\leq 6.2\TeV$ is derived for $\Delta sys =$0.5\%
and $4.5\TeV \leq m_{Z'}
\leq 5.9\TeV$ for $\Delta sys =$0.2, respectively.

It should be noted that a two-fold ambiguity in the signs of couplings
remains since all observables are bilinear products of $a'_f$ and $v'_f$.

\subsection[$W'$ limits]{$\bW'$ limits}
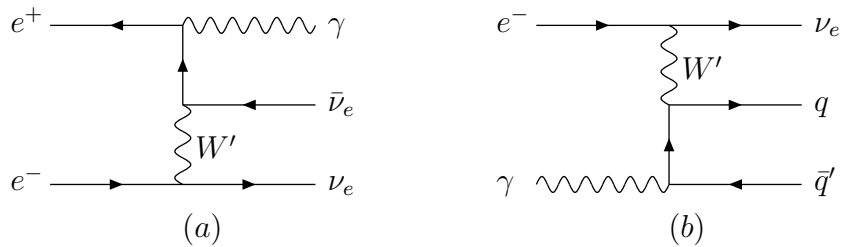
\begin{figure}[hbt]
\begin{picture}(100,85)(-80,10)
\ArrowLine(0,20)(50,20)
\ArrowLine(50,80)(0,80)
\ArrowLine(50,20)(100,20)
%\ArrowLine(100,80)(50,80)
%\Photon(50,80)(50,20){3}{6}
%\Photon(50,50)(100,50){3}{6}
%\put(55,60){$W'$}
%\put(105,48){$\gamma$}
%\put(105,78){$\bar \nu_e$}
\ArrowLine(50,50)(50,80)
\ArrowLine(100,50)(50,50)
\Photon(50,50)(50,20){3}{3}
\Photon(50,80)(100,80){3}{6}
\put(55,30){$W'$}
\put(105,78){$\gamma$}
\put(105,48){$\bar \nu_e$}
\put(-15,78){$e^+$}
\put(-15,18){$e^-$}
\put(105,18){$\nu_e$}
\put(50,0){$(a)$}
\end{picture}
\begin{picture}(100,85)(-160,10)
\ArrowLine(0,80)(50,80)
\ArrowLine(50,80)(100,80)
\ArrowLine(50,50)(100,50)
\ArrowLine(50,20)(50,50)
\ArrowLine(100,20)(50,20)
\Photon(50,80)(50,50){3}{3}
\Photon(0,20)(50,20){3}{6}
\put(-15,78){$e^-$}
\put(-15,18){$\gamma$}
\put(55,60){$W'$}
\put(105,78){$\nu_e$}
\put(105,48){$q$}
\put(105,18){$\bar q'$}
\put(50,0){$(b)$}
\end{picture} \\
\caption{\it \label{fg:wpdia}
Typical diagrams for the processes (a) $e^+e^-\to \nu\bar\nu \gamma$ and
(b) $e\gamma\to \nu q + X$.}
\end{figure}

The limits on extra charged gauge bosons shown here are based on 
the two reactions $e^+e^-\rightarrow \nu\bar\nu\gamma$ (see
Fig.~\ref{fg:wpdia}a) and 
$e\gamma\rightarrow \nu q + X$ (see Fig.~\ref{fg:wpdia}b) for three
different models: the SM-type heavy $W'$ (SSM $W'$), the left-right
model (LRM) and the SM-type Kaluza--Klein-excitation model (KK)
\cite{extg_prd61}.
The SM inputs $M_W = 80.33\GeV$, $M_Z = 91.187\GeV$, 
$\sin^2\theta_W = 0.23124$, $\alpha=1/128$ and $\Gamma_Z=2.49\GeV$ are 
used in the numerics.

Details of the $W'$ analysis based on $e^+e^-\rightarrow \nu\bar\nu\gamma$ can
be found in Ref.~\cite{extg_prd61}.
In order to take into account detector acceptance,
the photon energy, $E_\gamma$, and the angle of the photon with respect
to the beam axis, $\theta_\gamma$, are restricted to the ranges
$E_\gamma \geq 10 \,{\GeV}$ and $10^0 \leq \theta_\gamma \leq 170^0.$
These cuts also remove singularities arising for soft or collinear photons.
The photon's transverse momentum is restricted to 
$p_T^\gamma > \sqrt{s} \sin\theta_\gamma \sin\theta_v / 
(\sin\theta_\gamma + \sin\theta_v)$,
where $\theta_v$ is the minimum angle down to which the veto detectors
may observe electrons or positrons, here $\theta_v = 25$ mrad. 
This cut removes the largest background,
namely radiative Bhabha-scattering where the scattered $e^+$ and $e^-$
go undetected down the beam pipe.

\begin{figure}[hbt]
%-----------Fig.1-----------------------------
\vspace*{0.0cm}

\begin{minipage}[t]{7.2cm} {
\begin{center}
\hspace{-1.7cm}
\mbox{
\epsfysize=7.0cm
\href{pictures/7/wprer.pdf}{{\epsffile[0 0 500 500]{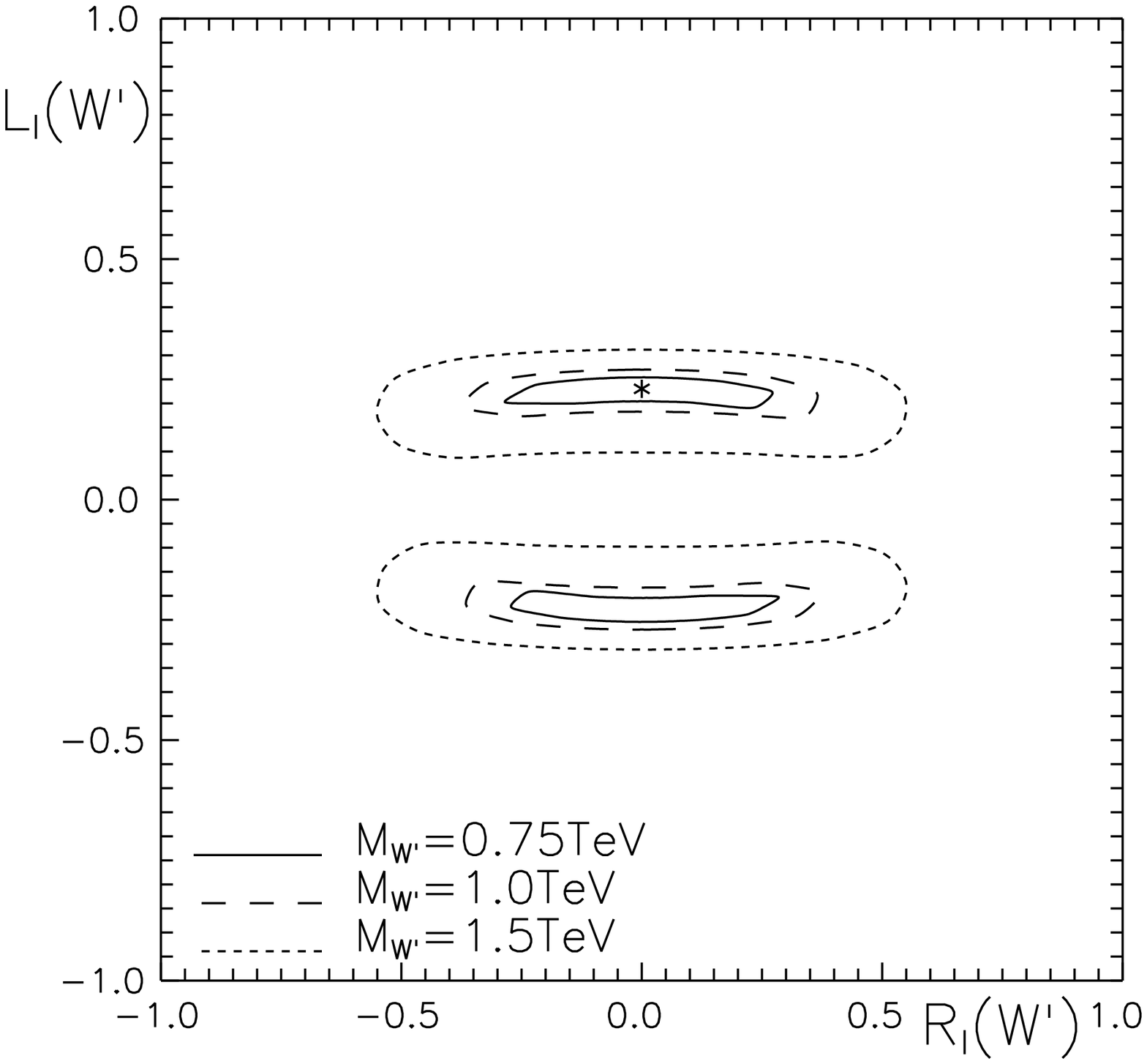}}}%
}
\end{center}
\vspace*{-1.2cm}
\caption{\it \label{fg:leike1}
95\% C.L. constraints from $e^+e^-\to \nu\bar\nu\gamma$ on couplings
of the SSM $W'$ indicated by a star for $\sqrt{s}=500$\TeV\ 
and $L_{int}=1000\,fb^{-1}$ with a systematic error of 0.5\% (0.25\%)
for $\sigma (A_{LR})$ for different $W'$ masses, see text.
}
}\end{minipage}
\hspace*{0.5cm}
% ----------------Fig. 2-----------------------
\begin{minipage}[t]{7.2cm} {
\begin{center}
\hspace{-1.7cm}
\mbox{
\epsfysize=7.0cm
\href{pictures/7/wppat.pdf}{{\epsffile[0 0 500 500]{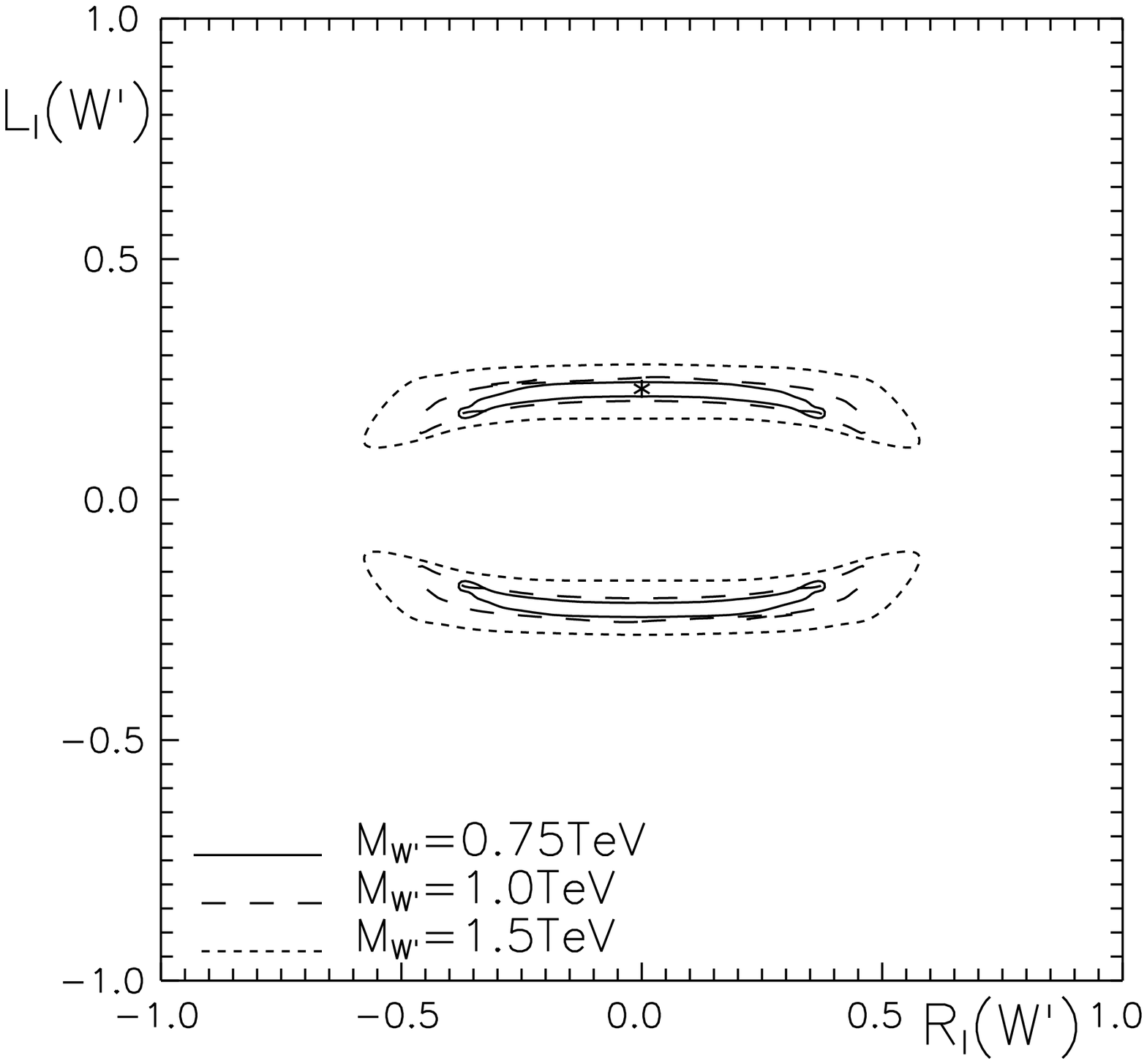}}}%
}
\end{center}
\vspace*{-1.2cm}
\caption{\it \label{fg:leike2}
95\% C.L. constraints from $e^+\gamma\to \bar \nu q+X$ on couplings
of the SSM $W'$ indicated by a star for $\sqrt{s}_{e^+e^-}=0.5\,\TeV\ $ and
$L_{int}=1000\,fb^{-1}$ with a 2\% systematic error for different $W'$ masses.
The results have been obtained by convolution with the spectrum of
Compton-backscattered laser photons.
}
}\end{minipage}
\end{figure}

Figure~\ref{fg:leike1} shows the possible constraints 
(95\% C.L.) on the right- and left-handed couplings of a $W'$ to fermions 
using the total cross section $\sigma$ and the left-right asymmetry $A_{LR}$
as observables. 
The assumed systematic errors for $\sigma (A_{LR})$ are 0.5\%(0.25\%).
80\% electron and 60\% positron polarization are assumed.
It is assumed in this figure that there exists a heavy SSM $W'$ and that
there is no signal from additional neutral gauge bosons.
The $W'$ couplings can only be constrained up to a two-fold ambiguity.
This ambiguity could be resolved by reactions where the $W'$ couples to a
triple gauge vertex.

Details of the $W'$ analysis based on  $e\gamma\rightarrow \nu q + X$ can
be found in Ref.~\cite{extg_prdxx}.
In  order to take into account detector acceptance, the angle $\theta_q$ 
of the detected quark relative to the beam axis is restricted to
$10^0 \leq \theta_q \leq 170^0$.
The quark's transverse momentum relative to the beam is restricted to
$p_T^q > 40\, (75) \GeV$ for $\sqrt{s}= 0.5(1.0)\TeV$.
This cut suppresses various SM backgrounds.

Figure~\ref{fg:leike2} shows the possible constraints 
(95\% C.L.) on SSM $W'$ couplings to fermions for backscattered laser photons. 
The best $W'$ limits come from the observable $d\sigma/dp_T^q$.
The assumed systematic error of 2\% dominates the statistical error,
thus eliminating the potential gain from high luminosities.
$W'$ limits from backscattered laser photons are considerably better than 
those from Weizs\"acker--Williams photons.
Polarized beams give only a minor improvement to $W'$ limits after
including systematic errors.
The $W'$ couplings can only be constrained up to a two-fold ambiguity.

Table~\ref{tb:leike} shows mass sensitivity limits (95\% C.L.) from both
reactions for a $W'$
predicted in the three models introduced above for different systematic errors.
All assumptions are the same as in Figures~\ref{fg:leike1} and
\ref{fg:leike2}.
The $e^+e^-$ limits on the SSM $W'$ do not improve with polarized beams.
The $e^+e^-$ limits on the $W'$ predicted in the LRM (KK) show a weak 
(considerable) improvement with polarized beams.
As mentioned before, the $e\gamma$ limits do not improve much with
polarized beams.
Backscattered laser photons give important complementary $W'$ limits 
relative to $e^+e^-$ scattering.
\begin{table}[hbt]
\begin{center}
\begin{tabular}{|l|lllllllllll|}
\hline
&\multicolumn{5}{c}{$\sqrt{s}=0.5\,\TeV, L_{int}=1000{\rm \, fb}^{-1}$} &
 \multicolumn{5}{c}{$\sqrt{s}=1\,\TeV,   L_{int}=1000{\rm \, fb}^{-1}$} &\\
&\multicolumn{3}{c}{$e^+e^-\rightarrow \nu\bar\nu\gamma$} 
&\multicolumn{2}{c}{$e\gamma\rightarrow \nu q + X$} &
 \multicolumn{3}{c}{$e^+e^-\rightarrow \nu\bar\nu\gamma$}
&\multicolumn{2}{c}{$e\gamma\rightarrow \nu q + X$} &\\
syst. error in \%&0.1&0.5&2.0&0.5&2.0&0.1&0.5&2.0&0.5&2.0&\\ 
\hline
Model &&&&&&&&&&&\\ \hline
SSM $W'$& 4.8 & 3.8 & 1.7 & 4.0 & 2.7 & 5.9 & 4.8 & 2.2 & 5.8 & 4.6 &\\
LRM     & 1.3 & 1.1 & 0.9 & 0.7 & 0.6 & 1.7 & 1.4 & 1.2 & 1.2 & 1.1 &\\
KK      & 5.0 & 4.0 & 1.8 & 5.7 & 3.8 & 6.4 & 5.1 & 2.3 & 8.2 & 6.5 &\\ \hline
\end{tabular}
\end{center}
\caption{\it \label{tb:leike} $W'$ sensitivity limits (95\% CL) in TeV,
see text.}
\end{table}

\subsection{SO(10) neutrinos and E$_6$ leptons}

\begin{figure}[hbt]
%\vspace*{-0.0cm}

%\hspace*{-1.6cm}
\begin{center}
%\hspace*{-0.6cm}
\begin{tabular}{cc}
\href{pictures/7/spira1.pdf}{{\includegraphics[height=7.5cm,angle=-90, bb= 29 175 546 695]{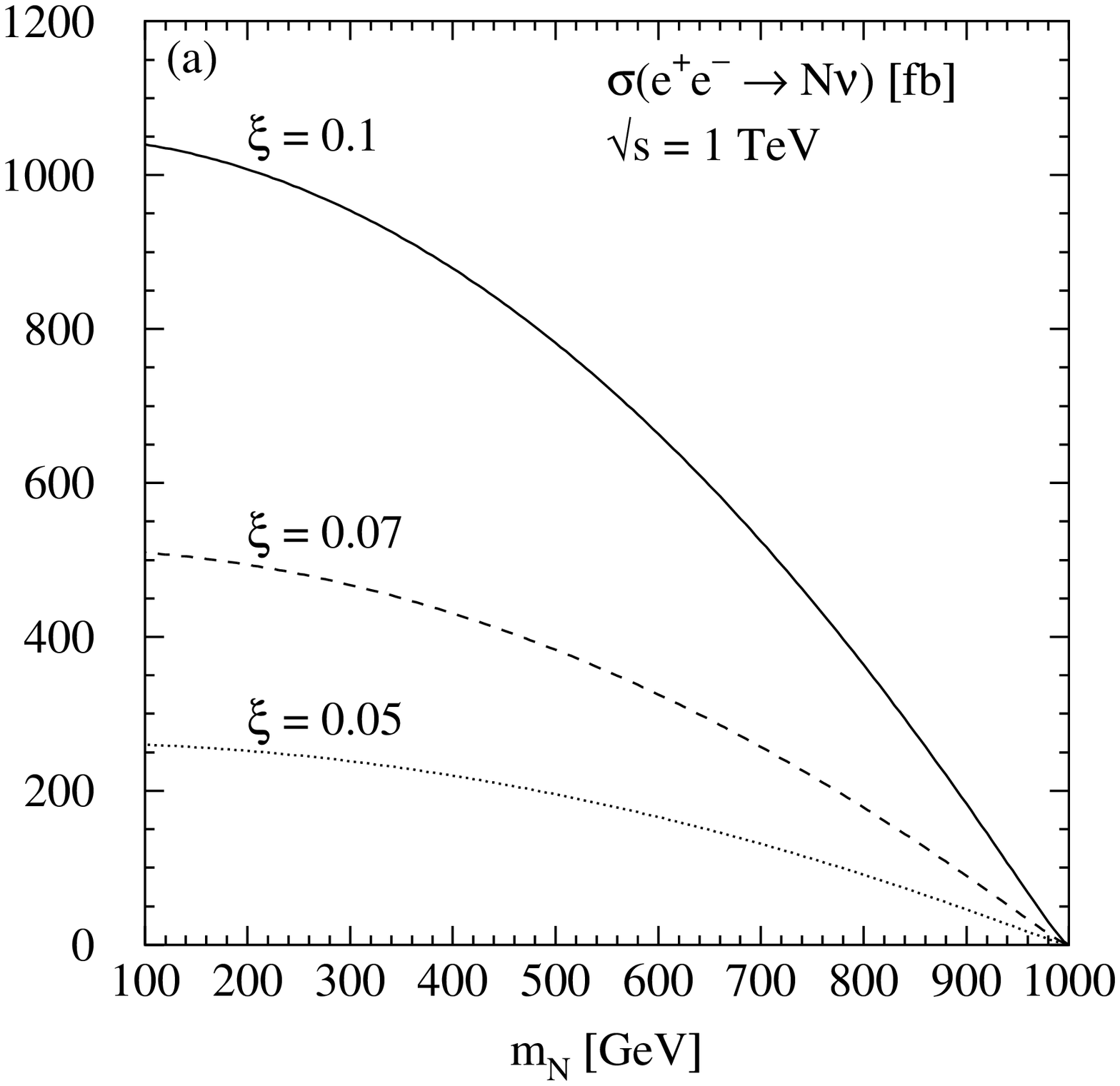}}}&
\href{pictures/7/spira2.pdf}{{\includegraphics[height=7.5cm,angle=-90, bb= 29 175 546 695]{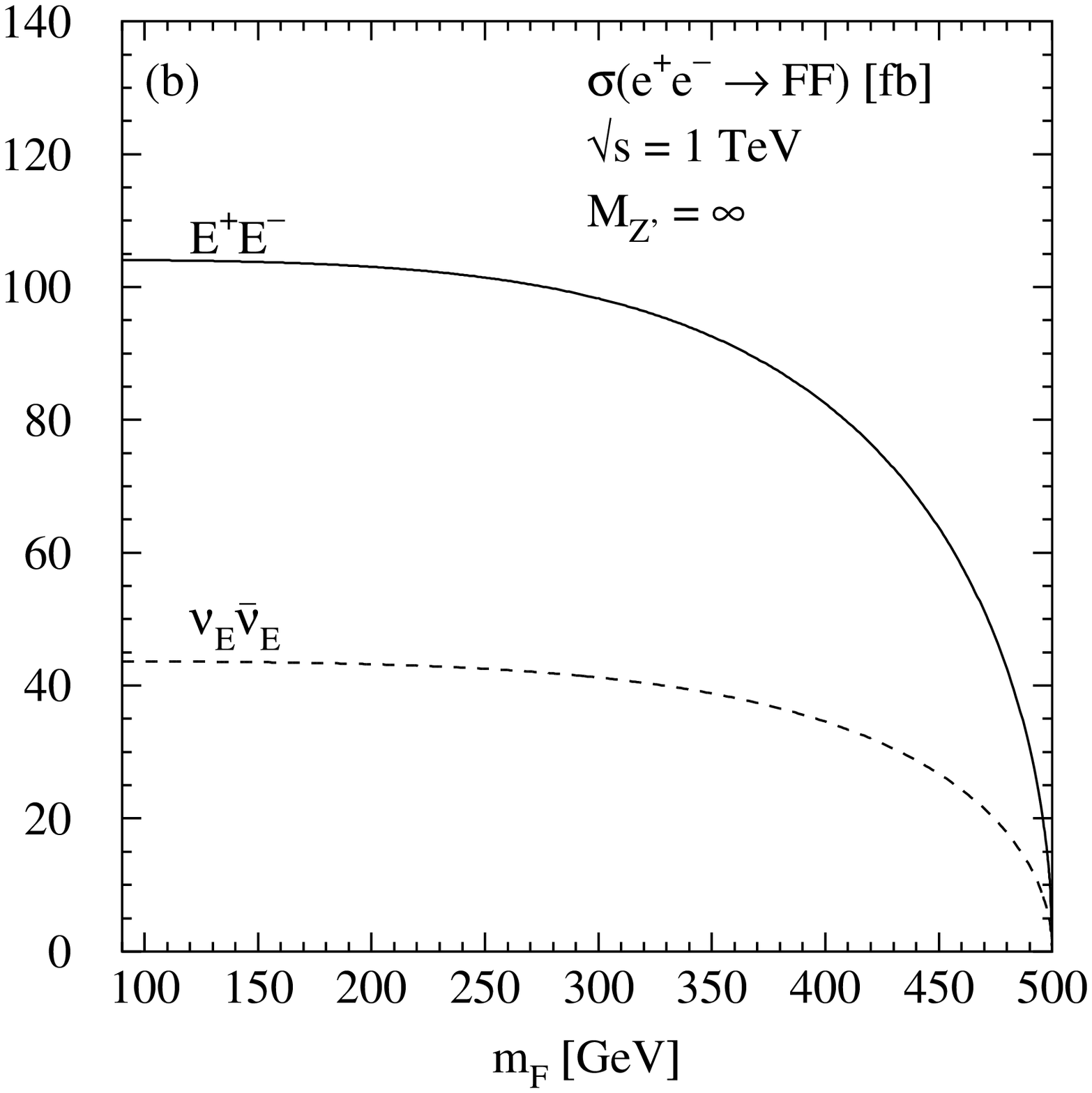}}}
\end{tabular}
\end{center}
%\begin{turn}{-90}%
%\epsfxsize=7cm \epsfbox{spira2.ps}
%\end{turn}
\vspace*{-0.5cm}

\caption[]{\label{fg:e6fer} \it Cross sections for (a) single SO(10)
Majorana neutrinos [for various mixing parameters $\xi$] and (b) heavy
$E_6$ lepton pair production [without $Z'$ exchange] at $\sqrt{s}=1$\TeV.}
\end{figure}

\noindent
{\bf i)} \underline{\it SO(10) neutrinos.}
The fundamental representation of the SO(10) gauge group contains 16
fermions, which consist of the 15 fermions of one family in the
Standard Model (SM) and a
right-handed neutrino, which is a singlet under the SM gauge group.
Mixing between ordinary and heavy right-handed neutrinos induces new
couplings, which allow for the single production of the latter. Due to
the large contribution of the $t$-channel $W$
exchange, single Majorana neutrinos can be produced with masses close to
the total c.m.~energy of the $e^+e^-$ collider; for mixing parameters not
too tiny, $\xi \gsim 10^{-2}$, the production rates are large enough for
the states to be detected \cite{extg_spira2}, see Fig.~\ref{fg:e6fer}\,a. \\

\noindent
{\bf ii)} \underline{\it E$_6$ leptons.}
Twelve new fermions are needed to complete the 27 dimensional
fundamental representation of E$_6$. They consist of two weak
isodoublet leptons, two isosinglet neutrinos, which can be either of the
Dirac or Majorana type, and an isosinglet quark with charge $-1/3$
appearing in a left- and a right-handed state:
\begin{equation}\nonumber
\left[ \begin{array}{c} \nu_E \\ E^- \end{array} \right]_L \qquad
\left[ \begin{array}{c} \nu_E \\ E^- \end{array} \right]_R \qquad
N \qquad N' \qquad D_L \qquad D_R
\end{equation}
Since the new fermions are either gauge singlets under the electroweak gauge
group or
vector-like there are no significant constraints on their masses and
couplings from precision data.
These particles can be pair produced via gauge boson exchange including
a new $Z'$ boson related to an additional abelian factor
in the gauge group structure at low energies. The cross sections for
pair production of the heavy charged and neutral isodoublet leptons are rather
large thus allowing for the discovery of these particles with masses
close to the beam energy \cite{extg_spira1}, as can be inferred from
Fig.~\ref{fg:e6fer}\,b.

\subsection[Heavy Majorana neutrinoc in $e^-e^-$ collisions]{Heavy Majorana neutrinos in $\bel^-\bel^-$ collisions}

    Throughout the TESLA physics studies, there has been the
consideration that, in its electron-electron version, this machine can
give unmistakable and irrefutable evidence for the existence of\TeV-level
Majorana neutrinos, should Nature have chosen to account in this way for
the deficiencies of the Standard Model neutral-lepton sector. Recent
evidence on possible neutrino oscillations has done little to clear up the
broader picture of {\it (i)} the Majorana vs. Dirac character of the neutrinos
and {\it (ii)} the disparity of masses of charged vs. neutral fermions in higher
symmetry schemes.

    The exchange of\TeV-level Majorana
neutrinos in high-energy scattering of left-handed electrons can lead to
clear signals in the reaction $e^-e^- \to W^-W^-$. In addition, the energy
dependence of those signals gives precise information on mass and
couplings of the exchanged Majorana neutrino. Assuming a discovery limit
of 8 events for an integrated luminosity of 500\,fb$^{-1}$ at $\sqrt{s} =
500$\GeV and 80\% polarized electron beams, the discovery potential of
TESLA ranges from about 1\TeV\  for smaller mixing
$\sim 3\cdot 10^{-4}$ to about 2\TeV\  for larger mixing $\sim 5\cdot 10^{-4}$,
corresponding to the range of upper bounds allowed by experimental
constraints \cite{extg_heusch1}. These numbers scale approximately with the
energy, i.e.~for $\sqrt{s} = 800$\GeV\  the discovery limits are raised to
1.6\TeV\  and 3.2\TeV, respectively.

It has to be examined whether {\it (a)} such a
process is excluded due to the non-observation of neutrinoless double beta
decay, and {\it (b)} the greatest sensitivity to the possible existence of heavy
right-handed singlets is offered by new and recently proposed experiments
involving large tanks of double-beta-decay candidates such as germanium.
Recently it has been demonstrated that the constraints on Majorana
neutrinos imposed by these experiments turn out to be very weak if color
effects and alternative contributions from e.g.~supersymmetric
particles or leptoquarks are properly taken into account \cite{extg_heusch1}.
TESLA is the one venue for cleanly interpretable experimentation, where
the energy and helicities of the incoming electrons provide for clean
identification and definition of the exchanged heavy neutral lepton.

\subsection{Conclusions}

Extensions of the SM gauge group $SU(3)\times SU(2)\times U(1)$ lead to
the existence of new heavy gauge bosons $Z'$ and $W'$ and novel heavy
fermions, the masses of which can be in the\TeV\  range. $Z'$ and $W'$
gauge bosons can be discovered with masses up to about 5\TeV\  at the LHC,
while TESLA exceeds the sensitivity to $Z'$ masses up to $\sim
15$\TeV\ 
and SSM $W'$ masses up to $\sim 6$\TeV\  via indirect virtual effects in
leptonic processes. The couplings of these novel gauge bosons can be
measured accurately at TESLA thanks to the
high luminosities available at TESLA. The LHC will be the
better environment for the search of new heavy quark states, while TESLA
will discover novel heavy leptons with masses up to the kinematical
limits. The comparison between TESLA and the LHC is summarized in
Table~\ref{tb:eelhcexgt}.
\begin{table}[hbt]
\renewcommand{\arraystretch}{1.5}
\begin{center}
\begin{tabular}{|l|l|l|}
\hline
Alternative & TESLA & LHC \\ \hline \hline
$Z'$ masses & $M_{Z'}\lsim 15$\TeV & $M_{Z'} \lsim 5$\TeV \\
$Z'$ couplings & ${\cal O}$(10\%) & ? \\
SSM $W'$ masses & $M_{W'}\lsim 6$\TeV & $M_{W'} \lsim 5$\TeV \\
$W'$ couplings & ${\cal O}$(10\%) & ? \\
SO(10), E$_6$ fermions & leptons & quarks \\ \hline
\end{tabular}
\end{center}
\renewcommand{\arraystretch}{1}
\caption{\label{tb:eelhcexgt} \it Comparison of TESLA and LHC for
several aspects of extended gauge theories.}
\end{table}

  %%%%%%%%%%%%%%%%%%%%%%%%%%%%%%%%%%%%%%%%%%%%
%  Chapter on TOP Physics for the TDR, 
% Version 0.6, 27.1. 2001, W. Bernreuther
%%%%%%%%%%%%%%%%%%%%%%%%%%%%%%%%%%%%%%%%%%%
%%%%%%
%%%%%%
%\newcommand{\STS}{\vspace{1mm}}
\newcommand{\STS}{\vspace{0mm}}

\newcommand{\GS}{\vspace{0mm}}

\renewcommand{\ee}{$e^+e^-$\ }
\newcommand{\VA}{V \! + \!A}
\providecommand{\tb}{\tan \beta }
\renewcommand{\LUM}{{\cal{L}}}
\newcommand{\NT}{\cdot }
\newcommand{\demi}{1\! /\! 2}
\renewcommand{\ttb}{t\overline{t}}
%%%%%
%%%%%
%\setcounter{page}{1}

\section{Top Quark Physics}
\label{physics_top}
The top quark is by far the heaviest fermion observed,
yet all the experimental results
tell us that it behaves exactly as would be expected for
a third generation Standard Model (SM) quark with charge +2/3.
In particular the direct measurement of the mass  of the top quark by the 
CDF and D0 collaborations at the
Tevatron, yielding a combined result  of $m_t = 174.3 \, \pm \,5.1$\GeV, 
is in striking agreement with the earlier SM electroweak analysis of 
data recorded at LEP and SLC \cite{top_N15}.

Its large mass, which is close to the scale of electroweak symmetry 
breaking, renders the top quark a unique object for studying 
the fundamental interactions in the attometer regime.
It is likely to play a key role in pinning down the origin of electroweak 
symmetry breaking and
in the search for clues to solve the flavour problem.
If the Higgs mechanism should be
verified then, for instance, 
 the measurement of the top Yukawa coupling  (see the section
on Higgs bosons) would help to discriminate between SM and non-SM scenarios.
High-precision measurements of the properties and interactions of top quarks 
are 
therefore mandatory at any future collider.

${\rm e}^+{\rm e}^-$ colliders are the most suitable instruments to study 
the properties of top quarks under clean 
experimental conditions. Operating the
machine at the $t\bar t$ threshold, the mass of the top quark 
can be determined with an accuracy that is an 
order of magnitude superior to measurements at hadron colliders. 
A further asset is the availability of beam polarisation which is a 
powerful  tool in precision studies of the neutral and charged 
current interactions
of the top quark, both at threshold and in the continuum.
These studies include  accurate determination of the ``static" properties 
of  top quarks,  its vector and axial vector couplings and 
its magnetic and
electric dipole moment, as well as  measurement of   
the charged-current couplings in the  main decay channel. 
Moreover, decays of the top quark into
novel particles, as predicted by extensions of the
Standard Model, for instance into 
charged Higgs bosons and/or stop particles
may be observed. The top-Higgs Yukawa
coupling is best determined  in the reaction $e^+e^- \to t {\bar t} H$
discussed in section 2.

Since the lifetime of the $t$ quark is much shorter than the typical 
hadronisation time
set by the  scale $\Lambda^{-1}_{QCD}$, 
top quark production and decay can be analysed within perturbative QCD
 \cite{top_Bigi:1986jk}.
Unlike the case
of light quarks the properties of the top quark, in particular its spin
 properties, are reflected directly
in the distributions of the jets, $W$ bosons, or leptons into which 
the $t$ and $\bar t$ decay. This additional
important distinctive feature of top quarks will open up the rich 
phenomenology refered to above.

\subsection{Profile of the top quark: decay modes}
\label{physics_top_profile}

\noindent
a) {\it The Dominant SM Decay.} The channel $t \rightarrow b + W^+$
is the dominant top quark decay mode, not only in the Standard 
Model but also in
extended scenarios.  In the SM  the total top width $\Gamma_t$ is, 
for all practical purposes,
equal to the partial width of this decay mode. To lowest order
\begin{equation}
\Gamma (t \rightarrow b + W^+) =
\frac{G_F {\mid V_{tb}\mid}^2 m^3_t}{8 \sqrt{2} \pi}
\left[ 1 - \frac{m^2_W}{m^2_t} \right]^2
\left[ 1 + 2 \frac{m^2_W}{m^2_t} \right] \, .
\end{equation}
A large fraction, $p_L = m^2_t / (m^2_t + 2 m^2_W) \approx
0.7$, of the decay $W$ bosons are longitudinally polarised.  The
proportionality of $\Gamma_t$ to the third power of $m_t$ is due
to the fact that in the SM  the longitudinal $W$ component, dominating 
for large $t$
masses, is to be  identified with the charged Goldstone boson, the
coupling of which grows with the $t$ mass.  The width of the top quark
is known to second-order QCD \cite{top_twqcd} and first-order
electroweak corrections \cite{top_twew}.   Numerically 
$\Gamma_t/{\mid V_{tb}\mid}^2 =
1.39 $\GeV\  for $m_t = 175$\GeV\  (pole mass).

The direct measurement of the top quark width is difficult.  The
most promising method appears to be the extraction of the width from 
 the
forward-backward asymmetry of $t$ quarks near the \ee production
threshold.  This asymmetry which is generated by the overlap of parity-even
$S$-- and parity-odd $P$--wave production channels  is 
sensitive to the width $\Gamma_t$.  Including the other threshold
observables, cross section and momentum distributions, a precision of about
10 \% can be expected for the measurement of $\Gamma_t$ in total
\cite{top_N18}. A more precise knowledge of $\Gamma_t$, which should
eventually be feasible,  would allow
an accurate determination of the CKM matrix elements $\mid V_{tq}\mid$
via measurement of the respective branching ratios \cite{top_LettMatt}.

\noindent 
{\it Chirality of the $(tb)$ decay current}.  The
precise determination of the weak isospin quantum numbers does not
allow for large deviations of the $(tb)$ decay current from the
standard $V \!\! - \!\! A$ structure.
Nevertheless, since
$\VA$  admixtures may grow with the masses of the quarks involved ($\sim
\sqrt{m_t / M_X}$ through mixing with heavy mirror quarks of mass
$M_X$, for instance), it is necessary to check the chirality of the
decay current directly.  The $l^+$ energy distribution in the
semileptonic decay chain $t \rightarrow W^+ \rightarrow l^+$ depends
on the chirality of the current. For $V \!\! - \!\! A$ couplings it is
given by $dN/dx_l \sim x^2_l (1 - x_l)$.  Another SM prediction, which is important for 
helicity analyses, is  
that the charged lepton in the semileptonic (or the d-type quark in non-leptonic)
decays of polarised top quarks is the best analyser of the top spin \cite{top_tspind}:
$dN_{pol}/dx_l d\Omega_l \sim dN/dx_l (1+{\bf s}_t\cdot {\bf \hat p}_l)$, 
where  ${\bf s}_t$ and ${\bf \hat p}_l$ are the top polarisation and 
the lepton direction of flight in the top rest frame, respectively.

A deviation from the
standard $V \!\! - \!\! A$ current would change this distribution;
in particular it would  stiffen the energy spectrum and it
would lead to a non-zero value  of the energy
distribution at the upper end-point. Extrapolating the analysis of \cite{top_18A}
to the present TESLA design luminosity gives an experimental 
 sensitivity to possible $V \!\! + \!\! A$
admixtures (corresponding to the form factor $F_{1R}^{W}$ that measures
 $(V \! + \! A)/(V \! - \! A)$)
which is listed  in {Table}~\ref{fig:top_topform}). 

\begin{figure}[ht]
\begin{center}
%\hspace*{-6cm}
\href{pictures/8/topfig1.pdf}{{\includegraphics[width=9cm]{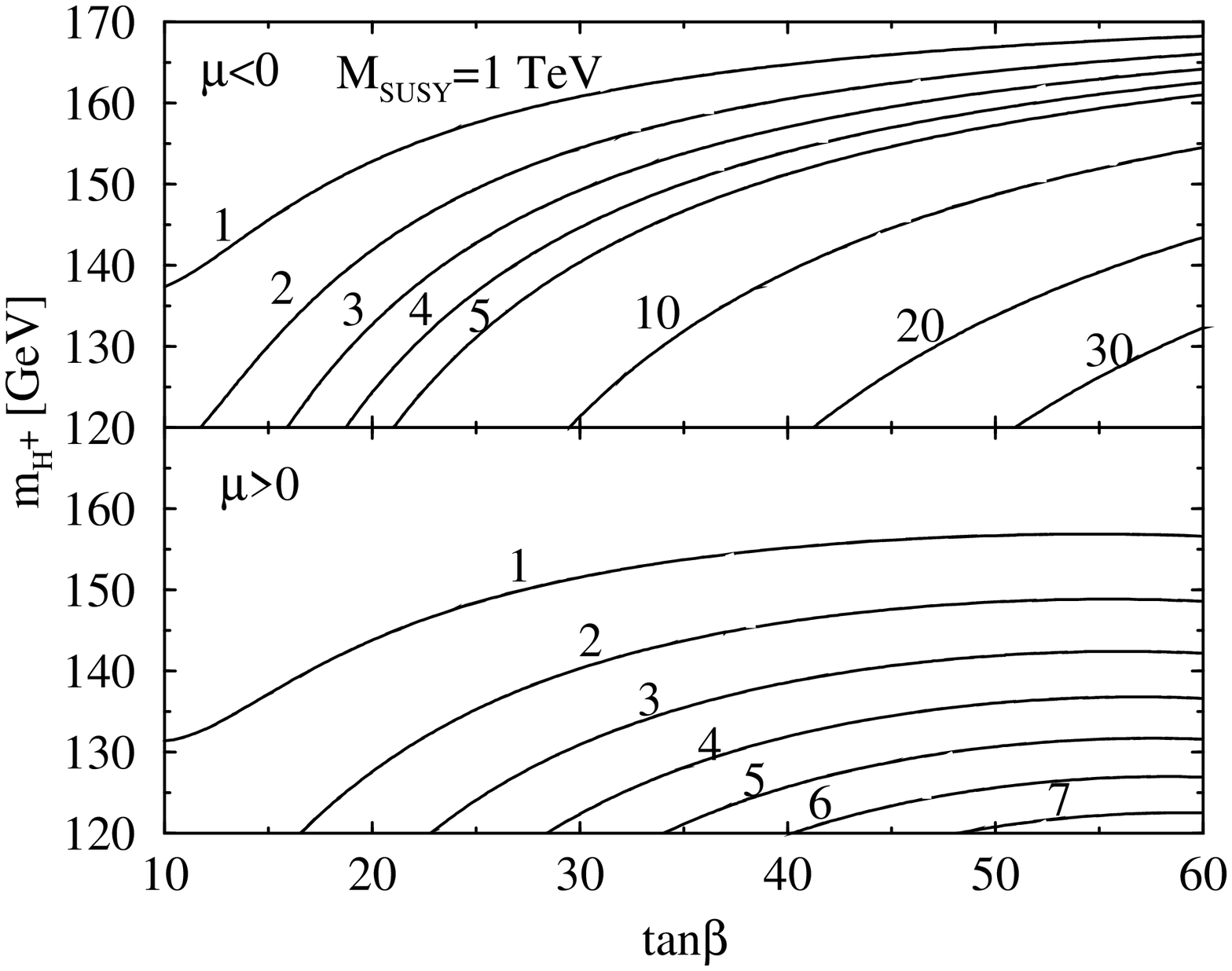}}} \\ 
\vspace*{8mm}
\href{pictures/8/topfig2.pdf}{{\includegraphics[width=8.8cm]{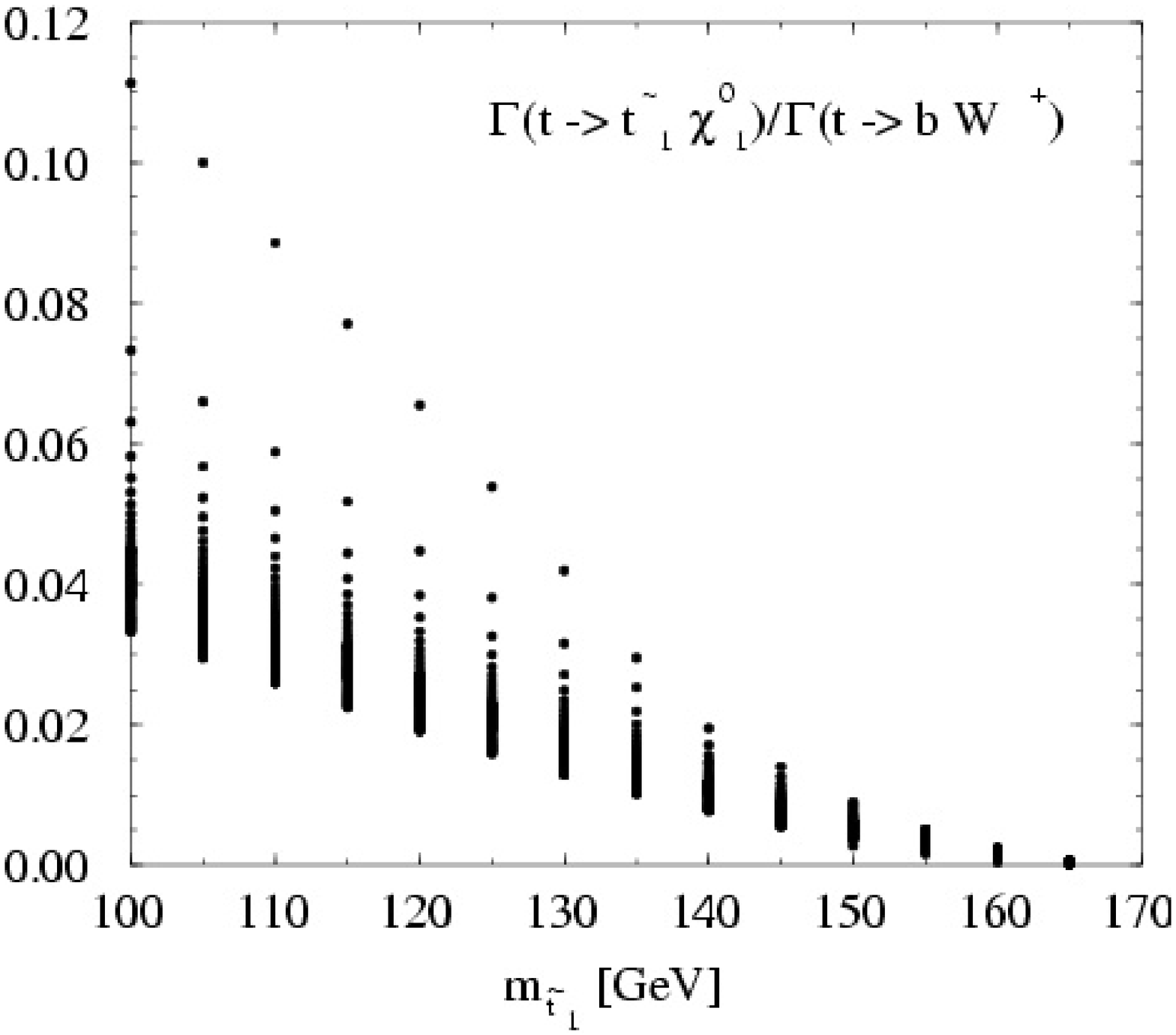}}}
%\epsfig{file=topfig1.eps,width= 6cm} 
%\vspace*{-12cm}
%\hspace*{-6cm}
%\epsfig{file=topfig2.eps,width= 6cm}
\end{center}
\vspace*{-.5cm}
\caption[]{ 
a) Contour of fixed branching ratios (in percent) of 
top quark decays
to charged Higgs bosons in supersymmetric theories, for two characteristic
sets of parameters \cite{top_N19}.
Also shown is  the range of charged Higgs boson masses 
as a function
of the coupling $\tan\beta$ that can be detected experimentally
for a given luminosity. b) Range of the branching
ratio of top quark decay  to a stop
particle and the lightest neutralino in supersymmetric theories \cite{top_N19}.
\protect\label{fig:top_f4}\label{fig:top_f5}}
\end{figure}

\GS
\noindent
b) {\it Non--Standard Top Decays.} 
Such decays could occur, for example, in supersymmetric extensions of
the Standard Model: top decays into charged Higgs bosons and/or into
stop particles and neutralinos:
\begin{equation}
t  \rightarrow  b + H^+ \hspace{+2mm} \, , \hspace{+2mm}
t \rightarrow  \tilde{t} + \tilde{\chi}^0_1 \,.
\end{equation}
If kinematically allowed, branching ratios for these decay modes could
be as large as ${\cal O}(10\%)$ for the Higgs and several percent
for the SUSY decay   Fig.~\ref{fig:top_f4} \cite{top_N19}, given the 
present constraints on  supersymmetric parameters.  The signatures for these
decay modes are very clear and they are easy to detect experimentally
\cite{top_N20}.  The subsequent decays of charged Higgs bosons $H^+$  
manifest themselves through 
decays to $\tau^+\nu_{\tau}$ and  $c\bar s$  with rates 
which are different from the universal $W$ decay rates in
the Standard Model, thus breaking $\tau$ {\it vs.} $e, \mu$
universality. If this decay exists it will first be seen at a hadron 
collider: perhaps at the Tevatron or eventually at the LHC. 
Nevertheless, at a Linear Collider additional important insight 
could be obtained
into the coupling strength of the charged Higgs boson and its properties 
by measuring
the branching ratio of this mode \cite{top_N20}. 

If neutralinos are the lightest
supersymmetric particles, they escape undetected in stop decays, so that a
large amount of missing energy would be observed in these decay modes. 
At a high luminosity linear collider this channel can be detected down to 
a branching fraction of slightly less than 1 percent \cite{top_N20}.

\par
 Besides breaking the $V \!\! - \!\! A$ law for the chirality of
the $ t \rightarrow bW$ decay current, mixing of the top quark with
other heavy quarks would break the GIM mechanism if the new quark species
do not belong to the standard doublet/singlet assignments of isospin
multiplets.  As a result, FCNC ($tc$) couplings of order $\sqrt{m_t
  m_c / M^2_X}$ may be induced.  FCNC $t$ quark decays, for example
$t \rightarrow c \gamma$ or $cZ$, may therefore occur at the level of
a few permille; down to this level they can be detected experimentally
\cite{top_20A}. The large number of top quarks produced at the LHC 
allows, however, to search for rare FCNC decays with clean signatures, 
such as $t \rightarrow cZ$, down to a branching ratio of less than
$10^{-4}$.

\STS
\subsection[Threshold production: the top mass]{Threshold production: 
the top mass}
\label{physics_top_threshold}

Quark-antiquark production near the threshold in \ee collisions is
of great interest as it offers a unique way to investigate the
bound-state dynamics of strongly interacting particles.
The long lifetime of the lighter quarks  allows the
strong interactions to build up rich structures of bound states and
resonances.  For the  top quark with its large mass and width, 
the picture is different: 
the decay time of the states is shorter than the revolution time of the
constituents so that toponium resonances can no longer form
\cite{top_Bigi:1986jk}.   Nevertheless, the remnants of the
toponium $S$-wave resonances induce a fast rise of the cross
section near the threshold.  The steep rise provides by far the best method
for high-precision measurements of the top quark mass. In comparison to
the reconstruction of the invariant  mass of jets originating from 
a single top quark  at future hadron
colliders the LC threshold method is superior by  an order of magnitude.
This method has the advantage that the cross section for the
production of a colour singlet $t \bar t$ state is analysed.
Infering the  mass of a coloured object like the top quark from the 
invariant mass of colour singlet final states recorded in the detector
necessarily has a larger uncertainty. 

 Why should it be desirable to measure the top mass with high
precision? Two immediate reasons can be given: \\
{$(i)$} The top mass is an important ingredient for the 
electroweak precision analyses at the quantum level \cite{top_N30}.
Suppose a Higgs boson has been found
at the Tevatron, the LHC
and/or the Linear Collider, and
its mass is  known from
 direct measurement.
If the $W$ and top quark masses are known to high precision then
the SM consistency checks, which at present provide an indirect  
 determination of
the mass of the Higgs particle,
will be substantially tightened. 
In the case of TESLA 
 the Higgs mass can finally be
extracted from the high-precision electroweak observables to an
accuracy of about 5\,\% as shown in Fig.~\ref{fig:gz_fit}.  
This would provide
the most stringent test of the Higgs mechanism
 at the quantum level.\\
{$(ii)$} The Standard Model provides no understanding
of the disparate quark and lepton mass spectra, 
and no answer to the question whether and how the fermion masses 
and mixing angles  are linked to each
other.  This deficiency
might be removed by a future theory of flavor dynamics, still to be 
discovered. The top quark, endowed with the heaviest
mass in the fermion sector, will very likely play a key role in this
context.  In the same way as present measurements test the relations
between the masses of the electroweak $W, Z$ vector bosons in the
Standard Model, similar relations between lepton and quark masses will
have to be scrutinized in the future. \\

The $t\bar t$ excitation curve can be predicted by
perturbative QCD \cite{top_N25,top_N26,top_N27} because  the rapid $t$ decay 
restricts the interaction region of the top
quark to small distances. The interquark potential is given
essentially by the short distance Coulombic part.

The excitation curve is built up primarily by the superposition
of the $nS$ states.  At leading order in the
non-relativistic expansion  of the total cross section
this sum can conveniently be performed by using
Green function techniques and the Schr\"odinger equation:
\begin{equation}
\sigma (e^+e^- \rightarrow t \overline{t})_{thr} =
\frac{6 \pi^2 \alpha^2 e^2_t}{m^4_t}
{\cal I}m \, G (\vec{x}= 0; E + i \Gamma_t) \, .
\end{equation}

\begin{figure}[ht]
\begin{center}
\href{pictures/8/topfig3.pdf}{{\includegraphics[height=14cm,bb= 80 260 430 720]{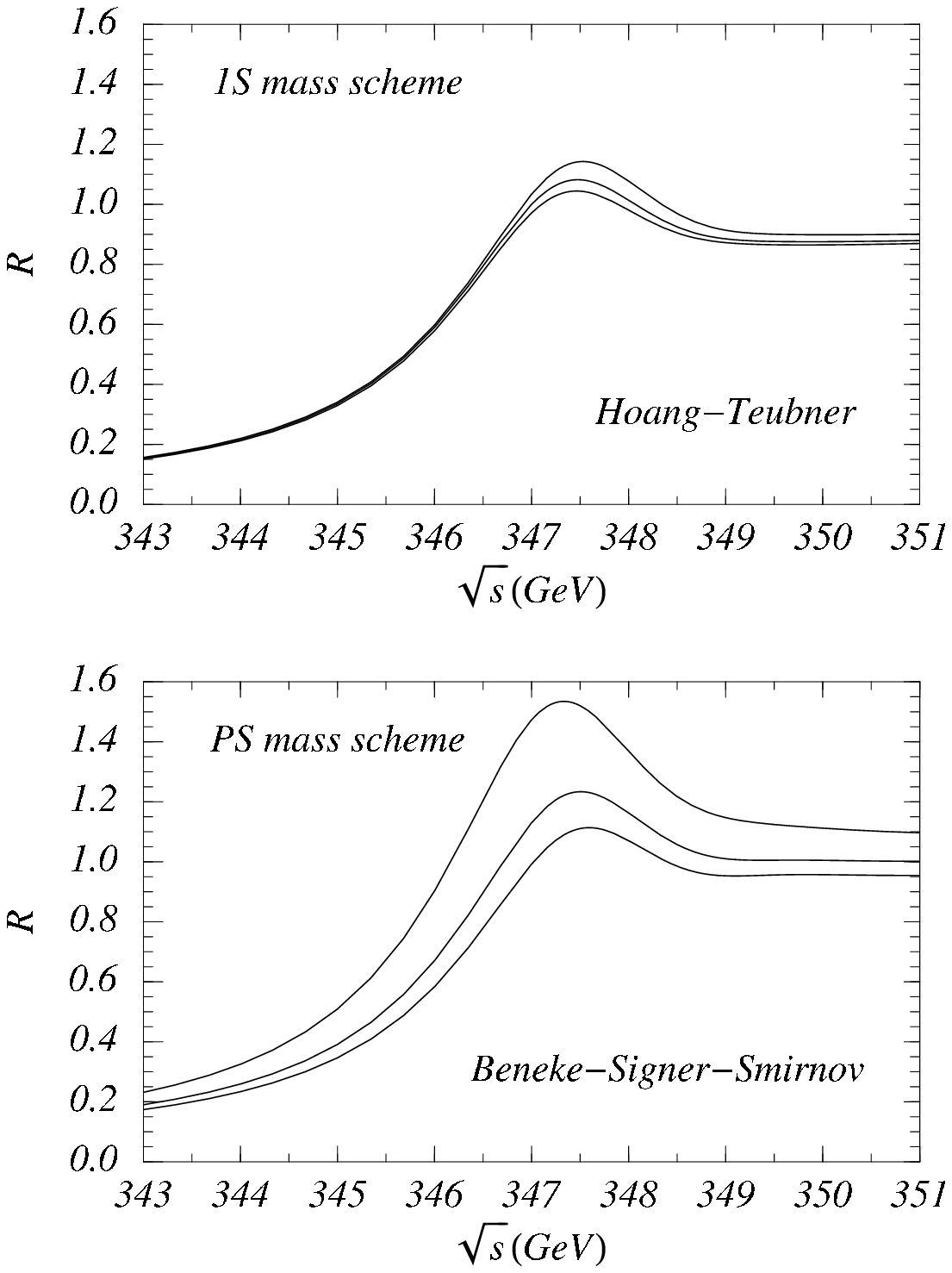}}}
\end{center}
\vspace*{-0.5cm}
\caption[]{\label{fig:top_f6}
The total normalised photon-induced
cross section $R$ at NNLO
for several top mass schemes  \cite{top_Hoang:2000yr}. For a given scheme
the curves refer to three different renormalisation scales 
 $\mu_{soft}$ = 15, 30, 60\GeV.}
\end{figure}

The form and the height of the excitation curve are very sensitive to
the mass of the top quark,
Fig.~\ref{fig:top_f6}a.  Since any increase of the $t$ quark 
(pole) mass can be
compensated by a rise of the QCD coupling, which lowers the energy
levels, the measurement errors of the two parameters are positively
correlated. 

\begin{figure}[ht]
\begin{center}
  \href{pictures/8/topfig4.pdf}{{\includegraphics[height=9cm]{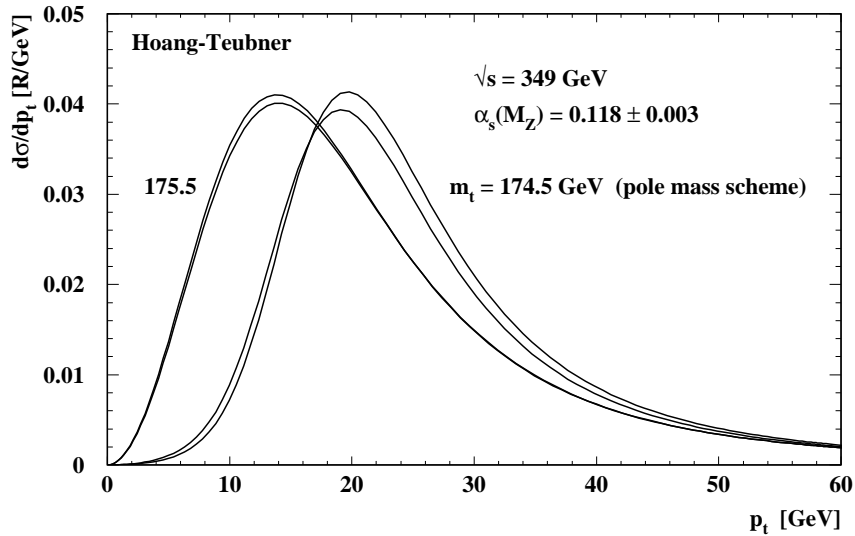}}}
%\hspace*{-6cm}
%\epsfig{file=topfig3.eps,width= 12cm} \\
%\vspace*{-5cm}
%\hspace*{-6cm}
%\epsfig{file=topfig4.eps,width= 9cm}
\end{center}
%\vspace*{-9.cm}
\vspace*{-0.5cm}
\caption{ The momentum spectrum of the top quarks near
  the threshold for a fixed total c.m. energy as
given in Ref. \cite{top_hoangTeu2}.  The momentum depends
 strongly on the top mass, yet less on the QCD coupling.
\protect\label{fig:top_sigdp}}
\end{figure}

The correlation between the top mass and the QCD coupling
can partially be resolved by measuring the
momentum of the top quark \cite{top_N27} which is reflected in the
momentum distribution of the decay $W$ boson. 
The $t$ momentum is
determined by the Fourier transform of the wave functions of the
overlapping resonances:
\begin{equation}
\frac{d \sigma}{d P_t} = \frac{3 \alpha^2 e^2_t}
{\pi s} \frac{\Gamma _t}{m^2_t}
| \hat{G} (P_t, E + i \Gamma_t) |^2
\end{equation}
The top quarks will have average
momenta of order  $\sim  \alpha_s m_t/2$. 
Together with the uncertainty $\sim \sqrt{\Gamma_t m_t}$ due to
the finite lifetime, this leads to average momenta $<P_t>$ of about 
15\GeV\  for $m_t \sim 175$\GeV, see Fig.~\ref{fig:top_sigdp}.

\begin{figure}[ht]
\begin{center}
  \href{pictures/8/topfig5.pdf}{{\epsfig{file=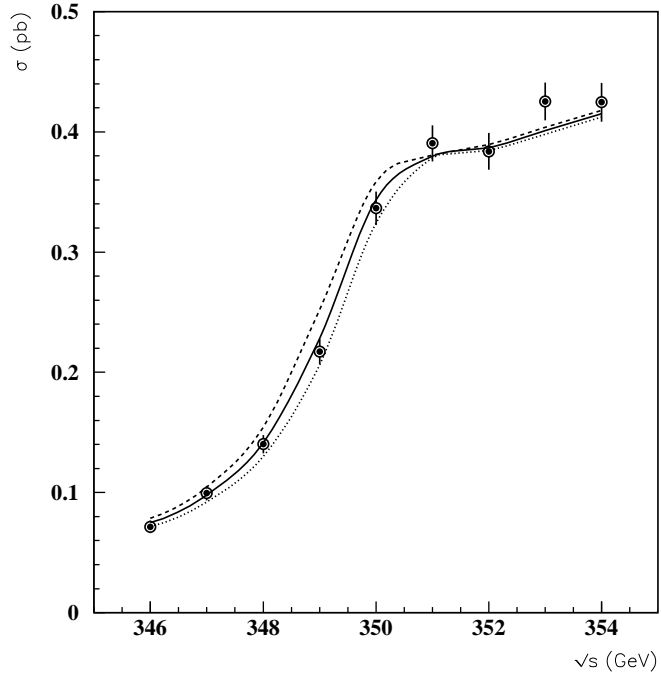,width=10cm}}}
\end{center}
\vspace*{-.5cm}
\caption[]{ 
  Excitation curve of $t\bar t$ quarks including initial-state
  radiation and beamstrahlung \protect\cite{top_excit}.  
  The errors of the data points
  correspond to an integrated luminosity of $\int \LUM$ = 100 fb$^{-1}$.
   The dotted curves indicate shifts of the top mass by
  $\pm$ 100\,MeV. \label{fig:top_f7}}
\end{figure}

Recently the next-to-next-to-leading order (NNLO) QCD corrections to the 
total cross section
were calculated by several groups \cite{top_NNLO}. 
The corrections to the location of the threshold and the shift in the
height of the cross section were found to be large. Top quark mass
definitions,  so-called threshold masses, were suggested \cite{top_Beneke98} 
in this context
that stabilise the location of the threshold with respect to the NNLO
corrections.
 These threshold masses can be extracted from  data 
with high accuracy and may 
then be converted into the frequently used  
$\overline{\mbox{MS}}$ mass parameter  of the top quark \cite{top_ChSMR}.
(For a detailed  comparison of the
different approaches and an assessment of theoretical uncertainties, 
see \cite{top_Hoang:2000yr}.) 

\begin{figure}[ht]
\begin{center}
\href{pictures/8/topfig6.pdf}{{\epsfig{file=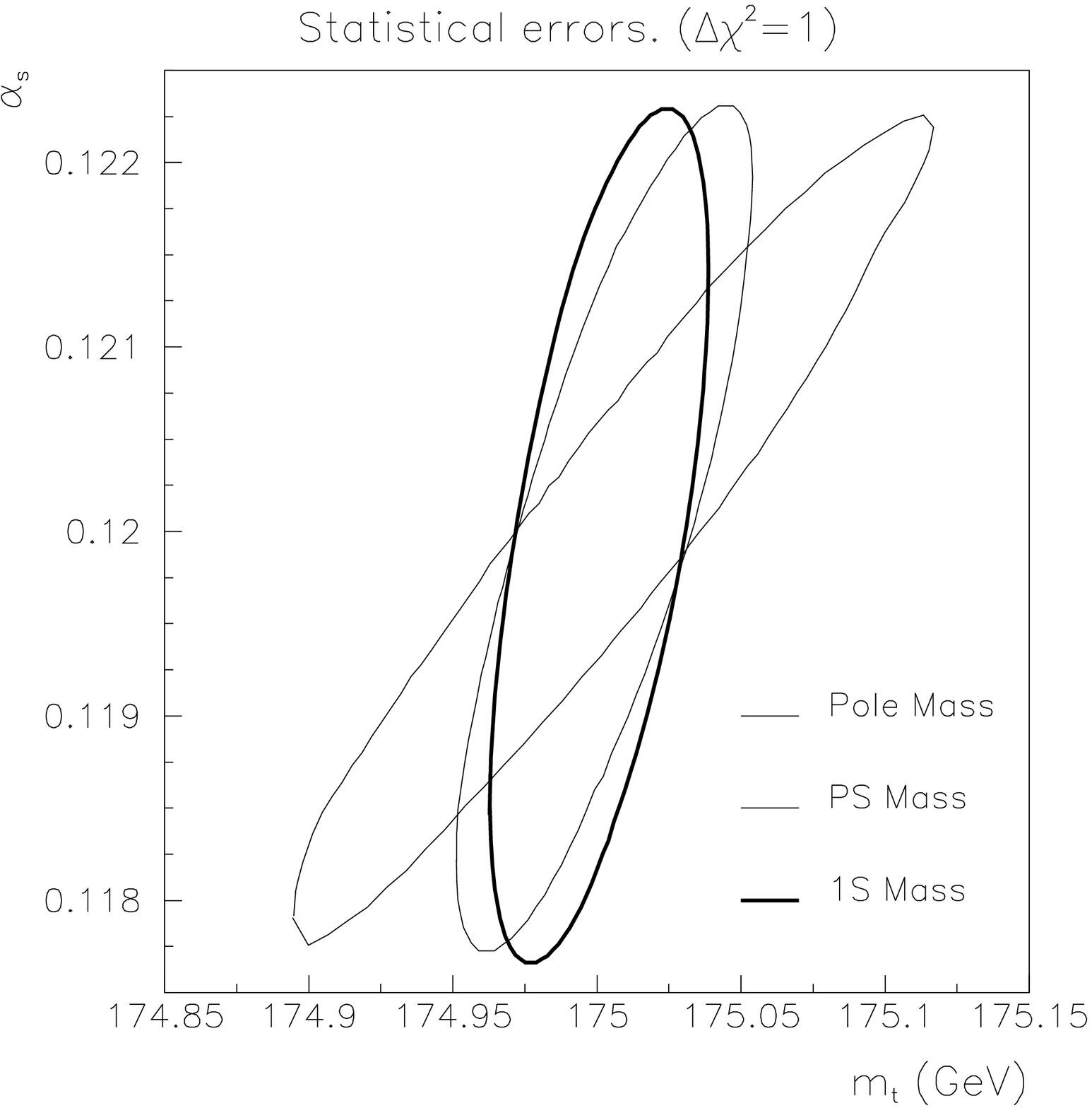,width=10cm}}}
\end{center}
\vspace{-.5cm}
\caption[]{ Statistical errors on $\alpha_s$ and the top mass 
 resulting from a 2-parameter fit to simulated data of the 
$t \bar t$ excitation curve, using the NNLO cross-section predictions with the
top threshold mass parameters indicated. Here $m_t$
denotes the top mass in one of the conventions given in the figure.
An integrated luminosity of 
100 fb$^{-1}$ was assumed.
From Ref. \protect\cite{top_excit}.  \protect\label{fig:top_fig7par}}
\end{figure}

Fig.~\ref{fig:top_f7} shows the simulation of a scan of the $t \bar t$ 
cross section
in the threshold region, including 
the effects of initial-state
radiation and beamstrahlung, with 9 energy locations \cite{top_excit}.
A two-parameter fit using the NNLO cross-section formulae yields
the experimental sensitivity to the top mass and to 
the QCD coupling shown in 
Fig.\ref{fig:top_fig7par}. 
This figure  shows that only when the top mass is defined
in the pole mass scheme there  is a strong correlation with $\alpha_s$ -- but not for 
threshold masses.
With the NNLO
formulae that use threshold mass parameters, the following 
statistical errors can be achieved:
$(\delta m_t)_{stat}  \approx  50\MeV$, 
$(\delta \alpha_s)_{stat}  \approx 0.0024 $.
These errors
 were derived for an integrated luminosity of $\int
{\LUM}$~$= 100\,\makebox{ fb}^{-1}$. The theory
errors 
in the determination
of $m_t$ and $\alpha_s$ were estimated 
by varying the renormalisation scale and by
comparing NNLO with NLO results \cite{top_excit}. This analysis
shows that
the two-parameter fit is  problematic in that
 $\alpha_s$ absorbs almost all the uncertainties in
the normalisation of $\sigma_{t\bar{t}}$, and these theoretical
uncertainties are large.
Hence determining  $\alpha_s$ from the $t\bar t$ cross section
at threshold appears not to be the best procedure for the time being.
However, a recent renormalisation-group improved calculation
of the threshold cross section \cite{top_hmst} leads to a considerable
reduction of the theoretical uncertainty in the normalisation of the cross 
section down to
2-3\%. The increase of the cross section from the exchange of a
115\GeV\  
Standard Model Higgs boson \cite{top_HarJeKu}
amounts to 5-8\% and would give direct access to the top-Higgs Yukawa
coupling.

Rather than using the top threshold scan to determine
$\alpha_s$ it is more effective to constrain its
value to the current world average and fit
the data to  $m_t$ alone.
With this strategy the statistical errors of the top mass determination
decrease, being now 30\,MeV for
the $m_t^{1S}$ and 40\,MeV for the $m_t^{PS}$ mass, while the theory
errors, estimated as above, 
increase to  110\,MeV for the $m_t^{1S}$ and to 180 MeV for the
$m_t^{PS}$ mass.  The larger theory error reflects the fact that
due to the correlation between 
$\alpha_s$ and $m_t$ the normalisation uncertainites are now shifted, into 
some extent, to the top mass.

A detailed  assessment of the theoretical error  must take into account
the  uncertainties due to  i) different methods used in the NNLO calculations, 
ii) the dependence of  a given NNLO cross-section formula, 
for a fixed convention with respect to  the QCD coupling,
on the definition of the
top mass parameter and on the renormalisation scale, iii)
 uncalculated higher order corrections.
Ref. \cite{top_Hoang:2000yr} estimates
that these uncertainties lead to an error on the 
$\overline{\mbox{MS}}$ mass of
$ (\delta m_t)_{th}  \approx  100\MeV$.

\STS For the top mass measurements at the LHC a sensitivity of 
about  1 -- 2\GeV\  was
estimated  \cite{top_lhc00}, based on 
the reconstruction of top quarks
from jet and lepton final states.  Smearing effects due to soft stray
gluons which are coherently radiated off the $t$ quark before 
the decay and off
the $b$ quark after the decay, add to the complexity of the
analysis.  Thus, \ee colliders will improve our knowledge on the
top-quark mass by at least an order of magnitude.

\vspace{1.cm}
\noindent {\it Polarised Beams and the Top Threshold} \cite{top_LCNoteKu}:
Close to threshold beam polarisation will be particularly useful to
determine the weak couplings responsible for top production and
decay. The left right asymmetry \cite{top_ARL} $A_{RL}$ can be
measured to an accuracy of 0.01, if an integrated luminosity of 
100\,fb${}^{-1}$ is distributed evenly among the four combinations of beam
polarisations ($|P_{e^-}|=0.8$ and $|P_{e^+}|=0.6$ will
be assumed throughout). If 300\,fb${}^{-1}$ 
are invested specifically into the L-R and R-L combinations of
electron-positron helicities, even $\delta A_{RL} =0.004$ can be
achieved. Such a measurement would determine the vector
coupling of top quarks to a relative precision of 2\% (or even 0.8\%)
and thus become sensitive to the quantum corrections \cite{top_27A}.  

Both top quarks and antiquarks will be highly polarised, with $P_t =
P_{\bar t} = 0.98$ and 0.88 respectively for the L-R and R-L helicity
combinations. For the decay channel $t\to b W(\to \bar\ell\nu)$
this allows to constrain the coefficient $\alpha$ of the
lepton angular distribution \cite{top_tspind} 
${\rm d} N \propto (1+\alpha {\bf s_t \cdot\hat p_\ell}) {\rm d}\Omega_\ell$ 
to better than 0.02 and 0.008, respectively, for the two assumptions 
on the luminosity. By the same line of reasoning the (energy dependent)
angular distribution of neutrinos can be analysed and the combination of
these measurements will lead to tight limits on anomalous couplings of
the top quark \cite{top_JezK94}.

Polarised beams would also play an important role in the study of the
angular distribution and the transverse and the normal polarisation of
the top quarks. These are sensitive to the axial coupling of the top
quark and to the $t\bar t$ potential through rescattering corrections
\cite{top_27A}.

Last but not not least, 
top quark polarisation will be extremely useful for the
analysis of non-standard decays. The parameter $\alpha_H$ in the angular
distribution of a charged Higgs boson \cite{top_JHKze88} 
${\rm d} N \propto (1+\alpha_H {\bf s_t \cdot \hat p_H}) {\rm d}\Omega_H$ 
from the decay $t\to H^+ b$ can be measured to 0.04 (or even
0.016) which will lead to important constraints on the handedness of the
Yukawa coupling.

\STS
\subsection[Continuum production and $t$ form factors]{Continuum 
production and $\bt$ form factors}
\label{physics_top_continuum}

\begin{figure}[ht]
%\vspace*{-2.cm}
%\vspace*{-5.cm}
\begin{center}
\href{pictures/8/topfig7.pdf}{{\epsfig{file= 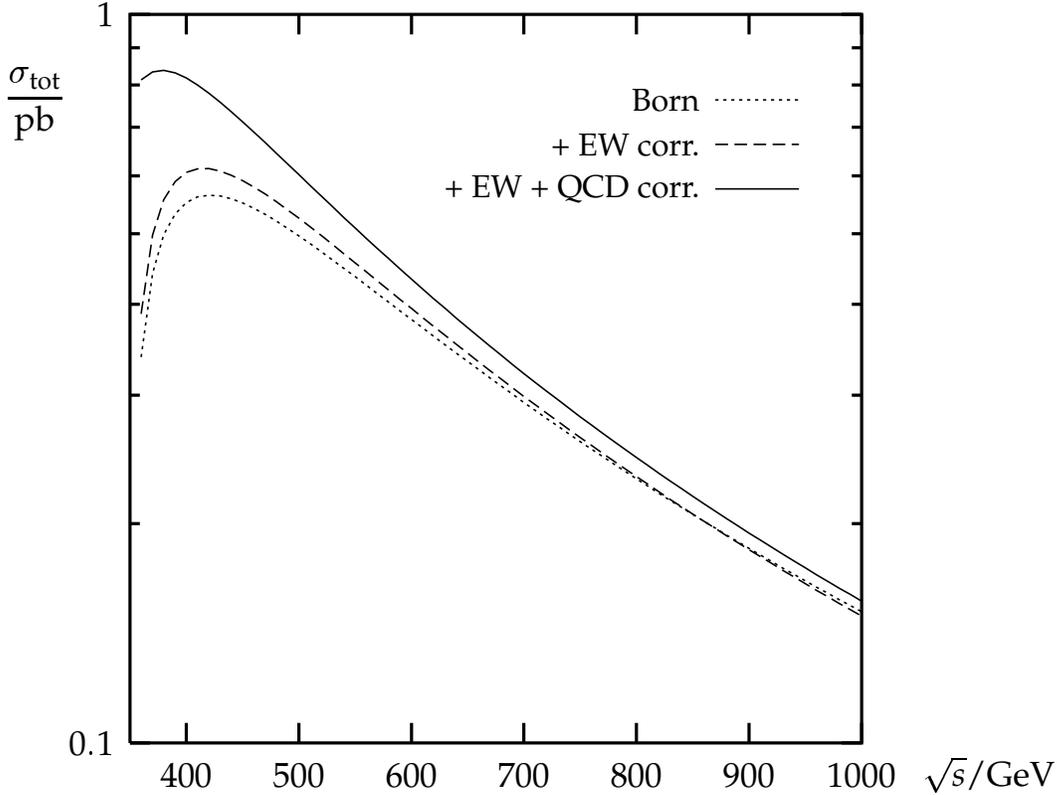, width=14cm}}}
\end{center}
%\vspace{-2.cm}
\caption[]{
  The cross section for the production of top-quark pairs in the
  continuum as a function of the c. m. energy to lowest order, to
 1-loop order in the electroweak (EW) corrections \cite{top_Beenakker:1991ca},
  and including the combined order $\alpha_s$ \cite{top_N21}, $\alpha_s^2$ 
\cite{top_KAal2} QCD and EW corrections \cite{top_KuHaHar}.
  The mass $m_t$ = 175\GeV\  (pole mass) is used.  \protect\label{fig:top_5t} }
\end{figure}

The main production mechanism for top quarks in \ee collisions is the
annihilation channel
\[
e^+e^- \stackrel{\gamma, Z}{\longrightarrow} t \overline{t} \, .
\]
Extensive theoretical knowledge has been gained about the total cross section
for this reaction: The QCD corrections were determined to 
order $\alpha_s^2$ \cite{top_KAal2}
and the electroweak SM corrections to 1-loop order \cite{top_Beenakker:1991ca}, including 
the hard photon corrections \cite{top_hardgcor}. The 1-loop
quantum corrections to the lowest-order cross section were also computed for 
the minimal supersymmetric extension of the SM \cite{top_Hollik:1999md}.
As shown in Fig.~\ref{fig:top_5t}, the cross section
 is of the order of 1 pb so that top quarks will be produced at
large rates in a clean environment at \ee linear colliders; about
300,000 pairs for an integrated luminosity of $ \int {\LUM} \sim$ $300
\makebox{ fb}^{-1}$.
  
\noindent
The production and decays of top quarks are not significantly affected by the
non-perturbative effects of hadronisation. Moreover, the
perturbative QCD corrections are small
for  the continuum production of  $t\bar t$ pairs
in a general spin configuration \cite{top_tspinp}, for the decay of
 polarized top
quarks \cite{top_tspind}, and for the QCD rescattering
corrections sufficiently away from threshold \cite{top_BeChap}. 
Therefore the helicities of the top
quarks can be determined from the distribution of the jets and leptons
in the decay chain $t \rightarrow b + W^+ \rightarrow b+ f \bar{f}'$ 
and the neutral and charged-current interactions  of the
top quark can be measured with good accuracy.

An obvious question is whether the top quark has non-standard
couplings to gauge bosons. Possible anomalous couplings of the top
quark to $\gamma,Z,W$ bosons can be parameterized by means of
an effective Lagrangian or, alternatively, by using form factors.
The form factors of the top quark 
in the electromagnetic
and the weak neutral currents are the Pauli--Dirac form factors
$F_{1V}^{\gamma, Z}$ and $F_{2V}^{\gamma, Z}$, the axial form 
factors $F_{1A}^{\gamma, Z}$,
and the CP-violating form factors $F_{2A}^{\gamma, Z}$ 
(see e.g. \cite{top_N22,top_Bernreuther:1992be}).
Suffice it to mention that the physical object is the
S-matrix element: in some models, for instance in supersymmetric SM 
extensions,
the new physics contributions to the
${e^+}{ e^-} \to t{\bar t}$ amplitude are not confined to the $Vt\bar t$
vertices  \cite{top_Hollik:1999md,top_Hollik:1999vz}. \\
 By convention, to lowest order in the SM
the chirality-conserving
form factors 
$F_{1V}^{\gamma, Z}$ and $F_{1A}^Z$ are 
normalized to unity, whereas $F_{2V}^{\gamma, Z}$ and $F_{2A}^{\gamma, Z}$, which 
are chirality-flipping,  
vanish to this order. Anomalous values could be a consequence of
strong-interaction-type electroweak symmetry breaking scenarios or of
composite quark structures. Detectable values of the  
electric-type dipole moments $F_{2A}^{\gamma, Z}$ would be 
evidence for a new CP-violating interaction beyond the
Kobayashi-Maskawa mechanism. 
Because the above form factors are functions of the time-like
energy variable  $s$ they can have, apart from dispersive also
non-zero absorptive parts if (new) physics thresholds are crossed. 
 For the $t\bar t$ production vertex
 this amounts to
real and imaginary parts, respectively. 
Depending on how the
form factors vary with $s$ their contribution to the
cross section or to other observables can increase  with the c.m. energy.\\
The general amplitude for the decay $t\to Wb$ contains
four (complex) form factors, two chirality-conserving
and two chirality-flipping ones, respectively. Here we consider only
the form factor $F_{1R}^{W}$ which signifies a 
$V$+$A$ admixture, and 
 ${\mbox Im} F_{2R}^{W}$ which, if found to be non-zero, would be
evidence for non-SM CP violation \cite{top_Bernreuther:1992be}. 
(For further studies, see \cite{top_tformf}.)\\
 Among the ``static'' parameters, that is, the above
form factors of the top quark which can be
determined only at \ee linear colliders, the following 
examples are of particular interest: \\
{\it $Z$ charges of the top quark}.  The form factors
$F_{1V}^Z, F_{1A}^Z$
can be determined from the $t\bar{t}$ production cross sections
with $e^-_L$ and $e^-_R$ beams and the left-right asymmetry
\cite{top_L27A,top_grahik}.  Moreover, the production of top quarks near the
threshold with longitudinally polarised beams leads to a sample of
highly polarised quarks.  The small admixture of transverse and normal
polarisation induced by $S$--wave/$P$--wave interference, is extremely
sensitive to the axial $Z$ charge $a_t$ of the top quark \cite{top_27A}.
Some models of strong-interaction-type electroweak symmetry breaking
predict rather large anomalous contributions, $\delta F_{1V}^Z$
up to 10 $\%$ \cite{top_Murayama:1996ec}. 
 \\
{\it Magnetic dipole moments of the top quark}.  These form
factors are generated already in the SM at the quantum level.
Gluon exchange induces a term 
$ F_{2V}^{\gamma, Z}\sim \alpha_s/\pi$
and the  interactions of a light Higgs boson would lead to 
a contribution of similar size. If the
electrons in the annihilation process $e^+e^- \rightarrow
t\overline{t}$ are left-handedly polarised, the top quarks are
produced preferentially as left-handed particles in the forward
direction while only a small fraction is produced as right-handed
particles in the backward direction \cite{top_N23}.  As a result of this
prediction in the Standard Model, the backward direction is most
sensitive to small anomalous magnetic moments of the top quarks.  The
anomalous magnetic moments can thus be searched for  by
measuring the angular dependence of the $t$ quark cross section
\cite{top_N23,top_18A}.

\begin{table}[ht!]
\begin{center}
\begin{tabular}{|l|c||cc|cc|} \hline
\rule[-3mm]{0mm}{8mm} Form factor
 & SM value  & \multicolumn{2}{|c|}{$\sqrt s$  = 500\GeV} &  
\multicolumn{2}{|c|}{$\sqrt s$ = 800\GeV }     \\
\hline
\hline
&\rule[-3mm]{0mm}{8mm} & $p$ = 0 & $p$ = -- 0.8 & 
$p$ = 0 & $p$ = -- 0.8\\
\hline
% $F_{1V}^{\gamma}$ &\rule[-3mm]{0mm}{8mm} 1 
% &  &   &   &    \\
$F_{1V}^{Z}$ &\rule[-3mm]{0mm}{8mm} 1
&  & 0.019 &  &     \\
$F_{1A}^{Z}$ &\rule[-3mm]{0mm}{8mm} 1
&  & 0.016  &  &     \\
% $F_{1A}^{\gamma}$ &\rule[-3mm]{0mm}{8mm} 0
% &  &  &   &     \\
$F_{2V}^{\gamma,Z}={(g-2)^{\gamma,Z}}_t$ &\rule[-3mm]{0mm}{8mm} 0
& 0.015 &0.011  & 0.011  & 0.008    \\
${\mbox Re}\, F_{2A}^{\gamma}$ &\rule[-3mm]{0mm}{8mm} 0
& 0.035  &0.007  & 0.015  &  0.004   \\
${\mbox Re}\, d_t^{\gamma}$ [10$^{-19}$ e cm] &\rule[-3mm]{0mm}{8mm} 0
& 20  & 4  & 8  &  2  \\
${\mbox Re}\, F_{2A}^{Z}$ &\rule[-3mm]{0mm}{8mm} 0
& 0.012  &0.008  & 0.008  &  0.007   \\
${\mbox Re}\, d_t^{Z}$ [10$^{-19}$ e cm] &\rule[-3mm]{0mm}{8mm} 0
& 7  & 5  & 5  &  4  \\
${\mbox Im}\, F_{2A}^{\gamma}$ &\rule[-3mm]{0mm}{8mm} 0
& 0.010  &0.008  & 0.006  &  0.005   \\
${\mbox Im}\, F_{2A}^{Z}$ &\rule[-3mm]{0mm}{8mm} 0
& 0.055  &0.010 & 0.037 &  0.007   \\
\hline
$F_{1R}^{W}$ &\rule[-3mm]{0mm}{8mm} 0
& 0.030 & 0.012 &   &     \\
${\mbox Im} F_{2R}^{W}$ &\rule[-3mm]{0mm}{8mm} 0
& 0.025  &0.010 &  &     \\
\hline
\end{tabular}
\caption[]{
{1 s.d. statistical sensitivities to some (non) SM form factors
in $t\bar t$ production \cite{top_18A,top_N24,top_rfrey} and in $t$ decay 
to $Wb$ 
\cite{top_18A,top_Bernreuther:1992be}. 
The second column contains
the respective SM value to lowest order, $p$ denotes the 
polarisation of the electron beam.
For the c.m. energy $\sqrt s$  = 500\GeV (800\GeV) an 
integrated luminosity of 300 fb$^{-1}$ 
(500 fb$^{-1}$) was used.
$F_{1R}^{W}$ measures $(V+A)/(V-A)$.} \label{fig:top_topform} }
\end{center}
\end{table}

\noindent
{\it Electric dipole moments of the top quark}.  Electric
dipole moments $d^{\gamma, Z}_t = e F_{2A}^{\gamma, Z}/2 m_t$
of detectable size can be
generated only by new  CP-violating interactions. If a light 
neutral Higgs boson ($m_h \lesssim$ 160\GeV) with
undefined CP parity exits, its reduced scalar and
pseudoscalar couplings to top quarks could be of order 1
which leads to CP-violating form factors that can be sizeable
not too far away from the $t\bar t$ threshold  
\cite{top_Bernreuther:1992dz}: at $\sqrt s$ = 370\GeV\ ,
Re$F_{2A}^{\gamma} \approx$ 
Im$F_{2A}^{\gamma}
\approx $ 2 - 3 $\%$, and $F_{2A}^{Z}\approx
0.34 F_{2A}^{\gamma}$. At a high luminosity LC
these effects could be measured with (optimized)
CP-odd observables \cite{top_Bernreuther:1992be,top_N24,top_jeznasu}.
The  exchange of supersymmetric particles involving 
new CP-violating phases \cite{top_N24,top_Abartl,top_Hollik:1999vz}
leads to smaller effects.

\STS The results of a number of sensitivity analyses
are given in {\it Table}\,\ref{fig:top_topform}. Most of these studies
were performed at the parton level. Only the dilepton and
single lepton  $t \bar t$ decay channels ($\ell = e,\mu$) were considered.
A possible strategy for performing these multi-parameter analyses
in future experiments is as follows. First one may measure CP-odd
angular correlations and asymmetries which are sensitive only to
CP-violating form factors. Observables were constructed which can be used 
to disentangle possible CP effects in $t\bar t$ production and decay
\cite{top_Bernreuther:1992be,top_N24}.
Once the values of -- or limits to -- these form factors are known,
one can proceed by measuring the CP-invariant moments  in
${\rm t}\bar {\rm t}$ production and decay by suitable distributions and asymmetries
(see above). In this way one probes for anomalous effects down to
length scales of {\it a few} $\times 10^{-19}$\,cm. 
It is worth emphasizing one aspect of such studies. 
 At \ee linear colliders  very clean
and sensitive searches can be made 
for new CP-violating sources  
which may only become visible
at high energies, in particular
in $t\bar t$ production and decay. Detection of such interactions in the
laboratory would have striking consequences for our attempts to
understand the matter-antimatter asymmetry of the universe.

\STS
\subsection[Complementarity with the LHC]{Complementarity with the LHC}
\label{physics_top_LHC}
Finally a brief comparison of some of the respective assets
of the proposed
TESLA collider (LC) and the LHC is in order. 
As far as as top quark physics is concerned these machines are complementary,
to a large extent, in their potential.
(For a recent compilation
of the perspectives of top quark physics at the LHC, see  \cite{top_lhc00}.)
Clearly, a remarkable feature of the LC is the possiblity to extract
the mass of the top quark from the threshold excitation curve
with an error estimated below 200\MeV. 
Also the total top decay width $\Gamma_t$ can be determined with
a relative error of about 10$\%$.  
At the LHC top mass measurements are
expected to be feasible with an accuracy of 
about  1 - 2\GeV. As far as  $\Gamma_t$ is concerned no sensible method
of extraction  is known. However, from single top quark
production at the LHC the measurement of the 
Kobayashi-Maskawa matrix element ${\mid V_{tb}\mid}^2$ to 10 $\%$
appears feasible.

At the LC very clean and accurate measurements  of the neutral current
couplings of the top quark to the photon and Z-boson -- its vector and
axial vector charges, anomalous magnetic and electric dipole moments --
can be made, both at threshold and in the continuum. A powerful tool for these
 studies and those of the charged current $t$ couplings 
will be the possiblity of  polarising the $e^{\pm}$ beams.

Owing to the expected production of $10^7$ or more 
top quark pairs  per year
the LHC  has a large discovery potential of a number
of non-standard rare top decays, such as the flavour-changing neutral
current reactions $t\to c + Z, c + \gamma$, down to branching ratios of about
$10^{-4}$ which is superior to the LC. If top decays into  charged Higgs bosons
exist they should be seen 
 first at hadron colliders. Then at the LC 
rather precise measurements of the branching fraction  and
clean polarisation studies can be made, offering insights into the 
Yukawa interactions at work in this mode.
Moreover, the LC allows for very sensitive searches of the
supersymmetric decay of the top quark into stop and neutralino particles. 

%\bibliography{physics_top}

%  \input{physics_top.bbl}
  \def\alpmz{\relax\ifmmode \alpha_s(M_Z)\else $\alpha_s(M_Z)$\fi\chkspace}
\def\z0{\relax\ifmmode Z \else {$Z$} \fi\chkspace}
\def\ep{{$e^+e^-$}\chkspace}
\def\alp{\relax\ifmmode \alpha_s\else $\alpha_s$\fi\chkspace}
\def\b_alp{\relax\ifmmode \balpha_s\else $\balpha_s$\fi\chkspace}

%new
\def\ggx{{$\gamma\gamma$}\chkspace}
\def\sigg{{$\sigma_{\gamma\gamma}^{tot}$}\chkspace}
\def\fg{{$F^{\gamma}_2$}\chkspace}
\def\xg{{$x_{\gamma}$}\chkspace}
\def\syy{{$\sqrt{s}_{\gamma\gamma}$}\chkspace}
\def\gstar{{$\gamma^*\gamma^*$}\chkspace}
\def\be{\begin{equation}}
\def\ee{\end{equation}}
                         \def\bearr{\begin{eqnarray}}
                         \def\eearr{\end{eqnarray}}
\def\benum{\begin{enumerate}}
\def\eenum{\end{enumerate}}
\def\bitem{\begin{itemize}}
\def\eitem{\end{itemize}}
%end new...

\section{Quantum Chromodynamics}
\label{tdr-phys-qcd}
\subsection{Introduction}

Strong-interaction measurements at TESLA
will form an important component of the physics 
programme. The collider 
offers the possibility of testing QCD~\cite{fritsch_QCD} at high energy scales
in the experimentally clean, theoretically tractable
\ep environment. In addition, virtual $\gamma\gamma$ interactions
will be delivered free by Nature, and a dedicated $\gamma\gamma$
collider is an additional option, allowing detailed measurements of
the relatively poorly understood photon structure.
The benchmark physics main topics are:

\noindent
$\bullet$
Precise determination of the strong coupling \alp.

\noindent
$\bullet$
Measurement of the $Q^2$ evolution of \alp and constraints on the GUT scale.

\noindent
$\bullet$
Measurement of the total $\gamma\gamma$ cross section and 
the photon structure function.

\subsection[Precise determination of \alp]{Precise determination of \b_alp}
The current precision of individual \alp
measurements is limited at best to several per cent~\cite{alphasrev}.
Since the uncertainty on \alp translates directly into an uncertainty 
on perturbative QCD (pQCD) predictions, especially for high-order multijet 
processes, it would be desirable to achieve much better precision. 
In addition, since the weak and electromagnetic couplings are known
with much greater relative precision, the error on \alp represents the
dominant uncertainty on our `prediction' of the scale for
grand unification of the strong, weak and electromagnetic forces~\cite{gut}.

Here we will refer to the conventional yardstick of
$\alpha_s$ quoted at the \z0 mass scale,
\alpmz, unless explicitly stated otherwise.
Several techniques for \alpmz determination will be available
at TESLA:

\subsubsection{Event shape observables}

The determination of \alpmz from event `shape' observables that are
sensitive to the 3-jet nature of the particle flow has been
pursued for two decades and is generally well understood~\cite{philalp}. 
In this method one usually forms a differential distribution, 
makes corrections for detector and hadronisation effects,
and fits a pQCD prediction to the data, allowing \alpmz to vary.
Examples of such observables are the thrust, jet masses and jet rates.

The latest generation of such \alpmz measurements, from SLC and LEP, has
shown that statistical errors below 0.001 can be obtained
with samples of a few tens of thousands of hadronic events.
With the current TESLA design luminosities of 
$3(5)\times10^{34}$/cm$^2$/s, at $Q$ = 500 (800)\GeV, 
hundreds of thousands of e$^+$e$^-$ $\rightarrow$ $q\overline{q}$ 
events would be produced each year,
and a statistical error on \alpmz below 0.0005 could
be achieved. 

Detector systematic errors, which relate mainly to
uncertainties on the corrections made for acceptance and resolution
effects and are observable-dependent, 
are under control in today's detectors at the $\Delta\alpmz$ = 
0.001--0.004 level~\cite{otmar}. 
If the TESLA detector is designed to
be very hermetic, with good tracking resolution and efficiency,
as well as good calorimetric
jet energy resolution, all of which are required for the search
for new physics processes, it seems reasonable to 
expect that the detector-related uncertainties can be beaten down to
the $\Delta\alpmz$ $\simeq$ 0.001 level or better.

$e^+e^-$ $\rightarrow ZZ, W^+W^-$, or $t\overline{t}$ events 
will present significant backgrounds to  $q\overline{q}$
 events for QCD studies, and the
selection of a highly pure $q\overline{q}$  event sample will not be quite
as straightforward as at the \z0 resonance. The application of kinematic cuts 
would cause a significant bias to the event-shape
distributions, necessitating compensating corrections at the level of
25\%~\cite{bias}. 
More recent studies have shown~\cite{schumm} that the majority of
$W^+W^-$ events can
be excluded without bias by using only events produced with 
right-handed electron beams 
for the \alpmz analysis. Furthermore, the application of
highly-efficient $b$-jet tagging can be used to reduce the $t\overline{t}$ 
contamination to the 1\% level. After statistical subtraction of the
remaining backgrounds (the \z0\z0 and $W^+W^-$ event properties
have been measured accurately at SLC and LEPI/II),
the residual bias on the event-shape distributions is expected to
be under control at the better than 0.001 level on \alpmz.

Additional corrections must be made for the effects of the smearing of the
particle momentum flow caused by hadronisation.
These are traditionally evaluated using Monte Carlo models.
The models have been well tuned at SLC and LEP and are widely 
used for evaluating systematic
effects. The size of the correction factor, and hence the uncertainty, 
is observable dependent, but the `best' observables measured at the
\z0 have 
uncertainties as low as $\Delta\alpmz$ $\simeq$ 0.001. Furthermore, one expects
the size of these hadronisation effects
to diminish with c.m. energy at least as fast as 1/$Q$.
Hence 10\%-level corrections at the \z0 should dwindle to 
1\%-level corrections at $Q$ $\geq$ 500\GeV, and the
associated uncertainties will be substantially below the 0.001 level
on \alpmz. This has been confirmed by explicit simulations using 
PYTHIA~\cite{otmar}.

Currently pQCD calculations of event shapes 
are available complete only
up to $O(\alpha_s^2)$, although resummed calculations are available for some
observables~\cite{resum}. 
One must therefore 
estimate the possible bias inherent in measuring
\alpmz using the truncated QCD series.
Though not universally accepted, it is customary to estimate this from
the dependence of the fitted \alpmz value on the QCD renormalisation scale,
yielding a large and dominant uncertainty of about $\Delta\alpmz$ $\simeq$
$\pm$0.006~\cite{philalp}.
Since the missing terms are $O(\alpha_s^3)$, and since 
\alp(500\GeV) is expected to be about 25\% smaller than \alpmz, 
one expects the uncalculated contributions to be
almost a factor of two smaller at the higher energy. However,
translating to the yardstick \alpmz yields an uncertainty of
$\pm$0.005, only slightly smaller than currently.
Therefore, although a 0.001-level \alpmz measurement is 
possible experimentally,
it will not be realised unless $O(\alpha_s^3)$ contributions are
calculated. There is reasonable expectation that this will be achieved within
the next 5 years~\cite{zvi}.
 
\subsubsection[The $t \overline t g$ System]{The $\bt\overline{\bt}(\bg)$ System}
The dependence of the $e^+e^-$ $\rightarrow$ $t\overline{t}$
production cross section, $\sigma_{t\overline{t}}$, on the top-quark mass, $m_t$,
and on \alpmz is discussed in section~\ref{physics_top_threshold}. 
In order to optimise the precision on the $m_t$ measurement near threshold
it is desirable to input a precise \alpmz measurement from elsewhere.
Furthermore, 
the current theoretical uncertainty on $\sigma_{t\overline{t}}$ translates into
$\Delta$\alpmz = $\pm 0.010$. Hence, although extraction of \alpmz from 
$\sigma_{t\overline{t}}$ near threshold may provide
a useful `sanity check' of QCD in the $t\overline{t}$  system,
it does not appear currently to offer the prospect of a competitive 
measurement.
A preliminary study has also been made~\cite{qcd_werner} of the determination
of \alpmz from $R_t$ $\equiv$ $\sigma_{t\overline{t}}/\sigma_{\mu^+\mu^-}$ 
{\it above} threshold.
For $Q$ $\geq$ 500\GeV the uncertainty on $R_t$ due to $m_t$ is around 0.0005. 
The limiting precision on $R_t$ will be given by the uncertainty on
the luminosity measurement. 
If this is as good as 0.5\% %~(see section~\ref{sec:gaugebosons})
then \alpmz could be determined with an experimental precision 
approaching 0.001, which would be extremely valuable as a complementary 
precision measurement from the $t\overline{t}$ system.

\subsubsection{A high-luminosity run at the Z resonance}

A Giga \z0 sample offers two additional options for 
\alpmz determination via measurements of the inclusive ratios 
$\Gamma^{had}_Z/\Gamma_Z^{lept}$ and
$\Gamma^{had}_{\tau}/\Gamma_{\tau}^{lept}$. Both are indirectly 
proportional to \alp, and hence require a very large event sample for 
a precise measurement. For example, the current LEP data sample of 16\,M
\z0 yields an error of 0.0025 on \alpmz from 
$\Gamma^{had}_Z/\Gamma_Z^{lept}$. The statistical error could,
naively, be pushed to below the $\Delta$\alpmz = 0.004 level, but systematic errors
arising from the hadronic and leptonic event
 selection will probably limit the precision
to 0.0008~(see section~\ref{sec:gaugebosons}). 
This would be a very precise, reliable 
measurement. In the case of 
$\Gamma^{had}_{\tau}/\Gamma_{\tau}^{lept}$ the experimental precision
from LEP and CLEO is already at the 0.001 level on \alpmz. However,
there has been considerable debate about the size of the theoretical
uncertainties, with estimates as large as 0.005 \cite{tau}. 
If this situation is clarified, and the theoretical uncertainty is small,  
$\Gamma^{had}_{\tau}/\Gamma_{\tau}^{lept}$ may offer a further
0.001-level \alpmz measurement. 

\subsection[$Q^2$ evolution of \alp]{$\bQ$ evolution of \b_alp}

In the preceeding sections we discussed the expected attainable precision
on the yardstick \alpmz. Translation of the measurements of \alp($Q$) 
($Q \neq M_Z$) to \alpmz requires the assumption that the `running' of
the coupling is determined by the QCD $\beta$ function.
However, since the logarithmic decrease of \alp with $Q$ is an essential
component of QCD, reflecting the underlying non-Abelian dynamics,
it is vital also to test this $Q$-dependence explicitly.
Such a test would be particularly interesting if new coloured particles were 
discovered, since deviations from QCD running would be expected at 
energies above the threshold for pair-production of the new particles.
Furthermore, extrapolation of \alp to very high energies of the 
order of $10^{15}$\GeV can be combined 
with corresponding extrapolations of the
dimensionless weak and electromagnetic couplings in order to
constrain the coupling-unification, or GUT, scale~\cite{gut}. 
Hence it would be desirable to measure \alp {\it in the same detector,
with the same technique, and by applying the same treatment to the
data} at a series of different energies $Q$, so as to maximise the
lever-arm for constraining the running.

\begin{figure}[htbp]
  \begin{center}
    \leavevmode 
    \centerline{\epsfxsize10cm\href{pictures/9/tdr_alphss.pdf}{{\epsfbox{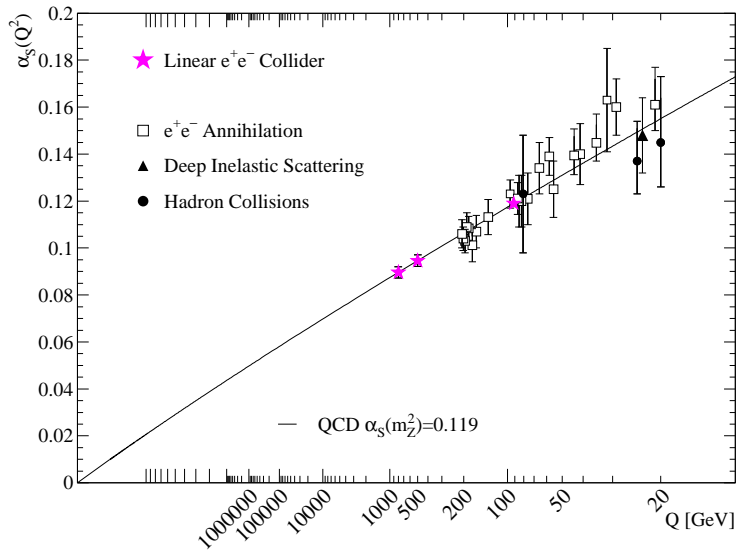}}}}
    \caption{The evolution of \alp with $1/\ln Q$ \protect{\cite{otmar}}; 
             sample $Q$ values (GeV) are indicated.}
    \label{tdr-phys-qcd:fig_alpha_s} 
  \end{center}
 \end{figure}

Simulated measurements of \alp($Q$) at $Q$ = 91, 500 and 800\GeV
are shown in Fig.~\ref{tdr-phys-qcd:fig_alpha_s}, together with
existing measurements which span the range $20\leq Q\leq 200$\GeV.
The highest-energy measurements are currently provided by LEPII. 
The point at $Q$ = 91\GeV is based on the $\Gamma_Z^{had}/\Gamma_Z^{lept}$
technique, and those at 500 and 800\GeV are based on the 
event shapes technique.  The last two include the current theoretical uncertainty,
which yields a total error on each point equivalent to $\Delta$\alpmz = 0.004.
It is clear that the TESLA data would add significantly to the
lever-arm in $Q$, and would allow a substantially improved extrapolation
to the GUT scale. 

% \begin{figure}[htbp]
%  \begin{center}
%    \leavevmode 
%    \centerline{\epsfxsize9cm\epsfbox{lc_2.eps}}
%    \caption{}
%    \label{fig:alpha_s_2}
%  \end{center}
%\end{figure}

\subsection{Further important topics}

Limited space allows only a brief mention
of several other important topics~\cite{previous}:

\noindent
$\bullet$
{\it Hard gluon radiation in $t\overline{t}$  events} would allow several 
tests of the strong dynamics of the top quark~\cite{arnd}:
test of the flavour-independence of strong interactions;
limits on anomalous chromo-electric and/or
chromo-magnetic dipole moments~\cite{rizzo};
determination of the running $m_t$.

\noindent
$\bullet$
{\it Soft gluon radiation in $t\overline{t}$ events} 
is expected to be strongly regulated by
the large mass and width of the top quark. Precise measurements of
gluon radiation patterns in $t\overline{t}g$  events would provide 
additional constraints on the top decay width~\cite{lynne}.

\noindent
$\bullet$
{\it Polarised electron (and positron) beams} can be exploited 
to test symmetries using multi-jet final states.
For polarized \ep annihilation to
three hadronic jets one can define ${\bf S}_e\cdot({\bf k}_1\times {\bf k}_2)$,
which correlates the electron-beam polarization vector ${\bf S}_e$
with the normal to the three-jet plane defined by
${\bf k}_1$ and ${\bf k}_2$, the momenta of the two quark jets.
If the jets are ordered by momentum (flavour)
the triple-product is CP even (odd) and T odd.
Standard Model T-odd contributions of this form are
expected~\cite{lance} to be immeasurably small, and limits 
have been set for the $b\overline{b}g$ system~\cite{sldtodd}.
At TESLA these observables will provide
an additional search-ground for anomalous effects in the $t\overline{t}g$
system.

\noindent
$\bullet$
{\it The difference between the particle multiplicity in heavy- ($b,c$)
and light-quark events} is predicted~\cite{doksh} to be independent of
c.m. energy. Precise measurements have been made at the \z0, but 
measurements at other energies are statistically limited in precision, 
rendering a
limited test of this important prediction. High-precision
measurements at TESLA would add the lever-arm for a powerful test.

\noindent
$\bullet$
{\it Colour reconnection and Bose-Einstein correlations}
are important to study precisely since
they may affect the precision with which the masses of heavy particles,
such as the $W^{\pm}$ and top-quark, can be reconstructed kinematically
via their multijet decays~\cite{torbj}.

\noindent
$\bullet$
{\it Hadronisation studies and renormalon physics}
can be explored via measurements of event-shape observables over a
range of $Q$ values.

\subsection{Two Photon physics}

Traditionally \ep colliders provide a wealth of two-photon data.
The photons are produced via bremsstrahlung~\cite{wws} from the
electron and positron 
beam, which  leads to a soft energy spectrum of the photons.
Such processes will also occur at future high energy 
\ep colliders.
Due to the single use of the colliding beams at these machines
 other operation modes become possible 
such as a $\gamma\gamma$ collider and e$\gamma$ 
collider~\cite{ginz,telnov}, where 
the  electron beam(s)  of a  linear  collider are 
converted into  photon
beams
via Compton laser backscattering.
This offers the exciting possibility to study
two-photon interactions at the highest possible energies
with high luminosity. A plethora of
QCD physics 
topics in two-photon interactions can be addressed 
with a linear \ep collider or $\gamma\gamma$ collider.
Furthermore,  good knowledge and understanding of two photon processes 
will be  essential for 
controlling background contributions to other processes.

\subsubsection{Total cross section}

\begin{figure}[htbp]
\begin{center}
\href{pictures/9/tdr-ggtot.pdf}{{\epsfig{figure=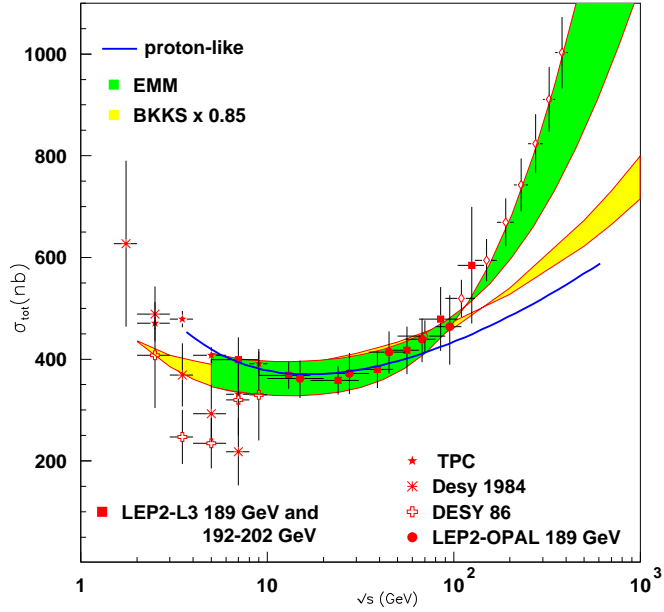,bbllx=0pt,bblly=150pt,bburx=540pt,bbury=680pt,width=9cm}}}
\caption{
The total \ggx cross-section as function of the collision energy, 
compared with model calculations:
BKKS band (upper and lower limit correspond to different   
photon densities \protect \cite{BKKS});
%SAS lines (Regge Pomeron exchange, upper and lower limits 
%as given by SAS); 
a proton-like model (solid line \protect \cite{aspen});
EMM band (Eikonal Minijet Model for total and 
inelastic cross-section, with different photon densities and 
different minimum jet transverse momentum \protect
\cite{pancheri}). The proton-like and BKKS
models have been normalized to the data, in order to show the 
energy dependence of the cross section.
}
    \label{tdr-phys-qcd:fig_ggtot}
\end{center}
\end{figure}

At a linear \ep collider and \ggx collider 
detailed properties of two-photon collisions
can be studied. A key example is the  total \ggx cross-section, which
is not yet understood from first principles.
Fig.~\ref{tdr-phys-qcd:fig_ggtot} shows 
  present photon-photon cross-section data
in comparison with recent phenomenological 
models~\cite{pancheri}.
All models predict a rise of the cross-section with the collision 
energy \syy. The
predictions for high   energies show dramatic
differences reflecting our present lack in understanding.
In proton-like models
(%dash-dotted~\cite{SAS,us1}, 
%dashed~\cite{ttwu},dotted~\cite{GLMN} and 
solid curve~\cite{aspen} ), the rise 
follows  closely that of the proton-proton cross-section, while in  QCD based 
models (upper~\cite{BKKS} and lower~\cite{pancheri} bands), the rise is 
obtained using the eikonalized pQCD jet cross-section. 
%Using an eikonal formalism to 
%unitarize the cross-section, one needs phenomenological inputs such as 
%the impact parameter distribution for partons in the photons, which are
%model dependent. 
%In general, in these models the curvatures is typically 
%$s^\epsilon$ with $\epsilon\approx 0.08\div 0.11$ which describes 
%the rise of both $pp$, $p {\bar p}$ and the older $\gamma p$ data
%obtained 
%from photoproduction.

%The figure demonstrates that large differences between the models 
%become apparent in the energy   
%of  a future 0.5-1\TeV \ep collider. 
A detailed comparison of the 
predictions reveals that in order to distinguish between 
all the models the cross-sections  need to be determined 
to a precision of better than 10\%~\cite{pancheri} 
at a future 0.5--1\TeV\  \ep collider.
This is difficult to achieve 
at an \ep collider, since the variable \syy needs to be 
reconstructed from the visible hadronic final state in the detector. 
At the highest 
energies the hadronic final state extends in pseudorapidity
$\eta = \ln\tan\theta/2$ in the region $-8 < \eta < 8$, while 
the detector covers roughly  the region $-3 < \eta <3$. 
Some information can be gained by measuring the total integrated
cross section above a value \syy, e.g. \syy $> 50$\GeV, for which
the total spread of the model predictions is 10-20\%~\cite{pancheri2}.

For a $\gamma \gamma$ collider the photon 
beam energy can be tuned with a spread of less than 10\%, such that
measurements
of \sigg can be made at a number of ``fixed'' energy values in e.g. the range
 $100 <$ \syy $< 400$\GeV, as shown in Fig.~\ref{tdr-phys-qcd:fig_ggtot}.
 The absolute precision with which
these cross-sections can be measured ranges from 5\% to 10\%,
where the largest contributions to the errors are due to 
the control of the diffractive component  of the cross-section,
Monte Carlo models used to correct for the event selection cuts,
the knowledge on absolute luminosity and  
shape of the luminosity spectrum~\cite{pancheri2}.
It will be necessary to constrain the diffractive component
in high energy two-photon data. A technique
to measure diffractive
contributions separately, mirrored to 
the rapidity gap methods used at HERA,
 has been proposed in \cite{engel}.

While the absolute cross-sections are measured with limited precision, the 
change of the cross-section with energy can be determined much more 
accurately. Fitting the  data of the collider 
to the Regge inspired form 
$s^{\epsilon}$ in the 
high energy region, one can determine $\epsilon $ with a precision 
of $\Delta \epsilon = 0.02$. The models show a variation between 
$\epsilon = 0.08$ and $\epsilon = 0.26$.

%While the absolute cross-sections are measured with limited precision, the 
%change of the cross-section with energy can be determined much more 
%accurately. Fitting the  data of the collider 
%to the Regge inspired form 
%$s^{\epsilon}$ in the 
%high energy region, one can determine $\epsilon $ with a precision 
%of $\Delta \epsilon = 0.02$. The models shown variation between 
%$\epsilon = 0.08$ and $\epsilon = 0.4$

%Another important aspect of two-photon interactions is its decomposition
%into
%diffractive and non-diffractive processes. 
%For a photon collider, the detector acceptance for elastic processes
%will be essentially zero. However the single diffractive and 
%double diffractive components can be measured, and thus the elastic
%part can be inferred from Regge factorization arguments. It would be
%advantageous however to test this factorization in \ggx at lower energy
%e.g. at LEP. 
%The available diffractive samples for photons in a wide range of 
%energies and virtualities 
%allow for 
% systematic studies of diffraction at high
%energies, currently of intense interest a HERA. 

\subsubsection{Photon structure}
The nature of the photon is complex. A high energy photon can fluctuate
into   a fermion pair or even into a bound state, i.e. a vector meson with
the
same quantum numbers as the photon $J^{PC} = 1^{--}$.
These  quantum fluctuations 
lead to the
so-called hadronic structure of the 
photon.
In contrast to  the proton, the structure 
function of the photon is predicted to rise linearly with the 
logarithm of the momentum transfer $Q^2$, and to increase with
increasing Bjorken-$x$~\cite{gg_zerwas}. The absolute magnitude 
of the photon structure function is asymptotically determined by 
the strong coupling constant~\cite{gg_witten}.

The classical way to study the structure of the photon is via
deep inelastic electron-photon scattering (DIS), i.e. two-photon interactions 
with one quasi-real (virtuality $Q^2 \sim 0$) and one virtual
($Q^2 >$ few\GeV$^2$) photon.
The unpolarised e$\gamma$ DIS cross-section is
\begin{equation}
\label{avs1}
  \frac{d \sigma (e \gamma \rightarrow eX)}{dQ^2 dx}
  \: = \:\frac{2\pi \alpha^2}{Q^4 x} 
  \:\cdot 
  \Big[ \big\{ 1 + (1-y)^2 \big\} F_2^{\, \gamma}(x,Q^2) 
  - y^2 F_L^{\, \gamma}(x,Q^2) \Big] \: , 
\end{equation}
where $F_{2,L}^{\, \gamma}(x,Q^2)$ denote the structure functions of the
real photon.
To leading order the structure function is given by the quark 
content, e.g.
\begin{equation}
F_2^{\gamma} = \Sigma_q e^2_q( xq^{\gamma}(x,Q^2)+ 
x\overline{q}^{\gamma}(x,Q^2)).
\end{equation}

\begin{figure}[htbp]
\begin{center}
\href{pictures/9/tdr-bl500m.pdf}{\epsfig{figure=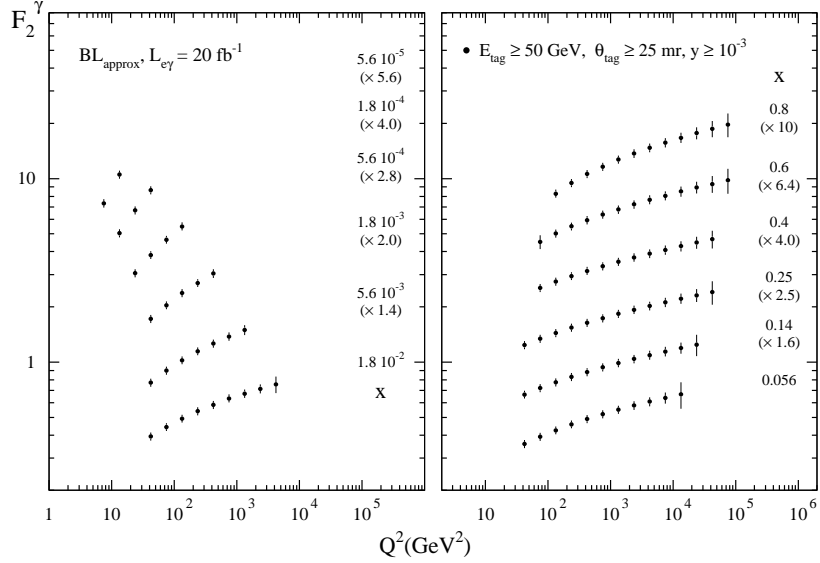,bbllx=80pt,bblly=30pt,bburx=530pt,bbury=720pt,angle=-90,width=11cm}}
\caption{The kinematic coverage of the measurement of \fg for the 
backscattered e$\gamma$ mode at a 500\GeV linear collider~\protect 
\cite{vogt}.}
    \label{tdr-phys-qcd:fig_ggf2}
\end{center}
\end{figure}

To measure \fg it is important to detect (tag) the scattered 
electron which has emitted the virtual photon. Background studies 
suggest that these electrons can be detected down to 25 mrad and down
to 50\GeV.  
%An important limiting factor for present measurements of \fg
%at LEP
%is the understanding and modelling of the hadronic final state, needed
%to reconstruct the kinematics of the events in the \ep collider mode, and this
% limitation will become even worse at higher energies  
%due to the increased rapidity span of the hadronic
%final state. 
For
e$\gamma$ scattering at an e$\gamma$ collider
the energy of the probed quasi-real photon  is 
known (within the beam spread of 10\%) and the systematic error can be
controlled to about 5\%.
Fig.~\ref{tdr-phys-qcd:fig_ggf2}
shows the measurement potential for an e$\gamma$ collider~\cite{vogt}.
The measurements are shown with statistical and (5\%)
systematical error, for 20\,fb$^{-1}$ e$\gamma$ collider luminosity, i.e.
about a year of data taking. Measurements can be made in the region 
$5.6 \cdot 10^{-5} < x < 0.56$,  in a region similar to the HERA proton
structure function measurements, and $10 < Q^2 < 8\cdot 10^{4}$\GeV$^2$.
The cross-sections at low-$x$ constitutes
 30-40\% of charm, and 
 corrections due to $F_L$ amount to more than 10\% for 
the region  $x < 5.6 \cdot 10^{-4}$, and can thus be studied.
For the \ep collider mode the hadronic final state needs to be measured
accurately in order to reconstruct $x$. This will limit
  the lowest reachable $x$ value to be 
around $10^{-3}$, but allows for measurements in the 
high $x$ ($0.1 < x < 0.8$) and high $Q^2$ ($ Q^2> 100$\GeV$^2$) 
range, for detailed \fg QCD evolution tests~\cite{nisius}.
%The $Q^2$ 
%evolution of the structure function
%at 
%large $x$ and $Q^2$ has also been often advocated as a clean measurement
%of $\alpha_s$. A 5\% change on $\alpha_s$ results however in a 
%3\% change in \fg only, hence such a 
%$\alpha_s$ determination  will require very precise \fg
%measurements.

%Hence the  data will allow
%to extract for the first time the longitudinal  structure functions
%of the photon at small $x$. 

At very high $Q^2$ values
($Q^2 \sim 10\,000$\GeV$^2$)
 also $Z$ and $W$
exchange will become important, 
the latter leading to charged current events~\cite{ridder}
yielding events with large missing transverse
momentum due to the escaping
neutrino. By
measuring the electroweak neutral and charged current structure functions,
the up and down type quark content of the photon can be determined
separately.

While e$\gamma$ scattering allows to measure the quark distributions
it only weakly constrains the gluon distribution via the QCD evolution
of the structure functions. Direct information on the gluon in the 
photon can however be obtained from measurements of jet~\cite{wengler}, open 
charm~\cite{jankowski}, and $J/\psi$~\cite{indumathi} production 
in $\gamma\gamma$ interactions
 at an \ep and $\gamma\gamma$ collider.
%The dominating diagrams are photon-gluon and 
%gluon-gluon fusion. 
%Charm cross-section measurements~\cite{jankowski} will
%give information on the gluon density in the photon down to 
%$x$ values of $10^{-3}$. 
%The event rates are large and will lead to
%statistical errors of the order of a few \% on differential distributions.
%A second handle on the gluon content in the photon is the measurement of 
%di-jet cross-sections~\cite{opaljets}. 
%These cross-sections are typically in the range of
%a few to a few hundred pb. 
%Dijet cross-sections are sensitive 
%to both the quark and gluon content of the photon.
%Fig.~\ref{tdr-phys-qcd:fig_ggjet} shows the Di-jet cross-section as 
%function of \xg $= x^{\pm}_{\gamma}=\Sigma_{jets}(E\pm p_z) /
%\Sigma_{hadrons}(E\pm p_z)$, with $p_z$ the longitudinal 
%momentum of a particle. This variable is closely related to the 
%true \xg at the parton level, and can be used to separate 
%resolved (e.g. $x^{\pm}_{\gamma}< 0.8$) from direct 
%(e.g. $x^{\pm}_{\gamma}> 0.8$) processes. 
%The \xg distribution is shown
%for two different assumptions of the parton distributions in dijet 
%production in
%Fig.~\ref{tdr-phys-qcd:fig_ggjet}. 
Values of $x$  down to a few
times 10$^{-3}$ can be reached with 
charm and di-jet measurements~\cite{wengler,jankowski}, 
a region where predicted gluon distributions 
typically differ by a factor of two or more.

% \begin{figure}[htbp]
%\begin{center}
%\epsfig{ggws_pdfcomp.ps,bbllx=30pt,bblly=0pt,bburx=410pt,bbury=380pt,width=8cm}
%\caption{Jet cross-sections versus \xg for the 
%backscattered $\gamma\gamma$ mode at a 500\GeV linear collider, 
%for two assumptions of parton distributions of the photon.}
%\label{tdr-phys-qcd:fig_ggjet}
%\end{center}
%\end{figure}

A linear collider also provides circularly polarised 
photon beams.
%, either from the polarised beams of the  \ep collider directly,
%or via polarised laser beams scattered on the polarised \ep drive beam.
This offers a unique opportunity to study the 
polarised parton distributions of the photon, for which to date no 
experimental data are available. 

%At the end of the '80s 
%the first precision data on the polarised structure of the proton 
%lead to the so called spin puzzle, i.e. the observation that
%quarks carry only a very small fraction of the proton spin.
%This puzzle has not been entirely resolved yet, and is presently 
%still subject 
%of extensive experimental and theoretical work.

\begin{figure}[htbp]
\begin{center}
\begin{picture}(500,200)(0,0)
\put(0,0){\href{pictures/9/tdr-strat2.pdf}{\epsfig{figure=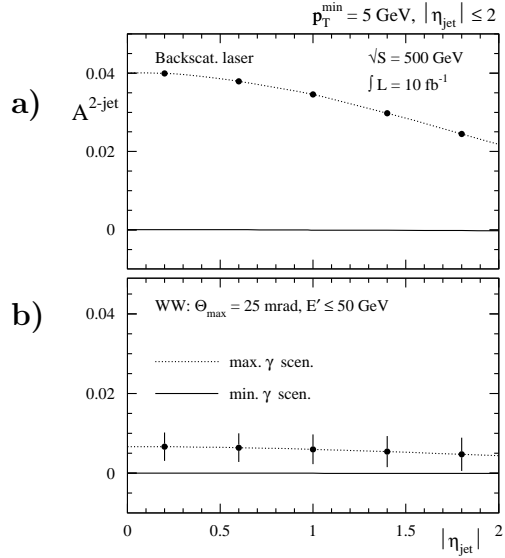,bbllx=-280pt,bblly=50pt,bburx=150pt,bbury=880pt,height=12cm}}}
\put(86,80){\bf b)}
\put(86,160){\bf a)}
\end{picture}

\caption{di-jet spin asymmetry for events with
$p^{jet}_T = 5$\GeV and $|\eta_{jet}|< 2$ for  collisions
at an e$\gamma$ collider (a) and
$\gamma\gamma$ (b) collisions at an \ep
collider\protect \cite{stratmann}. Predictions are shown for two different 
assumptions for the polarized parton distributions of the photon.
Only statistical errors are shown.
\label{tdr-phys-qcd:fig_ggpol}}

\end{center}
\end{figure}

Information on the spin structure of the photon can be obtained from 
inclusive polarised deep inelastic e$\gamma$ measurements and from 
jet and charm measurements~\cite{stratmann,g1_kwiecinski}
in polarised \ggx scattering. 
An example of a jet measurement is presented in 
Fig.~\ref{tdr-phys-qcd:fig_ggpol} which shows the asymmetry measured 
for dijet events, for the \ep and $\gamma\gamma$ collider modes 
separately. Two extreme
models are assumed for the polarised parton distributions in the 
photon.
Already with very modest luminosity significant measurements of the 
polarised parton distributions can be made at a linear collider.
The extraction of the polarised structure
function $g_1 (x,Q^2) =\Sigma_q e^2_q( \Delta q^{\gamma}(x,Q^2)+ 
\Delta\overline{q}^{\gamma}(x,Q^2))$, with $\Delta q$ the polarised
parton densities, can however be best done at an e$\gamma$  collider.
%for 
%similar reasons as outlined above for \fg measurements.
Measurements of $g_1$, particularly at low $x$,
 are  very important for studies of the high energy QCD
limit, or BFKL regime~\cite{bfkl}. 
Indeed, the most singular terms of the effects of small $x$ resummation
on $g_1(x,Q^2)$
behave like $\alpha_s^n\ln^{2n}1/x$, compared to $\alpha_s^n\ln^{n}1/x$
in the unpolarised case of \fg. 
Thus large $\ln 1/x$ effects are expected to set in 
much more rapidly for polarised than for unpolarised structure measurements.
Leading order calculations, which include kinematic constraints, 
show that  differences in predictions of $g_1$ 
with and without these large logarithms can be as large as a factor 
 3 to 4 for $x= 10^{-4}$ and can be measured with a few
years of data taking at a $\gamma\gamma$ collider.

 \begin{figure}[htbp]
\begin{center}
\href{pictures/9/tdr-kw3.pdf}{\epsfig{figure=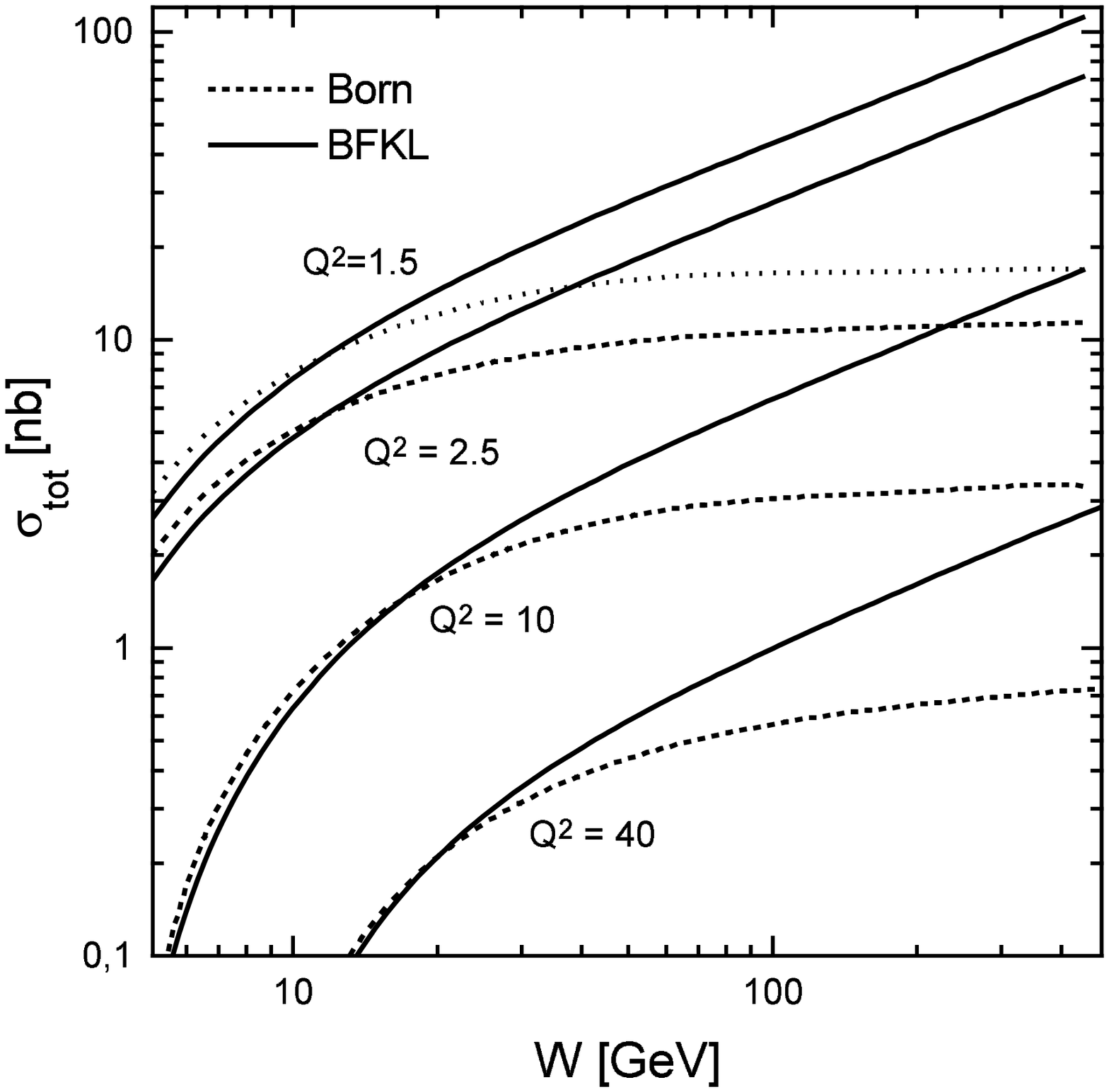,bbllx=0pt,bblly=220pt,bburx=600pt,bbury=830pt,width=9cm}}
\caption{Prediction of the $\sigma_{\gamma*\gamma*}(Q^2_1=Q_2^2,W^2)$ 
cross section 
(solid line)
and two gluon exchange cross section (dotted line) as function 
of $W^2$ for different $Q^2$ values, with $Q^2= Q^2_1= Q^2_2$
and $ Q^2_1,Q^2_2$ the virtualities of the two photons~\protect 
\cite{gammastar}.
}
\label{tdr-phys-qcd:fig_gsgs}
\end{center}
\end{figure}

\subsubsection{Testing of BFKL dynamics}
Dedicated measurements have been proposed for 
detecting and studying the large $\ln1/x$ logarithm 
resummation effects in QCD.
Experimentally establishing the BFKL effect in data is very 
important for the understanding of the high energy limit in QCD scattering. 
%These calculations, done in LO, underwent a revolution in the 1998 and 
%1999, when 
%it was pointed out that the NLO corrections could be very 
%large~\cite{lipatov}.
%The dust is settling, showing that the corrections to experimental
%variables are generally not as large as thought
%at first, and several methods have been developed to get improved 
%estimates\cite{nlobfkl}.

The most promising measurement for observing BFKL effects
is the total 
\gstar cross-section, i.e. the scattering of two virtual photons
with
 approximately equal virtualities~\cite{gammastar2}.
This cross section can be 
calculated entirely within pQCD and is found to be  
sufficiently large.
 The events are measured by tagging both scattered
electrons. At a 500\GeV \ \ep collider about 3000 events are expected
per year (200\,fb$^{-1}$) and a factor of 3 less in the 
absence of BFKL effects~\cite{gammastar}. The ability to
 tag electrons down to as low angles as possible
(e.g. 25 mrad) is essential for this measurement.
The growth of the cross section as function of 
$W^2$ due to the BFKL effect is shown in Fig.~\ref{tdr-phys-qcd:fig_gsgs},
(solid line) and compared with the cross section in absence of BFKL
(dashed line).

Closely related to the \gstar measurement is vector meson production,
e.g \ggx $\rightarrow J/\psi J/\psi$ or (at large $t$) \ggx 
$\rightarrow \rho\rho$,
where the hard scale in the process is given by the $J/\psi$ mass or the 
momentum transfer $t$. $J/\psi$'s can be detected via their decay 
into leptons, and separated from the background through a peak in the 
invariant mass. Approximately 100 fully 
reconstructed 4-muon events are expected for 200 fb$^{-1}$ of
luminosity for a 500\GeV\  \ep collider~\cite{KWADR}. 
For this channel it is crucial that the decay muons and/or electrons can be 
measured to angles below 10 degrees in the experiment.

Further processes which are strongly sensitive to BFKL effects 
include  e$\gamma$ scattering with associated jet production~\cite{contreras},
and $e^+e^- \rightarrow e^+e^-\gamma X$
and $\gamma\gamma \rightarrow \gamma X$~\cite{evanson}. 
%The gold plated method to look for BFKL effects at HERA is 
%the study of so-called forward jets and particles~\cite{forward}. 
%These are jets or particles which
%a $p_T$ similar to the virtuality of the $\gamma^*$ and
%very close to the direction of the outgoing proton beam at HERA.
%A process similar to the 'forward jets' at HERA 
% can be studied at a linear collider in e$\gamma$ 
%scattering, with a forward jet produced 
%in the direction of the real photon. 
%The BFKL effects,
%so far calculated in leading order,
% are indeed very large
% The measurements can reach out to smaller $x$ 
%values than 
%presently reachable at HERA, due to the more favourable kinematics
%of the final state
%~\cite{contreras}.
%Finally the processes $e^+e^- \rightarrow e^+e^-\gamma X$ and  
%$\gamma\gamma \rightarrow \gamma X$ have 
%been studied~\cite{evanson}, and found to be very sensitive to BFKL
%dynamics. Event rates for events with photons with energy larger than
%5\GeV and $p_T$ larger than 1\GeV are large. At an \ep 
%collider several thousand events will be collected per year, 
%while at a photon collider the event rate is about a factor ten larger.
In all, the study of all these processes will provide new fundamental
insight in small $x$ QCD physics.

\subsection{Complementarity of LHC}

QCD studies at the LHC will concentrate mainly on: jet studies, extracting
parton densities in the proton, hard diffraction and heavy 
quark studies~\cite{LHCQCD}. In principle QCD phenomena
can be studied at scales upto a few\,\TeV.
A precision measurement of \alp is foreseen via jet cross section 
measurements,
but the precision has not been quantified yet. 
BFKL phenomena can be studied mainly via di-jet production using jet pairs
with a large rapidity difference, which has an entirely different
systematics compared to the methods proposed for the linear collider.

Recently~\cite{piotrzkowski} it was proposed to study real two photon processes
at the LHC. 
If the technical challenges  to tag the outgoing protons
can be overcome, such data could allow for exploratory studies of 
quasi real two-photon physics in the high energy regime, such 
as the total $\gamma\gamma$ cross section and jet production
studies.

  \clearpage
  {\raggedright{%
% References: Standard Model Electroweak
%
% checked by Jakob Hauschild, Kristian Harder
% urls inserted by Jenny Boehme, Jakob Hauschildt
%

%
% remaining problems:
%
% PDG quoted twice ( ref:pdg and top_N15 )
% ref:gudrun :  not found, but confirmed to be announced... (hep-ph expected)
% top_N19 :     LC-Note number not yet assigned, although draft is
% submitted
% contreras,pancheri2: LC notes not yet available
%

%
}
\end{document}